\documentclass[11pt,a4paper]{article}
\pdfoutput=1
\usepackage{jheppub}
\usepackage[utf8]{inputenc} 
\usepackage{multirow}    
\usepackage{xcolor}
\usepackage{graphicx}
\usepackage{amsmath}
\usepackage{amssymb}
\usepackage{braket} 
\usepackage{hyperref}
\usepackage{frontespizio}
\usepackage{comment}  
\usepackage{epsfig}
\usepackage{float}
\usepackage{bbold}
\usepackage[title]{appendix}
\usepackage{multirow} 
\usepackage{tikz}
\usetikzlibrary{decorations.pathmorphing}
\usepackage{varwidth}
\usepackage{tikz}
\usetikzlibrary{backgrounds,patterns,calc}
\usepackage{xparse}  
\usepackage{subfigure}  
\usepackage{bbm}   
\usepackage{mathabx}

\usepackage{jheppub} 

\title{Dualities and Categorical Structures from 2D Up}

\author[]{Veronica Pasquarella}
\affiliation{Department of Applied Mathematics and Theoretical Physics (DAMTP)}
\affiliation{University of Cambridge, \\ Wilberforce Road, CB3 0WA, Cambridge, UK}

\emailAdd{vp360@damtp.cam.ac.uk}

\abstract{A fundamental step towards studying string theory vacua, and, ultimately, their stability, is that of understanding the underlying mathematical structure of the QFT resulting from its dimensional reduction on Calabi-Yau (CY) manifolds, the latter being complex manifolds admitting a category theory description. The present work aims at addressing this question in the case of lower-dimensional gravitating systems, and supersymmetric quiver gauge theories, highlighting the role played by dualities and categorical structures towards studying analytical structures of amplitudes. (Based on the author's PhD thesis).}

\keywords{factorisation homology, higher-categorical structures, geometric invariant theory, supersymmetric quiver gauge theories, lower-dimensional holographic duality and vacuum transitions}
\makeatletter
\gdef\@fpheader{}
\makeatother
\begin{document} 
\maketitle


%
%


\NewDocumentCommand{\dslash}{s}{%
  \IfBooleanTF{#1}
    {\big/\mkern-7mu\big/}
    {/\mkern-6mu/}%
}

\part{Introduction}

\section{What this work is about}

The present Thesis is, at the same time, a continuation of an original work I did with my supervisor, Professor Fernando Quevedo, \cite{DeAlwis:2019rxg}, and the birth of my research independence in the world of Physical Mathematics.

Our original aim was that of identifying a way for describing transitions in between different vacua within the String Theory Landscape, and ultimately using this as a criterion for establishing which String Theory models could more accurately describe the world in which we live. 

Identifying a rigorous formalism for describing vacuum decay processes has been a great matter of debate for almost half a century, and part of our joint collaboration was attempting to provide further insight into which prescription could better suit our purposes.  

As a step forward towards achieving such aim, we carried out an analytical study of the underlying mathematical structure of field theories. In doing so, the main tools adopted in our treatment are dualities and higher-categorical structures. 

This first section lays out the motivations for using these tools in addressing this question, as well as indicating future research directions that will be pursued by the author. We conclude the section outlining the content of the present work, highlighting some key findings, referring to their respective sections for detailed explanation.

\subsection{Motivations and Future Perspectives}

String theory\footnote{See for instance \cite{gsw1,gsw2,jp1,jp2,Green:1982ct,Schwarz:1982jn,Green:1984sg,Gross:1984dd,K.Wilson,Polchinski:1994mb}.} is a promising candidate unifying theory of fundamental interactions. In its fermionic formulation, it involves ten dimensions, of which nine are spatial and one is temporal. 
Crucially, such formulation strongly relies upon Supersymmetry (SUSY), \cite{Quevedo:2010ui}, which is a correspondence in between bosonic and fermionic particles. 

According to present-day evidence, SUSY is certainly broken in our world, and plenty of effort has been made in identifying traces of such symmetry breaking from the observational side\footnote{Among the most recent advancements in the field, we refer the interested reader to the following works \cite{Allanach:2005pv,LHCReinterpretationForum:2020xtr,Allanach:2021bbd,Banks:2020gpu,Allanach:2023bgg,Hammou:2023heg,Kassabov:2023hbm,Iranipour:2022iak,Nath:2010zj}.}, eventually leading to the extension of the Standard Model (SM), \cite{Weinberg:2004kv}, by encoding additional particles and interactions.  
Understanding how the Standard Model can be embedded in String Theory is certainly a step forward towards identifying patterns of SUSY breaking in our world. 
Albeit being broken, SUSY certainly plays a crucial role in understanding which String Theory model is most appropriate for the SM to arise as an effective field theory (EFT), precisely by studying its breaking.

String theory vacua\footnote{This has been a very active area of research on the past three decades. Some of the most relevant works in this regard are \cite{McAllister:2023vgy,Cicoli:2023njy,Cicoli:2023opf,Cicoli:2021dhg,Crino:2020qwk,AbdusSalam:2020ywo,Cicoli:2016olq,Cicoli:2015ylx,Aparicio:2015psl,Quevedo:2015ava,deAlwis:2014wia,AbdusSalam:2014uea,Quevedo:2014xia,deAlwis:2013gka,Burgess:2013sla,Font:2013hia,Cicoli:2012vw,Cicoli:2012fh,Cicoli:2012cy,Krippendorf:2010hj,Blumenhagen:2009gk,AbdusSalam:2009qd,Krippendorf:2009zza,Conlon:2008wa,Cicoli:2008gp,Conlon:2008qi,Conlon:2008cj,Cicoli:2008va,Burgess:2008ri,AbdusSalam:2007pm,Cicoli:2007xp,Conlon:2007xv,Cremades:2007ig,Conlon:2006wz,Burgess:2006mn,Conlon:2006tj,Avgoustidis:2006zp,Conlon:2006us}.}, where the SM would eventually be embedded, are parametrised by the so-called moduli space, and the latter is known to admit a categorical algebro-geometric description. Crucially, great advancements in the latter took place simultaneously to the birth and development of SUSY and String Theory. In recent years, joint collaboration in between pure mathematicians and mathematical physicists, primarily Freed, Hopkins, Lurie, Moore, Segal, and Teleman \footnote{For the purpose of our treatment, the most important works by these authors are the following: \cite{Freed:2019sco,Freed:2019jzd,Freed:2018cec,Freed:2014iua,Freed:2012hx,Freed:2010wsz,Freed:2009qp,Freed:2008jq,Freed:2006tm,Freed:2006ph,Freed:2005qu,Freed:2003qx,Freed:2002qp,Freed:2000ta,Freed:2000tt,Freed:1999vc,Deligne:1999qp,Freed:1999mn,Freed:1995ku,Freed:1994ad,Freed:1993wb,Freed:1992qb,Freed:1992vw,Freed:1992vf,Freed:1991bn,Freed:1984xe,Freed:2006ya,Segal:2002ei,Hitchin:1999at,Segal:1996ku,Pressley:1988qk,Segal:1987sk,Freed:2022qnc,Freed:2012bs}.}, shed light on new interesting perspectives which happen to be particularly relevant for the objectives outlined above. The main feature of their approach is that of applying the powerful tool of higher-categorical structures within the context of gauge theories and their supersymmetric counterparts, with the ultimate aim being that of explaining what is the underlying mathematical structure of a quantum field theory (QFT), \cite{W1,W2,W3}. Their work is based upon identifying underlying topological structures from which partition functions, and correlation functions can be derived. Furthermore, their formalism happens to be particularly efficient for keeping track of the spectrum (i.e. the particle-field content) of a given QFT, and how the latter changes under gauging of the global symmetries of the theory one started from.

Up to now, most of the mathematical control in describing higher-dimensional field theories has been for descendants of maximally-supersymmetric 6D ${\cal N}=(2,0)$ superconformal field theories (SCFTs), \cite{Witten:1995zh,Witten:2007ct,Witten:2009at,DBZ}, from which a rich web of dualities arises upon dimensional reduction to, either, 4 or 3 spacetime dimensions\footnote{See for instance \cite{Moore,Seiberg:1996nz,Gaiotto:2008ak,Gaiotto:2009gz,Gaiotto:2009jjh,Gaiotto:2008sa,Gaiotto:2008sd,Freed:2022yae,Neitzke:2014cja,Gaiotto:2011tf,Gaiotto:2010okc,Gaiotto:2009jjh,Tachikawa:2013hya,Alday:2009aq,Moore:2011ee}.}. Such descendants feature fewer supersymmetries with respect to their non-Lagrangian parent theory. In particular, the most studied are 4D ${\cal N}=2$ (also known as class ${\cal S}$ theories), and 3D ${\cal N}=4$ SCFTs, exhibiting exactly half of the amount of supersymmetry of their parent 6D theory.  

From the work of Alday, Gaiotto and Tachikawa (AGT), \cite{Alday:2009aq}, it is known that class ${\cal S}$ theories can be equally described in terms of a 2D topological field theory (TFT) living on the Riemann surface on which dimensional reduction has been performed. Such a 2D TFT is defined to be a functor in between 2-categories. Such 2-categories were first studied by Moore and Tachikawa, \cite{Moore:2011ee}, which, in turn, relied upon the pioneering work of Moore and Segal, \cite{Moore:2006dw}. The main point of their work is that, for a certain class ${\cal S}$ theory to be uniquely defined (i.e. for its spectrum of operators to be uniquely determined), one needs to fully specify the source and target 2-categories of the 2D TFT corresponding to the given 4D ${\cal N}=2$ SCFT in question. In technical terms, the objects of the source 2-category are circles, and the 1-morphisms between them are bordisms. The target 2-category is obtained by acting with the functor on these objects and 1-morphisms, resulting in the gauge group of the 4D dual theory, and the bordism operators (also known as the Moore-Segal bordism operators), with the latter being related to algebraic varieties defining moduli spaces in algebraic geometry, i.e. the Higgs and Coulomb branches.

Higgs and Coulomb branches of supersymmetric gauge theories are fascinating objects to explore, since they shed light on dualities characterising the theory they are associated with. From a pure mathematical point of view, they can be realised in terms of the so-called geometric invariant theory (GIT) construction, and quotienting\footnote{There is a vast literature in this regard, but we will mostly be focussing on the following works in due course \cite{FK,Nakajima:2022sbi,Teleman:2022oiu,Braverman:2016pwk,Braverman:2016wma,Teleman:2018wac,GIT,Braverman:2016pwk,Braverman:2017ofm}.}. For the case of 3D ${\cal N}=4$ SCFTs, exchange in between Higgs and Coulomb branches corresponds to a statement of 3D mirror symmetry between two theories (first studied by Intriligator and Seiberg from a pure theoretical-physics point of view, \cite{Intriligator:1996ex}). This 1-to-1 correspondence extends to the calculation of invariants, most importantly the Hilbert Series, accounting for all the gauge-invariant operators of the theory.

However, as also pointed out by Moore, Segal and Tachikawa, their analysis lacked a rigorous mathematical treatment, most importantly for the case in which the target category of the functor is a hyperk$\ddot{\text{a}}$hler quotient\footnote{Clarification on the terminology adopted will be provided in Part \ref{sec:V} of the present work.}, which is indeed the case when describing algebraic varieties such as the Higgs branch of a given theory. The work I have conducted so far, provides a step forward towards achieving a rigorous mathematical formulation of such 2-categorical structures for the reason that I will now outline.

As previously explained, TFTs play a crucial role in bridging the gap between QFT and Mathematics\footnote{For some of the pioneering, as well as most recent advancements in the field, we refer the interested reader to the following works \cite{Atiyah:2021hsc,Atiyah:2018ijp,Atiyah:2016vlv,Atiyah:2010qs,Atiyah:2007cka,Atiyah:2004jv,Segal:2021mpr,Segal:2010hbt,Allanach:2020zna,Allanach:2019gwp,Davighi:2022icj,Davighi:2020vcm,Davighi:2020nhv,Gripaios:2023ups,Gripaios:2022yjy,Gripaios:2022vvc,Allanach:2021bfe,Davighi:2018inx,Gripaios:2016ubw,Gripaios:2016xuo}.}. Because of this, the formalism outlined above suggests to always recast a given QFT to a TFT description. For example, 3D ${\cal N}=4$ SCFTs can be turned into 3D TFTs by performing a topological twist. Unlike ordinary 3D TFTs, which are usually trivial in nature in the sense that there are no local degrees of freedom, topologically-twisted TFTs come with nontrivial boundaries, called Lagrangian submanifolds. Importantly, the latter geometrise the boundary conditions of the field content of the theory in question, and, upon studying the intersection (i.e. the 1-morphisms) in between these Lagrangian submanifolds, one can reformulate the statement of 3D mirror symmetry of the parent 3D ${\cal N}=4$ SCFT as that of homological mirror symmetry, which is essentially a duality in between 2-categories. The steps described so far clearly show the advantage of bringing the description down to that the 2-category of Lagrangian submanifolds, which is exactly what we need to keep track of the field content of the theory we started with.  

The 3D mirror symmetry/homological mirror symmetry correspondence has been extensively studied for the specific case in which the correspondence in between the Coulomb and Higgs branch in 3D is 1-to-1, \cite{Kontsevich:1994dn,Teleman:2014jaa,CT}. However, recent developments within the context of supersymmetric quiver gauge theories, mostly led by Hanany and collaborators, \footnote{See for instance \cite{Bourget:2019aer,Bourget:2021siw,Bourget:2023cgs,Bourget:2019rtl,Hanany:2019tji,Cabrera:2018ann}.}, have shown that there are cases in which Higgs branches of fermionic theories can be broken down into several constituents, each one associated with the Coulomb branch of a certain 3D ${\cal N}=4$ SCFT (a.k.a. the cone structure), with the latter being described by the so-called magnetic quivers. Such correspondence was primarily identified by studying the Hilbert series as a geometric invariant. In one of my recent works, \cite{Pasquarella:2023exd}, I explain how the emergence of this multi-cone structure in the Higgs branch is equivalent to a generalised statement of 3D mirror symmetry, in turn descending form relaxing constraining assumptions on which homological mirror symmetry is based. As a follow up, in \cite{Pasquarella:2023ntw} and \cite{Pasquarella:2023exd}, I propose a generalisation of the Moore-Tachikawa varieties for the case in which the target category of the 2D TFT is a hyperk$\ddot{\text{a}}$hler quotient. As proved in the former of these two papers, achieving this requires generalising the bordism operators of Moore and Segal to the case involving lack of reparametrisation-invariance on the Riemann surface, ultimately enabling to relate this to the issue of defining a Drinfeld center for composite class ${\cal S}$ theories.

So far, my main focus has been on class ${\cal S}$ theories, and 3D ${\cal N}=4$ SCFTs, both featuring 8 supercharges. 
To what extent the generalised 3D mirror symmetry/homological mirror symmetry correspondence can be extended to algebraic varieties associated with theories featuring different amounts of supersymmetries is part of my currently ongoing research. The ultimate aim would be to extend this formalism to quiver gauge theories describing the SM and its possible extensions, eventually shedding light on the mathematical description of SUSY breaking mechanisms leading to observable EFTs.

In summary, the key take-home message of my work is that I have identified an essential question to be addressed, which can simply be stated as follows: generalising the mutual relation between homological mirror symmetry and 3D mirror symmetry, and its impact towards understanding the structure of quantum field theory, as well as the embedding of the SM in String Theory models, by means of categorical algebraic geometry. 

\medskip

\medskip

\subsection{Outline}

\begin{figure}[ht!]  
\begin{center}   
\includegraphics[scale=0.8]{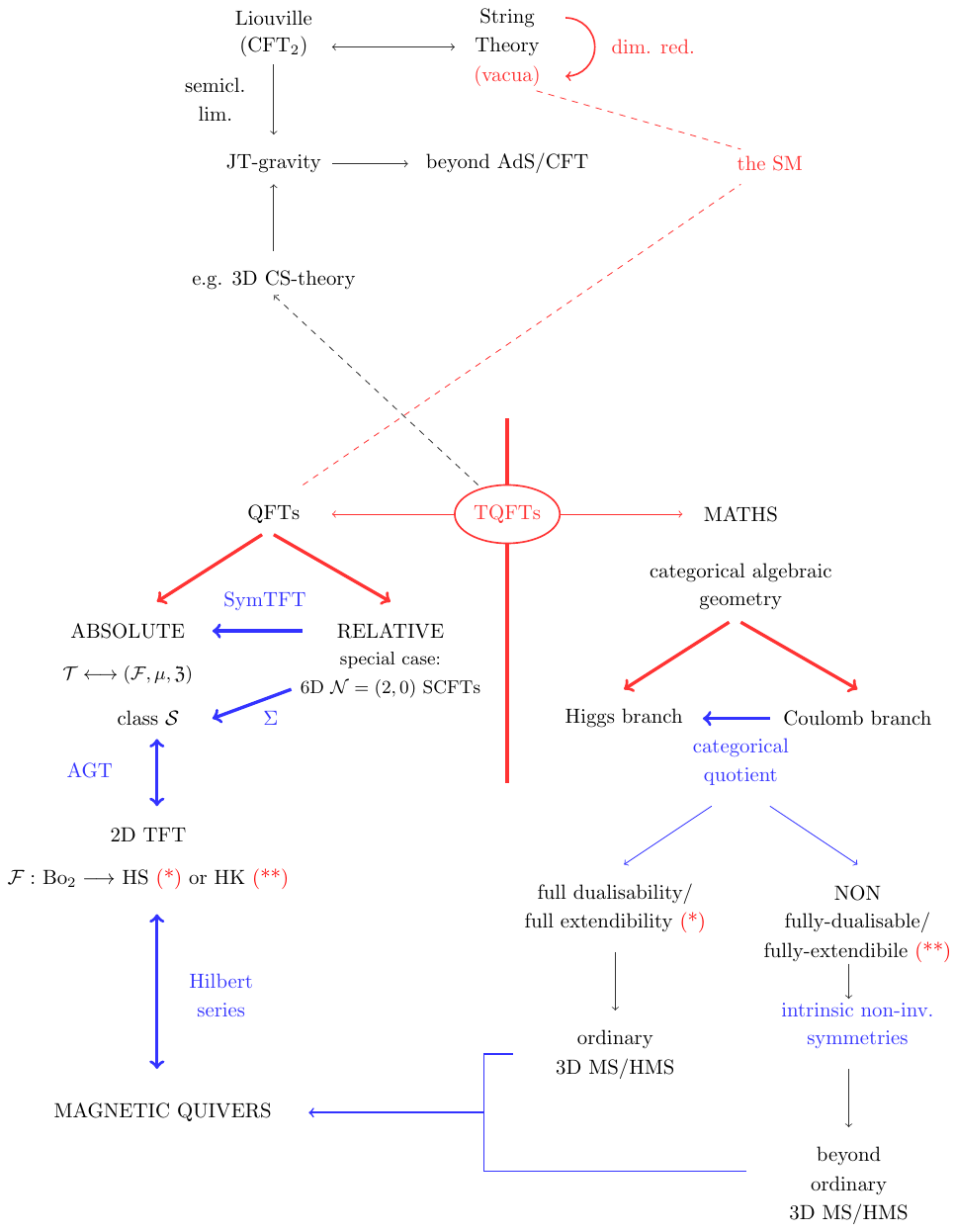}       
\caption{\small Summary diagram of the main components and motivations of this work. With it, we wish to highlight the top-down approach employed in our treatment, baring in mind the diagram on the left hand side of figure \ref{fig:SMSM12}.}   
\label{fig:SMSM12map}  
\end{center}  
\end{figure}

\begin{figure}[ht!]  
\begin{center}   
\includegraphics[scale=0.65]{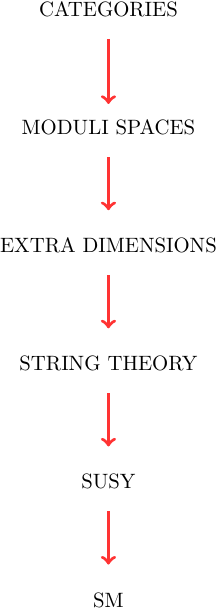}  \ \ \ \ \ \   \ \ \ \ \ \ \ \ \ 
\includegraphics[scale=0.85]{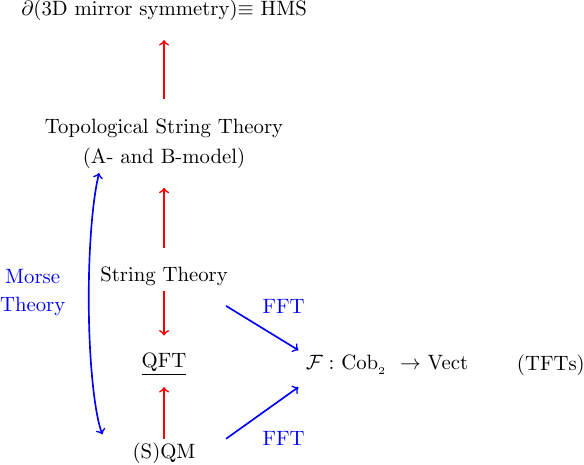}    
\caption{\small The diagram on the left hand side reproduces the top-down route connecting the main ideas underlying our treatment. The diagram on the right hand side, instead, outlines some more specific details of what we will be encountering in the following sections, and is mainly presented here as a guide the reader can refer to when delving in the more detailed analysis to come.}   
\label{fig:SMSM12}  
\end{center}  
\end{figure}

The concluding part of this first section outlines the roadmap that we will be taking in exploring the topics listed in the preliminary part of the Introduction. Given the richness and breadth of tools encountered in our journey, it would be rather challenging to sketch an accurate sequence of items in a unique diagram. We attempted to provide an oversimplified version of this in figures \ref{fig:SMSM12map} and \ref{fig:SMSM12}, but we wish to emphasise that this is by no means exhaustive. The reason for adding these diagrams at this stage is that we believe they might turn out useful to the reader as a guide to refer to when delving in the deeper analysis treated in the forthcoming sections.

\begin{figure}[ht!]  
\begin{center}  
\includegraphics[scale=0.85]{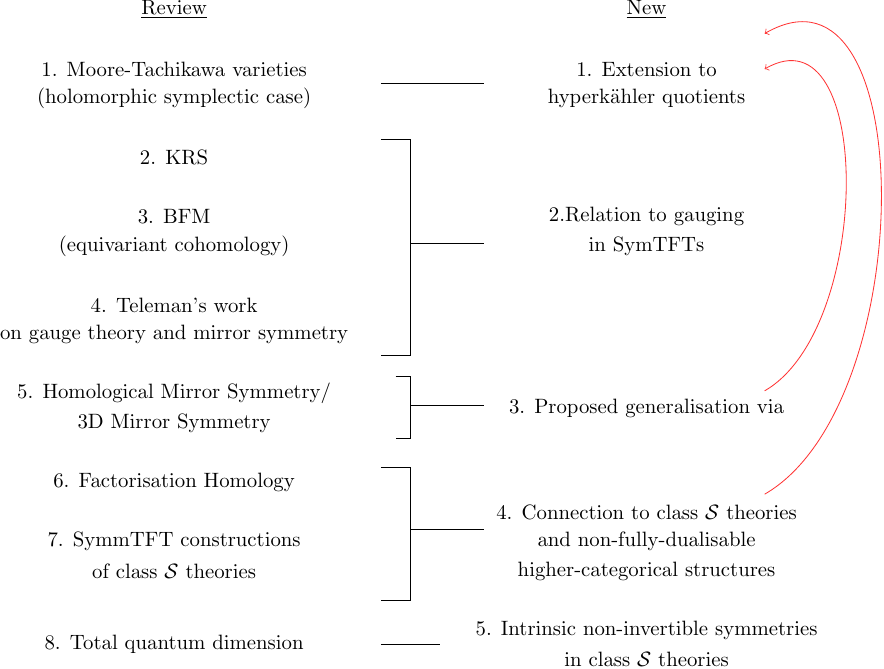}  
\caption{\small This list succinctly outlines the main achievements of Parts II, III, V of the present work. We kindly suggest the interested reader to refer to it while reading later sections. The important thing we wish to highlight is how all our major results are related to the main example described in the second half of section \ref{sec:rqft}. Please note that Part IV involves a separate treatment, analysing other aspects of lower-dimensional QFTs, mostly related to holographic duality and its generalisation, which are not listed in this table. }   
\label{fig:new}  
\end{center}  
\end{figure} 

As previously explained, our work is motivated by identifying a String Theory model that best suits the description of our world as an EFT. Albeit being a very promising candidate as a unifying theory, String Theory still features facets to be explored. Among the most pressing is its non-perturbative formulation. Notoriously, the latter was among the first motivations of the Holographic Principle, \cite{Bigatti:1999dp}, which relates strong and weak coupling limits of two different theories, one living at the boundary of the other. First inspired by black hole mechanics, the Holographic Principle relates the entropy of a certain volume, with the number of degrees of freedom living on its boundary. From statistical mechanics, we know that the entropy is nothing but a functional descending from extremisation of the partition function associated to a given theory.

The AdS/CFT correspondence is one of the most notorious realisations of the Holographic Principle, has often proved to be crucial in circumventing issues encountered in analytical calculations. 
According to the Anti-de Sitter/Conformal Field Theory (AdS/CFT) correspondence, one of the two theories in question is a theory of quantum gravity theory\footnote{Indeed, in Maldacena's first proposal, this corresponds to type-IIB string theory on an AdS$_{_5}\times S^{^5}$ background.}. Hence, studying the correspondence is expected to offer a promising way for describing a quantum gravity theory that most faithfully resembles our world. Clearly, the main shortcoming of the AdS$_{_5}\times S^{^5}$/4D ${\cal N}=4$ SYM correspondence, is that it features five extended spacetime dimensions. This prompted many subsequent formulations, attempting to explain what the correspondence should be when dealing with and AdS$_{_4}$ spacetime factor, instead. The pioneering works of Aharony, Bergman, Jafferis, and Maldacena (ABJM), showed that such a theory is dual to Chern-Simons (CS) matter theory.  

Regardless of the particular string theory model adopted, the vacuum descending from dimensional reduction of the 10-dimensional fermionic string, is an AdS spacetime, with uplifting techniques leading to, either, Minkowski or de Sitter (dS) solutions. From experimental observation, we know our universe is of the dS-type, and we therefore expect to be able to achieve this in a way which is compatible with the Holographic Principle. The only advancements that have been made so far in extending the AdS/CFT correspondence to positive signs of the cosmological constant, have been in lower-dimensional setups. Interestingly, it so too happens that proceedings (albeit not yet a full resolution) towards solving the information loss paradox, have also taken place in the setup of lower-dimensional gravity, specifically Jackiw-Teitelboim (JT), and its interplay with QM. The topological nature of JT-gravity makes it a very particular theory, and understanding to what extent its features can be extended to its 4D counterpart is still being investigated. 

Nevertheless, JT-gravity is also known to arise as the semiclassical limit of Liouville Theory, which is, in turn, a prototypical example of the bosonic string theory action, as well as being itself the dual of gravity in AdS$_{_3}$. One would therefore expect that results achieved in understanding this model from a holographic point of view should, in turn, be able to shed light on higher dimensional setups as well.

As already mentioned, the power of the Holographic Principle is in it being a duality mapping calculations in between different theories. Vacuum decay in between string theory vacua is a non-perturbative effect. Nonperturbative effects in string theory vacua, also known as \emph{instantons}, are defects interpolating between different vacua of the theory in question, and have been amply studied within the context of quantum mechanics (QM), \cite{W4}, QFT, and their 4D supersymmetric counterparts. Given the lack of a complete non-perturbative formulation of String Theory, it is reasonable to propose that one might address the same question from the perspective of a non-gravitating theory living a the boundary of the given quantum gravity (QG) model in question. 

Furthermore, the AdS/CFT correspondence strongly relies upon the boundary theories being supersymmetric. On the other hand, present-day observations clearly support the statement of SUSY being broken, and we would therefore expect that, if it were possible to assign a holographic dual to our 4-dimensional dS universe, then it must feature a non-supersymmetric QFT on its boundary. 

 \medskip    
   \medskip
\color{blue}

\noindent\fbox{%
    \parbox{\textwidth}{%
   \medskip    
   \medskip
   \medskip    
   \medskip
   \begin{minipage}{20pt}
        \ \ \ \ 
        \end{minipage}
        \begin{minipage}{380pt}
      \color{black}   \color{white}bbb\color{black}We are therefore drawn to the final standing question: what is a QFT? 
        \end{minipage}   
         \medskip    
   \medskip
        \\
    }%
}
 \medskip    
   \medskip     \color{black}

Our work provides some further advancements in answering this question. In exploring the importance of 2D structures towards describing higher-dimensional QFTs, we rely upon the joint role played by dualities and categories. We will now briefly explain what we mean by this in this introduction, while referring to the respective sections of this work for a more detailed outline. We believe that any advancement in explaining the decay of String Theory vacua, as well as identifying the most promising String Theory model completion of the SM requires extending a categorical algebraic geometry techniques to QFTs with reducing amounts of supersymmetries.

\begin{itemize}

\item  The dualities we will be dealing with are holographic dualities (and their generalisations), \cite{JM}, 3D mirror symmetry, \cite{Intriligator:1996ex}, and homological mirror symmetry, \cite{Kontsevich:1994dn}\footnote{Among the many relevant works on the topic, we mention the following \cite{hms1,hms2,hms3,hms4,hms5,hms6,hms7,hms8,hms9,hms10,hms11,hms12,hms13,hms14,hms15,hms16,hms17,hms18,hms19,hms20,hms21,Katzarkov:2008hs}.}.    

\item   The second crucial ingredient in our treatment are categories. In particular, we will be focussing on the importance of the formalism they provide for describing dualities.

\end{itemize}

The present work was structured in six parts:

\begin{enumerate} 

\item At first, we introduce the notion of relative QFTs, with detailed explanation for the case of sigma models, gauge theories, and 6D SCFTs. We then turn to the main case of interest to us, namely Coulomb branches of star-shaped quiver gauge theories, introducing the Kirwan map, and the notion of abelianisation, naturally paving the way to the introduction of category theory, which plays a crucial role in the remainder of our work.

\item Part II introduces the main ingredients of our treatment, namely categories and dualities. After having introduced higher-categorical structures, we specify how they are realised in the main example outlined in Part II thanks to the work of Moore and Tachikawa, \cite{Moore:2011ee}. Moreover, category theory is particularly useful for describing dualities, which is another crucial ingredient characterising this thesis. The dualities we encountered are: 

\begin{itemize}

\item Homological mirror symmetry.

\item 3D mirror symmetry. 

\item The Alday-Gaiotto-Tachikawa (AGT) correspondence.

\item Holographic duality. 

\end{itemize}   

Apart from revising well-known results in the literature, we highlight the importance of going beyond full categorical dualisability in the context of Moore-Tachikawa varieties, to which Part V was devoted. In doing so, we relate their formalism to that of SymTFT constructions of Freed, Moore and Teleman, \cite{Freed:2022qnc}, briefly overviewed in Part III.

\item  Part III explains how QFTs arise from knot invariants. Specifically, we overview the Symmetry Topological Field Theory (SymTFT) construction of Freed, Moore, and Teleman focussing on the dimensional reduction of non-Lagrangian theories, specifically 6D ${\cal N}=(2,0)$ SCFTs, \cite{Witten:1995zh,Witten:2007ct}, and the categorical structure of the resulting class ${\cal S}$ theories, \cite{Pasquarella:2023deo}. Making use of the AGT correspondence, \cite{Alday:2009aq}, we turn to a description in terms of 2D topological theories defined as functors between 2-categories. 

This Part mostly centers around: 

\begin{itemize} 

\item The algebro-geometric formulation of moduli spaces, and their construction by means of the Geometric Invariant Theory (GIT) construction. Importantly, we relate this construction to the abelianisation procedure outlined in our main example in section \ref{sec:main}, paving the way to further analysis in Part V.

\item Turning from QM to Symplectic Geometry by means of the WKB approximation. 

\end{itemize} 

Both topics have thoroughly been addressed in the literature, and we therefore devote some sections in the appendix to explaining them in greater depth.

\item Part IV builds from open questions pointed out in the first paper by the author and her supervisor, \cite{DeAlwis:2019rxg}. In addressing this topic, we restrict to lower-dimensional setups in order to deal with open issues usually arising in higher-dimensional settings, extensively addressed in the literature, \cite{Coleman:1980aw,Brown:1988kg,Fischler:1990pk}. In doing so, we built connection with setups involving the emergence of islands, as well as recent developments in the context of von Neumann algebras for calculating generalised entropies in gravitating systems and QFT, \cite{JM2, SWH, Banks:1983by, Giddings:1995gd, EW2, EW3, Witten:2021jzq}. The interesting connection with the remainder of our work is twofold: firstly, the need to resort to a certain dual picture in order to interpret the analytic results of the calculations performed; secondly, the richer structure resulting from the lower-dimensional analysis, emphasises the role played by internal structures in higher-dimensional theories, therefore suggesting alternative perspectives might be needed in order to address other open issues in QFT.

\item  Part V results from the merge of three single-authored papers, \cite{Pasquarella:2023exd,Pasquarella:2023ntw,Pasquarella:2023vks}. The key findings in this Part are:

\begin{itemize}

\item Moore-Tachikawa varieties without dualities are naturally realised in supersymmetric quiver gauge theories, and we therefore put forward a correspondence in between such theories, magnetic quivers, and Drinfeld centers, \cite{Pasquarella:2023exd}. As a result of the formalism adopted (namely the definition of Moore-Segal bordism operators, \cite{Pasquarella:2023ntw,Moore:2006dw}) applied to Moore-Tachikawa varieties without categorical duality, \cite{Moore:2011ee}, we provide further evidence of our conjectured relation in between homological mirror symmetry and 3D mirror symmetry, first presented in \cite{Pasquarella:2023exd}.

\item  The last paper written by the author prior to submitting this Thesis, \cite{Pasquarella:2023vlr}, nicely ties together the topics covered in \cite{Pasquarella:2023exd,Pasquarella:2023ntw}, while also preparing the stage for the future research the author will be conducting as a postdoc\footnote{Partly outlined in \cite{Pasquarella:2024mlr}.}. In particular, this part takes a more algebro-geometric approach in explaining the mathematical structure underlying supersymmetric quiver gauge theories, with a particular focus on dualities and their mutual relations in terms of higher-categories. As will be explained in depth, the crucial role is played by factorisation homology, \cite{FHP}. We will not attempt to provide an exhaustive definition of it in this introduction, and, instead, refer the reader to the related section in the core of this work. What we can anticipate, though, is that factorisation homology is related with the counting of ground state degeneracy, and, when applied to the AGT correspondence, provides a tool for evaluating the 4-sphere partition function of class ${\cal S}$ theories. Specifically, as we shall see, factorisation homology allows us to relate categorical dualisability and full-extendibility\footnote{And is therefore in line with the connection between dualities and categorical structures.}. Within the context of non-Lagrangian theories, such as 6D ${\cal N}=(2,0)$ SCFTs this is particularly relevant, since it introduces the notion of ring objects, of which Hilbert series are one of the examples of most interest to us. As an algebraic variety, we focus on the calculation of the Hilbert series on Coulomb and Higgs branches, with the latter being related by the geometric invariant theory (GIT) quotient construction, \cite{Deligne-Mumford}. As such, the GIT quotient provides a recipe for realising 3D mirror symmetry; we therefore show how it can be suitably extended for cases where mirror duals are not necessarily in a 1-to-1 correspondence. In relation to \cite{Cabrera:2019izd}, this nicely fits with the magnetic quiver prescription, for describing the Higgs branch of certain supersymmetric quiver gauge theories with non-primitive ideals\footnote{Namely the 2-cone structure, \cite{Ferlito:2016grh}.}. We then show how the 2D description of disk algebras can be suitably adopted for describing Hitchin systems of class ${\cal S}$ theories.  We concluded outlining future directions that are currently under investigation by the same author.  

\end{itemize}

\item Part VI closes outlining the main achievements of the present work, with an outlook on some key questions the author wishes to address in her future research.

\end{enumerate}

We believe our treatment opens the way to further investigating the mutual relation in between homological mirror symmetry and 3D mirror symmetry with the ultimate aim of aiding at understanding the underlying mathematical structure of QFT. These are currently under investigation by the same author, and we plan to report about any advancement in the field in the near future.

\section{Roadmap}

\subsection{Roadmap}    \label{sec:2}

\begin{figure}[ht!]  
\begin{center}  
\includegraphics[scale=1]{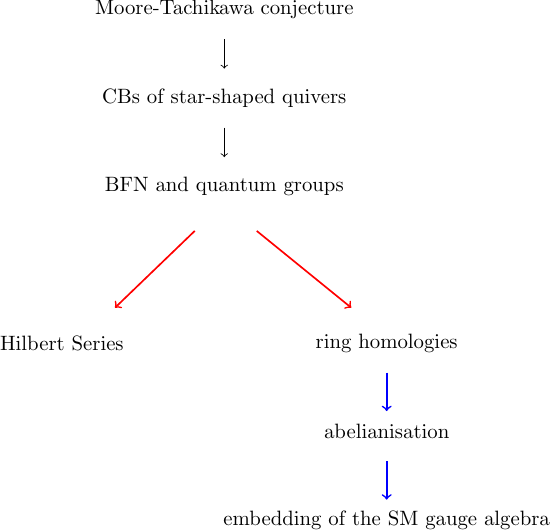}  
\label{fig:roadmap}  
\caption{\small Roadmap for the present and upcoming work.}  
\end{center}  
\end{figure}

This first section is devoted to outlining the path we will be adopting for addressing some of the issues mentioned in the Introduction. The diagram of figure \ref{fig:roadmap} simplifies this, and we will start addressing the topics outlined there starting from section \ref{sec:rqft}, continuing in an upcoming work \cite{VP}. For those who are already familiar with the terminology adopted in figure \ref{fig:roadmap}, the first few sections of this work are purely revision. However, they might still find some interesting connections, which, to the best of our knowledge, haven't been spelled out in the literature with the same emphasis and perspective.

Essentially, Part I can be thought of as an extremely concise summary of the basic knowledge gathered by a mathematical physicist in the attempt to explain a possible roadmap towards deepening the understanding of some major open questions in Theoretical Particle Physics from a Mathematician's point of view. 

As an advanced notice, we advice the reader not to be misguided by the oversimplified explanation of the first few pages of this work: technicalities will soon arise.

In this preliminary section, we introduce the unfamiliar reader to: 

\begin{enumerate}  

\item The basic building blocks of QFT\footnote{We strongly suggest the interested reader to refer to the work of pioneers in the field for a complete treatment of QFT, specifically \cite{W1,W2,W3}.}.  

\item Some key theoretical tools adopted by Phenomenologists for analysing them.   

\end{enumerate}

\subsubsection{Fields and Interactions}

Quantum fields are mutually-interacting entities exhibit wave-particle duality. QFT is arguably the most successful scientific theory so far, with the Standard Model (SM) of Particle Physics being a prototypical example. In some cases, QFT predicts to great accuracy observational outcomes, and has also been used in the context of pure mathematics. 

With the advent of the quantum revolution, fields were promoted to being fundamental constituents, whereas all fundamental particles were understood to arise as a quantum mechanical (QM) effect. This extends to all particles known so far, namely quarks, leptons, gluons, W- and Z-bosons, and the Higgs boson. 

The SM is an example of a QFT containing all the particles known up to now, with their associated three force fields\footnote{The matter fields comprised in it come in 3 families, plus the Higgs boson.} Extensions of the SM, referred to as Beyond the Standard Model (BSM) setups, are currently an active area of research, providing extensions of the SM by adding interactions and/or particles. 

A given QFT is defined in terms of a Lagrangian density, which, upon integrated over spacetime, defines the action, $S$, which is, essentially, a functional of the fields. The Variational Principle constraints the action in such a way that its extremisation leads to the equations of motion of the respective fields. Importantly, the action is needed for defining the partition function, i.e. the expectation value of the identity, as well as any other expectation value of operators\footnote{The latter are also known as correlation functions. } accounted for by a given effective field theory (EFT).

\subsubsection{Amplitudes and Hilbert series} 

\subsubsection*{Amplitudes}  

Scattering amplitudes are among the most important tools in Particle Physics. They are the building blocks for defining transition amplitudes of a certain set of states to another. In the context of collider physics, they are an essential theoretical background tool for predicting and analysing particle collisions and decays. Importantly, if new physics were to be discovered, it must be signalled by a mismatch in between experimental outcomes and scattering calculations, that could either feature as the detection of new particles, and/or the identification of a new interaction (and therefore a new fundamental force). 

\begin{figure}[ht!]  
\begin{center}  
\includegraphics[scale=0.8]{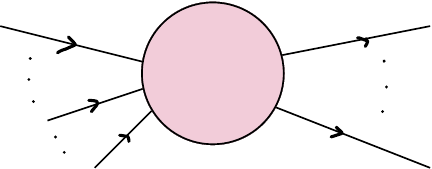}  
\label{fig:scatt}  
\caption{\small A simplified depiction of a scattering process. The $S$-matrix defines the amplitude from the incoming states (black lines on the left) to the outgoing states (black lines on the right). The shaded circle in the middle encodes the information regarding the interactions and operator content characterising a certain EFT.}  
\end{center}  
\end{figure}

In a nutshell, if our theoretical knowledge of particle physics were complete, we should be able to explain all experimental outcomes, recasting them to the Lagrangian theories at hand. However, it so happens that there are some crucial unanswered questions, that clearly signal the need for a deeper mathematical understanding of the models developed so far. Those that motivated our work in the first place are the following:  

\begin{enumerate}  

\item The mass hierarchy problem.   

\item The lightness of the Higgs mass. 

\end{enumerate} 

To help the unfamiliar reader, we will come back to explaining these issues in section \ref{sec:33}.

\subsubsection*{Hilbert series and effective operators}  

The Hilbert series, HS, is defined as follows

\begin{equation}  
\text{HS}[\phi]\ =\ \prod_{_{i=1}}^{^{n}}\ \int_{_{{\cal G}_{_{i}}}}d\mu_{_i}\ \text{PE}\ [\varphi, R]  
\label{eq:HShs}  
\end{equation}   
where ${\cal G}_{_i}$ denotes a given gauge group, $\varphi$ is a \emph{spurion} variable\footnote{A suprion could either be a scalar, a fermion, or a gauge field.}, $d\mu_{_i}$ denotes the Haar measure, and PE$[{\varphi, R}]$ the \emph{plethystic exponential}, which, in turn, is defined as follows\footnote{See for instance \cite{Anisha:2019nzx}.},

\begin{equation}  
\text{PE}\ [\phi, R]  =\exp\left[\sum_{r=1}^{\infty}\frac{\phi^{^r}\chi_{_R}(z_{_j}^{^r})}{r}\right]\ \ \ \ \ \ \ \ ,\ \ \ \ \ \ \ \ \text{PE}\ [\psi, R]  =\exp\left[\sum_{r=1}^{\infty}(-1)^{^{r+1}}\frac{\psi^{^r}\chi_{_R}(z_{_j}^{^r})}{r}\right]
\end{equation} 
for bosons and fermions, respectively\footnote{$\chi$ here denotes the character, and depends in the representation, $R$.}. We will be entering the details of the calculation of \eqref{eq:HShs} in an upcoming work, \cite{VP}. For the purpose of this introductory work, it is enough to know that HS is constructed in terms of the degrees of freedom (namely the particle content) and their transformation properties under the symmetries of a given theory. This is enough to say that, in order to build the Hilbert series, we need to know what the complete set of operators of a given theory is. Completeness of the spectrum of operators is a major sought-after objective in particle physics, since it paves the way towards identifying possible extensions of the fundamental laws of nature. Given the relation in between scattering amplitudes and HS, a mismatch with predictions for a given theory signal the need to extend the effective field theory of reference. If the theory in question is taken to be the Standard Model (SM) of Particle Physics, its resulting extensions are referred to as Beyond the Standard Model (BSM) effective field theories, \cite{Graf:2022rco,Bento:2023owf,Anisha:2019nzx}.

\medskip    

\medskip

\subsection{(Some) Open Questions in Quantum Field Theory}  \label{sec:33}  

Albeit having been successfully verified experimentally, the SM of Particle Physicsis clearly incomplete. The Higgs boson, \cite{Higgs:1964ia,Higgs:1966ev}, responsible for the mass of fundamental particles, was the last piece of a jigsaw to be found, yet the first of a more complicated one. The Higgs boson is a scalar responsible for the mass gain of fundamental particles, via the so-called Higgs mechanism\footnote{We refer the interested reader to the original work by Professor Peter Higgs for a detailed explanation of the process, \cite{Higgs:1964ia,Higgs:1966ev}.}. Undoubtedly one of the major theoretical physics achievements of the past century, the model is clearly incomplete. The two major issues we indicate are the mass hierarchy problem and the lightness of the Higgs boson mass. Each one of them is the object of active investigation by world-leading experts in the field, and by no means we are attempting to provide a full explanation of either of them in our treatment. Nevertheless, we consider them as part of the motivations leading to the need for furthering the understanding of the underlying mathematical structure of QFT.

\subsection*{The Mass Hierarchy Problem}  

Essentially, this is a statement regarding the unexplainable discrepancy between the energy scales at which the fundamental interactions decouple. A successful unifying theory should be able to explain why this symmetry breaking pattern. 

\subsection*{Lightness of the Higgs Boson and Supersymmetry}

Another open question highlighting the fact that the SM is still incomplete is that the Higgs mass is too light. Given that it is the only spinless particle in the SM, theorems imply it should be extremely massive. One of the possible reasons explaining the very light mass of the Higgs was thought to be Supersymmetry (SUSY), proposing that, in order to compensate the very light mass observed, new particles (referred to as superpartners) should be detected at the Large Hadron Collider (LHC). Up to now, though, these superpartners haven't been observed by collider experiments at the energies predicted by the theory. This inevitably leads to the question, to what extent is SUSY able to provide a consistent theory of everything, given that String Theory so strongly relies upon it?  

We believe that SUSY and String Theory are very promising candidates for a unifying theory of everything, but, in order to overcome the apparent shortcomings listed above, we need to address the deep question highlighted in the introduction, namely: "What is (the mathematical structure of) a QFT?". The present work is meant to provide further insights in addressing this question, by making use of mathematical advancements in the field of categorical algebraic geometry. Our findings shed light on a new perspective that is proposed to be a promising alternative top-down construction of QFTs. The ultimate aim would be that of identifying the mathematical structure underlying the SM, and, with it, its embedding in the most suitable string theory vacuum possible, to which we plan to report in upcoming work \cite{VP}.

Albeit being among the most successful accomplishments in theoretical physics, QFT, in its present-day formulation, is mathematically incomplete. To see why this is the case, let us take a step back, explaining a few preliminary features of the theory, and where the issues arise, \cite{DTong,BCA}.

In a QFT, at every point in spacetime one can think of there being an operator, i.e. an infinite-dimensional matrix, acting on a Hilbert space. The mathematical issue arises when attempting to reconcile this picture with the continuum of spacetime to describe our universe. In the quantum picture, the continuum limit is expected to arise upon bringing any two points infinitesimally-close to each other. From a computational perspective, though, this gives rise to divergences, that can only be overcome by supposing spacetime is discretised. In mathematical language, this discretisation turns spacetime into a \emph{lattice}, thereby enabling to keep a finite distance in between any two points in spacetime. However, according to the Nielsen-Ninomiya Theorem, this does not hold for any QFT, and, in particular, it doesn't hold for the SM. 

The great advantage of lattice field theories, is that they enable to have very good mathematical and analytic control, meaning that it is practically impossible for them to describe inconsistent theories. It is therefore tantalising to believe that, if we were able to place the SM on a lattice, we might be able to identify what it is that we are currently missing of QFT as it stands for it to be regarded to as a mathematically rigorous theory.

The reason why the SM cannot (yet) be reconciled with the lattice description is the fact that it is practically impossible to express parity violation in a mathematically rigorous way. In QFT, this is overcome by imposing the so-called \emph{anomaly cancellation}. Importantly, the latter are closely related to the topology of spacetime, and therefore can only arise in the continuum\footnote{For an alternative prescription, see \cite{Jansen:1994ym}.}.

Another open question highlighting the fact that the SM is still incomplete is that the Higgs mass is too light. Given that it is the only spinless particle in the SM, theorems imply it should be extremely massive. One of the possible reasons explaining the very light mass of the Higgs was thought to be Supersymmetry, proposing that, in order to compensate the very light mass observed, new particles should be detected at the LHC. Up to now, the predicted particles still haven't been observed by collider experiments at the energies predicted by the theory. This inevitably leads to the question, to what extent is SUSY able to provide a consistent theory of everything, given that String Theory so strongly relies upon it?  

We believe that SUSY and String Theory are very promising candidates for a unifying theory of everything, but, in order to overcome the apparent shortcomings listed above, we need to address the deep question highlighted in the introduction, namely: "What is (the mathematical structure of) a QFT?". The present work is meant to provide further insights in addressing this question, by making use of mathematical advancements in the field of higher-categorical structures. Our findings shed light on a new perspective that is proposed to be a promising alternative top-down construction of QFTs. The ultimate aim would be that of identifying the mathematical structure underlying the SM, and, with it, its embedding in the most suitable string theory vacuum possible.

\subsection{Where to next?}

Key advancements towards achieving a rigorous mathematical formulation of QFTs strongly relies upon a higher-categorical prescription. In particular, the symmetries of a QFT can be thought of as defects living in different categories connected by functors.

Overall, the present work clearly shows the richness and in-depth understanding that underlying mathematical structures and dualities are able to provide in, both, gravitating systems and QFTs alike. 

Part of this work explores the role played by black holes in transitions involving 2D theories of gravity. In particular, we build connection with the von Neumann algebras, recently put forward by Witten for calculating the entropy of states in QFT and gravity. We therefore show how generalisations of the holographic correspondence, and quantum information tools enable to overcome issues encountered in higher-dimensional setups. 

The complementary part of our treatment deals with interesting cases where lower-dimensional descriptions of higher-dimensional theories are needed. Focussing on dimensional reduction of non-Lagrangian theories, specifically 6D ${\cal N}=(2,0)$ SCFTs on the Riemann surface, and exploiting the Alday-Gaiotto-Tachikawa correspondence, we map the 4D description of class ${\cal S}$ theories to their 2D TFT duals. Viewing the 2D theory as a functor in between 2-categories, we propose a generalisation of Moore-Segal bordism operators as needed for describing Moore-Tachikawa varieties for the case not involving categorical duality, ultimately relating this to joined systems of class ${\cal S}$ theories joined by non-invertible (categorical) defects. By means of algebro-geometric tools, we show how this formalism nicely reconciles with the calculation of Hilbert series for theories admitting a magnetic quiver description. As a biproduct of the tools adopted, our findings provide further evidence of a generalised correspondence in between homological mirror symmetry and 3D mirror symmetry.

In our analysis, we encountered two major setups built upon the notion of higher-categories, namely symmetry TFTs (SymTFT) and topological orders (TO). Plenty of effort has been made towards building a correspondence in between the two, most recently in \cite{Ji:2019eqo}. Our findings provide further support towards strengthening the connection between the two description. 

En-passing, we commented on a proposal regarding the realisation of such composite absolute theories and the definition of a fiber functor intrinsically related to the notion of a partition function for a 3D theory rather than a 4D theory.

We conclude by stressing that our analysis is mostly motivated by furthering the understanding of  6D ${\cal N}=(2,0)$ SCFTs, from, both, a mathematical and physical point of view. To what extent these findings might be mapped to other setups is currently under investigation.

\section{Relative QFTs}  \label{sec:rqft}

The present section consists of two parts: 

\begin{enumerate}  

\item At first, we introduce the notion of relative QFT, and how this applies to sigma models, gauge theories, and 6D ${\cal N}=(2,0)$ SCFTs. This is essentially a brief overview of \cite{Freed:2012bs}.

\item Then, we turn to our main example in this context, namely Coulomb branches of star-shaped quivers\footnote{This terminology will be explained in due course.}, which will be useful to keep as a reference for later sections and Parts of this work. Most importantly, we introduce the notion of \emph{abelianisation}, which will be of great importance throughout our treatment. As we shall see, this naturally paves the way towards introducing category theory tools.   

\end{enumerate}

 \subsection{Rings, homologies, cohomologies, and modules}  \label{sec:prelim}

 In this section we start by introducing some preliminary notation that will be needed for the remainder of our treatment. Most of this is well-known to experts in the field, but we report the main definition just for completeness. 

\emph{Rings} are algebraic structures generalising fields, where multiplication need not be commutative, and inverses might not exist. It is a set endowed with two binary operations called addition and multiplication, such that the ring is an abelian group with respect to the addition operation, and has a multiplicative identity element.

\emph{Homologies} were originally defined in Algebraic Topology, as a general way of associating a sequence of algebraic objects, such as abelian groups or modules, with other mathematical objects, such as topological spaces.

\emph{Modules} generalise the notion of vector space, in which the field of scalars is replaced by a ring. A module also generalises the notion of abelian groups, since the latter are modules over the ring of integers. Modules are closely related to representation theory of groups, and are also widely used in homological algebra, Algebraic Geometry, and Algebraic Topology.

\subsection{Abstract topological symmetry data in field theory}  

To a given QFT, $F$, fixing $N\in\mathbb{Z}^{^{\ge 0}}$, one can assign topological data\footnote{For a brief overview of topological field theories (TFTs), we suggest the interested reader to appendix \ref{sec:TQFTs}, which is mostly overview of well-known background material.}, called quiche, namely a pair $(\sigma, \rho)$ in which

\begin{equation}   
\sigma:\text{Bord}_{_{N+1}}(F)\rightarrow{\cal C} 
\end{equation} 
is an $N+1$-dimensional TFT and $\rho$ is a right topological $\sigma$-module. The quiche is $N$-dimensional, hence it shares the same dimensionality as the theory on which it acts. Let $F$ be an $N$-dimensional field theory. A $(\sigma,\rho)$-\emph{module structure} on $F$ is a pair $(\tilde F, \theta)$, in which $\tilde F$ is a left $\sigma$-module and $\theta$ is an isomorphism 

\begin{equation}   
\theta\ : \rho\ \otimes_{_{\sigma}}\ \tilde F\xrightarrow{\ \ \simeq\ \ }\ F  
\label{eq:LHS}  
\end{equation}   
of absolute $N$-dimensional theories. The LHS of \eqref{eq:LHS} defines the dimensional reduction, i.e. the sandwich we were referring to earlier. $\sigma$ needs only be a \emph{once-categorified} $N$-dimensional theory, whereas $\rho$ and $\tilde F$ are relative theories.

Examples of relative theories are, \cite{Freed:2012bs}:  

\begin{enumerate}   

\item 2D chiral WZW CFTs based on a compact Lie group $G$ and level $k$, w.r.t. 3D topological CS theory associated to the $(G,k)$-data of its 2D boundary.

\item Gauge theories.

\item 6D ${\cal N}$=(2,0) SCFTs w.r.t. 7D TQFT in the bulk. 

\end{enumerate}          

More generally, \emph{relative theories}  are models whose fields form a fibration

\begin{equation}  
{\cal F}\ \longrightarrow\  {\cal M}\ \xrightarrow{\ \ \pi\ \ }\ {\cal B} 
\end{equation}   
where ${\cal B}$ and ${\cal F}$ denote the base and the fiber, respectively. Relative fields are the fibers of $\pi$. The path integral over the base reduces a finite sum. For the theory to be absolute, the same should hold for ${\cal F}$ as well. 

As originally outlined in \cite{Freed:2012bs}, \ QFT $F$ relative to $\alpha$ is a homomorphism 

\begin{equation}   
\tilde F:\ \tau_{_{\le n}}\alpha   \ \rightarrow\ \mathbf{1}\  
\end{equation}   
where $alpha$ is the bulk theory w.r.t. $F$.

\subsubsection{Fibering over \texorpdfstring{$BG$}{}}

Symmetries in field theory are expressed in terms of topological defects acting as operators on state spaces of the theory in question. The idea at the heart of this are the notions of abstract groups and abstract algebras. In their work, \cite{Freed:2022qnc}, Freed, Moore and Teleman propose an abstract symmetry structure in the context of field theory and its concrete realisation.

A field theory is analogous to a linear representation of a Lie group or a module over an algebra. Hence their definition of the boundary theory as a module. The essential content of their definition is that of expressing a QFT, $F$, as a sandwich $\rho\otimes_{\sigma}\tilde F$, with $\rho$ a regular Dirichlet BC (right boundary theory for $\sigma$) and $\tilde F$ the left boundary theory for $\sigma$ (typically not topological). 

Altogether, $(\sigma,\rho)$ constitutes the $n$-dimensional \emph{quiche}, defining the abstract symmetry structure of a given QFT. Defects away from $\tilde F$ are topological and belong to $(\sigma,\rho)$. $(\sigma,\rho)$-defects are analogous to elements of an abstract algebra (connection with GKSW). The novelty of the treatment by Freed, Moore and Teleman is the fact that the sandwich representation develops a calculus of topological defects acting on a QFT ( based on fully local TFT). The TFT imposes strong finiteness constraints on the QFT.   

Let $G$ be a finite group. A classifying space $BG$ is derived   from a contractible topological space $EG$ equipped with a free $G$-action by taking the quotient. The homotopy type of $BG$ is independent of choices. If $X$ is a topological space equipped with a $G$-action, then the Borel construction, \cite{BWB} is the total space of the fiber bundle.

\begin{equation}  
\begin{aligned}   
&X_{_{G}}\ \equiv\ EG\ \times_{_{G}}\ X  \\   
&\\    
&\downarrow\ \ \pi\\ 
&\\ 
&BG   
\end{aligned}   
\label{eq:fibration}
\end{equation}  
with fiber $X$. If $*\in BG$ is a point, and we choose a basepoint in the $G$-orbit in $EG$ labelled by $*$, then the fiber $\pi^{^-1}(*)$ is canonically identified with $X$. The abstract group symmetry data is the pair $(BG,*)$, and a realisation of the symmetry $(BG,*)$ on $X$ is a fiber bundle over $BG$ together with an identification of the fiber over $*\in BG$ with $X$.

\subsection{Preliminary examples: the \texorpdfstring{$\sigma$}{}-model and gauge theories}

In light of the techniques introduced in the following Parts of this work, we now introduce the notion of moduli space, as denoting the minima of the theory in question, namely the least energy field configurations\footnote{For a brief overview of moduli spaces in some basic examples, we refer the interested reader to appendix \ref{sec:sigmamodels}.}. Albeit the sample theories presented in this section are quite trivial, they are sufficient for building intuition for what lies ahead in our treatment. 
\subsection*{1) Sigma models}

Data: $G$ a finite group and $M$ a smooth manifold with a free left-acting $G$ and quotient $\bar M$ (equivalently the gauged $\sigma$ model on $M$).

Denoting with $BG(X)$ the collection of principal $G$-bundles over $X$, 

A principal $G$-bundle is a covering space $P\rightarrow X$ and a free-$G$ action on $P$ such that $P\rightarrow X$ is a quotient map for the $G$-action. We denote the collection of principal $G$-bundles over $X$ with $BG(X)$, with the latter being a grupoid, and, as such, is a category in which every morphism is invertible. 

A \emph{symmetry}

\begin{equation}  
\varphi: (P\ \rightarrow\ X)\ \rightarrow   \ (P^{\prime}\ \rightarrow\ X)
\end{equation}  is a diffeomorphism 

\begin{equation}  
\varphi: P\ \rightarrow\ P^{\prime}  
\end{equation}  
commuting with $G$ and covering id$_{_{X}}$. The automorphism group of $P\ \rightarrow X$ is the group of gauge transformations. 

The path-integral over $BG(Y)$ is an integral over the equivalence classes of $G$-bundles. 

Fixing $(P\ \rightarrow\ X)\ \in\ BG(X)$, a relative field over $(P\ \rightarrow\ X)$ is a pair $(f,\theta)$ consisting of a smooth map $f: X\rightarrow \bar M$ and an isomorphism $\theta:(P\ \rightarrow\ X)\rightarrow f^*(M\ \rightarrow\ \bar M)$ of $G$-bundles over $X$. Given that relative fields feature no automorphisms, they are \emph{rigid}, and therefore constitute a space rather than a category. 

Given a pair of theories, $(\alpha, F)$, with $\alpha$ topological and defined on all manifolds, and $F$ a Riemannian theory, using the knowledge of the $\sigma$-model $F$ to predict the structure of $\alpha$. 

The relative path-integral $F(X)$ on a closed $D$-dimensional manifold is an integral over relative fields, hence a function   

\begin{equation}  
F(X): BG(X)\ \rightarrow \ \mathbb{C}  
\label{eq:neqFX}   
\end{equation}   

As such, $F(X)$ is invariant under symmetries of $BG(X)$, and is therefore expressed as a function on equivalence classes, $H^{^{1}}(X;G)$, with the latter denoting the isomorphic classes if principal $G$-bundles over $X$. From this, \eqref{eq:neqFX} can be reexpressed as 

\begin{equation}  
F(X): H^{^{1}}(X;G)\ \rightarrow \ \mathbb{C}  
\label{eq:neqFX1}   
\end{equation}   

If $F$ is to be a QFT relative to a $D+1$-dimensional bulk $\alpha$, in the sense that 

\begin{equation}  
F(X):\ \mathbb{C}\ \rightarrow\ \alpha(X)   
\end{equation}  
then, $F(X)$ can be identified as an element of the vector space $\alpha(X)$, leading to the conclusion that $\alpha(X)$ id the free vector space on the finite set $H^{^{1}}(X;G)$.  

Next, considering a closed  $D-1$-dimensional manifold, $Y$, the relative canonical quantisation of $F(Y)$ is obtained by performing the quantisation on the fibers of the map 

\begin{equation}  
\text{Map}(Y,\bar M)\ \rightarrow\ BG(Y)  
\end{equation}    
such that the vector bundle is 

\begin{equation}  
F(Y)\ \rightarrow\ BG(Y)  
\label{eq:equivbund}
\end{equation}

\begin{figure}[ht!]   
\begin{center}
\includegraphics[scale=0.5]{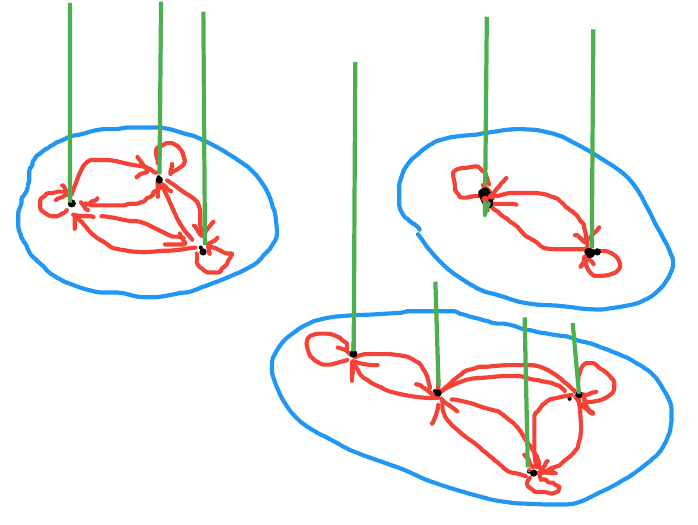} 
\label{fig:fiber}   
\caption{\small A depiction of the vector bundle.}
\end{center} 
\end{figure} 

The bundles are grouped into isomorphism classes. Points featuring in each isomorphism class denote the $G$-bundles $Q\rightarrow Y$, and the arrows the isomorphisms between the $G-bundles$. Selecting a representative bundle, $Q\rightarrow Y$, in each isomorphism class, the grupoid $BG(Y)$ corresponds to a simpler grupoid with a finite set of $H^{^{1}}(Y;G)$ and no morphisms between them.

If $m\ \in H^{^{1}}(Y;G)$ is the class of a $G$-bundle $Q\rightarrow Y$, then the automorphism group of $m$ is the group of gauge transformations of $Q\rightarrow Y$, and the equivariant bundle, \eqref{eq:equivbund}, decomposes into topological vector spaces indexed by pairs $(m,e)$ in which $m\ \in\ H^{^{1}}(Y;G)$ and $e$ is a complex irreducible representation of the automorphism group of $m$. The partition function on $Y$   

\begin{equation} 
F(Y):\ \mathbf{1}(Y)\ \rightarrow \ \alpha(Y)  
\end{equation}  
may be identified with an object belonging to the linear category $\alpha(Y)$, leading to the prediction that $\alpha(Y)$ is the free Vect$_{_{top}}$-module with basis pairs $(m,e)$.  

For the case in which $X$ is a 1D theory, namely $X\equiv\ S^{^{1}}$, $H^{^{1}}(S^{^{1}};G)$ is the set of \emph{conjugacy classes} in $G$. If $Y\equiv$pt, then $H^{^{1}}(\text{pt};G)$ has a single element, representing the trivial bundle with automorphism group $G$. Consequently, $\alpha(\text{pt})$ defines the category of representations of $G$. 

In this case, $\alpha$ is the finite 2D gauge theory with gauge group $G$, and can be defined on any manifold of $D\le 2$ by a finite path-integral. Given that the the theory has a vanishing Lagrangian, the partition function on a closed 2D manifold, $M_{_{2}}$, reduces to the weighted counting of $G$-bundles    

\begin{equation}  
\alpha(M_{_{2}})  
\ 
=   
\   
\sum_{[R\rightarrow\ M_{_{2}}]}\ \frac{1}{\ |\text{Aut}(R\rightarrow M_{_{2}})|\ }  
\end{equation}   

The finite path integral over $X\equiv S^{^{1}}$ is the space of invariant sections of the trivial equivariant line bundle $BG(S^{^{1}})$. On $Y\equiv$ pt, instead, the exponential of the action has constant value in the linear category Vect$_{_{top}}$, hence the path integral defines the subcategory of invariants under $G$, namely the category of representations of $G$.   

\subsection*{From relative to absolute}

An \emph{absolute} theory can be recovered from a relative one under gauging by a finite subgroup $G^{^{\prime}}\ \subset\ G$. In such case, the $\sigma$-model on a closed $D$-dimensional manifold can be expressed in terms of the relative theory as follows

\begin{equation}  
f_{_{G^{\prime}}}(X)   
\ 
=   
\   
\sum_{m^{\prime}\ \in\  H^{^{1}}(X;G^{\prime})}\ \frac{1}{\ |Z_{_{G^{\prime}}}(m^{\prime})|\ }\ F(X;\ m^{\prime})  
\end{equation} 
where the sum runs over the equivalence classes of $G^{\prime}$-bundles $P\rightarrow X$, and $Z_{_{G^{\prime}}}(m^{\prime})$ denotes the automorphism group of a representative of the equivalence class $m^{\prime}$. 

Following similar arguments, the quantum topological vector space on a closed $D-1$-dimensional manifold $Y$ can be expressed as

\begin{equation}  
f_{_{G^{\prime}}}(Y)   
\ 
=   
\   
\bigoplus_{m^{\prime}\ \equiv\ [Q^{\prime}\rightarrow Y]\in\  H^{^{1}}(X;G^{\prime})}\  F(Y;\ Q\rightarrow Y)^{^{\text{Aut}(Q^{\prime}\rightarrow Y)}}  
\end{equation}

\subsection*{2) Gauge theories}

A very similar analysis holds for the case of gauge theories, with all the fields' categorical levels enhanced by 1 unit, due to the presence of an additional symmetry. 

In this case, the theory is specified by the following data: a Lie algebra, $\mathfrak{g}$, a covering homomorphism $G\rightarrow\ \tilde G$ of compact connected Lie groups with kernel $G_{_{o}}$ a finite central abelian subgroup of $G$. For example, if $G_{_{o}}\ \subset\ G$ and $\tilde G\equiv G\diagdown G_{_{o}}$ is the adjoint group, this gives a canonical choice of data associated to a compact semisimple Lie algebra. 

Assuming $\tilde P\rightarrow X$ denotes a principal $\tilde G$ bundle, the obstruction to lifting to a principal $G$-bundle is measured by a $G_{_{o}}-gerbe$,  ${\cal G}(\tilde P)\rightarrow X$, and with $B^{^{2}}G_{_{o}}(X)$ the collection of $G_{_{o}}$-gerbes over $X$. $B^{^{2}}G_{_{o}}(X)$ is a 1-category. Its equivalence classes constitute the cohomology group $H^{^{2}}(X;G_{_{o}})$. The group of equivalence classes of automorphisms between objects is $H^{^{1}}(X;G_{_{o}})$, and the group of automorphisms of automorphisms is $H^{^{0}}(X;G_{_{o}})$. 

For gauge theories, a relative field can be defined as $({\cal G}\rightarrow X)\ \in\ B^{^{2}}G_{_{o}}(X)$. A relative field over $({\cal G}\rightarrow X)$ is a pair $(\tilde\Theta, \theta)$ consisting of a $\tilde G$-connection $\tilde \Theta$ and an isomorphism $\theta: {\cal G}\rightarrow{\cal G}(\tilde\Theta)$ of $G_{_{o}}$-gerbes. $(\tilde\Theta, \theta)$ form an ordinary grupoid, with no automorphisms between automorphisms. The fields are $\tilde G$-connections with $G$-gauge transformations.

The quantum theories $(\alpha, F)$ that can be built from these fields are a $D+1$ dimensional topological theory and a $D$-dimensional relative theory $F$. $\alpha$ is now defined as a path-integral over the $G_{_{o}}$-gerbes, specifically  

\begin{equation}  
\alpha(M_{_{2}})  
\ 
\equiv 
\    
\frac{\ |H^{^{2}}(X;G_{_{o}})| | H^{^{o}}(X;G_{_{o}})|\ }{|H^{^{1}}(X;G_{_{o}})| }     
\end{equation}  

Consequently, $\alpha(X)$ is the vector space of complex-valued functions on $B^{^{2}}(X)$. The relative theory $F$ is an element of $\alpha(X)$ since  

\begin{equation}
F(X)\equiv H^{^{2}}(X;G_{_{o}})\ \rightarrow \ \mathbb{C}  
\label{eq:neqFX11}   
\end{equation} 

 $\alpha(Y)$, instead, defines the linear category of vector bundles over $B^{^{2}}G_{_{o}}(Y)$. The relative theory $F$ determines a particular vector bundle $F(Y)\rightarrow B^{^{2}}G_{_{o}}(Y)$. 

 For any finite abelian group $AA$, its Pontryagin dual is defined as follows  

 \begin{equation}  
 A^{^{\text{V}}}\ \overset{def.}{=}\ \text{Hom}(A,G_{_{o}})   
 \end{equation}  

 Once having chosen the basepoints, $F(Y)\rightarrow B^{^{2}}G_{_{o}}(Y)$ determines the topological vector spaces $F(Y;m,e)$ for $m\in H^{^{^2}}(Y;G_{_{o}}), e\in H^{^{1}}(Y;G_{_{o}})^{^{\text{V}}}$, where $(m,e)$ denote the classes of discrete magnetic and electric fluxes. 

\subsection*{From relative to absolute}

 Once more, an absolute gauge theory can be obtained by gauging by a subgroup of $G$, leading to a relic gauge group $G\diagdown G_{_{o}}^{\prime}$, $G_{_{o}}^{\prime}\subset G_{_{o}}$, and its expression reads

 \begin{equation}  
f_{_{G_{_{o}}^{\prime}}}(X)   
\ 
=   
\   
\sum_{m^{\prime}\ \in\  H^{^{2}}(X;G_{_{o}})}\ \frac{\ |H^{^{0}}(X;G_{_{o}}^{\prime})|\ }{\ |H^{^{1}}(X;G_{_{o}}^{\prime})|\ }\ F(X;\ m^{\prime}).  
\end{equation}

\begin{figure}[ht!]   
\begin{center}
\includegraphics[scale=0.9]{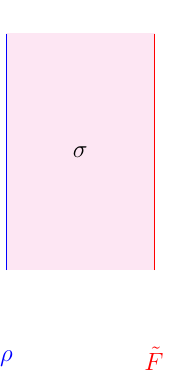}     
\label{fig:FrredMooreTeleman extended} 
\caption{\small Specifying the topological boundary conditions turns a relative QFT into an absolute one.}  
\end{center} 
\end{figure} 

\subsection{Theory \texorpdfstring{$X$}{}}\label{sec:X}

We now turn to the case of 6D ${\cal N}=(2,0)$ SCFTs. The data defining the theory in question are: 

\begin{enumerate}

\item    A reductive Lie algebra, $\mathfrak{g}$, with an invariant inner product, $<,>$, such that all coroots have square of length 2. A reductive real Lie algebra is defined as the direct sum 

\begin{equation}   
\mathfrak{g}=\mathfrak{z}\ \oplus\ \mathfrak{g}^{\prime},  
\end{equation}
where $\mathfrak{z}$ denotes its center, and $\mathfrak{g}^{\prime}\equiv[\mathfrak{g},\mathfrak{g}]$ its semisimple subalgebra. A Cartan subalgebra $\mathfrak{h}^{\prime}\ \subset\ \mathfrak{g}^{\prime}$ determines the coroot lattice, which is a full sublattice $\Gamma^{\prime}\ \subset\ \mathfrak{h}^{\prime}$.

\item A full lattice $\Gamma\ \supset\ \Gamma^{\prime}$ in $\mathfrak{h}=\mathfrak{Z}\ \oplus\ \mathfrak{h}^{\prime}$, such that the inner product is integral and even on $\Gamma$. 

\end{enumerate}

Given this data, one can define $\Lambda\ \supset\ \Gamma$ as the dual lattice to $\Gamma$ in $\mathfrak h$, with $\Lambda$ being the subset of vectors $\eta\ \in\ \mathfrak h$ such that

\begin{equation}   
<\eta, \Gamma>\ \subset \ \mathbb{Z}.  
\end{equation}

Then, the quotient 

\begin{equation}  
\pi\ \overset{def.}{=}\ \Lambda/\Gamma  
\end{equation}   
is a finite abelian group equipped with a perfect pairing

\begin{equation}  
\pi\ \times\ \pi\ \rightarrow\ \mathbb{Q}/\mathbb{Z}\ \subset\ \mathbb{T}  
\label{eq:pairing}
\end{equation}
induced from the inner product. This pairing induces an isomorphism   

\begin{equation}   
\pi\ \simeq\ \pi^{\text{v}},      
\end{equation}          
implying $\pi$ is Pontrjagin self-dual. 

The expectation outlined by Freed and Teleman on the basis of Witten's pioneering work, is therefore that, given   

\begin{equation}   
\left(\mathfrak{g}, <,>,\Gamma\right),     
\end{equation}   
there exists a finite 7D TQFT   

\begin{equation}  
\alpha_{_{\mathfrak{g}}}\ \overset{def.}{=}\ \alpha\left(\mathfrak{g}, <,>,\Gamma\right),    
\end{equation}   
and a 6D QFT

\begin{equation}  
X_{_{\mathfrak{g}}}\ \overset{def.}{=}\ X_{_{\left(\mathfrak{g}, <,>,\Gamma\right)}},    
\end{equation}  
relative to $\alpha_{_{\mathfrak{g}}}$, with $\alpha_{_{\mathfrak{g}}}$ a bordism mutli-category of manifolds with unspecified tangential structure. This theory is constitutes a 7D analogue of CS theory for torus groups. On the other hand, the 6D theory, $X_{_{\mathfrak{g}}}$, is meant to be a geometric bordism category with a conformal structure.

\begin{figure}[ht!]   
\begin{center}
\includegraphics[scale=1]{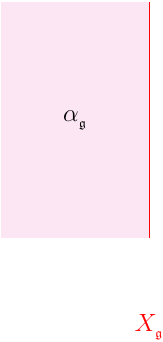}   
\caption{\small Theory $X$, here denoted $X_{_{\mathfrak{g}}}$, lives on the boundary of a 7D bulk TFT, $\alpha_{_{\mathfrak{g}}}$. No topological boundary theory can be defined to make the theory absolute.}
\label{eq:mew}   
\end{center}
\end{figure}

For $X$ a closed oriented 6-manifold, the cup product and pairing \eqref{eq:pairing} combine to give a non-degenerate skew-symmetric pairing

\begin{equation}   
H^{^3}(X;\pi)\ \times\ H^{^3}(X;\pi)\ \rightarrow\ \mathbb{T}  
\end{equation}  
on the finite group $H^{^3}(X;\pi)$. Upon taking $Y$ to be an oriented a 5-dimensional manifold, $H^{^3}(Y;\pi)$ defines the group of self-dual fluxes, and, from the pairing and Poincare'-Pontrjagin duality, the following isomorphism follows   

\begin{equation}  
H^{^3}(Y;\pi)\ \simeq\ H^{^3}(Y;\pi^{\text{v}})\ \simeq\ H^{^2}(Y;\pi)^{\text{v}},    
\end{equation}
where self-duality identifies the electric and magnetic groups. Then, the linear category $\alpha_{_{\mathfrak{g}}}$ is the free Vect$_{_{\text{top}}}$-module with basis $H^{^3}(Y;\pi)$, and, to each self-dual flux, $\sigma\ \in\ H^{^3}(Y;\pi)$, should correspond a quantum topological vector space, $X_{_{\mathfrak{g}}}(Y;\sigma)$.

\subsection{Main Example: the Coulomb branch chiral algebra}  \label{sec:main}   

An example of \eqref{eq:fibration} is 

\begin{equation}  
\begin{aligned}   
&{\cal M}_{_C}\ \longleftarrow\ T^{^{\text{V}}}_{_{\mathbb{C}}}  \\   
&\\    
&\downarrow\ \ \pi\\ 
&\\ 
&\mathfrak{t}_{_{\mathbb{C}}}   /\ W
\end{aligned}   
\end{equation} 
where $\mathfrak{t}_{_{\mathbb{C}}}$ denotes the complexified Cartan subalgebra of $G$, $\mathfrak{t}_{_{\mathbf{C}}}\ \subset\ \mathfrak{g}_{_{\mathbf{C}}}$, $W$ is the Weyl group, and $T^{^{\text{V}}}_{_{\mathbb{C}}}$ is the complexified dual of the maximal torus. $T^{^{\text{V}}}_{_{\mathbb{C}}}\simeq\mathbb{C}^{^{\text{rank}(G)}}$ is a holomorphic Lagrangian torus with respect to the holomorphic symplectic structure. The base $\mathfrak{t}_{_{\mathbb{C}}}   /\ W$ is parametrised by the diagonal expectation value of a complex sclar vectormultiplet, \cite{hkq}, 

\begin{equation}  
\varphi\ \in\ \mathfrak{t}_{_{\mathbb{C}}}  \ \subset\ \mathfrak{g}_{_{\mathbb{C}}}   .  
\end{equation}  

Global coordiantes on the base are given by polynomials in $\varphi$ which are gauge-invariant operators of a non-abelian theory. $\varphi$ is referred to as a \emph{generic} point of the Coulomb branch if: 

\begin{enumerate}  
\item It fully breaks gauge symmetry to the torus (practically turning the W-boson massive), 

\begin{equation}  
M_{_{\alpha}}\ \equiv\ <\alpha,\varphi>\ \neq \ 0,  
\end{equation}  
for all $\alpha$, roots of $G$.

\item Giving a non-zero effective mass to every hypermultiplet,  

\begin{equation}  
M_{_{\lambda}}\ \equiv\ <\lambda, \varphi>\ \neq \ 0,  
\end{equation}   
for all $\lambda$, weights of $R$.

\end{enumerate}  

Hence, a generic point of $\mathfrak{t}_{_{\mathbb{C}}}$ lies in the complement of all weight and root hyperplanes, \cite{hkq}.  

${\cal M}_{_C}$ is a complex symplectic manifold, on which one can calculate the chiral ring $\mathbb{C}[{\cal M}_{_C}]$ and its Poisson structure. The chiral-ring operators lie in the cohomology of a topological supercharge, and therefore lie within the topological subsector of the 3D gauge theory in question. Consequently, the product of chiral operators is topologically protected and can be calculated by means of TQFT techniques.    

The chiral algebra ${\cal A}$ can be defined as the local chiral ring in the neighbourhood of a generic point $\varphi$ on the base of the Coulomb branchwhen W-bosons and hypermultiplets become massless, 

\begin{equation}  
{\cal M}_{_C}^{^{\text{abel.}}}\ \overset{def.}{=}\ \pi^{^{-1}}\left(\left(\mathfrak{t}_{_C}\backslash\ \Delta\cup\Delta_{_R}\right)/W \right)\ \subset\ {\cal M}_{_C}     
\label{eq:abelalg}    
\end{equation}  
where  
\begin{equation}  
\Delta\ \overset{def.}{=}\ \bigcup_{_{\text{roots $\alpha$}}}\ \left\{ M_{_{\alpha}}(\varphi)=0\ \right\}\ \subset\ \mathfrak{t}_{_C}\ \ \  \ , \ \ \ \ \Delta_{_R}\ \overset{def.}{=}\ \bigcup_{_{\text{weights $\lambda$}}}\ \left\{ M_{_{\lambda}}(\varphi)=0\ \right\}\ \subset\ \mathfrak{t}_{_C}.
\end{equation}  

Form this, it follows that the algebra ${\cal A}$ is defined from the covering of \eqref{eq:abelalg} as        

\begin{equation}   
{\cal A}\ \overset{def.}{=}\ \mathbb{C}\left[\tilde{\cal M}_{_{\mathbf{C}}}^{^{\text{abel}}}\right] 
\end{equation}  
from which it follows that 

\begin{equation}  
\mathbb{C}[{\cal M}_{_C}]\ \hookrightarrow\ {\cal A}.  
\end{equation}    

The generators of ${\cal A}$ come in two main types, \cite{hkq}:  

\begin{enumerate}

\item Polynomials in $\varphi, M_{_{\alpha}}^{^{-1}}, M_{_{\lambda}}^{^{-1}}$.

\item Abelian monopole operators for any cocharacter $A\ \in\ \text{Hom}(U(1), T)\ \simeq\ \mathbb{Z}^{^{{\text{rank}(G)}}}$.

\end{enumerate}   

Gathering these properties together, the abelianised embedding of the chiral ring operators can succinctly be denoted as follows 

\begin{equation}  
{\cal A}\ =\ \mathbb{C}\left[\ \varphi, \left\{M_{_{\alpha}}^{^{-1}}\right\}, \left\{M_{_{\lambda}}^{^{-1}}\right\}, \{v_{_A}\}\ \right]\ /\ \text{rel.}  
\end{equation}

\subsubsection{Star-shaped quivers}  

\subsubsection*{The Coulomb branch}

For the case of interest to us, we focus on star-shaped quivers, conventionally denoted by ${\cal T}_{_{N, \mathbf{k}}}$, where $N$ dictates the number of gauge nodes in the quiver, and $\mathbf{k}$ the number of legs.  To see how we have outlined above practically leads to the chiral ring defining the Coulomb branch, lats assume that $N=2$ and $\mathbf{k}=3$. In this case the Coulomb branch is flat 

\begin{equation}  
{\cal M}_{_C}\ \simeq\ \mathbb{C}^{^8}.  
\end{equation}   

This is dual to a 4D class ${\cal S}$ theory with $A_{_1}$ structure, namely the Grassmannian. The gauge grouup of the theory in question, ${\cal T}_{_{2, 3}}$, reads  

\begin{equation}  
G\ =\ \left[U(2)\times U(1)^{^3}\right]\ /\ U(1)_{_{\text{diag}}}.     
\end{equation}

Monopole charges are labelled by, \cite{hkq},  

\begin{equation}  
\text{cochar} (G)\ =\ \text{Hom} (U(1), T)\ =\ \mathbb{Z}^{^5}\ /\ \mathbb{Z}_{_{\text{diag}}}     
\end{equation}
assigning 5 integers  as follows

\begin{equation}    
\left(A_{_1}, A_{_2}; B_{_1}, B_{_2}, B_{_3}\right)\ \in\ \text{cochar}(U(2))\times \text{cochar}\left(U(1)^{^3}\right)\ /\ A_{_{\text{diag}}}.      
\end{equation}    

By definition, two cocharacters are equivalent if they differ by an integer multiple of $A_{_{\text{diag}}}$. On the other hand, the weights are defined as follows   

\begin{equation}  
\lambda\ =\ \text{weights}(G)\ \overset{def.}{=}\ \text{Hom} (T, U(1)))  
\end{equation}  
such that

\begin{equation}  
\text{weights}(G)\times\text{cochar} (G)\ \longrightarrow\ \mathbb{Z}\ \ \ \ \ \ \ \ \ \text{and}\ \ \ \ \ \ \ \ <\lambda, A_{_{\text{diag}}}>\ =\ 0\  \ \ \ \ \ \ \ \ \forall\ \lambda.     
\end{equation}    

The matter representation is

\begin{equation}    
{\cal R}\ =\ R\ \oplus\ R^{^*}.  
\end{equation}

${\cal T}_{_{2,3}}$ has an $SU(2)^{^3}$ flavour symmetry acting on its Coulomb branch, and a corresponding $SL(2, \mathbb{C})^{^3}$ symmetry in the chiral ring. This should be generated by three $\mathfrak{sl}(2, \mathbb{C})^{^*}$-valued complex moment maps $\mu_{_a}$, $a=1,2,3$.  

This is a particular example, where the Coulomb branch of the star-shaped quiver coincides with the Higgs branch of the 4D  class ${\cal S}$ theory ${\cal T}_{_2}\left[\Sigma_{_{0,3}}\right]$, hence it simply reads  

\begin{equation}  
{\cal M}_{_H}\ \simeq\ \mathbb{C}^{^8}. 
\end{equation}

\subsubsection*{The Higgs branch}   

$\mathbf{k}$ denotes the number of punctures on sphere. The Higgs branch is defined once having identified all the pair of pants decompositions of $\Sigma_{_{0,\mathbf{k}}}$, namely  

\begin{equation}   
\Sigma_{_{0,\mathbf{k}}}\ \simeq\ \bigcup_{_{\alpha=1}}^{^{k-2}}\ \Sigma_{_{0,\mathbf{3}}}^{^{(\alpha)}},      
\end{equation}  
from which the Higgs branch is extracted as a holomorphic symplectic quotient  

\begin{equation}  
{\cal M}_{_H}^{^{4D}}\ \simeq\ \left[\ \prod_{_{\alpha=1}}^{^{k-2}}\ \mathbb{C}^{^8}\ \right]\ \dslash\ SL(2, \mathbb{C})^{^{k-3}}.  
\end{equation}   

However, the disadvantage of this case is that the decomposition is not unique, and the operators and relations are not manifest. 

\subsubsection*{1- and 2-legged star-shaped quivers}  

For $\mathbf{k}=1,2$, the Coulomb branches simply read  

\begin{equation}  
{\cal M}_{_C}\ \simeq\ T^{^*}SL(2,\mathbb{C})  
\ \ \ \ \ \ ,\ \ \ \ \ \ {\cal M}_{_C}\ \simeq\ T^{^*}SL(2,\mathbb{C}) //_{_{\psi}}\ \mathfrak{N}  
\end{equation}  
where $T^{^*}SL(2,\mathbb{C})$ is the cotangent bundle of the complex group $SL(2,\mathbb{C})$, whose points are parametrised by a triple $\left(\mu_{_L}, g, \mu_{_R}\right)$, such that   

\begin{equation}  
T^{^*}SL(2,\mathbb{C})\ \equiv\ \left\{\left(\mu_{_L}, g, \mu_{_R}\right)\ \bigg|\ \mu_{_{L,R}}\ \in\ \mathfrak{sl}(N,\mathbb{C})^{^*}\ , \ g\in SL(N,\mathbb{C})\ ,\ \mu_{_L}g=g\mu_{_R}\right\},     
\label{eq:tripleeq}   
\end{equation}
with $\mu_{_{L,R}}$ complex moment maps for the cation of the left and right multiplication of $SL(N, \mathbb{C})$ on itself, extending to $T^{^*}SL(2,\mathbb{C})$ as complex Hamiltonian actions\footnote{Essentially, $\mu_{_{L,R}}$ respectively denote the left- and right-invariant trivialisation of the cotangent bundle.}.       

\subsection{Lagrangian submanifolds of a symplectic manifold}    

We now outline some key terminology needed for later purposes in our treatment. In particular we introduce Lagrangian submanifolds of a symplectic manifold in the context of quiver gauge theories. The remainder of this section is mostly a review of \cite{hkq}.

\subsubsection{Hyperk\texorpdfstring{$\ddot{\text{a}}$}{}hler quotients}  

Let $X$ be a K$\ddot{\text{a}}$hler manifold presented as a K$\ddot{\text{a}}$hler quotient of $\mathbb{C}^{^n}$ by the linear action of a compact group $G$. The hyperk$\ddot{\text{a}}$hler analogue, $\mathfrak{M}$, of $X$ is a hyperk$\ddot{\text{a}}$hler quotient of the cotangent bundle $T^{^*}\mathbb{C}^{^n}$ by the induced $G$-action. Special instances of this construction include hypertoric and quiver varieties, \cite{Moore:2011ee}.     

Let $M$ be a hyperk$\ddot{\text{a}}$hler manifold equipped with a hyperhamiltonian action of a compact Lie group $G$, and moment maps $\mu_{_a}$, $a=1,2,3$. $M$ is a symplectic manifold, and $X$ is a Lagrangian submanifold with respect to $M$, \footnote{More of this will be explained in the context of \ref{sec:KRS}.}

Suppose that 

\begin{equation}  
\xi\ =\ \xi_{_{\mathbb{R}}}\ \oplus\ \xi_{_{\mathbb{C}}}  
\end{equation}  
is a central regular value of $\mu_{_{\text{HK}}}^{^{-1}}(\xi)$. Then there is a unique hyperk$\ddot{\text{a}}$hler structure on the hyperk$\ddot{\text{a}}$hler quotient    

\begin{equation}   
\mathfrak{M}\ =\ M\dslash_{_{\xi}}G\ \overset{def.}{=}\ \mu_{_{\text{HK}}}^{^{-1}}(\xi)/G,       
\end{equation}   
which is the hyperk$\ddot{\text{a}}$hler analogue of a k$\ddot{\text{a}}$hler quotient    

\begin{equation}  
X\ =\ \mathbb{C}^{^n}\dslash\  G,    
\end{equation}       
where   

\begin{equation}   
\mu_{_{\text{HK}}}\ \equiv\ \mu_{_{\mathbb{R}}}\ \oplus\ \mu_{_{\mathbb{C}}}\ :\ M\ \rightarrow\ \mathfrak{g}^{^*}\ \oplus\ \mathfrak{g}_{_{\mathbb{C}}}^{^*}   
\label{eq:muhk}    
\end{equation}    
with associated symplectic and holomorphic symplectic forms, $\omega_{_{\mathbb{R}}}^{^{\xi}}$ and $\omega_{_{\mathbb{C}}}^{^{\xi}}$, such that they pull back to the restrictions of $\omega_{_{\mathbb{R}}}$ and $\omega_{_{\mathbb{C}}}$ to $\mu_{_{\text{HK}}}^{^{-1}}(\xi)$.

Restricting to the case in which $M=T^{^*}\mathbb{C}^{^n}$, \eqref{eq:muhk} reduces to    

\begin{equation}   
\mu_{_{\mathbb{R}}}\ \oplus\ \mu_{_{\mathbb{C}}}\ :\ T^{^*}\mathbb{C}^{^n}\ \rightarrow\ \left(\mathfrak{su}(2)^{^*}\ \oplus\ \left(\mathfrak{t}^{^n}\right)^{^*}\right)\ \oplus\   \left(\mathfrak{sl}\left(2, \mathbb{C}\right)^{^*}\ \oplus\ \left(\mathfrak{u}(1)^{^n}\right)_{_{\mathbb{C}}}^{^*}\right),   
\label{eq:muhk1}    
\end{equation}   
in which case the hyperpolygon space is the hyperk$\ddot{\text{a}}$hler quotient   

\begin{equation}   
\mathfrak{M} (\alpha)\ \overset{def.}{=}\ T^{^*}\mathbb{C}^{^n}\dslash_{_{(\alpha,0)}}G\ =\ \left(\mu_{_{\mathbb{R}}}^{^{-1}}(\alpha)\ \cap \ \mu_{_{\mathbb{C}}}^{^{-1}}(0)\right)/G,       
\end{equation}    
which is a smooth, noncompact hyperk$\ddot{\text{a}}$hler manifold of complex dimension $2(n-3)$.

\subsection{Abelianisation}

To calculate the circle-equivariant cohomology of a hyperk$\ddot{\text{a}}$hler space $\mathfrak{M}$, we need the Kirwan map   

\begin{equation}   
\kappa_{_{G}}\ :\ H^{^*}_{_{S^{^1}\times G}}(M)\ \longrightarrow\ H^{^*}_{_{S^{^1}}}(M\dslash\ G) 
\end{equation}

\begin{equation}    
\hat H_{_{S^{^1}}}^{^*}(M\dslash\ G)\ \simeq\ \frac{\hat H^{^*}_{_{S^{^1}}}(M\dslash\ T) ^{^W}}{\text{ann}(e)},    
\end{equation}     
where $T$ is the maximal torus, $W$ is the Weyl group of $G$, and

\begin{equation}    
e\ =\ \prod_{_{\alpha\in A}}\alpha(x-\alpha)\ \in\ (\text{Sym}\ \mathfrak{t}^{^*})^{^W}\ \otimes\ \mathbb{Q}[x]\ \simeq\ H_{_{S^{^1}\times T}}(\text{pt})^{^W}\ \subseteq\ \hat H_{_{S^{^1}\times T}}(\text{pt})^{^W}.      
\end{equation}

The main goal of his work is that of stating and proving an abelianisation theorem for hyperk$\ddot{\text{a}}$hler quotients. This features two main obstacles:

\begin{enumerate}   

\item  Hyperk$\ddot{\text{a}}$hler quotients are rarely compact.

\item  Surjectivity of the Kirwan map from $H^{^{\bullet}}_{_G}(X)$ to $H^{^{\bullet}}(X\dslash\ G)$.  

\end{enumerate}     

The analogous map for circle compact hyperk$\ddot{\text{a}}$hler quotients is conjecturally surjective, but only a few cases are known to be so. They adopt a different method, instead, weaker than surjectivity. But this is true for the case of Nakajima quiver varieties:    

\begin{equation}  
\kappa_{_G}:\ H^{^{\bullet}}_{_{S^{^1}\times G}}(M)\ \rightarrow\ H^{^{\bullet}}_{_{S^{^1}}}(M\dslash G).     
\end{equation}  

Suppose that $M\dslash G$ and $M\dslash T$ are both circle compact. If $\gamma\ \in\ \hat H_{_{S^{^1}\times G}}^{^{\bullet}}(X)$, then integration over the Kirwan map leads to

\begin{equation}  
\int_{_{X\dslash G}}\hat\kappa_{_{G}}(\gamma)\ =\ \frac{1}{|W|}\int_{_{X\dslash T}}\hat\kappa_{_{G}}(\gamma)\ \circ\ r^{^G}_{_T}(\gamma) \cdot e,   
\label{eq:intkm}   
\end{equation}           
where 

\begin{equation}    
r_{_T}^{^G}:\ \hat H^{^{\bullet}}_{_{S^{^1}\times G}}(M)\ \rightarrow\ \hat H_{_{S^{^1}\times T}}(M)^{^W}.  
\end{equation} 

Importantly in this case, the rational Kirwan map is surjective, implying the following chain of isomorphisms  

\begin{equation}    
\hat H^{^{\bullet}}_{_{S^{^1}}}(M\dslash\ G)\ \simeq\ \frac{\hat H^{^{\bullet}}_{_{S^{^1}}}(M\dslash\ T)^{^W}}{\text{ann}(e)}\ \simeq\ \left(\frac{\hat H^{^{\bullet}}_{_{S^{^1}}}(M\dslash\ T)}{\text{ann}(e^{^\prime})}\right)^{^W}   
\end{equation}   
where     

\begin{equation}    
e\ =\ \prod_{_{\alpha\in A}}\alpha(x-\alpha)\ \in\ (\text{Sym}\ \mathfrak{t}^{^*})^{^W}\ \otimes\ \mathbb{Q}[x]\ \simeq\ H_{_{S^{^1}\times T}}(\text{pt})^{^W}\ \subseteq\ \hat H_{_{S^{^1}\times T}}(\text{pt})^{^W}.      
\end{equation}

\begin{equation}    
e^{\prime}\ =\ \prod_{_{\alpha\in \Delta^{^-}}}\alpha\ \prod_{_{\alpha\in \Delta}}(x-\alpha)\ \in\ (\text{Sym}\ \mathfrak{t}^{^*})^{^W}\ \otimes\ \mathbb{Q}[x]\ \simeq\ H_{_{S^{^1}\times T}}(\text{pt})^{^W}\ \subseteq\ \hat H_{_{S^{^1}\times T}}(\text{pt})^{^W}.      
\end{equation}

\begin{equation}  
M\dslash G\ =\ \mu_{_G}^{^{-1}}(\xi, 0)/ \ G\ \ \ , \ \ \ M\dslash T\ =\ \mu_{_T}^{^{-1}}(\text{pr}(\xi),0)/\ T  
\label{eq:hkqs}    
\end{equation}  
and    
\begin{equation}    
\text{pr}:\ \mathfrak{g}^{^*}\ \rightarrow\ \mathfrak{t}^{^*}    
\end{equation}      
is the natural projection. Then $T$ acts on $M$ with hyperk$\ddot{\text{a}}$hler moment map   

\begin{equation}   
\mu_{_T}\ =\ \text{pr}\ \circ\ \mu_{_{\mathbb{R}}}\ \oplus\ \text{pr}_{_{\mathbb{C}}}\ \circ\ \mu_{_{\mathbb{C}}}\ :\ M\ \rightarrow\ \mathfrak{t}^{^*}\ \oplus\ \mathfrak{t}_{_{\mathbb{C}}}^{^*}.   
\end{equation}    

Let $\xi\ \in\ \mathfrak{g}^{^*}$ be central elements such that $(\xi,0)$ is a regular value of $\mu_{_G}$ and $(\text{pr}(\xi),0)$ is a regular value of $\mu_{_T}$. By further assuming that $G$ acts freely on $\mu_{_G}^{^{-1}}(\xi, 0)$, and $T$ acts freely on $\mu_{_T}^{^{-1}}(\text{pr}(\xi), 0)$. Upon taking \eqref{eq:hkqs} to be hyperk$\ddot{\text{a}}$hler quotients of $M$ by $G$ and $T$, respectively, since $\mu_{_G}$ and $\mu_{_T}$ are circle equivariant, then the action of $S^{^1}$ on $M$ descends to the action of $S^{^1}$ on $M$ descends to the action on the hyperk$\ddot{\text{a}}$hler quotients. $M\dslash\ T$ also inherits an action of the Weyl group $W$ of $G$.

\subsubsection{Towards Category Theory}

Understanding the impact of what we have outlined up to now in the context of this project, we need a crucial ingredient, namely that of higher-categorical structures, to which Part II of this work is devoted. As we shall see there, they provide a natural language for describing interesting dualities arising in supersymmetric quantum field theories. In particular, we will be focussing on the category of circle compact manifolds.

\part{Higher-Categories and Dualities}

\section{Higher-categorical structures }  \label{sec:hcandd}

Part II of our treatment introduces two main tools needed in the remainder of our work, namely higher-categorical structures and dualities. Given the breadth of topics either terms encompass, we now pause momentarily to briefly outline what are the types of most interest to us, with the purpose of guiding the reader through the remaining sections of this Part. Importantly, this will build understanding of the main tools needed for addressing questions raised in later sections.

\medskip 

\medskip 

\begin{figure}[ht!]  
\begin{center}   
\includegraphics[scale=1]{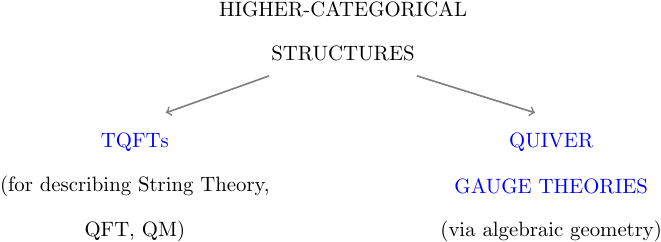}   
\caption{\small Main cases of higher-categorical structures, and tools adopted for describing them. Each one of them will arise in due course in the remainder of our treatment.}   
\label{fig:SMSM}  
\end{center}  
\end{figure}

\medskip 

\medskip

Figure \ref{fig:SMSM} summarises the key examples we are interested in: 

\begin{enumerate}

\item  Higher-categorical structures describing String Theory, QFT, and QM, mostly relying upon functorial field theory (FFT), and algebro-geometric tools.

\item   Holographic dualities (such as the AdS/CFT correspondence, CS/WZW-models, wedge holography, and generalisations thereof, in turn accounting for the emergence of islands), and categorical dualities (3D mirror symmetry, and homological mirror symmetry).  

\end{enumerate}

As explained in the introductory part of this work, understanding the underlining mathematical structure of QFT is one of the core open questions in the realm of physical mathematics. Higher-categorical structures have proved being extremely efficient in achieving this for gauge theories and their supersymmetric counterparts. In particular, this formalism happens to be particularly efficient for keeping track of the spectrum (i.e. the particle-field content) of a given QFT, and how the latter changes under gauging of the global symmetries of the theory one started from.

Up to now, most of the mathematical control in describing higher-dimensional field theories has been for descendants of maximally-supersymmetric 6D ${\cal N}=(2,0)$ superconformal field theories (SCFTs), from which a rich web of dualities arises upon dimensional reduction to, either, 4 or 3 spacetime dimensions. Such descendants feature fewer supersymmetries with respect to their non-Lagrangian parent theory. In particular, the most studied are 4D ${\cal N}=2$ (also known as class ${\cal S}$ theories), and 3D ${\cal N}=4$ SCFTs, exhibiting exactly half of the amount of supersymmetry of their parent 6D theory. In this concluding part of our work, we will be restricting to these, \cite{Witten:1995zh,Witten:2007ct,Moore,Witten:2009at,DBZ}. 

In a later section of this Part, we will be focusing on the applicability of category theory to the specific case of theories descending from dimensional reduction of 6D ${\cal N}=(2,0)$ SCFTs, specifically class ${\cal S}$ theories, which, according to the Alday-Gaiotto-Tachikawa (AGT)-correspondence, can be equivalently described in terms of a 2D TFT living on the Riemann surface with respect to which dimensional reduction has been performed. Because of this, class ${\cal S}$ theories constitute a prototypical example of theories bridging between the two upper nodes of the diagram in figure \ref{fig:SMSM}. This part of our work also introduces TQFTs and varieties arising in algebraic geometry as two main examples of higher-categorical structures.

In section \ref{sec:rqft}, we have gathered sufficient evidence supporting the utility of applying category theory for describing simple examples of QFT models. 
We now switch gears, and turn to a more algebro-geometric analysis of the setup outlined so far. In doing so, we will mostly be relying on \cite{Braverman:2017ofm,Braverman:2016wma,Dimofte:2018abu,Bullimore:2015lsa}, shedding new light and emphasising techniques that should be further developed and applied to Beyond the Standard Model setups\footnote{More details in this regard will appear in an upcoming work by the same author, \cite{VP}.}.

\subsection{Higher-categorical structures }  \label{sec:BMTCs}

For a QFT to be well defined, its spectrum of allowed operators must be compatible with the symmetries of the given theory, \cite{Aharony:2013hda}. This statement has been a very active area of research, and mostly motivated furthering a full mathematical formulation of QFTs. Category theory has been pointed out as a promising candidate in pursuing such aim. 

Following suite, the present work aims at exploring the higher-categorical theory of and descending from 6D ${\cal N}=(2,0)$ SCFTs with the aim of identifying a categorical theory quantity enabling to distinguish between intrinsic and non-intrinsic non-invertible symmetries in class ${\cal S}$ theories obtained by dimensional reduction of their parent theory. 

Before delving into the specifics of the 6D theory, to which section \ref{sec:Freed-Moore-Teleman}    is devoted, the current section aims at providing a brief overview of some essential tools we will need in the remainder of the paper, as well as further motivating the use of such mathematical tools within the following theoretical physics treatment. This section is structured as follows:

\begin{itemize}  

\item Section \ref{sec:BMTCs} provides a brief overview of some key features of higher-categorical structures that are relevant for our analysis. In the first part of this section, we briefly recall the relation between the total quantum dimension, fusion and braiding structures.

\item As an example of a higher-categorical structure, we first describe the Chern-Simons theory for a flux-charge-quasiparticle setup, and how gauging a subgroup of the anyonic symmetry affects the total quantum dimension. The section concludes with an explanation for how to gauge a higher-categorical structure by means of algebraic gauging, \cite{TJF}.  This is really meant to be a warm-up example to the follow-up example of interest to us, namely string theory. For the purpose of our work, its functorial field theory formulation turns out being particularly significant. As we shall see, our analysis extends that of Moore and Segal, \cite{Moore:2006dw}, and Moore and Tachikawa, \cite{Moore:2011ee}.

\end{itemize}

\subsubsection{Defect fusion categories}   \label{sec:dfc}

\emph{Categories} are defined as sets of objects related to each other by morphisms. A $2$-category is a collection of categories related to each other by \emph{functors}. Groups constitute a special type of categories, where only one element is present, and every morphism is an isomorphism. 

$(n,r)$-categories are generalisations of, both, $n$-categories and $n$-grupoids. As $n$ increases, there are many possibilities, until there are infinitely many of $(\infty,r)$-categories. An $(\infty,1)$-category is am $\infty$-category in which all $n$0morphisms for $n\ge 2$ are equivalences. This is the joint generalisation of the notion of category and $\infty$-grupoid.  

Among all $(n,r)$-categories, $(\infty,1)$-categories are special in that they are the simplest structures that at the same time admit a higher version of category theory, and encode information about higher-equivalences. In general, to understand $(n,r)$-categories, one requires the notion of the underlying $(n,1)$-, hence, the $(\infty,1)$-category, which already captures most of the information of interest.

A \emph{higher}-category, or $n$-category, is an $(\infty,1)$-category enriched with $(n-1)$-categories, with $n$ denoting the dimensionality.

A category, ${\cal C}$, is \emph{monoidal} (MC) if it comes equipped with a bi-functor   

\begin{equation}    
\otimes\ :\ {\cal C}\times{\cal C}\ \rightarrow\ {\cal C}.
\label{eq:mult}    
\end{equation}  

For ${\cal C}$ semisimple, the tensor product of any two simple objects $x,y\in{\cal C}$ is 

\begin{equation}  
x\ \otimes\ y\ \simeq\ \bigoplus_{z\in\text{Irr}({\cal C})}\ N_{_{xy}}^{^{z}}\cdot z     
\ \ \ \ 
,\ \ \ \ N_{_{xy}}^{^{z}}\in\mathbb{N},
\end{equation}  
where $\text{Irr}({\cal C})$ denotes the irreducible representations, $\left\{N_{_{xy}}^{^{z}}\right\}_{_{x,y,z\in\text{Irr}({\cal C})}}$ denote the \emph{fusion rules} of ${\cal C}$. Equivalently, the isomorphism classes of simple objects ${\cal C}$ generate the \emph{fusion ring} or \emph{Grothendieck ring} of ${\cal C}$, where multiplication is given by \eqref{eq:mult}, and $N_{_{xy}}^{^{z}}$ as structure constants.

Given $x, y\in\ {\cal C}$ 

\begin{equation}   
\text{Hom}:\ x,y\ \ \rightarrow\ \text{Hom}(x,y)   
\end{equation}  
is an $(n-1)$-category, and for $\mathbf{1}\in{\cal C}$  

\begin{equation}   
\text{Hom}:\ \mathbf{1}, \mathbf{1}\ \ \rightarrow\ \text{Hom}(\mathbf{1}\ ,\mathbf{1}) \ \overset{def.}{=}\ \Omega\ {\cal C},        
\end{equation}  
$\Omega\ {\cal C}$ is a monoidal $(n-1)$-category. More generally, an $(n,r)$-category is a higher-category such that:   

\begin{enumerate}   

\item All $k$-morphisms for $k>n$ are trivial.  

\item All $k$-morphisms for $k>r$ are invertible.  

\end{enumerate}

\medskip  

\medskip

\subsection*{Braiding and ribbon structure}

\medskip  

\medskip 

A modular category (MC) is \emph{braided} if, in addition to \eqref{eq:mult} it is also equipped with a natural isomorphism 

\begin{equation}  
c_{_{x,y}}: x\otimes y\ \rightarrow\ y\otimes x,  
\label{eq:braiding}   
\end{equation} 
called \emph{braiding}. Given two simple objects $x,y\in{\cal C}$, the \emph{S-matrix} is defined as

\begin{equation} 
\boxed{\ \ S\ \overset{def.}{=}\ \left(S_{_{xy}}\right)_{_{z\in\text{Irr}({\cal C})}},  \color{white}\bigg]\ \ }     
\label{eq:SM1}
\end{equation} 
with components defined from \eqref{eq:braiding}

\begin{equation} 
S{_{xy}}\ \overset{def.}{=} \ \text{Tr}\ \left(c_{_{y,x^{*}}}\ \circ\ c_{_{x^{*},y}}\right)\ \equiv\ \text{Tr}\ \left(c_{_{y^{*},x}}\ \circ\ c_{_{x,y^{*}}}\right).
\label{eq:SM}
\end{equation}

\begin{figure}[ht!]  
\begin{center}   
\includegraphics[scale=1]{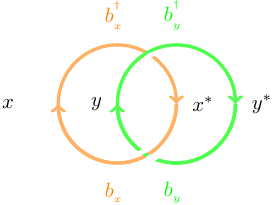}  
\ \ \ \ \ \ \ \ \ \ \ \ \ \ \ \ \ \ \ \ 
\includegraphics[scale=1]{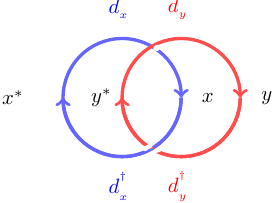}  
\caption{\small The S-matrix corresponds to the Hopf link, and encodes the information of mutual-statistics. As shown in these pictures, the linking between particle worldlines encodes information of mutual-statistics.}   
\label{fig:SM}  
\end{center}  
\end{figure}

\begin{figure}[ht!]  
\begin{center}   
\includegraphics[scale=1]{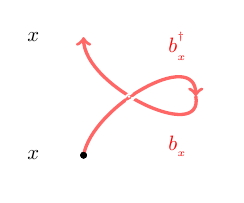}  
\ \ \ \ \ \ \ \ \ \ \ \ \ \ \ \ \ \ \ \ 
\includegraphics[scale=1]{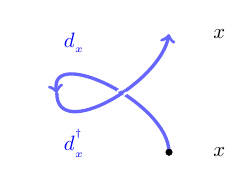}  
\caption{\small The ribbon structure describing self-statistics is here drawn as the worldline of a particle excitation $x$ winding around itself. Such information is encoded in the T-matrix.}   
\label{fig:ribbonT}  
\end{center}  
\end{figure}

For ${\cal C}$ a braided fusion category and $x\in {\cal C}$, the 2 morphisms

\begin{equation}  
x\ \xrightarrow{\text{id}_{_{x}}\otimes b_{_{x}}}\ x\otimes x\otimes x^*\ \xrightarrow{c_{_{x,x}}\otimes \text{id}_{_{x^*}}}\ x\otimes x\otimes x^*\xrightarrow{\text{id}_{_{x}}\otimes b_{_{x}}^{\dag}}\ x  
\end{equation}    

\begin{equation}  
x\ \xrightarrow{d_{_{x}}^{\dag}\otimes\text{id}_{_{x}}}\ x^*\otimes x\otimes x\ \xrightarrow{\text{id}_{_{x^*}}\otimes c_{_{x,x}}}\ x^*\otimes x\otimes x\xrightarrow{b_{_{x}}\otimes\text{id}_{_{x}}  }\ x  
\end{equation}  
are both equivalent to 

\begin{equation}  
\theta_{_{x}}:\ x\rightarrow x,  
\end{equation}  
defining the \emph{twist} or \emph{topological spin of x}. Importantly, $\theta_{_{x}}\neq\text{id}_{_{x}}$, since the worldline should be viewed as a ribbon, hence a line equipped with a framing. Both processes described in figure \ref{fig:ribbonT} are therefore equivalent to twisting the ribbon counterclockwise. The twisting rotates the topological defect by 360$^{\circ}$, reason why $\theta_{_{x}}$ is called the topological spin of x.

$\forall x\in {\cal C}$, since Hom$_{_{{\cal C}}}(x,x)\simeq\mathbb{C}$, 

\begin{equation}  
\theta_{_{x}}\equiv T_{_{x}}\cdot \text{id}_{_{x}}  
\ \ \ 
, \ \ \ 
T_{_{x}}\in\mathbb{C}, 
\end{equation}  
where the $T$-matrix of the unitary braided fusion category is defined as

\begin{equation}  
T\overset{def.}{=}
\left(T_{_{x}}\delta_{_{xy}}\right)_{_{x,y\in\text{Irr}({\cal C})}},
\label{eq:T}
\end{equation}  
encoding the information of self-statistics.

\subsection*{Quantum double category of a finite group }

$\forall$ finite group $G$, there is a unitary MTC ${\cal D}_{_{G}}$ called \emph{double quantum category of G}, \cite{Kong:2022cpy}, such that:  

\begin{enumerate}  

\item An object in ${\cal D}_{_{G}}$ is a vector space $V$ equipped with a $G$-action 

\begin{equation}   
\rho:\ G\ \longrightarrow\ GL(V),      
\end{equation}  
and a $G-grading$ $V\equiv \underset{g\in G}{\bigoplus} V_{_{g}}$ such that 

\begin{equation}  
\rho(g): V_{_{h}}\ \longrightarrow\ V_{_{ghg^{^{-1}}}}\ \ \ ,\ \ \ \forall g,h\ \in\ G.  
\end{equation}  

\item A morphism in ${\cal D}_{_{G}}$ is a $\mathbb{C}$-linear map that is, both, a morphism in Rep$(G)$ and a morphism in Vec$(G)$. 

\item  Given $g\in G$, we denote its conjugacy class as $[g]$, and its centraliser as $Z(g)$. If $\pi$ is an irreducible $Z(g)$-representation, the induced representation Ind$_{_{Z(g)}}^{^{G}}(\pi)\overset{def.}{=} X_{_{(g,\pi)}}$ is a $G$-representation admitting a canonical $G/Z(g)\simeq[g]$-grading. $X_{_{(g,\pi)}}$ is a simple object in ${\cal D}_{_{G}}$.  

\item The isomorphism classes of objects in ${\cal D}_{_{G}}$ are labelled by pairs $([g], [\rho])$ where $[g]$ is the conjugacy class in $G$ and $[\rho]$ is an isomorphism class of irreducible representations of $Z(g)$, where $Z[g]$ only depends on the conjugacy class of $g$.

\end{enumerate}

The properties of the $S$ and $T$ matrices defined in \eqref{eq:SM1} and \eqref{eq:T}, and their generalisations to the case of quantum double categories have been extensively described in previous works (see, for example, \cite{Kong:2022cpy}). For the purpose of the present work, it is important to stress that the $S$ and $T$ matrices are related to the generators of the $SL(2, \mathbb{Z})$ group 

\begin{equation} 
s\overset{def.}{=}\left(\begin{matrix}0\ \ -1\\
1\ \ \ \ 0\\ \end{matrix}\right)\ \ \ ,\ \ \ t\overset{def.}{=}\left(\begin{matrix}1\ \ \ 1\\
0\ \ \ 1\\ \end{matrix}\right),   
\end{equation}  
such that 

\begin{equation} 
(st)^{^3}\ \equiv\ s^{^2}\ \ \ ,\ \ \ s^{^4}\equiv 1.  
\end{equation}   
under a projective $SL(2, \mathbb{Z})$-representation

\begin{equation}  
s\ \mapsto\ \frac{S}{\ \sqrt{\text{dim}({\cal C})\ } } \ \ \ ,\ \ \ t\ \mapsto\ \ T.  
\label{eq:proj}   
\end{equation}  

The quantity ${\cal D}\overset{def.}{=}$dim$({\cal C})$ is the \emph{total quantum dimension} of the category ${\cal C}$. For the case in which ${\cal C}$ is made up of purely simple objects, the total quantum dimension of a given superselection sector simply reads  

\begin{equation}  
{\cal D}_{_{x}}\overset{def.}{=}\sqrt{\ \sum_{i}\ \text{dim}^{^2}(x_{_{i}})\ } =\sqrt {\ \sum_{i}\ S_{_{\mathbf{1}x_{i}}}^{^2}\ },
\end{equation}   
where $x_{_i}$ are the elements of the category and $S_{_{\mathbf{1}x_{i}}}^{^2}$ the square of the $S$-matrix evaluated between the identity element, $\mathbf{1}$, and any other non-trivial element, $x_{_i}$.

 \subsubsection{Example of higher-categorical structures}    \label{sec:excatstr}

\section*{CS theory}
  
We now turn to describing the main example of MTC of interest to us. We will assume that the total category associated to a theory ${\mathfrak T}$ with gauge group $G$, ${\cal C}_{_{\mathfrak T}}$ is given by

\begin{equation}   
{\cal C}_{_{\mathfrak T}}\ \overset{def.}{=}\{{\cal C}_{_{point}}, {\cal C}_{_{flux}}, {\cal C}_{_{Hopf}}\ \},       
\label{eq:supersel}     
\end{equation}  
where each subcategory is a different superseclection sector of the theory, defined as follows:

\begin{enumerate}   

\item   ${\cal C}_{_{point}}\ \equiv \ \{1,a,...\}$, $a\equiv R\in\ (G)_{_{IRREP}}$, are point excitations, corresponding to the irreducible representations of the group $G$. The quantum dimension of each element of the category is $d_{_{a}}\ \overset{def.}{=}\ \text{dim}\ R$.

\item ${\cal C}_{_{flux}}\ \equiv\   \{1,\mu,...\}$ is the superselection sector of pure fluxes, and therefore corresponds to the conjugacy classes of the finite group $G$, $C\in\ (G)_{_{cj}}$. Their quantum dimensions are $d_{_{\mu}}\ \overset{def.}{=}\ \sqrt{|C|\ }$, with $C$ denoting the number of group elements in a given conjugacy class.

\item  ${\cal C}_{_{Hopf}}\ \equiv \ \{1,\eta,...\}$ is the collection of Hopf link excitations, with $\eta\overset{def.}{=}\ \left(C_{_{(g,h)}}, R\right), gh\equiv hg, C_{_{(g,h)}} \equiv\ \{(tgt^{-1}, tht^{-1}|t\in G\}, R\in\ E_{_{(g,h)}}\ \equiv\ \{t\in G\ |(g,h)\equiv \left(tgt^{-1}, tht^{-1}\right)\}$, and quantum dimensions 
\begin{equation}   
d_{_{\eta}}\ \overset{def.}{=}\ \frac{|G|}{\ |E_{_{(g,h)}}|\ }\ \text{dim}\ R   
\end{equation}   

\end{enumerate}

The 1-morphisms between these categories are

\begin{equation}  
\varphi\ :\ {\cal C}_{_{flux}}\ \hookrightarrow\ {\cal C}_{_{loop}}\ \subset\ {\cal C}_{_{Hopf}}  
\label{eq:varphi}
\end{equation}

\begin{equation}  
\phi\ :\ {\cal C}_{_{point}}\ \hookrightarrow\ {\cal C}_{_{loop}}\ \subset\ {\cal C}_{_{Hopf}}
\label{eq:phi}
\end{equation}

\begin{equation} 
v\ :\ {\cal C}_{_{Hopf}}\ \xrightarrow{\color{white}aaaa\color{black}}\ {\cal C}_{_{Hopf}}   \ \ \ , \ \ \ v(\eta)\ \equiv \eta^{^{\text{V}}}   
\label{eq:v}
\end{equation}

The aforementioned properties suggest the physical interpretation of the superselection sectors in \eqref{eq:supersel} as the electric, magnetic and dyonic charges characterising the spectrum of the theory $\mathfrak{T}$. As a result, the objects living in ${\cal C}_{_{\mathfrak T}}$ are defined by a 3-tuple $\chi\equiv([M],\lambda,\rho)$, corresponding to a flux-quasiparticle-charge composite.

As mentioned earlier on in the introduction, application of categories in physics began in the study of 2D CFTs and $D+1$ TQFTs. Their mathematical formulation was originally motivated by the idea of formalising the factorisation property of path-integrals as a certain monoidal functor defined on a cobordism category. Thanks to the pioneering work of Moore and Seiberg on \emph{modular tensor categories} in 2D RCFTs, Rashetikhin and Turaeev reformulated their discovery in present-day categorical formulation, and used it to give the 2+1 D RT TQFT formulation, eventually leading to the study of topological excitations (or anyons) in a 2D topological order. Anyons are quasiparticles, and as such can be thought of as being the physical counterpart of \eqref{eq:supersel}. Systems exhibiting anyon symmetries with gauge group $G$ in a given topological phase efficiently be related to Chern-Simons (CS) theory.

For such correspondence to hold, the CS theory needs to account for the topological information defined by the braiding statistics and quasiparticle fusion rules (cf. section \ref{sec:dfc}). For example, an abelian phase in (2+1)D can be characterised by a QFT with partition function defined as follows

\begin{equation}  
{\cal Z}[{\cal J}]\ \equiv\ \int\left[{\cal D}\alpha(\overset{\rightarrow}{r})\right]\ \exp\left(iS[{\cal J}] \right), 
\end{equation}  
involving an $N$-component set of $U(1)$ gauge fields $\alpha\equiv(\alpha_{_{1}},...,\alpha_{_{N}})$ with action

\begin{equation}  
S[{\cal J}] \equiv \frac{1}{4\pi}\ \int\left(K_{_{IJ}}\alpha^{^{I}}\wedge d\alpha^{^{J}}+\alpha^{^{I}}\ {\cal J}_{_{I}}\right).  
\label{eq:action}    
\end{equation}  

Quasiparticles, $\psi^{^{\overset{\rightarrow}{a}}}$, are sources for ${\cal J}_{_{1}}^{^{a_{_{1}}}},...,{\cal J}_{_{N}}^{^{a_{_{N}}}}$, labelled by $\overset{\rightarrow}{a} \equiv(a_{_{1}},...a_{_{N}})$ on a lattice $\Gamma^*\equiv\mathbb{Z}^{^N}$. At long distance, nearby quasiparticles can form a single entity, leading to the definition of a fusion structure. The $K$-matrix in \eqref{eq:action} dictates the braiding statistics of quasi-particles  

\begin{equation}
{\cal D}\ S_{_{ab}}\ \equiv\ e^{^{2\pi i a^{^T}K^{-1}b}}  
\ \ \ 
,   
\ \ \ 
{\cal D}\equiv\sqrt{|\text{det}(K)|\ }  \equiv\sqrt{{\cal A}\ }  
\end{equation} 
with ${\cal D}$ the total quantum dimension. The exchange statistics, instead, is defined by 

\begin{equation}  
\theta_{_{a}}\ \equiv\ e^{^{2\pi i a^{^T}K^{-1}b}},  
\end{equation} 
corresponding to the spin of the quasi particle. 
The topological phase generated by gauging the anyonic symmetry, is referred to as \emph{twisted liquid}. The latter are generalisations of (2+1)D discrete gauge theories. Their quasiparticles are compositions of fluxes and charges associated to the gauged anyonic symmetry, as well as a superselection sector of the original topological state. Their total quantum dimensions are related as follows 

\begin{equation}  
{\cal D}_{_{TL}}\ \equiv\ {\cal D}_{_{o}} \ |G|,  
\label{eq:DTL}   
\end{equation}  
with $|G|$ the order of the anyonic symmetry group $G$ being gauged, and ${\cal D}_{_o}$ the quantum dimension of the pre-gauged theory. 

An important remark is of order, though. A CS description of a topological phase (cf. \eqref{eq:action}) is not unique. This follows from the fact that the $K$ matrix encodes the same fusion and braiding structure even after undergoing a basis transformation. The set of automorphisms, Aut$(K)$, preserving $K$ correspond an anyonic symmetry operation permuting the anyons with the same fusion properties and spin-statistics. As such, it classifies the global symmetries of the TQFT associated to the CS action, $S_{_{CS}}$. Once having modded-out the trivial relabellings, Inn$(K)$, we are left with the outer automorphisms 

\begin{equation}  
\text{Outer}(K)\ \overset{def.}{=}\ \frac{\text{Aut}(K)}{\text{Inner}(K)}, 
\end{equation}  
which will play a key role in the remainder of the present work.

These results about the total quantum dimension in gauged topological orders have been thoroughly investigated in recent years. In the next section, we shall see how the presence of intrinsic non-invertibles arising in the gauge theory lead to a different expression for the total quantum dimension, thereby signalling that the multiplicity of some superselection sectors can be greater that unity.

\subsection{String Theory and Quantum Mechanics (QM) as Functorial Field Theories (FFTs)}

As previously mentioned, the previous section is really meant to be a warm-up example to the follow-up example of interest to us, namely string theory. For the purpose of our work, its functorial field theory formulation turns out being particularly significant. As we shall see, our analysis extends that of Moore and Segal, \cite{Moore:2006dw}, and Moore and Tachikawa, \cite{Moore:2011ee}.

String Theory should be interpreted as a functor

\begin{equation} 
{\cal F}:\ \text{Cob}_{_2}\ \rightarrow\ \text{Vect},  
\end{equation}
where vector spaces describe the quantum states of strings and the operators describe how the possible interaction processes that can happen between strings affect the quantum state. Cob$_{_2}$ denotes the geometry of spacetime, whereas Vect denotes the linear algebra of quantum physics. 

\begin{figure}[ht!]  
\begin{center}
\includegraphics[scale=0.6]{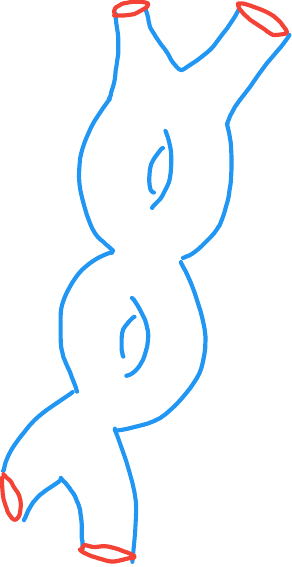}  \ \ \ \ \ \ \ \ \ \ \ \ \ \ \ \ 
\includegraphics[scale=0.8]{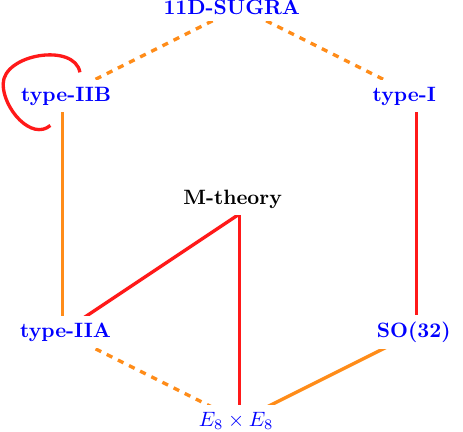} 
\caption{\small String Theory as cobordism functor (on the LHS). On the RHS, instead, the different string theory sectors emerging as, either, low-energy limits (for 11D SUGRA) or from different compactifications of M-theory (R.H.S.). The different sectors differ by means of the p-forms that can be defined in the 10 or 11D spacetimes, thereby constraining the types of objects to which they couple, i.e. Dp-branes, which can really be thought of as the string theory analogues of the Wilson and 't Hooft operators, Some sectors are related by $S$ and $T$-duality, denoted by the red and orange continuous lines, respectively. }  
\label{fig:FNQV}  
\end{center}  
\end{figure}

The functorial approach can similarly be extended to QFT (with QM as a particular case). The most subtle part being that of completely identifying the exact structure of, both, the source and target category.

Usually, QM is described by: 

\begin{enumerate} 

\item  A set, $V$, of possible quantum states of a particle (described by vectors in a complex inner product space).

\item  A set, $U_{_t}$, of operators on $V$, one for each $t \in  \mathbf{R}_{_{\ge 0}}$, equipped with an identity element, $U_{_o}\equiv I$, and group composition law

\begin{equation}  
U_{_{t+s}}=U_{_t}\ \circ\ U_{_s},  
\end{equation}
with $U_{_t}$ denoting the evaluation of a state from time 0 to time $t$.

\end{enumerate}

\subsubsection{Calabi-Yau (CY) categories}  \label{sec:CY}

Calabi Yau categories were originally studied as an abstract version of the derived category of coherent sheaves on a CY manifold, \cite{STY,PC1,PC2}.  We now provide a short digression explaining this terminology. 

An algebraic scheme is \emph{locally noetherian} if it can be covered by a family $U_{_{\alpha}}$ of open subsets of the form  

\begin{equation}  
U_{_{\alpha}}    
\ =\ 
\text{Spec}\ R_{_{\alpha}},     
\end{equation}  
where $R_{_{\alpha}}$ are noetherian rings. 

When $X$ is a noetherian scheme, one has the full \emph{abelian subcategory of coherent sheaves}, corresponding to locally finitely-generated modules, \cite{dccs},

\begin{equation}  
\text{Coh}\left(O_{_X}\right)\ \subset\ Q\text{Coh}\left(O_{_X}\right).    
\end{equation}

One can define from these the triangulated subcategory

\begin{equation}  
D\left(\text{Coh}\left(O_{_X}\right)\right)\ \subset\ D\left(Q\text{Coh}\left(O_{_X}\right)\right).   
\end{equation} 

Denoting with $D^{^b}\left(\text{Coh}\left(O_{_X}\right)\right)$ the bounded derived category of Coh$\left(O_{_X}\right)$, the fully-faithful functor

\begin{equation}  
D^{^b}\left(\text{Coh}\left(O_{_X}\right)\right)\ \hookrightarrow\ D\left(\text{Mod}\left(O_{_X}\right)\right),   
\end{equation}
identifies $D^{^b}\left(\text{Coh}\left(O_{_X}\right)\right)$ with the full subcategory $D\left(\text{Mod}\left(O_{_X}\right)\right)$ of borderedcomplexes whose cohomology objects are coherent sheaves.

A CY category is defined as a Vect-enriched\footnote{Given a monoidal category, $K$, i.e. one such that 

\begin{equation}   
\circ:\ \text{Hom}(y,z)\ \otimes\ \text{Hom}(x,y)\ \rightarrow\ \text{Hom}(x,z)  
\end{equation}  
an \emph{enriched category}, ${\cal C}$ over $K$ features a collection of objects ob(${\cal C}$), such that, for any pair $x,y\in$ ob(${\cal C}$), Hom($x,y$)$\in K$.} category ${\cal C}$ equipped for each object $c\in{\cal C}$ with a trace-like map  

\begin{equation}  
\text{Tr}_{_{{\cal C}}}:\ {\cal C}(c,c) \rightarrow\ k  
\end{equation}  
to the ground field, $k$, such that, for any $d\in{\cal C}$, the induced pairing  

\begin{equation}  
<-,->_{_{c,d}}:\ {\cal C}(c,d)\ \otimes\ {\cal C}(d,c)\ \rightarrow\ k  
\end{equation}
given by  

\begin{equation}  
<f,g>\ =\ \text{Tr}\ (g\circ f)  
\end{equation}  
is symmetric and non-degenerate.   

Examples of CY-categories are:    

\begin{enumerate}   

\item   From CY varieties, \cite{STY,PC1,PC2}.

\item  Fukaya categories associated with a symplectic manifold, $X$, \cite{hms6}.  

\end{enumerate}  

We will be coming back to these in section \ref{sec:hms}. For the moment, we will simply say that the latter have been used to classify topological conformal field theories (TCFTs), originally arising as an abstract version of CFTs constructed from sigma-models whose tagrets are CY-manifolds.

\section{Main Example: Cochain level theories} \label{sec:clt}   

We now start putting to practice the formalism outlined in section \ref{sec:hcandd} to our main example, namely the algebraic varieties introduced in section \ref{sec:rqft}. The reason for doing so is to explain the relation in between them and Moore-Tachikawa varieties. As noted by the authors themselves, \cite{Moore:2011ee}, though, their treatment calls for a categorical description of the hyperk$\ddot{\text{a}}$hler generalisation of the target category. We anticipate that the answer to such generalisation can be achieved by making use of equivariant cohomology as in the approach of Teleman, \cite{Teleman:2014jaa}. In Part III we will be connecting this setup to the emergence of peculiar higher-categorical structures in classs ${\cal S}$ theories and their 2D TFT duals.

The present section is structured as follows:  

\begin{enumerate}  

\item  At first, we overview some key features of equivariant cohomology and the convolution product, both within the context of quiver gauge theories, \cite{Braverman:2017ofm,Braverman:2016wma}

\item Then, we introduce Moore-Tachikawa varieties and their 2D TFT counterpart\footnote{For completeness, we provide some further details in appendix \ref{sec:mt}.}, connecting categorical structures and quiver varieties, \cite{Moore:2011ee}. Making use of \cite{Dimofte:2018abu}, we emphasise how and when 3D mirror symmetry manifests from a categorical perspective, specifically in terms of bordisms, making use of the Kostant-Whittaker symplectic reduction, \cite{Braverman:2017ofm}.

\item  Last but not least, we close the section highlighting the importance of abelianisation for embedding the ring homology of the theory in question. In Part III, we will see how this is in turn related to (un)gauging categorical SymTFTs.

\end{enumerate}  

\begin{figure}[ht!]  
\begin{center}  
\includegraphics[scale=1]{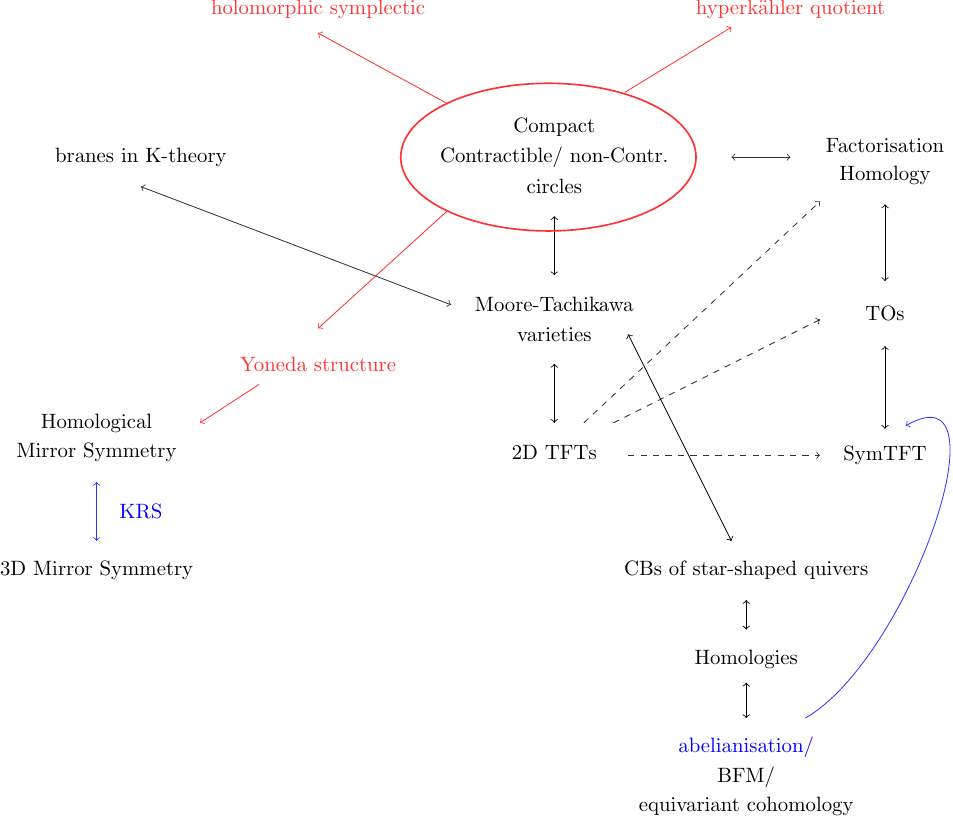}  
\caption{\small This diagram shows how the main example introduced in section \ref{sec:main} fits within the 2-categorical description of Moore-Tachikawa varieties, together with their generalisation, and the gauging prescription within the SymTFT setup.}   
\label{fig:cm}  
\end{center}  
\end{figure}

\subsection{Cochain complexes}

A \emph{cochain complex}, $(A^{^{\tiny\bullet}}, d^{^{\tiny\bullet}})$, is  an algebraic structure that consists of a sequence of abelian groups (or modules), $A^{^{\tiny\bullet}}$, and a sequence of homomorphisms between consecutive groups, $d^{^{\tiny\bullet}}$, such that the image of each homomorphism is included in the kernel of the next. To a chain complex, $(A_{_{\tiny\bullet}}, d_{_{\tiny\bullet}})$, there is an associated homology, which describes how the images are included in the kernels. A cochain complex is similar to a chain complex, except that its homomorphisms are in the opposite direction. The homology of a cochain complex is called its \emph{cohomology}

\begin{equation}  
H({\cal C})\ \overset{def.}{=}\ \text{Ker}({\cal Q})/ \ \text{Im}({\cal Q}).
\end{equation}  

The n$^{th}$ homology group, $H_{_n}(H^{^0})$ is 

\begin{equation}  
H_{_n}\ \overset{def.}{=}\ \text{Ker}\ d_{_n}/ \ \text{Im}\ d_{_{n+1}}.
\end{equation} 

The central object in closed string theory is the vector space ${\cal C}\equiv{\cal C}_{_{S^{^1}}}$ of states of a single parametrised string. ${\cal C}$ denotes the cochain complex in this case, \cite{Moore:2006dw}. The latter comes equipped with a grading given by the ghost number, and an operator ${\cal Q}:\ {\cal C}\ \rightarrow\ {\cal C}$ called the BRST operator, raising the ghost number by 1, and such that ${\cal Q}^{^2}\equiv 0$.

\subsection{Moore-Segal bordism operators}

\subsubsection{Cochain level theories}   \label{sec:Moore-Segal}  

Cochain level theories constitute the most important generalisation of the open and closed TFT construction, reason why we will be devoting the present section to briefly overviewing some of its key features that are mostly needed for our treatment. The reason for doing so is that these mathematical structures enable to determine the set of possible D-branes given a closed string background, \cite{Moore:2006dw}. In their work, \cite{Moore:2006dw} address this problem from the point of view of a 2D TFT, \eqref{eq:etaGC11}, where the whole content of the theory is encoded in a finite-dimensional commutative Frobenius algebra. 






The present section is therefore structured as follows:   

\begin{enumerate}  

\item At first, we briefly overview cochain complexes as the essential mathematical tools needed for translating the setup of our previous work, \cite{Pasquarella:2023deo}, in the formalism of Moore and Segal. 

\item We then turn to highlighting the construction of cobordism operators, \cite{Moore:2006dw}, emphasising its dependence on the conformal structure of the Riemann surface.

\end{enumerate}

\subsection{The Moore-Segal setup}   

The most general finite-dimensional commutative algebra over the complex numbers is of the form

\begin{equation}  
{\cal C}\ \overset{def.}{=}\ \underset{x}{\bigoplus}\ {\cal C}_{_{x}}  \ \ \ , \ \ \ x\in\ \text{Spec}\ ({\cal C}),  
\end{equation}
with 

\begin{equation}  
{\cal C}_{_{x}}\ \overset{def.}{=}\ \mathbb{C}\ {\cal E}_{_{x}}\ \oplus\ m_{_{x}},    
\end{equation}  
where ${\cal E}_{_{x}}$ is an idempotent, and $m_{_{x}}$ a nilpotent ideal. If ${\cal C}$ is a Frobenius algebra, then so too is each ${\cal C}_{_{x}}$.

In their treatment, \cite{Moore:2006dw} restrict to the semisimple\footnote{Despite appearing quite restrictive, committing to semisimplicity is enough to shed light on the essential structure of the theory. According to \cite{Moore:2006dw}, to go beyond it, the appropriate objects of study, are cochain-complex valued TFTs rather than non-semisimple TFTs in the usual sense.}  case. Semisimplicity admits many equivalent definitions:

\begin{enumerate}   

\item  The presence of simultaneously-diagonalisable fusion rules.   

\item  There exists a set of basic idempotents ${\cal E}_{_{x}}$ such that 
\begin{equation}  
{\cal C}\ \overset{def.}{=}\ \underset{x}{\bigoplus}\ \mathbb{C}\ {\cal E}_{_{x}}  \ \ \ , \ \ \ x\in\ \text{Spec}\ ({\cal C}),\ \ \ \text{with}\ \ \ \ \   
{\cal E}_{_{x}}{\cal E}_{_{y}}\ \equiv\ \delta_{_{xy}} {\cal E}_{_{y}}.   
\end{equation}

\item   ${\cal C}$ is the algebra of complex-valued functions on the finite set of characters of ${\cal C}$, $X\in\ \text{Spec}\ ({\cal C})$.   

\end{enumerate}

For any pair of boundary conditions, $a,b$, the corresponding cochain complex for a semisimple category is defined as follows, \cite{Moore:2006dw},

\begin{equation}  
{\cal O}_{_{aa}}\ \simeq\ \underset{x}{\bigoplus}\ \text{End}\left(W_{_{x,a}}\right),     
\end{equation}

\begin{equation}  
{\cal O}_{_{ab}}\ \simeq\ \underset{x}{\bigoplus}\ \text{Hom}\left(W_{_{x,a}}; W_{_{x,b}}\right),      
\end{equation}   
where $W_{_{x,a}}$ is a vector space associated to every idempotent ${\cal E}_{_{x}}$.

\subsection*{Cobordism operators}   

We now turn to the key elements for our analysis, namely cobordism operators. In doing so, we will be using the definition provided by \cite{Moore:2006dw}, highlighting the crucial property that will be mostly needed in sections \ref{sec:333} and \ref{sec:last}. A cobordism $\Sigma$ from $p$ circles to $q$ circles gives an operator   

\begin{equation}   
U_{_{\Sigma, \alpha}}:\ {\cal C}^{^{\otimes p}}\ \rightarrow\ {\cal C}^{^{\otimes q}},  
\label{eq:U}  
\end{equation}   
which depends on the conformal structure $\alpha$ on $\Sigma$. This operator, \eqref{eq:U}, is a cochain map, but its crucial feature is that, changing the conformal structure $\alpha$ on $\Sigma$, changes $U_{_{\Sigma, \alpha}}$ only by a cochain homotopy.   

To describe how $U_{_{\Sigma, \alpha}}$ varies with $\alpha$, if ${\cal M}_{_{\Sigma}}$ is the moduli space of conformal structures on the cobordism $\Sigma$ which are the identity on the boundary circles, there is a resulting cochain map    

\begin{equation}  
U_{_{\Sigma}}:\ {\cal C}^{^{\otimes p}}\ \rightarrow\ \Omega\left({\cal M}_{_{\Sigma}}; {\cal C}^{^{\otimes q}}\right),     
\end{equation}  
with the target denoting the de Rham complex of forms on ${\cal M}_{_{\Sigma}}$ with values in ${\cal C}^{^{\otimes q}}$. 

An alternative, totally equivalent, definition is that of the following cochain map

\begin{equation}  
U_{_{\Sigma}}:\ C_{_{_{_{\tiny\bullet}}}}\ \left( {\cal M}_{_{\Sigma}}\right)\ \rightarrow\ \left({\cal C}^{^{\otimes p}}\right)^{^*}\ \otimes\ {\cal C}^{^{\otimes q}}.   
\end{equation} 

In the next sections, we will show that lack of reparametrisation-invariance on the Riemann surface implies interesting mathematical and physical features of the resulting theory of interest.

\section*{Key points}    

The main points to keep in mind throughout the reminder of our treatment are the following:  

\begin{itemize}  

\item Cochain level theories provide the natural mathematical formalism for describing absolute theories obtained by partial gaugings of the SymTFT in the Freed-Moore-Teleman setup. 

\item The definition of cobordism operators associated to such complex cochain structure follows from the assumption that the Riemann surface is reparametrisation-invariant. 

\end{itemize}

\section*{Quivers}

Equipped with the tools revised in the previous sections, we are now ready to begin the analysis outlined in the roadmap illustrated in figure \ref{fig:roadmap}. The top entry of our diagram reads an example of a 2-category, namely that of Moore-Tachikawa varieties\footnote{To facilitate the reader, we have added an appendix with further details to be read, albeit we highly recommend to go back to the original paper \cite{Moore:2011ee} for a more thorough overview.}. The reason why we are referring to them as our main example is their relation to Coulomb branches of star-shaped quivers, \cite{Dimofte:2018abu}, to which we can apply the analysis of section \ref{sec:rqft}. The setups we are dealing with are highly supersymmetric, thereby describing theories that are very far from the SM of particle physics, even in its minimally supersymmetric realisation (MSSM). However, the purpose of this article is that of highlighting features in categorical structures of supersymmetric gauge theories that could come to aid in furthering the understanding of extensions of the SM to possible completions in unifying theories, in turn descending from a superymmetric parent setting, for the reasons outlined in section \ref{sec:X}. Furthermore, it is important to remark that, the quiver description of the SM and its extensions is an essential step towards embedding them in any string theory setup, and we refer the reader to some key works on this regard, such as \cite{Cicoli:2021dhg, Uranga}.

\subsection{Moore-Tachikawa varieties}   \label{sec:mt}

Having outlined the importance of reparametrisation-invariance in the definition of bordism operators, \cite{Moore:2006dw}, we now turn to the particular application in describing maximal dimensional Higgs branches of class ${\cal S}$ theories, as first proposed by \cite{Moore:2011ee}. Our major contribution in the present section will be highlighting where upgrades to the categories defined in \cite{Moore:2006dw} are needed for dealing with setups as the ones associated to the correspondence depicted in figure \ref{fig:correspondence}, namely those leading to the emergence of composite class ${\cal S}$ theories separated by a non-invertible defect. 

This section is structured as follows:   

\begin{enumerate}

\item  At first, we briefly overview the source and target categorical structure proposed in \cite{Moore:2011ee} assuming duality.   

\item  We then explain what categorical duality means from an algebraic perspective.

\item  We conclude the section indicating the relation between Moore-Tachikawa varieties and Coulomb branches of quiver gauge theories as an interesting realisation of 3D mirror symmetry, and how the categorical generalisation proposed in this work suggests interesting applications to quiver varieties that will be addressed in more detail in Part III.

\end{enumerate}

\begin{figure}[ht!]   
\begin{center}  
\includegraphics[scale=0.9]{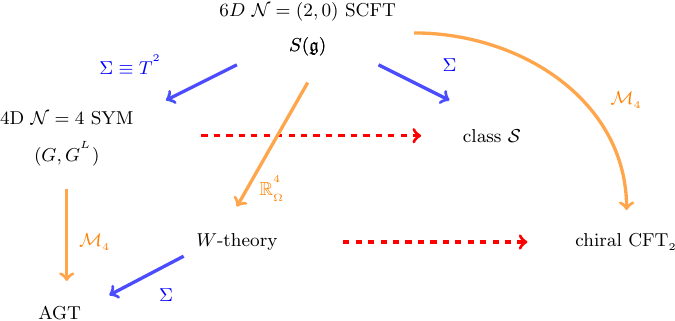}   
\caption{\small Partial reproduction of a diagram displayed in \cite{Moore}. The first part of our treatment focuses on the functorial field theory description of class ${\cal S}$ theories and their Higgs branches in terms of 2D TFT cobordism constructions.}    
\label{fig:classS}    
\end{center}  
\end{figure}

\subsubsection{Categorical structure assuming duality}  \label{sec:MooreTach}   

According to the AGT correspondence, to a given class ${\cal S}$ theory, one can assign a 2D TFT valued in a symmetric monoidal category, \cite{Moore:2011ee},

\begin{equation}   
\boxed{\ \ \ \eta_{_{G_{\mathbb{C}}}}:\ \text{Bo}_{_2}\ \rightarrow\ \text{HS} \color{white}\bigg]\ \ \ }   
\label{eq:etaGC}
\end{equation}

The existence of this 2D TFT relies on the the source and target categories satisfying a certain list of properties, \cite{Moore:2011ee}. We will not reproduce all of them in our treatment, and refer the interested reader to the original work of Moore and Tachikawa for a detailed explanation. In this first part of the section, we will only point out some of the crucial assumptions made in their work for reasons that will become clear in the following pages. 

\section*{Duality}

\begin{figure}[ht!]       
\begin{center}  
\includegraphics[scale=1]{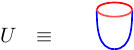}   \ \ \  \ \ \ 
  \includegraphics[scale=1]{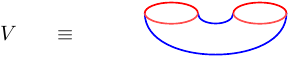} \ \ \   \ \ \ 
\includegraphics[scale=1]{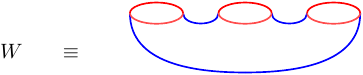}   
\caption{\small Basic bordisms assuming duality of, both, the source and target categories leading to the definition of the identity element, $V_{_{G_{_{\mathbb{C}}}}}$, and the maximal dimensional Higgs branch, $W_{_{G_{_{\mathbb{C}}}}}$.}    
\label{figure:UW}    
\end{center}  
\end{figure}

For the purpose of our work, the crucial assumption made in \cite{Moore:2011ee} is the duality structure of the source category Bo$_{_2}$. As explained in \cite{Moore:2011ee}, duality implies that the 2-category Bo$_{_2}$ is fully specified by its objects, $S^{1}$, and 1-morphisms, namely the bordisms depicted in figure \ref{figure:UW}. The middle bordism, i.e. the one labelled $V$, is the identity bordism. One can easily see this by noticing that $V$ is topologically equivalent to a cylinder whose edges are the red circles, i.e. the object of 2-category Bo$_{_2}$ (the closed string we were referring to in section \ref{sec:Moore-Segal}).

For $\eta_{_{G_{\mathbb{C}}}}$ to be well defined, the source and target categories are required to satisfy certain sewing relations, \cite{Moore:2006dw,Moore:2011ee}. This practically means that, compositions between morphisms should close. In particular, the identity itself can be defined in terms of composite homomorphisms as follows,

\begin{equation}             
\boxed{\ \ \ U_{_{G_{\mathbb{C}}}}\ \circ\ W_{_{G_{\mathbb{C}}}}\ \equiv\ T^{^*}G_{_{\mathbb{C}}} \color{white}\bigg]\ },  
\label{eq:defofid}     
\end{equation} 
where $T^{^*}G_{_{\mathbb{C}}}\ \equiv\ V{_{G_{\mathbb{C}}}}$.
Indeed, one can easily see that combining the first and third bordisms in figure \ref{figure:UW}, is topologically equivalent to $V$.\footnote{Indeed, $V$ is topologically equivalent to the cylinder, i.e. the cobordism between $S^{^1}$ and itself.} 

We therefore wish to highlight the following  

 \medskip    
   \medskip
\color{blue}

\noindent\fbox{%
    \parbox{\textwidth}{%
   \medskip    
   \medskip
   \begin{minipage}{20pt}
        \ \ \ \ 
        \end{minipage}
        \begin{minipage}{380pt}
      \color{black}  \underline{Main point:} Duality ensures the presence of an identity associated to a certain gauge group, $G_{_{\mathbb{C}}}\equiv\eta_{_{G_{\mathbb{C}}}}\ (S^{^1})$, \eqref{eq:gaugegroup}.
        \end{minipage}   
         \medskip    
   \medskip
        \\
    }%
}
 \medskip    
   \medskip     \color{black}

\subsection*{Key axiom}  

\eqref{eq:defofid} is essential for us in relating the formalism of \cite{Moore:2006dw,Moore:2011ee} to the setup of figure \ref{fig:correspondence}. In particular, it is what leads to the definition of the triple featuring on the RHS of \eqref{eq:HB1}. To see this explicitly, let us recall a crucial axiom required to be satisfied by \eqref{eq:etaGC}, and, therefore, in turn by \eqref{eq:defofid}, \cite{Moore:2011ee}.

For $X\in$\ Hom\ $(G_{_{\mathbb{C}}}^{^{\prime}}, G_{_{\mathbb{C}}})$ and $Y\in$\ Hom\ $(G_{_{\mathbb{C}}}, G_{_{\mathbb{C}}}^{^{\prime\prime}})$, their composition 

\begin{equation}   
Y\ \circ\ X\ \in\ \text{Hom}\ (G_{_{\mathbb{C}}}^{^{\prime}}, G_{_{\mathbb{C}}}^{^{\prime\prime}}) 
\label{eq:etaGC1}
\end{equation}    
is identified with the holomorphic symplectic quotient   

\begin{equation}   
\begin{aligned}
Y\ \circ\ X\ &\overset{def.}{=}\ X\ \times\ Y\ //\ G_{_{\mathbb{C}}}   \\
&\ =\{(x,y)\ \in\ X\times Y|\ \mu_{_X}(x)+\mu_{_Y}(y)=0\}\ /\ G_{_{\mathbb{C}}},     
\end{aligned}
\label{eq:etaGC1}
\end{equation}  
where

\begin{equation}   
\mu_{_X}:\ X\ \longrightarrow\ \mathfrak{g}_{_{\mathbb{C}}}^{*}\ \ \ , \ \ \ \mu_{_Y}:\ Y\ \longrightarrow\ \mathfrak{g}_{_{\mathbb{C}}}^{*}
\label{eq:etaGC1b}
\end{equation}  
are the moment maps of the action of $G_{_{\mathbb{C}}}$ on $X$ and $Y$, with $\mathfrak{g}_{_{\mathbb{C}}}$ the Lie algebra associated to $G_{_{\mathbb{C}}}$.  The identity element

\begin{equation}   
T^{^*}G_{_{\mathbb{C}}}\ \overset{def.}{=}\ \text{id}_{_{G_{_{\mathbb{C}}}}}\ \in\ \text{Hom}\ (G_{_{\mathbb{C}}}, G_{_{\mathbb{C}}}) 
\label{eq:etaGC2}
\end{equation}           
comes with a Hamiltonian $G_{_{\mathbb{C}}}\ \times\ G_{_{\mathbb{C}}}$ action. As also explained in \cite{Moore:2011ee}, to see that $T^{^*}G_{_{\mathbb{C}}}$ acts as the identity, it is enough to consider a composition of homomorphisms, $T^{^*}G_{_{\mathbb{C}}}\ \circ\ X$. Identifying $T^{^*}G_{_{\mathbb{C}}}\ \simeq\ G_{_{\mathbb{C}}}\ \times\ \mathfrak{g}_{_{\mathbb{C}}}$, and an element of $T^{^*}G_{_{\mathbb{C}}}$ as $(g,a)$.   The moment map condition, \eqref{eq:etaGC1}, reduces to   

\begin{equation}   
a+\mu(x)=0,      
\label{eq:etaGC3}
\end{equation}  
from which $a$ can be removed. Consequently, the induced 2-form on the solution space is $G_{_{\mathbb{C}}}$-invariant and basic. Upon taking the quotient with respect to $G_{_{\mathbb{C}}}$, we can gauge $g$ to 1, leading to a holomorphic isomorphism with the original $X$ space with its symplectic form. Note that this exacly like the constraint found in our main example \eqref{eq:tripleeq} within the context of Coulomb branches of star-shaped quivers.

The categorical quotient taken in defining the composition \eqref{eq:etaGC1} is equivalent to the one that a given absolute theory should be equipped with to potentially gauge away its entire operator content, while leaving only the identity in the spectrum\footnote{cf. explanation in the Introduction. This is basically what leads to the definition of the fiber functor, moment map and Drinfeld center.}. Indeed, this is true as long as the embeddings of the subalgebras associated to $G_{_{\mathbb{C}}}^{^{\prime}}$ and $G_{_{\mathbb{C}}}^{^{\prime\prime}}$ are subsets of each other within the mother algebra $\mathfrak{g}_{_{\mathbb{C}}}$. However, we are interested in describing more general setups, where the embeddings of the algebras are intersecting albeit not one included within the other. In the remainder of our treatment, we will explain that, for this to be described in the formalism of \cite{Moore:2006dw,Moore:2011ee}, the standard identity element associated to the gauge group $G_{_{\mathbb{C}}}$ and embedding Lie algebra $g_{_{\mathbb{C}}}$ needs to be removed from Bo$_{_2}$, while being replaced by a new composite bordism, and propose the definition of a new functor.

\subsubsection{Duality from an algebraic perspective} \label{sec:od}      

Before turning to explaining what are the changes that the source and target categories should undergo, we will briefly pause for a digression explaining how reparametrisation-invariance on the Riemann surface involved in the definition of the bordism operators, outlined in section \ref{sec:Moore-Segal}, is strongly related to the aforementioned duality assumption.

As explained in \cite{Moore:2011ee}, the crucial point is that, thanks to the duality propriety of the source 2-category Bo$_{_2}$, the identity element in the target category, $T^{^*}G_{_{\mathbf{C}}}$, is reparametrisation-invariant.  In particular, one could compose the identity morphisms as follows\footnote{Making use of the axiom \eqref{eq:etaGC1}.}  

\begin{equation}  
T^{^*}G_{_{\mathbf{C}}}^{a^{\prime}}\ \circ\ T^{^*}G_{_{\mathbf{C}}}^{^a}\equiv\left(T^{^*}G_{_{\mathbf{C}}}^{^a}\ \times\ T^{^*}G_{_{\mathbf{C}}}^{a^{\prime}} \right)\  \dslash\   G_{_{\mathbf{C}}},         
\label{eq:composition}  
\end{equation}
where $a, a^{\prime}$ denote the choice of the conformal structure. In particular, one can think of them as denoting the area of the bordism.

From the considerations made above, it therefore follows that one could rephrase \eqref{eq:composition} as the definition of the Drinfeld center for the composite system made up of two class $\cal S$ theories (associated to the two gauge groups involved) separated by an invertible defect, with the latter ensuring reparametrisation invariance of the Riemann surface \cite{Moore:2006dw}.   

Concretely, under the duality assumption, one could gauge away either of the two groups, while being left with the following

\begin{equation}  
\mathbf{1}_{_{G_{_{\mathbf{C}}}^{a^{\prime}}}}\ \circ\ T^{^*}G_{_{\mathbf{C}}}^{^a}\equiv\left(T^{^*}G_{_{\mathbf{C}}}^{^a}\ \times\ \mathbf{1}_{_{G_{_{\mathbf{C}}}^{a^{\prime}}}} \right)\ \dslash\  G_{_{\mathbf{C}}}.      
\end{equation}

If the conformal structure were the same on the two sides, then it would be the same, either with or without the composition rule. If 

\begin{equation}  
G_{_{\mathbf{C}}}^{^a}\ \times \ G_{_{\mathbf{C}}}^{a^{\prime}}\ \equiv\ G_{_{\mathbf{C}}}^{^{a+a^{\prime}}}\ \equiv\ G_{_{\mathbf{C}}}, \ \ \ \forall\  a, a^{\prime},    
\end{equation}   
then \eqref{eq:composition} can be recast to the following

\begin{equation}  
\boxed{\ \ \ T^{^*}G_{_{\mathbf{C}}}^{^a}\equiv T^{^*}G_{_{\mathbf{C}}}^{^a}\ \dslash  \ G_{_{\mathbf{C}}} \color{white}\bigg]\ \ } \ ,      
\label{eq:composition1}  
\end{equation}  
which is equivalent to a statement of S-duality\footnote{Note that the conformal structure $a$ is the same on both sides.}. Once more, we highlight that this is possible because the source category for the 2D TFT associated to the group $G_{_{\mathbf{C}}}$ contains the identity element. But, in case this is not true\footnote{Such as the case in which the Riemann surface is no longer reparametrisation-invariant.}, \eqref{eq:composition1} needs to be changed accordingly, which one could think of as a generalisation of an S-duality statement. Indeed, if the group composition rules do not hold, 

\begin{equation}  
G_{_{\mathbf{C}}}^{^{a+a^{\prime}}}\ \neq\ G_{_{\mathbf{C}}}, \ \ \ \forall a, a^{\prime},  
\end{equation}
we get something that is not simply the ordinary S-dual theory, \eqref{eq:composition1}.

The main purpose of our work is basically to go backwards, starting from the LHS of \eqref{eq:composition} and determining what the RHS should be. Most importantly, we need to:   

\begin{enumerate}  

\item Identify $G_{_{\mathbf{C}}}$ in the new theory obtained by composing the two theories on the LHS, each one characterised by a different choice of conformal structure on the Riemann surface.  

\item  Equivalently to 1., reconstruct $T^{^*}G_{_{\mathbf{C}}}$, namely the identity of the composite theory.

\item  We highlight that the most important generalisation of the 2D TFT \eqref{eq:etaGC1} that one should really be using for the case of interest to us is instead the following

\begin{equation}   
\boxed{\ \ \ \tilde\eta_{_{G_{\mathbb{C}}}}:\ \text{Bo}_{_2}\backslash\ V \rightarrow\ \text{HK} \color{white}\bigg]\ \ \ } 
\label{eq:newetaGC}   
\end{equation}    
which, as already pointed out in \cite{Moore:2011ee}, requires removing the identity element from the source category. Its effect on the target is to turn it into a hyperk$\ddot{\text{a}}$hler quotient. Its connection with theoretical physics\footnote{Already presented in \cite{Pasquarella:2023exd}.} is the main focus of section \ref{sec:last}.   

\end{enumerate} 

From step 1., an important observation is in order. $G_{_{\mathbf{C}}}$ acts on the two factors on the RHS of \eqref{eq:composition} in separate ways. This is part of the meaning of the generalisation of S-duality that we were previously referring to. Indeed, the categorical quotient $//G_{_{\mathbf{C}}}$ tells us what the identity is as a result of gauging a certain subalgebra. This is obtained by taking the 1-morphisms on either side of the correspondence and taking their nontrivial composition w.r.t. $T^{^*}G_{_{\mathbf{C}}}$, with the latter being the identity in the target category. But the latter was assumed to be removed. The immediate suggestion to circumvent this shortcoming is that the functors defining the identity element of each individual theory on the LHS of \eqref{eq:composition} is different w.r.t. the one on the RHS. In section \ref{sec:last} we therefore propose the generalisation of 3D mirror symmetry as the need to define two different 2D TFT functors associated to the left and right hand sides of \eqref{eq:composition}.

\subsection*{Algebraic Varieties} \label{sec:FBFN}

In the concluding part of this section, we highlight an interesting application of \eqref{eq:newetaGC} in the context of quiver gauge theories, which will be explained in more detail in the concluding sections of this work, in particular leading towards a generalisation 3D mirror symmetry.

The Higgs branches described by Moore and Tachikawa are known to have been reproduced by \cite{Braverman:2017ofm} as the Coulomb branches of 3D ${\cal N}=4$ supersymmetric quiver gauge theories. Such correspondence is therefore equivalent to a statement of 3D mirror symmetry. The purpose of \cite{Pasquarella:2023ntw} is to explain what the 3D dual of a theory described by 
\eqref{eq:newetaGC} is in terms of Coulomb branches of 3D ${\cal N}=4$ quiver gauge theories. In this way, we expect to be able to prove the statements made in \cite{Pasquarella:2023exd}.

If $V$ has been removed from the source, one should expect there to be more than one 2D TFT 
 of the kind \eqref{eq:etaGC} associated to two different gauge groups whose embedding in the gauge group associated to the original TFT with identity element $V$ is not simply a cochain complex. Correspondingly, this also means that there is more than one 1-morphism $U_{_{G_{\mathbb{C}}}}$.

Given that the identity of the embedding theory is defined as follows

\begin{equation}             
\boxed{\ \ \ U_{_{G_{\mathbb{C}}}}\ \circ\ W_{_{G_{\mathbb{C}}}}\ \equiv\ T^{^*}G_{_{\mathbb{C}}} \color{white}\bigg]\ }\ \ \ \ , \ \ \ \  
\boxed{\ \ \ \eta_{_{G_{\mathbb{C}}}}(V)\ \equiv\ T^{^*}G_{_{\mathbb{C}}}\color{white} \ \ }.    
\end{equation} 
and that 

\begin{equation}             
U_{_{G_{\mathbb{C}}}}\ \overset{def.}{=}\ G_{_{\mathbb{C}}}\times S_{_{n}}\ \subset\ G_{_{\mathbb{C}}}\ \times\ \mathfrak{g}_{_{\mathbb{C}}}\ \simeq\ T^{^*}G_{_{\mathbb{C}}},    
\end{equation} 
with $S_{_n}$ is the Slodowy slice at a principal nilpotent element $n$. The physical theories of class ${\cal S}$ predict the existence of a variety $W_{_{G_{_{\mathbb{C}}}}}$ satisfying the properties needed to define a TFT, $\eta_{_{G_{\mathbb{C}}}}$.  From the duality assumption, it follows that the dimensionalities of the two varieties are related as follows

\begin{equation}             
\text{dim}_{_{\mathbb{C}}}\ U_{_{G_{\mathbb{C}}}}\ \overset{def.}{=}\ \text{dim}_{_{\mathbb{C}}}\ G_{_{\mathbb{C}}}\ +\ \text{rank}\ G_{_{\mathbb{C}}}.  
\end{equation} 

\begin{equation}             
\text{dim}_{_{\mathbb{C}}}\ W_{_{G_{\mathbb{C}}}}\ \overset{def.}{=}\ 3\ \text{dim}_{_{\mathbb{C}}}\ G_{_{\mathbb{C}}}\ -\ \text{rank}\ G_{_{\mathbb{C}}}. 
\label{eq:dimhb}   
\end{equation} 

However, if the identity needs to be removed from Bo$_{_2}$, $T^{^*}G_{_{\mathbb{C}}}$ is not the identity and, in particular \eqref{eq:dimhb} needs to be redefined precisely because the source is no longer a dual category.

As a concluding remark to what we have just said, in \cite{Moore:2006dw} they conjecture the following property for the moment maps associated to the $G^{^3}$ action on the Higgs branch $W_{_{G_{_{\mathbb{C}}}}}$

\begin{equation}   
\boxed{\ \ \ \mu_{_i}:\ W_{_{G_{_{\mathbb{C}}}}}\rightarrow \mathfrak{g}_{_{\mathbf{C}}}^{^*}\ \ \ ,\ \ \ i=1,2,3.  \color{white}]\ \ }  
\label{eq:HBB}  
\end{equation}   

This is crucial to our analysis since \eqref{eq:HBB} can be inverted to obtain the Higgs branch as a hyperk$\ddot{\text{a}}$ler quotient 

\begin{equation}   
\boxed{\ \ \ W_{_{G_{_{\mathbb{C}}}}}\ \equiv \ \mu^{^-1} /G^{^3}\color{white}\bigg]\ \ }.  
\label{eq:HBtrue}  
\end{equation}

However, for the case in which the identity is removed from the source category, \eqref{eq:HBtrue}  does not hold anymore precisely because of the lack of permutational symmetry arising in the quotient. In section \ref{sec:last} we will be explaining how to define $W_{_{G_{_{\mathbb{C}}}}}$ and its dimensionality for the case involving categories without a duality structure. 

\subsection{Examples}

These are exactly the examples we encountered in section \ref{sec:main}, when dealing with Coulomb branches of star-shaped quivers.

\section*{Key points}    

The main points to bare in mind from this section are the following:  

\begin{itemize}  

\item Categorical duality ensures the presence of an identity object.

\item 3D mirror symmetry requires reparametrisation-invariance.

\end{itemize}

\section{Dualities among Algebraic Varieties}\label{sec:otherdualities}     

In this section we describe how abelianisation fits in the algebro-geometric description of 3D mirror symmetry. In doing so, we explicitly explain the analogy in between different treatments, \cite{Moore:2011ee}. The main aim for doing so at this stage is that of introducing the reader to some crucial notions that will later be developed in more exotic setups, to which Part III and V are devoted.   

The present section is structured as follows:  

\begin{enumerate}  

\item  At first, we overview some key features of equivariant cohomology and the convolution product, both within the context of quiver gauge theories, \cite{Braverman:2017ofm,Braverman:2016wma}. 

\item Then, making use of \cite{Dimofte:2018abu}, we emphasise how and when 3D mirror symmetry manifests from a categorical perspective, specifically in terms of bordisms, making use of the Kostant-Whittaker symplectic reduction, \cite{Braverman:2017ofm}.

\item  Last but not least, we close the section highlighting the importance of abelianisation for embedding the ring homology of the theory in question, explaining how this is in turn related to (un)gauging the categorical SymTFT introduced in section \ref{sec:4}. In doing so, we also overview the work of Teleman, \cite{Teleman:2014jaa}, thereby connecting it with Moore-Tachikawa varieties, and Coulomb branches ofstar-shaped quivers.         

\end{enumerate}

\subsection{Equivariant cohomology, fixed point localisation and convolution product}

Equivariant cohomology, \cite{BFM}, also known as Borel-Moore cohomology, \cite{BWB}, is a cohomology theory from algebraic topology which applies to topological spaces with a group action. It can be viewed as a generalisation of group cohomology. More explicitly, the equivariant cohomology ring of a space $X$ with action of a topological group $G$ is defined as the ordinary cohomology ring with coefficient ring $\Lambda$ of the homotopy quotient $EG\times_{_G}X$  

\begin{equation}  
H^{^{\bullet}}_{_G}(X;\Lambda)\ =\ H^{^{\bullet}}\left(EG\times_{_G} X;\Lambda\right).    
\end{equation}  

\subsection*{The Kirwan map}

In section \ref{sec:rqft} we saw that the Kirwan map, first introduced in \cite{FK}, states 

\begin{equation}  
H^{^{\bullet}}_{_G}(M)\ \rightarrow\ H^{^{\bullet}}\left(M\dslash _{_p} G\right), 
\end{equation}  
where $M$ is a Hamiltonian $G$-space, i.e. a symplectic manifold acted on by a Lie group $G$ with a moment map 

\begin{equation}     
\mu:\ M\ \rightarrow\ \mathfrak{g}^{*}.  
\end{equation}   

$H^{^{\bullet}}(M)$ is an equivariant cohomology ring of $M$, i.e. the cohomology ring of the homotopy quotient $EG\times_{_G}M$ of $M$ by $G$.

\begin{equation}  
M\dslash_{_p}G\ =\ \mu^{^{-1}}(p)/G  
\end{equation}   
is the symplectic quotient of $M$ by $G$ at a regular central value $p\in Z(\mathfrak{g}^{*})$ of $\mu$. It is defined as the map of equivariant cohomology induced by the inclusion 

\begin{equation}    
\mu^{^{-1}}(p)\ \hookrightarrow\ M   
\end{equation}
followed by the canonical isomorphism  

\begin{equation}  
H^{^{\bullet}}_{_G}\left(\mu^{^{-1}}(p)\right)\ \simeq\ H^{^{\bullet}}\left(M\dslash_{_p} G\right).
\end{equation} 

\section*{Fixed point localisation and convolution product}   

In BFN, the embedding of algebras is performed under the so called \emph{convolution product}. To see how this is defined, we start off with the convolution diagram for the affine Grassmannian, \cite{Braverman:2017ofm, Braverman:2016wma} 

\begin{equation}  
\text{Gr}_{_G}\times\text{Gr}_{_G}\ \overset{p}{\leftarrow}\ G_{_{\cal K}}\ \times\ \text{Gr}_{_G}\ \overset{q}{\rightarrow}\ \text{Gr}_{_G}\tilde\times\ \text{Gr}_{_G}\ \overset{m}{\rightarrow}\ \text{Gr}_{_G},   
\end{equation}   
where $p,q$ are projections, $m$ is a multiplication, and

\begin{equation}  
\text{Gr}_{_G}\tilde\times\ \text{Gr}_{_G}\ =\ G_{_{\cal K}}\ \times_{_{G_{_{\cal O}}}}\ \text{Gr}_{_G}. 
\label{eq:convoldiagr}     
\end{equation}

Hence, given any two $G_{_{\cal O}}$-equivariant perverse sheaves ${\cal A}_{_1}, {\cal A}_{_2}$, the pullback  

\begin{equation}  
p^{^*}\left({\cal A}_{_1}\boxtimes {\cal A}_{_2}\right) \ \equiv\ {\cal A}_{_1}\ \tilde\boxtimes\ {\cal A}_{_2}   
\end{equation}  
descends to $\text{Gr}_{_G}\ \tilde\times\ \text{Gr}_{_G}$ by equivariance, and we can define the convolution product as follows  

\begin{equation}  
{\cal A}_{_1}\star {\cal A}_{_2}  \ =\ m_{_*}\left({\cal A}_{_1}\ \tilde\boxtimes\ {\cal A}_{_2}\right).  
\end{equation} 

The latter defines a symmetric monoidal structure on the category of $G_{_{\cal O}}$-equivariant perverse sheaves on $\text{Gr}_{_G}$, and is equivariant to the monoidal category of finite-dimensional representations of the Langlands dual of $G$.   

\eqref{eq:convoldiagr} is the convolution diagram for the case in which $G$ is a complex reductive group and representation $\mathbf{N}=0$. For arbitrary $\mathbf{N}$, we can generalise the convolution diagram, achieving a noncommutative algebraic structure. In order to do so, we define a triple ${\cal R}=({\cal P},\varphi, s)$, where ${\cal P}$ is a $G$-bundle on a formal disk $D=\text{Spec}\ \mathbf{C}[[z]]$, $\varphi$ is its trivialisation over the punctured disk $D^{^*}=\text{Spec}\ \mathbf{C}((z))$, and $s$ is a section of the associated vector bundle ${\cal P}\times_{_G}\mathbf{N}$ such that it is sent to a regular section of the trivial bundle under $\varphi$.

\section*{Coulomb branches} 

Given a complex vector space, $V$, on a connected reductive algebraic group, $G$, with a fixed faithful linear action on $V$, the \emph{Higgs branch} is defined as follows  

\begin{equation}  
{\cal M}_{_H}\ =\ \mu^{^{-1}}(0)\ \dslash\ G\ =\ \text{Spec}\left(\mathbf{C}\left[\mu^{^{-1}}(0)\right]^{^G}\right)  
\end{equation}   
with 

\begin{equation}   
\mu:\ T^{^*}V\ \rightarrow \ \mathfrak{g}  
\end{equation}  
denoting the moment map, whereas the \emph{Coulomb branch}, ${\cal M}_{_C}$, is the spectrum of a ring constructed as a convolution algebra in the homology of the affine Grassmannian, where the choice of the representation $V$ is incorporated as certain quantum corrections to convolution in homology, which are kept track of by an auxiliary vector bundle, \cite{Braverman:2017ofm, Braverman:2016wma}. ${\cal M}_{_H}$ and ${\cal M}_{_C}$ are conjectured to be symplectic duals to each other, with the most important relation in between them being Koszul duality between the categories of the quantised varieties, \cite{Webster:2016rhh}.

Let $G$ be a simply-connected complex semisimple group, and let $G_{_{ad}}$ be its adjoint form. The group $G_{_{ad}}$ acts on $G$ by conjugation, and $G$ contains a transverse slice $\Sigma$ for this action, which was first introduced by Steinberg, \cite{stein}. The resulting multiplicative universal centraliser is the smooth affine variety   

\begin{equation}    
Z_{_{G}}\ =\ \left\{(a,h)\ \in\ G_{_{ad}}\times\Sigma\ \bigg|\ a\in G_{_{ad}}^{^h}\right\}.      
\end{equation}   

When $G$ is simply-laced, Bezrukavnikov, Finkelberg, and Mirkovíc, \cite{BFM}, showed that its coordinate ring is isomorphic to the equivariant K-theory of the affine Grassmannian of the Langlands dual group $G^{^{\text{V}}}$, therefore, in this case, $Z_{_{G}}$ is an example of a Coulomb branch as defined by BFN \cite{Braverman:2017ofm, Braverman:2016wma}.

\subsection{Mirror symmetry in Moore-Tachikawa varieties}  

We now turn to the specific case of interest to us, namely Moore-Tachikawa varieties, and their relation with the work of \cite{Freed:2022qnc,Dimofte:2018abu,Teleman:2014jaa,Bullimore:2015lsa}. First and foremost, we will briefly overview how the Higgs and Coulomb branches arise in this setup.

Given a reductive connected group, $G$, there is a functor defined as follows   

\begin{equation}  
\eta_{_{G_{_{\mathbf{C}}}}}:\ \ \text{Bo}_{_2}\ \rightarrow\ {\cal C} 
\end{equation} 
where ${\cal C}$ is a hyperbolic symplectic manifold, whose objects are affine algebraic groups, and whose morphisms 

\begin{equation}   
\text{Hom}_{_{\cal C}}(G_{_1}, G_{_2})  
\end{equation}    
are isomorphism classes of symplectic\footnote{By this we mean affine, Poisson, operationally symplectic, with a Hamiltonian chosen moment map.} varieties with Hamiltonian $G_{_1}\times G_{_2}$-action. $\eta_{_{G_{_{\mathbf{C}}}}}$ is fully specified by its action on objects and morphisms, namely

\begin{equation}   
\begin{aligned}
&\eta_{_{G_{_{\mathbf{C}}}}}\left(S^{^1}\right)\ \mapsto\ G\ \ \ \ \ \ \ \ ,  \ \ \ \ \ \ \ \ \eta_{_{G_{_{\mathbf{C}}}}}\left(S^{^2}\right)\ \mapsto\ Z_{_G}\\  
&\\
&\ \ \ \ \ \ \ \ \ \ \ \ \ \ \ \ \eta_{_{G_{_{\mathbf{C}}}}}\left(S^{^1}\times S^{^1}\right)\ \mapsto\ T^{^*}G\\   
\end{aligned}   
\label{eq:maps}   
\end{equation}  
where $Z_{_G}$ is the \emph{universal centraliser}, consisting of all the pairs $(x,g)$ with  $x\in\mathfrak{g}$, $\mathfrak{g}$ Lie, regular, $g\in G$ and $ad(x)_{_{\mathfrak{g}}}=x$. Given $K\subset\mathfrak{g}$ a Kostant slice, and $(e, h,f)$ a principal $\mathfrak{sl}_{_2}$-triple in $\mathfrak{g}$, the Kostant slice can be rewritten as follows  

\begin{equation}  
K=e+Z(f)   
\end{equation}  
with $Z(f)$ the centraliser of $f$ in $\mathfrak{g}$, $\mathfrak{g}\simeq \mathfrak{g}^{^*}$. For a given Kostant slice, the universal centraliser can therefore be redefined as follows  

\begin{equation}  
Z_{_G}=\left\{(g, x)|g\in G, x\in K, \text{ad}_{_{\mathfrak{g}}}(x)=x\right\}.  
\end{equation}  

What we have outlined so far can be suitably adapted to the case of interest to us, namely when $\Sigma$ is a Riemann surface whose boundary has $n$ components, from which one gets a symplectic variety $X$ with a $G^{^n}$-action. This symplectic variety, $X$, is often times referred to as the \emph{Higgs branch}.

\subsubsection{The Kostant-Whittaker symplectic reduction}

The action of the functor explicited in \eqref{eq:maps}, however, is not exhaustive. Indeed, there is one more piece of information that is needed in order for the 2D TFT to be fully determined, and this is its action on the bordism corresponding to the three-punctured sphere.  

To see how this relates to quivers, let us consider the case in which $X$ is a symplectic variety with $G$-action. Let $N\subset G$ be a maximal map, and 

\begin{equation}         
\chi:\ \ N\ \longrightarrow\ \ \mathbf{C}  
\end{equation}    
a non-degenerate homomorphism such that 

\begin{equation}  
N\ \bigg/\ [N,N]\ \simeq\ \bigoplus_{_{{\alpha}_{_i}}}\ \mathbf{C}_{_{\alpha_{_i}}},  
\end{equation}   
where $\alpha_{_i}$ denote the simple roots. Then, the Kostant-Whittaker reduction on the symplectic variety corresponds to the quotient   

\begin{equation}  
KW(X)\ =\ \mu^{^{-1}}_{_N}(\chi)\ \bigg/\ N,  
\end{equation}
and the universal centraliser becomes   

\begin{equation}  
Z_{_G}\ \simeq\ KW_{_{G\times G}}\left(T^{^*}G\right).
\end{equation}    

Furthermore, we can take 

\begin{equation}  
G\times K = KW_{_{G^{^\text{V}}}}\left(T^{^*}G\right)  \ \ \ \ \ , \ \ \ \ \ KW_{_G}(G\times K)\ =\ Z_{_G},  
\end{equation}  
meaning that filling in a hole on the sphere corresponds to performing a $KW_{_G}$ reduction. Note that this is a statement of 3D mirror symmetry.

According to the Moore-Tachikawa conjecture, overviewed in slightly more detail in appendix \ref{sec:mt}, a 2D TFT is an example of a 2-categorical structure, that is the Alday-Gaiotto-Tachikawa dual of a supersymmetric 4D theory arising from dimensional reduction of 6D ${\cal N}=(2,0)$ SCFTs of type $A_{_{N-1}}$ on a sphere with $N$ maximal punctures. These theories, oftentimes denoted ${\cal T}_{_{N}}\left[\Sigma_{_{0,k}}\right]$, upon being further dimensionally reduced on a circle, ${\cal T}_{_{N}}\left[\Sigma_{_{0,k}}\times S^{^1}\right]$, are known to be related to Sicilian 3D ${\cal N}=4$ theories, the so-called star-shaped quivers, upon taking the $\underset{S^{^1}\rightarrow0}{\lim}$, \cite{Dimofte:2018abu}.  

In \cite{Cremonesi:2014vla}, the authors calculated the Hilbert series of the chiral ring 

\begin{equation}         
\mathbf{C}\left[{\cal M}_{_{C}}\right]\ \simeq\ \mathbf{C}\left[{\cal M}_{_{H}}^{^{4D}}\right]  
\end{equation}  
exaclty with the intention of obtaining new information about the structure of the Higgs branches of the Higgs branch of the 4D theory, ${\cal M}_{_{H}}^{^{4D}}$, via a direct analysis of the corresponding Coulomb branches ${\cal M}_{_{C}}$ of ${\cal T}_{_{N,k}}$, for general $N$ and $k$. In addition to this, our work wishes to emphasise the importance of the calculation of the Hilbert series together with the correct identification of the ring elements and relations in between them, which plays a crucial role in developing connection with Particle Physics and calculations therein. In achieving the second objective, a crucial role is played by the so-called \emph{abelianisation} procedure, first introduced by \cite{Bullimore:2015lsa}, and subsequently emphasised also in \cite{Dimofte:2018abu}.

\subsection{Homological mirror symmetry} \label{sec:hms}

We now turn to a more general setup, first outlined by \cite{Teleman:2014jaa}. In our brief overview, we highlight the importance of its relation with our main example, \ref{sec:rqft}, and how it fits within the context of abelianisation. As we shall see, the key expression that we need to consider is \eqref{eq:tripleeq}. 

At first, let us briefly overview the notion of homological mirror symmetry in its first formulation, \cite{Kontsevich:1994dn}.

Homological mirror symmetry, consists in a proposed agreement between two categories, namely:

\begin{itemize}  

\item  Fukaya's $A_{_{\infty}}$-category, ${\cal F}(X)$, on the symplectic side,

\item The derived category with Yoneda structure, $D^{^b}\mathfrak{Coh}(X^{^{\text{V}}})$, on the complex side. 

\end{itemize} 

For completeness, we now turn to explain this terminology.

\subsection*{Fukaya categories}

A symplectic manifold, $X$, is a smooth manifold of even dimension equipped with a non-degenerate symplectic form $\omega\ \in\ \Omega^{^2}_{_{cl}}(X)$. A Fukaya category, \cite{hms6}, of a symplectic manifold, $X$, is an $A_{_{\infty}}$-category with Lagrangian submanifolds $L\ \subset\ X$ as objects, whose intersections define the Hom-space. 
An $A_{_{\infty}}$-category is a category with associativity condition ($A)$ relaxed without bound on degrees of homotopies ($\infty$). They are linear categories, i.e. their Hom-objects are chain complexes.

\subsection*{Yoneda structure}  

Yoneda structure, \cite{Yonedastructure}, provides in a 2-categorical setting the axiomatic description of the formal properties of the usual presheaf construction and Yoneda embedding of locally smaller categories. 

The axioms of Yoneda structure capture the properties of the presheaf construction with CAT replaced by a general 2-category, ${\cal K}$. In order to handle size issues, a class of admissible or legitimate 0-cells is singled-out in $|{\cal K}|$. In order to handle size issues, as well as a class of 1-cells that behave well w.r.t. this class and the presheaf construction. In fact, it suffices to describe the admissible 1-cells, since one can identify the admissible 0-cells with the admissible identity 1-cells.  

Let ${\cal K}$ be a 2-category, and $\mathbb{A}$ a class of 1-cells. The 1-cells of $f\ \in\ \mathbb{A}$ are called \emph{admissible} if for all $f\ \in\ \mathbb{A}$ and composable 1-cells $g\in{\cal K}$, $f\circ g\in\mathbb{A}$. In this setup, a 0-cell $C\in|{\cal K}|$ is admissible if id$_{_C}$ is.  

A \emph{Yoneda structure} is therefore defined as a pair $({\cal P}, \mathbb{A})$, where ${\cal P}$ is a pre-sheaf construction for $\mathbb{A}$ assigns to every admissible object $A\in|\mathbb{A}|$ and object ${\cal P}A\in|{\cal K}|$ called its object of presheaves, and an admissible 1-cell

\begin{equation}   
y_{_A}:\ A\ \rightarrow\ {\cal P}A  
\end{equation}  
called its Yoneda morphism subject to three conditions:

\subsubsection*{Relation to Factorisation Homology}  

As we shall see later on in section \ref{sec:cd}, this formulation is compatible with that of factorisation homology for class ${\cal S}$ theories, first outlined in our work, \cite{Pasquarella:2023vks}.

\subsubsection{Relation to our main example}

The main point, to which we will be coming back later on in Part V is that the morphisms in the category of compact contractible circles encountered in Moore-Tachikawa varieties are the 1-cells featuring in the Yoneda structure description. 

\subsubsection{Lagrangian submanifolds}

A Lagrangian submanifold of a symplectic manifold\footnote{Note that here we are referring to Calabi-Yau 3-folds, hence $X$ is of even dimensions.}, $L\ \subset\ X$, is a submanifold which is a maximal isotropic submanifold on which $\omega=0$. They constitute the leaves of real polarisations, and are, therefore, crucial elements of symplectic geometry.

There is an equivalence of categories between the category of $\mathbb{C}$-linear representations of a quiver ${\cal Q}$ and the category of left $\mathbb{C}{\cal Q}$-modules.

\begin{figure}[ht!]  
\begin{center} 
\includegraphics[scale=1]{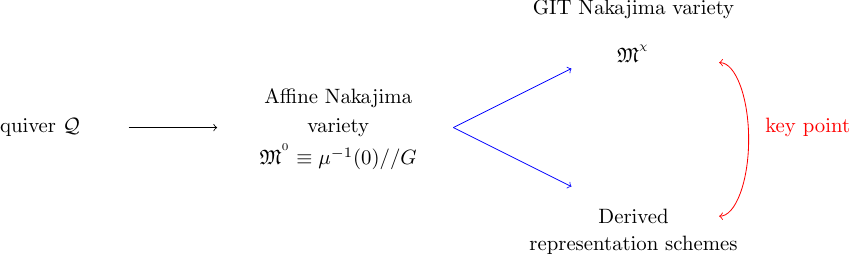} 
\caption{\small The upper and lower blue arrows correspond to geometric and algebraic resolutions of ${\mathfrak{M}^{0}}$, respectively. }  
\label{fig:GIT}  
\end{center}    
\end{figure}

    \subsection*{ Moment map and higher homologies}

In general, homologies of derived representation schemes can be highly nontrivial. However, in this particular case, one can identify a necessary and sufficient condition for the vanishing of the higher homologies based on the flatness of $\mu$, \cite{DAlesio:2021hlp}. In particular, in such reference, it was shown that the derived representation scheme DRep$_{_{v,w}}({\cal A})$ has vanishing higher homologies if and only if $\mu^{^{-1}}(0)\ \subset\ M({\cal Q},v,w)$ is a complete intersection, which happens only if the moment map is flat, \cite{mommap}. As we shall see, the requirement of the algebraic variety to be a complete intersection is crucial for the purpose of our treatment. In particular, it ensures the emergence of a 2-categorical structure, whose importance will be the core topic of section \ref{sec:3}. 

Prior to that, we devote the following subsection to explain the role played by magnetic quivers for describing generalisations of homological mirror symmetry in quiver gauge theories with 8 supercharges.

\subsection{3D mirror symmetry}

We now briefly overview one of the main dualities of interest to us, namely 3D mirror symmetry. The literature exploring this topic is extremely vast, and we are aware of the fact that our reference list is by no means complete. TO the extent possible, we will be referring to the works that are more closely related to the purpose and aims of our work. From the physical point of view, the pioneering work dates back to the joint work of Intriligator and Seiberg, \cite{Intriligator:1996ex}, which in turn is related to that of CY-manifolds, developed in the mathematical community. 

As initially quoted in section \ref{sec:CY}, CY-manifolds are the target of functors defining CY-categories.

The aim of this and the next section is that of geometrising the 3D mirror symmetry outlined in our main example in section \ref{sec:main} as a duality exchanging Coulomb and Higgs branches of 3D ${\cal N}=4$ SCFTs. This aim was already achieved by Kapustin, Rozansky, and Saulina, \cite{Kapustin:2008sc}, who were the first to address this in the context of Rozansky-Witten (RW) theories, \cite{Rozansky:1996bq}. 

The simplest topological BCs in the RW model with target $X$ correspond to complex Lagrangian submanifolds of $X$ corresponding to the fact that one associates a boundary condition to a fibration over a complex Lagrangian submanifold whose fibration is a CY-manifold. The set of BCs for the RW model with target $X$ should be thought of as a set of objects of a 2-category.

\subsubsection{Categorical formulation}

Gromov-Witten (GW) theory is among the most studied 2D TQFTs, \cite{hms8}. It assigns to a compact symplectic manifold $X$ a space of states $H^{\bullet}(X)$. Surfaces with points labelled by states give correlators counting pseudo-holomorphic maps to $X$ with coincidence conditions. 

Mirror symmetry reduces GW theory to calculations in the complex geometry of a mirror manifold, $X^{^V}$, introduced by Kontsevich, \cite{Kontsevich:1994dn}. 

The key idea is that, including all surfaces with corners, and BCs (branes) forming a linear category with structure, this category should be able to determine all invariants.  

Mathematically, the description starts with a group action acting on the generating category (the Fukaya category), which is factored through the topology of the group. Their mirror description involves holomorphic symplectic manifolds and Lagrangians related to the Langlands dual group. As an application: the complex mirrors of flag varieties proposed by Marsh and Rietsch, \cite{bmodel}, \footnote{We refer to section \ref{sec:main} for preliminary definitions. }. 

The aim is that of studying equivariance in the higher-algebra surrounding TQFTs. The homological information lies in the neighborhood of a Lagrangian within $X$ (a symplectic manifold), whereas most of the interesting physics happens elsewhere. The emerging geometric picture is the following: representations admit a character theory, but characters are now coherent sheaves on a manifold related to the conjugacy classes, instead of functions. The manifold in question is the Bezrukavnikov-Finkelberg-Mirkovic (BFM) space of the Langlands dual Lie  group, $G^{^V}$, and is closely related to the space of conjugacy classes in the complex group $G_{_{\mathbb{C}}}^{^V}$.   Multiplicity spaces of $G$-invariant maps between linear representations are now replaced by multiplicity categories, whose dimensions are the Hom-spaces in the category of coherent sheaves.  

There is a preferred family of simple representations that foliate the BFM space. Every such representation is symplectically-induced from a 1D representation of a certain Levi subgroup of $G$: it is the Fukaya category of a flag variety of $G$ (similarly to holomorphic induction, the Borel-Weil construction of irreducible representations of $G$). This is the same governing structure emerging in KRS, where BCs of the 3D TQFT are associated to a holomorphic Lagrangian submanifold.

\medskip   

\medskip 

\underline{Meta-statement, \cite{Teleman:2014jaa}:} Pure topological gauge theory in 3D for a compact Lie group $G$ is equivalent to the Rozansky-Witten theory for the BFM space of the Langlands dual Lie group $G^{^V}$.

\medskip   

\medskip 

2D TQFT can be thought of as a higher analogue of cohomology (Fukaya-Floer theory of a symplectic manifold is like a refinement of  ordinary cohomology), according to which gauged TQFTs are described in terms of equivariant cohomology. This describes categorification of representation theory of a compact Lie group $G$, with representations being topological.

In the RW model, one must describe geometrically two functors from the 2-category of linear categories with $G$-action to linear categories: 

\begin{enumerate}  

\item The forgetful functor, remembering the underlying category (describing the pre-gauged TQFT). This is represented by a regular representation of $G$, i.e. categories of coherent sheaves over the Lagrangians Spec$H_{*}(\Omega G)$. 

\item The invariant category, generating the gauged TQFT, represented by a trivial representation of $G$, $\mathfrak{t}_{_{\mathbb{C}}}/W\equiv T^*_{_{1}}T^{^V}_{_{\mathbb{C}}}/W$.  

\end{enumerate}

\subsection{From Homological mirror symmetry to 3D mirror symmetry}  \label{sec:KRS}

The concluding part of this section is a very succinct overview of some key features emerging in the work of \cite{Teleman:2014jaa}, which is of particular importance to us. As we shall see, his work is deeply related to that of Moore and Tachikawa, thanks to their relation to Coulomb branches of star-shaped quivers, briefly overviewed in section \ref{sec:main}.   

To the best of our knowledge, the correspondence in between these three setups has only been spelled out in our work, \cite{Pasquarella:2023exd}. The crucial step in understanding these different setups is looking at how mirror symmetry is realised in each case. As we shall see, this builds a dictionary that can be used for furthering the study of more exotic underlying mathematical structures.

\subsection*{The 2-category of KRS and the Bezrukavnikov-Finkelberg-Mirkovic (BFM) space}

The image of a point $Z$(pt) is an object in the 3D bordism 3-category. Lurie's generator for pure 3-dimensional gauge theory should have categorical depth 2. His proposal for such generator is a 2-category associated to a certain holomorphic symplectic manifold, whose existence had already been conjectured by KRS and KR. When $X$ is compact, this 2-category should generate the RW-theory of $X$. In particular, its Hochschild cohomology (a 1-category with braided tensor structure), should be a dg-refinement of the derived category of coherent sheaves on $X$, $\sqrt{\mathfrak{Coh}(X)\ }\equiv$ KRS$(X)$. Among objects in KRS$(X)$ are smooth holomorphic Lagrangians $L\subset X$. More general objects are coherent sheaves of ${\cal O}_{_{L}}$-linear categories on such $L$. Given two Lagrangians, $L,L^{\prime}\in X$, Hom$(L,L^{\prime})$ is a sheaf of categories supported on $L\cap L^{\prime}$, and a $(\mathfrak{Coh}(L),\otimes)-(\mathfrak{Coh}(L^{\prime}),\otimes)$ bimodule. 

\begin{figure}[ht!]   
\begin{center}
\includegraphics[scale=0.9]{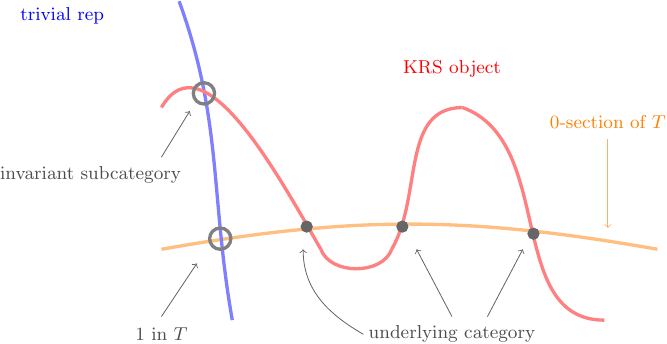}    
\caption{\small BFM space for an abelian gauge group, \cite{Teleman:2014jaa}.}
\label{fig:ct1}  
\end{center} 
\end{figure}

Localising at $L$, we choose a formal neighbourhood identified symplectically with $T^*L$ such that we can regard (locally) as the graph of a differential d$\Psi$ for a potential function $\Psi:L\rightarrow\mathbb{C}$. Locally, when this identification is valid, Hom$(L,L^{\prime})\equiv$MF$(L,\Psi)$. The corresponding figure is \ref{fig:ct1}.

\begin{figure}[ht!]   
\begin{center}
\includegraphics[scale=0.9]{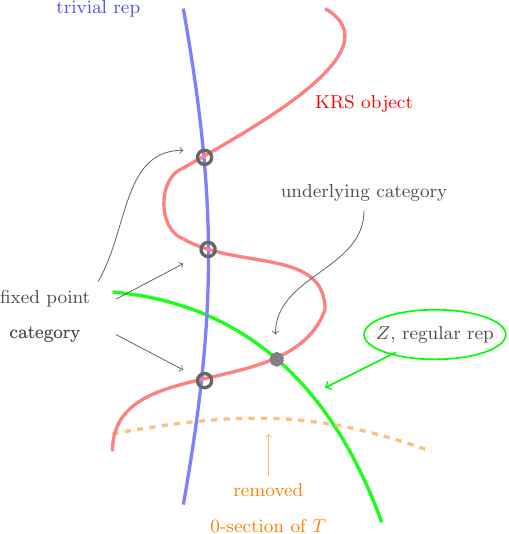}       
\caption{\small BFM space for a non-abelian gauge group, \cite{Teleman:2014jaa}.}
\label{fig:ct2}  
\end{center} 
\end{figure}

For the non-abelian case, instead, this description in terms of complex manifolds and potentials is inadequate, since the BFM space is not quite a cotangent bundle. The corresponding figure is \ref{fig:ct2}.

\subsubsection{Relation to our main example}   

The Hom category Hom$(L,L^{\prime})$ for two Lagrangians $(L,L^{\prime})\subset X$ with finite intersection supplies a 2D TFT for framed surfaces (following from matrix factorisations). $X$ itself aims at defining RW-theory, and each $L,L^{\prime}$ is a BC for it.

\begin{equation}  
L,L^{\prime}:\text{Id}\rightarrow \text{RW}_{_{X}}
\end{equation}    
are morphisms from the trivial 3D theory Id to RW$_{_{X}}$ viewed as functors from Bord$_2$ to the 3-category of linear 2-categories and Hom($L,L^{\prime})$ of natural transformations between these morphisms is the generator of the theory in \cite{Teleman:2014jaa}. Note that this is equivalent to the definition of the triple \eqref{eq:tripleeq} in our main example in section \ref{sec:rqft}, since the $RW_{_X}$-theory lives on a Calabi-Yau manifold, and the Lagrangians $L, L^{^{\prime}}$ act as the left- and right- moment maps projecting the identity element Id. This basically shows that the method put forward by \cite{Teleman:2014jaa} is an abelianisation procedure enabling to foliate the BFM space for any given pair of Lagrangian submanifolds.

The Hochschild homology

\begin{equation} 
\text{HH}_{_{\text{dim} L}}\text{Hom}(L,L^{\prime})\simeq\Gamma\left(L\cap L^{\prime};(\omega\otimes\omega^{\prime})^{^{1/2}}\right) 
\end{equation}

A domain wall between different TQFTs generalises the notion of BCs. This is an adjoint pair of functors between the TQFTs meeting certain dualisability conditions. A BC is a domain wall with a trivial TQFT. Just as a holomorphic Lagrangian is expected to define a BC for RW$_{_{X}}$, a holomorphic Lagrangian correspondence 

\begin{equation}  
X\ \leftarrow\ C\ \rightarrow\ Y  
\end{equation}   
should define a domain wall between RW$_{_{X}}$ and RW$_{_{Y}}$. This can be used for comparing gauge theories of different groups. 

\begin{equation}  
\text{BFM}(G^{^V})\ \longleftarrow\ \text{BFM}(G^{^V})\times_{\mathfrak{t}_{_{\mathbf{C}}}/W}\mathfrak{t}_{_{\mathbb{C}}}\ \longrightarrow\ T^*T^{^V}_{_{\mathbb{C}}}     
\end{equation}

The correspondence matches an adjoint pair of restriction-induction functors between categorical $T$- and $G$-representations. Induction from a category ${\cal C}$ with topological $T$-action is affected by string topology with coefficients in the flag variety $G/T$: 

\begin{equation} 
\text{Ind}({\cal C})\equiv C_{_{*}}\Omega G\otimes_{_{C_{_{*}}\Omega T}}\ {\cal C}  
\end{equation}  

Restricting the $G$-action to $T$, the flag manifold $G/L$ is a transformation from $L$-gauge theory to a $T$-gauge theory, given by composition of the symplectic induction and string topology domain walls 

\begin{equation}  
T(L^{^V})\ \overset{\text{SInd}}{\hookrightarrow}\ T(G^{^V})\ \underset{\text{Toda}}{\overset{~}{\rightarrow}}
 \text{BFM}(G)\ \overset{\text{ST}}{\rightarrow} \ \text{BFM}(T^{^V})\ \equiv\ T^*T^{^V}_{_{\mathbf{C}}}      
 \end{equation}

 The flag manifold $G/L$ is therefore a transformation from $L$-gauge theory to $T$-gauge theory given by the composition of symplectic induction and string topology domain walls.

One of the main aims of the work by KRS was that of furthering the understanding of 3D mirror symmetry in terms of homological mirror symmetry between categories of Lagrangian submanifolds. In the concluding part of this work, we show that this correspondence can equivalently be realised in terms of the magnetic quiver prescription, which can be further extended to the case involving non-invertible defects in presence of two cones\footnote{We will come back to this in section \ref{sec:333}.}.

\section{Holographic Dualities and Beyond } 

\begin{figure}[ht!]  
\begin{center}   
\includegraphics[scale=1]{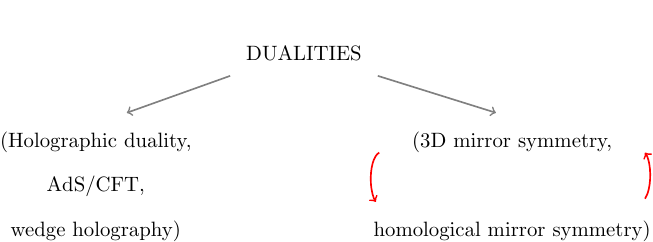}  
\caption{\small Main types of dualities arising in due course in the remainder of our treatment.}   
\label{fig:SMSM}  
\end{center} 
\end{figure}   

\subsection{Holographic duality}

Inspired by black hole mechanics, the Holographic Principle was first proposed by Gerard 't Hooft, and subsequently given a more precise string-theoretic interpretation by Susskind. It is a property of String Theory as well as a supposed property of QG, stating that, the description of a volume of space can be encoded in its boundary. The AdS/CFT correspondence provides a non-perturbative formulation of String Theory with certain boundary conditions, and constitutes one of the most successful realisations of the Holograpic Principle.

One of the original motivations for studying the AdS/CFT correspondence was developing a non-perturbative formulation of string theory.

\subsection*{The 4D \texorpdfstring{${\cal N}=4$}{} SYM/ string theory on AdS\texorpdfstring{$_{_5}\times S^{^5}$}{} case} 

For the particular case of 4D ${\cal N}=4$ SYM/ string theory on AdS$_{_5}\times S^{^5}$, symmetries play a crucial role in establishing the holographic correspondence. Specifically, the isometry invariance group of AdS$_{_5}$, namely $SO(4,2)$, corresponds to the symmetry group of CFT$_{_4}$s, whereas the isometry group of $S^{^5}$, $SO(6)\simeq SU(4)$, corresponds to the R-symmetry group of ${\cal N}=4$ SYM. 

\begin{figure}[ht!]  
\begin{center}   
\includegraphics[scale=1]{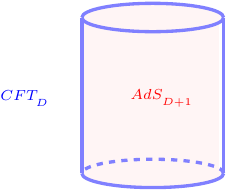}  
\caption{\small }   
\label{fig:feynman}  
\end{center} 
\end{figure}

This statement can be generalised to arbitrary dimensions, in the sense that the isometry group of AdS$_{_{D+1}}$ is $SO(D,2)$, corresponding to the conformal group in $(D-1,1)$ dimensions. The boundary of AdS$_{_{D+1}}$ is $\mathbb{R}^{^D}$ in Poincare' coordinates, and $\mathbb{R}_{_t}\times S^{^{D-1}}$ in global coordinates, which is conformal to $\mathbb{R}^{^D}$. Figure \ref{fig:feynman} shows this.

A crucial feature of the correspondence is the fact that AdS acts as a quantum mechanical box, meaning light can travel from the bulk to the boundary in finite time, implying the two are in causal contact with each other. Consequently to this, unlike the case of Minkowski spacetime, the notion of S-matrix falls short from being well-defined in AdS. On the other hand, the well-defined observables at hand are correlation functions of fields, with sources placed on the conformal boundary.   

For the purpose of our treatment, we only need to state what the key points of the holographic correspondence imply from the point view of the bulk-boundary interaction, which is what we will now turn to. 

In its original formulation, one expects there to be a correspondence in between the theory of open strings living on $N$ $D$-branes, i.e. ${\cal N}=4$ SYM with $SU(N)$ gauge group, and the gravitating theory corresponding to fields of the Hawking radiation existing in the background curved by the D-branes. Indeed, the Hawking radiation process plays a crucial role in the AdS/CFT correspondence. Considering the $D$-branes as endpoints of open strings, then string theory with D3-branes has three ingredients: 

\begin{enumerate} 

\item    The open strings existing on the D3-branes, giving a theory that reduces to 4D ${\cal N}=4$ SYM in the low-energy limit.

\item   The closed strings existing in the bulk spacetime, giving a theory that is SUGRA coupled to the massive modes of the string. In the low-energy limit, only SUGRA will remain.    

\item  The interactions in between the two theories, of which Hawking radiation is a manifestation.

\end{enumerate}

\begin{figure}[ht!]  
\begin{center}   
\includegraphics[scale=1]{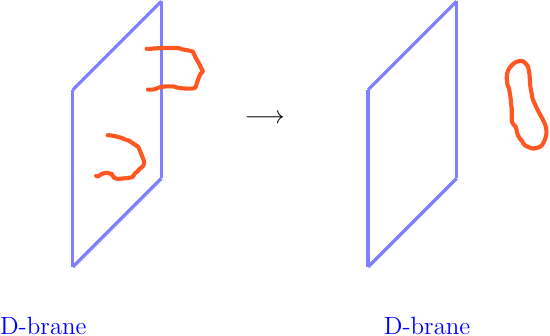}  
\caption{\small }   
\label{fig:hawking}  
\end{center} 
\end{figure}

\subsection*{Generalisations}  

To better understand the quantum aspects of our 4D world, some have considered lower-dimensional mathematical models, where spacetime is assumed to be 3-dimensional. The advantage of such a lower-dimensional setup is the fact that the description of the gravitational field is greatly simplified, and one can therefore use standard QFT techniques, circumventing the need to resort to string theory in 4D. 

Two of the most important proposals for such lower-dimensional correspondence are the following: 

\begin{enumerate}     

\item   In 1986, Brown and Henneaux, \cite{Brown:1986nw}, realised that quantum gravity in 3D is closely related to a particular CFT$_{_2}$, namely Liouville Field Theory.

\item  Edward Witten related 3D gravity in AdS to a CFT with monster group theory.   

\end{enumerate}   

Both cases provide examples of the AdS/CFT correspondence that do not require the full apparatus of string or M-theory.  

\subsubsection{Liouville Theory}   

Liouville Theory describes the dynamics of a field, $\varphi$ on a 2D space, subject to the potential   

\begin{equation}  
V(\varphi)=e^{^{2b\varphi}},      
\end{equation}   
where $b$ plays the role of the coupling constant. Its action reads   

\begin{equation}  
S[\varphi]=\frac{1}{4\pi}\int d^{^2}x\sqrt{g\ }\left(g^{^{\mu\nu}}\partial_{_{\mu}}\varphi\partial_{_{\nu}}\varphi+Q{\cal R}\varphi+\lambda^{^{\prime}}e^{^{2b\varphi}}\right),  
\label{eq:liouville}
\end{equation}   
where $g^{^{\mu\nu}}$ is the 2D metric, ${\cal R}$ the Ricci scalar, and $\lambda^{^{\prime}}$ is often referred to as the cosmological constant. 

The equations of motion from \eqref{eq:liouville} read   

\begin{equation}  
\frac{1}{\sqrt{g\ }}\partial_{_{\mu}}\left(\sqrt{g\ }g^{^{\mu\nu}}\partial_{_{\nu}}\right)\varphi=\frac{1}{2}Q{\cal R}+\lambda^{^{\prime}}be^{^{2b\varphi}}.   
\label{eq:liouville}
\end{equation} 

In Euclidean signature, \eqref{eq:liouville} reduces to

\begin{equation}  
\left(\frac{\partial^{^2}}{\partial x^{^2}}+\frac{\partial^{^2}}{\partial y^{^2}}\right)\varphi=\frac{1}{2}Q{\cal R}+\lambda^{^{\prime}}be^{^{2b\varphi}}.   
\label{eq:liouvilleeucl}
\end{equation}  

Assuming the Euclidean metric takes the following form  

\begin{equation}  
ds^{^2}=g_{_{\mu\nu}}dx^{^{\mu}}dx^{^{\nu}} = dzd\bar z,  
\label{eq:liouvilleeuclmetric}
\end{equation} 
the components of the energy-momentum tensor read 

\begin{equation}  
T_{_{z\bar z}}=T_{_{\bar z z}}=0
\end{equation} 
and

\begin{equation}  
T_{_{z z}}=\left(\partial_{_z}\varphi\right)^{^2}+Q\partial_{_z}^{^2}\varphi 
\ \ \ \ \ \ \ \ , \ \ \ \ \ \ \ \  
T_{_{\bar z \bar z}}=\left(\partial_{_{\bar z}}\varphi\right)^{^2}+Q\partial_{_{\bar z}}^{^2}\varphi,  
\label{eq:nonv}
\end{equation} 
with 

\begin{equation}  
\partial_{_{\bar z}} T_{_{z z}}=0
\ \ \ \ \ \ \ \ , \ \ \ \ \ \ \ \  
\partial_{_z}T_{_{\bar z \bar z}}=0.     
\end{equation} 

Each one of the non-vanishing components \eqref{eq:nonv} generates a Virasoro algebra with central charge

\begin{equation}  
c=1+6Q^{^2}.    
\end{equation} 

For such Virasoro algebras, $e^{^{2\alpha\varphi}}$ is a primary field with conformal dimension   

\begin{equation}  
\Delta=\alpha(Q-\alpha).  
\end{equation}  

For the theory \eqref{eq:liouville} to be conformally-invariant, the field $e^{^{2b\varphi}}$ must be exactly marginal, i.e. it must have conformal dimension such that

\begin{equation}  
\Delta(b)=1\ \ \ \ \ \ \Rightarrow\ \ \ \ \ \ Q=b+\frac{1}{b}.  
\label{eq:confinv}
\end{equation} 

The last condition in \eqref{eq:confinv} ensures $(c, \Delta)$ is invariant under the duality    

\begin{equation}  
b\ \longrightarrow\ \frac{1}{b}.  
\label{eq:confinvduality}
\end{equation}  

\subsection*{Relation to other theories}  

Liouville theory admits two different supersymmetric extensions, namely with ${\cal N}=1$ and ${\cal N}=2$. Furthermore, it can be obtained from the WZW model. 

Liouville theory also appears in the context of bosonic string theory. In its semiclassical limit it coincides with Jackiw-Teitelboim (JT)-gravity, which is another formulation of 2D gravity. To see how JT-gravity can be recovered from Liouville theory, we start from a double Liouville theory built combining a spacelike and a timelike Liouville theory. 

To begin with, we start by defining the following CFT action

\begin{equation}  
S_{_L}=\int d^{^2}x\sqrt{g\ }\left[\left(\partial_{_{\mu}}\phi\right)^{^2}-\left(\partial_{_{\mu}}\chi\right)^{^2}+\mu e^{^{2b\phi}}-\mu e^{^{-2b\chi}}\right] +2\int d\theta\  (Q\phi+q\chi) {\cal K}^{^{(0)}}.   
\label{eq:doubleliouville}    
\end{equation}

Under the field redefinition

\begin{equation}  
\Phi\overset{def.}{=}-\frac{1}{2b}(\phi+\chi)\ \ \ ,\ \ \ \rho\overset{def.}{=}\frac{b}{2}(\phi-\chi),    
\end{equation}
\eqref{eq:doubleliouville} reduces to

\begin{equation}  
S=\int d^{^2}x\sqrt{g\ }\left[-4\left(\partial\rho\right)\left(\partial\Phi\right)-2\mu e^{^{2\rho}}-\sinh\left(2b^{^2}\Phi\right)\right] -4\int d\theta\  \Phi{\cal K}^{^{(0)}}.   
\label{eq:doubleliouville1}    
\end{equation}

Taking $\rho$ to be the conformal factor of the JT-gravity metric

\begin{equation}  
ds_{_{JT}}^{^2}=e^{^{2\rho}}\left(dr^{^2}+r^{^2}d\theta^{^2}\right),  
\end{equation}    
\eqref{eq:doubleliouville1} ultimately becomes

\begin{equation}  
S=-2\int d^{^2}x\sqrt{g_{_{JT}}\ }\left[\Phi{\cal R}+\mu e^{^{2\rho}}-\sinh\left(2b^{^2}\Phi\right)\right] -4\int d\theta\ \sqrt{\gamma_{_{JT}}\ } \Phi{\cal K}, 
\label{eq:JTfromL}    
\end{equation}  
which in the semiclassical limit $\left(\text{namely}\  \underset{b\rightarrow 0}{\lim}\  \text{with}\  \mu b^{^2}=1 \right)$ is exactly equivalent to JT-gravity.

\subsection{The Chern-Simons/Wess-Zumino-Witten (CS/WZW) correspondence}  

One realisation of the holographic principle in QFT between 3D $G$-CS theory as the bulk field theory, and the 2D WZW model on a suitable Lie group $G$ as the boundary theory. This case was known and understood well before the holographic principle was first formulated.  

Unlike the AdS/CFT correspondence, this is an actual theorem, stating the normal equivalence in between the space of quantum states of CS theory on a surface $\Sigma$, and the space of conformal blocks on the WZW model.  

The FRS construction/classification provides a rigorous classification of all rational full CFT$_{_2}$ s, where full means that these theories are defined at all genera, namely on all 2D cobordisms. 

The construction of CFT$_{_2}$s relies on the following:  

\begin{enumerate}  

\item A local geometric aspect. 

\item A global topological aspect. Taking in the spaces of conformal blocks one element such that this choice globally fits together under gluing of 2D cobordisms defined the sewing constraint.

\end{enumerate}

Because of this, the FRS formalism provides a mathematical formulation of the holographic correspondence. 

Given a vertex operator algebra (VOA) that controls the CFT locally, then, for RCFTs, their representation category is a MTC (in turn obtained from, either, a Hopf algebra, or loop groups representations), and the Reshetikhin-Turaev or Turaev-Viro construction defines the 3D TQFTs, namely the 3D CS theory.   

Full CFT$_{_2}$ s are relevant in String Theory, since these correspond to the vacua that string perturbation theory series are perturbing about. The FRS classification is part of the mathematical identification of the landscape of string theory vacua, 

RCFT$_{_2}$s are a small part of all possible CFT$_{_2}$s (indeed, usually, any deformed CFT$_{_2}$ is non-rational). 
However, RCFT$_{_2}$s (namely the WZW-model) plays an important role as the internal parts of string theory KK-compactifications.

\medskip 
\medskip 

\subsection{Jackiw-Teitelboim (JT)-gravity from CS-theory}  

In its first-order formulation, the Einstein-Hilbert action for 3D $SL(2,\mathbf{R})$ CS gravity is defined as the difference of two CS actions 

\begin{equation}   
S_{_{EH}}\ =\ S_{_{CS}}[A]-S_{_{CS}}[\bar A].  
\label{eq:ehcs3d}
\end{equation}

For AdS$_{_3}$ in Lorentzian signature, the connections $A$ and $\bar A$ are $\mathfrak{sl}(2,\mathbf{R})$-valued 1-forms. In the fundamental representation, the generators of the Lie algebra $\mathfrak{sl}(2,\mathbf{R})$ read   

\begin{equation}  
L_{_o}=\left(\begin{matrix}
    \frac{1}{2}\ \ \ \ \ 0\\
    0\ \ -\frac{1}{2}\\
    \end{matrix}\right)\ \ \ \  ,  \ \ \ \    L_{_1}=\left(\begin{matrix}
    \ 0\ \ \ \ \ \ \ 0\\
    -1\ \ \ \ \ \ 0\\
    \end{matrix}\right)\ \ \ \  ,  \ \ \ \ L_{_{-1}}=\left(\begin{matrix}
    0\ \ \ \ 1\\
    0\ \ \ \ 0\\
    \end{matrix}\right)\ \ \ ,
\end{equation}   
obeying the commutation relation

\begin{equation}  
\left[L_{_m}, L_{_n}\right]=(m-n)\ L_{_{m+n}}.   
\end{equation}  

The e.o.m. for the connections in \eqref{eq:ehcs3d} read   

\begin{equation}  
F=dA+A\wedge A=0\ \ \ \ \ \ \ ,\ \ \ \ \ \ \ \bar F=d\bar A+\bar A\wedge \bar A =0 
\end{equation}  
whose solution is pure gauge  

\begin{equation}  
A=g^{^{-1}}dg\ \ \ \ \ ,\ \ \ \ \ \bar A=\bar g^{^{-1}}d\bar g  \ \ \ \ \ \, \ \ \ \ g, \bar g\in\ \mathfrak{sl}(2,\mathbf{R}). 
\label{eq:solAAbar}
\end{equation}  

In terms of $(t,z,r)$-coordinates, where $r$ denotes the holographic direction, upon choosing suitable boundary conditions, \eqref{eq:solAAbar} reads as follows  

\begin{equation}  
A=b(r)(d+a(z))b(r)\ \ \ \ \ \ \ ,\ \ \ \ \ \ \ \ b(r)=r^{^{L_{_{o}}}} 
\end{equation}  

\begin{equation}  
\bar A=\bar b(r)(d+\bar a(z))\bar b(r)^{^{-1}}\ \ \ \ \ \ \ ,\ \ \ \ \ \ \ \bar b(r)=r^{^{\bar L_{_{o}}}}. 
\end{equation} 

In presence of a boundary, the CS action needs to be equipped with a suitable boundary term, ensuring \label{eq:ehcs3d} admits a well-defined variational principle  

\begin{equation}   
\delta S_{_{EH}}\ =\ \delta S_{_{CS}}[A]-\delta S_{_{CS}}[\bar A]    
\end{equation}  

\begin{equation}  
\Rightarrow\ \delta S_{_{EH}}+\delta S_{_{bdy}}\ =\ \frac{1}{4\pi G}\int_{_{\partial_{_{M_{_{3}}}}}}\ \epsilon_{_{ab}}f^{^a}\wedge\delta e^{^b},     
\end{equation}     
with $e^{^a}$ the source and $f^{^a}$ the expectation value of the dual operator. In particular, the operator dual to the boundary vielbein is the stress tensor   

\begin{equation}  
\delta S\ =\ 4\int_{_{\partial_{_{M_{_{3}}}}}}\ d^{^2}x\ (\text{det}\  e)\ T^{^i}_{_a}\delta e^{^b}.    
\end{equation}   

Interestingly, upon choosing a particular boundary term and subsequently taking the dimensional reduction of either CS action, we end up with a BF-theory

\begin{equation}  
\begin{aligned}
8\pi G S_{_{CS}}\ &=\ \frac{1}{2}\int_{_{M_{_{3}}}}d^{^3}x\ \epsilon^{^{\mu\nu\rho}}\ \text{Tr}\left(A_{_{\mu}}\partial_{_{\nu}}A_{_{\rho}}+\frac{2}{3}A_{_{\mu}}A_{{\nu}}A_{_{\rho}}\right) \\
&=\int_{_{M_{_{3}}}}d^{^3}x\ \text{Tr}\left(A_{_{\varphi}}F_{_{tr}}+A_{_r}\partial_{_{\varphi}}A_{_t}\right)+\frac{1}{2}\oint_{_{\partial M_{_{3}}}}\text{Tr} A_{_t}^{^2}. 
\label{eq:bdyth}
\end{aligned}
\end{equation} 

The corresponding 2D theory living on the conformal boundary is obtained upon taking the $\underset{r\rightarrow 0}{\lim}$ of \eqref{eq:bdyth}. Imposing the following boundary conditions   

\begin{equation}  
A_{_t}=A_{_{\varphi}}\bigg|_{_{\partial M_{_{3}}}}\ \ \ \ \text{such that}\ \ \ \phi\equiv A_{_{\varphi}}  \ \ \ \text{and}\ \ \ \partial_{_{\varphi}}=0,   
\end{equation}  
\eqref{eq:bdyth} turns into 

\begin{equation}   
\boxed{\ \ S_{_{BF}}\ =\ \int_{_{M_{_{2}}}}\ \text{Tr}\left(\phi F\right) +\frac{1}{2}\oint_{_{\partial M_{_{2}}}}\text{Tr} \left(\phi^{^2}\right)\color{white}\bigg]\ },   
\label{eq:bdythBF} 
\end{equation} 
where the first term is the usual BF-term for the gauge group $G=SL(2, \mathbf{R})$, whereas the second term is the Schwarzian-type boundary term. 

The degrees of freedom in this resulting action, \eqref{eq:bdythBF}, are: a gauge field $A_{_{\mu}}$, with associated field strength $F_{_{\mu\nu}}$, along with an $SL(2, \mathbf{R})$-valued scalar field $\phi$. The latter corresponds to a boundary degree of freedom (also known as the particle-on-a-group).

\=subsection{Branes in QFTs and String Theory}   

\subsection*{D-branes}   

In the cobordism formulation of a higher-dimensional QFT, a D-brane is a type of data assigned to the QFTs boudaries of cobordisms.

Key examples include the following: 

\begin{enumerate}   

\item   RCFT$_{_2}$s (the full rational CFT$_{_2}$, are classified using the FRS-formalism\footnote{FRS stands for Fuchs, Runke, and Schweigert, who provided a rigorous construction and classification of all rational full 2D CFTs, where full means that these theories are defined at all genera, namely on all 2D cobordisms.}, by means of:  

\begin{itemize}  

\item A MTC, ${\cal C}$, the category of representations of the VOA of the CFT$_{_2}$.  

\item A special symmetric Frobenius algebra object, $A$, internal to ${\cal C}$.

\end{itemize}  

Hence, a D-brane of the theory is an $A$-module in ${\cal C}$.

\item   The $(\infty,1)$-categorical formulation of TFT$_{_2}$s, leads to the A- and B-models. The A-model's branes form the Fukaya category of the target space. The B-model's branes form a stable $(\infty,1)$-category of complexes. The WZW-model is a particular example of this, where the target space is a simple Lie group $X\equiv G$, and the background field is a circle 2-bundle with connection on $G$, representing the background field, known as the Kapustin-Rozansky (KR) field. 
The study of Kontsevich's homological algebra reformulation of mirror symmetry, and the study of derived D-brane categories constituted a field of pure maths on its own.

\item    In type-IIA SUGRA, they are the D0-, D2-, D4-, D6-, D8-, F1-, NS5-branes.

\item   In type-IIB SUGRA, they are D1-, D3-, D5-, D7-, F1-, NS5-, $(p,q)$-5, $[p,q]$-7 branes.

\item In M-theory, M2-, M5-, M9-, M-wave.

\end{enumerate}

The topological part of D-brane actions behave as generalised symmetry operators. In particular, they exhibit:  

\begin{enumerate} 

\item Confinement/deconfinement transition, via the Hanany-Witten effect via D3-brane nucleation.

\begin{figure}[ht!]    
\begin{center} 
  \includegraphics[scale=0.7]{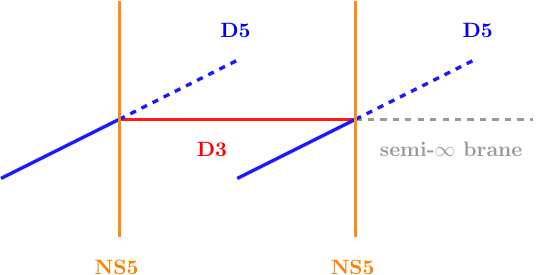}
\label{fig:lr1}     
\caption{\small}
\end{center}      
\end{figure}   

The Hanany-Witten mechanism, \cite{Hanany:1996ie}, is realised in brane web setups. As shown in figure \ref{fig:lr1}, it consists in the extension of F1-branes between D3- and D5-branes as they pass across each other. This behaviour is naturally arising within the context of 4D QFTs with mixed t'Hooft anomaly. We will briefly outline this example in Part III of the present treatment.  

\item Fusion rules via the Myers effect. This is an additional feature connecting braneweb constructions with symmetry generators in 4D QFTs. In particular, the Myers effect, well known in the context of D0-D2 branes, is analogous to the fusion rules arising from gauging higher-form symmetries in QFT. 

\end{enumerate}

\subsubsection{(P)SU(N) in 4D \texorpdfstring{${\cal N}=1$}{} Super Yang Mills}  \label{sec:braneex}    

To put in practice what we have said up to now, we briefly reproduce an example outlined in \cite{Apruzzi:2022rei}, in the context of symmetries of 4D QFTs in relation to D-branes structures. The 4D theory is a gauge theory with a 0-form symmetry, $\mathbb{Z}_{_M}$, with background field $A_{_1}$, and a 1-form symmetry, $Z_{_M}$, with background field $B_{_2}$, related by a mixed 't Hooft anomaly  

\begin{equation}  
{\cal A}=-\frac{2\pi}{M}\int A\cup\frac{{\cal P}(B_{_2})}{2},   
\label{eq:anomaly}    
\end{equation}    
where ${\cal P}(B_{_2})$ is the Pontrjagin square of $B_{_2}$.  

The generators of the 0- and 1-form symmetries are generated by 3D and 2D topological defects $D_{_3}^{^g}(M_{_3}), D_{_2}^{^h}(M_{_2})$, respectively, such that their fusion rules obey group-like structure relations      

\begin{equation}   
D_{_p}^{^{g_{_1}}}(M_{_p})\ \otimes\ D_{_p}^{^{g_{_2}}}(M_{_p}) = D_{_p}^{^{g_{_1} g_{_2}}}(M_{_p}).      
\end{equation}    

Due to the anomaly \eqref{eq:anomaly}, the generators of the 0-form symmetry transform nontrivially, namely

\begin{equation}  
D_{_3}^{^g}(M_{_3})\ \rightarrow\ D_{_3}^{^g}(M_{_3})\ \exp\left(\int_{_{M_{_4}}}-\frac{2\pi i}{M}\frac{{\cal P}(B_{_2})}{2}\right),       
\label{eq:transform0}   
\end{equation}    
with $\partial M_{_4}=M_{_3}$. In particular, when gauging the 1-form symmetry, \eqref{eq:transform0} is inconsistent. To solve this problem, we need to dress $D_{_3}(M_{_3})$ with a minimal TQFT ${\cal A}^{^{M,p}}$ featuring a 1-form symmetry $Z_{_{M}}$ that cancels the anomaly. For this example, this is a $U(1)_{_M}$ theory, and the dressed defect is redefined as follows  

\begin{equation}   
{\cal N}_{_3}^{(1)}\ =\ D_{_3}^{^{(1)}}\ \otimes\ {\cal A}^{^{M,1}},     
\end{equation}   
with non-invertible fusion rules  

\begin{equation}  
{\cal N}_{_3}^{^{(1)}}\ \otimes\ {\cal N}^{^{(1)}}\ =\ {\cal A}^{^{M,2}}{\cal N}_{_3}^{^{(2)}}  \ \ \ \ \ \ \ ,\ \ \ \ \ \ \ {\cal N}_{_3}^{^{(1)}}\ \otimes\ {\cal N}^{^{(1)}\dag}\ =\ \sum_{_{M_{_2}\in H_{_2}(M_{_3},\mathbb{Z}_{_M})}}\frac{(-1)^{^{Q(M_{_2})}}D_{_2}(M_{_2})}{|H^{^0}(M_{_3},\mathbb{Z}_{_M})|},      
\end{equation}  
where 

\begin{equation}  
D_{_2}(M_{_2})\ =\ e^{^{i 2\pi\ \int_{_{M_{_2}}} \frac{b_{_2}}{M}}},    
\end{equation}   
with $b_{_2}$ the gauge field for the 1-form symmetry. In relation to type-IIB string theory, these symmetry operators ${\cal N}_{_3}^{^{(1)}}$ are associated to D5-branes, whereas 't Hooft operators are associated to D3-branes, \cite{Apruzzi:2022rei}. We will be referring to this example for later convenience in Part III when introducing Symmetry Topological Field Theories (SymTFTs).

\subsection{From brane webs to quiver gauge theories}

\begin{figure}[ht!]    
\begin{center} 
\includegraphics[scale=1]{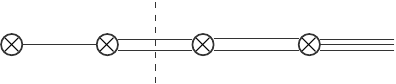} \ \ \ \ \ \ \ \ \ \ \ \ \ \ \ \ \ \ \ \ \ \ \ \ 
\includegraphics[scale=1]{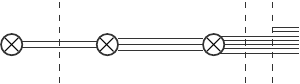} 
\label{fig:lr}   
\caption{\small}
\end{center}      
\end{figure}

Quivers encode the Lagrangian of a supersymmetric gauge theory. They take different forms according to the number of supercharges. For the case in which there are 8 of them, quiver diagrams encode the gauge symmetries, flavour symmetries, and matter field content. The fields featuring in a certain theory are contained in vector multiplets and hypermultiplets. The field content changes according to the dimensionality of the theory in question.  

The quiver diagram encodes the representations of the gauge and matter fields under which they transform. Specifically, the vector multiplets transform under the adjoint representation of the gauge group representation, whereas the hypermultiplets transform under the bifundamental representation of the gauge group and the flavour group.

\begin{figure}[ht!]    
\begin{center} 
\includegraphics[scale=1]{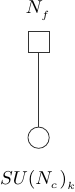} \ \ \ \ \ \ \ \ \ \ \ \ \ \ \ \ \ \ \ \ \ \ \ \ 
\includegraphics[scale=1.1]{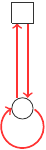} 
\label{fig:lrw}   
\caption{\small}
\end{center}      
\end{figure}

Importantly, from the quiver, we can read the superpotential. In figure \ref{fig:lrw} we show the example of a framed Nakajima quiver variety. In this case, the corresponding superpotential reads as follows

\begin{equation}   
{\cal W}=\text{Tr}\left(A^{^{\alpha}}_{_i}\Phi^{^i}_{_j}B^{^{j\beta}}\right),    
\end{equation}  
where $A^{^{\alpha}}_{_j}, \Phi^{^i}_{_j}$ and $B^{^{j\beta}}$ denote the chiral multiplets, the anti-chiral multiplets, and the adjoint chirals, respectively. The indices $i,j$ are related to the gauge group, whereas the indices $\alpha, \beta$ are related to the flavour group.

\subsection{Beyond holographic duality: adding boundaries}  

\subsection*{Boundary Conditions (BCs) of \texorpdfstring{${\cal N}=4$}{} SCFTs}   

$AdS_{_4}\times S_{_2}^{^{(A)}}\times S_{_2}^{^{(B)}}$ fibered over a Riemann surface, $\Sigma$, the corresponding metric solution reads

\begin{equation}   
ds^{^2}=f_{_4}^{^2}ds^{^2}_{_{AdS_{_4}}}+f_{_1}^{^2}ds^{^2}_{_{S_{_2}^{^{(A)}}}}+f_{_2}^{^2}ds^{^2}_{_{S_{_2}^{^{(B)}}}}+4\rho^{^2}\left(dr^{^2}+r^{^2}d\theta^{^2}\right),      
\end{equation}   
where the metric functions $f_{_i}, \rho$ are functions of $h_{_1}, h_{_2}$ defined on $\Sigma$, via $(r,\theta)$. As explained in \cite{Raamsdonk:2020tin}, the field theory in question is uniquely defined by the following set of parameters  

\begin{equation}  
N_{_{D5}}^{^{(A)}}\overset{def}{=}\frac{1}{\sqrt{g\ }}c_{_A}\ \ \ \ \ \ \ \ \ \ \ \ , \ \ \ \ \ \ \ \ \ \ \ \ N_{_{BS5}}^{^{(B)}}\overset{def}{=}\sqrt{g\ }d_{_B}
\end{equation}

\begin{equation}  
L_{_A}\overset{def}{=}\sqrt{g\ }l_{_A}+\frac{2}{\pi}\sum_{_B}N_{_{D5}}^{^{(B)}}\tan^{^{-1}}\left(\frac{l_{_A}}{k_{_B}}\right)\ \ \ , \ \ \ K_{_B}\overset{def}{=}\frac{1}{\sqrt{g\ }}k_{_B}+\frac{2}{\pi}\sum_{_A}N_{_{NS5}}^{^{(B)}}\tan^{^{-1}}\left(\frac{k_{_B}}{l_{_A}}\right).
\end{equation}  

At this point, one could further redefine $\hat L_{_i}$ as the number of D3-branes ending from the right on the $i^{^{th}}$ D5-brane plus the number of NS5-branes to the left of this D5-brane, and with $\hat K_{_i}$ the number of D3-branes ending on the $i^{^{th}}$ NS5-brane plus the number of D5-branes to the left of this NS5-brane, leading to the brane-web intersection reported in figure \ref{fig:lr}.

\begin{figure}[ht!]    
\begin{center} 
\includegraphics[scale=1]{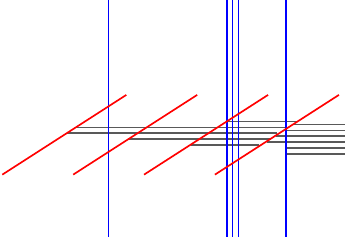} \ \ \ \ \ \ \ \ \ \ \ \ \ \ \ \ \ \ \ \ \ \ \ \ 
\includegraphics[scale=1]{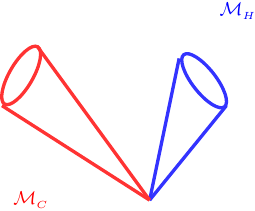} 
\label{fig:lr}   
\caption{\small}
\end{center}      
\end{figure} 

In a SCFT with ${\cal N}=4$ supersymmetry, the Higgs and Coulomb branch are both conical and meet at their common origin. In general, there might also be mixed branches.

\subsection*{\texorpdfstring{$\frac{1}{2}$}{}-BPS BCs in \texorpdfstring{${\cal N}=4$}{} SYM}    

$\frac{1}{2}$-BPS BCs in ${\cal N}=4$ SYM with gauge group $G$ admit a general classification in terms of a triple, $(\rho, H, {\cal B})$, \cite{Gaiotto:2008sa}, where 

\begin{equation}   
\rho:\ \mathfrak{su}(2)\ \rightarrow\ \mathfrak{g}  
\label{eq:rho}
\end{equation}
denotes the embedding of the Lie algebra of $SU(2)$ in the Lie algebra of $G$, $H$ is a subgroup of $G$ resulting from the symmetry breaking induced by \eqref{eq:rho}, whereas ${\cal B}$ denotes the 3D boundary field theory with symmetry $H$.

\subsubsection{Islands in JT-gravity}
The holographic description of the entanglement entropy for Hawking radiation in arbitrary spacetime dimensions can be described in 3 different ways. This follows from the extension of the 2d JT-gravity model to arbitrary dimensions relying upon the Randall-Sundrum model with a brane embedded in AdS$_{d+1}$. Maldacena et al.'s treatment \cite{JM2} is equivalent to the brane (EFT) description. The bulk perspective shows that it is indeed reasonable to extend JT-gravity results to higher-dimensions relying upon dualities. In particular, wedge holography \cite{Fujita:2011fp} looks very promising for addressing vacuum transitions in higher dimensions.

\begin{enumerate} 

\item{ \underline{Boundary description}

 \begin{figure}[ht!] 
  \begin{center} 
\includegraphics[scale=0.7]{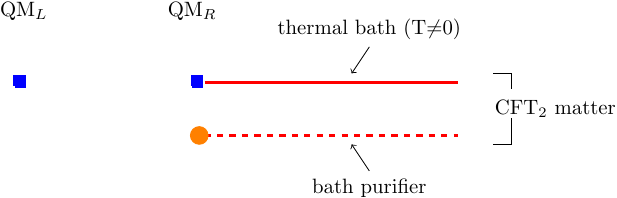}
\caption{Both of the QMs and field theory systems are prepared in independent thermofield double states.} 
\end{center} 
\end{figure} 

} 

\item{\underline{Brane description}

 \begin{figure}[ht!] 
  \begin{center}
\includegraphics[scale=0.7]{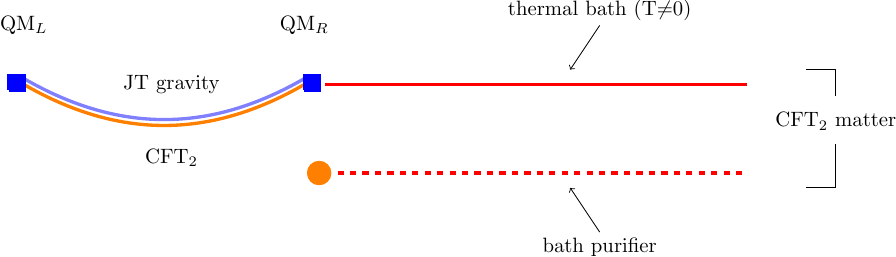}
\caption{Introducing a 2D holographic geometry (JT-gravity), dual to the entangled state of QM$_{L}$ and QM$_{R}$. The CFT$_{2}$ coupling to JT-gravity is the same as the one featuring in the bath.} 
\end{center} 
\end{figure} 
}

\item{\underline{Bulk description}

 \begin{figure}[ht!] 
  \begin{center}
\includegraphics[scale=0.6]{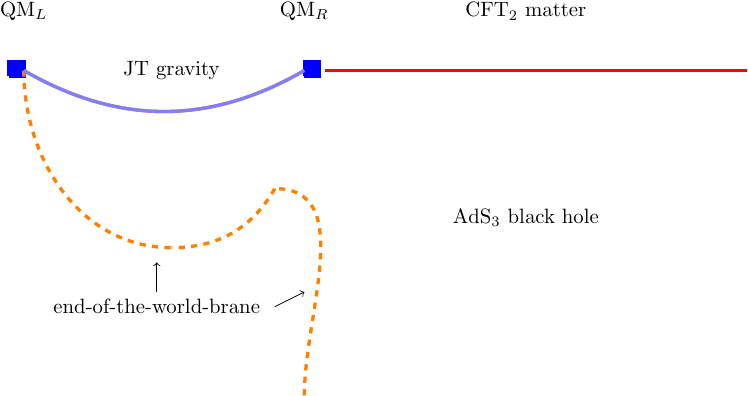}    
\caption{The doubly-holographic description, where the holographic CFT$_{2}$ is replaced by a bulk region with geometry of an AdS$_{3}$ black hole. }    
\end{center} 
\end{figure} 

} 
\end{enumerate}

\subsubsection{Conformal interfaces as boundary states} 

The idea of the replica trick, at first applied within the context of statistical field theory and condensed matter physics, basically consists in taking $n$-copies of a certain system and relating its behaviour to that of the original system, i.e for $n=1$. The reason for doing so is that, in principle, there might be noncommutative behaviour emerging from permutations of different subregions of the $n$-copied system, which in turn would highlight the presence of interesting entangling features that are in turn scale-dependent, i.e. their behaviour is determined by the size of one copy. 

Translating this idea in a suitable language for our purposes, i.e. describing quantum transition in between different spacetimes,\footnote{and as it will become clear later on in the treatment.} there is one additional key point to bare in mind. Quantum transitions are really meant to describe \emph{local} spacetime transformations. As an example, describing uptunneling from AdS would really mean considering a portion of the domain at a given instant of time (in global coordinates) and then defining its corresponding density matrix by post-selection according to Takayanagi et al.'s prescription discussed in the previous section. The final state $|\varphi>$ will be defined such that the covariant holographic entanglement entropy satisfies the 2-step minimisation-extremisation (min-extr) of the expectation value of $Area(\Gamma_{A})$. $\rho_{A}$ is therefore the density matrix associated to the states defined on such subregion at the boundary of AdS. Being this a portion of the entire domain at $r_{\infty}$, we can think of $A$ as arising by quotienting w.r.t. $Z_{n}$-symmetry, with $n$ determined by the particular choice of the domain $A$. However, as also explained in the appendix, even for the simplest $U(1)$-case, an anomaly emerges, since the symmetry is broken in the quantum theory by the state and the partition function. Restoring it at the level of the partition function requires adding Wilson loops, i.e. additional operators to the quantum theory. In turn, these operators can be thought of as the lower-dimensional counterparts of background fields living in one spatial dimension higher. Their coupling will ultimately lead to the definition of Chern-Simons terms, i.e. defects added to the bulk theory in order to compensate for the anomaly featuring in the original QM theory. This procedure is known as the \emph{anomaly-inflow mechanism}. 

Our main interest is addressing the issue form the quantum gravity perspective. In order to do so, we rely upon the holographic dictionary of the AdS$_{d+1}$/CFT$_{d}$ correspondence, at first starting with $d=2$, from which some major considerations can be extracted and later on extended to higher-dimensional setups. 

We start by considering 2 CFTs separated by a conformal interface, which basically provide a special example of extended operators, such as Wilson lines, D-branes, or other extended sources. Upon exchanging the role of space and time, these conformal boundaries can be mapped to boundary states, following the so-called \emph{folding trick}. As a result of this procedure, we end up with a BCFT. The boundary conditions defined at the interface can be suitably chosen accordingly to the setup we wish to describe. 2 particular examples are provided by the totally-reflecting and the topological BCs. 

The boundary arising from the folding trick can really be thought of as a special case of a \emph{defect}, for which an entropy can be derived from the entanglement entropy of a $(1+1)$ CFT. Specifically, the entanglement entropies with and without the boundary defect resepctively read 

\begin{equation} 
S_{1} 
\ 
= 
\ 
\frac{c}{6}\ln\frac{L}{\epsilon}\ +\ \ln(g_{B})\ +\ \alpha_{0} 
\ \ \ 
, 
\ \ \ 
S_{2} 
\ 
= 
\ 
\frac{c}{6}\ln\frac{L}{\epsilon}\ +\ \alpha_{0}    
\label{eq:entent}    
\end{equation} 
where $c$ is the central charge of the CFT$_{_2}$, $\epsilon$ is the UV cutoff introduced for regularisation purposes, $L$ is the length of the spatial region on which the entanglement entropy is being evaluated, $\alpha_{_0}$ is an overall constant, and $g_{_B}$ is the ground state degeneracy. Upon taking the difference in between the two quantities in \eqref{eq:entent}, we are left with the entropy of the defect   

\begin{equation} 
S_{bndry} 
\ 
= 
\ 
S_{1}\ -\ S_{2} 
\ 
= 
\ 
\ln(g_{B}) 
\ \ \ 
. 
\end{equation} 

The defect, therefore, contributes with an additional term to the partition function, which asymptotically reads 

\begin{equation} 
S 
\ 
= 
\ 
\ln\ g_{B}  
\ \ \ 
, 
\ \ \ 
g_{B} 
\ 
= 
\ 
<1> 
\ 
= 
\ 
<0|\ D\ |0> 
\ \ \ 
\end{equation}

with $D$ defining the \emph{defect operator}

\begin{equation} 
D 
\ 
= 
\ 
g|0><0|\ +\ ... 
\ \ \ 
. 
\end{equation} 

In string theory, e.g., the role of the $g$-factor is played by the tension of a D-brane. It is interesting to note that this is exactly the term entering in the Hamiltonian formulation of quantum transitions by implementing the junction conditions, with the brane tension separating different spacetimes. 

The \emph{degeneracy} of the ground state is therefore defined by the overlap between the boundary state and the vacuum

\begin{equation} 
g_{B} 
\ 
= 
\ 
<0|B> 
\ \ \ 
. 
\end{equation} 

The \emph{g-theorem} states that the value of $g_{B}$ must decrease under the boundary RG-flow \cite{Azeyanagi:2007qj}.

Given $g_{B}$, the \emph{distance} in between the two original CFTs is defined as follows 

\begin{equation} 
d(1,2) 
\ 
= 
\ 
\text{min}\sqrt{\ln g_{B}} 
\ \ \ 
. 
\end{equation}

From this, we therefore understand that BCFTs, such as the case of interest for us being $\rho_{A}=\text{Tr}_{B}\rho$, are incredibly rich in structure and carry many interesting and hidden features that will turn out essential in the forthcoming sections. In particular, the \emph{boundaries} of the BCFT carry d.o.f. associated to the degeneracy of the ground state and and therefore signal whether a defect needs to be added to the path-integral evaluation when taking n-copies of the given subregion where the BCFT lives. Such defect is, in turn associated to an anomaly induced by the anomaly-inflow mechanism of a bulk background field which needs to be added to the quotiented geometry. 

In terms of the von Neumann formulation, the entanglement entropy reads 

\begin{equation}
S_{A} 
\ 
= 
\ 
    - \text{Tr}_{A}\ (\rho_{A}\ln(\rho_{A}) )
    \ 
    = 
    \ 
    - 
    \ 
    \lim_{n\rightarrow 1}\ \frac{d}{dn}\text{Tr}_{A}(\rho_{A}^{n}) 
    \ \ \ 
    , 
    \label{eq:VNE} 
\end{equation} 

where in the last step of (\ref{eq:VNE}) we implemented the \emph{replica trick}. 

For integer values of $n$-replicas, $\text{Tr}(\rho_{A}^{n})$ is the partition function on a Riemann surface, i.e. 

\begin{equation} 
\text{Tr}_{A}(\rho_{A}^{n}) 
\ 
\sim
\ 
{\cal Z}_{n} 
\ 
= 
\ 
\int\ [d\varphi_{1}...d\varphi_{n}]_{{\cal M}_{n}}\ \exp\left[-\int_{{\cal M}_{n}}\ d^{2}x\ {\cal L}^{(n)}[\varphi_{1},...,\varphi_{n}](x)\right] 
\ \ \ 
. 
\label{eq:pf} 
\end{equation} 

Invariance of the boundary theory under $Z_{n}$-symmetry corresponds to the Lagrangian being preserved under permutations of the n-copies of the QFT, i.e. 

\begin{equation} 
{\cal L}^{(n)}[\varphi_{1},...,\varphi_{n}](x)...,\sigma\varphi_{n}](x) 
\ 
= 
\ 
{\cal L}[\varphi_{1}](x)\ +\ ...\ +\ {\cal L}[\varphi_{n}](x) 
\ \ \ 
, 
\end{equation} 
with $\sigma\varphi_{i}\ =\ \varphi_{i+1 \text{mod n}}$. The action of $\sigma$ belonging the symmetry group ${\cal Z}_{n}$ and its inverse $\sigma^{-1}$ are associated to a \emph{twist fields}, whose correlation function is proportional to (\ref{eq:pf}). The partition function of the n-copied system can therefore be used for redefining the entanglement entropy in the $\lim_{n\rightarrow 1}$, and (\ref{eq:VNE}) therefore reduces to 

\begin{equation} 
S_{A} 
\ 
= 
\ 
- 
\ 
\lim_{n\rightarrow 1} \ \frac{d}{d n}\ {\cal Z}_{n} 
\ \ \ 
. 
\end{equation}

With this setup, we are now ready to derive the generalised expression for the entanglement entropy, where the contribution from quantum extremal surfaces adds to the von Neumann entropy.

\part{From knot invariants to QFTs}

\section{Summary and Motivations of Part III}

\subsection{Motivations}

Since first encountered, \cite{Witten:1995zh}, 6D ${\cal N}=(2,0)$ SCFTs, also referred to as \emph{theory X},\footnote{Please see appendix \ref{sec:rqft} for some brief overview of its importance in pure mathematics and the Geometric Langlands Program.}, have attracted great interest in, both, the mathematics and high-energy physics community alike. From the high-energy physics point of view, they are essential for deriving lower-dimensional SCFTs, \cite{Witten:2007ct,Chacaltana:2012zy,Tachikawa:2013hya,Moore}, whose features are much better understood\footnote{We review some key features of its 4D descendants in appendix \ref{sec:6d4d}.}. On the other hand, from a purely mathematical perspective, they provide an extremely interesting and rich setup where to exploit the present-day knowledge of representation theory, \cite{Witten:2009at,DBZ}.

In recent years, the two communities have been brought together by fascinating developments on either side, \cite{Bah:2022wot}, highlighting the importance of bridging the gap in between different formalisms, with category theory playing a key role in pursuing this task. Indeed, the pioneering works of \cite{Moore:1988qv,Moore:1988ss,Verlinde:1988sn} led to the discovery of modular tensor categories in anyon gauging theory. Since then, application of such deep mathematical language inspired applicability to higher-dimensional QFTs, by identifying their symmetries with topological defects, \cite{Gaiotto:2014kfa}. Of particular importance to us is the richness of higher-categorical theory applied to topological orders (TOs), \cite{Kong:2020cie,Kong:2022cpy,Kong:2019byq,Kong:2019cuu,Kong:lastbutone,Gaiotto:2019xmp,Johnson-Freyd:2021tbq,Johnson-Freyd:2020usu,MYu}, and symmetry topological field theories (SymTFTs), \cite{Freed:2012bs,Freed:2022qnc}. 

Embracing the thrust of recent developments in the field, the present work is meant to be the first of a series of papers where the author applies the techniques of category theory to furthering the understanding of 6D ${\cal N}=(2,0)$ SCFTs and theories derived from them. In doing so, we will attempt, wherever possible, to provide an understanding of the procedure as well as of the outcome of our findings, from, both, a mathematician's and a theoretical physicist's point of view. Hopefully, this will further motivate both communities to benefit form each other's influence. 

This part of our work aims at furthering the understanding of the categorical structure arising from dimensional reduction of 6D ${\cal N}=(2,0)$ SCFTs. In particular, our aim is that of identifying a criterion to distinguish between intrinsic and non-intrinsic non-invertible symmetries\footnote{See, for example, \cite{Bashmakov:2022jtl,Bashmakov:2022uek,Bhardwaj:2022kot,Bhardwaj:2022yxj,9,Kaidi:2021xfk,Kaidi:2022uux,Kaidi:2022cpf,Choi:2022zal,Choi:2021kmx} for explanation on the current state of the art.} in terms of the quantum dimension\footnote{The total quantum dimension is defined from the braiding and fusion rules of the underlying categorical structure of the topological symmetries of the theory in question. Further explanation will be provided in due course.}. In doing so, we rely upon their description in terms of relative field theories, as described in \cite{Bashmakov:2022uek} in terms of the Freed-Moore-Teleman setup, \cite{Freed:2012bs,Freed:2022qnc}.

As previously argued, class ${\cal S}$ theories are special in the sense that they are absolute, and admit a quiver gauge theory description. Their TFT counterparts are the so-called minimal models, which, in turn are characterised by a higher-categorical structure, as reviewed in section \ref{sec:2}.
The multiplicities of the different superselection sectors might exceed unity, thereby signalling the possibility of information storage, correspondingly leading to a decrease in ${\cal D}$. Given that the definition of the entanglement entropy involves a partial tracing of the total boundary state, it might be that the R\'enyi entropy shows a characteristic periodicity $\mathfrak{p}$ signalling that the state defining the reduced density matrix can be factorised under suitable redefinition of the modular operators involved in the definition of the link state itself.

Our main proposal is that the total quantum dimension of the gaugeable algebra implementing the gauging in the bulk Sym TFT is a key quantity to distinguish in between non-invertibility being intrinsic or non-intrinsic. Specifically, for the intrinsic case, multiplicity is greater w.r.t. the non-intrinsic case, thereby signalling the possibility for additional d.o.f. to be stored in certain superselection sectors of the resulting absolute theory. 

Our results extend arguments proposed in \cite{Kaidi:2022cpf}, where the authors also proposed a way of distinguishing intrinsic from non-intrinsic non-invertibility in 2D by means of the quantum dimension of the non-invertible defect.

\subsection{Outline and main results}

The present Part contains the main findings first presented in an original work by the same author, and serves as a bridge between the preliminary sections of the present work and the remainder of this thesis. The content of the following two sections is as follows:  

\begin{enumerate}

\item   In section \ref{sec:symtft} we introduce the symmetry topological field theory (SymTFT) construction proposed by Freed, Moore, and Teleman for describing supersymmetric gauge theories descending from dimensional reduction of 6d ${\cal N}=(2,0)$ SCFTs. In particular, we focus on the non-invertible defects arising in between class ${\cal S}$ theories. Our main focus is on the following: 

\begin{itemize}   

\item We highlight the importance of defining the Drinfeld center for the case in which non-invertibility is non-intrinsic, and relate this to the need of generalising the 3D mirror symmetry/homological mirror symmetry correspondence explained in section \ref{sec:otherdualities}. This prepares the stage for the more detailed analysis to which sections \ref{sec:cd}, \ref{sec:clt}, \ref{sec:333}, and \ref{sec:last} are devoted. 

\item Exploiting the relation in between the fusion rules of such defects and the partition function on the absolute 4D theory, we show that the quantum dimension of the gaugeable algebra in the intrinsic case always exceeds that featuring in the non-intrinsic case. We thereby propose this as a general criterion to be applied to class ${\cal S}$ theories descending from 6D ${\cal N}=(2,0)$ SCFTs.    

\end{itemize}

\item   In section \ref{sec:inst}, we then turn to one of the key topics bridging with the remainder of our treatment, namely instantons. After having overviewed how they arise in QQM, QFT, 4D supersymmetric gauge theories, and oriented string theory, we explain their relation to the ADHM construction, and the Higgs branch moduli space. In the concluding part of the same section, we then introduce the WKB approximation as a method used for determining approximate solutions to wall-crossing setups, with particular focus on two settings of interest for us, namely Hitchin systems and vacuum transitions.

\end{enumerate}

\section{Quantum Field Theory from Category Theory} \label{sec:symtft}

We now turn to explaining how QFT can be mathematically described in terms of Category Theory. 

To begin with, we start off by listing some of the ingredients that are most relevant to our analysis\footnote{As we shall see, more richness will be added in due course in our treatment. The aim of the table is simply that of helping the reader in gaining intuition for the setup and the motivations behind the procedure adopted.}, briefly outlined in the table below.

\begin{equation}
\begin{aligned}
&\text{\underline{Particle Physics}} \ \ \ \color{white}spacespacespacespace\color{black} \ \ \text{\underline{Category Theory}} \\
&\\
&\text{Gauge Theory}\ \ \ \color{white}spacespacespacespaces\color{black} \ \ \text{Algebras}\\
&\text{\color{red}Operators}\ \ \ \color{white}spacespacespacespacespace\color{black} \ \ \text{Homologies}\\
&\text{Quivers}\ \ \ \color{white}spacespacespacespacespacesp\color{black} \ \ \text{Quivers/Algebraic varieties}\\
&\text{Hilbert Series}\ \ \ \color{white}spacespacespacespacece\color{black} \ \ \text{Hilbert Series}\\
&\text{\color{red}SUSY}\ \ \ \color{white}spacespacespacespacecespacesp\color{black} \ \ \text{SUSY}\\
\end{aligned}
\nonumber
\end{equation}  

\medskip 

\medskip 

This shows some quantities/entities encountered in the realm of Particle Physics and their Category Theory counterpart. We have highlighted two of them in red. The reason for this will be more clear in Part V, but the familiar reader will certainly notice that some motivation was already hinted at in section \ref{sec:rqft}. For the moment, we simply recall what was mentioned in the Introduction, namely that the Hilbert series and ring homologies are key complementary ingredients towards understanding a given QFT. Among the two, though, the latter is the one that is more directly related to understanding the underlying categorical structure of the theory in question, and therefore constitutes the ultimate objective one wishes to obtain. On the other hand, the Hilbert series, is obtained from ring homologies, but hides some key categorical features that are only accessible via the homological perspective.\footnote{Indeed, it is known that different ring homologies could lead to the same Hilbert series, \cite{Cremonesi:2014vla}.}

This intermediate section is mostly an overview of \cite{ Freed:2022qnc,Freed:2012bs,Freed:2022qnc}, and is needed in order to understand the analysis carried out in the concluding part of our work, \ref{sec:5}. This section is structured as follows:  

\begin{enumerate}  

\item At first, we review mathematical gauge theory and the definition of moduli spaces therein, \cite{Freed:2022qnc}. 

\item Then, we recall the notion of relative field theories, \cite{Freed:2012bs}, which can also be adapted to gauge theories. 

\item In conclusion, we explain how this fits with the higher-categorical prescription of Freed, Moore and Teleman, \cite{Freed:2022qnc}, and how gauging can be implemented in terms of gaugeable homomorphisms in the Symmetry Topological Field Theory categorising the symmetries of a given QFT, \cite{TJF}. 

\end{enumerate}  

For reasons that will become clear by the end of this section, the crucial part of the higher-categorical perspective can be summarised as follows: 

\begin{itemize}  

\item Keeping track of the identity. 

\item Embedding the gauged algebra. 

\end{itemize}  

Most importantly, they both lie within the notion of \emph{abelianisation}. In section \ref{sec:5} we will see explicitly how they arise within the context of supersymmetric quiver varieties, \cite{Dimofte:2018abu,Bullimore:2015lsa}.


\subsection{The Freed-Moore-Teleman construction}

\subsubsection{Gauging in relative QFTs}  \label{sec:FMT}

\begin{figure}[ht!]  
\begin{center}
\includegraphics[scale=0.9]{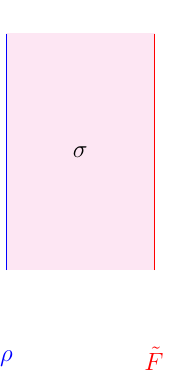} 
\ \ \ \ \ \ \ \ 
\includegraphics[scale=0.9]{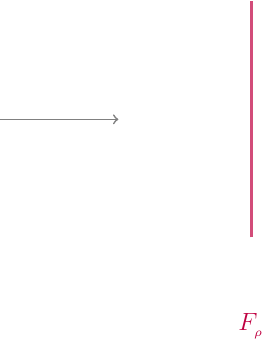}  
\caption{\small The Freed-Moore-Teleman setup, with $\tilde F$ denoting a relative QFT. Specifying the topological data $(\sigma,\rho)$, the resulting theory, $\tilde F_{_{\rho}}$ is absolute.}
\label{fig:FMT}    
\end{center}  
\end{figure}

In the formulation of \cite{Freed:2012bs}, a \emph{relative} field theory, $\tilde F$, requires additional topological data in order to be fully specified. Such data is encoded in a pair $(\sigma,\rho)$, referred to as \emph{quiche}. $\sigma$ is the \emph{symmetry topological field theory} (SymTFT), whereas $\rho$ geometrises the choice of boundary conditions for the fields defining the relative theory $\tilde F$. The overall system, depicted in figure \ref{fig:FMT}, gives rise to an \emph{absolute} QFT, $F_{_{\rho}}$.

In mathematical terms, the description outlined above can be formulated in terms of bordism in the following way. Fixing $N\in\mathbb{Z}^{^{\ge 0}}$, then a quiche is a pair $(\sigma, \rho)$ in which $\sigma:\text{Bord}_{_{N+1}}(F)\rightarrow{\cal C}$ is an $N+1$-dimensional TFT and $\rho$ is a right topological $\sigma$-module. The quiche is $N$-dimensional, hence it shares the same dimensionality as the theory on which it acts. Let $F$ be an $N$-dimensional field theory. A $(\sigma,\rho)$-\emph{module structure} on $F$ is a pair $(\tilde F, \theta)$, in which $\tilde F$ is a left $\sigma$-module and $\theta$ is an isomorphism 

\begin{equation}   
\theta\ : \rho\ \otimes_{_{\sigma}}\ \tilde F\xrightarrow{\ \ \simeq\ \ }\ F_{_{\rho}},  
\label{eq:LHS}  
\end{equation}   
of absolute $N$-dimensional theories, with \eqref{eq:LHS} defining the dimensional reduction leading to the absolute theory. $\sigma$ needs only be a \emph{once-categorified} $N$-dimensional theory, whereas $\rho$ and $\tilde F$ are relative theories.

In theories admitting a higher-categorical structure as \eqref{eq:supersel}, topological defects in different superselection sectors have different dimensionality. Consequently, the fusion rules are categorical, in the sense that they usually don't follow group-like composition laws. Because of this these defects are also called categorical or non-invertible. 

\begin{figure}[ht!]   
\begin{center}
\includegraphics[scale=0.7]{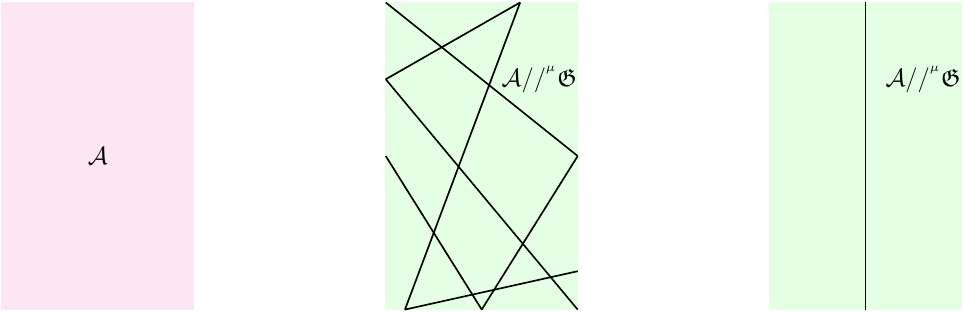}     
\caption{\small Gauging corresponds to condensing an algebra in a TFT. Idempotency ensures the resulting theory can be effectively thought of as featuring a unique defect, as shown on the RHS.}
\label{fig:codensationTJF}  
\end{center} 
\end{figure}

Prior to delving into the specific case of interest to us, namely relative QFTs, we first wish to provide some further explanation for what gauging a categorical structure actually means. In doing so, we refer to the work of many experts in the field, and, in particular \cite{TJF}. As explained in such reference, for any fusion n-category $\mathfrak{G}$, any fiber functor

\begin{equation} 
{\cal F}:\ \mathfrak{G}\ \rightarrow\ \text{nVec},  
\label{eq:functor}   
\end{equation}   
selects nVec as the image of a gaugeable algebra living in $\mathfrak{G}$, corresponding to a projection on the identity. The gauging process, can therefore be defined as a map

\begin{equation} 
\mu:\ \mathfrak{G}\ \rightarrow \ {\cal A}_{_C}, 
\label{eq:mu}   
\end{equation}   
with ${\cal A}_{_C}$ the algebra of invertible topological operators in $\mathfrak{G}$. In terms of the example reproduced in section \ref{sec:braneex}, ${\cal A}_{_C}$ refers to the algebra that dresses the symmetry operators once having implemented the gauging to ensure gauge invariance. Hence, $\mu$ really corresponds to the projection of the identity to a subalgebra of the total algebra, ${\cal A}$, featuring in the pre-gauged theory, with ${\cal A}_{_C}\ \subset\ {\cal A}$. For later purposes, we highlight the relation in between this description with that introduced in section \ref{sec:rqft} in relation to Moore-Tachikawa varieties. As also defined there, the crucial role from the category theory point of view really consists in defining the cotangent bundle as a triple of elements, two of which are left- and right- acting moment maps, mutually constrained by the main axiom outlined by Moore and Tachikawa. As explained in section \ref{sec:MooreTach}, this follows from full-dualisability of the 2-categories involved in defining the 2D TFT associated to a given class ${\cal S}$ theory\footnote{As explained in Part V, this follows from the fact that the category of contractible cycles associated to a given algebraic variety admits a Kirwan map ensuring abelianisation in terms of equivariant cohomology.}. Given \eqref{eq:mu}, the norm element

\begin{equation} 
N\overset{def}{=}\bigoplus_{g\in\mathfrak{G}}\ \mu(g).  
\label{eq:idemp}   
\end{equation}    
carries the structure of an n-categorical idempotent, also known as \emph{gaugeable algebra}, depicted in black in figure \ref{fig:codensationTJF}. The requirement for \eqref{eq:idemp} to be a higher-idempotent is needed to ensure the flooding doesn't depend on the specific features of the network being adopted to perform the gauging. The algebra of topological operators that are left are denoted by ${\cal A}//^{^{\mu}}\mathfrak{G}$. The equivalence of the second and third picture from the left in figure \ref{fig:codensationTJF} follows from $N$ being a higher-gaugeable algebra. As we shall see, this pattern emerges when gauging the 5D SymTFT of class ${\cal S}$ theories, where the objects of the 2-category in question will be Wilson surfaces charged under the 1-form symmetry being gauged.

\begin{figure}[ht!]   
\begin{center}
\includegraphics[scale=0.7]{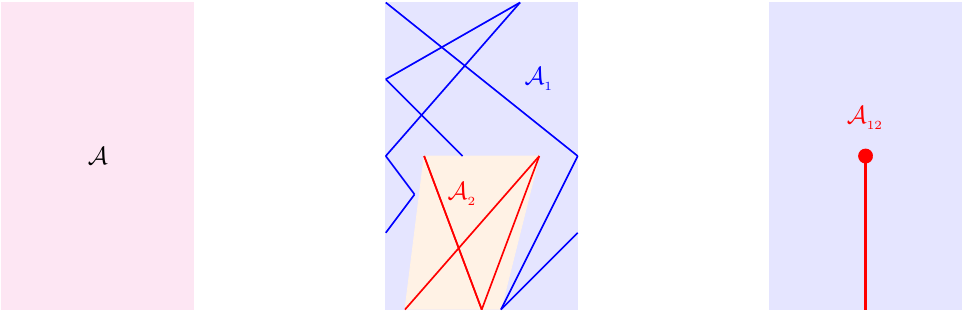}     
\caption{\small Condensing two different subalgebras, ${\cal A}_{_1}, {\cal A}_{_2}\ \subset\ {\cal A}$, the resulting theory corresponds to one with a changed phase with a gauged defect resulting from a relative gaugeable algebra, ${\cal A}_{_{12}}$ ending in the bulk. The defect at the endpoint is nontrivial, and can therefore be thought of as a Hom$(\mathbf{1}_{_{{\cal C}}},{\cal A}_{_{12}})$.}  
\label{fig:2codensationsTJF}  
\end{center} 
\end{figure}

Unlike \ref{fig:codensationTJF}, \ref{fig:2codensationsTJF} does not admit a straightforward expression as \eqref{eq:functor}.

\begin{figure}[ht!]   
\begin{center}
\includegraphics[scale=0.9]{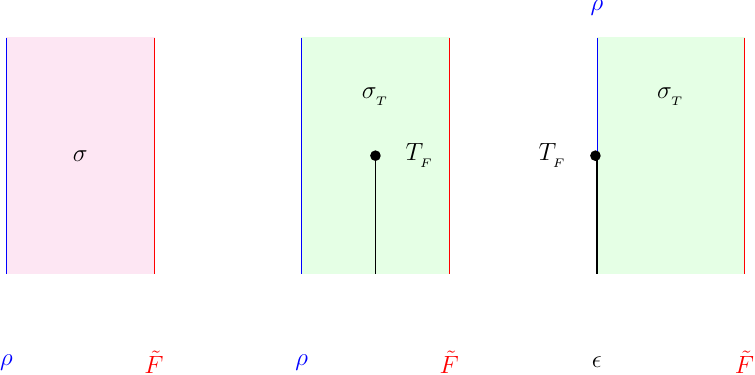}  \ \ \ \ \ \ \ \ \ \ \ \ \ \ \ \ \ \ 
\includegraphics[scale=1]{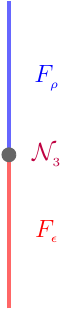} 
\caption{\small Adaptation of Freed-Moore-Teleman to the case involving twisted gauged defects. The figure on the far right corresponds to the case of interest to us, namely a configuration involving two different absolute 4D gauge theories separated by a defect. As we shall see, such defect is intrinsically non-invertible, corresponding to the presence of a relative uncondensed subalgebra, dressing ${\cal N}_{_3}$ in a nontrivial way. In higher-categorical terms, it corresponds to a fusion tensor category implementing the morphisms between the operators charged under the gauged symmetry.}
\label{fig:FrredMooreTeleman extended}  
\end{center} 
\end{figure}

\subsubsection{Absolute 4D theories from 6D \texorpdfstring{${\cal N}=(2,0)$}{} SCFTs}    \label{sec:4dfrom6d}

The main features of 6D ${\cal N}$=(2,0) SCFTs are that:  

\begin{itemize}  

\item They admit no Lagrangian formulation, \cite
{Witten:2009at, Freed:2012bs}.  

\item The theory depends on the Lie algebra rather than the Lie group. 

\item Ordinary\footnote{By this we mean that the lower-dimensional theories depend on the specific choice of the Lie group.} 5D and 4D QFTs can be obtained by dimensional reduction on a Riemann surface, $\Sigma_{_{g,n}}$.  

\item It is a \emph{relative} w.r.t. an \emph{extended} QFT in 1-dimension higher. 

\end{itemize}

Examples of relative theories are models whose fields form a fibration, \cite
{Freed:2012bs},

\begin{equation}  
{\cal F}\ \longrightarrow\  {\cal M}\ \xrightarrow{\ \ \pi\ \ }\ {\cal B} 
\end{equation}   
where ${\cal B}$ and ${\cal F}$ denote the base and the fiber, respectively. Relative fields are the fibers of $\pi$. The path integral over the base reduces a finite sum. For the theory to be absolute, the same should hold for ${\cal F}$ as well.

When compactifying the 6D theory on $\Sigma_{_{g,n}}$ without punctures:  

\begin{enumerate}   

\item If $g\equiv 1$ $\Sigma_{_{g}}\equiv T^{^{2}}$, the resulting class ${\cal S}$ theories are ${\cal N}$=4 SYM with gauge algebra $\mathfrak{su}(N)$.

\item For $g>1$, the resulting theories are models with an $S$-duality frame, interpreted as conformal gaugings of $2g-2$ trinion $T_{_{N}}$ theories coupled to $3g-3$ gauge groups with $\mathfrak{su}(N)$ gauge algebras.  

\end{enumerate}

\subsubsection{From relative to absolute}

The procedure described in section \ref{sec:FMT} cannot be applied to the case of maximally supersymmetric 6D theories. The reason being that, while a bulk SymTFT can be defined, thanks to the Milnor-Moore theorem, \cite{MilnorMoore}, specifying any topological boundary conditions would inevitably imply supersymmetry breaking\footnote{To the best of our knowledge, the only exception to this is dealt with in \cite{Frey:2018vpw}.}, \cite{Freed:2012bs}. Nevertheless, such 6D theories provide the natural realm from which one can realise lower-dimensional absolute theories, as well as constituting an extremely rich setting where to explore and put to practice mathematical techniques provided by representation theory, \cite{DBZ}.

In this subsection we briefly overview the work of \cite{Bashmakov:2022uek}, and propose a criterion for distinguishing intrinsic from non-intrinsic non-invertible defects.

Assuming the 6D ${\cal N}=(2,0)$ SCFT is defined on $Y=X\times \Sigma_{_{g}}$, with $X$ a compact 4D spacetime, torsion-free, and with trivial 1$^{^{st}}$ cohomology, $H^{^{1}}(X,\mathbb{Z}_{_{N}})$, the 6D self-dual fluxes belong to\footnote{$\mathbb{Z}_{_{N}}$ denotes the \emph{defect group} of the 6D theory.}

\begin{equation}  
H^{^{3}}(Y,\mathbb{Z}_{_{N}})\ \simeq\ H^{^{1}}(\Sigma_{_{g}},\mathbb{Z}_{_{N}})\ \otimes\ H^{^{2}}(X,\mathbb{Z}_{_{N}}).   
\end{equation}     

As far as the 6D theory is concerned, this is the only information at our disposal, given that the theory is non-Lagrangian, \cite{Freed:2012bs, Witten:2009at}. However, for the 4D theory to be fully defined, additional data is needed. In particular, one needs to identify a \emph{maximal isotropic sublattice} $L$ of $H^{^{1}}(\Sigma_{_{g}},\mathbb{Z}_{_{N}})$ w.r.t. the canonical pairing induced by the intersection pairing on $\Sigma_{_{g}}$, thereby leading to 

\begin{equation}  
{\cal L}\ \simeq\ L\ \otimes\ H^{^{2}}(X,\mathbb{Z}_{_{N}})   
\label{eq:maxissub}    
\end{equation} 
as the maximal isotropic lattice of $H^{^{3}}(Y,\mathbb{Z}_{_{N}})$. \eqref{eq:maxissub} enables to distinguish between different global forms of class ${\cal S}$ theories. Having fixed ${\cal L}$, the fluxes in its complement

\begin{equation} 
{\cal L}^{\perp}\  \overset{def.}{=}\ H_{_{3}}(X_{_{6}},\mathbb{Z}_{_{N}})/{\cal L}  
\end{equation} 
parametrise the possible partition functions for the 4D theory with inequivalent 1-form symmetry backgrounds along $X$. 

The 6D ${\cal N}$=(2,0) theory, assigns to the 6D manifold $Y$ a partition vector rather than a partition function, with the former belonging to the Hilbert space constituting a representation of a Heisenberg algebra of non-commuting discrete 3-form fluxes in $\mathbb{Z}_{_{N}}$.  

For fixed ${\cal L}$, the partition vector takes the form

\begin{equation}  
|{\cal Z}(Y)>\ \equiv\ \sum_{v\ \in\ {\cal L}^{^{\perp}}}\ {\cal Z}_{_{v}}(Y)\ |{\cal L}; v>,   
\label{eq:chofb}   
\end{equation}    
with ${\cal Z}_{_{v}}(Y)$ denoting the 6D conformal blocks, which also encode the \emph{global structures} of the 4D theory. Under a change of basis, \eqref{eq:chofb} turns into   

\begin{equation}  
|{\cal Z}(Y)>\ \equiv\ \sum_{v^{\prime}\ \in\ {\cal L}^{^{\prime\perp}}}\ {\cal Z}_{_{v}}(Y)\ \sum_{v\ \in\ {\cal L}^{^{\perp}}}\ R_{_{v}}^{^{v^{\prime}}}\ |{\cal L}; v>,  
\label{eq:chofb1}   
\end{equation} 
leading to the following relation between conformal blocks

\begin{equation}  
{\cal Z}_{_{v}}(Y)\ \equiv\ \sum_{v^{\prime}\ \in\ {\cal L}^{^{\prime\perp}}}\ R_{_{v}}^{^{v^{\prime}}}\ {\cal Z}_{_{v}}(Y).     
\label{eq:chofb2}   
\end{equation}

\subsection*{The 7D and 5D SymTFTs}

The Milnor-Moore Theorem states that, the universal enveloping algebra of a Lie algebra is a Hopf algebra whose primitive elements are the elements of the original Lie algebra. A Hopf algebra on an associative algebra turns the category of modules into a monoidal category equipped with a fiber functor. This is a statement of Tannaka duality. 

Putting this together for the case of 6D ${\cal N}=(2,0)$ SCFTs, the bulk SymTFT is a pointed braided tensor category where the braiding is in bijection with 3-cocycles.

Following the arguments outlined in section \ref{sec:excatstr}, one can take the 7D TQFT in the bulk of the relative 6D ${\cal N}=$(2,0) theory, to be a CS theory with action

\begin{equation}  
S_{_{7D}}=\frac{N}{4\pi}\ \int_{_{W_{_{7}}}}\ dc\ \wedge\ c\ \ \ ,\ \ \ c\in H^{^{3}}(W_{_{7}},U(1)), \label{eq:7D}   
\end{equation}   
and Wilson surfaces  

\begin{equation}  
\Phi_{_{q}}({\cal M}_{_{3}})\overset{def.}{=}\ e^{^{iq\oint_{_{{\cal M}_{_{3}}}} c}}\ \ \ ,\ \ \ q\in\mathbb{Z}_{_{N}}\ \ \ ,\ \ \ {\cal M}_{_{3}}\ \in H^{^{3}}(W_{_{7}},U(1)). 
\end{equation}   

Taking two such Wilson surfaces $\Phi_{_{q}}({\cal M}_{_{3}}), \Phi_{_{q^{\prime}}}({\cal M}_{_{3}}^{\prime})$, with ${\cal M}_{_{3}}, {\cal M}_{_{3}}^{\prime}$ forming a Hopf link in $W_{_{7}}$, and inserting them in the path integral, amounts to changing the action \eqref{eq:7D} by adding the holonomy terms associated to the operator insertions

\begin{equation}  
\begin{aligned}
S_{_{7D}}&=\frac{N}{4\pi}\ \int_{_{W_{_{7}}}}\ dc\ \wedge\ c  + q\int_{_{{\cal M}_{_{3}}}} c\ +q^{\prime} \int_{_{{\cal M}_{_{3}}^{\prime}}} c\\  
&=\frac{N}{4\pi}\ \int_{_{W_{_{7}}}}\ dc\ \wedge\ c +\ \int_{_{W_{_{7}}}}\ \left(q\omega_{_{{\cal M}_{_{3}}}}+q^{\prime}\omega_{_{{\cal M}_{_{3}}^{\prime}}}\right)\ \wedge\ c,
\label{eq:CS7D}    
\end{aligned}
\end{equation} 
with $\omega_{_{{\cal M}_{_{3}}}}, \omega_{_{{\cal M}_{_{3}}^{\prime}}}$ denoting the Poincar\'e duals of ${\cal M}_{_{3}}, {\cal M}_{_{3}}^{\prime}$, respectively. Integrating-out $c$, and imposing 

\begin{equation}  
dc\equiv-\frac{2\pi}{N}\left(q\omega_{_{{\cal M}_{_{3}}}}+q^{\prime}\omega_{_{{\cal M}_{_{3}}^{\prime}}}\right), 
\end{equation} 
defining $V_{_{4}}$ such that 

\begin{equation} 
\partial V_{_{4}}\equiv\left(q\omega_{_{{\cal M}_{_{3}}}}+q^{\prime}\omega_{_{{\cal M}_{_{3}}^{\prime}}}\right),  
\end{equation} 
implies

\begin{equation}   
c\equiv -\frac{2\pi}{N}\ \text{PD}(V_{_{4}}).  
\end{equation}

From this, \eqref{eq:CS7D} reduces to 

\begin{equation}  
\begin{aligned}
S_{_{7D}} 
&=\frac{2\pi}{N}\ q q^{\prime}\ \text{link}\left({\cal M}_{_{3}}, {\cal M}_{_{3}}^{\prime}\right),
\label{eq:linkact7}  
\end{aligned}
\end{equation}
where the linking is the Hopf link occurring in $W_{_{7}}$. It features in \eqref{eq:linkact7} due to the fact that $V_{_{4}}$ is a Seifert surface for the combination $q{\cal M}_{_{3}}+q^{\prime}{\cal M}_{_{3}}^{\prime}$, meaning that, every time ${\cal M}_{_{3}}$ pierces $V_{_{4}}$, it links ${\cal M}_{_{3}}^{\prime}$ once. It thereby follows that, the Hopf link of Wilson surfaces supported on ${\cal M}_{_{3}}$ and ${\cal M}_{_{3}}^{\prime}$ in $W_{_{7}}$ can be thought of as an operator insertion in the 7D path integral, 

\begin{equation}  
<\Phi_{_{q}}({\cal M}_{_{3}})\Phi_{_{q^{\prime}}}({\cal M}_{_{3}}^{\prime})...>\equiv e^{^{\frac{2\pi i}{N}q q^{\prime}\text{link}({\cal M}_{_{3}},{\cal M}_{_{3}}^{\prime})}}\ <...>,  
\end{equation}  
where the phase on the RHS being the analog of the $S$-matrix element $S_{_{ab}}$ between anyons of charges $a,b$ in $U(1)_{_{k}}$ CS theory in 3D, namely $S_{_{ab}}\equiv e^{^{2\pi iab/k}}$.

For a given $X_{_{6}}\equiv \Sigma_{_{g,n}}\times X_{_{4}}$, there is a particular class of 7-manifolds $W_{_{7}}$ obtained by taking $W_{_{7}}\equiv V_{_{g,n}}\times X_{_{4}}$, with $V_{_{g,n}}$ a 3-manifold with $\partial V_{_{g,n}}\equiv\Sigma_{_{g,n}}$. 

For any Riemann surface $\Sigma_{_{g,0}}$, there are many inequivelent 3-manifolds with $\Sigma_{_{g,0}}$ as its boundary. One such example are handlebodies. To construct a handlebody, choose a set $g$ of \emph{meridians}, $\{\mu_{_{i}}\}, i\equiv1,...g$, i.e. a set of generators of $H_{_{1}}(\Sigma_{_{g,0}}, \mathbb{Z})$ which become trivial as elements of $H_{_{1}}(V_{_{g,0}}, \mathbb{Z})$.  The remaining $g$ generators of $H_{_{1}}(\Sigma_{_{g,0}}, \mathbb{Z})$ lift to generators of $H_{_{1}}(V_{_{g,0}}, \mathbb{Z})$, and are referred to as \emph{longitudes}, $\{\lambda_{_{i}}\}, i\equiv1,...g$. Not every choice of $g$ generators gives rise to a legitimate set of meridians (i.e. vanishing cycles). For a specific choice of meridians to be legitimate, it must correspond to a maximal isotropic sublattice   

\begin{equation}  
L\ \in\ H_{_{1}}(\Sigma_{_{g,0}},\mathbb{Z}). 
\end{equation}  

The handlebody specified by the choice of meridians $L$ will be denoted by $V_{_{g,0}}^{^{L}}$. Given $V_{_{g,0}}^{^{L}}$, there are multiple choices of longitudes, differing by shifts in meridians   

\begin{equation}  
\lambda_{_{i}}^{\prime}\equiv\lambda_{_{i}}+\sum_{j=1}^{3}k_{_{ij}}\mu_{_{j}}\ \ \ ,\ \ k_{_{ij}}\in\mathbb{Z},
\end{equation}   
satisfying the following constraint 

\begin{equation}
<\mu_{_{i}},\lambda_{_{j}}>\equiv-<\lambda_{_{j}},\mu_{_{i}}>\equiv\delta_{_{ij}} . 
\end{equation}   

7D CS theory on a handlebody can be formulated by starting with a definition of the fields which reads 

\begin{equation}   
b_{_{i}}\overset{def.}{=}\oint_{_{\mu_{_{i}}}} c\ \ \ ,\ \ \ \hat b_{_{i}}\overset{def.}{=}\oint_{_{\lambda_{_{i}}}} c\ \ \ ,\ \ \ i=1,...,g.  
\end{equation}  

The meridians are naively contractible by definition. However, this does not necessarily imply the vanishing of $b_{_{i}}$, since there might be Wilson surfaces supported on $\lambda_{_{i}}$ and $\mu_{_{i}}$ that are nontrivially linking, effectively making $\mu_{_{i}}$ non-contractible. 

The Wilson lines wrapping meridians and longitudes can be expressed as follows  

\begin{equation}   
\Phi_{_{i}}({\cal M}_{_{2}})\ \overset{def.}{=} \Phi_{_{i}}({\cal M}_{_{2}}\times\mu_{_{i}})\equiv e^{^{\frac{2\pi i}{N}\oint_{_{{\cal M}_{_{2}}}}b_{_{i}}}}  
\ \ \ ,\ \ \ \hat \Phi_{_{i}}({\cal M}_{_{2}})\ \overset{def.}{=} \hat \Phi_{_{i}}({\cal M}_{_{2}}\times\lambda_{_{i}})\equiv e^{^{\frac{2\pi i}{N}\oint_{_{{\cal M}_{_{2}}}}\hat b_{_{i}}}},
\end{equation}   
which might be nontrivially linking in the 5D theory resulting from dimensional reduction of the 7D TFT on $\Sigma_{_{g,0}}$. Compactifying the theory on $\Sigma_{_{g,0}}$ in presence of a series of Wilson 3-surfaces on $\lambda_{_{j}}\times {\cal M}_{_{2,j}}^{\prime}$, the linking equation becomes an operator equation in 4D 

\begin{equation}  
\Phi_{_{i}}({\cal M}_{_{2}})\ \equiv\ e^{^{\frac{2\pi i}{N}\sum_{j=1}^{g}<\mu_{_{i}},\lambda_{_{i}}>\oint_{_{{\cal M}_{_{2}}}}B_{_{j}}}},  
\label{eq:extraphase}
\end{equation}  
where $B_{_{j}}\in H^{^{2}}(X_{_{4}},\mathbb{Z}_{_{N}})$ denotes the Poincar\'e dual of ${\cal M}_{_{2,j}}^{\prime}$. Hence, in any 4D calculation involving $\Phi_{_{i}}$, the latter can be moved to the right of $\hat\Phi_{_{j}}$, and subsequently be shrunk to a point by introducing the phase \eqref{eq:extraphase}. Hence, the insertion of longitudinal Wilson lines allows us to make $b_{_{i}}\neq 0$, while still keeping it constant, thereby defining a background field.

At this point, the Sym TFT can be obtained by splitting the handlebody in the following way:

\begin{equation}  
V_{_{g,0}}\ \overset{def.}\ \begin{cases} V_{_{g,0}}^{^{L,in}}\ \ \ \text{for}\ \ \ y\ge y_{_{*}}\\ 
\\
V_{_{g,0}}^{^{ext}}\overset{def}{=}\Sigma_{_{g,0}}\times[0,y_{_{*}}]\ \ \ \text{for}\ \ \ 0\le y\le y_{_{*}}\\ 
\end{cases}.  
\label{eq:splitting}  
\end{equation}  

\begin{figure}[ht!]   
\begin{center}  
\includegraphics[scale=0.7]{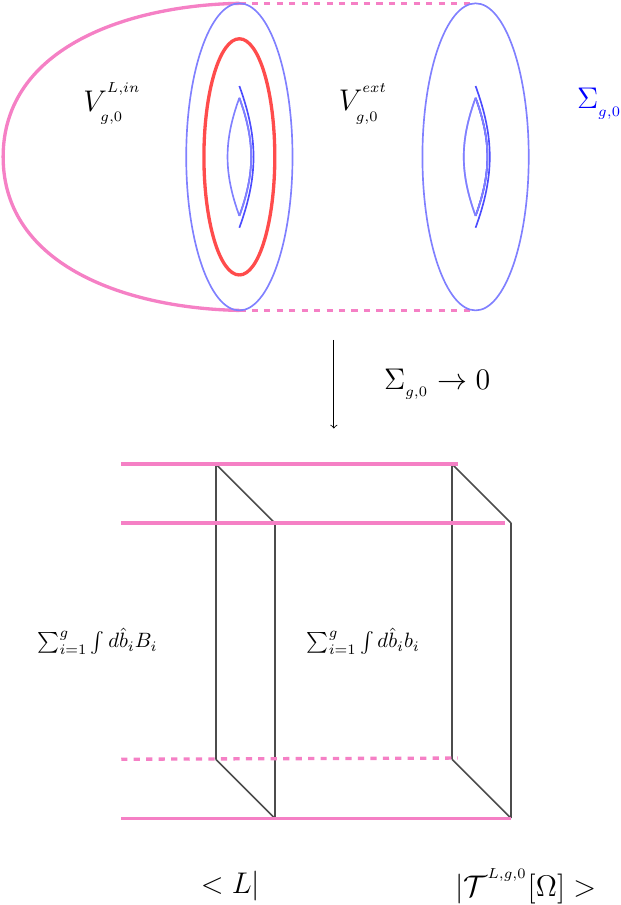}   
\caption{\small In presence of the splitting \eqref{eq:splitting}, the state $<L|$ separates the anomaly TFT from the Sym TFT. The picture above is valid for $\forall g$, but has been drawn with $g=1$ for simplicity.}    
\label{fig:shrsigma}  
\end{center}    
\end{figure}

Importantly, it features two boundaries (cf. figure \ref{fig:shrsigma}): 

\begin{itemize} 

\item One, at $y=0$, is where the 6D ${\cal N}=$(2,0) theory lives. 

\item The other, at $y=y_{_{*}}$, is where the longitudinal Wilson surfaces fixing the background fields are placed. 

\end{itemize}

\subsection{Fiber functors from non-invertible symmetries} \label{sec:Freed-Moore-Teleman}   

The main motivation of the present section is that of studying the defects interpolating different 4D class ${\cal S}$ theories descending from dimensional reduction of 6D ${\cal N}$=(2,0) SCFTs. In particular, explaining how such defects emerge under gauging of multiple subalgebras, we propose the quantum dimension of the relative gaugeable subalgebra as the quantity needed to distinguish whether such interpolating defect is intrinsically or non-intrinsically non-invertible, thereby proposing the 4D counterpart of arguments proposed in \cite{Kaidi:2022cpf} for the 2D case.

The present section is therefore structured into 4 main parts:  

\begin{itemize} 

\item At first, we describe the procedure, \cite{TJF,Kong:2013aya,Yu:2021zmu}, and its adaptation to the Freed-Teleman, \cite{Freed:2012bs}, and Freed-Moore-Teleman construction, \cite{Freed:2022qnc}, of relative field theories and topological symmetries of QFTs, respectively, with particular emphasis on the case of multiple simultaneous gauging of different subalgebras. 

\item We then briefly overview the procedure adopted by \cite{Bashmakov:2022uek} for realising absolute 4D SCFTs from ${\cal N}=(2,0)$ SCFTs, explicitly rephrasing it in terms of the multiple gauging procedure described in section \ref{sec:FMT}. 

\item When describing such interpolating defects in terms of gauging of homomorphisms, they naturally admit an interpretation in terms of fusion tensor categories interpolating between different 2-categories, to a total quantum dimension can be assigned. Section \ref{sec:2} closes with the relation between the issue in defining a fiber functor in presence of non-invertible defects interpolating between different absolute theories.

\item Making use of the gauging procedure explained in section \ref{sec:FMT}, we propose a criterion for distinguishing between the two types of non-invertible defects in terms of the total quantum dimension of the relative gauged algebra implementing the gauging. 

\end{itemize}   

  \subsubsection{Intrinsic non-invertible symmetries and fiber functors} \label{sec:2}

On the handlebody, one can perform two different operations:  

\begin{enumerate}  

\item A \emph{modular transformation} of $\Sigma_{_{g,0}}$, acting on the entire handlebody and generically changing the period matrix of $\Sigma_{_{g,0}}$ at each cross section. The theory in the bulk of the handlebody is topological, implying only $\Sigma_{_{g,0}}$ is affected by such transformations, which is where the 6D ${\cal N}=$(2,0) theory lives. 

\item Excising an \emph{inner} handlebody, in the setup where the splitting \eqref{eq:splitting} takes place. The operation of interest consists in gluing together the 2 parts $V_{_{g,0}}^{^{L,in}}$ and $V_{_{g,0}}^{^{ext}}$ with a nontrivial element of the modular group $Sp(2g, \mathbb{Z}_{_{N}})$, corresponding to a surgery operation on the 3-manifolds. 

\end{enumerate}    

Given that the latter takes place in the interior of the handlebody, it doesn't affect $\Sigma_{_{g,0}}$ where the 6D theory lives, hence, the period matrix remains unchanged in the region the theory is sensitive to. However, this surgery operation changes the global form of $L$, since it changes which are the contractible cycles in the handlebody. Hence, surgery is implemented by an element $F$ of the modular group which only affects the inner boundary of the handlebody, thereby changing the global form. This operation may in turn be combined with a modular transformation acting on the entire handlebody, reverting the interior of the handlebody back to its original form, but now changing the period matrix of the Riemann surface. 

Choosing a period matrix that remains invariant under the action of $F$, the geometry with the combined surgery plus modular transformation has the same boundaries as the original geometry. Importantly, though, the full geometry is no longer the same, due to the internal twist.

Under dimensional reduction on $\Sigma_{_{g,0}}$, \eqref{eq:CS7D} becomes    

\begin{equation}  
S_{_{5D}}\equiv\frac{2\pi}{N}\sum_{i=1}^{g}\int_{_{X_{_{4}\times[0,y_{_{*}}]}}}b_{_{i}}\ \cup\ \delta\hat b_{_{i}}. 
\label{eq:newBFnew}   
\end{equation}

The fact that $b_{_{i}}, \hat b_{_{i}}$ are dynamical, implies $\lambda_{_{i}}, \mu_{_{i}}$ are both nontrivial in $V_{_{g,0}}^{^{ext}}$. This follows from the fact that $L$ specifies which directions of the Riemann surface become trivial in the handlebody, namely the meridians, and which of the dynamical fields of the $BF$-theory become background. To each $L$ corresponds a different choice of Dirichlet boundary conditions for different sets of fields in the $BF$-theory. As explained in \cite{Bashmakov:2022uek}, the 7D CS theory on a Riemann surface can only capture the full Sym TFT iff all the non-invertible defects are non-intrinsic.  

The 5D Wilson lines are defined from the 7D Wilson surfaces as follows

\begin{equation}  
\Phi_{\overset{\rightarrow}{n}}\left({\cal M}_{_{2}}\right)\ \overset{def.}{=}\  \Phi\left({\cal M}_{_{2}}\times\gamma_{_{\overset{\rightarrow}{n}}}\right),
\label{eq:phim2}   
\end{equation}
with 1-cycles 

\begin{equation}  
\gamma_{_{\overset{\rightarrow}{n}}}  \ \overset{def.}{=}\ \sum_{j=1}^{g} e_{_{j}}\lambda_{_{j}}+\sum_{j=1}^{g}\ m_{_{j}}\mu_{_{j}}\ \ \ ,\ \ \ \overset{\rightarrow}{n}\ \overset{def.}{=}\ \left(e_{_{1}},..., e_{_{g}};m_{_{1}},...,m_{_{g}}\right).
\end{equation}

Explicitly, \eqref{eq:phim2} can be expressed as follows  

\begin{equation}  
\Phi_{\overset{\rightarrow}{n}}\left({\cal M}_{_{2}}\right)\ \equiv\  e^{^{\frac{2\pi i}{N}\ \frac{1}{2}e\cdot m({\cal M}_{2},{\cal M}_{2})}}\ \prod_{i=1}^{g}\Phi^{^{e_{_{i}}}}\left({\cal M}_{_{2}}\times\gamma_{_{\overset{\rightarrow}{n}}}\right) \ \ \prod_{i=1}^{g}\Phi^{^{m_{_{i}}}}\left({\cal M}_{_{2}}\times\gamma_{_{\overset{\rightarrow}{n}}}\right),  
\label{eq:phim21}   
\end{equation}
with each electric and magnetic component, in turn descending from a 3D Wilson surface in 7D associated to a cycle $\gamma_{_{\overset{\rightarrow}{n}}}$. 

Surgery defects in 7D correspond to loci across which $F$ acts on $\gamma_{_{\overset{\rightarrow}{n}}}$ as follows 

\begin{equation}   
F\left(\gamma_{_{\overset{\rightarrow}{n}}}\right)\ \equiv\ \gamma_{F\overset{\rightarrow}{n}}.  
\end{equation}

In 5D, they reduce to gaugeable defects implementing morphisms in a higher-categorical structure

\begin{equation}  
{\cal C}_{_{F}}(X_{_{4}})\ :\ \Phi_{\overset{\rightarrow}{n}}\left({\cal M}_{_{2}}\right)\ \mapsto\ \Phi_{F\overset{\rightarrow}{n}}\left({\cal M}_{_{2}}\right)\ \ \ \ ,\ \ \ {\cal M}_{_{2}}\ \subset\ X_{_{4}}.  
\label{eq:morphism}  
\end{equation}

\begin{figure}[ht!]   
\begin{center}
\includegraphics[scale=0.8]{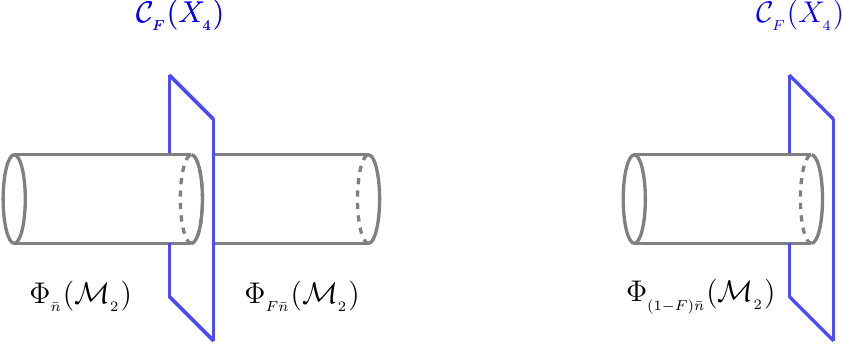}  
\caption{\small The picture on the LHS shows the identification of a morphism between objects as a defect in a higher-categorical structure. On the RHS, the interface turns into a boundary defect once having performed the folding trick.}
\label{fig:jphi}  
\end{center} 
\end{figure}

\begin{figure}[ht!]    
\begin{center}   
\includegraphics[scale=0.8]{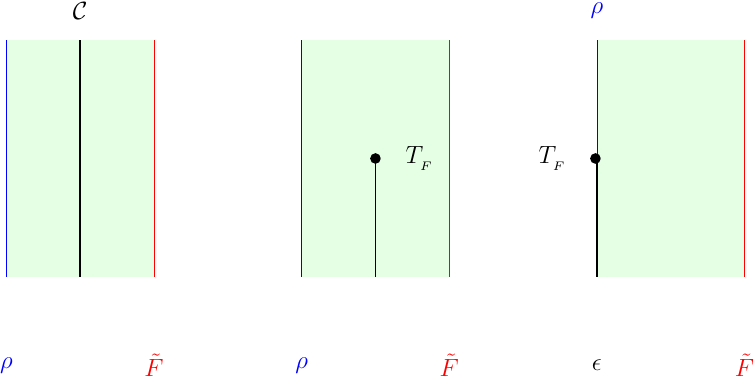}  
\caption{\small Gauging the defects living in the SymTFT upon implementing an outer automorphisms, changes the boundary conditions of the resulting absolute theory. In presence of a unique gauging, the resulting configuration is depicted on the LHS. On the other hand, upon condensing two different subalgebras of the original bulk theory, leads to a bulk gauging defect effectively ending in the bulk, as shown in the figure in the middle. This is equivalent to realising a configuration comprising two different boundary conditions separated by a topological defect (as shown on the RHS).}   
\label{fig:newidentity}   
\end{center}   
\end{figure}

As suggested by \eqref{eq:morphism}, $F$ acts on the labellings of the objects living in the theory. These objects are dyonic, with nontrivial linking. An explicit expression for the morphism \eqref{eq:morphism} as a defect follows by performing the folding trick on the surgery point in the 5D configuration depicted in figure \ref{fig:jphi}

\begin{equation}  
{\cal C}(X_{_{4}}) \ =\ \frac{\ |H^{^{0}}(X_{_{4}},\mathbb{Z}_{_{p}})|\ }{|H^{^{1}}(X_{_{4}},\mathbb{Z}_{_{p}})|}\ \sum_{_{{\cal M}_{_{2}} \in H_{_{2}}\left(X_{_{4}}, \mathbb{Z}_{_{p}}^{^{2g}}\right)}}\ \exp\left(\frac{2\pi i}{p}\ <F{\cal M}_{_{2}},{\cal M}_{_{2}}>\right) \Phi \left((\mathbb{1}-F){\cal M}_{_{2}} \right),    
\label{eq:conddef}  
\end{equation} 
implying that, $\forall \ F\in\ Sp(2g, \mathbb{Z}_{_{p}})$, there is a codimension-1 defect implementing the symmetry associated to $F$ in the bulk 5D TQFT. Such defect is invertible, and therefore its fusion rules simply read

\begin{equation}   
\boxed{\ \ \ {\cal C}(X_{_{4}})\ \times\ \bar {\cal C}(X_{_{4}})\ \equiv\ {\cal A}\ \equiv\ \mathbf{1}\color{white}\bigg]\ \ }. 
\label{eq:identity}
\end{equation}
where ${\cal A}$ corresponds to the condensed algebra implementing the gauging. Notice that the defect \eqref{eq:conddef} is equivalent to the one arising from gauging by algebraic gauging in a topological order, as explained in section \ref{sec:FMT} and depicted in figure \ref{fig:codensationTJF}. The crucial meaning of equation \eqref{eq:identity} is the fact that, gauging is effectively acting as a redefinition of the identity from the mother theory to the gauged one. In particular, the last equality claims that the condensed algebra in the mother theory is projected to the identity of the gauged theory. Pictorially, this corresponds to the configuration on the LHS of \ref{fig:newidentity}.

Gauged defects associated to $F$, can in turn give rise to $|F|$-ality defects in the boundary 4D gauge theory. Such defects can be constructed from \eqref{eq:conddef} by admitting Dirichlet boundary conditions for the gauged defects, allowing them to terminate in the bulk, \cite{Kaidi:2022cpf, Teo:2015xla, Barkeshli:2014cna}. Practically, this corresponds to a 3D-manifold attached to a 4D manifold, which, combined together, give rise to a \emph{twist defect}, $T_{_{F}}({\cal M}_{_{3}}, {\cal M}_{_{4}})$, whose expression is obtained from \eqref{eq:conddef} once rewritten in terms of relative cohomology. From the perspective of the gauging procedure, this amounts to considering simultaneous gauging of two different subalgebras within the gaugeable subalgebra of the mother theory. In terms of its SymTFT characterisation, this would correspond to the gaugeable defect effectively ending in the bulk SymTFT as shown in the middle of figure \ref{fig:newidentity}. Equivalently, this is equivalent to describing a relative field theory with two boundary conditions interpolated by a topological defect.

\begin{figure}[ht!]    
\begin{center}   
\includegraphics[scale=0.8]{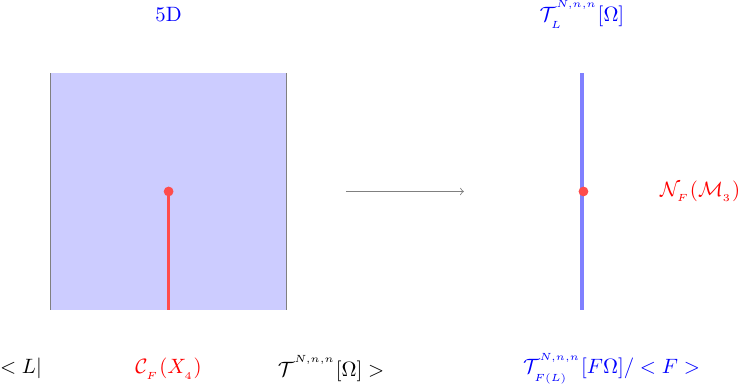}  
\caption{\small The gauged defect ${\cal C}_{_{F}}(X_{_{4}})$ implements the outer automorphism in the 5D TQFT. Applying twisted BCs, the gauged defect can effectively terminate in the 5D bulk, and is therefore topological. At this point, the width of the 5D theory can be taken to vanish, resulting in the configuration on the RHS, with the two phases of the 4D theory separated by a topological non-invertible defect, ${\cal N}_{_{F}}({\cal M}_{_{3}})$.}   
\label{fig:condensationdefect}   
\end{center}   
\end{figure}

The gauged defect can be reexpressed in terms of the quantum dimensions of the underlying anyon description by noticing that the action of $F$ is equivalent to that of the outer automorphism keeping track of the anyonic symmetry being gauged\footnote{As explained in section \ref{sec:excatstr} and \ref{sec:FMT}.}. This follows from the fact that the elements of the sum appearing in the gauged defect, \eqref{eq:conddef}, can be broken down into two main parts, namely     

\begin{equation}  
\exp\left(\frac{2\pi i}{p}\ <F{\cal M}_{_{2}},{\cal M}_{_{2}}>\right) \Phi \left((\mathbf{1}-F){\cal M}_{_{2}} \right)\ \equiv\ S_{_{\overset{\rightarrow}{a}\ F\overset{\rightarrow}{a}}}\ {\cal G}_{_{F\overset{\rightarrow}{a}}},  
\end{equation}  
with $S_{_{\overset{\rightarrow}{a},\ F\overset{\rightarrow}{a}}}$ and ${\cal G}_{_{F\overset{\rightarrow}{a}}}$ denoting the $S$-matrix entries for the associated anyonic charges, and the exceptional fibration operator, respectively. Extending these arguments, the sum over the 2-manifolds ${\cal M}_{_{2}}$ can be replaced by a sum over the anyonic charges compatible with the symmetries of the given theory. This way, \eqref{eq:conddef} can be rewritten as follows

\begin{equation}  
T_{_{F}}({\cal M}_{_{3}}, {\cal M}_{_{4}}) \ \overset{def.}{=}\ \frac{\ |H^{^{0}}(X_{_{4}},\mathbb{Z}_{_{p}})|^{^{2g}}\ }{|H^{^{1}}(X_{_{4}},\mathbb{Z}_{_{p}})|^{^{2g}}}\ \sum_{_{\overset{\rightarrow}{a}}} \ S_{_{\overset{\rightarrow}{a},\ F\overset{\rightarrow}{a}}}\ {\cal G}_{_{F\overset{\rightarrow}{a}}},
\label{eq:conddef1}  
\end{equation} 
where $\overset{\rightarrow}{a}\ \overset{def.}{=}\ \left([a], \rho(a)\right)$, with $[a], \rho(a)$ denoting the conjugacy class and the representation of the corresponding anyon charges, respectively. 
Given that the defect is now topological, it can be moved to the 4D boundary, defining an interface for the 4D gauge theory

\begin{equation}  
{\cal N}({\cal M}_{_{3}})\ \overset{def.}{=}\ \underset{y\rightarrow0}{\lim}\ T_{_{F}}({\cal M}_{_{3}}, {\cal M}_{_{4}}),
\end{equation}   
obeying non-invertible fusion rules. From the 4D perspective of the absolute theory on the RHS of figure \ref{fig:condensationdefect}, ${\cal N}({\cal M}_{_{3}})$ is \emph{intrinsically}-non-invertible, namely, it cannot be rendered invertible by acting with $SL(2,\mathbb{Z})$-transformations. This is equivalent to stating that the action of $F$ cannot be reversed simply by acting with a modular transformation. Non-invertibility implies its fusion rules now read 

\begin{equation}  
\boxed{\ \ \ {\cal N}({\cal M}_{_{3}})\ \times \bar{\cal N}({\cal M}_{_{3}})\ =\ {\cal A}_{_{X_{_{4}}}}\color{white}\bigg]\ \ },   
\label{eq:NNA}   
\end{equation}
with ${\cal A}_{_{X_{_{4}}}}$ denoting the relative gaugeable algebra in the absolute theory. 

Crucially, the RHS of equation \eqref{eq:NNA} is not simply the identity of a 4D absolute theory, corresponding to the fact that there is no direct counterpart of equation \eqref{eq:functor} upon performing a double gauging involving two distinct subalgebras of the gaugeable algebra characterising the mother SymTFT. Given that the latter is basically equivalent to the existence of a well-defined partition function for the resulting absolute theory, we claim that, if might be able to define a counterpart of the functor \eqref{eq:functor} for the case depicted at the centre and RHS of figure \ref{fig:newidentity}, by composing the following chain of functors

\begin{figure}[ht!]    
\begin{center}   
\includegraphics[scale=1.25]{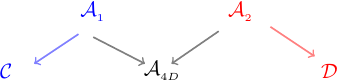}  
\caption{\small A qualitative diagram depicting the counterpart of the fiber functor's domain for the case of figure \ref{fig:newidentity}. The black arrows indicate the choice of different gaugeable subalgebras, leading to different 4D absolute theories, labelled by 2-categories ${\cal C,D}$.}   
\label{fig:3fibers}   
\end{center}   
\end{figure}  
where ${\cal A}_{_{4D}}$ is the gaugeable algebra leading to a single 4D absolute theory, whereas ${\cal A}_{_1}$ and ${\cal A}_{_2}$ correspond to different subalgebras of ${\cal A}_{_{4D}}$ leading to different 2-categories of underlying charges associated to the gauged symmetry characterising the two distinct 4D absolute theories. This is current work in progress by the same author, and we plan to report about it in a forthcoming work. The main message we want to convey with this is that the fiber functor that can be defined for the configuration at the centre and RHS of figure \ref{fig:newidentity} leads to the partition function of a 3D theory.

\subsubsection{Quantum dimension from relative gaugeable algebra}  \label{sec:2.4}  

In the concluding part of this section, we make use of the tools outlined in the previous parts of the present treatment, leading to the proposal of a probe quantity enabling to distinguish whether the defect interpolating in between class ${\cal S}$ theories is intrinsically or non-intrinsically non-invertible. The starting point will be the setting of section \ref{sec:FMT}, where we revised the gauging procedure, implemented by flooding the topological order corresponding to the 2-category of Wilson surfaces living in the bulk SymTFT with a tensor network of an idempotent element, namely the norm element \eqref{eq:idemp}. 

\begin{figure}[ht!]    
\begin{center}   
\includegraphics[scale=1]{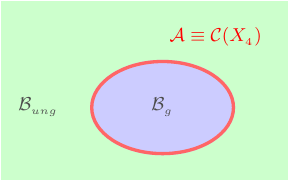}  
\caption{\small This picture shows the categories of anyons associated to the ungauged and gauged theories, denoted by ${\cal B}_{_{ung}}$ and ${\cal B}_{_g}$, respectively. Gauging can be visualised as the domain wall effectively separating two different topological phases.}   
\label{fig:condensing algebra}   
\end{center}   
\end{figure}

Practically, the two phases (prior and after gauging) can be thought of as arising from a phase transition in between two MTCs, ${\cal B}_{_{ung}}$ and ${\cal B}_{_g}$, corresponding to the ungauged and the gauged phases, respectively, \cite{Kong:2013aya}. Imposing natural physical requirements, it is possible to derive a relation between the anyons in the ${\cal B}_{_{g}}$ phase and those in the ${\cal B}_{_{ung}}$ phase. The vacuum, or tensor unit ${\cal A}$ in ${\cal B}_{_g}$ is necessarily a connected commutative separable algebra in ${\cal B}_{_{ung}}$, and the category ${\cal B}_{_g}$ is equivalent to the category of local ${\cal A}$-modules as MTCs. This gauging produces a gapped domain wall (DW) with wall excitations given by the category of ${\cal A}$-modules in ${\cal B}_{_{ung}}$. The domain wall separating them, is, in turn a category, ${\cal A}\equiv{\cal C}(X_{_4})$, as shown in figure \ref{fig:condensing algebra}. 

For any pair of elements, $M, N$, 

\begin{equation}  
\text{Hom}_{_{{\cal B}_{_g}}}(M,N)\ \hookrightarrow \ \text{Hom}_{_{{\cal B}_{_{ung}}}}(M,N).
\end{equation}  

Assuming vacuum degeneracy being trivial, 

\begin{equation}  
\text{Hom}_{_{{\cal A}}}(\mathbf{1}_{_{\cal A}},\mathbf{1}_{_{\cal A}}) \ \simeq \ \mathbb{C},  
\end{equation}    
${\cal A}$ must be a UFC with a unique spherical structure. All objects in ${\cal A}$ should come from objects in ${\cal B}_{_{ung}}$, and should satisfy the following properties:  

\begin{enumerate}

\item ${\cal A}$ is a subcategory of ${\cal B}_{_{ung}}$   

\item \begin{equation}  
\text{Hom}_{_{{\cal A}}}(M,N)\ \hookrightarrow\   \text{Hom}_{_{{\cal B}_{_{ung}}}}(M,N)
\end{equation} 

\item    $\mathbf{1}_{_{ung}}$ should condense into $B$, such that 

\begin{equation} 
\iota_{_B}:\ \mathbf{1}_{_{{\cal B}_{_{ung}}}}\ \rightarrow\ B\ \in\ {\cal B}_{_{ung}}  
\end{equation}   

$\mathbf{1}_{_{{\cal B}_{_g}}}\equiv{\cal A}$ should fuse into the vacuum on the wall when moving ${\cal A}$ close to the wall, such that, in ${\cal B}_{_{ung}}$, the corresponding map reads

\begin{equation} 
\iota_{_A}^{^B}:\ A\ \rightarrow\ B  
\end{equation}  

\end{enumerate}

What we have just said can be succinctly re-expressed as follows  

\begin{equation}  
\boxed{\ \ \mathbf{1}_{_{{\cal B}_{_g}}}\ \equiv\ {\cal A}\ \subset\ {\cal B}_{_{ung}} \color{white}\bigg]\ \ }. 
\label{eq:identityoperator}
\end{equation}

After what we have just said, the key point to focus on is the fact that the gauging procedure, the identity operator changes, in the sense that, from being a simple element in the category ${\cal B}_{_{ung}}$, it is nontrivial in the category of the gauged theory, ${\cal B}_{_g}$, since $\mathbf{1}_{_{{\cal B}_{_g}}}\equiv{\cal A}$. 

The analogy we have drawn so far with the example shown in figure \eqref{eq:identityoperator} taken from \cite{Kong:2013aya}, corresponds to the case of a single gauging of the 5D SymTFT, leading to the LHS of our figure \ref{fig:newidentity}. For the case involving a double gauging, instead, such prescription needs to be suitably adapted. The motivation still follows the arguments presented in \cite{Kong:2013aya}. We now argue why this is indeed the case specifying to the example described in section \ref{sec:4dfrom6d}.

Recall that ${\cal C}(X_{_4})$ is the operator implementing the outer automorphism on the category of defects defining the 5D SymTFT, and corresponds to the idempotent object that is constituted of the algebra elements that are flooding the SymTFT in order to implement the gauging. Denoting by ${\cal C}_{_g}, {\cal C}_{_{ung}}$ the 2-category of defects characterising the 4D absolute theories, if the two phases are related by a single bulk gauging, it follows that

\begin{equation}  
\text{dim}_{_{{\cal C}_{_g}}}\ \mathbf{1}_{_{{\cal C}_{_g}}}\ \equiv\ \frac{\ \text{dim}_{_{{\cal C}_{_{ung}}}} {\cal A}_{_{4D}}\ }{\text{dim}_{_{{\cal C}_{_{ung}}}} {\cal A}_{_{4D}}}\ \equiv\ 1.  
\label{eq:BBungg5}   
\end{equation}

But this is only from the point of view of the gauged theory, ${\cal C}_{_g}$. On the other hand, ${\cal A}_{_{4D}}$ is a graded object in ${\cal C}_{_{ung}}$, and therefore 

\begin{equation}  
\text{dim}_{_{{\cal C}_{_{ung}}}}\ {\cal A}_{_{4D}}\ \neq\ 1.     
\label{eq:BBungg6}   
\end{equation}

However, our aim is that of describing a composite configuration for the absolute theory of the kind depicted on the LHS of figure \ref{fig:condensing algebra2}, where, as explained in sections \ref{sec:4dfrom6d} and \ref{sec:2}, both phases really descend form the gauging of different subalgebras within the gaugeable algebra of the original ungauged SymTFT\footnote{This is our proposal on the basis of \cite{TJF,Kong:2013aya}.}. The different 4D absolute gauge theories are interpolated by a non-invertible defect defined as 

\begin{equation} 
{\cal N}({\cal M}_{_3})\ \overset{def.}{\equiv}\ {\cal D}_{_3}\ {\cal A}_{_{\epsilon\rho}} 
\label{eq:noninv}   
\end{equation}   
where ${\cal A}_{_{\epsilon\rho}}$ plays the role of a relative gaugeable algebra that cannot be identified with the vacuum on either side.

\begin{figure}[ht!]    
\begin{center}   
\includegraphics[scale=1]{fnf.pdf}  
\ \ \ \ \ \ \ \ \ \ \ \ \ \ \ \ \ \ \ \ \ \ \ \ \ \ \ 
\includegraphics[scale=1]{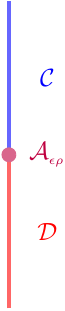}  
\caption{\small  Composite 4D absolute theories separated by an intrinsic non-invertible defect on the LHS. Rephrased in terms of the higher-categorical structure underlying the resulting theory, the algebra dressing the interpolating defects ${\cal N}_{_3}$ is a homomorphism between the objects of a 2-category.}   
\label{fig:condensing algebra2}   
\end{center}   
\end{figure}

Consistency of the configuration associated to the LHS of figure \ref{fig:condensing algebra2}, is mapped to a statement of Witt equivalence between the underlying 2-categories depicted on the RHS of the same figure, \cite{Kong:2013aya},

\begin{equation} 
{\cal C}\ \boxtimes\ \bar{\cal D}\ \simeq\ \mathfrak{Z}({\cal {\cal A}_{_{\epsilon\rho}}}), 
\label{eq:inclusion}   
\end{equation}   
where ${\cal A}_{_{\epsilon\rho}}$ denotes the relative gaugeable algebra featuring in \eqref{eq:noninv}. 

The relative gaugeable algebra, ${\cal A}_{_{\epsilon\rho}}$, constitutes a fusion tensor category\footnote{With ${\cal A}_{_1}$ and ${\cal A}_{_0}$ denoting the object and the morphisms, respectively.} defining morphisms between any pair of elements $L, L^{\prime}\ \in\ {\cal C,D}$.

\begin{equation}    
{\cal A}_{_{\epsilon\rho}}^{^{L,L^{^{\prime}}}} \ \overset{def.}{=}\ \left\{{\cal A}_{_1}^{^{L,L^{^{\prime}}}}, {\cal A}_{_0}^{^{L,L^{^{\prime}}}}\ \right\}, 
\end{equation}
whose total quantum dimension can be compared to that of the identity, i.e. of the algebra that has been condensed to achieve the reference gauge theory, with 2-category ${\cal C}$.


\begin{figure}[ht!]    
\begin{center}   
\includegraphics[scale=1]{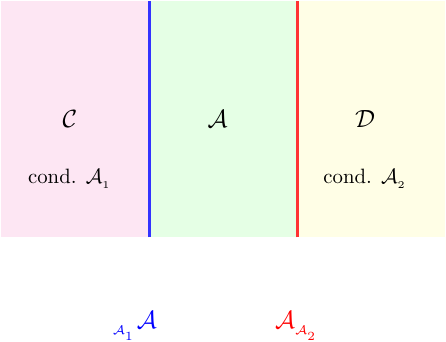}  
\caption{\small The algebra for the underlying category, denoted by ${\cal A}$ can be partially condensed, under suitable choice of subalgebras ${\cal A}_{_1}, {\cal A}_{_2}\ \subset\ {\cal A}$, such that they lead to different gauge theories, whose 2-categories are labelled by ${\cal C, D}$.}   
\label{fig:condensing algebra1}   
\end{center}   
\end{figure}

\begin{figure}[ht!]    
\begin{center}   
\includegraphics[scale=1]{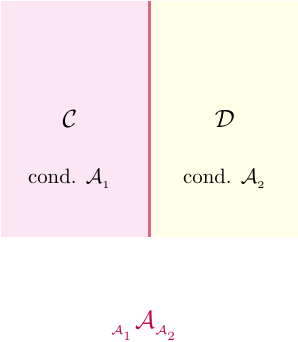}  
\caption{\small The composite system \eqref{fig:condensing algebra1} can be thought of as describing a thickened domain wall, $_{_{{\cal A}_{_1}}}{\cal A}_{_{{\cal A}_{_2}}}\ \overset{def.}{\equiv}\ {\cal A}_{_1}\ \otimes_{_{\cal A}}\ {\cal A}_{_2}$.}   
\label{fig:condensing algebra2}   
\end{center}   
\end{figure}

From the point of view of the absolute 4D theory, the non-invertible defect can be thought of as a thickened domain wall obtained by condensing two different subalgebras within the original condensing algebra ${\cal A}$, as shown in figure \ref{fig:condensing algebra2}. Given that the target absolute theory we want to achieve is that of figure \ref{fig:condensing algebra1}, then the corresponding 2-categories of defects can be obtained starting from 

\begin{equation}   
{\cal A}\ \equiv\ {\cal D}\ \boxtimes\ \mathfrak{Z}({\cal E}),  
\end{equation}

\begin{equation}  
G:\ {\cal C}\ \boxtimes\ \mathfrak{Z}({\cal D})\ \simeq\ {\cal D}\ \boxtimes\ \mathfrak{Z}({\cal E}), 
\end{equation}  
and the  fiber functors 

\begin{equation}  
{\cal F}_{_{\cal E}}\ :\ \mathfrak{Z}({\cal E})\ \rightarrow\ {\cal E}  
\ \ \ ,\ \ \ 
{\cal F}_{_{\cal D}}\ :\ \mathfrak{Z}({\cal D})\ \rightarrow\ {\cal D}   
\end{equation}   
with corresponding duals ${\cal F}_{_{\cal E}}^{^V}, {\cal F}_{_{\cal D}}^{^V}$.






The fact that ${\cal N}({\cal M}_{_{3}})$ is intrinsically non-invertible corresponds to saying that the outer automorphism implemented by the action of $F$ cannot be undone by a combination of modular transformations, as described in section \ref{sec:2}. In particular, ${\cal A}_{_{\epsilon\rho}}$ is different from the identity on either side of the defect, and, therefore, ${\cal D}_{_{{\cal A}_{_{\epsilon\rho}}}}\neq {\cal D}_{_{\mathbb{1}}}$. 

\section*{Proof}    

Take ${\cal B}$ to be a semisimple abelian category over $\mathbb{C}$; assuming it is a rigid braided tensor category. $I$ is the set of isomorphism classes of irreducible objects in ${\cal B}$, with representatives $V_{_i}, \ \forall i\in I$. Assuming the spaces of morphisms are finite-dimensional. $\mathbb 1\in {\cal B}$ is assumed to be a simple object corresponding to the labelling $i\equiv0\in I\ \Rightarrow\ V_{_0}\overset{def.}{\equiv}\mathbf{1}$. A ${\cal B}$-algebra is an object ${\cal A}\in{\cal C}$ with morphisms $\mu:\ {\cal A}\otimes{\cal A}\rightarrow{\cal A}$ and $\iota_{_{{\cal A}}}:\mathbf{1}\hookrightarrow{\cal A}$, with 

\begin{equation} 
\text{dim \ Hom}_{_{{\cal B}}}(\mathbf{1},{\cal A})\ \equiv\ \mathbf{1}.  
\end{equation}

Given ${\cal B}$ and a ${\cal B}$-algebra ${\cal A}$, Rep${\cal A}$ is defined as pairs $(V,\mu_{_{V}})$, where $V\in{\cal B}$ and $\mu_{_{V}}:{\cal A}\otimes V\rightarrow V$ morphisms in ${\cal B}$. 

For example, if $G$ is a finite group and ${\cal B}$ is the category of finite-dimensional complex representations of $G$, semisimple ${\cal B}$-algebras correspond to different semigroups in $G$ (this is what we need to implement gauging in the Freed-Moore-Teleman setup). 

Here, in Rep${\cal A}$, ${\cal A}$ is the unit object. For the case where there are two different absolute theories separated by an intrinsically non-invertible defect, the setup of joint absolute theories results from a double gauging procedure [Liang Kong] and it makes sense to consider the composite object $X\otimes_{_{{\cal A}}}Y$ with ${\cal A}$ the gaugeable algebra in ${\cal C}$; so, we have that $X,Y\in$ Rep${\cal A}$. But, at the same time, $X$ and $Y$ are also elements of two different subalgebras within ${\cal A}$ corresponding to two different gaugings. E.g. $X\in$Rep${\cal A}_{_1}$ and $Y\in$Rep${\cal A}_{_2}$, with ${\cal A}{_{_1}}, {\cal A}_{_{2}}\subset{\cal A}$. In turn, ${\cal A}_{_1}$ is the unit element in Rep${\cal A}_{_1}$ and ${\cal A}_{_2}$ id the unite element in Rep${\cal A}_{_2}$. 

Hence, if we consider the original mother algebra ${\cal A}$ with its characteristic elementary objects $V_{_i}$, and assuming every one of them has multiplicity $1$ within ${\cal A}$, we get 

\begin{equation}  
\text{dim}_{_{\cal A}}(X\otimes_{_{{\cal A}}}Y) \ \equiv\ \text{dim}_{_{\cal A}}(X)\ \text{dim}_{_{\cal A}}(Y)
\label{eq:first}
\end{equation}  
and 

\begin{equation}  
\text{dim}_{_{\cal A}}(X) \ \equiv\ \frac{\text{dim}_{_{\cal B}}({ X})}{\text{dim}_{_{\cal B}}({\cal A})}.   
\label{eq:second}
\end{equation} 

Furthermore, we have that

\begin{equation}  
\text{dim}_{_{\cal A}}(X\otimes_{_{\cal A}}Y) \ \equiv\ \frac{\text{dim}_{_{\cal B}}({ X})\ \text{dim}_{_{\cal B}}({ Y})}{\text{dim}_{_{\cal B}}({\cal A})}.   
\label{eq:third}
\end{equation}

Extending equation \eqref{eq:third} to the case in which $X, Y$ are replaced by the ${\cal A}_{_1}, {\cal A}_{_2}$ subalgebras themselves, we can look at the following ratio

\begin{equation}  
\frac{\text{dim}_{_{\cal B}}({\cal A}_{_1}\otimes_{_{\cal A}}{\cal A}_{_2})}{\text{dim}_{_{\cal B}}({\cal A}_{_1})}\ \equiv\ \frac{{\cal D}^{^{intr.}}_{{\cal A}}}{{\cal D}^{^{non-intr.}}_{{\cal A}}},  
\label{eq:third1}
\end{equation} 
where on the RHS we identified the numerator and denominator with the quantum dimension of algebras associated to the intrinsic or non-intrinsic non-invertible defects separating absolute theories resulting from the Freed-Moore-Teleman construction discussed in the first part of section \ref{sec:FMT}. Making use of \eqref{eq:first} and \eqref{eq:second}, \eqref{eq:third1}, we get

\begin{equation}  
\frac{\text{dim}_{_{\cal A}}({\cal A}_{_1}\otimes_{_{\cal A}}{\cal A}_{_2})\ \text{dim}_{_{\cal B}}({\cal A})}{ \text{dim}_{_{\cal A}}({\cal A}_{_1})\ \text{dim}_{_{\cal B}}({\cal A})}\ \equiv\ \frac{\text{dim}_{_{\cal A}}({\cal A}_{_1})\ \text{dim}_{_{\cal A}}({\cal A}_{_2})}{\text{dim}_{_{\cal A}}({\cal A}_{_1})}\ \equiv\ \text{dim}_{_{\cal A}}({\cal A}_{_2}).      
\label{eq:third1}
\end{equation} 

Given that 

\begin{equation} 
\text{dim}_{_{\cal A}}({\cal A}_{_2})\ \overset{def.}{=}\ \sum_{_i}d_{_i}V_{_i}  
\end{equation} 
with $V_{_i}\in {\cal A}_{_2}$ being simple objects in ${\cal A}_{_2}$ (similar arguments hold for ${\cal A}_{_1}$, since both are assumed to be subalgebras in ${\cal A}$, whereas 

\begin{equation}  
d_{_i}\ \overset{def.}{=}\ \text{dim}_{_{\cal A}}V_{_i},   
\end{equation}
it therefore follows that 

\begin{equation}  
\boxed{\ \ \ \frac{\text{dim}_{_{\cal B}}({\cal A}_{_1}\otimes_{_{\cal A}}{\cal A}_{_2})}{\text{dim}_{_{\cal B}}({\cal A}_{_1})}\ \equiv\ \frac{{\cal D}^{^{intr.}}_{{\cal A}}}{{\cal D}^{^{non-intr.}}_{{\cal A}}}\ >\ 1\color{white}\bigg]\ \ }.  
\label{eq:third2}
\end{equation}

For the configuration of interest, namely two different absolute theories obtained by gauging of two different subalgebras ${\cal A}_{_1}, {\cal A}_{_2}\subset{\cal A}$, separated by an intrinsic non-invertible defect, the numerator in \eqref{eq:third2} corresponds to the nontrivial fusion. 

We are therefore lead to conclude that the total quantum dimension of the relative gaugeable algebra ${\cal A}_{_{\epsilon\rho}}$ is greater w.r.t. that of the condensed algebra leading to the single gauged phase, ${\cal C}$, and we propose the following

\begin{equation} 
{\cal N}({\cal M}_{_{3}})\ \times\ \bar {\cal N}({\cal M}_{_{3}}) 
\equiv{\cal A}_{_{4D}}    
\end{equation}  
for class ${\cal S}$ theories, whereas the denominator in \eqref{eq:third2} corresponds to the quantum dimension of the algebra that has been condensed in the SymTFT to obtain a specific absolute theory, namely

\begin{equation} 
{\cal D}_{_3}\ \times\ \bar {\cal D}_{_3} \equiv\ \mathbf{1}\ 
\equiv\ {\cal A}_{_1}    
\end{equation}

Gathering these results together, we can rewrite equation \eqref{eq:third2} in terms of the notation adopted in the first part of this section, leading to the following result

\begin{equation} 
\begin{aligned}    
\boxed{\ \ \ \frac{<{\cal N}({\cal M}_{_{3}})\ \times\ \bar {\cal N}({\cal M}_{_{3}}) >_{_{{\cal B}}}}{<{\cal D}_{3}\ \times\ \bar{\cal D}_{3}>_{_{{\cal B}}}}\  \equiv\ \frac{ <{\cal A}_{_{\epsilon\rho}} >_{_{{\cal B}}} }{ \ <\mathbf{1}_{_{{\cal B}}} >_{_{{\cal B}}} \ }\ \equiv\ \left(\frac{{\cal D}^{^{intr.}}_{{\cal A}}}{{\cal D}^{^{non-intr.}}_{{\cal A}}} \right)_{_{X_{4} }}\ >\ 1\color{white}\bigg]\ \ },      
\label{eq:4DNN3}
\end{aligned}  
\end{equation} 
with the last relation indicating that such ratio can be used as a parameter probing whether the two gauge theories are separated by an intrinsic non-invertible symmetry. Consequently, the following relation 

\begin{equation}  
\boxed{\ \ \ {\cal D}_{{\cal A}_{_{\epsilon\rho}}{(X_{_{4}})}}^{^{intr.}} \ >  \ {\cal D}_{{\cal A}_{_{\epsilon\rho}}{(X_{_{4}})}}^{^{non-intr. }}\color{white}\bigg]\color{black}\ \  }   
\label{eq:BFE}  
\end{equation} 
implies that the fusion category associated to intrinsic non-invertible defects is characterised by a bigger quantum dimension, implying such configuration (arising from double gauging) is able to store a higher number of degrees of freedom w.r.t. the case in which there is a unique gauging taking place\footnote{Note that this criterion is perfectly in agreement with Remark 2.5 in \cite{JP}.}.

\section*{Key points}  

\begin{enumerate}  

\item For class ${\cal S}$ theories, ${\cal D}_{{\cal A}_{_{\epsilon\rho}}}$ enables to probe whether a defect is intrinsically or non-intrinsically non-invertible.  

\item ${\cal D}_{{\cal A}_{_{\epsilon\rho}}{(X_{_{4}})}}^{^{intr.}} \ >  \ {\cal D}_{{\cal A}_{_{\epsilon\rho}}{(X_{_{4}})}}^{^{non-intr. }}$, and therefore the former is a theory equipped with a higher-categorical structure enabling to store more d.o.f. in certain superselection sectors. 

\item For theories descending from 6D ${\cal N}=(2,0)$ SCFTs, if the gauging procedure leads to the presence of intrinsic non-invertible defects, the total quantum dimension of the algebra implementing the non-invertibility of such defects exhibits an increase in its total quantum dimension, signalling the relative gaugeable algebra ${\cal A}_{_{\epsilon\rho}}$ is characterised by superselection sectors admitting a richer structure w.r.t. the condensed algebra leading to a single absolute 4D gauged theory.

\end{enumerate}

At first sight, the title chosen for this article might appear provocative, and the length of the paper is certainly not fair towards the rich and deep advancements made by the phenomenological community so far\footnote{For a detailed overview of the state of the art and major developments in the field, we refer the interested reader to the rich literature at our disposal, including \cite{Nilles:1995ci,Csaki:1996ks,Slavich:2020zjv,LHCReinterpretationForum:2020xtr,Allanach:2021bbd,Banks:2020gpu,Allanach:2023bgg,Hammou:2023heg,Kassabov:2023hbm,Iranipour:2022iak,Nath:2010zj}, and \cite{Allanach:2005pv,Cicoli:2021dhg,Krippendorf:2010hj,AbdusSalam:2009qd,AbdusSalam:2007pm,Conlon:2007xv,Cremades:2007ig} for slightly more theoretically-oriented perspectives in this regard, especially in relation with string theory, \cite{gsw1,gsw2,jp1,jp2,Green:1982ct,Schwarz:1982jn,Green:1984sg,Gross:1984dd,Polchinski:1994mb}.}. However, we stress that our aim is that of introducing important open questions in the realm of Particle Physics in a mathematical language approachable by categorical algebraic geometers.

Actually, this work is really meant to be the first of a series of papers by the same author continuing from past developments, bridging the gap between Pure Mathematics and Theoretical Particle Physics achievements. The author's engagement with experts from both lines of research is what motivated this project in the first instance. In upcoming work, \cite{VP}, we will be developing this description in terms of algebraic and Hilbert series calculation in greater detail. For the moment, we anticipate that the purpose of this introductory article is mainly fourfold:  

\begin{itemize}  

\item   To illustrate (some) key fundamental problems in Theoretical Particle Physics necessitating further mathematical understanding.   

\item   Opening the scene to well-established techniques in representation theory and categorical algebraic geometry to help in this regard.    

\item   For those who are not experts in the field, serve as an oversimplified explanation of the powerful formalism put forward by Braverman, Finkelberg, and Nakajima (BFN), \cite{Braverman:2017ofm,Braverman:2016wma}. In doing so we will explicitly show the relation in between their work, that of Freed, Moore, and Teleman (FMT), \cite{Freed:2022qnc}, and Teleman's, \cite{Teleman:2014jaa}, which, to the best of our knowledge, has only previously been referred to in \cite{Pasquarella:2023exd}. 

\item Motivating Phenomenologists to engage more with Pure Mathematicians, fostering mutual inspiration for common aims.

\end{itemize}

The main objective of this paper is that of explaining the higher categorical structures that are needed for describing the invariants associated to specific supersymmetric quiver gauge theories, with a particular focus on dualities and their mutual relations in terms of higher-categories.  We believe that deepening its understanding and generalisations thereof, could lead to uncharted corners of mathematics that could be used for embedding of the Standard Model (SM) of Particle Physics, \cite{Weinberg:2004kv} in possible supersymmetric unifying theories, \cite{Quevedo:2010ui}.

The present work is structured as follows: 

\begin{enumerate}

\item Section \ref{sec:2}  explains the motivations and outline of the present and forthcoming papers \cite{VP}. In doing so, we briefly introduce scattering amplitudes as crucial tools within the context of particle physics, explaining their connection with Hilbert series. We then briefly explain some of the key open questions in Theoretical Particle Physics necessitating further mathematical understanding, specifically the hierarchy problem, and the lightness of the Higgs mass (and its relation to Supersymmetry (SUSY), \cite{Quevedo:2010ui}).

\item     Section \ref{sec:4}   is mostly a revision of \cite{Freed:2022qnc,Freed:2012bs}, and we use it at this stage for multiple reasons. First of all because it allows to introduce gauge theories and Lagrangians in a mathematical language, which in turn is suitable to make connection with the higher-categorical Symmetry Topological Field Theory (SymTFT) prescription put forward by Freed, Moore, and Teleman, (FMT), \cite{Freed:2022qnc}, in the context of relative quantum field theories (QFTs).  

\item     The natural way of describing gauge theories is by means of quivers. The Standard Model itself admits such a representation. It is also the case, though, that quivers are crucial for describing certain algebraic varieties associated with highly supersymmetric setups. Section \ref{sec:5} is therefore devoted to explaining in a simplified way the well-established BFN construction leading to ring homologies. The theories in question are 3D ${\cal N}=4$ supersymmetric quiver gauge theories. This section highlights the importance that categorical algebraic geometry plays in identifying ring homologies accounting for the complete spectrum of a given theory, therefore giving a complementary insight with respect to the Hilbert series alone. In so doing, we will highlight the importance of \emph{abelianisation}, \cite{Dimofte:2018abu,Bullimore:2015lsa}.  In doing so we will explicitly show the relation in between their work, that of Freed, Moore, and Teleman (FMT), \cite{Freed:2022qnc}, and Teleman's, \cite{Teleman:2014jaa}, which, to the best of our knowledge, has only previously been mentioned in \cite{Pasquarella:2023exd}. 

\item From the discussion of the previous section, the concluding section, \ref{sec:6}, paves the way to forthcoming work \cite{VP}, where a more detailed analysis with regard to the algebraic structure and Hilbert series of (Beyond) the Standard Model scenarios will be carried out.  
Additional complementary directions addressed by the same author will be developing ring homologies for Moore-Tachikawa varieties, \cite{Moore:2011ee}, beyond categorical duality\footnote{Thereby following up from the analysis carried out in \cite{Pasquarella:2023ntw}.}, and understanding their implications for Koszul duality, \cite{Webster:2016rhh}. We plan to report of any advancements in this regard in the near future, \cite{VP}.

\end{enumerate}

\subsection{Abelianisation (once more)}  

The basic idea of abelianisation is to embed the Coulomb branch chiral ring $\mathbf{C}[{\cal M}_{_C}]$ of a non-abelian theory into a larger, semisimple abelian algebra, ${\cal A}$,   

\begin{equation}  
\boxed{\ \ \ \mathbf{C}[{\cal M}_C]\ \hookrightarrow\ {\cal A}\ \ \color{white}\bigg]}.    
\label{eq:emb}
\end{equation}

Essentially, abelianisation corresponds to fixed-point localisation in the equivariant homology, which we will briefly explain momentarily. The embedding \eqref{eq:emb} allows to:    

\begin{enumerate}    

\item  Verifying relations among elements of $\mathbf{C}[{\cal M}_{_C}]$.

\item Identifying the Poisson structure on $\mathbf{C}[{\cal M}_{_C}]$ and its deformation quantisation.  

\item Extend the algebra $\mathbf{C}[{\cal M}_{_C}]$ over twistor space, ultimately enabling to access the hyperk$\ddot{\text{a}}$hler structure on ${\cal M}_{_C}$.   

\end{enumerate}

In the context of BFN, this arises when considering a 3D ${\cal N}=4$ gauge theory on $\mathbb{C}\times\mathbb{R}$ spacetime with $\frac{1}{2}$-BPS boundary conditions ${\cal B}$ near spatial infinity on $\mathbb{C}$.

\begin{figure}[ht!]  
\begin{center}  
\includegraphics[scale=0.8]{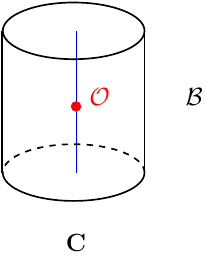}  \ \ \ \ \ \ \ \ \ \  \ \ \ \ \ \ \ \ \ \
\includegraphics[scale=0.8]{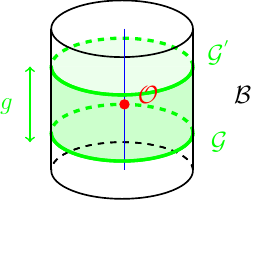} 
\caption{\small Picture reproduced from the work of \cite{Dimofte:2018abu}, encoding the key idea outlined in their treatment, and from which our work was in turn inspired.}  
\label{fig:cyl}   
\end{center} 
\end{figure}

Working in the cohomology of the Rozansky-Witten theory, ${\cal Q}_{_{RW}}$, the fields may be localised in the QM to sections of bundles that solve the ${\cal Q}_{_{RW}}$ BPS equations. This yields the gauged ${\cal N}=4$ QM on   

\begin{equation}  
{\cal M}_{_{[\mathbf{C}]}}=\left\{\ (E,X)\ \right\},  
\end{equation}  
where $E$ is a holomorphic $G_{_{\mathbf{C}}}$-bundle on $\mathbf{C}$, and $X$ is a holomorphic section of an associated $R$-bundle with gauge group, ${cal G}$ is the holomorphic $G_{_{\mathbf{C}}}$ gauge transformation on $\mathbf{C}$.  

The Hilbert space of the space of the QMs in the ${\cal Q}_{_{RW}}$ twist should be the ${\cal G}$-equivariant de Rham cohomology of ${\cal M}_{_{[\mathbf{C}]}}$. Turning on the Omega background further, corresponds to working equivariantly w.r.t. the $U(1)_{_{\epsilon}}$ spatial rotation group of $\mathbf{C}$, which is an ordinary symmetry of the moduli space, the Hilbert space is   

\begin{equation}   
{\cal H}\ \overset{def.}{=}\ H^{^{\bullet}}_{_{{\cal G}\times U(1)_{_{\epsilon}}}}\left({\cal M}_{_{[\mathbf{C}]}}\right).  
\end{equation}

${\cal M}_{_{[\mathbf{C}]}}$ is contractible to a point where $E$ is the trivial bundle and $X$ is the zero-section. ${\cal G}$ is contractible to $G$, therefore the Hilbert space reduces to   

\begin{equation}  
{\cal H}\ \simeq\ H^{^{\bullet}}_{_{{\cal G}\times U(1)_{_{\epsilon}}}}\left(\text{pt.}\right) =\mathbf{C}[\varphi,\epsilon]^{^W},   
\end{equation}    
namely Weyl-invariant polynomials in the equivariant weights $\varphi\in\mathfrak{t}_{_{\mathbf{C}}}$ and $\epsilon$.   

We can therefore write the BFN construction, which essentially reads as follows  

\begin{equation}  
\mathbf{C}_{_{\epsilon}}\left[{\cal M}_{_C}\right]\ \equiv\ H^{^{\bullet}}_{_{{\cal G}\times {\cal G}^{^{\prime}}\times U(1)_{_{\epsilon}}}}\left({\cal M}_{_{[\mathbf{C}\cup\mathbf{C}]}}\right)   \simeq H^{^{\bullet}}_{_{G\times U(1)_{_{\epsilon}}}}\left({\cal M}_{_{[\mathbf{C}\cup\mathbf{C}]}}\ \bigg /\ {\cal G}^{^{\prime}}\right),   
\label{eq:space}   
\end{equation}   
where ${\cal G,G}^{^{\prime}}$ are the groups of regular holomorphic gauge transformations on the top and bottom copies of $\mathbf{C}$, and $U(1)_{_{\epsilon}}$ is the usual spatial rotation group. Importantly, this space, \eqref{eq:space}, is non-contractible, and has highly nontrivial topology since it contains monopole operators. OPEs of local operators in the algebra $\mathbf{C}_{_{\epsilon}}[{\cal M}_{_{\epsilon}}]$ naturally correspond to the convolution product rather than the cup product.     

We are now ready to explain a particularly useful tool in the context of equivariant cohomology, namely fixed-point localisation. Letting $T\subset G$ denote the maximal torus, one finds that $T\times U(1)_{_{\epsilon}}$ fixed points of ${\cal M}_{_{[\mathbf{C}\cup\mathbf{C}]}}\bigg / {\cal G}^{^{\prime}}$ are isolated points described as pairs $(E, X), (E^{^{\prime}}, X^{^{\prime}})$ where $X=X^{^{\prime}}=0$ are zero sections, $E$ is trivial, and $E^{^{\prime}}$ is obtained from $E$ by a gauge transformation  

\begin{equation}  
g(z)=z^{^{A}}\ \ \ ,\ \ \ A\in\ \text{cochar}(G),  
\end{equation}  
meaning fixed points are labelled by cocharacters corresponding to ableian monopole operators. 

If ${\cal F}$ is the fixed point set of the $T\times U(1)_{_{\epsilon}}$ action on ${\cal M}_{_{[\mathbf{C}\cup\mathbf{C}]}}\bigg / {\cal G}^{^{\prime}}$, ${\cal F}\simeq\text{cochar}(G)$ is isomorphic to the cocharacter lattice. 

The equivariant cohomology of the fixed point set contains a copy of

\begin{equation}   
H^{^{\bullet}}_{_{T\times U(1)_{_{\epsilon}}}} (\text{pt})=\mathbf{C}[\varphi, \epsilon]  
\end{equation}  
for every point in ${\cal F}$, i.e.  

\begin{equation}   
H^{^{\bullet}}_{_{T\times U(1)_{_{\epsilon}}}} ({\cal F})\simeq\mathbf{C}\left[\varphi, \epsilon, \{v_{_{A}}\}_{_{A\ \in\ \text{cochar}(G)}}\right].  
\end{equation}  

In its localised version, one gets that 

\begin{equation}   
{\cal A}_{_{\epsilon}}\ \subset\ H^{^{\bullet}}_{_{T\times U(1)_{_{\epsilon}}}} ({\cal F})^{^{\text{loc}}}.   
\label{eq:aeht}   
\end{equation}  

The crucial difference in between the two sides in \eqref{eq:aeht} is that the localised cohomology indiscriminately inverts all weights, whereas in ${\cal A}_{_{\epsilon}}$ we only invert part of them. The localisation theorem provides the following algebraic embedding 

\begin{equation}  
\mathbf{C}_{_{\epsilon}}\left[{\cal M}_{_{[\mathbf{C}]}}\right]=H^{^{\bullet}}_{_{G\times U(1)_{_{\epsilon}}}}\left({\cal M}_{_{[\mathbf{C}\cup\mathbf{C}]}}\bigg /\ {\cal G}^{^{\prime}}\right)\ \hookrightarrow\ H^{^{\bullet}}_{_{T\times U(1)_{_{\epsilon}}}}\left({\cal F}\right)^{^{\text{loc}}},    
\label{eq:emb1}
\end{equation}  
which is the abelianisation we introduced in \eqref{eq:emb}. Importantly, the RHS of \eqref{eq:emb1} is the mother algebra we were looking for, the un-gauged algebra referred to in section \ref{sec:4}, and the key entity encoding essential information about the QFT in question. In conclusion to this section, we wish to highlight that, for any successful QFT to be fully specified, one should really be able to determine such an embedding. Any shortcomings in achieving it, should be taken as signalling a yet incomplete understanding of the underlying mathematical structure of the theory in question.

\medskip   

\medskip 

\section{Instantons and Scattering Amplitudes}    \label{sec:inst}

\subsection{Instantons}  

\emph{Instantons} are solutions to the classical equations of motion in Euclidean spacetime with a finite non-zero action. They arise in, both, QFT and QM, and can be thought of as critical points of the action. 

In QFT, instantons: 

\begin{enumerate}

\item Appear in the path-integral as leading quantum corrections to the classical behaviour of the system, and 

\item Can be used to study tunnelling in various systems, such as Yang-Mills (YM) theory. 

\end{enumerate}

In mathematical terms, a YM instanton is a self-dual or anti-self-dual connection in a principal bundle over a 4D Riemann manifold, and are topologically nontrivial solutions to the YM equations minimising the energy functional.

\subsubsection{Instantons in QM}  

An example of a system with an instanton effect is the double-well potential, for example 

\begin{equation} 
V(z)=\left(z^2-1\right)^2, 
\label{eq:dwp}   
\end{equation}
in which case the classical minima lie at $z=\pm1$, corresponding to the two lowest energy states. In QM, instead, the energy eigenstates are found by solving the Schr$\ddot{\text{o}}$dinger equation

\begin{equation}   
-\frac{\hbar^{^2}}{2m}\frac{\partial}{\partial z^{^2}}\psi+V\psi=E\psi.  
\end{equation}

In this case, though, the lowest-energy eigenvalue is unique, since the ground state wavefunction localises at both classical minima, $z=\pm1$, due to quantum interference, which justifies introducing the notion of tunnelling in the first place. In particular, instantons enable to explain why this is so in the semi-classical approximation of the path-integral formulation in Euclidean time.

\begin{figure}[ht!] 
\begin{center}   
\includegraphics[scale=1]{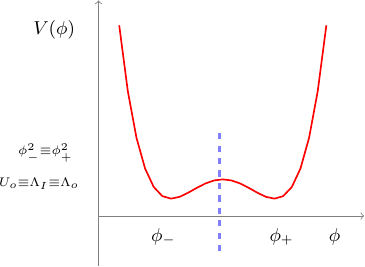}
\label{fig:dwp}  
\caption{\small}  
\end{center}  
\end{figure}

In the path-integral formulation of the tunnelling process, the transition amplitude reads as follows    

\begin{equation}  
\begin{aligned}
K\left(z_{_1}, z_{_2};t\right)&=\bigg<z=z_{_1}\bigg|\ e^{^{-\frac{iHt}{\hbar}}}\ \bigg|z=z_{_{2}}\bigg> \\
&=\int\ d[z(t)]\ e^{^{\frac{iS}{\hbar}}}.       
\end{aligned}
\end{equation}  

Under Wick-rotation, 

\begin{equation}  
\begin{aligned}
\tilde K\left(z_{_1}, z_{_2};t\right)&=\bigg<z=z_{_1}\bigg|\ e^{^{-\frac{H\tau}{\hbar}}}\ \bigg|z=z_{_{2}}\bigg> \\
&=\int\ d[z(t)]\ e^{^{-\frac{S_{_E}}{\hbar}}},          
\end{aligned}
\end{equation} 
with Euclidean action

\begin{equation}  
\begin{aligned}
S_{_E}=\int_{_{\tau_{_1}}}^{^{\tau_{_2}}}\ \left(\frac{1}{2}m\left(\frac{dz}{d\tau}\right)^{^2}+V\right)\ d\tau.     
\label{eq:nes1}
\end{aligned}
\end{equation} 

Specifying to the double-well potential \eqref{fig:dwp}, setting $z_{_2}=-z_{_1}=1$, \eqref{eq:nes1} becomes

\begin{equation}  
\begin{aligned}
S_{_E}&=\int_{_{\tau_{_1}}}^{^{\tau_{_2}}}\ \left[\frac{1}{2}\left(\left(\frac{dz}{d\tau}\right)^{^2}-\sqrt{2 V\ }\right)^{^2}\ +\sqrt{2\ }\frac{dz}{d\tau}\sqrt{V\ }\right]d\tau\\
&=\int_{_{\tau_{_1}}}^{^{\tau_{_2}}}\ \frac{1}{2}\left(\left(\frac{dz}{d\tau}\right)^{^2}-\sqrt{2 V\ }\right)^{^2}d\tau\ +\ \sqrt{2\ }\int_{_{-1}}^{^1}\ \left(1-z^{^2}\right) dz\\
&\ge\ \frac{4\sqrt{2\ }}{3}.
\label{eq:nes2}
\end{aligned}
\end{equation} 

The inequality in the last line of \eqref{eq:nes2} is saturated by the solution of the following differential equation (arising by setting the first integral in \eqref{eq:nes2} to zero)

\begin{equation}   
\frac{dz}{d\tau} =\sqrt{2 V\ }   
\end{equation}   
subject to the boundary conditions  

\begin{equation}   
z\left(\tau_{_1}\right)=-1\ \ \ \ \ \ \ \ \ \ \ , \ \ \ \ \ \ \ \ \ \ \ z\left(\tau_{_2}\right)=1. 
\label{eq:bcs}   
\end{equation}   

The instanton solution interpolating between the two classical minima therefore reads

\begin{equation} 
z(\tau)\ =\ \tanh\left(\tau-\tau_{_o}\right),    
\end{equation} 
where $\tau_{_o}=$ const., and interpolates between the two asymptotic values set by the boundary conditions \eqref{eq:bcs}.

\begin{figure}[ht!] 
\begin{center}   
\includegraphics[scale=1]{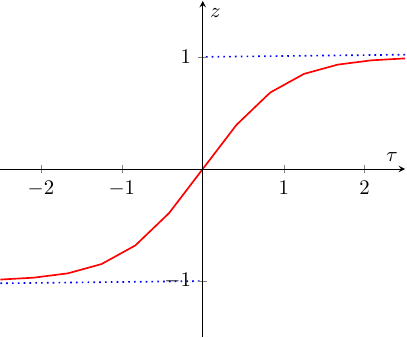}
\label{fig:tanh}  
\caption{\small}  
\end{center}  
\end{figure}

\subsubsection{Instantons in QFT}

Instantons also arise when studying  the vacuum structure of QFTs, where the true vacuum can be defined as an overlap of several topologically-inequivalent sectors (also known as topological vacua). Among the most known examples are QFTs with non-abelian gauge group, i.e. Yang-Mills theories. In such case, the inequivalent topological sectors are classified by means of the third homotopy group of $SU(2)$, 

\begin{equation}  
\pi_{_3} \left(S^{^3}\right)\ =\ \mathbb{Z},  
\label{eq:inflymany}
\end{equation}  
whose group manifold is the 3-sphere, $S^{^3}$. From \eqref{eq:inflymany}, it therefore follows that there are infinitely-many topologically-inequivalent vacua, labelled by the Pontrjagin index, $N$.  

In this case, an instanton is interpreted as a tunnelling effect between these inequivalent topological vacua. More concretely, if an instanton is assigned a Pontrjagin index $Q$, then it mediates between the topological vacua $|N>$ and $|N+Q>$, whereas the true vacuum of the theory would be expressed as the following linear supersposition  

\begin{equation}  
|\theta>\ \overset{def.}{=}\ \sum_{_{N=-\infty}}^{^{+\infty}}\ e^{^{i\theta N}}\ |N>.   
\end{equation}

\subsubsection{Instantons in 4D supersymmetric gauge theories}

Supersymmetric gauge theories often obey nonrenormalisation theorems, with the latter restricting the allowed kinds of quantum corrections. Many such theorems only apply to perturbative corrections, reason why instantons, that are non-perturbative in nature, turn out being the only corrections possible.   

Many instanton calculations were performed in supersymmetric theories during the 1970s and 1980s\footnote{See for instance \cite{Belavin:1975fg}.}.

According to the amount of supersymmetry that is present, different corrections can be considered:  

\begin{enumerate}

\item   ${\cal N}=1$ SUSY gauge theories: instantons can modify the superpotential, potentially lifting the vacua.

\item    ${\cal N}=2$: the superpotential receives no quantum correction, but the corrections to the metric of the space of vacua from instantons can be nontrivial.

\item    ${\cal N}=4$: instantons lead to no quantum corrections, neither in the superpotential, nor the K$\ddot{\text{a}}$hler metric.

\end{enumerate}

\subsubsection{Instantons in oriented string theory: the ADHM construction}  

A D$p$-brane is a gauge theory instanton in the world volume $(p+5)$-dimensional $U(N)$ gauge theory on a stack of $N$ D$(p+4)$-branes.

YM-instantons are used for studying the vacuum structure of YM theory and classifying 4-manifolds. A method for studying self-dual YM-instantons is due to ADHM, who first performed this calculation for classical gauge groups, $SU(N), SO(N)$, and $Sp(N)$. Douglas, \cite{Douglas:1996uz}, and Witten, \cite{Witten:1995gx}, found that the ADHM construction can be realised in string theory. In particular, the moduli space of instantons on $\mathbb{R}^{^4}$ is equivalent to the Higgs branch of supersymmetric gauge theories on a system of D$p$-D$(p+4)$ branes. These are quiver gauge theories with 8 supercharges (e.g. 4D ${\cal N}=2$ for $p=3$).

The moduli space of instantons is the Higgs branch of SUSY gauge theories. Gauge theories for classical gauge groups can be embedded as worldvolume theories of D$p$-branes in backgrounds of D$(p+4)$-branes. 

Self-dual instantons on $\mathbb{CP}^{^2}$ or $\mathbb{C}^{^2}$ are described an ADHM-like construction allowing to calculate the Hilbert series of the moduli space\footnote{For a more detailed explanation and overview of the Hilbert series in this context, please see appendix \ref{sec:magneticquivers}.}.

\subsection{The WKB approximation}

The aim is that of understanding the geometric structures naturally arising when trying to solve the Schr$\ddot{\text o}$dinger equations. It involves symplectic geometry, spectral theory, and microlocal analysis.

The WKB method is a technique adopted for finding approximate solutions to the Schwarzschild equation. More sophisticated developments of this have led to other techniques such as microlocal analysis, \cite{RichardM}.  

For a given Hamiltonian

\begin{equation} 
H(q,p)=\frac{p^{2}}{2m} +V(q)  
\end{equation}   
the Hamiltonian flow equations read

\begin{equation} 
\dot q=\frac{p}{m}\ \ \ ,\ \ \ \dot p=-V^{^{\prime}}(q).
\end{equation}  

Fixing $\hbar\ \in\ \mathbf{R}_{_+}$, the Schr$\ddot{\text{o}}$dinger equation reads as follows

\begin{equation}    
i\hbar\ \frac{\partial\psi}{\partial t}\ =\ \hat H\ \psi\ \ \ \ \text{with}\  \hat H=-\frac{\hbar^{^2}}{2m}\ \frac{\partial^{^2}}{\partial x^{^2}}+m.  
\end{equation}   

Looking for stationary solutions of the form

\begin{equation} 
\psi(x,t)=\varphi(x)\ e^{^{-i\omega t}}, 
\end{equation}
and assuming 

\begin{equation}   
\varphi(x)=e^{^{ix\xi}},    
\end{equation}
namely, as $V$ varies with $x$, so too does $\xi$, the WKB approximation leads to a solution of the form  

\begin{equation}  
\varphi(x)=e^{^{iS(x)/\hbar}},     
\end{equation}
with $S(x)$ the phase function solving the Hamilton-Jacobi (HJ) equations obtained from

\begin{equation} 
(\hat H-E)\varphi=\left[\frac{(S^{^{\prime}}(x)}{2m}+(V-E)-\frac{i\hbar}{2m}S^{^{\prime\prime}}(x)\right]e^{^{iS(x)/\hbar}}
\end{equation}  
assuming $\hbar<<1$, 

\begin{equation}  
H(x, S^{^{\prime}}(x))=\frac{S^{^{\prime}}(x)}{2m}+V(x)=E  
\end{equation}

\begin{equation}   
\Rightarrow\ \ \ S^{^{\prime}}(x)=\pm\sqrt{2m(E-V(x))\ }. 
\label{eq:almostbounce}
\end{equation}

Given the classical phase plane $\mathbb{R}^{^2}\simeq T^{^*}\mathbb{R}$ with coordinates $(q,p)$, the differential $dS=S^{^{\prime}}dX$ can be viewed as a mapping

\begin{equation}   
dS:\ \mathbf{R}\ \rightarrow\ T^{^*}\mathbf{R}, 
\end{equation}  
where $p\equiv S^{^{\prime}}$. Then, $S$ satisfies the HJ equation if and only if the images of $dS$ lie in the level manifold $H^{^{-1}}(E)$.   

These arguments can be extended to $\mathbf{R}^{^n}$, at which point we get 

\begin{equation}   
S:\ \mathbf{R}^{^n}\ \rightarrow\ \mathbf{R}  
\end{equation}
as a solution of the HJ equation.

In the remainder of this section, we briefly overview the applicability of the WKB approximation to two setups that are most important for our work, namely Hitchin systems and vacuum transitions.

\subsubsection{WKB for Hitchin systems}    

Given the construction of the coordinates\footnote{We refer the reader to the respective section in the appendix for a more detailed analysis of the specific setup, which is a review of the original work by Gaiotto, Moore, and Neitzke.} of appendix \ref{sec:metrics} associated to the metric on the Riemann surface for class ${\cal S}$ theories, one can look for particular asymptotic solutions of the connection ${\cal A}$ such that 

\begin{equation} 
\underset{\zeta\rightarrow0}{\lim}\ {\cal A}=\frac{R\varphi}{\zeta}.  
\end{equation}  

At every point of the Riemann surface, $\varphi$ takes two eigenvalues, $\pm\lambda$, with $\lambda$ being a multi-valued 1-form on the Riemann surface. The WKB approximation then states that, if one chooses a gauge such that 

\begin{equation}  
\varphi=\left(\begin{matrix}
    &\lambda\ \ 0\\
    &0\ \ -\lambda\\
\end{matrix}\right),
\end{equation}  
then there are two independent approximate ${\cal A}$-flat sections of the type

\begin{equation}
\psi^{^{(1)}}\ \sim\ \left(\begin{matrix}
    &e^{^{-\frac{R}{\zeta}\int\lambda}}\\
    &0\\
\end{matrix}\right)\ \ \ ,\ \ \ \psi^{^{(2)}}\ \sim\ \left(\begin{matrix}
&0\\
    &e^{^{-\frac{R}{\zeta}\int\lambda}}\\
\end{matrix}\right),
\end{equation}
and one would therefore expect that, in the $\underset{\zeta\rightarrow0}{\lim}$, the calculation reduces to that of periods of the 1-form $\lambda$ on the Riemann surface\footnote{More detailed explanation on this terminology is provided in section \ref{sec:metrics}.}.

\subsubsection{WKB for vacuum transitions}

The treatment of the first part of this subsection extends to the case of QM tunnelling between different minima of a given theory. To see this explicitly, it is sufficient to consider the simplest setting possible, where the field theory in question is defined by the action

\begin{equation}  
S=\int d^{^4}x\ \left(-\frac{1}{2}\partial_{_{\mu}}\varphi\partial^{^{\mu}}\varphi-V(\varphi)\right). 
\label{sec:actionforwkb}
\end{equation}

The corresponding energy of the system therefore reads  

\begin{equation}   
E[\varphi(x)]=\int d^{^3}x\left(\frac{1}{2}(\nabla\varphi)^{^2}+V(\varphi)\right).  
\end{equation}  

The specific profile of the potential, $V(\varphi)$, is obviously going to dictate the types (and number) of vacua available in a give theory.

\begin{figure}[ht!]    
\begin{center}   
\includegraphics[scale=1]{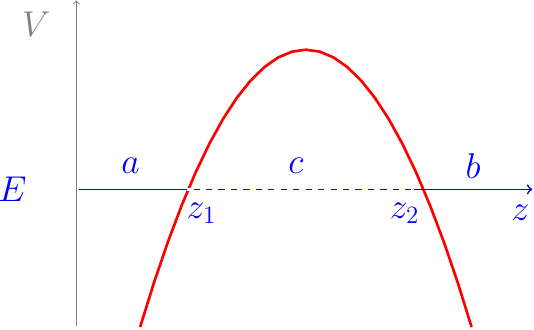}  
\caption{\small }   
\label{fig:wkb}   
\end{center}   
\end{figure}

Albeit \eqref{sec:actionforwkb} explicitly being higher-dimensional, under the assumption of spherical symmetry, the nucleation process can be recast to that of QM tunnelling, whose unique degree of freedom is the radius of the new vacuum bubble emerging in the background vacuum. 
Figure \ref{fig:wkb} is useful in understanding what we have just said. In the under-the-barrier region, the solution to the HJ equations, \eqref{eq:almostbounce}, becomes imaginary, leading to the definition of the \emph{bounce}

\begin{equation}   
\boxed{\ \ \ B\ \overset{def.}{=}\ iS= 2i\int_{_{z_{_1}}}^{^{z_{_2}}}\sqrt{2m(E-V(x))\ }\color{white}\bigg]\ \ }. 
\label{eq:almostbounce}
\end{equation}

In the following Part, we will be analysing in great depth the case of vacuum transitions for 2D field theories with gravity, where the procedure just outlined is also known as the Coleman-De Luccia vacuum decay description. For reasons that we will be explaining in due course, though, we will also be outlining the importance of identifying alternative methods for calculating decay rates. The need to do so is multifold. If we were to pick one of the most crucial points to be addressed, is the absence of a closed-form expression for the String Theory Landscape potential. More of this will be motivated in section \ref{sec:prelchapt}, as introductory to Part \ref{sec:IV}.

\section{Conclusions and Outlook of Part III}

Key advancements towards achieving a rigorous mathematical formulation of QFTs strongly relies upon a higher-categorical prescription. In particular, the symmetries of a QFT can be thought of as defects living in different categories connected by functors.  

In this Part of our work, we identified a criterion to distinguish between intrinsic and non-intrinsic non-invertible symmetries arising from the categorical structure of 6D ${\cal N}=(2,0)$ SCFTs. In particular we were able to derive a relation in terms of the quantum dimension of the condensing algebra implementing the gauging in the bulk SymTFT. In doing so, we relied upon the description in terms of relative field theories, as described in \cite{Bashmakov:2022uek} in terms of the Freed-Moore-Teleman setup, \cite{Freed:2012bs,Freed:2022qnc}, together with the gauging prescription, \cite{TJF}. For the intrinsic non-invertible case, multiplicity is greater w.r.t. the non-intrinsic case, thereby signalling the possibility for additional d.o.f. to be stored in certain superselection sectors of the resulting absolute theory. 

Our results extend arguments proposed in \cite{Kaidi:2022cpf}, where the authors also proposed a way of distinguishing intrinsic from non-intrinsic non-invertibility in 2D by means of the quantum dimension of the non-invertible defect. 

In our analysis, we encountered two major setups built upon the notion of higher-categories, namely symmetry TFTs (SymTFT) and topological orders (TO). Plenty of effort has been made towards building a correspondence in between the two, most recently in \cite{Ji:2019eqo}. Our findings provide further support towards strengthening the connection between the two description. 

We conclude this Part by stressing that our analysis is mostly motivated by furthering the understanding of  6D ${\cal N}=(2,0)$ SCFTs, from, both, a mathematical and physical point of view. To what extent these findings might be mapped to other setups is currently under investigation. In an upcoming work \cite{VP}, we will report on further advancements building on these findings.

\part{Holographic interpretation of 2D vacuum transitions}  \label{sec:IV}

\section{Summary and Motivations of Part IV} \label{sec:prelchapt}   

\subsection{Motivations}

\begin{figure}[h!] 
\begin{center}
\includegraphics[scale=0.45]{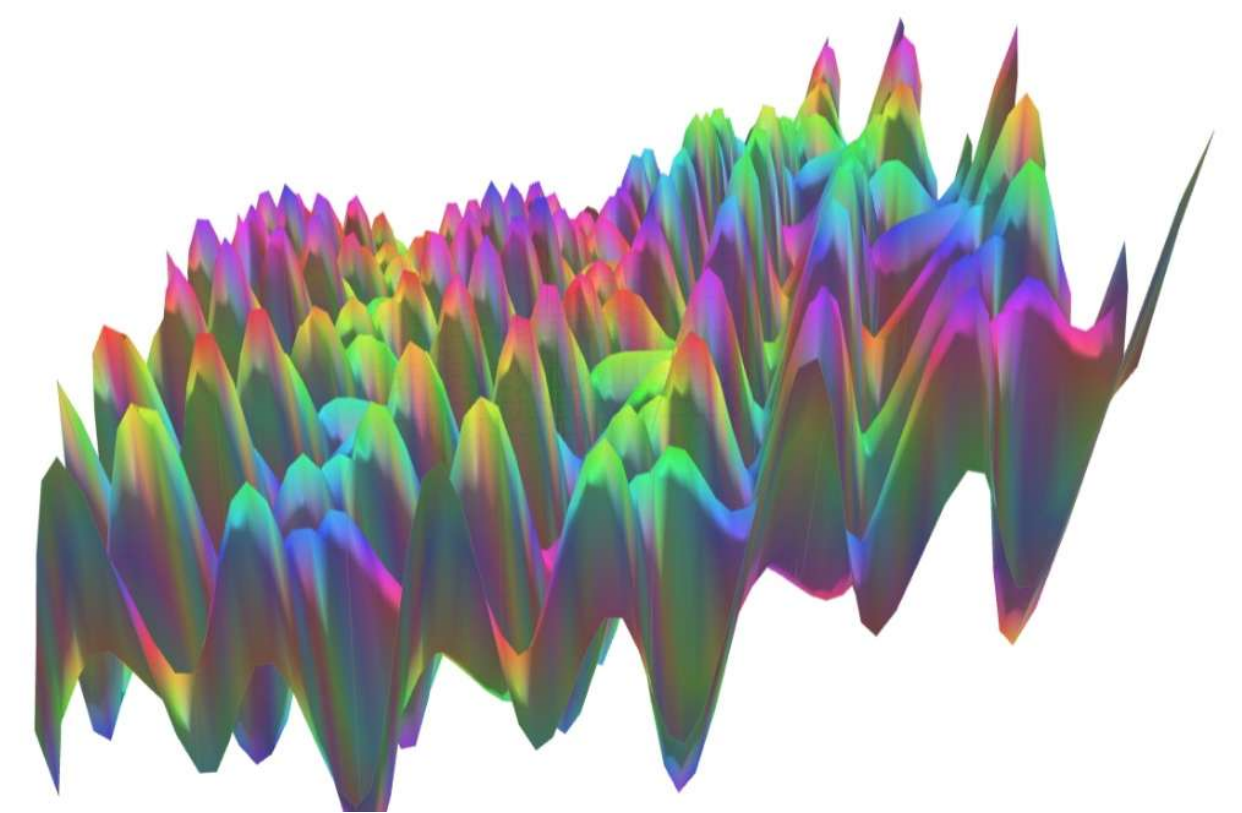}
\caption{The string theory landscape.}
\label{fig:landscape}
\end{center}
\end{figure}

As already mentioned in the introductory part of the present work, the original motivation to our work was that of studying the stability of vacua arising in the string theory landscape. The main reason for this being that of identifying a string theory model that would better suit the description of our universe in its present-day status, as well as describing its evolution according to well-established cosmological models.

A faithful description of the stability of such vacua, requires identifying the fundamental degrees of freedom characterising the minima of the string theory landscape, and plenty of effort has been made so far in doing so, by means of, both, analytic and numerical techniques\footnote{Among the most important works in this regard, we refer the interested reader to the following \cite{McAllister:2023vgy,Cicoli:2023njy,Cicoli:2023opf,Cicoli:2021dhg,Crino:2020qwk,AbdusSalam:2020ywo,Cicoli:2016olq,Cicoli:2015ylx,Aparicio:2015psl,Quevedo:2015ava,deAlwis:2014wia,AbdusSalam:2014uea,Quevedo:2014xia,deAlwis:2013gka,Burgess:2013sla,Font:2013hia,Cicoli:2012vw,Cicoli:2012fh,Cicoli:2012cy,Krippendorf:2010hj,Blumenhagen:2009gk,AbdusSalam:2009qd,Krippendorf:2009zza,Conlon:2008wa,Cicoli:2008gp,Conlon:2008qi,Conlon:2008cj,Cicoli:2008va,Burgess:2008ri,AbdusSalam:2007pm,Cicoli:2007xp,Conlon:2007xv,Cremades:2007ig,Conlon:2006wz,Burgess:2006mn,Conlon:2006tj,Avgoustidis:2006zp,Conlon:2006us}.}. However, to the best of our knowledge, no systematic method has yet been identified as effectively describing the decay of such vacua into others, and the work in collaboration with my supervisor constitutes a step forward towards this aim.

Figure \ref{fig:landscape} was purposely meant to show the randomness with which different vacua build the string theory landscape, highlighting the lack of a unique potential accounting for all of them at once.

Identifying a rigorous formalism towards describing vacuum decay processes has been a great matter of debate for almost half a century, \cite{Coleman:1980aw,Brown:1988kg,Fischler:1990pk}, and part of our joint collaboration was attempting to provide further insights into which description could better suit our purposes.

Together with the quantum study of black hole physics, the study of vacuum transitions provides a rich arena to explore quantum aspects of gravity, and may shed some light towards a proper understanding of quantum gravity. Recent progress in the study of black hole information (for a review see \cite{JM2}) has been achieved, in great part, studying concrete examples in 2D, and it is natural to explore if the concepts used there, such as quantum extremal surfaces, generalised entropies and islands  may also play a role in addressing questions regarding other physical systems involving quantum aspects of gravity, namely, early universe cosmology and vacuum transitions.

The following four sections of this work aim at providing a general perspective towards understanding  vacuum transitions in 2D. The reason for doing so is multi-fold. Among the most important, we highlight the following:  

\begin{enumerate} 

\item To understand, from the simple 2D set-up, the stability of vacua with different values and signs of the vacuum energy, in particular having in mind the string theory landscape.

\item To study the possible emergence of unitarity-violating behaviour, such as encountered in the black hole information paradox \cite{SWH, Banks:1983by,Giddings:1995gd}, in a different physical system.   

\item To identify potential holographic interpretations of vacuum transitions, \cite{Maldacena:2010un, JM}, and their generalisations to higher-dimensional holographic embeddings and defect field theories. 

\item To explore the possibility of assigning an effective entropy to transition amplitudes for spacetimes with arbitrary sign of the cosmological constant. The main motivation for this comes from recent developments towards studying the emergence of spacetime from entanglement \cite{RT, BB14, VanRaamsdonk:2016exw, VanRaamsdonk:2020ydg}.

\end{enumerate}

\subsection*{Different approaches, same aim}

Throughout the last four decades, the study of quantum transitions in 4D has been addressed by means of three main approaches, differing in terms of the way the cosmological constants are being defined, and the use of Euclidean or Lorentzian techniques. In order of appearance in the literature, they are:

\begin{itemize} 

\item Coleman-de Luccia (CDL) \cite{Coleman:1980aw}, describing transitions between different  local minima in a scalar field potential, following a Euclidean approach.

\item Brown-Teitelboim (BT) \cite{Brown:1988kg}, Euclidean vacuum transitions mediated by brane nucleation. 

\item Fischler-Morgan-Polchinski (FMP) \cite{Fischler:1990pk}, transitions between two spacetimes with different cosmological constants,  by means of the Hamiltonian formalism in Lorentzian signature. 

\end{itemize}

In each case, the quantity of interest is the transition amplitude, $\Gamma$. In the Euclidean methods, this is obtained from the \emph{bounce}, $B$, defined as the difference  between the Euclidean action evaluated on the instanton $S_{E}|_{inst} $ and on the background  $S_{E}|_{bckgr} $: 

\begin{equation}   
\Gamma\ \sim\   \exp\left(-B\right)\ \ \ , \ \ \ B\overset{def.}{=}S_{E}|_{inst}-S_{E}|_{bckgr}.
\label{eq:gamma}   
\end{equation}

The main motivation for this approach follows from the relation between the WKB approximation and  vacuum transitions, with the latter being viewed as the quantum mechanical process of a particle crossing a potential barrier. 

The Lorentzian method of FMP, instead, proposes an alternative definition for $\Gamma$; the transition amplitude from a spacetime ${\cal M}_1$ to another ${\cal M}_2$, is defined as the relative ratio of two probabilities,

\begin{equation}
\Gamma_{1\rightarrow 2}
\ 
\overset{\text{def.}}{=}    
\ 
\frac{P_{{\text{nothing}}\rightarrow {{\cal M}_1}{\text{/Wall}}/{\cal M}_2}}{P_{\text{nothing}\rightarrow {{\cal M}_1}}} 
\ 
= 
\ 
\frac{|\Psi_{\text{nothing}\rightarrow {{\cal M}_1}{\text{/Wall/}}{\cal M}_2}|^{2}}{|\Psi_{\text{nothing}\rightarrow {{\cal M}_1}}|^{2}} 
\ \ \ 
, 
\label{eq:Gamma0}
\end{equation}
with both nucleations out of nothing being identified with Hartle-Hawking (HH)-like states \cite{Hartle:1983ai}. This is a conditional probability, with the numerator and denominator corresponding to the end point of the transition and the original background spacetime, respectively.

In an earlier paper, \cite{DeAlwis:2019rxg} (see also \cite{Bachlechner:2016mtp, BBOC}), we recovered the original BT result for dS$_{4}\rightarrow$dS$_{4}$ \footnote{The outer and inner vacua are denoted by $o,I$, respectively.}, by means of the FMP method, with total action given by

\begin{equation}       
\log \Gamma_{\text{dS}\rightarrow \text{dS}}  
= 
\frac{\eta\pi}{2G}\left[\frac{[(H_{o}^{2}-H_{I}^{2})^{2}+\kappa^{2}(H_{o}^{2}+H_{I}^{2})]R_{o}}{4\kappa H_{o}^{2}H_{I}^{2}} + \frac{1}{2H_{o}^{2}} - \frac{1}{2H_{I}^{2}}\right] \ \ \ 
, 
\label{eq:bck} 
\end{equation}

Here $H_o, H_I$ refer to the Hubble parameter for the spacetimes outside and inside the bubble, $\kappa$ is the tension of the bubble, $R_o$ its radius at nucleation and $\eta=\pm 1$.
Under mutual exchange of the 2 vacua, the 1$^{st}$ term in (\ref{eq:bck}), coming from the brane, is symmetric, whereas the other 2 terms just flip their sign. From this follows that, the ratio between the direct and reverse transition reads

\begin{equation}   
\frac{\Gamma_{o\rightarrow I}}{\Gamma_{I\rightarrow o}}= e^{\frac{\eta\pi}{2G}\left(\frac{1}{H_o^2}-\frac{1}{H_I^2}\right)}=e^{\eta(S_o-S_I)}
\label{eq:db1}    
\end{equation}    
For $\eta=1$, since  $S_o$ and $S_I$ are the entropies of the corresponding de Sitter spacetimes, this is a statement of \emph{detailed balance}. According to \cite{LW}, such entropic argument cannot be extended to spacetimes other than dS\footnote{The definition of the entropy for the static dS patch is a longstanding matter of debate. Motivated by our findings, we will be arguing that the entropy definition arising from the study of vacuum transitions is only meaningful when considering a finite portion of the dS patch. Our findings are therefore in agreement with those of others, such as, e.g. \cite{Banihashemi:2022htw}, where the authors outline the need to introduce a boundary to dS for assigning a notion of entropy to it.}, due to the topological change the spacetime would undergo as a consequence of the change in sign of its curvature. Furthermore, upon taking a vanishing background cosmological constant, \eqref{eq:bck} diverges, thereby forbidding up-tunnelling from a flat spacetime. In such case, \eqref{eq:db1} would thereby suggest that the background entropy is divergent. However, this is clearly in contrast with the common lore that the Minkowski vacuum should have vanishing entropy instead.

Following the steps of \cite{Fischler:1990pk}, whose work was in turn motivated by \cite{Farhi:1989yr}\footnote{The configuration described by Farhi, Guth and Guven, \cite{Farhi:1989yr}, has recently been addressed in the literature, \cite{Susskind:2021yvs}, and claimed to be unsuitable for the application of detailed balance. However, we will be able to prove that the claims made in \cite{Susskind:2021yvs} are not applicable in our setup given the locality of the processes being described.}, we showed in \cite{DeAlwis:2019rxg} that this issue can be addressed by  introducing a Schwarzschild black hole in the background, and realising Minkowski as the limit of the vanishing mass of the black hole itself. Indeed, the total action, in such case, reads

\begin{equation}  
S_{\text{Sch}\rightarrow \text{dS}}  
= 
\frac{\eta\pi}{2G}\left[S_{bounce} - S_{bckg} \right] 
= 
\frac{\eta\pi}{2G}\left[S_{bounce} - 4G^{2}M^{2} \right]   
\label{eq:bhds}  
\end{equation}  
and clearly remains finite in the formal limit \footnote{Note that this limit is only an approximation since the lower bound on a black hole mass is the Planck scale. } $\underset{M\rightarrow0}{\lim}$.

The question of whether, and how, we can possibly reconcile these apparently contrasting behaviours partly  motivated the present work, with the possibility to correctly assign an entropy to the spacetimes involved. Before embarking in the analysis outlined in the present work, our original interpretation was that transitions were taking place in \emph{local} regions of spacetime; as such, topologically-inspired arguments for the inapplicability of detailed balance could be dropped. But still, a probably more interesting question emerged: can we assign an entropic interpretation to the direct amplitude itself and not only to the ratio of the amplitudes, independently of detailed balance? As we shall see, the answer will turn out to be affirmative in 2D. 

In achieving this result, holography will turn out to play a key role. Since its 1$^{st}$ formulation, \cite{JM}, there have been many generalisations: AdS/BCFT, \cite{Fujita:2011fp}, wedge holography (codim-2), \cite{BB-1}, $T\bar T$, \cite{BB901, BB902, BB70, BB48, BB47}, the dS/CFT , \cite{BB20}, and dS$_{_{D+1}}$/dS$_{_{D}}$ correspondence, \cite{BB22}, etc. Last but not least, the \emph{island}, \cite{JM1, Penington:2019npb, JS}, pushing the holographic duality to its currently most extreme formulation, since it involves the  entanglement of spatially-disconnected regions. Throughout our treatment we will encounter applications\footnote{For other recent applications of entanglement and islands within a cosmological setup, please see, e.g. \cite{Antonini:2022xzo, Antonini:2022blk, VanRaamsdonk:2020tlr, MVRHFC, iic, BBTH, Geng:2021iyq, Geng:2021wcq, Geng2:2021wcq,Geng:2022slq,  Langhoff:2021uct, Betzios:2019rds, Betzios:2021fnm, Bousso:2022gth, dadg,daeh,Chen:2020tes}.} of all such cases within the context of 2D vacuum transitions, with the different methods showing complementary holographic interpretations, thereby consistently reconciling all predictable disagreements.\footnote{This shows that the holographic obstructions towards realising specific cosmological and gravitational processes (as in \cite{Freivogel:2005qh, BBHEP, Fu:2019oyc}), can be overcome.}         
   
Our main result in this regard is having found expressions of the form \eqref{eq:bck} for vacuum transition rates in the 2-dimensional case. We will perform the calculation in all three formalisms outlined above (CDL, BT and FMP), providing their corresponding holographic interpretation. Given that the BT case has already been addressed by the authors in their original work, \cite{Brown:1988kg}, we will be using it as a ``standard ruler'' to check the results obtained by other means. In doing so, however, we will not simply take it for granted, but, rather, derive it from the suitably adjusted setup of CDL in 2D, which, to the best of our knowledge, has not been dealt with explicitly in the literature. As outlined in section \ref{sec:4}, the common starting point for both Euclidean methods will be the Almheiri-Polchinski action, \cite{AP},

\begin{equation}   
S_{2D}   
= 
\int\ d^2 x\ \sqrt{-g\ }\ \left[\ \phi^2\  {\cal R} + \lambda\ \phi^{\prime2}-U(\phi)\ 
\right]  
\end{equation} 
which  resembles the starting point of \cite{Coleman:1980aw} for decays in the presence of gravity. 

Due to the specific features of the 2D setup, though, the Ricci scalar or the scalar kinetic term can be eliminated by a suitable rescaling of the metric and of the potential. As we shall see, removing the $\phi^{\prime\ 2}$-term we can calculate the transition amplitude either with or without the \emph{thin-wall} approximation, proper to CDL. On the other hand, upon removing the Ricci scalar term, and adding suitable boundary terms ensuring the action satisfies a variational principle, we recover the gravitational action of BT. The importance of the addition of the boundary terms will be highlighted in due course, but it is worth stressing that, for the case in which the kinetic term is suppressed, they trivially vanish in either vacuum. Correspondingly, this is mapped to the fact that the calculation by means of the CDL method in 2D  describes transitions in \emph{absence} of gravity. From the holographic interpretation, we will see that, indeed, the total bounce is reminiscent of the entropy of an \emph{internal} CFT$_2$, which geometrises the RG-flow interpolating between different values of the dilaton.

For the case of BT, instead, the addition of the boundary terms, accounting for the presence of gravity, is encoded in boundary entropies of defect CFT$_1$ s placed at the endpoints of the \emph{bulk} CFT$_2$, which in turn are dual to the end-of-the-world (ETW) branes, \cite{ATC1, ATC}, where the spacetimes undergoing the transition live (cf. figure \ref{fig:plot2f1}). A more detailed explanation of figure \ref{fig:plot2f1} will be provided in section \ref{sec:4}.

\begin{figure}[h!]    
\begin{center}    
\includegraphics[scale=1]{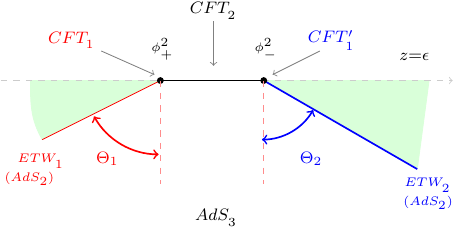}    
\caption{\small The CFT$_2$ geometrises the RG-flow defining $B_{wall}$, and in this sense is \emph{internal}. The CFT$_1$s are the duals of the AdS$_2$ spacetimes involved in the transition living on the 2 ETW branes. Their opening angles, $\Theta_{1,2}$, and corresponding tensions, result in the value of $\Lambda_{\pm}$ on the 2 sides of the wall. A more detailed explanation is provided in section \ref{sec:4}}    
\label{fig:plot2f1} 
\end{center}      
\end{figure}

\medskip

\subsection{Outline and main results}

In summary, Part IV of this thesis is structured as follows: in section \ref{sec:4} we extend the CDL formalism to 2D, which, to the best of our knowledge, has not been explicitly done in the literature. In doing so, we use the JT-gravity theory analysed by Almheiri and Polchinski \cite{AP}. Thanks to the nature of the 2D setup, under  suitable  Weyl transformation, we can trade the kinetic term for the dilaton with the Ricci scalar. We thereby perform the calculation in both ways, showing that the case without the kinetic term describes transitions in absence of gravity. Upon removing the Ricci term, instead, the action can be brought back to the same form as in BT, up to a boundary term whose role is essential to account for the gravitational interaction. The second part of the section provides a brief overview of the 2D treatment of BT, as already addressed by the authors in their original work, \cite{Brown:1988kg}, also highlighting some key features that will turn  out useful in following sections. 

In section \ref{sec:2}, we turn to the Hamiltonian formalism of \cite{Fischler:1990pk} applied to 2D JT-gravity \cite{Jackiw:1984je,Teitelboim:1983ux, BB21}, and determine the transition rates in absence of black holes, finding that uptunnelling from and downtunnelling to Minkowski spacetimes are not allowed. 

In section \ref{sec:3}, we generalise the results of section \ref{sec:2} by adding a constant term in the action that allows for non-extremal black holes to emerge, as long as their mass lies within a certain range to be specified in due course. The range emerges from requiring the existence of 2 distinct phisical turning points. We conclude that, upon adding a black hole of suitable mass on either side, the flat spacetime limit does not violate unitarity, and we are still left with a well defined transition amplitude. 

In section \ref{sec:4.4} we provide a possible holographic interpretation of the total bounces and actions calculated throughout our work, showing mutual compatibility and complementarity. Altogether, this leads to an exhaustive explanation of figure \ref{fig:plot2f1}. In particular, we find that the corresponding expression for the transition rate in presence of gravity, and in absence of black holes, is given by the difference of entropies of $T\bar T$-deformed CFTs, hence proving the locality of the nucleation process. Upon adding black holes, within a suitable mass range, instead, the total action can be expressed as the difference of generalised entropies, with an island emerging beyond a critical value of the black hole mass. In particular, we show that, whenever an island is present, uptunnelling is always possible. Furthermore, the results obtained by means of the FMP method are found to agree with the expression provided by \cite{MVR} for describing mutual approximation of boundary states belonging to different CFTs under suitable parametric redefinition. The BT results, which in section \ref{sec:3} were proved to be equivalent to the FMP results in absence of black holes, can be expressed in terms of entropies of BCFT$_2$s with 2 nontrivial boundary conditions dual to ETW branes. Last but not least, the CDL result can be recast in the form of an entropy product of a CFT$_2$, thereby showing agreement with the expectations following from the analytic behaviour encountered in section \ref{sec:4}. These findings further support the interpretation of the dilatonic field as playing the role of an \emph{internal} CFT$_2$ on which the entanglement entropy is being evaluated. 

One of the key results of our work can therefore be synthesised by saying that the total action (or bounce) associated to the direct vacuum transition process carries an \emph{internal} entropic interpretation. In particular, for the BT and FMP cases, they can always be expressed as the difference of generalised entropies. However, only the formalism of \cite{Fischler:1990pk} provides the right setup for an island to emerge. Following a brief summary of our findings, at the end of section \ref{sec:4.4}, we also point out: 

\begin{enumerate}    

\item  The emergence of an entropic hierarchy, similar to the one featuring among the von Neumann algebras recently analysed by Witten et al. within the context of the algebra of observables, \cite{EW2, EW3},  

\item The importance of scales throughout our analysis, by making a comparison with recent developments in the literature, specifically Susskind \cite{Susskind}, Schlenker and Witten, \cite{EW1}.   

\end{enumerate}

\section{Euclidean transitions in 2D }\label{sec:4}

In the first section of this Part, we analyse 2D vacuum transitions by means of the Euclidean methods of CDL and BT, with the latter being essentially already known from the original work of \cite{Brown:1988kg}. Due to the specifics of the 2D setup, both results can be obtained from the same formalism, namely that of Almheiri and Polchinski, \cite{AP}. However, along the way, we encounter significant differences, resulting in different final expressions for the bounces. A physical explanation for this will be outlined throughout the treatment, leaving a more detailed justification to section \ref{sec:4.4}, where, by means of holographic arguments, we will be showing the complementarity of the methods applied.

\subsection{Vacuum transitions in 2D without gravity}

\begin{figure}[ht!]
\begin{minipage}[c]{0.45\textwidth}    
\caption{\footnotesize  The transition between the minima of the potential is independent of the specific profile of $V$, and is only a function of $\Lambda_{I,o}$ and the tension of the wall. }  
\label{fig:cdl}  
\end{minipage}\hfill   
\begin{minipage}[c]{0.45\textwidth}      
\includegraphics[scale=0.9]{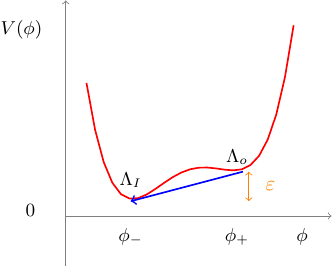} 
\end{minipage} 
\end{figure}

Let us start considering the simplest case of vacuum transitions in 2D without including gravity. As already argued by the authors of \cite{Coleman:1980aw}, the final amplitude does not depend on the particular profile of the potential. However, their calculation still relies upon the potential exhibiting specific features. In particular, for the purpose of describing the process, it must have at least two minima, corresponding to the vacua involved in the transition. The simplest starting point is therefore to assume a double well potential, such as the one depicted in figure \ref{fig:cdl}, whose analytic expression can be split into two parts as follows, 

\begin{equation} 
V(\phi)\ \overset{def.}{=} \ V_o + V_{\varepsilon},  \label{eq:V}
\end{equation} 
where $V_{_{o}}$ is the ordinary degenerate double well potential satisfying the following properties

\begin{equation}    
V_o(\phi_+)\equiv V_o(\phi_-)\ \ \ ,\ \ \ \frac{d V_o}{d\phi}\bigg|_{\phi_{\pm}}=0, 
\end{equation}
with an additional symmetry breaking term proportional to the energy difference between the minima, 
\begin{equation} 
\varepsilon\overset{def.}{=} V(\phi_+)-V(\phi_-)\ \equiv\ \Lambda_{o}-\Lambda_{I}.
\end{equation}
The interpolation between the two minima is mediated by the scalar field, as depicted in figure \ref{fig:cdl}, since the latter is characterised by a nontrivial spatial profile

According to the Euclidean procedure, the decay rate, $\Gamma$, is defined from the bounce, $B$, as follows 
\begin{equation}
\Gamma\sim e^{-B } 
\ \ \ 
, 
\ \ \ 
B\overset{def.}{=}S_{E}|_{inst}-S_{E}|_{bckgr},
\end{equation}    
where $S_{E}$ denotes the Euclidean action associated to a given theory. Specifically, there are 3 main contributions,
\begin{equation}  
\begin{aligned}
B_{tot}^{CDL}
&=& 
B_{out}+B_{wall}+B_{inside}. 
\label{eq:BTOT1}    
\end{aligned} 
\end{equation}
For a scalar field admitting an O(2)-symmetric instanton solution   
$ds^2    
=    
d\rho^2+\rho^2d\theta^2,   $
the corresponding Euclidean action reads:

\begin{equation}   
S_E    
=
\int\ dx^2\ \left[\frac{1}{2}(\partial_{\mu}\phi)^2+V(\phi)\right]       
=
\pi\int\ d\rho\ \rho\ \left[\frac{1}{2}(\phi^{\prime})^2+V(\phi)\right].
\end{equation}  
where $^{\prime}\equiv\frac{d}{d\rho}$. The equation of motion for $\phi$, in the thin-wall approximation, reduces to:    

\begin{equation}   
\phi^{\prime}\phi^{\prime\prime}    
-V_o^{\prime}=\left(\frac{1}{2}(\phi^{\prime})^2-V_o\right)^{\prime}=0,
\label{eq:eomnew}
\end{equation}
which is exactly integrable once having chosen suitable boundary conditions, $\phi(\rho=\infty)=\phi_+$.

\begin{equation}    
\phi^{\prime} 
=    
\sqrt{2(V_o(\phi)-V_o(\phi_{\pm}))\ }\ \ \ \Rightarrow\ \ \ \rho-\bar\rho 
=    
\int_{\bar\phi}^{\phi}\frac{d\phi}{\sqrt{2(V_o(\phi)-V_o(\phi_{\pm}))\ } }  
\label{eq:eomnew1}
\end{equation}   
with $\bar\phi\overset{def.}{=}\frac{\phi_++\phi_-}{2}$. Equation \eqref{eq:eomnew1} defines the turning point $(\bar\rho,\bar\phi)$.  
The total bounce is simply given by the interior and the wall

\begin{equation}  
\begin{aligned}
B_{tot}   
=&    
B_{in}+B_{wall}  
=   
-\pi\bar\rho^2\ \varepsilon+2\pi\bar\rho S_1,   
\label{eq:btotcdlnog}
\end{aligned}
\end{equation}
where, making use of \eqref{eq:eomnew1}, the tension of the wall is defined as

\begin{equation}    
S_1    
=     
\int d\rho\ \left[2(V_o(\phi)-V_o(\phi_+))\right] .  
\end{equation}
Equation \eqref{eq:btotcdlnog} is extremised at $\bar\rho   
=    
\frac{S_1}{\varepsilon} $, at which the total extremised bounce reads:

\begin{equation}   
\boxed{\ \ B_{tot}^{\ \ extr} 
=     
\frac{\pi\ S_1^2}{\varepsilon}    \ \ }
\label{eq:Schw}   
\end{equation}   
The presence of a single spatial dimension allows for a direct identification of the nucleation process described by \eqref{eq:Schw}    with the Schwinger pair production, \cite{Schwinger:1951nm}, whose bounce reads    

\begin{equation}   
B_{Schw}  
=   
\frac{\pi\ m^2}{|eE_{on}|} 
=\pi m\bar\rho_{_{BT}},
\label{eq:schpr}       
\end{equation}
with $\bar\rho_{_{BT}}\overset{def.}{=}\frac{m}{|eE_{on}|}$ being the value of the turning point extremising the bounce of the corresponding process. Equation \eqref{eq:schpr} suggests the following parametric identification:

\begin{equation}   
m\ \ \leftrightarrow\ \ S_1\ \ \ ,\ \ \ \varepsilon\ \ \leftrightarrow\ \ |e E_{on}|, 
\end{equation}  
where $m$ is the mass of the particles, whereas $|eE_{on}|$ is the difference in energy provided by the electric field prior and after pair creation, and can therefore be associated to the energy difference between the minima in $V$. 

\begin{figure}[ht!]    
\begin{minipage}[c]{0.6\textwidth}    
\caption{\footnotesize Brane nucleation in one spatial dimension, with $\Lambda_{\pm}\overset{def.}{=}V(\phi_{\pm})$ is analogous to the Schwinger process upon interpreting the brane with tension $\sigma$ as the particle/antiparticle pair (denoted by the two red dots).}    
\label{fig:Schw}      
\end{minipage}\hfill  
\begin{minipage}[c]{0.3\textwidth}       
\includegraphics[scale=1]{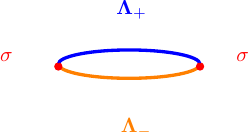}    
\end{minipage}    
\end{figure}

\subsection{Coleman-De Luccia from the Almheiri-Polchinski setup }\label{sec:4.1}

We will now turn to the case with gravity. The Euclidean action for gravity coupled to a scalar field should contain a potential and a kinetic term. However, in 2D gravity, the Ricci scalar needs to be coupled to the dilaton to ensure that the theory is nontrivial, as first formulated by Jackiw and Teitelboim in \cite{Jackiw:1984je}. This is a first major difference with respect to the 4D treatment, from which a dynamical coupling between $\phi$ and $\rho$ is expected to arise at the level of the equations of motion.

\subsubsection*{Light-cone and cartesian coordinates}

Given the conformal metric 

\begin{equation}   
ds^2 
= 
-e^{2\omega}\ dx^+dx^-    
= 
4\ e^{2\omega}\left(-dt^2+dz^2\right)      
\label{eq:lineelemlineelem}    
\end{equation} 
with $x^{\pm}\overset{def.}{=}t\pm z$ and $2\ \partial_{\pm}=\partial_t\pm\partial_z$, the Ricci scalar in 2D is given by the following expression:
\begin{equation} 
\begin{aligned}
{\cal R}    
&=        
-4\  e^{-2\omega}\left(\ \partial_{+}\partial_{-}\omega\ -  2\ \partial_+\omega\ \partial_-\omega\ \right)     \nonumber   \\
&=       
-2 \left[\ \partial_t\left(e^{-2\omega}\  \partial_t\ \omega   \right)- \partial_z\left(e^{-2\omega}\  \partial_z\ \omega   \right)\ \right].
\label{eq:Ricci}               
\end{aligned} 
\end{equation}

In Lorentzian signature, the 2D action in terms of $x^{\pm}$ reads

\begin{equation}   
S_{2D}   
= 
\int\ d^2 x\ \sqrt{-g\ }\ \left[\ \phi^2\  {\cal R} + \lambda\  \partial_{+}\phi\ \partial_{-}\phi-U(\phi)\ 
\right].
\label{eq:Se}    
\end{equation}

Given the particular nature of the 2D setup, we can proceed in two ways: either removing the kinetic term or performing a Weyl rescaling of the metric and setting ${\cal R}=0$. Both procedures will be explored in turn in the present section.


\subsubsection*{Removing the kinetic term}

The kinetic term can be removed under suitable rescalings

\begin{equation}  
\rho\rightarrow\phi^{-\lambda/4}\rho   
\ \ \ 
,    
\ \ \ 
U\rightarrow\phi^{-\lambda/2}U\overset{def.}{=}U_o,   \ \ \    
\label{eq:fredef}    
\end{equation}  
such that the action resembles that of JT-gravity. Once integrated by parts, equation \eqref{eq:Se} is simply

\begin{equation} 
\begin{aligned}
S_{2D}       
&=        
-8\ \left(\phi^2\partial_-\omega\right)\bigg|_{x^+=\text{const.}}-\int d^2 x \ \left[8\left(\partial_+\phi^2\right)\left(\partial_{-}\omega\right)     +U_o e^{2\omega}\right],
\label{eq:newacr}   
\end{aligned} 
\end{equation}
from which the e.o.m. for $\phi^2$ and $\omega$ can be extracted

\begin{equation}    
4\partial_+\partial_-\omega    
= 
\frac{e^{2\omega}}{2}\partial_{\phi^2}U_o \equiv   T_{+-},
\ \ \ \ \ \ \ \ \ \ \ \    \ \ \ \ \ \ \ \ \ \ \ \ 
4\partial_-\partial_+\phi^2    
=   
U_oe^{2\omega}   \equiv T_{-+},
\label{eq:SE1}   
\end{equation}   
corresponding to the off-diagonal components of the energy momentum tensor $T_{\mu\nu}$. The remaining components, instead, read:

\begin{equation}   
T_{\pm\pm}    
= 
0 
\ \ \ 
\Rightarrow    
\ \ \ 
\partial_{\pm}\left(e^{-2\omega}\ \partial_{\pm}\phi^2\right) =0.  
\label{eq:thc1}
\end{equation}    

In cartesian coordinates, the Lorentzian action reads:  
\begin{equation} 
\begin{aligned}
S_{2D}         
&=&         
S_{bdy}-\int\ dt\ dz\ \left[\ 2\phi^2\ \partial_z^2\ \omega   +U_o e^{2\omega}\ \right], 
\label{eq:newact2}   
\end{aligned} 
\end{equation}
where $\omega=\omega(z), \phi^2=\phi^2(z)$. \eqref{eq:newact2} corresponds to the starting point for the extension of CDL to 2D, and the constraints \eqref{eq:thc1} reduce to \cite{AP}, 
\begin{equation} 
\partial_z\left(e^{-2\omega}\ \partial_z\left(\phi^2\right)\right)    
=     
0.   
\label{eq:3rdcinz}    
\end{equation}

\subsubsection*{Wick rotation and polar coordinates}

Following \cite{Coleman:1980aw}, we turn to Wick-rotated and  polar coordinates:
\begin{equation}   
ds^2 
= 
e^{2\omega}\ \left(dt^2+dz^2\right) 
= 
\rho^2\ \left(dr^2+r^2d\theta^2\right),    
\label{eq:lineelem}    
\end{equation} 
 where,
\begin{equation}   
\begin{cases}   
t
= 
r\ \cos\theta 
\\    
z=r\ \sin\theta
\end{cases}   
\ \ \ 
,    
\ \ \ 
r\overset{def.}{=}\sqrt{t^2+ z^2\ }\ \ \ , \ \ \ \theta\overset{def.}{=}\tan^{-1}\frac{\ z\ }{t}.   
\label{eq:newvar4}   
\end{equation}  
The Euclidean action reads:   
\begin{equation} 
\begin{aligned}
S_E         
&=  
4\pi\ \phi^2r \omega^{\prime}\bigg|_{r_i}^{r_f}\  -2\pi\ \int\ dr \left[  2\left(\phi^2\right)^{\prime}\ r \omega^{\prime}-r U_o e^{2\omega}\right], 
\label{eq:Lagr1}    
\end{aligned} 
\end{equation}
where the $2\pi$ factor comes from the integration over $\theta$ and $^{\prime}\overset{def.}{=}\partial_r$, with the equations of motion (e.o.m.)  for $\phi^2$ and $\rho$ now being:
\begin{equation}    
2 \left(r\ \omega^{\prime}\right)^{\prime} = -r\ e^{2\omega} \ \partial_{\phi^2}U_o, 
\ \ \ \ \ \ \ \ \ \ \ \    \ \ \ \ \ \ \ \ \ \ \ \    
( r(\phi^2)^{\prime})^ {\prime}    
=   
- r\ e^{2\omega}    \ U_o.
\label{eq:er}
\end{equation}     
Notice that the latter is exactly the e.o.m. for the scalar field obtained by CDL in the thin-wall approximation. Indeed, the only term that is missing w.r.t. the full equations is the $\frac{\rho^{\prime}\phi^{\prime}}{\rho}$-term that can be tuned to zero in the thin-wall approximation. Assuming $\theta$-independence of $\phi$ and $\omega$, \eqref{eq:3rdcinz} reads
\begin{equation} 
\begin{aligned}
\partial_{r}\ \left(\frac{e^{-2\omega}}{r}\ \partial_{r}\ \phi^2 \right)  =0.
\label{eq:3rdcinr}    
\end{aligned} 
\end{equation}
To obtain an explicit expression  for \eqref{eq:BTOT1}, we need to specify what we mean by the minima of the potential and the wall separating them within the Almheiri-Polchinski setup. We now turn to describing both, one at a time.

\subsubsection*{\texorpdfstring{${\left({\phi}^2\right)^{\prime}= 0}$}{} : defining the vacua\ }

For $\phi^2$=const., \eqref{eq:er} implies $U_o$=0 and \eqref{eq:3rdcinr} is trivially satisfied. Redefining $u\overset{def.}{=}\ln r$, \eqref{eq:er} becomes    

\begin{equation}    
2\ \ddot\omega = -r^2\ e^{2\omega} \ \partial_{\phi^2}U_o=- \partial_{\phi^2}U_oe^{2(\omega+u)} ,
\label{eq:ep1}
\end{equation}
where $^{\cdot}\overset{def.}{=}\partial_u$. Redefining $\omega+u\overset{def.}{=}f$, equation \eqref{eq:ep1} becomes     

\begin{equation}    
2\ \ddot f  =- \partial_{\phi^2}U_oe^{2f}.
\label{eq:ep2}
\end{equation}
The first equation in \eqref{eq:er} shows that $\partial_{\phi^2}U_o \overset{def.}{=}2\Lambda=$ constant. It defines the cosmological constant of the 2D spacetime. We can therefore determine the solutions corresponding to dS and AdS by respectively taking positive or negative values of such constant:


\medskip   

\underline{\ 1) $\Lambda>0$ \ } \ equation \eqref{eq:ep2} is solved by
\begin{equation}  
 \rho 
= 
e^{\omega}    
= 
e^{f-u}    
= 
\frac{2a}{\sqrt{ 2\Lambda\  }\ \cosh(au+b) \ r\ } . 
\label{eq:case1}    
\end{equation}



\begin{figure}[ht!]    
\begin{center}    
\includegraphics[scale=0.5]{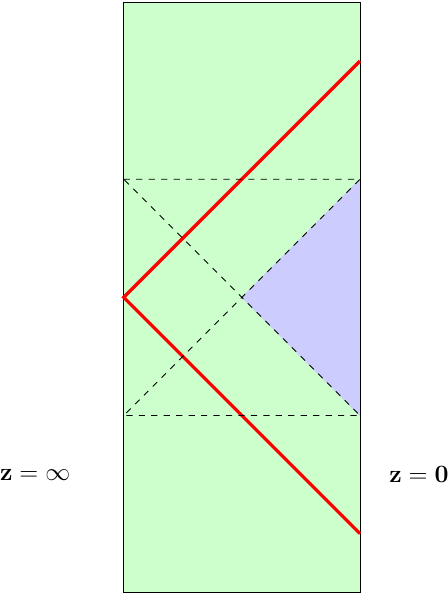} 
\ \ \     
\includegraphics[scale=0.5]{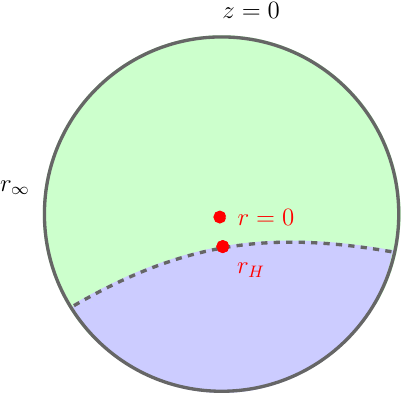}   
\ \ \ 
\includegraphics[scale=0.5]{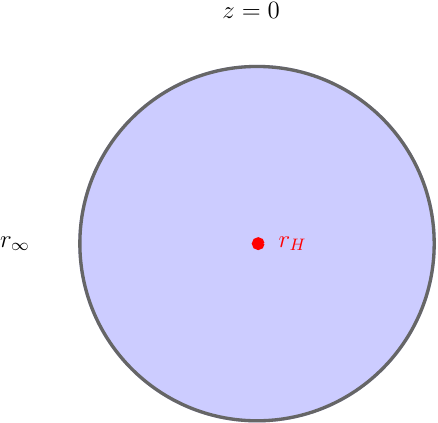}
\caption{\footnotesize The picture on the LHS shows the different regions of AdS$_2$ covered by the global (the whole rectangular strip), Poincarè (on the RHS of the red lines) and BTZ black hole (the shaded blue region). For the purpose of interest to us, we will only be using the latter, given that the first two are associated to the infinite regions of spacetime. The figures in the middle and on the right, show that the BTZ metric effectively covers only a portion of the Poincarè disk.   }   
\label{fig:3pict3} 
\end{center}   
\end{figure}

\underline{\ 2) $\Lambda<0$ \ }  \ equation \eqref{eq:ep2} is solved by  


\begin{equation}  
\rho    
=    
e^{\omega}   
=    
e^{f-u}   
= \begin{cases}
\frac{a}{\sqrt{2\Lambda\ }\ (au+b)\ r} \\       

\frac{a}{\ \sqrt{2\Lambda\ }\ \sin(\text{h})  (au+b) \ r}  \\    
\frac{a}{\sqrt{2\Lambda\ }\ \sin  (au+b) \ r}  . 
\end{cases}    
\label{eq:cases2}    
\end{equation}  
The three cases outlined in \ref{eq:cases2} correspond to the Poincar\'e patch of AdS$_2$, global AdS$_2$ and an AdS$_2$ black hole, respectively. Figure \ref{fig:3pict3} helps visualising the regions covered by the different coordinate systems. 
For either sign of the cosmological constant, the radial coordinate can be redefined, such that the line element takes the following form: 
\begin{equation}
ds^2    
=    
d\hat r^2+f(\hat r)\ d\theta^2 .    
\end{equation}  
The corresponding redefinitions are   
\begin{equation} 
r 
= 
\left(\ \frac{\ 1\ }{K}\ \tan\ \left(\frac{\sqrt{2\Lambda\ }\ \hat r}{2}\ \right)\ \right)^{1/a}    \ \ \  \text{for } \Lambda>0, \ \ \ \    
\label{eq:hatr1}
\end{equation}  
and    
\begin{equation} 
\frac{\ \ln\left|a\ \ln|r|+b\right|\ }{\ \sqrt{ 2\Lambda\  } \ }     
=         
\hat r   
\ \ \ ,\ \ \ r 
= 
\left(\ \frac{\ 1\ }{K}\ \tanh\ \left(\frac{\sqrt{2\Lambda\ } \hat r}{2}\ \right)\ \right)^{1/a}     \ \ \  \text{for } \Lambda<0, \ \ \ \   
\label{eq:hatr}     
\end{equation}    
for the $1/r\ \ln r$ and $\frac{1}{\sinh}$ cases, respectively.

For $a=1, b=0$, \eqref{eq:case1} and \eqref{eq:cases2} reduce to: 

\begin{equation}    
\boxed{\ \ 
\rho_{\text{dS}_2} 
=    
\frac{2}{\ \sqrt{2\Lambda\ }\ (r^2+1)\ } 
\ \ \ 
,    
\ \ \ 
\rho_{\text{AdS}_2} 
=    
\begin{cases}    
\frac{a}{\sqrt{2\Lambda\ }\  \ln|r|\ r \ }  \\    
\frac{2i}{\ \sqrt{2\Lambda\ }\ (r^2-1)\ } \\    
\frac{2i}{\ \sqrt{2\Lambda\ }\ (r^{2i}-1). \ }    
\end{cases}    \  }    
\label{eq:casesmetric}    
\end{equation}    
Under a suitable coordinate transformation, these results are equivalent to the ones obtained by Almheiri and Polchinski.
The first and third solutions in \eqref{eq:cases2} are not relevant for our purposes, given that they correspond to the Poincarè patch, and global AdS. In figure \ref{fig:3pict3}, they correspond to the region inside the red triangle and the green shaded strip, respectively. Instead, we will restrict to the $\rho \sim \frac{1}{\sinh}$ case, corresponding to the BTZ black hole, \cite{Banados:1992wn}, (the region outside the black hole is shaded in blue), which is a quotient of the Poincarè patch.

\subsubsection*{\texorpdfstring{$(\phi^2)^{\prime}\neq 0$}{} : defining the wall }

We now turn to the case in which $(\phi^2)^{\prime}\neq0$. Integrating over \eqref{eq:3rdcinr} twice, we find, \cite{AP}
\begin{equation}      
(\phi^2)^{\prime} 
= 
r\ c_1\ \ e^{2\omega}     
\ \ \ 
\Rightarrow   
\ \ \ 
\phi_+^2-\bar \phi^2=\frac{c_1}{2}\int_{0}^{\bar r} dr\ r\ e^{2\omega},      
\label{eq:follident122}    
\end{equation}  
where $\bar\phi^2\overset{def.}{=}\frac{\phi_{+}^2+\phi_{-}^{2}}{2}$ is the mean value of the field between the two minima. This defines the turning point ($\bar\phi^2,\bar r$) in analogy with the Euclidean formalism of \cite{Coleman:1980aw}. Up to now, the constant $c_1$ is unspecified. However, its importance is central in our treatment, as will be explained later.

The radius of the nucleated bubble is parameterised by the $\hat r$-variable, defined in \eqref{eq:hatr1}  and \eqref{eq:hatr} for dS$_2$ and AdS$_2$, respectively. Setting $a=1, b=0$, it is a monotonic function and sgn\ $\hat r\equiv$ sgn$ r$. It therefore follows that we can trade the integration over $\hat r$ with that over $r$ without loss of generality.
From the equations of motion, we know that $U_o=0$ at any minimum, but different values of $\phi^2$ come with a corresponding value of $c_1$ triggering the flow described by \eqref{eq:follident122}. 
Equation \eqref{eq:follident122}  describes the flow from one value of the dilaton to that at the turning point   
\begin{figure}[ht!]          
\begin{minipage}[c]{0.3\textwidth}        
\includegraphics[scale=2.5]{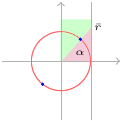}   
\end{minipage}\hfill
\begin{minipage}[c]{0.6\textwidth}
\caption{\footnotesize The brane in 2D is mapped to a particle/antiparticle pair (blue dots). The angle $\alpha\overset{def.}{=}\sqrt{\frac{\Lambda}{2}\ }\ \hat r$, denotes the location of the brane separating the background (green) from the interior of the bubble (purple). The corresponding value of $r$ is obtained by the intersection shown in the figure, namely from $\tan$(h)$\alpha$. Note that $\bar r=1$ corresponds to $\alpha=\frac{\pi}{4}$.}     \label{fig:2p}   
\end{minipage}    
\end{figure}

\begin{equation} 
\begin{aligned}
\phi_+^2-\phi_-^2   
&=    
c_1^+\int_0^{\bar r}\ dr\ r\ e^{2\omega}    
=      
\frac{c_1^+}{\Lambda_+}\ \int_0^{\bar r}\ dr\ \frac{2r}{ \ (r^2-1)^2\ }   =      
-\frac{c_1^+}{\Lambda_+}\ \left[\ \frac{1}{\ \bar r^2-1\ } +1\ \right].  
\label{eq:RGflow}      
\end{aligned} 
\end{equation} 

Equation \eqref{eq:RGflow} can also be interpreted as describing the running of the cosmological constant along the RG-flow originating from $\phi_+^2$ and parameterised by $c_1^+$.
Upon integrating over the reverse flow, defined by $c_1^-\equiv -c_1^+$, we get

\begin{equation} 
\begin{aligned}
\phi_-^2-\phi_+^2   
&=       
-\frac{c_1^-}{\Lambda_-}\ \left[\ \frac{1}{\ \bar r^2-1\ } +1\ \right].  
\label{eq:RGflow4}     
\end{aligned} 
\end{equation}
Consistency between \eqref{eq:RGflow}and \eqref{eq:RGflow4} leads to a constraint for the turning point $\bar r$, which in turn places constraints on the parameters of the theory of either vacua. A similar analysis can be carried out for dS and mixed AdS/dS transitions, and the respective constraints arising from this procedure can be summarised as follows

\begin{equation}  
\boxed{\ \ \ \begin{aligned}       
&\bar r^2 \bigg|_{  \text{AdS}_2\rightarrow\text{AdS}_2}
=    
1+\frac{\Lambda_-}{\Lambda_+} \ \ \ \ \ \ \  ,\ \ \ \ \  \ \    \bar r^2  \bigg|_{ \text{dS}_2\rightarrow\text{dS}_2} 
\frac{\Lambda_+}{\Lambda_-} -1  \\    
\\    
&\bar r^2   \bigg|_{\text{AdS}_2\leftrightarrow\text{dS}_2}    
=   
\frac{1}{2}\ \left(-\frac{\Lambda_++\Lambda_-}{\Lambda_+-\Lambda_-}\pm\sqrt{ \left(\frac{\Lambda_++\Lambda_-}{\Lambda_+-\Lambda_-}\right)^2+8\ \ }\right).  
\label{eq:constrads} 
\end{aligned}\ \ }         
\end{equation}  
For dS$_2\rightarrow$dS$_2$ processes, the constraint $\bar r>0$ requires $\Lambda_+>\Lambda_-$, implying only downtunnellling is allowed, wheras for transitions of the AdS$_2\leftrightarrow\ $dS$_2$ kind, only the the $+$ solution ensures $\bar r$ is physical.

\subsection*{Minkowski vacua     }

Before turning to the evaluation of the bounces, we wish to highlight that the system of equations \eqref{eq:er} and \eqref{eq:3rdcinr}, together with the following expression valid at local minima of the potential

\begin{equation}   
U_o\bigg|_{min}    
=    
\phi^2\ \partial_{\phi^2}\ U_o+2c_1         
=   
0     
\label{eq:defuo}        
\end{equation}     
implies that the Minkowski vacuum cannot provide a stable solution originating a nontrivial flow. Indeed, from \eqref{eq:defuo}, for $\Lambda=0$, it follows that $c_1=0$, implying the flow generated by such vacuum defines a flat direction in the space of metrics and does not end in a vacuum other than Minkowski itself. Because of this, at least for the moment, we will not be analysing any transitions involving such spacetimes, and we will only be looking at the flat limit of the bounces by taking $\underset{\Lambda_{\pm}\rightarrow0}{\lim}$.

\subsubsection*{The total bounce}

The total bounce is the result from the sum of 2 contributions,    

\begin{equation} 
\begin{aligned}
B_{tot} 
= 
B_{wall}+B_{in}, 
\label{eq:tbe}   
\end{aligned} 
\end{equation}
given that $B_{out}=0$. The scalar field is constant in a local minimum of the potential, and therefore $U_o=0$ due to \eqref{eq:er}. From \eqref{eq:er}, the Euclidean actions for (A)dS$_2$, read

\begin{equation} 
\begin{aligned}
S_E\bigg|_{\text{dS}_2}   
&=      
-2\pi\ \phi_-^2\int_{0}^{\bar r}\ dr\ r\ e^{2\omega}\ \partial_{\phi^2}U_o  =      
4\pi\ \phi_-^2\ \left[\ \frac{1}{\ \bar r^2+1\ }-1\ \right],
\label{eq:dS2}   
\end{aligned} 
\end{equation}

\begin{equation} 
\begin{aligned}
S_E\bigg|_{\text{AdS}_2}   
&=          
-2\pi\ \phi_-^2\int_{0}^{\bar r}\ dr\ r\ e^{2\omega} \ \partial_{\phi^2}U_o  =     
4\pi\ \phi_-^2\ \left[\ \frac{1}{\ \bar r^2-1\ }+1\ \right],
\label{eq:AdS2}   
\end{aligned} 
\end{equation}
from which the inner bounces associated to any possible transition between (A)dS$_2$ spacetimes, can be evaluated.
Inside the wall, the bounce reads   

\begin{equation} 
\begin{aligned}
B_{wall}    
&=     
2\pi\ \int_{0}^{\bar r} d\ r\ r\ e^{2\omega}\left[-\phi^2\ \partial_{\phi^2}\ U_o+U_o(\phi)+\phi_+^2\ \partial_{\phi^2}\ U_o(\phi_+)-U_o(\phi_+)\right] \nonumber\\   
&=    
-2\pi\ \phi^2\ U_o\bigg|_{\phi_+^2}^{\phi_-^2}+\frac{4\pi}{c_1}\ \int_{\phi_-^2}^{\phi_+^2}d  \left(\phi^2\right)\ \left[U_o(\phi)+\phi_+^2\ \Lambda_+\right]\nonumber\\  
&=        
\frac{4\pi}{c_1}\ S_1+4\pi\ \left(\phi_-^2-\phi_+^2\right),    
\label{eq:wallB}    
\end{aligned}   
\end{equation}
where 

\begin{equation}    
S_1    
\overset{def.}{=}    
 \int_{\phi_+^2}^{\phi_-^2}d  \left(\phi^2\right)\ U_o(\phi)    \ \ \ \text{with} \ \ U_o   
=   
-\frac{e^{-2\omega}}{r}\ \left(r(\phi^2)^{\prime}\right)^{\prime},   
\label{eq:tensionT}    
 \end{equation}   
where the last expression follows from the equation of motion for $\phi^2$. According to the specific configuration being studied, there are 4 possible cases:

\begin{equation} 
\begin{aligned}
B_{wall}        
&=   4\pi\Delta\phi^{^2}\cdot \begin{cases}  
 \left(2\frac{\phi_-^2}{\phi_+^2}+1\right)    \\
 \\
 \left(2\frac{\phi_-^2}{\phi_+^2}-1\right)    \\
 \\
\left(2\frac{\phi_-^2}{\phi_+^2}+1\right)     \\
\\
 \left(2\frac{\phi_-^2}{\phi_+^2}-1\right)    
    \end{cases}    
\Rightarrow \boxed{ B_{tot}= 4\pi \Delta\phi^{^2}\cdot \begin{cases}  
     \left(\frac{\phi_-^2}{\phi_+^2}+2\right)   \ \ \ [\text{AdS}_2\rightarrow \text{AdS}_{2}]\\
\\
 \left(3\frac{\phi_-^2}{\phi_+^2}-2\right)  \ \   [\text{dS}_2\rightarrow \text{dS}_{2} ] \\
\\
\left(3\frac{\phi_-^2}{\phi_+^2}+2\right)   \ \   [\text{AdS}_2\rightarrow \text{dS}_{2}]\\
\\
\left(\frac{\phi_-^2}{\phi_+^2}-2\right)    \ \ \ [\text{dS}_2\rightarrow \text{AdS}_{2}]
    \end{cases} }
\label{eq:wallnew1}    
\end{aligned} 
\end{equation}
where $\Delta\phi^{^2}\overset{def.}{=}\phi_-^{^2}-\phi_+^{^2}$.  

Note that, upon taking $\underset{\Lambda_{\pm}\rightarrow 0}{\lim}$, all transition amplitudes vanish. In particular, there is no 2D counterpart for, either, dS$_4\rightarrow$\ Mink$_4$ nor Mink$_4\rightarrow$\ AdS$_4$ processes, which were originally addressed in \cite{Coleman:1980aw}.

\subsection*{CDL in the thin-wall approximation} \label{sec:thinwall}

In the above derivation of the total bounces,  a closed form expression could be found without needing to resort to the thin-wall approximation. It is therefore interesting to see what the result might be for the case in which we were to apply the standard procedure as in \cite{Coleman:1980aw}. 
The starting point is still an O(2)-symmeytric instanton solution 

\begin{equation} 
\begin{aligned}
ds^2=d\xi^2+\rho^2(\xi)d\theta^2.
\label{eq:O2}
\end{aligned}
\end{equation}
The radial coordinate used in our derivations, $r$,  is related to $\xi$ in \eqref{eq:O2} as follows\footnote{Notice that fo AdS the same arguments follow, the only difference being that trigonometric functions turn into hyperbolic functions due to the negative sign of $\Lambda$.}
\begin{equation} 
\begin{aligned}
\hat r
=\sqrt{\frac{2}{\Lambda}\ }\tan^{-1} r  
\ \ \ 
\Rightarrow    
\ \ \ 
d\hat r   
=     
\sqrt{\frac{2}{\Lambda}\ }\ \frac{dr}{1+r^2}
=     
\sqrt{\frac{2}{\Lambda}\ }\ \cos^2\left(\sqrt{\frac{\Lambda}{2}\ }\xi\right)\ dr   
=d\xi,
\end{aligned}
\end{equation} 
and to the function $\rho_{_{CDL}}$ in \eqref{eq:O2} as 
\begin{equation}
    \rho_{_{CDL}}    
    =\sqrt{\frac{2}{\Lambda}\ }\ \frac{r}{1+r^2}.
\end{equation}
In particular, we will be looking at the case of dS$_2\rightarrow$Mink$_2$ transitions.
In order to do so, we need to re-express the Euclidean action

\begin{equation}   
S_E  
=2\pi\int_{0}^{\bar r} dr\ \left[2\phi^2(r\omega^{\prime})^{\prime}+rU_oe^{2\omega}\right],   
\end{equation}   
in terms of the new variables $(\hat r, \rho)$ or $(\xi,\rho)$. 
For Minkowski, $\hat r\equiv r\equiv \xi\equiv \rho$. The Euclidean action therefore simply vanishes on the inside. For dS, we know that
$e^{2\omega}=\frac{2}{\Lambda(r^2+1)^2}$, 
as previously derived. On the inside, the dilaton is constant, and the Euclidean action  reduces to
\begin{equation} 
S_{in}  
= 
2\pi\phi^2r\omega^{\prime}   
\ \ \ \ \ \ \ \ \text{with} \ \ \ \ \ \ \ \  
\omega^{\prime}
= 
\frac{(e^{\omega})^{\prime}}{e^{\omega}}     =-2\rho_{_{CDL}}\sqrt{\frac{\Lambda}{2}\ }.  
\label{eq:rasrho}    
\end{equation}    
For ease of notation, we will simply write the radial coordinate as $\rho$. Inverting the last equality in \eqref{eq:rasrho}, we get     

\begin{equation} 
r 
= 
\frac{1-\sqrt{1-2\Lambda\rho^2\ }}{\sqrt{2\Lambda}\rho}   
\ \ \ \ \ \ \  
\Rightarrow\ \ \ \ \ \ \ S_{in}  
= 
-2\pi\phi^2\left(1-\sqrt{1-2\Lambda\bar\rho^2\ }\right).
\label{eq:Sin1}    
\end{equation} 
The wall's bounce, instead, reads  

\begin{equation}   
\begin{aligned}   
B_{wall}   
=&    
\pi\bar\rho\ \bar\phi^2\ S_1+4\pi(\phi_-^2-\phi_+^2),
\label{eq:newea}    
\end{aligned}   
\end{equation}
where we made use of the thin-wall approximation for extracting $\bar\rho$ from the integral and have also defined the tension of the wall in terms of the rescaled potential in the Einstein frame, $U_o^{^{new}}$ 
\begin{equation}    
S_1   =    
2\int d\xi \ U_o^{^{new}}  =    2\int d\xi \ \left[\phi^{-2}U_o-\partial_{\phi^2}U_o\right].   
\end{equation}   
For the case of a dS$_2\rightarrow$Mink$_2$ transition, the total bounce therefore reads   
\begin{equation}   
B_{tot}    
= 
-2\pi\phi_+^2\left(1+\sqrt{1-2\Lambda\bar\rho^2\ }\right)+\pi\bar\rho\ \bar\phi^2\ S_1 ,  
\label{eq:Btotdsmink} 
\end{equation}    
and it is extremised at
\begin{equation}      
\bar\rho=\frac{S_1\ \bar\phi^2}{4\Lambda\ \phi_+^2\sqrt{1+\frac{S_1^2\ \bar\phi^4}{8\Lambda\phi_+^4}\ }} .  
\label{eq:tp}
\end{equation}

\begin{equation}   
\Rightarrow\ \ \ \boxed{\ \ B_{tot}^{\ \text{dS}_2\rightarrow\text{Mink}_2}\bigg|_{extr}  
= 
2\pi\phi_+^2\ \sqrt{1+\frac{S_1^2\ \bar\phi^4}{8\Lambda\phi_+^4}\ }\left(1-\frac{3}{1+\frac{S_1^2\ \bar\phi^4}{8\Lambda\phi_+^4}}\right)  .}    
\end{equation}   
In a similar fashion, for the generic (A)dS$_2\rightarrow$(A)dS$_2$ case, the total bounce reads   

\begin{equation}   
\boxed{\ \ \ B_{tot}^{\ \text{(A)dS}_2\rightarrow\text{(A)dS}_2}    
= 
2\pi\phi_-^2\left(1+\sqrt{1-2\Lambda_-\bar\rho^2\ }\right)-2\pi\phi_+^2\left(1+\sqrt{1-2\Lambda_+\bar\rho^2\ }\right)+\pi\bar\rho\ \bar\phi^2\ S_1 \textcolor{white}{\Biggl [} \color{black} .}    
\label{eq:dstodstw}
\end{equation}

\subsection{Brown-Teitelboim in 2D from Almheiri-Polchinski}  \label{sec:BTFAP}

We now turn to the alternative path for determining the bounce describing 2D transitions, namely by removing the Ricci scalar and restoring the kinetic term in the AP action:

\begin{equation}   
S_{AP} 
=\int dx^2\ \left[\phi^2{\cal R}+U_o\right]e^{2\omega}. 
\label{eq:AP}   
\end{equation}   
The $\phi^{\prime\ 2}$-term can be restored by reabsorbing $\phi^2$ in the metric. In particular, under the rescalings $\rho\rightarrow\phi\rho$ and $U_o\rightarrow \phi^{-2}U_o\overset{def.}{=}16\Lambda_{_{BT}}$, \eqref{eq:AP} turns into

\begin{equation}   
\begin{aligned}
S_{AP}  
&=8\pi\int dz\ \left[-u^{\prime\ 2}+4\Lambda_{_{BT}}e^{2f}\right], 
\label{eq:AP1}   
\end{aligned}
\end{equation} 
with $f\overset{def.}{=}\ln\phi+\omega\overset{def.}{=}u+\omega$. 
Before delving into further calculations, some important remarks are in order. The equations of motion undoubtably play a key role within the CDL formalism, as can be appreciated by direct inspection of their original work in 4D, \cite{Coleman:1980aw}. When performing the calculation for transition amplitudes within the AP setup (once having removed the kinetic term for the scalar field), the crucial relation that allowed us to circumvent the need to resort to the thin-wall approximation (upon which \cite{Coleman:1980aw} rely) was the change of variables resulting from requiring vanishing diagonal components for the energy momentum tensor, \eqref{eq:thc1}. As previously argued, in polar coordinates, they reduce to   

\begin{equation} 
\partial_r\left(\frac{e^{-2\omega}}{r}\ \partial_r\phi^2\right)   
=0 
\ \ \ 
\Rightarrow 
\ \ \ 
\partial_z\left( u\ \partial_z u\right)   =   (uu^{\prime})^{\prime}
=0,
\label{eq:encons}
\end{equation}
where $z\overset{def.}{=}\ln r$. At the same time, the term on the RHS in \eqref{eq:encons} also emerges upon integrating by parts the  kinetic term in  \eqref{eq:AP1}. Correspondingly, we are led to consider the following quantity, $M_{_{ADM}}\overset{def.}{=}uu^{\prime}\bigg|_{bdy}$, which can be identified with the ADM mass of the spacetime. When removing the kinetic term in ${\cal L}_{_{AP}}$, the dilaton is constant in each vacuum solution, therefore $M_{_{ADM}}$ is trivially vanishing to start with, as follows from the equation of motion\footnote{As we shall see in greater detail in section \ref{sec:4.4}, this is consistent with the fact that such types of transitions take place in absence of gravity.} On the other hand, when performing the Weyl transformation of the metric leading to \eqref{eq:AP1}, this is no longer true. Indeed, when using \eqref{eq:AP1} to describe transitions, $\Delta M_{ADM}\overset{def.}{=}uu^{\prime}\bigg|_{bck}^{inst}\neq 0$. We thereby need to add such boundary term by hand to ensure energy conservation throughout the nucleation process as a whole.

From the considerations we have just made, \eqref{eq:AP1} should therefore be upgraded in the following way

\begin{equation}   
\begin{aligned}
S_{BT}^{\ grav} 
&=S_{AP}+16\pi uu^{\prime}\bigg|_{bdy}\\  
&=8\pi\int dz\ \left[-u^{\prime\ 2}+4\Lambda_{_{BT}}e^{2f}\right]  +16\pi uu^{\prime}\bigg|_{bdy},
\label{eq:BT1}   
\end{aligned}
\end{equation}
which is exactly equivalent to the BT action describing brane nuclaetion in 2D in presence of gravity, in which case the total bounce in conformal gauge (with  metric function parameterised by $e^{\varphi}$) reads, \cite{Brown:1988kg},  

\begin{equation}  
B_{_{TOT}} 
= 
2\pi\ m \bar\rho+\frac{\pi}{2k}\ \int_{\bar r_{o}}^{{\infty}}\ dr\ r\ \left[-\left(\varphi_{,r}\right)^2\bigg|^{inst}_{bck}+4\Lambda_{I}e^{\varphi_{inst}}-4\Lambda_{o}e^{\varphi_{bck}}\right]+\frac{\pi}{k}r\varphi_{,r}\varphi\bigg|^{inst}_{bck},
\label{eq:origBT1}    
\end{equation}  
where the \emph{effective} cosmological constants are defined as follows

\begin{equation} 
\Lambda_{o,I} 
\ 
\overset{def.}{=}      
\ 
\lambda+\frac{1}{2} k\ E_{o,I}^{2}.
\label{eq:Leff}
\end{equation} 
In BT's notation, $k$ is proportional to $G_{_{N}}$ and is kept fixed throughout their treatment. The dynamics is thereby carried by the change in the electric field, as follows from \eqref{eq:Leff}. On the other hand, on the case of AP, as discussed earlier on in the section, we can think of $k$ as being the running coupling ensuring the interpolation in between the two vacua parameterised by different values of the dilatonic field. Indeed, in this sense we argued earlier on that transitions in AP are equivalent to RG-flows.

The equations of motion for $\phi$ is:   

\begin{equation}   
-u^{\prime\prime}=4\Lambda_{_{BT}}\ e^{2\omega+2u}.  
\label{eq:eompbt}
\end{equation}  
It admits  a solution of the kind

\begin{equation}     
u(z)=    
-\ln\left|\ \cos\text{(h)}\left(-2\sqrt{\Lambda_{_{BT}}}e^{2\omega}z+\text{const.}\right)\ \right| . 
\end{equation}  
Fixing $\omega\equiv 0$ and const.$\equiv 0$, the profile for the scalar field reads: 

\begin{equation}   
\phi(z)=\frac{1}{\ \cos\text{(h)}\left(-2\sqrt{\Lambda_{_{BT}}\ } \ z\right)\ },
\label{eq:eompbt1}
\end{equation}  
which is either a trigonometric or hyperbolic function, according to the sign of $\Lambda_{_{BT}}$.
Upon integrating the first term in \eqref{eq:BT1}    by parts, and using \eqref{eq:eompbt}, we get

\begin{equation} 
\begin{aligned}   
S       
&=8\pi\int_{\bar z}^{z_{_{h}}}\ dz \ u^{\prime\prime}\ (2u-1), 
\label{eq:CDLAP1}    
\end{aligned}
\end{equation}   
where $\bar z$ and $z_{_{h}}$ denote the location of the turning point and the horizon, respectively. Specifically, for AdS, the latter is simply $z_{_{h}}\equiv 0$, whereas for dS it is finite. 
For the instanton to exist, $\bar z>z_{_{_h}}$ is needed, and \eqref{eq:CDLAP1} therefore reads:

\begin{equation}   
\begin{aligned}
S_{tot}   
&=-16\pi\ \left[\sqrt{\Lambda_{_{BT}}}\tan\text{(h)}\left(-2\sqrt{\Lambda_{_{BT}}}  z\right)\ln\left|\ \cos\text{(h)}^2\left(-2\sqrt{\Lambda_{_{BT}}}\  z\right)\right|+\right.\\
&\ \ \ \ \ \ \ \ \ \left.+4\Lambda_{_{BT}}\   z+3\sqrt{\Lambda_{_{BT}}}\tan\text{(h)}\left(-2\sqrt{\Lambda_{_{BT}}}\  z\right)\right]\bigg|_{\bar z}^{z_{_{h}}}.
\label{eq:stotbtfromap}    
\end{aligned}
\end{equation}
Equation \eqref{eq:stotbtfromap}  holds for, both, the instanton and background solutions, the only difference being the value of $\Lambda_{_{BT}}$. From now on it is convenient to analyse the AdS and dS cases separately. 
\underline{1)\ \ AdS$_2\rightarrow$AdS$_2$}

\medskip

In this case $z_{_{h}}\equiv0$. Redefining 

\begin{equation}   
\sqrt{\Lambda_{_{\pm}}\ }\  \sqrt{1\pm\phi_\pm^2\ } \overset{def.}{=}\phi_\pm^{new}   
\ \ \ 
\Rightarrow    
\ \ \ 
\phi_{\pm}^2= \pm\left(1-\frac{\phi_\pm^{\ new\ 2}   }{\Lambda_{\pm}}\right),  
\label{eq:sametp}    
 \end{equation}    
 ensures that now $\phi_{tp}$ is the same on either side at $\bar z$, hence

\begin{equation} 
\begin{aligned}   
B_{in}^{\ \text{AdS}_2\rightarrow\text{AdS}_2}   
&= 16\pi \left[\phi_{tp}^{new} \ln\left|\ \frac{1-\frac{\phi_{tp,-}^{new\ 2}}{\Lambda_{_{-}}}}{1-\frac{\phi_{tp,+}^{new\ 2}}{\Lambda_{_{+}}}}\  \right|+3(\Lambda_{_{-}}-\Lambda_{_{+}})\ \bar z\right]. 
\label{eq:AP111}  
\end{aligned}   
\end{equation} 
Which follows from $\phi_{tp,-}^{new}\ \equiv\ \phi_{tp,+}^{new}$, by means of \eqref{eq:sametp}, whereas the wall's bounce reads:

\begin{equation} 
\begin{aligned}   
B_{wall}       
&=8\pi\int dz\ \left[- u^{\prime\ 2}+4\Lambda_{_{BT}}e^{u2}\right] \overset{def.}{=}32\pi\ \sigma   =  32\pi\ m\ \bar z,    
\label{eq:AP1111}  
\end{aligned}   
\end{equation} 
where $\sigma$ denotes the tension of the brane and $m\overset{def.}{=}\Lambda_+-\Lambda_-$. The total bounce therefore reads:

\begin{equation} 
\begin{aligned}   
\boxed{\ \ \ B_{tot}^{\ \text{AdS}_2\rightarrow\text{AdS}_2}       
=16\pi \left[\phi_{tp}^{new} \ln\left|\ \frac{1-\frac{\phi_{tp,-}^{new\ 2}}{\Lambda_{_{-}}}}{1-\frac{\phi_{tp,+}^{new\ 2}}{\Lambda_{_{+}}}}\  \right|-\frac{1}{3}m\  \bar z\right]. \color{white}\bigg].\color{black}\ \ }
\label{eq:AP111adsads}  
\end{aligned}   
\end{equation} 
\underline{2)\ \ dS$_2\rightarrow$dS$_2$}   
Similar considerations can be performed for the case of dS. The main difference is that now the horizons contribute with nontrivial terms to the actions. In particular, being $z_{_{h}}\neq0$, we get 

\begin{equation} 
\boxed{\ \ \ 
\begin{aligned}   
B_{tot}^{\ \text{dS}_2\rightarrow\text{dS}_2}              
&= 16\pi \left[\phi_{tp}^{new} \ln\left|\ \frac{1-\frac{\phi_{tp,-}^{new\ 2}}{\Lambda_{_{-}}}}{1-\frac{\phi_{tp,+}^{new\ 2}}{\Lambda_{_{+}}}}\  \right|+3\left(\Lambda_{_{-}}-\Lambda_{_{+}}+\frac{m}{6}\right)\ \bar z\right.+\\
&\left.+\phi_{h,-}^{new}\left(2-3\sqrt{1-\frac{\phi_{h,-}^{new\ 2}}{\Lambda_-}\ }\right)-\phi_{h,+}^{new}\left(2-3\sqrt{1-\frac{\phi_{h,+}^{new\ 2}}{\Lambda_-}\ }\right)\right]. 
\label{eq:AP111adsads1}  
\end{aligned}  \ \ }    
\end{equation} 
Once more, the terms linear in $\bar z$ provide an energy conservation relation and are therefore vanishing. The terms featuring in the last line, instead, provide the contributions from the horizons. The results \eqref{eq:AP111adsads}, \eqref{eq:AP111adsads1} agree with the expression of the total bounce for type-1 instantons in \cite{Brown:1988kg}

\begin{equation} 
\boxed{\ \ \ B 
= 
2 \pi m\bar\rho-\frac{4\pi}{k}\ \ln\frac{r_i}{r_o}  + \frac{2\pi\bar\rho}{k}\left(\Lambda_i r_i-\Lambda_o r_o\right)  \textcolor{white}{\Biggl [} \color{black} .\ \ }
\label{eq:BT}
\end{equation} 
prior to extremising with respect to $\bar\rho$. In \eqref{eq:BT}, we made use of the same notation as can be found in the literature, with

\begin{equation}  
\bar r_{_{I,o}}  
\overset{def.}{=}  
\frac{2}{\Lambda_{_{I,o}}\bar\rho}\ \left[1-\sqrt{1-\Lambda_{_{I,o}}\bar\rho\ }\ \right].   
\end{equation}   
Importantly, the last 2 terms in \eqref{eq:BT} arise only for the dS$_2\rightarrow $ dS$_2$ case, and are absent for AdS$_2\rightarrow$AdS$_2$, consistently with the results obtained in \eqref{eq:AP111adsads1} and \eqref{eq:AP111adsads}, respectively. Compatibility with our results follows from the following identifications

\begin{equation}   
\bar z\longleftrightarrow\ \bar\rho\ \ \ ,\ \ \ \phi_{_{tp}}^{^{new}}\overset{def.}{=}\frac{1}{k}=\frac{E_{o,I}^{2}}{2\left(\Lambda_{o,I}-\lambda\right)}\ \ \ ,\ \ \ \frac{\phi_-^{new\ 2}}{\Lambda_{_{-}}}\overset{def.}{=}1+\frac{\bar r_{i,o}^2\Lambda_{\pm}}{4} .   
\end{equation}

\subsection*{Key points and remarks}

\begin{itemize}

\item  Unlike the ordinary Schwinger process, in presence of gravity, the Euclidean action needs to be dressed with a boundary term corresponding to the ADM mass
\begin{equation}   
\begin{aligned}
S_{BT}^{\ grav} 
&=S_{AP}+16\pi uu^{\prime}\bigg|_{bdy} .
\label{eq:B1T}   
\end{aligned}
\end{equation}

\item In BT, the dynamical process of brane nucleation is encoded in the change in the electric field, while keeping $k$ (i.e. $G_{_N}$) fixed, as follows from \eqref{eq:Leff}. On the other hand, on the case of AP, $k$ is related to the running coupling, $\phi$, ensuring the interpolation in between the two vacua parameterised by different values of the dilatonic field.

\item Up-tunneling falls into the forbidden region of parameter space of the type-1 instanton, since, this configuration fails to satisfy the energy conservation relation associated to the brane nucleation process.

\item 
Interestingly, for either sign of the cosmological constant, the flat limit cannot be taken. This follows from the fact that such processes correspond to topology change, and would thereby fall within a different instanton type. The reason for this can be traced back to the extremality of the spacetimes involved, due to the definition of effective cosmological constants, 

\begin{equation} 
\Lambda_{o,I} 
\ 
\overset{def.}{=}      
\ 
\lambda+\frac{1}{2} k\ E_{o,I}^{2}.
\label{eq:Leff}
\end{equation}

\item As already highlighted in \cite{Brown:1988kg}, there are two different ways of calculating the bounce action. Either substituting the singular solutions to the field equations in the Euclidean action, or, alternatively, integrating over the 2 bulk spacetimes separately, in which case the $\delta$-function in the fields would be replaced by the Gibbons-Hawking boundary terms, with the membrane providing the natural boundary for both spacetimes. The 2$^{nd}$ procedure agrees with the first upon  extremising the bounce with respect to the membrane size, $\bar\rho$. According to \cite{Brown:1988kg}, this is preferable, since it ensures that the extremisation of the bounce on the classical instanton solution is not an artefact of the formalism\footnote{In section \ref{sec:2} we will argue that the starting point in the Hamiltonian formalism of FMP is precisely the first of the two methods outlined in \cite{Brown:1988kg}.}.

\end{itemize}   


\section{Lorentzian transitions in 2D}   \label{sec:2}
  
  
As proved in an earlier work, \cite{DeAlwis:2019rxg}, vacuum transitions in 4D can be described in three equivalent ways, namely by means of the CDL, BT and FMP formalisms. The importance of their agreement in the well known case of dS$_4$ spacetime, is what led us to explore to what extent the obstruction to other types of transitions may be overcome. In particular, the role plawed by black holes in \cite{DeAlwis:2019rxg}, inspired by the original work of FMP, which had no counterpart in the BT and CDL methods, turned out to be crucial for enabling to define transitions involving Minkowski spacetimes. 

Given these motivations, we now turn to extending the Hamiltonian method of Fischler, Morgan and Polchinski (FMP) \cite{Fischler:1990pk} to its 2D counterpart, namely Jackiw-Teitelboim (JT)-gravity \cite{Jackiw:1984je,Teitelboim:1983ux}, whose action, up to boundary terms, reads

\begin{equation} 
S_{_{JT}} 
\ 
= 
\ 
\frac{\phi_{o}}{2\kappa^{2}}\ \int\ d^{2}x\ \sqrt{-g}\ {\cal R}+\frac{1}{2\kappa^{2}}\ \int\ d^{2}x\ \sqrt{-g}\ \phi( {\cal R}-2\Lambda),
\label{eq:JTact} 
\end{equation} 
with $\phi_{o}$ being a constant. The coupling of the dilaton $\phi$ to the Ricci scalar ensures gravity is nontrivial. We can think of (\ref{eq:JTact}) as being the 2D version of the Einstein-Hilbert action with a cosmological constant describing gravity in arbitrary dimension. $\phi$ plays the role of a Lagrange multiplier and ensures curvature is fixed. From a Field Theory point of view, the expression  (\ref{eq:JTact}) is that of a renormalised theory, with the dilaton accounting for the renormalisation of Newton's constant.


The setup consists of pure gravity with spherical symmetry and a wall (the bubble's surface) at $z=\hat z$, separating two spacetimes with two different cosmological constants $\Lambda_{+}, \Lambda_{-}$, similarly to Brown and Teitelboim \cite{Brown:1988kg}.
Given the above considerations, our proposed initial Lagrangian density, thereby takes the general form:

\begin{equation} 
\begin{aligned}
  {\cal L} 
&
   = 
   \sqrt{-g}\phi\left[{\cal R}-2\Lambda_{+}\Theta(z-\hat z)-2\Lambda_{-}\Theta(\hat z-z)\right]-\delta(z-\hat z)\sigma\sqrt{(N^{t})^2+L^{2}(N^{z}+ {\dot{\hat{z}}})^{2}} \ +    \nonumber \\
  &
+  \sqrt{-g}\left[ \phi_{o}\chi_{_{{\cal M}}}-{\cal B}_{+}\Theta(z-\hat z)-{\cal B}_{-}\Theta(\hat z-z)\right],
   \label{eq:LagrTh1}  
\end{aligned} 
\end{equation}
i.e. the action for JT-gravity coupled to matter, with the latter being provided by the brane term, the two cosmological constants $\Lambda_{+},\Lambda_{-}$  and two additional constants\footnote{These act as deformation parameters in the 2D theory, contributing to the Hamiltonian of the system in the form of constant energy terms. As we shall see in section \ref{sec:3}, they play the role of black hole masses, and will be subject to constraints specified in due course.}, ${\cal B}_{\pm}$. Following FMP we start with the metric:

  \begin{equation} 
  g_{\mu\nu}   
  \ 
  = 
  \ 
  \left( 
  \begin{matrix} 
  -(N^{t})^{2}+(N^{z})^{2}\ \ \  & \ \ \  LN^{z}\\
  LN^{z} \ \ \  &  \ \ \  L^{2}\\ 
  \end{matrix} 
  \right) .
  \\  
  \\    
  \label{eq:mans} 
  \end{equation}

   
\subsection{Actions with vanishing constant energies}


First, we will be considering the ${\cal B}_\pm=0$ case. The ${\cal B}\neq 0$ case will be dealt with in section \ref{sec:3}.
Substituting (\ref{eq:mans}) in (\ref{eq:LagrTh1}) with ${\cal B}_{\pm}=0$ and integrating by parts, we get\footnote{Notice that in the calculations we are omitting the topological term but we will add it back to the final expression for the action.}

\begin{equation} 
\begin{aligned}
  {\cal L} 
  &
   = 
 {\cal L}_{_{BTs}}+2\phi\left(\frac{N^{t\prime}}{L}\right)^{\prime}-\frac{2\dot\phi\dot L}{N^{t}}-\frac{2\phi^{\prime}N^{z}(N^{z\prime}-\dot L)}{LN^{t}}+\nonumber \\
  & -2\Lambda_{+}\Theta(z-\hat z)-2\Lambda_{-}\Theta(\hat z-z)-\delta(z-\hat z)\sigma\sqrt{(N^{t})^2+L^{2}(N^{z}+ {\dot{\hat{z}}})^{2}} \nonumber\\
  &= 
   \ 
      {\cal L}_{_{BTs}}+ \pi_{L}\dot L+\pi_{\phi}\dot\phi-N^{t}{\cal H}_{t}-N^{z}{\cal H}_{z} ,
   \end{aligned} 
   \end{equation}
   where the total derivative terms read 
   
   \begin{equation}
    {\cal L}_{_{BTs}}    
    \ 
    = 
    \ 
    -2\left(\frac{\phi N^{t\prime}}{L}\right)^{\prime} + \left(\frac{2\phi\dot L}{N^{t}}\right)^{.}+\left(\frac{2 N^{z} (N^{z \prime}-\dot L)}{L N^{t}}\right)^{\prime} .   
    \end{equation} 
 The conjugate momenta are: 
 
   \begin{equation}
   \begin{aligned}    
&\pi_{L} 
\ 
= 
\ 
2\left(\frac{\phi^{\prime}N^{z}}{LN^{t}}-\frac{\dot\phi}{ N^{t}}\right), 
\ \ 
\ \ 
\pi_{\phi} 
\ 
= 
\ 
\frac{2(N^{z\prime}-\dot L)}{N^{t}},
\ \  
\ \  
\pi_{\hat z} 
\ 
= 
\ 
- \frac{\sigma L^{2}(N^{z}+{\dot{\hat{z}}})\delta(z-\hat z)}{\sqrt{(N^{t})^2+L^{2}(N^{z}+ {\dot{\hat{z}}})^{2}}},
\ \ \   
\end{aligned}
\end{equation} 
The Hamiltonian and total momentum are:

\begin{equation}
\begin{aligned}
   -{\cal H}_{t} 
   &
   =  
   \frac{\delta {\cal L}}{\delta N^{t}} 
= 
   -2\left(\frac{\phi^{\prime}}{L}\right)^{\prime}-\frac{2\dot\phi}{N^{t}}(N^{z\prime}-\dot L)+\frac{2\phi^{\prime}N^{z}(N^{z^{\prime}}-\dot L)}{N^{t2}L}+  \\ 
   & -2\phi L \left[\Lambda_{+}\Theta(z-\hat z)+\Lambda_{-}\Theta(\hat z-z)\right]-\frac{\sigma\ N^{t}\delta(z-\hat z)}{\sqrt{(N^{t})^2+L^{2}(N^{z}+ {\dot{\hat{z}}})^{2}}},\\    
   -{\cal H}_{z} 
   &
   =  
\pi^{\prime}_{L}-\phi^{\prime}\frac{\pi_{\phi}}{L}-\delta(z-\hat z)\pi_{\hat z}.
\label{eq:Hzc}
\end{aligned}   
\end{equation}
Away from the wall, the momentum constraint reads 

\begin{equation} 
\pi^{\prime}_{L} 
\ 
= 
\ 
\phi^{\prime}\frac{\pi_{\phi}}{L} ,
\ \ \ 
\end{equation} 
in terms of which ${\cal H}_t$ can be re-expressed\footnote{Imposing the Hamiltonian constraint is equivalent to selecting a specific energy for the configuration described by the Lagrangian of the theory. Because of this, the Lorentzian approach of FMP is naturally providing a microcanonical description of vacuum decays. The improvement of their method with respect to those outlined in section \ref{sec:4} is comparable to the one in \cite{Marolf:2022jra}, where the authors provide a microcanonical generalisation of the canonical treatment analysed in \cite{Marolf:2022ntb}.}, thereby leading to the following expression 
 
 \begin{equation} 
\boxed{\ \ \ 
 \left(\pi_{L}^{2}\right)\bigg|_{\hat z\pm\epsilon} 
 \ 
 = 
 \ 
4 \left[\left(\frac{\phi^{\prime}}{L}\right)^{2}\bigg|_{\hat z\pm\epsilon} + {\cal C}_{\pm} +\Lambda_{\pm}\phi^{2}\right],  \textcolor{white}{\Biggl [} \color{black} \ \ } 
 \label{eq:scale} 
 \end{equation} 
with ${\cal C }_{\pm}$ denoting an integration constants. A key feature emerging from this analysis is that (\ref{eq:scale}) is a dimensionful relation. This follows from the fact that $[\phi]=0$ in 2D, hence $[{\cal C}_{\pm}]=2$.  
 The junction conditions can be extracted upon integrating (\ref{eq:Hzc}) across the brane. In the rest frame of the latter, we get

\begin{equation} 
\begin{aligned}
\begin{cases}
 -\int_{\hat z-\epsilon}^{\hat z+\epsilon} dz\ {\cal H}_{z} 
 \ 
= 
 \ 
\pi_{L}\bigg|_{\hat z-\epsilon}^{\hat z+\epsilon} - \hat\pi_{\hat z}
 \ 
 = 
 \ 
 0 \nonumber \\ 
 -\int_{\hat z-\epsilon}^{\hat z+\epsilon} dz\ {\cal H}_{t} 
 \ 
=
 \ 
 -2\frac{\phi^{\prime}}{L}\bigg|_{\hat z-\epsilon}^{\hat z+\epsilon} -\frac{\sigma N^{t}}{\sqrt{(N^{t})^2+L^{2}(N^{z}+ {\dot{\hat{z}}})^{2}}} 
 \ 
 = 
 \ 
 0 
 \end{cases}\ \ \ \ \overset{N^{t}=1\ ,\ N^{z}=0}{\longrightarrow}\ \ \  \begin{cases}\frac{\phi^{\prime}}{L}\bigg|_{\hat z_{_{-}}}^{\hat z_{_{+}}}    
 \ 
 = 
 \ 
 -\frac{\sigma}{2} 
\\ 
\pi_{L}\bigg|_{\hat z_{_{-}}}^{\hat z_{_{+}}} 
 \ 
 = 
 \ 
 0,  
 \end{cases}
 \label{eq:jc}
 \end{aligned} 
 \end{equation}
where the extrinsic curvatures on either side read

\begin{equation}  
\boxed{\ \ \ \frac{\phi^{\prime}}{L}\bigg|_{\hat z_{_{\pm}}} 
\ 
= 
\ 
\frac{(\Lambda_{+}-\Lambda_{-})\phi^{2}}{\sigma}\mp\frac{\sigma}{4} +\frac{{\cal C}_{+}-{\cal C}_{-}}{\sigma}. \textcolor{white}{\Biggl [} \color{black} \ \ } 
\ \ \ 
\label{eq:genexpr}    
\end{equation} 
Equating the new expression for $\pi_{L}$ with the one obtained from the variation of ${\cal L}$, we get

\begin{equation} 
\pi_{L}\bigg|_{\hat z_{_{+}}} 
\ 
= 
\ 
-2\dot\phi.
\end{equation} 
The constants ${\cal C_{+}}, {\cal C}_{-}$ can be absorbed by a rescaling of $\phi$ and $\pi_L$ . We will partially use this freedom to choose  ${\cal C_{+}}={\cal C}_{-}$ . Squaring both sides and
redefining, where the $b$-subindex denotes the value of the dilaton at the boundary. $\Phi_{b}=\phi_{b}/\sqrt{|{\cal C}|}$ and solving this quadratic  equation with respect to $\dot\Phi_{b}^{2}$  leads to the energy conservation relation 

\begin{equation} 
\boxed{\ \ \ \dot\Phi_{b}^{2}+V_{eff}
 \ 
 = 
 \ 
-1, \textcolor{white}{\Biggl [} \color{black} \ \ } 
 \end{equation} 
with effective quadratic  potential defined as

 \begin{equation} 
 \boxed{\ \ \ V_{eff} 
 \ 
 = 
 \ 
-\ \frac{1}{4}\left[\left( \frac{\Lambda_{+}-\Lambda_{-}-\frac{\kappa^{2}}{4}}{\kappa\sqrt{|{\cal C}|}}\right)^{2}+\Lambda_{+}\right]\Phi_{b}^{2}-\frac{3}{4},\textcolor{white}{\Biggl [} \color{black} \ \ }
\end{equation} 
where we have used that $\sigma=\kappa\phi_{b}$. The value of the dilaton at the turning point is found by setting $\dot\Phi_{_{b}}=0$, i.e. when the classical motion is reversed, and solving for the unique turning point 

\begin{equation} 
V_{eff} 
\ 
= 
\ 
-1 
\ \ \ 
\Rightarrow 
\ \ \ 
\Phi_{o}
\ 
= 
\ 
\left[\left(\frac{\Lambda_{+}-\Lambda_{-}-\frac{\kappa^{2}}{4}}{\kappa\sqrt{|{\cal C}|}}\right)^{2}+\Lambda_{+}\right]^{-\frac{1}{2}}. 
\label{eq:phio}    
\end{equation} 
Notice the similarity with the 4D case from FMP where the dilaton is playing the role of the compactification radius.    


\subsubsection*{Metric and dilaton profiles} \label{subsec:422a} 


We now determine the solutions to the equations of motion arising from the variational principle. These are needed for evaluating the bulk and boundary actions, addressed next.     
In conformal gauge, the constraint ${\cal R}=2\Lambda$ can be solved in order to recover an expression for $L$ in terms of the $z$-coordinate. In the Lagrangian density we started from, $\phi({\cal R}-2\Lambda_{\pm})$ plays the role of a perturbation around the background action which, for the 2D case is purely topological. For the given gauge choice, the metric and dilatonic profile can be expressed as

\begin{equation} 
 ds^{2} 
\ 
= 
\ 
\frac{1}{z^{2}}\left(-dt^{2}+dz^{2}\right) 
\ \ \ 
, 
\ \ \ 
\text{with} 
\ \ \ L 
\ 
= 
\ 
\frac{1}{z}     
\ \ \ 
, 
\ \ \ 
\phi 
\ 
= 
\ 
\text{const.} 
\label{eq:Pcads}    
\end{equation} 
The e.o.m. arising from ${\cal L}_{_{JT}}$, i.e. 

\begin{equation} 
\phi\left(-\frac{L^{\prime\prime}}{L}+\frac{2L^{\prime2}}{L^{2}}\right)    
\ 
= 
\ 
2\phi\Lambda_{\pm},     
\ \ \ \ \ \ \ \ 
\ \ \ \ \ \ \ \ 
\phi^{\prime\prime} 
\ 
= 
\ 
2\phi \Lambda_{\pm} L^{2},
\label{eq:1eom}    
\end{equation} 
are solved by  

\medskip 

\begin{equation}      
\boxed{\ \ \ L 
\ 
= 
\ 
\frac{c}{\sinh(bz)} 
\ \ \ 
, 
\ \ \ 
\phi 
 \ 
 = 
 \ 
 2 a\coth(bz),\textcolor{white}{\Bigg [} \color{black}  \ \ } 
 \ \ \ 
 \text{with } 
 \ \ \ 
 \Lambda_{\pm}    
 \ 
\overset{\text{def.}}{=}     
 \ 
\frac{b^{2}}{c^{2}}.
\label{eq:sol1}     
\end{equation}

\medskip 
 For the case of  pure AdS$_{2}$ in Poincarè coordinates, (\ref{eq:Pcads}), the metric diverges at the location of the conformal boundary, placed at $z=0$. The latter can be mapped to the horizon at $\rho=\gamma$ via the following coordinate transformation      

\begin{equation} 
\boxed{\ \ \ z 
\ 
\overset{\text{def.}}{=}     
\ 
\frac{1}{b}\coth^{-1}\left(\frac{\rho}{\gamma}\right) \textcolor{white}{\Biggl [} \color{black} \ \ } 
\ \ \ \Rightarrow 
\ \ \ 
dz 
\ 
= 
\ 
\frac{\gamma}{b}\frac{d\rho}{\gamma^{2}-\rho^{2}},     
\label{eq:coordtr1} 
\end{equation} 
whereas the dilatonic profile becomes linearly-dependent on the redefined spatial coordinate 

\begin{equation}
\boxed{\ \ \ \phi 
\ 
= 
\ 
2a\frac{\rho}{\gamma}.\textcolor{white}{\Biggl [} \color{black}  \ \ }    
\end{equation} 
Notice that, going from $\rho$ to $z$ in (\ref{eq:coordtr1}), is analogous to the analytic extension of the metric beyond the horizon. In particular, such transformation takes place by introducing an additional scale, $\mu$.
Upon substituting (\ref{eq:coordtr1}) in the original line element (\ref{eq:Pcads}), we get

\begin{equation} 
ds^{2} 
\ 
= 
\ 
\frac{\rho^{2}-\gamma^{2}}{\gamma^{2}}\left(-dt^{2}+\frac{\gamma^{2}}{b^{2}}\frac{d\rho^{2}}{(\rho^{2}-\gamma^{2})^{2}}\right).    
\label{eq:newle} 
\end{equation} 
The main feature of this result is the explicit emergence of a horizon at $\rho=\gamma$ if $\gamma^{2}>0$. For the $\gamma^{2}<0$ case, instead, there is no horizon. Rewriting (\ref{eq:newle}) in terms of $\phi$, with $\gamma=1/b$, we get    

\begin{equation} 
\boxed{\ \ ds^{2} 
\ 
= 
\ 
\left(\frac{\phi^{2}}{4a^{2}}-1\right)\left(-dt^{2}+\frac{1}{4a^{2}b^{2}}\frac{d\phi^{2}}{\left(\frac{\phi^{2}}{4a^{2}}-1\right)^{2}}\right). \ \ } 
\label{eq:newlee} 
\end{equation} 
The vanishing of the conjugate momentum with respect to $L$ imply:   

\begin{equation}
\ \ \ \ \boxed{\ \ \ \phi^{2} 
\ 
= 
\ 
-\frac{{\cal C}}{\Lambda_{\pm}} \ \overset{def.}{=}\ \mu_{\pm}^2, \color{white}\bigg]\color{black}\ \ }    
\label{eq:pil3} 
\end{equation}  
where the system is understood to be evaluated at the turning point in the $N^{z}=0$ gauge.
Equation \eqref{eq:pil3} enables to identify the parameter in (\ref{eq:newlee}) $2a=\sqrt{-{\cal C}/\Lambda_{\pm}}=\mu_{\pm}$ as the cosmological horizon in a dS$_{2}$ metric. Furthermore, from (\ref{eq:sol1}), we get: 

\begin{equation} 
\phi^{\prime} 
\ 
=  
 \ 
 2ab\left(1-\coth^{2}(bz)\right) 
\ 
=  
 \ 
 2ab-\frac{b}{2a}\phi^{2} 
 \ 
 = 
 \ 
 \frac{b}{2a}\left(4a^{2}-\phi^{2}\right),
 \label{eq:2v}    
 \end{equation} 
 from which it can be appreciated that the change in sign of the extrinsic curvature ($\phi^{\prime}$) takes place when crossing the cosmological horizon at $\phi=2a$. 


\section*{Bulk and boundary actions}   


In the Hamltonian formalism, the transition rate is defined in terms of the extremised total action including, both, bulk and wall terms 
\begin{equation}
\Gamma(A\rightarrow A\oplus W\oplus B)= \exp\left[-S_{tot}+S_{bckgr}\right],
\end{equation}
where $A,B$ denote the spacetimes involved separated by the brane $W$, and  

\begin{equation} 
S_{tot}^{\ \text{A}/\text{B}}  
\ 
= 
\ 
S_{top}+S_{bulk}+S_{brane} \ \ \ ,\ \ \  S_{bckgr} 
\ 
 \overset{def.}{=} 
\ 
 \frac{\phi_{bckgr}}{16\pi G} \chi_{_{\cal M}}.
\label{eq:totAB}
\end{equation}
The topological term $S_{top}$ cancels with the background contribution
when evaluating the transition rate. 
The extremised action is obtained by integrating over the Hamilton-Jacobi equations. By imposing the secondary constraints derived above, the variation of the bulk action therefore reads 

\begin{equation} 
\delta S_{bulk} 
\ 
= 
\ 
\int dz\left[\delta L(z)\pi_{L}+\delta\phi(z)\pi_{\phi}\right] 
\ 
= 
\ 
2\int dz\ \delta L(z)\pi_{L} 
\label{eq:var} 
\end{equation} 

\begin{equation} 
\begin{aligned}  
\Rightarrow 
\ \ \ 
S_{bulk}^{(+)} 
\ 
&=    
\ 
\frac{2\eta}{G}\int dz\ \int dL \sqrt{\left(\frac{\phi^{\prime}}{L}\right)^{2}+{\cal C}+\Lambda_{+}\phi^{2}} \\    
&= 
\frac{2\eta}{ G}\int_{0}^{\hat z} dz\ \left[ L\  \sqrt{\left(\frac{\phi^{\prime}}{L}\right)^{2}+{\cal C}+\Lambda_{+}\phi^{2}}  -\phi^{\prime}\ \cosh^{-1}\left(\frac{\phi^{\prime}}{L\sqrt{-{\cal C}-\Lambda_{+}\phi^{2}}}\right)  \right].
\label{eq:bulk1} 
\end{aligned}   
\end{equation}
Similarly, for the other vacuum, 
\begin{equation} 
S_{bulk}^{(-)} 
\ 
= 
\ 
\frac{2\eta}{ G}\int_{\hat z}^{\infty} dz\ \left[ L\  \sqrt{\left(\frac{\phi^{\prime}}{L}\right)^{2}+{\cal C}+\Lambda_{-}\phi^{2}}  -\phi^{\prime}\ \cosh^{-1}\left(\frac{\phi^{\prime}}{L\sqrt{-{\cal C}-\Lambda_{-}\phi^{2}}}\right)  \right].
\label{eq:bulk2} 
\end{equation}

The bulk action is extremised when the integrand featuring in the first term of (\ref{eq:bulk1}) and (\ref{eq:bulk2}) vanishes, therefore leaving with the corresponding contribution from the value of the dilaton at the turning point and at the horizons characterising spacetimes being involved. For such values, the argument of $\cosh^{-1}$ is imaginary, therefore, we can rewrite it as $\cos^{-1}$, which provides an overall $\pi$ factor. The extremes of integration for, both, (\ref{eq:bulk1}) and (\ref{eq:bulk2}) are dictated by the relative position of the innermost horizon in a given spacetime with respect to the turning point $\phi_{o}$ , whose expression was explicitly derived in the previous subsection, cf. eq. (\ref{eq:phio}). From equations (\ref{eq:2v}) we therefore get the following contributions

\begin{equation} 
S_{bulk}^{(-)} 
\ 
= 
\ 
\frac{2\pi\eta}{ G}\int_{z_{tp}}^{z_{h,-}} dz\ \phi^{\prime} 
\ 
= 
\ 
\frac{2\pi\eta}{ G}\int_{\mu_{-}}^{\phi_{o}\ \epsilon_{-}} d\phi\ 
\ 
= 
\ 
\frac{2\pi\eta}{ G}\left[\phi_{o}\ \Theta\left(-\frac{\phi^{\prime}}{L}\bigg|_{-}\right)-{\mu_{-}}\right],   
\label{eq:bulk3} 
\end{equation} 
with $\Theta\left(-\frac{\phi^{\prime}}{L}\bigg|_{-}\right)$ being the $\Theta$-function that is non-vanishing only if the extrinsic curvature on the inner side of the brane is negative, thereby indicating that the turning point is placed beyond the horizon. The same argument follows for the bulk integral over the outer vacuum.For the case involving pure AdS, instead, due to the absence of horizons, only the $\phi_{o}$-terms will be contributing as relevant extremes of integration. Furthermore, precisely because of this, the extrinsic curvature on either side will also preserve the sign over the entire spacetime, thereby leading to the mutual cancellation of the $\Theta$-terms. Combining the claims just made, the corresponding result for mixed AdS/dS configurations follows.  
For the 3 cases of interest, the total bulk action therefore reads 

\begin{equation} 
\boxed{\ \ \ \begin{aligned}
&S_{bulk}^{\  \text{dS}\rightarrow\text{dS}} 
\ 
= 
\ 
S_{bulk}^{(+)}+S_{bulk}^{(-)} 
\ 
= 
\ 
\frac{2\pi\eta}{G}\left[\phi_{o}\ \Theta\left(-\frac{\phi^{\prime}}{L}\bigg|_{-}\right)-{\mu_{-}}-\phi_{o}\ \Theta\left(-\frac{\phi^{\prime}}{L}\bigg|_{+}\right)+{\mu_{+}}\  \right]   \\
\\    
&S_{bulk}^{ \ \text{AdS}\rightarrow\text{AdS}} 
\ 
= 
\ 
S_{bulk}^{(+)}+S_{bulk}^{(-)} 
\ 
= 
\ 
0\\
\\    
& S_{bulk}^{\  \text{dS}\rightarrow\text{AdS}} 
\ 
= 
\ 
S_{bulk}^{(+)}+S_{bulk}^{(-)} 
\ 
= 
\ 
\frac{2\pi\eta}{G}\left[\phi_{o}\ \ \Theta\left(-\frac{\phi^{\prime}}{L}\bigg|_{-}\right)-\phi_{o}\ \Theta\left(-\frac{\phi^{\prime}}{L}\bigg|_{+}\right)+{\mu_{+}}\  \right].
\label{eq:actdsds}   
\end{aligned}\ \ \ } 
\end{equation}

The brane action is obtained by considering the variation of  (\ref{eq:var}) at the location of the wall $z=\hat z$. From the junction conditions, $\hat \pi_{L}\bigg|_{+}=\hat \pi_{L}\bigg|_{-}$, it follows that only the second term in, both,  (\ref{eq:bulk1}) and  (\ref{eq:bulk2}) contributes nontrivially  

\begin{equation} 
\begin{aligned}
S_{brane} 
&
= 
\frac{2}{G}\int_{0}^{{\phi}_{o}}\ d\phi_{b}\left[\cosh^{-1}\left(\frac{\phi_{b}^{\prime}(\hat z-\epsilon)}{L\sqrt{-{\cal C}-\Lambda_{-}\phi_{b}^{2}}}\right)-\cosh^{-1}\left(\frac{\phi_{b}^{\prime}(\hat z+\epsilon)}{L\sqrt{-{\cal C}-\Lambda_{+}\phi_{b}^{2}}}\right)   \right]\nonumber \\
&
= 
\frac{2}{G}\int_{0}^{{\phi}_{o}}\ d\phi_{b}\left[\cosh^{-1}\left(\frac{A_{1}\phi_{b}}{\sqrt{1-\frac{\Lambda_{-}}{|{\cal C}|}\phi_{b}^{2}}}\right)-\cosh^{-1}\left(\frac{A_{2}\phi_{b}}{\sqrt{1-\frac{\Lambda_{+}}{|{\cal C}|}\phi_{b}^{2}}}\right)   \right] ,
\label{eq:brane2in1}
\end{aligned} 
\end{equation}
where

\begin{equation}   
A_{1} 
\ 
\overset{\text{def.}}{=}
\ 
\frac{\Lambda_{+}-\Lambda_{-}+\frac{\kappa^{2}}{4}}{\kappa\sqrt{-{\cal C}}} 
\ \ \ 
, 
\ \ \ 
A_{2} 
\ 
\overset{\text{def.}}{=} 
\ 
\frac{\Lambda_{+}-\Lambda_{-}-\frac{\kappa^{2}}{4}}{\kappa\sqrt{-{\cal C}}}.
\end{equation} 
The brane action associated to the cases listed in \eqref{eq:actdsds} reads:

\begin{equation} 
\begin{aligned}
S_{brane}^{\  \text{dS}/\text{dS}} 
&
= 
\frac{2\pi\eta}{G}\left[\phi_{o}\ \Theta\left(-\frac{\phi^{\prime}}{L}\bigg|_{-}\right)-\phi_{o}\ \Theta\left(-\frac{\phi^{\prime}}{L}\bigg|_{+}\right)+\right.\\
& \left.+\frac{{\mu_{-}}}{2\pi} \ln\bigg|\frac{{\mu_{-}}A_{1}+1}{{\mu_{-}}A_{1}-1}\bigg| -\frac{{\mu_{+}}}{2\pi} \ln\bigg|\frac{{\mu_{+}}A_{2}+1}{{\mu_{+}}A_{2}-1}\bigg| \right],  \nonumber
\label{eq:vr} 
\end{aligned} 
\end{equation}

\begin{equation}    
S_{brane}^{\  \text{AdS}/\text{AdS}}
\ 
= 
\     
\frac{\eta}{G}\bigg[{\mu_{-}} \ln\bigg|\frac{{\mu_{-}}A_{1}+1}{{\mu_{-}}A_{1}-1}\bigg| -{\mu_{+}}\ln\bigg|\frac{{\mu_{+}}A_{2}+1}{{\mu_{+}}A_{2}-1}\bigg| \bigg] .    
\label{eq:vr1} 
\end{equation}   
 The $\Theta$-terms in (\ref{eq:vr1}) cancel each other out for the same argument outlined when dealing with the bulk action. For the case of mixed transitions of the dS$\rightarrow$AdS-kind, the brane action would be structurally equivalent to that of the dS$\rightarrow$ dS case, once having suitably accounted for the difference in sign of the cosmological constants involved.


\subsection{Transitions among de Sitter, anti-de Sitter, and Minkowski vacua}


In summary we can write the transitions so far as:
\begin{equation}  
\boxed{\ \ \ \begin{aligned}    
\\
&\ln\Gamma_{_{\ \text{AdS}\rightarrow\text{AdS}}  }    
\ 
= 
\ 
\frac{\eta}{G}\left[{\mu_{-}} \ln\left|\frac{{\mu_{-}}A_{1}+1}{{\mu_{-}}A_{1}-1}\right| -{\mu_{+}} \ln\left|\frac{{\mu_{+}}A_{2}+1}{{\mu_{+}}A_{2}-1}\right| \right]  \\ 
\\
&  \ln\Gamma_{_{\ \text{dS}\rightarrow\text{dS}} }     
\ 
= 
\ 
\frac{2\pi\eta}{G}\left[{\mu_{+}}-{\mu_{-}}+\frac{{\mu_{-}}}{2\pi} \ln\left|\frac{{\mu_{-}}A_{1}+1}{{\mu_{-}}A_{1}-1}\right| -\frac{{\mu_{+}}}{2\pi} \ln\left|\frac{{\mu_{+}}A_{2}+1}{{\mu_{+}}A_{2}-1}\right| \right]\\
\\
& \ln\Gamma_{_{\ \text{dS}\rightarrow\text{AdS}}  }    
\ 
= 
\ 
\frac{2\pi\eta}{G}\left[{\mu_{+}}+\frac{{\mu_{-}}}{2\pi} \ln\left|\frac{{\mu_{-}}A_{1}+1}{{\mu_{-}}A_{1}-1}\right| -\frac{{\mu_{+}}}{2\pi} \ln\left|\frac{{\mu_{+}}A_{2}+1}{{\mu_{+}}A_{2}-1}\right| \right].
\\
\label{eq:joined1} 
\end{aligned} \ \ }    
\end{equation} 
Their equivalence with the corresponding expressions in BT becomes manifest under the following parametric redefinitions

\begin{equation} 
\begin{aligned}
\bar z  
&=-\frac{1}{4\sqrt{\Lambda_+\ }}\ \ln\left|\ 1-\frac{\phi_{tp,+}^{new\ 2}}{\Lambda_{_{+}}}\ \right|\ \ \ ,\ \ \  
\phi_{tp}^{new\ 2}    
\overset{def.}{=}    
\Lambda_{\pm}A_{_{1,2}}\mu_{\pm}  
= 
A_{_{1,2}}\sqrt{\frac{-{\cal C}}{\Lambda_\pm}\ }\ \Lambda_{_{BT}}.\   
\end{aligned}
\end{equation}

\subsubsection*{Comparing dS, AdS and Minkowski transitions}

Some important observations can be drawn when comparing (\ref{eq:joined1}). 
\begin{itemize}

\item{Horizons' contribution }

The first main difference between the AdS and dS transitions is the horizon contribution to the amplitude, which is crucial when considering the flat limit. A detailed explanation and interpretation of these findings will be outlined in section \ref{sec:5}.

    \item{Bounds on wall tension} 
    
    For the AdS$_2\rightarrow$AdS$_2$ case, the classical turning point $\phi_{o}$ is real if and only if the tension of the brane is constrained within the following range

\begin{equation} 
\boxed{\ \ \    
\kappa      
\ 
<    
\ 
 2\left|\sqrt{-\Lambda_{+}}-\sqrt{-\Lambda_{-}}\right| .   \textcolor{white}{\Biggl [} \color{black}         
\ \ }    
\label{eq:range}    
\end{equation}

\item{The flat limit from AdS} 

Interestingly, upon taking, either, $\Lambda_{\pm}=0$ in AdS$_2\rightarrow$AdS$_2$ processes, the amplitude is still finite

\begin{equation} 
\boxed{\ \ \ \begin{aligned} 
&S_{tot}^{\ \text{Mink.}\rightarrow \text{AdS}} 
\ 
= 
\ 
\frac{\eta}{G}\ {\mu_{-}}\ \ln\bigg|\frac{{\mu_{-}}A_{1}+1}{{\mu_{-}}A_{1}-1}\bigg|   \\   
\\    
& S_{tot}^{\ \text{AdS}\rightarrow \text{Mink}} 
\ 
= 
\ 
-\frac{\eta}{G}\ {\mu_{+}}\ \ln\bigg|\frac{{\mu_{+}}A_{2}+1}{{\mu_{+}}A_{2}-1}\bigg|
\label{eq:mads} 
\end{aligned}
\ \ } 
\ \ \ \ \text{with}\ \ \ A_{1} 
\ 
\overset{\text{def.}}{=}
\ 
\frac{-\Lambda_{-}+\frac{\kappa^{2}}{4}}{\kappa\sqrt{-{\cal C}}} \ \ ,\ \ A_{2} 
\ 
\overset{\text{def.}}{=}
\ 
\frac{\Lambda_{+}+\frac{\kappa^{2}}{4}}{\kappa\sqrt{-{\cal C}}}.
\end{equation} 
Therefore up and down-tunneling are allowed in this case. Finiteness follow from the fact that the brane action vanishes on the side where the cosmological constant is turned off.



\item{Detailed balance}

Under the exchange $\mu_{+}\leftrightarrow\mu_{-}$, with ${\cal C}_{+}={\cal C}_{-}$,  we get $A_1 \leftrightarrow A_2$ therefore, from  (\ref{eq:vr}), we obtain:

\begin{equation}
\boxed{\ \ \ \frac{\Gamma_{1\rightarrow 2}}{\Gamma_{2\rightarrow 1}}= e^{\frac{8\eta\pi}{G}\left(\phi_{+}^{h}-\phi_{-}^{h}\right)}=e^{\eta(S_1-S_2),}  \textcolor{white}{\Biggl [} \color{black} \ \ }    
\label{eq:detbal}
\end{equation}   
proving that the principle of detailed balance is satisfied, once we identify   the value of the dilaton at the horizon with the entropy. It is interesting to notice that detailed balance is still working even in this setup, thereby consistently proving that the value of the dilaton at the horizon is playing the role of the black hole entropy. 
However, the particular nature of the cosmological horizon leads to unitarity issues upon taking the flat spacetime limit from a pure dS setup.


\item{The flat limit from dS}

The main difference with respect to the AdS case is that, this time, the definition of the turning point, $\phi_o$, is not subject to any parametric constraint, since $\Lambda_{\pm}>0$. However, the presence of an additional term proportional to the cosmological horizon in $S_{_{TOT}}$, inevitably leads to a divergent action upon taking the flat limit on eiter side of the brane

\begin{equation} 
\boxed{\ \ \ 
\begin{aligned}
&S_{tot}^{\ \text{Mink.}\rightarrow \text{dS}} 
\ 
= 
\ 
\frac{2\eta\pi}{G}\left[\ -{\mu_{-}}+\frac{{\mu_{-}}}{2\pi}\ \ln\bigg|\frac{{\mu_{-}}A_{1}+1}{{\mu_{-}}A_{1}-1}\bigg|+{\mu_{+}}\right] \rightarrow \infty \\
\\
&S_{tot}^{\ \text{dS}\rightarrow \text{Mink.}} 
\ 
= 
\ 
\frac{2\eta\pi}{G}\left[\ {\mu_{+}}-\frac{{\mu_{+}}}{2\pi}\ \ln\bigg|\frac{{\mu_{+}}A_{2}+1}{{\mu_{+}}A_{2}-1}\bigg|-{\mu_{-}}\right] \rightarrow -\infty, 
\end{aligned}\ \ }    
\end{equation} 
for up and down-tunneling, respectively. In particular, this proves that there is no 2D counterpart of the CDL transition for dS$\rightarrow$ Mink. It is an important feature of the 2D theory that the divergence does not cancel, thereby signalling the need for a deeper understanding. Indeed, when taking the $\underset{\mu_{+}\rightarrow \infty}{\lim}$, we are really moving very far away with respect to the perturbative regime where the near-horizon approximation of JT-gravity is valid. As the horizon moves deeper inside the IR bulk, so too does the turning point. However, $\phi_{o}$ will at most asymptote to a finite value, ultimately reemerging from the horizon in the form of a naked singularity. At this point, the causal structure of the spacetime prevents the nucleation of a new bulk phase.  





\end{itemize}


\section{Lorentzian transitions with black holes}\label{sec:3}


\subsection{Actions with non-vanishing constant energies} \label{sec:aed0}

We will now turn to the more general case in which a constant term (with respect to $\phi$) is being added to the Lagrangian density, which now reads

\begin{equation}
\begin{aligned}
  {\cal L} 
& 
= 
   \sqrt{-g}\phi\left[{\cal R}-2\Lambda_{+}\Theta(z-\hat z)-2\Lambda_{-}\Theta(\hat z-z)\right]-\delta(z-\hat z)\sigma\sqrt{(N^{t})^2+L^{2}(N^{z}+ {\dot{\hat{z}}})^{2}}\nonumber\\
&+
\sqrt{-g}\left[ \phi_{o}\chi_{_{{\cal M}}}-{\cal B}_{+}\Theta(z-\hat z)-{\cal B}_{-}\Theta(\hat z-z)\right].
\end{aligned} 
\end{equation}

The additional ${\cal B}$-terms play the same role as the black hole mass in the 4D analysis carried out in \cite{Fischler:1990pk}. Indeed, such term contributes to the 2D metric function by means of a linear term in $\phi$, therefore featuring the appropriate scaling for a 2D black hole. However, unlike the 4D case, where $M$ emerges as an integration constant, the lower dimensionality of the current analysis requires it to be defined as a parameter of the theory. 

The procedure for determining the total action  is very similar to the one carried out in section \ref{sec:2}, hence we will just highlight the main differences. The Hamiltonian constraint picks up additional ${\cal B}_{\pm}$ terms, whereas the momentum constraint remain unaltered. This follows from the fact that the determinant of the metric is $N^{z}$-independent. In particular, the latter ensures that the relation in between the conjugate momenta $\pi_{L}$ and $\pi_{\phi}$ is preserved. Upon integrating the Hamiltonian constraint across the wall, the conjugate momentum with respect to the metric reads

 \begin{equation} 
 \boxed{\ \ \left(\pi_{L}^{2}\right)\bigg|_{\hat z_{_{\pm}}} 
 \ 
 = 
 \ 
 4\ \left[\left(\frac{\phi^{\prime}}{L}\right)^{2}\bigg|_{\hat z_{_{\pm}}}  + {\cal C}_{\pm} + {\cal B}_{\pm}\phi +\Lambda_{\pm}\phi^{2}\right], \ \ } 
 \label{eq:dual}    
 \end{equation} 
and the extrinsic curvatures 

\begin{equation} 
\frac{\phi^{\prime}}{L}\bigg|_{\hat z_{_{\pm}}} 
\ 
= 
\ 
\frac{(\Lambda_{+}-\Lambda_{-})\phi^{2}}{\sigma}\mp\frac{\sigma}{4} +\frac{{\cal C}_{+}-{\cal C}_{-}}{\sigma} +\ \frac{({\cal B}_{+}-{\cal B}_{-})\phi}{\sigma} 
\ \ \ 
,
\label{eq:genexpr1}    
\end{equation} 
from which the junction conditions follow
similarly to the ${\cal B}_{\pm}=0$ case. Upon performing the same procedure outlined in section \ref{sec:2}, and still requiring ${\cal C}_{+}={\cal C}_{-}$, the effective potential reads 

 \begin{equation} 
 V_{eff} 
 \ 
 = 
 \ 
-\ \frac{1}{4}\left[\left(\left( \frac{\Lambda_{+}-\Lambda_{-}-\frac{\sigma^{2}}{4}}{\sigma} \right)\Phi_{b}^{2}+\frac{({\cal B}_{+}-{\cal B}_{-})\Phi_{b}}{\sqrt{|{\cal C}|}\sigma} \right)^{2}-1+\frac{{\cal B}_{+}}{\sqrt{|{\cal C}|}}\Phi_{b}+\Lambda_{+}\Phi_{b}^{2}\right]-1.
\end{equation} 
We will restrict to the case ${\cal B}_{+}={\cal B}_{-}={\cal B}$. The general case  ${\cal B}_{+}\neq{\cal B}_{-}$ can be considered without changing the conclusions. 
 For ${\cal B}_{+}={\cal B}_{-}$, the turning points are

\begin{equation}     
\boxed{\ \ \  \Phi_{1,2}   
\ 
\overset{\text{def.}}{=}     
\ 
-\frac{{\cal B}\Phi_{o}^{2}}{2\sqrt{|{\cal C}|}}\ \left[\ 1\pm\sqrt{1+\frac{4{\cal C}}{{\cal B}^{2}\Phi_{o}^{2}}}\ \right],  \textcolor{white}{\Biggl [} \color{black}\ \ }    
\label{eq:phi12}    
\end{equation} 
which, in the ${\cal B}\rightarrow0$ limit, consistently reduces to $|\phi_{1}|=|\phi_{2}|=\phi_{o}$, thereby correctly recovering the result obtained in section \ref{sec:2}. From (\ref{eq:phi12}), the existence of 2 physical turning points is ensured upon requiring the following parametric constraints 
 
\begin{equation} 
\boxed{\ \ \ -\frac{1}{4}<\frac{{\cal C}}{{\cal B}^2\Phi_{o}^2}<0, \ \ \ \ \ \ \text{with}\ \ \ {\cal B}, {\cal C}<0, \textcolor{white}{\Biggl [} \color{black}\ \ }   
\label{eq:constrm}
\end{equation} 
which can also be rewritten as follows   

\begin{equation} 
\boxed{\ \ \ -2\sqrt{|\Lambda{\cal C}|\ }<{\cal B}<0.  \textcolor{white}{\Biggl [} \color{black}\ \ }   
\label{eq:constrm1}
\end{equation}

In particular, \eqref{eq:constrm1} defines a bound for ${\cal B}$, which, as will be argued in section \ref{sec:fl}, plays an essential role in our analysis. Indeed, as shown next, this parameter is related to the black hole mass.  


\subsubsection*{Metric and dilaton profiles}


Let us generalise the results of section  \ref{subsec:422a}  to the case ${\cal B}\neq 0$.  Here, the dilatonic e.o.m. pick up an additional constant term, and, consequently, so too does the field, 

\begin{equation} 
 \phi^{\prime\prime} 
\ 
= 
\ 
2\phi L^{2}\Lambda_{\pm} \pm  {\cal B}L^{2} 
\ \ \ 
\Rightarrow 
\ \ \ 
\phi 
\ 
= 
\    
2a\coth(bz)\pm\frac{{\cal B}}{2\Lambda_{\pm}}. 
\label{eq:2eom}        
\end{equation}    
Under the coordinate transformation (\ref{eq:coordtr1}), the line element turns into 

\begin{equation} 
\boxed{\ \ ds^{2} 
\ 
= 
\ 
\frac{\left(\phi\pm\frac{{\cal B}}{2\Lambda_{\pm}}\right)^{2}-4a^{2}}{4a^{2}}\left(-dt^{2}+\frac{4a^{2}}{b^{2}}\frac{ d\phi^{2}}{\left(\left(\phi\pm\frac{{\cal B}}{2\Lambda_{\pm}}\right)^{2}-4a^{2}\right)^{2}}\right), \ \ } 
\label{eq:newlee1} 
\end{equation} 
which is exactly of the S(A)dS-kind. The parameters featuring in (\ref{eq:newlee1}) can be related to the ones we have and will be using in the expressions for the total actions via the following identification 

\begin{equation} 
\frac{{\cal B}^{2}}{4\Lambda_{\pm}^{2}}-4a^{2}  
\ 
= 
\ 
\frac{{\cal  C}}{\Lambda_{\pm}} 
\ \ \ 
\Rightarrow 
\ \ \ 
{\cal B} 
\ 
= 
\ 
\pm 2\Lambda_{\pm}\sqrt{4a^{2}+\frac{\cal C}{\Lambda_{\pm}}}    
\ \ \ 
. \label{eq:conv}    
\end{equation}      
Note that, in this case,  the metric has 2 horizons, associated to the roots of the metric function in (\ref{eq:newlee1}) that read

\begin{equation} 
\phi_{h,1,2}^{\pm}   
\ 
 = 
 \ 
-\frac{|{\cal B}|}{2\Lambda_{\pm}}\left[\text{sign}_{\pm}({\cal B})\pm\frac{4a\sqrt{\Lambda_{\pm}}}{{\cal B}}\right]    
\ 
 \overset{(\ref{eq:conv})}{=}   
 \ 
-\frac{|{\cal B}|}{2\Lambda_{\pm}}\left[\text{sign}_{\pm}({\cal B})\pm\sqrt{1-\frac{4{\cal C}\Lambda_{\pm}}{{\cal B}^{2}}}\right].
\ \ \ 
\end{equation}

This shows the importance of introducing $\cal B$. Which sign is being identified with the innermost or outermost horizon of the given spacetimes entirely depends on the signs of the parameters of the theory $({\cal B}, \Lambda_{\pm})$. In particular, ${\cal B}$ will have opposite sign on the 2 sides to ensure the presence of 2 horizons for the $\Lambda_{\pm}>0$ case. Setting the outer horizon as a reference, the innermost horizon for ${\cal B}<0$\footnote{This condition follows from requiring both turning points to be physical, see eq. (\ref{eq:phi12}).} is given by the choice of the $+$ root. As a consequence of this, the outermost horizon on the inner vacuum is given by the$-$ sign, with ${\cal B}>0$. The change in sign for ${\cal B}$ in the 2 vacua ensures that we can sistematically use the same coordinate $\phi$ on both sides of the brane. Indeed, this is not something surprising, given that the value of the dilaton changes sign in different causal patches of a maximally-extended metric, and therefore it is in agreement with the description outlined above given that the nucleation process requires crossing the horizon of a given spacetime in static coordinates.

In the 4D case, FMP showed that the black hole mass arises as an integration variable, while in the 2D case we show that it is part of the theory, since it features at the level of the Lagrangian density. However, one should not be misled by this, in the sense that, unlike the Euclidean formalisms referred to in section 2, the main feature of the Hamiltonian method is precisely that of implementing constraints at the level of the Lagrangian, such that the total action describing the transition is effectively accounting for a joined system of two vacua separated by a wall. The black hole mass will therefore enter in the joined system's Lagrangian inside the Hamiltonian constraint, and the matching conditions will therefore act as a selection rule assigning to each spacetime a notion of state labelled by their respective value of the black hole mass, ${\cal B}_{\pm}$.


\subsection{Transitions with conical defects} 


The evaluation of the extremised total action can be carried out in complete analogy with the procedure applied in \ref{sec:2}. The bulk and brane actions read 

\begin{equation}  
\begin{aligned}
S_{bulk}^{^{(\pm)}} 
\ 
= 
\ 
\frac{2\eta}{ G}\int dz\ \left[ L\  \sqrt{\left(\frac{\phi^{\prime}}{L}\right)^{2}+{\cal C}+{\cal B}\phi+\Lambda_{\pm}\phi^{2}}  -\phi^{\prime}\ \cosh^{-1}\left(\frac{\phi^{\prime}}{L\sqrt{-{\cal C}-{\cal B}\phi-\Lambda_{\pm}\phi^{2}}}\right)  \right],
\nonumber
\end{aligned} 
\end{equation}

\begin{equation} 
\begin{aligned}
S_{brane} 
&
= 
\frac{2\eta}{G}\int_{\phi_{1}}^{\phi_{2}}\ d\phi_{b}\left[\cosh^{-1}\left(\frac{\phi_{b}^{\prime}(\hat z-\epsilon)}{L\sqrt{-{\cal C}-{\cal B}\phi_{b}-\Lambda_{-}\phi_{b}^{2}}}\right)-\cosh^{-1}\left(\frac{\phi_{b}^{\prime}(\hat z+\epsilon)}{L\sqrt{-{\cal C}-{\cal B}\phi_{b}-\Lambda_{+}\phi_{b}^{2}}}\right)   \right]. 
\nonumber
\end{aligned} 
\end{equation}

Once more, extremisation of the bulk actions occurs at the horizons according to the range of the mass parameter. This time, we have two horizons on either side, defined by 
\begin{equation} 
\phi_{h,1,2}^{+} 
\ 
\overset{\text{def.}}{=} 
\ 
-\frac{|{\cal B}|}{2\Lambda_{+}}\left[\text{sgn}({\cal B})\mp\sqrt{1-\frac{4\Lambda_{+}{\cal C}}{{\cal B}^{2}}}\right]    
\ \ \ 
, 
\ \ \ 
\phi_{h,1,2}^{-} 
\ 
\overset{\text{def.}}{=} 
\ 
-\frac{|{\cal B}|}{2\Lambda_{+}}\left[\text{sgn}({\cal B})\mp\sqrt{1-\frac{4\Lambda_{-}{\cal C}}{{\cal B}^{2}}}\right],
\label{eq:hor2}
\end{equation} 
with ${\cal C}_{\pm}<0$. Given that $\Lambda_{\pm}\ge 0$, for both horizons to be physical, it must be that $\text{sgn}({\cal B})=-1$. The bulk action is extremised at the outermost and innermost horizons on the 2 sides of the brane, respectively, (\ref{eq:hor2}), and at the turning point $\phi_{2}$, provided the argument of $\cosh^{-1}$ changes sign in $S_{bulk}$. 

The total bulk action reads
\begin{equation} 
\begin{aligned}
S_{bulk} 
=     
\frac{2\eta}{G}\ \left[\ -\phi_{h,2}^{-}+\phi_{h,1}^{+}+\phi_{2}\ \left[\Theta\left(-\frac {\phi^{\prime}}{L}\bigg|_{-}\right)-\Theta\left(-\frac {\phi^{\prime}}{L}\bigg|_{+}\right)\right]\right],   
\ \ \ 
\end{aligned} 
\end{equation}
and therefore we can write the transitions as:
 
\begin{equation}
\boxed{\ \ \      
\begin{aligned}      
S_{_{tot}}^{\text{(A)dS}\rightarrow\text{(A)dS}} 
& =
\frac{2\eta}{G} \left[ -\frac{{\cal B}}{2\Lambda_{+}} \left[1- \sqrt{1-\frac{4{\cal C}\ \Lambda_{+}}{{\cal B}^{2}}}-\ln\left|\frac{y_{4}}{y_{3}}\sqrt{\frac{1-y_{3}^{2}}{1-y_{4}^{2}}}\right| \right] +\right. 
\nonumber\\    
&   
+ 
\frac{{\cal B}}{2\Lambda_{-}}\ \left[1- \sqrt{1-\frac{4{\cal C}\ \Lambda_{-}}{{\cal B}^{2}}}-\ln\left|\frac{y_{2}}{y_{1}}\sqrt{\frac{1-y_{1}^{2}}{1-y_{2}^{2}}}\right|\right] +               
\nonumber\\  
& 
-  
\phi_{2}\ln\left|\frac{A_{2}\ y_{2}}{A_{4}y_{4}}\sqrt{\frac{1-y_{4}^{2}}{1-y_{2}^{2}}}\right|+\phi_{1}\ln\left|\frac{A_{2}\ y_{1}}{A_{4}y_{3}}\sqrt{\frac{1-y_{3}^{2}}{1-y_{1}^{2}}}\right| + 
\label{eq:totadsads} \\     
&    
-     
\left.\frac{1}{2}\left[\sqrt{\frac{\frac{{\cal B}^{2}}{4\Lambda_{-}}-{\cal C}}{\Lambda_{-}}}\ \ln\ \left|\frac{1-y_{2}}{1+y_{2}}\frac{1+y_{1}}{1-y_{1}}\right| - 
\sqrt{\frac{\frac{{\cal B}^{2}}{4\Lambda_{+}}-{\cal C}}{\Lambda_{+}}}\ \ln\ \left|\frac{1-y_{4}}{1+y_{4}}\frac{1+y_{3}}{1-y_{3}}\right|\right] \right],        
\end{aligned}       \  \ }
\end{equation}

where $y_{1,2} 
\equiv y_{3,4} 
\ 
= 
\ 
y(\phi_{1,2})$, 

\begin{equation} 
y 
\ 
\overset{\text{def.}}{= }    
\ 
\sqrt{\frac{\Lambda}{\frac{{\cal B}^{2}}{4\Lambda}-{\cal C}}}\left(\frac{{\cal B}}{2\Lambda}+\phi_{b}\right) 
\ \ \ 
,   \ \ \  \phi_{1,2}    
\ 
= 
\ 
-\frac{{\cal B}\phi_{o}^{2}}{2\sqrt{|{\cal C}|}}  \left[1\pm\sqrt{1+\frac{4|{\cal C}|}{{\cal B}^{2}\phi_{o}^{2}}}\right] 
\end{equation} 
and, for pure notational convenience, the following parametric redefinitions were performed

\begin{equation} 
\begin{aligned}
&A_{1} 
\ 
= 
\ 
\frac{\Lambda_{+}-\Lambda_{-}+\frac{\kappa^{2}}{4}}{\kappa\sqrt{{\cal C}-\frac{{\cal B}_{-}^{2}}{4\Lambda_{-}}}}, 
\ \ \  
\ \ \ 
A_{3} 
\ 
= 
\ 
\frac{\Lambda_{+}-\Lambda_{-}-\frac{\kappa^{2}}{4}}{\kappa\sqrt{{\cal C}-\frac{{\cal B}_{+}^{2}}{4\Lambda_{+}}}}  \\
\\
&A_{2}
\ 
\overset{\text{def.}}{= } 
\ 
\frac{A_{1}}{\sqrt{\Lambda_{-}}}\sqrt{\frac{{\cal B}^{2}}{4\Lambda_{-}}-{\cal C}},  
\ \ \  
\ \ \ A_{4}
\ 
\overset{\text{def.}}{= } 
\ 
\frac{A_{3}}{\sqrt{\Lambda_{+}}}\sqrt{\frac{{\cal B}^{2}}{4\Lambda_{+}}-{\cal C}}.  
\ \ \ 
\ \ \ 
\label{eq:a1a3}    
\end{aligned}
\end{equation} 
Setting ${\cal B}=0$, (\ref{eq:totadsads}) reduces to the results obtained in section \ref{sec:2}. Upon exchanging $\Lambda_{-}\longleftrightarrow \Lambda_{+}$, the turning points on the 2 sides are correspondingly mapped to $y_{2}\rightarrow -y_{4}$ and $y_{1}\rightarrow -y_{3}$. From this follows that the arguments of all the $\ln$-terms are inverted. The ratio of direct and reverse processes, lead to an expression for detailed balance which reads

\begin{equation}         
\boxed{\ \ \ln \frac{\Gamma_{\Lambda_{+}\rightarrow\Lambda_{-}}}{\Gamma_{\Lambda_{-}\rightarrow\Lambda_{+}}}
\ 
=    
\ 
\frac{4\eta}{G}\ \left[\ \phi_{h}^{+} -\phi_{h}^{-}\ \right],    \ \ }       
\label{eq:next101} 
\end{equation}
 with $\phi_{h}^{\pm}$ denoting the value of the dilaton at the BH horizons on either side. Notice that this holds for, either, (A)dS$\rightarrow$(A)dS.

 
\subsection{Flat limit}   \label{sec:fl}


In the last part of this section, we will show that, upon taking the flat limit on either side of the brane, the 2D transition amplitude is still well defined. Let us consider the $\Lambda_-\equiv 0$ case. In the flat limit, the value of the dilaton at the outermost horizon on the inner vacuum's side becomes constant

\begin{equation}
\underset{\mu_{-}\rightarrow \infty}{\lim}\ \frac{{\cal B}}{2\Lambda_{-}}\ \left[1-\sqrt{1-\frac{4{\cal C}\Lambda_{-}}{{\cal B}^{2}}}\right]   
\ \equiv\ 
\boxed{\ \ \ \frac{{\cal C}}{{\cal B}}     
\overset{\text{def.}}{=}    
\phi_{f}.    \textcolor{white}{\Biggl [} \color{black}   \ \ } 
\label{eq:constphi}   
\end{equation}   
The constraint on the black hole mass ensuring the existence of 2 physical turning points, \eqref{eq:constrm}, (derived in section \ref{sec:aed0}) can be re-expressed in terms of \eqref{eq:constphi} as follows 

\begin{equation}  
\boxed{\ \ \ 0>{\cal B}>-\frac{\phi_f}{\ 4\Phi_{o}^{2\ }}.\textcolor{white}{\Biggl [} \color{black}   \ \ } 
\label{eq:ncm}   
\end{equation}
In the $\underset{{\cal B}\rightarrow 0}{\lim}$, (\ref{eq:constphi}) diverges and therefore we correctly recover the behaviour of the divergent dilaton at the horizon encountered in section \ref{sec:2}. The rate  (including background subtraction) becomes:

\begin{equation}
\boxed{\ \ \    
\begin{aligned}          
B_{tot}^{\text{\ \ Mink.}\rightarrow\text{dS}} 
& 
=      
\frac{2\eta}{G}\ \left[\ -\phi_{h,2}^{-}+\phi_{f}- \sqrt{\frac{\frac{{\cal B}^{2}}{4\Lambda_{-}}-{\cal C}}{\Lambda_{-}}}\ \left[\ y_{2} \ln\left|\frac{A_{2}y_{2}}{\sqrt{1-y_{2}^{2}}}\right| - y_{1} \ln\left|\frac{A_{2}y_{1}}{\sqrt{1-y_{1}^{2}}}\right| \right]+ \right.
\nonumber\\     
&      
-        
\left. 
\sqrt{\frac{\frac{{\cal B}^{2}}{4\Lambda_{-}}-{\cal C}}{\Lambda_{-}}}\ \left[ 
\tanh^{-1} y_{2} -
\tanh^{-1} y_{1}\right] \right],       
\label{eq:eq13} 
\end{aligned}\ \ }    
\end{equation}       
and is symmetric under the exchange of inner and outer vacua. Detailed balance is still satisfied:

\begin{equation}         
\boxed{\ \ \ \ln \frac{\Gamma_{\text{Mink.}\rightarrow\text{(A)dS}}}{\Gamma_{\text{(A)dS}\rightarrow \text{Mink.}}}
\ 
=    
\ 
 \frac{4\eta}{G}\ \left[\ \phi_{f}-\phi_{h}\ \right].  \textcolor{white}{\Biggl [} \color{black}    \ \ }       
\label{eq:next1011} 
\end{equation}

\subsection*{Summary of results and comparison with the 4D case} 

The above calculations show that:

\begin{itemize} 

\item The constant ${\cal B}$-term added to ${\cal L}$ is playing the role of the black hole mass. This introduces a second turning point from the 2D dS point of view, as long as \eqref{eq:constrm} is satisfied.

\item The flat spacetime limit for (A)dS$\rightarrow$ (A)dS transitions, in the presence of a black hole, is finite. The dilaton at the horizon reaches a constant value, consistent with unitarity.

\item In particular, the Mink$\rightarrow$AdSBH transition we obtain is in agreement with arguments provided in \cite{Maldacena:2010un}, supporting compatibility of the process with the holographic principle. 

\end{itemize} 
The key differences with respect to the 4D case analysed in \cite{DeAlwis:2019rxg} are:

\begin{itemize} 

\item The divergence of dS$_{2}\rightarrow$ Mink$_{2}$ actions in absence of black holes 

\item The black hole mass is a parameter of the theory, explicitly featuring in the Lagrangian density, and is not an integration constant, albeit subject to the constraint \eqref{eq:constrm}.

\end{itemize}

\section{Holographic interpretation}\label{sec:4.4}

The importance played by black holes in ensuring the finiteness of transition amplitudes upon taking the flat limit, motivates seeking for a deeper physical  interpretation of our findings. In doing so we rely upon holography, with the main reason for this being that holographic techniques in 2D have played a key role for addressing the information loss paradox within the context of black hole physics, \cite{JM1}.
The importance of understanding decay processes involving Minkowski and AdS spacetimes from a holographic perspective was already addressed in \cite{Maldacena:2010un}, and one of our  findings is proving agreement between the arguments outlined in  \cite{Maldacena:2010un} and ours in the lower-dimensional setup.

In the present section, we argue that 2D vacuum transitions can be embedded within a similar holographic setup as the one describing black hole evaporation in braneworld models. The main reason being the role played by JT-gravity in both settings. In pursuing such task, we argue that:

\begin{itemize}

\item An entropy interpretation may  be assigned to the total action or bounce and not only to the ratio of decay rates as in detailed balance.

\item Unitarity issues emerging in the flat limit can be overcome in the Hamiltonian formalism in presence of non-extremal spacetimes in a way which is analogous to the island proposal. This follows from the fact that the total action can be rewritten as a difference of generalised entropies, each one defined as \cite{JM1, JM2}\footnote{This expression turns out to be applicable to dS$_2$ spacetimes as well, and our results are therefore in agreement with those of \cite{iic}.}

\begin{equation}    
S_{\text{gen}} (R) 
=
\underset{\partial I}{\text{{min\ extr}}}\ \left[\frac{\text{Area}(\partial I)}{4G_{N}}\ +\ S_{_{EE}}(I\cup R)\right].  
\label{eq:genentr1}   
\end{equation}

\item Transitions described by means of the BT and FMP methods are actually \emph{local}.

\item For the case of vacuum transitions, the entanglement in \ref{eq:genentr1} is \emph{internal}.

\item From the boundary perspective, a vacuum transition in the bulk may correspond to a phase transition in the boundary such as a deconfinement to confinement transition.

\end{itemize}

\subsection{Islands and horizons in holographic black hole evaporation}

\small{We will start by briefly reviewing the recent work on black holes and the information paradox that will be useful for addressing the vacuum transitions in a similar way. Our first motivation in searching for a holographic embedding of vacuum transitions, is the importance of the role played by horizons, with the Holographic Principle being its prototypical example. Before delving into a detailed analysis of our findings, we first provide some further arguments in support of the similarity between the information loss paradox, and the diveregence of the entropy associated to the cosmological horizon in the static dS patch. To the best of our knowledge, the arguments we propose provide an original motivation towards applying new holographic techniques, as outlined from section \ref{sec:KRvt} onwards.

\subsubsection*{Holographic black hole evaporation and gravitating baths }    \label{sec:KR}

In the first formulation of the information paradox, namely within the context of the evaporating (1-sided) black hole, spacetime is asymptotically flat. In its corresponding holographic embedding, this assumption is mapped to having a \emph{non-gravitating bath}. In the KR/HM-construction realising it,\cite{JM2,JM1,ATC1} (as depicted on the RHS of figure \ref{fig:RS}), the bath (denoted in blue) lives on the conformal boundary at $z=0$, and, in this sense, is non-gravitating.

\begin{figure}[ht!]        
\begin{center} 
\includegraphics[scale=0.95]{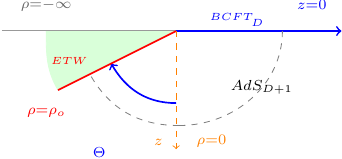} 
\ \ \ \ \ \ \ \ \ \ \ 
\includegraphics[scale=0.95]{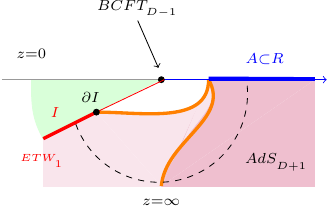} 
\caption{\footnotesize The dashed vertical line (on the LHS) denotes the \emph{Cardy branes} within the KR/HM-construction, \cite{JM1,JM2,ATC1}, separating the causal wedge of the \emph{bulk} CFT$_{D}$ from that of the black hole QMs living on the defect BCFT$_{D-1}$. The purple shaded regions on the RHS denote the entanglement wedges for the radiation region $A\subset R$ as time evolves. The orange curves on the RHS correspond to the RT-surfaces asociated to the choice of the boundary subregion $A$. }    
\label{fig:RS}    
\end{center} 
\end{figure}

The KR/HM-construction is able to simultaneously realise three different, albeit physically equivalent, descriptions of the same system, \cite{JM1}, namely:

\begin{enumerate} 
\item{From the \emph{bulk}, the system is described by an AdS$_{_{D+1}}$ with an ETW brane.} 

\item{From the \emph{brane}, it is given by a CFT$_{D}$ with a UV-cutoff + gravity on the ETW brane (asymptotically AdS$_{_{D}}$) which in turn is coupled with transparent boundary conditions to a BCFT$_{_{D}}$. }

\item{From the \emph{boundary}, it is a BCFT$_{_{D}}$ with nontrivial BCs. } 

\end{enumerate}

The key quantity to keep track of is the ratio between the central charges of the bulk and defect CFTs defining the BCFT$_D$, \cite{BB3456}, 

\begin{equation} 
\boxed{\ \ \ F 
\ 
\overset{\text{def.}}{=} 
\ 
\frac{c_{bdy}}{c_{bulk}} 
\ 
= 
\ 
\frac{6\ln g}{c_{bulk}}, \textcolor{white}{\Biggl [} \color{black}\ \ } 
\label{eq:F}   
\end{equation}
from which the tension of the brane, $T$, and brane angle, $\Theta$, can be determined as follows, \cite{BB3456}, 

\begin{equation} 
e^{F} 
\ 
= 
\ 
\frac{1+T}{\sqrt{1-T^{2}}} 
\ 
= 
\ 
\frac{1+\sin\Theta}{\cos\Theta} 
\ \ \ 
,   
\ \ \ 
1-e^{-2F}
\ 
= 
\ 
\frac{2\sin\Theta}{1+\sin\Theta},  
\label{eq:coordf}    
\end{equation} 
with BH evaporation being described by the mutual exchange of degrees of freedom between the bath and the black hole. Constraints on the CFT parameters are holographically dual to the allowed ranges for $T$ and $\Theta$ by virtue of \eqref{eq:coordf}.  From this follows that $t_{Page}$ is mapped to a critical value of $F$, beyond which an island is expected to arise. The second orange line in figure \ref{fig:lr}, which joins $R$ with $I$, becomes the dominant channel\footnote{The study of phase transitions in holography initiated from the first work of \cite{Witten:1998qj}.} for $t>t_{Page}$. The endpoint of $I$ is the quantum extremal surface (QES) contributing with the area law term to $S_{gen}$

\begin{equation}    
\boxed{\ \ \ S_{\text{gen}} (R) 
=
\underset{\partial I}{\text{{min\ extr}}}\ \left[\frac{\text{Area}(\partial I)}{4G_{N}}\ +\ S_{_{EE}}(I\cup R)\right].  \textcolor{white}{\Biggl [} \color{black} \ \ } 
\label{eq:genentr} 
\end{equation}

The main caveat featuring in the non-gravitating bath setup, namely the presence of a massive graviton \cite{BBK}, $m_{_{grav}}^{2}\sim c_{bulk}/c_{bdy}$, can be overcome by introducing a gravitating bath, as shown on the RHS of figure \ref{fig:lr}. The 2 configurations can be smoothly interpolated under the action of an RG-flow driving the conformal boundary at a finite cutoff in the AdS bulk, thereby rendering it gravitating. In terms of the parameter $F$, \eqref{eq:F}, the RG-flow corresponds to the limit in which $c_{_{bdy}}>>c_{_{bulk}}$ such that, to sufficiently good approximation, the BCs of the BCFT$_D$ encode all the degrees of freedom of the boundary theory. Correspondingly, the only part of the conformal boundary that is left is a codimension-2 theory, CFT$_{D-1}$, located at $z=\epsilon$, as shown on the RHS of figure \ref{fig:lr}, where the issue of the \emph{flat entanglement spectrum} can be circumvented if calculating the entanglement on a subsystem of the relic codimension-2 boundary theory, \cite{BBK}.

\begin{figure}   
\begin{center} 
\includegraphics[scale=1]{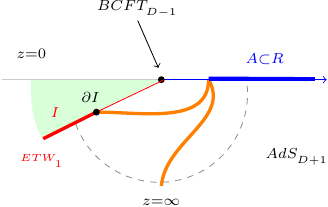}
\includegraphics[scale=1]{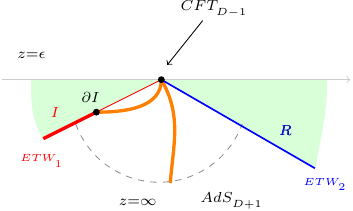}      
\caption{Non-gravitating (left) and a gravitating bath (right) KR/HM-constructions.} 
\label{fig:lr}   
\end{center}     
\end{figure}

\section*{Cosmological horizon}

\begin{figure}[ht!]      
\begin{center} 
\includegraphics[scale=0.55]{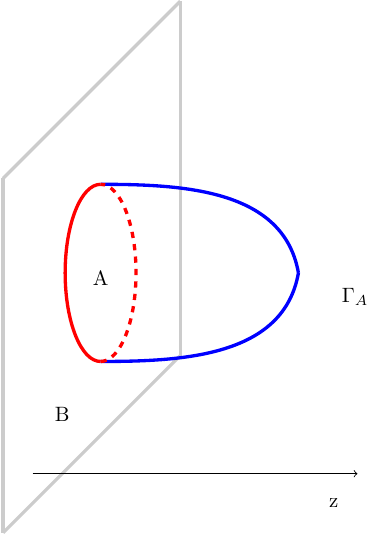}
\ \ \ \
\includegraphics[scale=1.2]{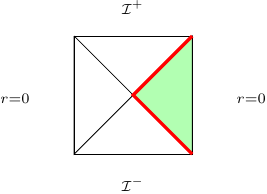} 
\caption{\footnotesize The picture on the left is at a fixed time slice. The plane is where the CFT lives and the bulk is AdS, with $z$ realised holographically. The subregion of bulk spacetime lying between $A$ and $\Gamma_{A}$ defines the \emph{entanglement wedge} of $A$. Global dS$_{_{D}}$ for $D>2$ (on the RHS). An observer at $r=0$ can only access degrees of freedom within the cosmological horizon $r=1/H$. } 
\label{fig:dS}
\end{center} 
\end{figure}

Equipped with this insight, we now turn to the case of the cosmological horizon, the main point of this digression being that of outlining the similarity between the unitarity issues in global dS and black hole evaporation in asymptotically flat spacetimes.

The cosmological horizon is placed at a finite distance\footnote{Each point along the red line in figure \ref{fig:dS} is a codimension-2 surface of radius $r_{h}$.}, $r_{h}=1/H$, with respect to an observer located at the origin of the static patch, thereby implying that there is only a finite amount of spacetime that she/he can be in causal contact with. Because of this, one might be led to conclude that the area of the cosmological horizon should somehow be related to the entropy of the static patch itself, as the Holographic Principle would suggest. However, it is also known that the dS Hilbert space is infinite-dimensional, \cite{B}. The reason for this follows from the fact that the dS isometry group is noncompact, and therefore has no finite dimensional unitary representations. This provides one of the main obstacles towards extending the holographic dictionary of \cite{JM} to the case of a positive cosmological constant.

The configuration just described seems to share similar unitarity issues as encountered for the case of the evaporating black hole. Assuming the horizon surrounding the static patch arises from partial tracing over the global spacetime, c.f. figure \ref{fig:dS},  when taking $\underset{\Lambda\rightarrow 0}{\lim}$, the cosmological horizon will correspondingly be pushed towards ${\cal I}^{+}$, at which the entropy should vanish. Intuitively, this is in agreement with the partial-tracing argument outlined above, since, as the cosmological horizon is pushed further away from the observer in the static patch, the degrees of freedom which were previously ``hidden'' (i.e. traced-over), are expected to re-enter the horizon, thereby becoming accessible to the observer. However, this falls short from being true given the divergence of the entropy defined by the area-law, which, e.g., in 4D scales as $\sim{\cal O}\left(\frac{1}{H^2}\right)$. The area of the codimension-2 surface located at such distance can be interpreted by holographic arguments as the entropy of the static patch, which, in the von Neumann formulation, follows from having traced over the spacetime beyond the cosmological horizon, in analogy with the black hole picture. As such, it is not a QES. 

Given the monotonic behaviour of, both, the background action for dS$\rightarrow$dS transitions (as found in \cite{DeAlwis:2019rxg}) and the von Neumann entropy, we suggest that the divergence issue arising from the flat spacetime limit of de Sitter might be solved by introducing a more suitable definition of the de Sitter entropy, in a similar fashion as $S_{gen}$ enables to recover unitarity within the context of black hole evaporation. As anticipated at the beginning of this section, this identification turns out to be possible\footnote{Interesting new developments towards realising a dS$_4$/CFT$_3$ correspondence have recently been presented in \cite{Cotler:2023xku}, where the interpretation of the maximal entanglement between the north and south pole in global dS shares the same motivation as the argument presented in our work.}.


\subsection{Holographic embedding of vacuum transitions}    \label{sec:KRvt}

Here we present our proposal for the holographic description of the vacuum transitions. As highlighted throughout the previous sections, the three methods used for deriving transition amplitudes exhibit unexpected behaviours. In the present section, we wish to provide a unifying holographic framework describing all processes, mostly motivated by the fact that:

\begin{enumerate}  
\item The action for the system of 2D spacetimes joined by a wall is analogous to the JT-gravity coupled to a CFT$_2$ setup adopted to formulate the island proposal within the context of the information paradox. 

\item The \emph{bulk-brane-boundary} description of the same phenomenon in the KR/HM setup accounts for complementary features of the same underlying process.

\end{enumerate}

It is indeed in the spirit of the BCFT formulation of the information paradox, that we propose a holographic embedding of the transitions described analytically in the first three sections of the present work, after having implemented suitable adaptations due to the specifics of the configurations being analysed.

\begin{figure}[ht!]
\begin{center}
\includegraphics[scale=0.9]{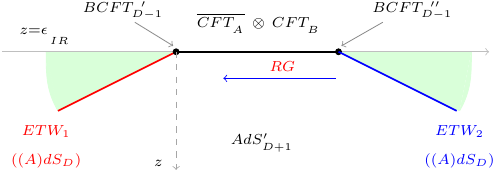}  
\ \ \  
\includegraphics[scale=0.9]{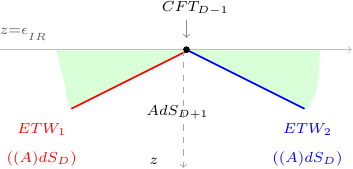}
\caption{\footnotesize The LHS shows a generalisation of figure \ref{fig:lr} for vacuum transitions. A deformation drives the RG-flow on the boundary, assuming $c_{bulk}<<c_{bdy}$ at both endpoints. In the IR of the RG-flow (as shown on the RHS), the vacuum transition takes place in between subregions of different $AdS_{_{D}}$'s, each one lying on a different ETW. More details will be outlined in the concluding part of this section.  } 
\label{fig:foldtr221} 
\end{center} 
\end{figure}

Figure \ref{fig:foldtr221} shows our proposal for the holographic embedding of 2D vacuum transitions:
\begin{itemize} 

\item Given that the process involves two different spacetimes, each one being associated to a different JT-action, the holographic embedding of the transition can be described by two ETW branes (each one corresponding to one of the two vauca, $ETW_{_{1,2}}$) with the brane separating them (mediating the decay process) being geometrically realised by a composite CFT (denoted by $\overline{CFT}_{_{A}}\otimes CFT_{_{B}}$ in figure \ref{fig:foldtr221}), which can be achieved by performing the folding trick.

\item The radiation region lies on the conformal boundary (denoted by the tensor product of CFTs arising from the folding trick), which in turn is brought at a finite cutoff, $\epsilon_{_{IR}}$, with nontrivial BCs, the latter denoted by $BCFT_{_{D-1}}, BCFT_{_{D-1}}^{\prime}$. The theory living on the gravitating bath interpolates between the values of the $D$-dimensional cosmological constants. As such, the wall plays the role of the entangling surface in between the spacetimes involved, with the entanglement being \emph{internal}.

\item  Attached to the radiation region are the two ETW branes. Each one of them accommodates one of the two spacetimes, (A)dS$_{_{D}}$.

\item The composite CFT on the gravitating bath (depicted on the LHS of figure \ref{fig:foldtr221}) is ultimately driven by an irrelevant operator to the configuration on the RHS of figure \ref{fig:foldtr221}, which is reminiscent of the wedge holographic construction, \cite{BB-1}.  

\item As explained at multiple stages throughout our treatment, the total bounce and action for a given vacuum transition, result once having suitably accounted for background subtraction. As we shall see, it is particularly instructive to assign a corresponding holographic realisation of the background configuration as well. Our proposal is represented in figure \ref{fig:foldtr2211} on the RHS. We now turn to explaining it in more detail, albeit most arguments follow through from the ones outlined when explaining the RHS of figure \ref{fig:foldtr221}. The background action is a pure CFT. As such, it can be equally described independently of the choice of the UV cutoff. Hence, starting from a pure AdS$_{_{D+1}}$/CFT$_{_{D}}$ setup, with the conformal boundary defined on the whole real line at $z=0$, we can perform a suitable conformal transformation enabling to restrict the boundary domain to an interval with BCs dual to two ETW branes characterised by the same $D$-dimensional cosmological constant. Given that the theory is still conformal, the width of the segment can be arbitrary; for the purpose of interest to us, we need to bring the conformal boundary at the same IR cutoff as the two-spacetime joined system on the RHS of figure \ref{fig:foldtr221}, such that the background subtraction exactly overlaps with the instanton.

 \end{itemize}   

\begin{figure}[ht!]
\begin{center}
\includegraphics[scale=0.9]{deconfined.pdf}  
\ \ \  
\includegraphics[scale=0.9]{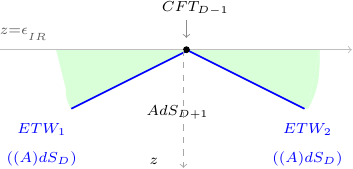}
\caption{\footnotesize The background configuration, with respect to which the decay rate is normalised, is drawn on the right. From the boundary theory, CFT$_{D-1}$, this corresponds to the deconfined phase of the theory. }
\label{fig:foldtr2211} 
\end{center} 
\end{figure}

\subsubsection*{Concrete proposal for the vacuum transition from the CFT side}

If there is a holographic dual formulation of vacuum transitions, a natural question to ask is: what is the corresponding physical effect that happens on the CFT side  that describes the vacuum transition in the bulk. Here we proposed that on the CFT side, the vacuum transition corresponds to a field theoretical phase transition such as a confinement/de-confinement phase transition. For this we need to identify the dual of the background spacetime and the dual of the final composite spacetime. For concreteness let us concentrate on a bulk AdS.
As we described before the vacuum transition  corresponds not to the decay of a full AdS but a portion of an AdS. For the dual we may represent this in terms of a double-well scalar potential that originally has two vacua related by $\phi \rightarrow -\phi$ separated by a domain wall. Modding out by the $\phi \rightarrow -\phi$ symmetry this reduces the bulk spacetime to a portion  with a boundary. This can provide a description of the dual of the original AdS. The description of the final spacetime can be seen in this way as a similar scalar potential but with the two minima being non-degenerate corresponding to two different spacetimes joined by the boundary. The question is how are the theories in the two boundaries related.

In figure \ref{fig:foldtr2211}  the background  can be thought of as being given by a degenerate double well potential, with the two vacua identified. In their holographic realisation in terms of ETW branes (as shown on the RHS of figure \ref{fig:foldtr2211}), only one of them will be undergoing the transition, leading to the configuration depicted on the RHS of figure \ref{fig:foldtr221}.
 
The interpolation between the deconfined and the confined phase of the theory living on the boundary can be realised by adding to the deconfined phase corresponding to figure \ref{fig:foldtr2211}, a  deformation, leading to degeneracy breaking between the two vacua, as shown on the RHS of figure \ref{fig:SSB}. Given that our analysis focuses on 2D, $T\bar T$-deformations will turn out being the relevant ones. 

The reason why we can effectively interpret vacuum transitions in terms of a deconfinement/confinement transition can be understood making use of the  analysis carried out in \cite{Klebanov:2007ws} and \cite{Komargodski:2020mxz}. In the former, it is shown that the difference between the two phases is dictated by the different dependence of the entanglement entropy on $N$, which is the key parameter in the holographic dictionary\footnote{Indeed, $N$ is related to the central charge and the AdS radius.}. In particular, following the RT prescription, this amounts to a diffference in the behaviour of the expectation value of the Wilson line connecting two arbitrary points on the conformal boundary: for the deconfined phase, $S_{_{EE}}\sim {\cal O}(N^{^2})$, wheras for the confined phase, $S_{_{EE}}\sim$ const. What this practically means is that, the former is actually associated to a conformal theory at $z=0$, for which the Wilson line diverges unless a UV-cutoff is introduced. Bulk reconstruction, in such setup, allows to recover the entire (infinite) AdS$_{_{D+1}}$ bulk volume upon removing the cutoff. On the other hand, the confined phase is characterised by a constant value of the entanglement entropy with respect to $N$, meaning that bulk reconstruction can effectively account for a fixed AdS$_{_{D+1}}$ bulk volume. In figure \ref{fig:foldtr2211}, this corresponds to having brought the conformal bundary at $\epsilon_{_{IR}}$, resulting only in a finite (A)dS$_{_{D}}$ volume being involved in the process.  

In a similar, and somewhat complementary fashion, the CFT description of the processes dealt with in our treatment find realisation in a recent work \cite{Komargodski:2020mxz}, where scalar potentials of the kind depicted in figure \ref{fig:SSB} have been identified with degenerate and non-degenerate vacua in studies of 2d QCD. In \cite{Komargodski:2020mxz} the degenerate case can be identified with a deconfinement phase, whereas the non-degenerate case with a confined phase\footnote{See also \cite{MVRHFC} for an interesting discussion of confinement and cosmology.}. In their treatment, the expectation value of the Wilson loop separating different vacua exhibits, either, an area- or perimeter-law like behaviour according to, whether, the system is in the confined or deconfined phase, respectively. More efficiently, such classification can be further performed by analysing the ratio of partition functions of the gauged and ungauged theory in the infinite volume\footnote{Namely the large-$N$ limit.} limit. If such ratio is order unity, it means the system is deconfined. On the other hand, if the ratio vanishes it is confined.

\begin{figure}[ht!]
\begin{center}
\includegraphics[scale=0.9]{deconfined.pdf}  
\ \ \ \ \ \ \ \ \ \ \ \  \ \ \ \ \ \ \ \ \ \ \ \  
\includegraphics[scale=0.9]{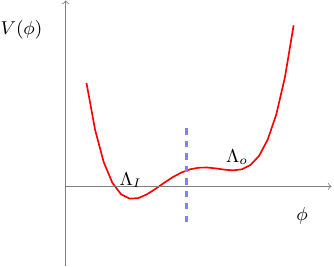}
\caption{\footnotesize{Under the action of an irrelevant operator, the boundary theory can be smoothly interpolated from the deconfined (left) to the confined phase (right). The latter is the field theory realisation of the RHS of figure \ref{fig:foldtr221}. In such description, the interpolating scalar, $\phi$, plays the role of a monopole field in a deconfinement/confinement transition, in a similar fashion as in a previous work by Komargodski et al.} } 
\label{fig:SSB} 
\end{center} 
\end{figure}

In drawing a comparison between the processes dealt with in the present work and those of Komargodski et al., it is important to notice that, in such reference, their analysis embraces, both, vacua and universes. While often used interchangeably, the main difference between vacua and universes is that the latter are separated by infinite barriers, whereas the former might allow finite tension domain walls interpolating between different superselection sectors of the same universe, and are therefore the ones relevant for our treatment. The systems analysed in Komargodski et al.'s work exhibit multiple vacua, distributed between different universes, and study confinement and deconfinement of Wilson lines interpolating between them. Their counterpart for transitions of the kind associated to the RHS of figure \ref{fig:SSB},  are therefore those involving confined Wilson lines interpolating between vacua belonging to the same universe. 

Clearly more work needs to be done in order to fully describe the vacuum transitions from the CFT side.


\subsubsection{Local transitions,  entanglement and islands in wedge holography}  \label{sec:5}     


As a first concrete application of the tools outlined in the first part of this section, we now turn to the holographic interpretation of the FMP results obtained in sections \ref{sec:2} and \ref{sec:3}. In particular, we prove that:

\begin{itemize} 

\item Under suitable parametric redefinition, the results obtained by means of the FMP method are found to agree with the expression provided by \cite{MVR} for describing mutual approximation of boundary states belonging to different CFTs.
\item 
The corresponding expression for the transition rate in presence of gravity, and in absence of black holes, is given by the difference of entropies of $T\bar T$-deformed CFTs, hence proving the locality of the nucleation process. Given its interpretation as being an internal entanglement, such transitions provide an example of an AdS$_2$/CFT$_1\subset$ AdS$_3$/CFT$_2$. \footnote{The lower-dimensional holographic setup involved in these processes shares similar features as to the one involving spacetimes emerging from matrix QMs, as analysed, e.g. in \cite{Anous:2019rqb}, and their higher dimensional counterparts in \cite{VanRaamsdonk:2021duo}.}

\item Upon adding black holes, instead, the total action can be expressed as the difference of generalised entropies, with an island emerging beyond a critical value of the black hole mass. As such, this is an example of an  AdS$_2$/CFT$_1\not\subset$ AdS$_3$/CFT$_2$.     

\end{itemize}


\subsection*{ICFTs and wedge holography} \label{subsec:3}


In this subsection, we prove that quantum transitions involving different subregions of spacetime can be described by means of dual CFTs interacting via an interface, \cite{MVR}. The authors of the latter describe mutual approximation of boundary states, $\Psi_{_{A,B}}$, belonging to the Hilbert spaces of different CFTs, (as shown in figure \ref{fig:compcft}).

The correspondence adopted in \cite{MVR}, specifying to the case of AdS$_3$/CFT$_2$, relates the following set of parameters on either side of the interface

\begin{equation}    
\boxed{\ \ \ L_{1}\ ,\ L_{2}\ ,\ \kappa 
\ \ \ 
\longleftrightarrow 
\ \ \ 
c_{_{A}}\ ,\ c_{_{B}},\ c_{bdy}\sim\ln g,\color{white} \textcolor{white}{\Biggl[} \color{black} \ \ } 
\end{equation}
with $L_{1,2}$ being the AdS radii, $\kappa$ the tension of the wall separating the bulk spacetimes, $c_{_{A,B}}$ the central charges, and $\ln g$ denoting the entropy of the ICFT, $S_{ICFT}$, with $g$ defining the degeneracy of the ground state.

\begin{figure}[ht!]      
\begin{center} 
\includegraphics[scale=0.8]{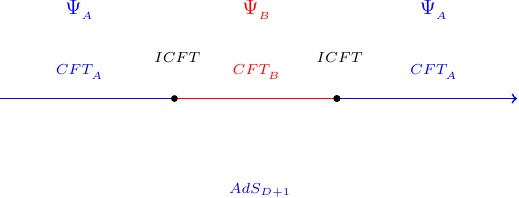}   
\caption{\footnotesize Replacing a small portion of the background CFT$_{_{A}}$ with a different theory, CFT$_{_{B}}$ implies a change in the energy spectrum. Because of this, for a given boundary state, the corresponding $S_{_{EE}}$ leads to a different effective bulk-reconstruction of the AdS dual. }   
\label{fig:compcft}  
\end{center}     
\end{figure}

Our first finding is that AdS$_2\rightarrow$AdS$_2$ transitions are equivalent to the case where the mutual approximation described in \cite{MVR} only involves the ground states. Indeed, under suitable parametric redefinition, the action for the transition coincides with the one obtained by \cite{MVR}, as explicitly shown below

\begin{equation} 
\begin{aligned}
S_{tot}^{\ \text{AdS}\rightarrow\text{AdS}}     
&
= 
\frac{\eta}{G}\left[\ {\mu_{-}}\ln\bigg|\frac{{\mu_{-}}A_{1}+1}{{\mu_{-}}A_{1}-1}\bigg|-\mu_{+}\ln\bigg|\frac{{\mu_{+}}A_{2}+1}{{\mu_{+}}A_{2}-1}\bigg|\ \right] 
\nonumber \\
& = 
\frac{2\eta}{G}\bigg[\left({\mu_{+}}-{\mu_{-}}\right)\tanh^{-1}\left(\frac{\Delta_{-} }{\kappa}\right) -\left({\mu_{+}}+{\mu_{-}}\right)\tanh^{-1}\left(\frac{\kappa}{\Delta_{+}  }\right)     
\bigg]  \nonumber \\ 
&
=   
2\eta\ S_{_{ICFT}}    
\ 
= 
\ 
2\eta\ \ln g (\kappa) \ =\ 
-2\eta\ F_{\partial},   
\label{eq:bricft}
\end{aligned} 
\end{equation}
where we defined

\begin{equation} 
\Delta_{\pm}  
\ 
\overset{\text{def.}}{= } 
\ 
2\left(\sqrt{-\Lambda_{+}}\pm\sqrt{-\Lambda_{-}}\right),   
\end{equation} 
\begin{figure}[ht!] 
\begin{center} 
\includegraphics[scale=0.8]{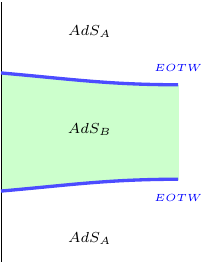}     
\ \ \ \ \ \ \ \ \ \ \ \ \ \ \ \ \ \ 
\includegraphics[scale=0.8]{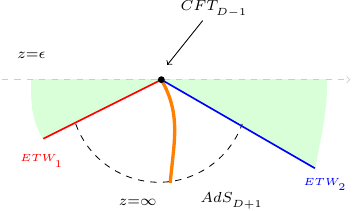} 
\caption{\footnotesize{As shown in \cite{MVR}, the trajectory of the interface separating pure AdS$_{3}$ phases has no turning point (LHS), and, correspondingly, the degeneracy of the ground state is a monotonic function of the tension $\kappa$, as found in section \ref{sec:4}. No island emerges in absence of conical defects (cf. picture on the RHS). The bulk spacetime is completely determined by means of wedge holography from the $CFT_{_{D-1}}$. $S_{_{EE}}$ is evaluated on the codimension-2 surface at the conformal boundary.  }}    
\label{fig:lngk1}
\end{center} 
\end{figure} 
and $g (\kappa)$ denotes the ground state degeneracy and $F_{\partial}$ the \emph{boundary free energy}, which is a monotonic function of $\kappa$. The entire transition is hence described by means of a codimension-2 theory as depicted in figure \ref{fig:lngk1}, with the bulk emerging via wedge holography.

Making use of the following parametric redefinitions

\begin{equation} 
\mu 
\ 
\overset{def.}{=}     
\ 
\sqrt{\frac{\lambda c}{12}\  }
\ \ \ 
, 
\ \ \ 
\phi_{o}   
\ 
\overset{def.}{=}    
\ 
r
\ \ \ 
, 
\ \ \ 
\lambda   
\ 
\overset{def.}{=}   
\ 
2\sqrt{-{\cal C}\Lambda,\ }      
\label{eq:2termssttb}    
\end{equation} 
the total action can be rewritten as

\begin{equation} 
\boxed{\ \ S_{tot}^{\ \text{(A)dS}\rightarrow\text{(A)dS}}
\ 
= 
\ 
2\pi\eta\left[\ S_{_{T\bar T}}^{^{-}}- S_{_{T\bar T}}^{^{+}}   \ \right]_{_{\text{univ.}}}   
\ 
= 
\ 
-2\pi\eta\   F_{\partial} 
= 
\ 
2\pi\eta\ S_{_{ICFT}}, \textcolor{white}{\Biggl[} \color{black} \ \ }     
\label{eq:chofrel}     
\end{equation} 
where    
\begin{equation} 
\begin{aligned}
S_{_{T\bar T}} 
\ 
&= 
\ 
\left(1-\frac{r}{2}\frac{\partial}{\partial r}\right) \ln {\cal Z}_{S^{2}}^{^{\ \ AdS, T\bar T}} 
\ 
= 
\ 
\pi \frac{c}{3}\  \sinh^{-1}\ \left(\sqrt{\frac{12}{c\lambda}}\ r\ \right)    \\
S_{_{T\bar T}} 
&=   
\left(1-\frac{r}{2}\ \frac{\partial  }{\partial r}\right)\ln {\cal Z}_{S^{2}}^{^{\ \ dS,  T\bar T}} 
\ = 
 \ 
 \frac{c}{6}+\frac{c}{3}\ \ln\left|\ \sqrt{\frac{\lambda c}{12 }}\frac{1}{r}\ \right|+\frac{c}{3}\ \tan^{-1}\ \left(\frac{1}{\sqrt{\frac{\lambda c}{12r^{2}}-1}  }\right)   +\text{const.}   \nonumber\\  
&=
 S_{\epsilon} + S_{_{\text{univ.}}},   
 \label{eq:sttds}    
  \end{aligned}
\end{equation}    
for the duals of AdS$_3$ and dS$_3$, respectively, \cite{BB70, BB48}. $S_{_{\text{univ.}}}$ and $S_{\epsilon}$ denote the \emph{universal} and \emph{cutoff-dependent} parts of \eqref{eq:sttds}, respectively, with

\begin{equation}   
 S_{\epsilon}   
 \overset{def.}{=}   
 \frac{c}{3}\ \ln\left|\ \sqrt{\frac{\lambda c}{12 }}\frac{1}{r}\ \right|,
 \ \ \ \ \ \ 
 \text{and}\ \ \ \epsilon\overset{def.}{=}\sqrt{\frac{\lambda c}{12 }.\ }    
 \end{equation}
According to the Casini-Huerta-Myers prescription, the universal part is the one contributing to the definition of the boundary free energy for a BCFT, and therefore this is the only term we need to retain for our purposes. Indeed, $S_{\epsilon}$ can be removed by simply setting $r=\epsilon$, namely choosing the cutoff to be equal to the localisation radius $r$.
From \eqref{eq:chofrel}, we deduce that:    

\begin{enumerate} 

\item This chain of equivalence relations, \eqref{eq:chofrel}, suggests an interesting direct correspondence between the mutual approximation of CFT states, \cite{MVR}, and $T\bar T$-deformed CFTs.
By virtue of the identification with $S_{gen}$, equation (\ref{eq:bricft}) implies the absence of a QES, and therefore of an island, consistently with the fact that the spacetimes involved in the transitions have no event horizons. 

\item The direct relation between the extremised action and $S_{_{T\bar T}}$ in 2D, ensures the locality of the process being described. However, as also suggested from \eqref{eq:2termssttb}, upon taking the flat limit on either side, the cutoff radius diverges ($r\rightarrow\infty$), and so too does the turning point, meaning the process is forbidden. We therefore conclude that, as long as the parameters of the theory are kept constant, no issue arises, and the transition remains local.

\item The cutoff nature of the cosmological horizon suggested by \eqref{eq:chofrel}, proves that its contribution to $S_{gen}$ should be understood as being part of $S_{_{T\bar T}}$. The motivation that led us to connect this issue with the one encountered in the black hole information paradox, precisely resided in the need to correctly define the dS entropy. Indeed, its divergence resembles that of the monotonic von Neumann entropy in the BH evaporation process, and, holographically, this corresponds to the standard divergence of $S_{_{EE}}$ upon removing the UV-cutoff. Just as in the black hole evaporation process the event horizon is not the QES extremising $S_{gen}$, so too can be claimed for the cosmological horizon, therefore proving the absence of islands in pure de Sitter spacetimes. 

\item Localisation techniques play a key role in evaluating the partition function for a given CFT. Given its key role in determining $S_{_{EE}}$, the extremisation procedure rooted in the formalism is compatible with the method defining $S_{_{TOT}}$ by means of the FMP method. In this section, we prove that these quantities are indeed proportional to each other.

\end{enumerate}

\subsubsection*{The emergence of the island}  \label{sec:5.5}

In presence of a non-extremal black hole, the total action for AdS$_2\rightarrow$AdSBH$_2$ reads: 

\begin{equation}   
\begin{aligned}
S_{tot}^{\text{\ AdS$\rightarrow$AdSBH}} 
&      
=         
\frac{2\pi\eta}{G}\ \left[\   
\frac{{\cal B}}{2\Lambda_{-}}\ \left[1-\sqrt{1-\frac{4{\cal C}\ \Lambda_{-}}{{\cal B}^{2}}}- \ln\left|\frac{y_{2}}{y_{1}}\sqrt{\frac{1-y_{1}^{2}}{1-y_{2}^{2}}}\right| \right]+\  \right.    
\nonumber\\ 
& 
- 
\phi_{2}\ln\left|\frac{A_{2}\ y_{2}\sqrt{1-y_{4}^{2}}}{A_{4}\ y_{4}\sqrt{1-y_{2}^{2}}}\right|+\phi_{1}\ln\left|\frac{A_{2}\ y_{1}\sqrt{1-y_{3}^{2}}}{A_{4}\ y_{3}\sqrt{1-y_{1}^{2}}}\right| +\nonumber\\ 
& 
- 
\left.\frac{1}{2}\sqrt{\frac{\frac{{\cal B}^{2}}{4\Lambda_{-}}-{\cal C}}{\Lambda_{-}}}\ \ln\ \left|\frac{1-y_{2}}{1+y_{2}}\frac{1+y_{1}}{1-y_{1}}\right| +\mu_{+}\ \ln\ \left|\frac{1-y_{4}}{1+y_{4}}\frac{1+y_{3}}{1-y_{3}}\right| \  \right],      
\label{eq:next2} 
\end{aligned}  
\end{equation}
where the notation is the same as in section \ref{sec:3}. \eqref{eq:next2} coincides with the result by \cite{MVR} for the case of an excited state of the new CFT approximating the ground state of the background theory under
the following identifications

\begin{equation} 
\begin{aligned}
S_{brane} 
& 
=  
(1-\mu)\Delta t_{2} -\Delta t_{1} \nonumber\\   
&   
=     
- \frac{1}{2}\sqrt{\frac{\frac{{\cal B}^{2}}{4\Lambda_{-}}-{\cal C}}{\Lambda_{-}}}\ \ln\ \left|\frac{1-y_{2}}{1+y_{2}}\frac{1+y_{1}}{1-y_{1}}\right|    +\mu_{+}\ \ln\ \left|\frac{1-y_{4}}{1+y_{4}}\frac{1+y_{3}}{1-y_{3}}\right| ,    
\\ 
S_{hor}^{^{(-)}} 
&      
=       
(1-\mu) \frac{\beta}{2}
\ 
= 
\    
- \frac{1}{2}\sqrt{\frac{\frac{{\cal B}^{2}}{4\Lambda_{-}}-{\cal C}}{\Lambda_{-}}} ,  
\label{eq:shordt}   
\end{aligned}  
\end{equation}
with the latter, \eqref{eq:shordt},  being the 2$^{nd}$ term featuring in the total action \eqref{eq:next2}. The terms proportional to $\phi_{1,2}$ are $\sim{\cal O}({\cal B})$, and so too are the first and third terms featuring in the first line of the expression for the total action. The results in \cite{MVR} are in agreement with ours upon expanding the total action with respect to the deformation parameter, ${\cal B}$. As also mentioned by the authors of \cite{MVR}, their expression does not contain terms that vanish upon removing the cutoff, thereby explaining the missing terms in performing the comparison. We also made a consistency check with the results obtained in section \ref{sec:3}, and found that subleading terms in the l.o. ${\cal B}$-expansion vanish in the $\underset{{\cal B}\rightarrow 0}{\lim}$.

\begin{figure}[ht!]     
\begin{center}
\includegraphics[scale=0.8]{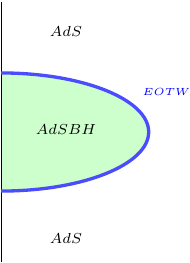} 
\ \ \ \ \ \ \ \ \ \ \ \ \ \ \ \ \ \ 
\includegraphics[scale=0.8]{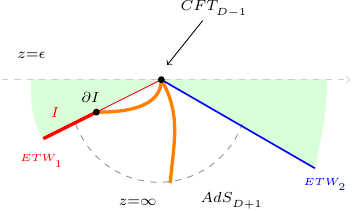} 
\caption{\small{By adding a black hole in the 3D bulk (LHS), instead, a turning point emerges, resulting in the ETW brane closing up and rejoning the conformal boundary. Correspondingly, an island emerges in the wedge holographic picture (RHS). This is the key difference with respect to figure \ref{fig:lngk1} on the LHS, where there is no bulk turning point, and therefore, the RT surface doesn't intersects the ETW brane.  }}    
\label{fig:VR1}    
\end{center} 
\end{figure}

The integration variables in the descriptions outlined in section \ref{sec:3} and \cite{MVR} are related as follows    

\begin{equation} 
\boxed{\ \ y 
\ 
\overset{\text{def.}}{=}  
\ 
\sqrt{\frac{1-\mu}{\lambda}}\ r. \ \ }    
\label{eq:newident1}   
\end{equation}
In terms of which, the 3D BTZ geometry, defining an orbifold of the universal cover of the Poincarè metric, can be associated with the line element

\begin{equation} 
ds^{2} 
\ 
= 
\ 
-\frac{r^{2}-r_{h}^{2}}{L^{2}}\ dt^{2}+\frac{L^{2}}{r^{2}-r_{h}^{2}}\ dr^{2}+\frac{r^{2}}{L^{2}}\ dx^{2}, 
\label{eq:BTZM3}      
\end{equation}
and becomes 2D conformally flat at the cutoff $r=r_{c}$, where the $T\bar T$-deformed CFT of inverse temperature $\beta\overset{def.}{=}2\pi L^{2}/r_{h}$ is located. Notice that $\underset{r_{h}\rightarrow 0}{\lim}\ \beta=\infty$, as expected for the T=0 theory. 
Recall that the variable $y$, featuring in \eqref{eq:newident1}, introduced in section \ref{sec:3}
parametrises the value of the dilaton, $\phi_b$, on the brane interpolating between the 2D vacua involved in the decay process. As previously argued, in the holographic interpretation of the process, the 1D brane is mapped to a bulk CFT$_2$ on which $S_{_{T\bar T}}$ is evaluated. 

In the same fashion, one might be led to believe that the role played by the cutoff $r_c$ is analogous to $\phi_b$, since all other parameters in the definition of $y$ are held fixed by the formalism. However, we will now be arguing that extra care is needed in building the correspondence between the 2 setups. For ${\cal B}=0$ , $y=\frac{\phi_b}{\mu}    
$, and, due to the absence of horizons, $
y_{tp} 
\ 
=1$ is unique. On the other hand, in presence of a black hole, the following identification needs to be made
\begin{equation} 
\tanh x 
\ 
\overset{\text{def.}}{=}     
\ 
y 
\ \ \ 
\Rightarrow 
\ \ \ 
\sinh x 
\     
= 
\ 
\frac{r_{h}}{r_{c}}    
\ 
= 
\ 
\frac{\sqrt{1-y^{2}} }{y}    
\ \ \ \Rightarrow\ \ \ 
 r_{c} 
\overset{\text{def.}}{= } 
\ 
{r_{h}}\frac{y}{\sqrt{1-y^{2}} }  
\ \ \ 
\overset{y\rightarrow 1}{\longrightarrow } 
\ \ \    
\infty,    
\label{eq:divergcutoff}    
\end{equation}
from which the following information can be extracted:     

\begin{enumerate}   
\item 
If ${\cal B}=0$, then $y_{tp}=1$ is unique, and $r_c=\infty$. Correspondingly, \eqref{eq:BTZM3} reduces to a pure AdS$_3$ metric, and the bulk configuration is the same as the one drawn on the LHS of figure \ref{fig:lngk1}. 

\item As proved in section \ref{sec:fl}, $y_{tp}=1$ is also the limiting value of the turning points upon taking the flat limit on either side of the 2D transition. 
The last passage in \eqref{eq:divergcutoff} thereby shows that the effect of removing the cutoff in \cite{MVR} is equivalent to taking $\underset{{\cal B}\rightarrow 0}{\lim}$ given the identification of the turning point $y_{tp}$ with $r_{c}$. 

\item These observations imply the need to add a black hole in the 3D bulk in order to to keep $r_{c}$ finite, thereby ensuring the ETW brane separating the two phases can actually close up, as shown on the RHS of figure \ref{fig:VR1}. 

\end{enumerate}

In order to provide further justification for the claims made so far, we will now show that: 

\begin{itemize} 

\item Transition in between pure (A)dS spacetimes described by \eqref{eq:chofrel} can also be obtained in presence of a suitable value of ${\cal B}\neq 0$, thereby justifying the identification of the black hole mass with the $T\bar T$-deformation ensuring the locality of the transition taking place\footnote{Our finding is hence in agreement with \cite{Morvan:2022ybp}, where the authors show that dS black holes behave as localized, constrained, states, and therefore feature a lower entropy with respect to pure dS.}.

\item For arbitrary values of ${\cal B}$, the brane action undergoes a phase transition when the black hole mass decreases below a certain critical value. 

\end{itemize}

\section*{How the number of turning points depends on ${\cal B}$}

For particular values of ${\cal B}$, the turning point reduces to

\begin{equation}   
   \boxed{\ \ \  \phi_{2} 
 \ 
 = 
 \ 
    \begin{cases}
\ 0\ \ \ \ \ \text{for  ${\cal B}=0$ \ \ \ \ \ \ \ \ \ \ \ \ \ $\equiv$ \ \ pure dS, 1 horizon, $y_{2}=0$} \\
\\
\ \mu_{\pm}\ \ \   \text{for ${\cal B}=-2\sqrt{|{\cal C}|\Lambda}$ \ \ $\equiv$ \ \ extremal dSBH, 2 degenerate horizons, $y_{2}=1$} \\    
\\
\ \in\ \ ]0,\mu_{\pm}[\   \text{for  $-2\sqrt{|{\cal C}|\Lambda}<{\cal B}<0$  $\equiv$  non-extremal dSBH, 2 horizons, $y_{2}\neq\{0,1\}$.} \\
    \end{cases} \textcolor{white}{\Biggl[} \color{black}}    
    \nonumber
    \label{eq:cases1}   
\end{equation}    
This shows that, in terms of the coordinate $y$, there is no turning point in absence of a conformal symmetry breaking term in ${\cal L}$, thereby forbidding the transition to take place. On the other hand, for ${\cal B}_{\text{dS}}=-2\sqrt{|{\cal C}|\Lambda}$, the turning point is unique, and therefore coincides with the results obtained in section \ref{sec:2}. For any other value ranging in between, i.e. ${\cal B}\ \in\ ]-2\sqrt{|{\cal C}|\Lambda},0[$, there are 2 physical turning points in terms of $\phi_b$, consistently with the constraint derived in section \ref{sec:3}, \eqref{eq:constrm1}.

Having said this, we therefore conclude that:    

\begin{itemize} 

\item The role of ${\cal B}$ is effectively equivalent to that of an irrelevant operator on a CFT$_2$.   

\item The results obtained in section \ref{sec:2} should be understood of as arising from the IR limit of the RG-flow of an irrelevant operator, whose imprint in the transition amplitude is given by the finiteness of the turning point at which nucleation takes place. 

\item The cosmological horizons and the cutoff radius at which localisation takes place are proportional to each other, implying the flat limit is equivalent to removing the cutoff altogether; this explains the divergences encountered in section \ref{sec:2}. 

\item According to the value of ${\cal B}$, the nucleation process features a different number of physical turning points. Only in presence of two distinct physical values of the turning points, the flat limit can be taken (as explained in section \ref{sec:fl}). 

\item In light of further considerations made in the remainder of our work, it is important to notice that, due to the specifics of the formalism, the parameter ${\cal B}$ is not simply part of the theory, but, rather specifies the state involved in the transition. This follows from the Hamiltonian constraint and the definition of the transition amplitude. 

\end{itemize}

\section*{Relation to \texorpdfstring{$S_{gen}$}{} and phase transition of \texorpdfstring{$S_{brane}$}{}}

The systems described in section \ref{sec:3} experience a phase transition as the black hole mass decreases, thereby providing further evidence of the fact that, in some cases, 2D vacuum tranistion can only take place in presence of islands.
The actions obtained in sections \ref{sec:2} and \ref{sec:3} can be recast to the difference of two generalised entropies, \eqref{eq:genentr}, 

\begin{equation}
\boxed{\ \ \ \ S_{tot}^{\text{\ \ (A)dS}\rightarrow\text{(A)dS}} 
\ 
= 
\ 
2\pi\eta\ \left[S_{gen}-S_{gen}\right]
\ 
= 
\ 
2\pi\eta\ \left[\frac{\phi_{h,+}}{G}-\frac{\phi_{h,-}}{G}+S_{_{EE,-}}^{^{(A)dS, T\bar T}}-S_{_{EE,+}}^{^{(A)dS, T\bar T}}\right].  \textcolor{white}{\Biggl[} \color{black}  \ \ }    
\label{eq:phirho14}      
\end{equation}

In agreement with arguments outlined in \cite{EW1}, the island only emerges beyond a certain threshold value of the black hole mass, which, as previously shown, is accounted for by the parameter ${\cal B}$. Each part of the brane action is an entanglement entropy of two disjoined intervals on a  CFT$_2$ with deformation parameter ${\cal B}$. Expanding to leading order in ${\cal B}$, both terms can be brought to the following form

\begin{equation} 
\begin{aligned}
S_{_{EE}}  
&=
S_{_{EE}}^{^{(2)}}\bigg|_{{\cal B}=0}+\delta S_{_{EE}}^{\ s}\nonumber\\   
&\sim 
 -\frac{8c_-}{3}\ln\left|\frac{1}{\epsilon}\ \sinh\left(\frac{2\pi r_-}{\beta_-}\right)\right|-\left(1-\frac{\phi_{o}^2}{\mu_{-}^2}\right)\frac{{\cal B}c_{-}^2}{9\Lambda_-}\coth^2\left(\frac{2\pi r_-}{\beta_-}\right)+\frac{\frac{c_-}{3}}{\sinh^2\left(\frac{2\pi r_-}{\beta_-}\right)}.   
 \label{eq:sl}  
 \end{aligned}
\end{equation}    

In the second line of \eqref{eq:sl}, we redefined the parameters featuring in our original expressions in order to make the comparison with the standard CFT calculation more explicit. Notice that this interpretation is fully legitimate, since the parameter ${\cal B}$ specifies the states on either vacuum being involved in the transition process.

\begin{figure}[ht!] 
\begin{center} 
\includegraphics[scale=0.3]{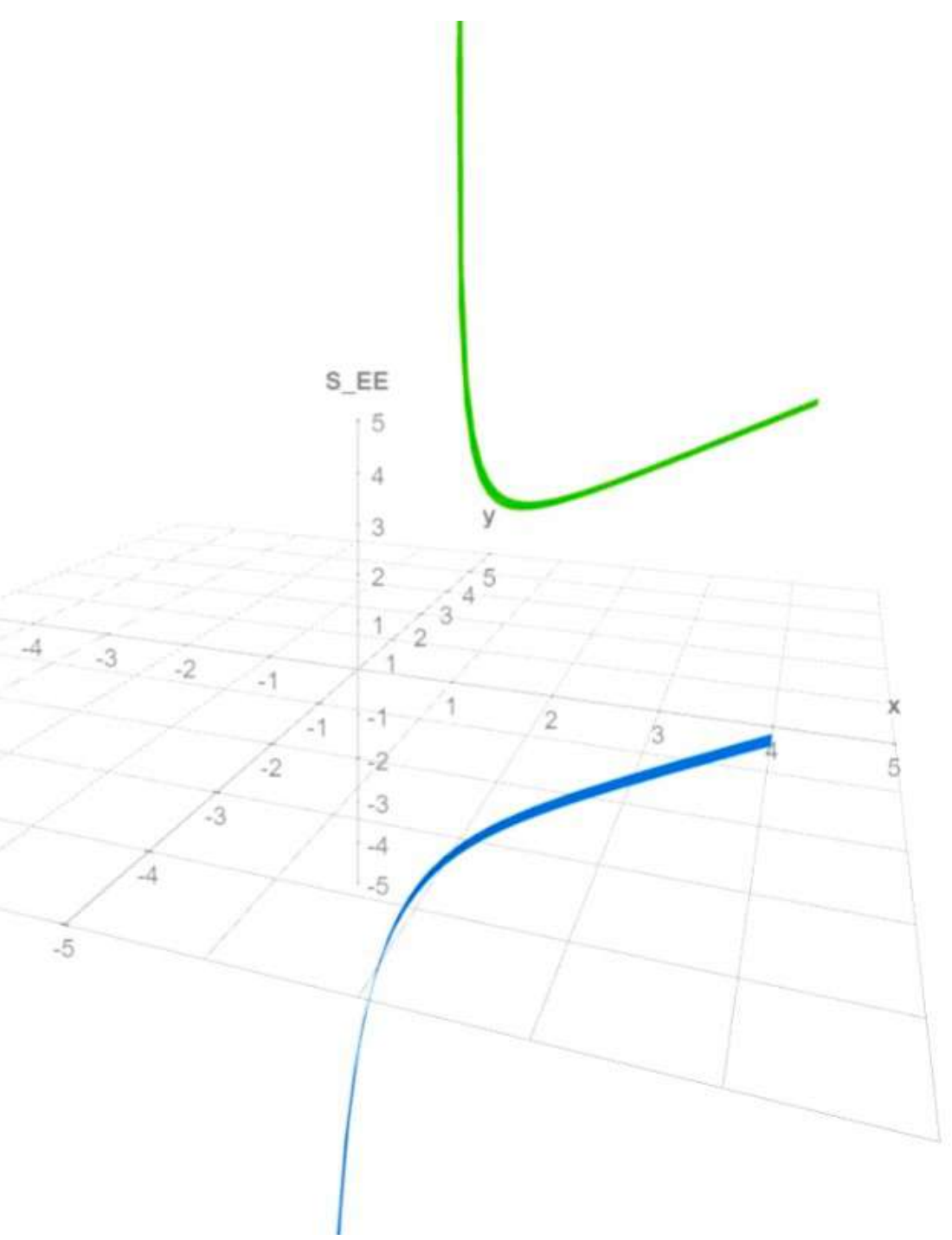}      
\ \ \ \ \ \ \ \ \ \ \ \ \ \    
\includegraphics[scale=0.28]{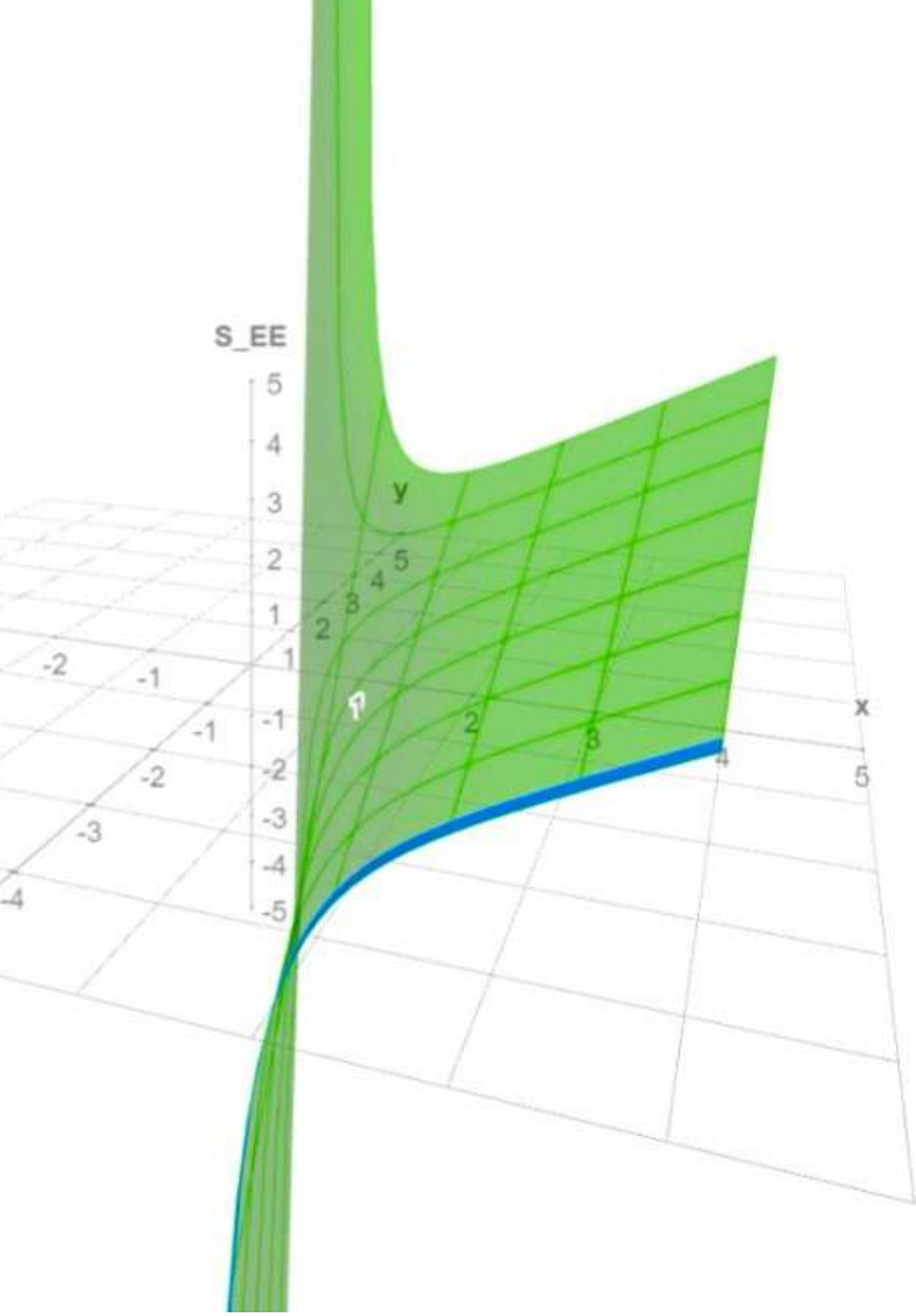} 
\caption{\footnotesize{A 3D plot of $S_{_{EE}}({\cal B}, x)$. $x$ denotes the portion of the CFT$_2$ on which $S_{_{EE}}$ is evaluated. As ${\cal B}$ changes, so too does the slope of \eqref{eq:sl}, as shown on the LHS. The figure on the RHS shows the smooth interpolation in between the monotonic behaviour of $S_{_{EE}}$ as one would expect if evaluated on the disconnected channel over 2 intervals, to a profile admitting a minimum. The phase transition occurs at ${\cal B}<0$, corresponding to a upper bound to the black hole mass below which an island needs to be taken into account.  }}    
\label{fig:lngk}
\end{center} 
\end{figure}

In section \ref{sec:3}, we showed that ${\cal B}<0$ and, from section \ref{sec:2}, we know that $\phi_{o}^2>\mu_{-}^2$, therefore $y>0$ is the range of interest for the decay process described by \eqref{eq:phirho14}. As shown on the RHS of figure \ref{fig:lngk}, the entanglement entropy, $S_{_{EE}}$, experiences a phase transition, for $y>0$. In particular, the latter takes place at $y=1$. It is possible to argue that this is indeed consistent with our finding in section \ref{sec:3} that the black hole mass features in the Lagrangian density as a $T\bar T$-deformation, which needs to be constrained between a certain range, to ensure the existence of two turning points, \eqref{eq:constrm}. 

Indeed, figure \ref{fig:lngk} further supports our argument that, the phase tranistion undergone by the brane action corresponds to a Page transition beyond a suitable value of ${\cal B}$, with the latter playing a similar role as $t_{Page}$ within the context of the information loss paradox. Consequently, this justifies the emergence of the island in \eqref{eq:phirho14} upon interpreting the total action as being the difference of generalised entropies.

\subsection*{An emergent holographic embedding of 2D vacuum transitions with gravity}

In summary, from the holographic interpretation of the FMP results we proved that:

\begin{itemize}

\item The flexibility of the FMP method relies upon the fact that such kinds of \emph{exactly integrable deformations} ($T\bar T$s) can be accounted for at the level of the Lagrangain density, as explicitly outlined in section \ref{sec:3}. Because of this, this setup can potentially accommodate an island, as long as the black hole mass lies within a certain range. 

\item In absence of black holes, the total action for the transition is proportional to the defect entropy, $S_{_{ICFT}}$, which is codimension-2 with respect to the auxiliary AdS$_3$ bulk. This can also be re-expressed as a difference of $S_{_{EE}}^{^{T\bar T}}$s, thereby ensuring the locality of the transition as well as the fact that $S_{_{TOT}}$  is given in terms of \emph{internal} entanglement entropies evaluated on a gravitating bath.

\item In presence of black holes, $S_{_{brane}}$ behaves as a quantity experiencing a phase transition beyond a certain value of ${\cal B}$, which can be thought of as playing the role of $t_{_{Page}}$ within the island formulation. Furthermore, given the fact that the $S_{_{EE}}$s defining $S_{_{brane}}$ are basically describing an \emph{internal} entanglement, due to the role of the dilaton in the given setup, our findings are compatible with the claims of \cite{BB44}.

\end{itemize}

Having said this, we are now able to further motivate our proposal, i.e. figure \ref{fig:foldtr221}, as being an appropriate adaptation of figure \ref{fig:lr} for describing vacuum transitions in presence of gravity.  

\begin{itemize}   

\item    The total action for AdS$_2\rightarrow$AdS$_2$ transitions calculated by means of the FMP method, is a codimension-2 quantity with respect to its AdS$_3$ embedding, and therefore is to be asssociated to the IR-limit of an RG-flow, in agreement with wedge-holography. The same picture can be drawn for transitions involving dS spacetimes, given their realisation via $T\bar T$-deformations in the IR of an AdS.

\item   The portions of spacetime involved in the transition lie on different ETW branes, which, in turn, are holographically dual of the defects BCFT$_{D-1}$s. The interval in between the two defect theories now accommodates a composite CFT$_{D}$ with nontrivial BCs provided by the defect theories themselves. 

\item  Under the assumption that $c_{bdy}>>c_{bulk}$, the RG-flow will drive the pair of defect CFT$_{D-1}$s in the UV to a single CFT$_{D-1}$ in the IR. An island emerges if the resulting CFT$_{D-1}$ alone is unable to realise the bulk AdS$_3$ by wedge holograhy.

\end{itemize}

\subsubsection*{Implications for up-tunnelling}  \label{sec:5.4}

Given the chain of relations \eqref{eq:chofrel}, which in turn relies upon the parametric identification \eqref{eq:2termssttb}, we can now appreciate the fact that the obstructions to up-tunneling outlined in section \ref{sec:2} share a common origin. In particular, we notice that:     

\begin{itemize} 

\item The vanishing brane action in the flat limit, is compatible with the fact that $S_{_{EE}}^{^{T\bar T}}\rightarrow 0$ as $\epsilon\equiv\frac{r}{\sqrt{c\lambda}}\rightarrow0$, thereby further justifying \eqref{eq:chofrel}.

\item The divergence of the bulk action in the flat limit of dS can be recast to the need for redefining the universal part of $S_{_{EE}}^{^{T\bar T}}$ for dS including the $\frac{c}{6}$ term, which, for the decay process analysed in section \ref{sec:2} corresponds to $\mu$, namely the value of the dilaton at the horizon. Given the cutoff nature of the cosmological horizon for pure dS spacetimes (due to \eqref{eq:chofrel}), it therefore follows that, upon taking the flat limit, the transition is no longer local.

\item AdS$_2\rightarrow$AdS$_2$ up-tunneling, AdS$_2\rightarrow$Mink$_2$ and Mink$_2\rightarrow$dS$_2$ can only be achieved as long as we are able to ensure the locality of the decay processes, or, equivalently, the finiteness of the turning point associated to them. For this to happen, the localisation radius, $r$, and the cosmological constant of a given vacuum need to disentangle. This can be achieved by introducing an additional scale, ultimately ensuring the finiteness of the transition under arbitrary change of the cosmological constant.

\item This new scale, namely the black hole mass, is responsible for an additional contribution to the total action which cannot be reabsorbed within the definition of $S_{_{EE}}^{^{T\bar T}}$, thereby signalling the emergence of a QES in \eqref{eq:phirho14}, and proving the need for an island to be present for up-tunneling to take place.

\end{itemize}

\subsubsection{A BCFT\texorpdfstring{$_2$}{} perspective: transitions between end-of-the-world branes}\label{sec:ETW}

In section \ref{sec:3} we proved agreement between the BT and FMP results in absence of black holes. This subsection is meant to provide further supportive evidence of their matching from a holographic point of view, showing that transitions described via the BT formalisms provide an example of an AdS$_2$/CFT$_1 \subset\ $AdS$_3$/CFT$_2$. In doing so, we will be focussing on AdS$_2\rightarrow$ AdS$_2$ processes, showing that the total bounce can be mapped to the entanglement entropy on a BCFT$_2$.

As already explained in section \ref{sec:4}, in the BT setup, the starting point is analogous to that of a type-II Einstein-Maxwell-dilaton gravity theory in 2D, which was proved to arise from dimensional reduction of a BTZ black hole in \cite{AA}. 

In performing this comparison, the key quantity of interest is the boundary free energy, $F_{\partial}$, \cite{CHM}, which, for a BCFT$_2$ reads

\begin{equation}    
-F_{\partial} 
= 
S_{CFT_{_{2}}}+S_{bdy} 
= 
\frac{T}{\ \sqrt{1-T^2\ }\ } - \ln\left|\frac{\ 1-T\ }{\ 1+ T\ }\right| ,  
\label{eq:freeE}    
\end{equation}    
where $T$ is the tension of the ETW brane and is related to its opening angle with respect to the Cardy brane as $\Theta\overset{def.}{=}\tan^{-1} T$, as depicted on the LHS of figure \ref{fig:plot2f}. 

\begin{figure}[ht!]        
\begin{center}   
\includegraphics[scale=0.8]{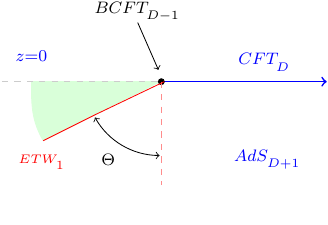}  
\ \ \ \ \ \ \ \ 
\includegraphics[scale=0.9]{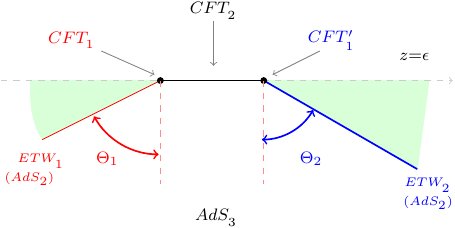}   
\caption{\footnotesize  The CFT$_1$s are dual to the AdS$_2$ spacetimes involved in the transition living on the two ETW branes and the wedges of opening angles $\Theta_{1,2}$. Their corresponding tensions, result in the value of $\Lambda_{\pm}$ on the two sides of the wall.}    
\label{fig:plot2ff}    
\end{center}    
\end{figure}

For type-1 instantons, \eqref{eq:sbdy} is nontrivial for both vacua, i.e. $T,\Theta\neq0$, and the total bounce can be re-expressed as follows:

\begin{equation} 
\boxed{\ \ \ B_{type-1} 
= 
2\pi\ S_{BCFT_2}=2\pi\ \left[S_{CFT_2}+S_{bdy}^{-}-S_{bdy}^{+}\right] \textcolor{white}{\Biggl [},\ \ }    
\label{eq:BThol}    
\end{equation}    
where

\begin{equation}    
\begin{aligned}
r_o^{-1}\overset{def.}{=}\bar\rho\ \ \ ,\ \ \  r_H\overset{def.}{=}\sqrt{\Lambda_{o,I}\ }\ \ \ ,\ \ \ 
T    
\overset{def.}{=}    
\sqrt{1-\frac{\ r_H^2\ }{r_o^2}\ }    
\ \ ,\ \ \            
S_{bdy}     =
- 2\ln\left|\frac{1-\sqrt{1-\Lambda\ \bar\rho^2\ }  }{\ \sqrt{\Lambda\ }\ \bar\rho\ }\right|.\nonumber  
\label{eq:sbdy}      
\end{aligned}  
\end{equation}

Equation \eqref{eq:BThol} indicates that either vacuum can be associated to a defect CFT$_1$ living at the endpoints of a bulk CFT$_2$. The overall system therefore defines a BCFT$_2$ as depicted on the RHS of figure \ref{fig:plot2ff}, with $S_{bdy}$ being evaluated on the lower-dimensional defect CFTs.  In particular, there are two different values of $T$, one for each vacuum. The mismatch in between the two gives rise to the holographic central charge, which counts the degrees of freedom of the ground state of the codimension-2 theory living on the boundary.

\subsubsection{An internal CFT\texorpdfstring{$_2$}{}  }\label{sec:2.31}

We now turn to the holographic interpretation of the results obtained in section \ref{sec:4} by means of the CDL method, further supporting the emergent analytic complementarity between the CDL and BT formalisms in 2D. In doing so, two main features will become manifest: 

\begin{enumerate} 

\item The entanglement along the dilatonic direction is effectively \emph{internal}. 

\item The CDL method (in 2D) describes vacuum decays in \emph{absence} of gravity.

\end{enumerate}

A CFT$_2$ is uniquely defined by its central charge and the conformal dimensions of the operators living in the theory, $(c,\Delta_i)$. Combined together, they define the energy spectrum of the states of a given CFT. 

From the fundamental parameters of the CFT$_2$, the entropy of an excited state is defined in terms of the \emph{Cardy formula},

\begin{equation}  
S_{_{Cardy}  }    
=      
2\pi \sqrt{ \frac{c}{6}\ \left(\Delta-\frac{c}{24}\right),\ } 
\label{eq:CardyS}    
\end{equation}    
where $\Delta$ and $\Delta_o=\frac{c}{24}$ denote the black hole and ground state energies, respectively. 


For the case of a CFT$_2$ where the degeneracy of the ground state is 1, the following definitions hold

\begin{equation}     
S_{CFT_2} 
= 
\frac{c}{3}\ \sqrt{\mu-1\ }, 
\ \ \  
\ \ \ 
E_o    
= 
\frac{c}{12}\ \mu, 
   \ \ \ \ \ \ 
c       
= 
6\frac{\ \partial S_{CFT}^2\ }{\partial E_o},
\label{eq:entr1}    
\end{equation}  
with $\mu\overset{def.}{=}\frac{\Delta}{c}$ being related to the black hole mass in the excited state. 
Focussing on the AdS$_2\rightarrow$ AdS$_2$ process, with associated extremised bounce that can be rewritten as follows

\begin{equation}   
\begin{aligned}
B_{tot}^{\ \text{AdS}_2\rightarrow \text{AdS}_{2}}    
&=      4\pi\ S_\pm^2 =
4\pi\ c_{hol}^{\ {^{CFT_2}}}\ M_{_{BH}},  
\label{eq:fc1}    
\end{aligned}   
\end{equation}

\begin{equation}    
c_{hol}^{\ ^{CFT_2}} 
\overset{def.}{=}         
\phi_-^2+2\phi_+^2 ,
\ \ \ \ \ \ 
M_{_{BH}}    
\overset{def.}{=}  
\frac{\phi_-^2}{\phi_+^2}-1,
\ \ \ \ \ \     
S_{\pm}
\overset{def.}{=}     
\sqrt{\phi_-^2+2\phi_+^2 \ }\ \sqrt{\frac{\phi_-^2}{\phi_+^2}-1\ },     
\label{eq:ep11}    
\end{equation}    
with $M_{_{BH}}  $ being a conserved charge along the flow and $c_{hol}^{\ ^{CFT_2}}$ its conjugate chemical potential. The last passage in \eqref{eq:fc1} follows from the relation between the entropy product $S_+S_-$ and the central charge.  Our result \eqref{eq:ep11} goes beyond the original formulation of $c$ by Cardy because it is associated to the defect lying in between the two vacua.

A key feature of \eqref{eq:fc1} is the fact that it defines an entropy product, and (apparently) not a difference in between entropies, as one might have expected to start with. The main reason for this is that the background configuration in CDL is providing a reference state, the \emph{ground state} of the CFT$_2$, with respect to which the newly-nucleated spacetime behaves as an excited state, namely an extremal black hole. Comparison with \eqref{eq:entr1} enables us to deduce that

\begin{equation}    
\boxed{\ \ \ \frac{\phi_-^2}{\phi_+^2}\overset{def.}{=}\mu,\textcolor{white}{\Biggl[}\ \ \ }    
\label{eq:boxed1}    
\end{equation}   
indicating that the newly nucleated vacuum is to be identified with an extremal black hole spacetime embedded in the background vacuum. 
Extremality follows from the fact that the mass of the BH is related to the cosmological constant itself. Indeed, from \eqref{eq:ep11}, $M_{_{BH}}$ is a function of $\Lambda_{\pm}$.

Relying upon arguments inspired by black hole microstate cosmology, \cite{MVR1}, a further remark is in order. We have shown that transitions of the kind AdS$_2\rightarrow$Mink$_2$ are forbidden. This is perfectly in agreement with the identification we have just made, namely \eqref{eq:boxed1}, since, from the CFT point of view, we can think of Mink$_2$ as being obtained from $\underset{c\rightarrow \infty}{\lim}$ of a CFT$_2$. The nucleation process would therefore correspond to the ground state of a CFT$_2$ insertion attempting to mimic an excited state of the background CFT$_2$. However, by definition, there is no horizon in the new vacuum that could attempt to screen the infinite number of degrees of freedom associated to a divergent central charge, and therefore the relative entropy \eqref{eq:entr1} cannot be defined. In conclusion, due to the extremality nature of the spacetime suggested by \eqref{eq:boxed1}, such transition is forbidden, since it corresponds to an overcounting of the microstates defining the entropy.

\section*{Final remarks}

To finish this section  we briefly collect some relevant points and mention the potential relation of our results  with other recent developments.

\begin{itemize} 

\item

In summary, in absence of black holes, the three methods lead to the following results in 2D

\begin{equation}                             
\begin{cases}            
B_{_{CDL/AP}}^{\ \ wo.kt}   
=    
2\pi\ S_{_{CFT_2}}^2  
=    
2\pi\ c_{hol}^{\ {^{CFT_2}}}\ M_{_{BH}}\\    \\    
S_{_{FMP}}^{^{\ {\cal B}=0}}=B_{_{BT} }   
= 
B_{_{CDL/AP}}^{\ \ w.kt}+M\bigg|^-_+   
=   
2\pi\ S_{_{BCFT_2}}
=2\pi\ \left[S_{CFT_2}+S_{bdy}^{-}-S_{bdy}^{+}\right]
\end{cases}     ,
\label{eq:BT1BT1}    
\end{equation}              
where $B_{_{CDL/AP}}^{\ \ wo.kt}$ and $B_{_{CDL/AP}}^{\ \ w.kt}$ denote the bounces calculated in absence and in presence of the kinetic term for $\phi$, respectively, whereas 

\begin{equation}   
S_{bdy}    
=    \ln\  <0|{\cal B}>\overset{def.}{=}    
\ln g= 
\ln\left|\frac{1+T}{1-T}\right|  ,
\end{equation}  
in turn highlighting the consistency and mutual complementarity of their holographic interpretations.

\item The identifications featuring in \eqref{eq:BT1BT1}, signal the emergence of an \emph{entropic hierarchy}, resulting in an interesting correspondence between our findings and those of \cite{EW2,EW3} within the realm of von Neumann algebras (VNAs)\footnote{See also \cite{Penedones:2023uqc} for an alternative perspective in the study of the Hilbert space of dS in arbitrary dimensions, mostly motivated by the bootstrap program. We thank the anonymous referee for bringing this article to our attention.}. In particular, the fact that the CDL calculation describes transitions in absence of gravity and the total bounce can be recast in the form of a relative entropy, is compatible with the fact that a type-III VNA only admits a relative entropy definition, rather than a von Neumann entropy. Turning on gravity, namely going beyond large-$N$, we have the BT and FMP results (in absence of black holes), therefore corresponding to type-II$_{\infty}$ VNAs. Upon adding a black hole, instead, which at the level of the Lagrangian density corresponds to turning on an irrelevant deformation, the UV theory is different with respect to that of the case with ${\cal B}=0$, leading us to conclude that the projection operator ensuring the reduction from type- II$_{\infty}$ to type-II$_1$ is mapped to the choice of the deformation in the FMP setup, hence to the choice of the black hole mass.

\item Vacuum transitions are defined in terms of amplitudes, and we have gathered supportive evidence throughout this work that such processes actually resemble a statement of mutual approximation of CFT states, \cite{MVR}. We emphasised that the emergence of the island can only be achieved in presence of non-extremal black holes, and that, because of this, upon taking the flat limit, the corresponding action remains finite and experiences a phase transition beyond a critical value of the black hole mass. Rephrasing this in the language \cite{Susskind,EW1}, the flat limit corresponds to the large-$N$ limit of the dual gravitational theory at hand. Inspired by the arguments of \cite{EW1}, regarding the importance of scale separation for reconciling connected amplitudes from disconnected boundaries (CADBs) with holography, we therefore claim that, below a critical value of the black hole mass, the flat limit leads to a divergent action due to the lack of scale separation between the cosmological constant and the black hole mass.

\end{itemize}


\section{Conclusions and Outlook of Part IV}

Part IV of the present work was devoted to understanding the subject of vacuum transitions in the simple setting of 2-dimensional gravity. Due to the peculiarities of the latter, the general analysis is not a straightforward extension of the 4D cases. Our main results are the calculation of transition amplitudes obtained following three known approaches, namely the two Euclidean methods of CDL and BT,  and the Lorentzian one of FMP. We presented explicit expressions for the transition rates in each method, and for the different signs of the vacuum energies, and took the first steps towards  understanding the results. Interestingly, in comparing the Euclidean and Lorentzian prescriptions, we find many similarities, as well as differences. 

Specifying to the case of (A)dS$_2\rightarrow$(A)dS$_2$ transitions, we derived the corresponding expressions for the total bounce or action in the AP setup following the CDL method in thick, \eqref{eq:wallnew1}, and thin-wall approximation, \eqref{eq:dstodstw}, the BT, \eqref{eq:AP111adsads}, and FMP method without black holes, \eqref{eq:joined1}. As emphasised in section \ref{sec:2}, the BT and FMP results agree under suitable parametric redefinition, thereby proving expected agreement in between the 2 methods. Such redefinition, though, is highly nontrivial, and can be recast to the different way in which the cosmological constants are being defined in the 2 setups, which in turn is due to the difference in between the Lagrangian densities defining the 2 theories. 

On the other hand, the result obtained by means of the CDL setup in the thin-wall approximation, \eqref{eq:dstodstw}, is equivalent to part of the BT result. In particular, all terms can be identified (under suitable redefinition), apart from the ones arising from the boundary terms associated to the ADM mass. As also argued in section \ref{sec:4}, these provide a nontrivial contribution to the total bounce upon turning on gravity. However, as a consequence of the particular nature of the 2D theory, the equation of motion for the AP setup (when removing the kinetic term for the dilaton from the Lagrangian density), imply $M_{_{ADM}}=0$. Therefore, 2D transitions described by means of the thin-wall approximated CDL method, should really be interpreted as describing nucleation processes in absence of gravity.

\begin{figure}[h!]    
\begin{center}    
\includegraphics[scale=0.9]{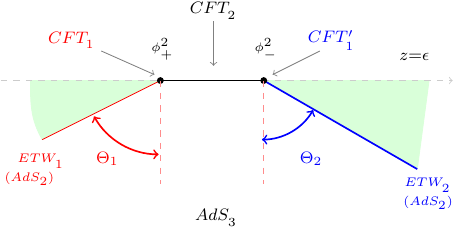}    
\caption{\small The 3 holographic descriptions are complementary to each other.}    
\label{fig:plot2f} 
\end{center}      
\end{figure}

Furthermore, the thick-wall analysis leads to an expression for $B_{_{TOT}}$ with unique features, in the sense that there seems to be no suitable reparametrisation enabling to recast it in any of the expressions obtained by means of other procedures. However, as argued in section \ref{sec:2}, under suitable parametric redefinitions, we can still provide a holographic interpretation of our result, which is complementarily in agreement with the others. 

We may summarise the holographic interpretation of our results as follows:

\begin{itemize}

\item  In section \ref{sec:4.4} we provide a possible holographic interpretation of the total bounces and actions calculated throughout our work, showing mutual compatibility and complementarity. Figure \ref{fig:plot2f} shows how these processes can be understood  as taking place in a KR/HM setup with two ETW branes and one \emph{internal} gravitating bath, geometrising the brane mediating the nucleation process. In particular, we find that the corresponding expression for the transition rate in presence of gravity, and in absence of black holes, is given by the difference of entropies of $T\bar T$-deformed CFTs, hence illustrating the \emph{locality} of the nucleation process. 

\item  Upon adding black holes of suitable mass, \eqref{eq:constrm}, instead, the total action can be expressed as the difference of generalised entropies, with an island emerging beyond a critical value of the black hole mass. Furthermore, under suitable parametric redefinition, the results obtained by means of the FMP method are found to agree with the expressions derived in \cite{MVR} for describing mutual approximation of boundary states belonging to different CFTs.  

\item We propose that the vacuum transitions in the bulk correspond to a deconfinement/confinement transition on the CFT side although further study needs to be made towards this correspondence.

\item   The BT results, which were shown to be equivalent to those obtained through the FMP method in absence of black holes, can be expressed in terms of entropies of BCFT$_2$'s with 2 nontrivial boundary conditions dual to ETW branes. Furthermore, the CDL result can be recast in the form of an entropy product of a CFT$_2$ thereby showing agreement with the expectations following from the analytic behaviour encountered in section \ref{sec:4}. 

\item  One of our  results is that the total action (or bounce) associated to the decay process carries an \emph{internal} entropic interpretation. In particular, for the BT and FMP cases, they can always be expressed as the difference of generalised entropies. Only the latter, however, provides the right setup for an island to emerge. In particular, the black hole mass, which in 2D is an explicit parameter of the theory, is, in a way, playing the role of time in the Hawking evaporation process, and is responsible for $S_{_{brane}}$ undergoing a phase transition, as explained in section \ref{sec:5}. 


\end{itemize}

Understanding the behaviour of the same physical process by means of different formalisms is equally important in all fields of research, and appears to be particularly promising within the context of quantum gravity. Recent progress towards understanding the behaviour of the partition function for QG within, either, the canonical or microcanonical ensemble, \cite{Marolf:2022jra, Marolf:2022ntb}, as well as the study of the phases of gauge theories and their holographic duals, \cite{Dias:2022eyq}, are certainly among the most promising directions deserving more investigation. Clearly our work leaves many open questions. We hope that our results  will be useful for addressing further questions related to vacuum transitions, early universe cosmology, and holography.

Understanding the behaviour of the same physical process by means of different formalisms is equally important in all fields of research, and appears to be particularly promising within the context of quantum gravity. Recent progress towards understanding the behaviour of the partition function for QG within, either, the canonical or microcanonical ensemble, \cite{Marolf:2022jra, Marolf:2022ntb}, as well as the study of the phases of gauge theories and their holographic duals, \cite{Dias:2022eyq}, are certainly among the most promising directions deserving more investigation. We hope the results found through this investigation will be useful for addressing further questions related to vacuum transitions, early universe cosmology, and holography.

\part{Beyond Cetegorical Dualisability}  \label{sec:V}

\section{Summary and Motivations of Part V}

\subsection{Motivations}

For a theory, ${\cal T}$, to be absolute, the following triple needs to be defined

\begin{equation}   
\boxed{\ \ \ {\cal T}\ \longleftrightarrow\ ({\cal F}, \mu, \mathfrak{Z})\color{white}\bigg]\  },    
\label{eq:HB1}  
\end{equation}     
where ${\cal F}$ is the fiber functor, $\mu$ the moment map, and $\mathfrak{Z}$ the Drinfeld center\footnote{We refer to section \ref{sec:symtft} for a detailed explanation of this terminology.}. Consistency of the underlying mathematical structure requires these three quantities to be mutually related. Indeed, upon defining any one of them, the other two should automatically follow. The purpose of \cite{Pasquarella:2023deo,Pasquarella:2023exd} and the present work is to show that an apparent shortcoming in defining such triple corresponds to the emergence of interesting physics, rather than being a fault of the sought after absolute theory. In particular, we will show that this can be used to explain the emergence of non-invertible symmetries separating different class ${\cal S}$ theories, \cite{Pasquarella:2023deo,Bashmakov:2022jtl,Bashmakov:2022uek,9,Kaidi:2021xfk,Choi:2022zal,Choi:2021kmx}. 


The crucial references we rely upon are the works of Moore and Segal, \cite{Moore:2006dw}, and Moore and Tachikawa, \cite{Moore:2011ee}. As briefly reviewed in the following sections, \cite{Moore:2011ee} proposes a redefinition of class ${\cal S}$ theories (cf. figure \ref{fig:classS}) in terms of a 2D TFT, namely the functor

\begin{equation}   
\boxed{\ \ \ \eta_{_{G_{\mathbb{C}}}}:\ \text{Bo}_{_2}\ \rightarrow\ \text{HS} \color{white}\bigg]\ \ \ }   
\label{eq:etaGC11}
\end{equation} 
with Bo$_{_2}$ and HS denoting the bordism 2-category and the holomorphic symplectic 2-category, respectively\footnote{For more details, we refer the reader to section \ref{sec:BMTCs}.}, associated to a given 4D ${\cal N}=2$ SCFT. The definition of \eqref{eq:etaGC11} strongly relies upon assuming, both, the source and target categories, enjoy a duality structure which, in turn follows from the presence of an identity element in both categories. In \cite{Moore:2011ee}, the authors show that, under the duality assumption, for the categories in \eqref{eq:etaGC11} to be well-defined, it is enough to specify their objects and 1-morphisms. Essentially, the objects of Bo$_{_2}$ are circles, $S^{^1}$, and the 1-morphisms are cobordisms between different disjoint unions of circles and the empty set. Their respective counterpart on the holomorphic symplectic side correspond to the gauge group, \cite{Moore:2011ee},

\begin{equation}    
\eta_{_{G_{_{\mathbb{C}}}}}\ \left(S^{^1}\right)\ \overset{def.}{=}\ G_{_{\mathbb{C}}},  
\label{eq:gaugegroup}   
\end{equation}  
and the cobordism operators, \cite{Moore:2011ee},
\begin{equation}    
\eta_{_{G_{_{\mathbb{C}}}}}\ \left(\text{Hom}\ \left(S^{^1}, \emptyset\right)\right)\ \overset{def.}{=}\ U_{_{G_{_{\mathbb{C}}}}}
\label{eq:UG1} 
\end{equation}  
\begin{equation}
\eta_{_{G_{_{\mathbb{C}}}}}\ \left(\text{Hom}\ \left(S^{^1}\ \sqcup\ S^{^1}, \emptyset\right)\right)\ \overset{def.}{=}\ V_{_{G_{_{\mathbb{C}}}}}  
\label{eq:VG1} 
\end{equation}  

\begin{equation} 
\eta_{_{G_{_{\mathbb{C}}}}}\ \left(\text{Hom}\ \left(S^{^1}\ \sqcup\ S^{^1}\ \sqcup \ S^{^1},\emptyset\right)\right)\ \overset{def.}{=}\ W_{_{G_{_{\mathbb{C}}}}}.    
\label{eq:HBCS}   
\end{equation}  
respectively.

Importantly for us, the Moore-Tachikawa varieties described by \eqref{eq:gaugegroup}, \eqref{eq:UG1}, \eqref{eq:VG1}, and \eqref{eq:HBCS} constitute the quiver gauge theory\footnote{Among the cobordism operators is the Higgs branch of class ${\cal S}$ theories, \eqref{eq:HBCS}.} realisation of \cite{Moore:2006dw}, where Moore and Segal propose the mathematical formalism needed for addressing the following question: given a certain closed string theory background, what is its corresponding D-brane content\footnote{Note that this is essentially the same question addressed in \cite{Aharony:2013hda}.}?

Our aim is that of explaining how and why one needs to generalise the construction of \cite{Moore:2011ee} from a higher-categorical point of view, and what its implications are on the theoretical physics side. In doing so, we highlight the crucial properties and axioms satisfied by the cobordism operators outlined in \cite{Moore:2006dw,Moore:2011ee}, and how they can be generalised to account for more general setups from, both, the mathematical and theoretical physics perspectives. 

\begin{figure}[ht!]   
\begin{center}  
\includegraphics[scale=0.75]{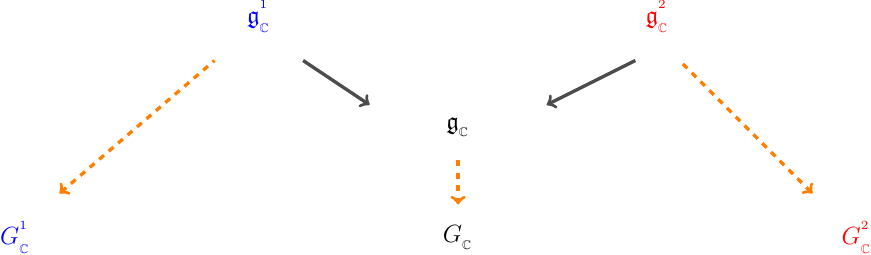}   
\caption{\small Adaptation of a correspondence first proposed in \cite{Pasquarella:2023deo} playing a key role towards generalising \cite{Moore:2011ee} to the hyperk$\ddot{\text{a}}$hler target category case. As explained in the following sections, this also requires the generalisation of cobordism operators, \cite{Moore:2006dw}, due to the lack of reparametrisation-invariance on the Riemann surface on which the compactification of the 6D ${\cal N}=(2,0)$ SCFT is performed.}    
\label{fig:correspondence}    
\end{center}  
\end{figure}

In particular, we emphasise the dependence of \eqref{eq:etaGC11} on the conformal structure of the Riemann surface on which the compactification of the 6D ${\cal N}=(2,0)$ SCFT has been performed to achieve a certain class ${\cal S}$ theory and how lack of reparametrisation invariance, corresponding to the absence of the identity element in its source and target categories, \cite{Moore:2011ee}, signals the presence of (non-invertible) categorical symmetries separating different absolute theories. 

At the heart of this is the correspondence sketched in figure \ref{fig:correspondence}.

In \cite{Pasquarella:2023deo} we explained how gauging a Symmetry Topological Field Theory (SymTFT) enables to change the boundary conditions of the fields living in the absolute field theory resulting from the Freed-Moore-Teleman construction, \cite{Freed:2012bs,Freed:2022qnc,TJF,Kong:2013aya}. To each gauging corresponds a choice of triples, \eqref{eq:etaGC11}, and, for any absolute theory, it is enough to define one of the three entries on the RHS of \eqref{eq:etaGC11} to determine the other two. For the purpose of this article, we will mostly focus on the second, namely the moment  map, defined as follows

\begin{equation} 
\mu:\ \mathfrak{G}\ \rightarrow\ {\cal A},      
\end{equation} 
where $\mathfrak{G}$ is an n-categorical structure, and ${\cal A}$ is the algebra of invertible topological defects associated to the action $\mu\ (\mathfrak{G})$. Gauging means taking the categorical quotient with respect to $\mathfrak{G}$ (in notation $//_{_{\mu}}\ \mathfrak{G}$), and projecting its image under $\mu$ to the identity\footnote{We thank Nathan Seiberg for instructive discussion regarding the appropriateness of the terminology to be adopted in describing this formalism.}. Practically, one could perform a total gauging of the theory by choosing ${\cal A}$ such that the overall spectrum of the theory of the gauged theory is only the (new) identity element. For the purpose of our work, instead, we are interested in understanding mathematical structures arising by gauging with respect to different subalgebras within ${\cal A}$, that are mutually intersecting, albeit not contained within each other. The ultimate aim is that of explaining the emergence of (non-invertible) categorical symmetries in certain supersymmetric quiver gauge theories once described in terms of Coulomb branches of magnetic quivers of 3D ${\cal N}=4$ quiver gauge theories, which is the main focus of an upcoming paper by the same author, \cite{VP}. In such analysis we will be applying some of the findings of \cite{Braverman:2017ofm,Benini:2010uu}.

In \cite{Pasquarella:2023exd}, we argued in favour of the following statements:  

\begin{itemize}  

\item Lack of reparametrisation-invariance of bordism operators signals the presence of intrinsic non-invertible defects separating different class ${\cal S}$ theories.

\item Defining the Drinfeld center for a system of composite class ${\cal S}$ theories separated by non-invertible defects constitutes a nontrivial generalisation of an S-duality statement.

\end{itemize}

Mostly relying upon \cite{Moore:2006dw}, we highlighted the importance of the construction of cobordism operators, emphasising their dependence on the conformal structure of the Riemann surface. We then turned to the discussion of a particular 2D TFT valued in a symmetric monoidal category, namely the maximal dimension Higgs branch of class ${\cal S}$ theories. After briefly reviewing the properties outlined in \cite{Moore:2011ee}, we proposed their generalisation for the case in which the target category of the $\eta_{_{G_{_{\mathbb{C}}}}}$ functor is a hyperk$\ddot{\text{a}}$hler quotient. We concluded outlining the possible extension of this treatment towards a mathematical formulation of magnetic quivers within the context of Coulomb branches of 3D ${\cal N}=4$ quiver gauge theories.

In this new work, we take a more algebro-geometric approach in explaining the mathematical structure underlying supersymmetric quiver gauge theories, with a particular focus on dualities and their mutual relations in terms of higher-categories. 

As will be explained in depth, the crucial role is played by factorisation homology. We will not attempt to provide an exhaustive definition here, and refer the reader to the related section in the core of this work, instead. What we can anticipate, though, is that factorisation homology is related with the counting of ground state degeneracy, and, when applied to the AGT correspondence, provides a tool for evaluating the 4-sphere partition function of class ${\cal S}$ theories. Specifically, as we shall see, factorisation homology allows us to relate categorical dualisability and full-extendibility\footnote{And is therefore in line with the connection between dualities and categorical structures.}. Within the context of non-Lagrangian theories, such as 6D ${\cal N}=(2,0)$ SCFTs, this is particularly relevant since it introduces the notion of ring objects, of which Hilbert series are one of the examples of most interest to us.

As an algebraic variety, we focus on the calculation of the Hilbert series on Coulomb and Higgs branches, with the latter being related by the geometric invariant theory (GIT) construction, \cite{Deligne-Mumford}. GIT being a recipe for realising 3D mirror symmetry, \cite{Intriligator:1996ex}, we show how it can be suitably extended for cases where mirror duals are not necessarily in a 1-to-1 correspondence. In relation to Part II, this nicely fits with the magnetic quiver prescription for describing the Higgs branch of certain supersymmetric quiver gauge theories with non-primitive ideals\footnote{Namely the 2-cone structure, \cite{Ferlito:2016grh}.}. 

We then show how the 2D description of disk algebras can be suitably adopted for describing Hitchin systems associated to joint class ${\cal S}$ theories.  We conclude outlining future directions that are currently under investigation by the same author.

Following our treatment in \cite{Pasquarella:2023exd} we explain the higher categorical structures that are needed for describing the invariants associated to specific supersymmetric quiver gauge theories admitting a magnetic quiver description.  

The key tool we will be using in achieving such aim is factorisation homology, of which a more detailed definition will be provided in the core of the paper. For the purpose of the introduction, we might just say that its role is that of assigning invariants to topological manifolds.




\subsection{Outline and main results}

Part V of this work is structured as follows: 

\begin{enumerate}

\item section \ref{sec:cd} introduces factorisation homology and its applicability within the context of class ${\cal S}$ theories by means of the AGT correspndence. The section ends recapitulating the interplay between dualisability and full-extendibility of TQFTs, with a particular focus of Moore-Tachikawa varieties\footnote{In section \ref{sec:clt}, we briefly overviewed cochain level theories as the most important generalisation of the open and closed TFT construction. Mostly relying upon \cite{Moore:2006dw}, we highlighted the importance of the construction of the cobordism operator, highlighting their dependence on the conformal structure of the Riemann surface. We then turn to the discussion of a particular 2D TFT valued in a symmetric monoidal category, namely the maximal dimension Higgs branch of class ${\cal S}$ theories.}.

 \item Section \ref{sec:333} explains the relation in between the Drinfeld center and magnetic quivers, combining sections \ref{sec:clt} and \ref{sec:symtft}.

\item In section \ref{sec:last} we propose the generalisation of Moore-Tachikawa varieties for the case in which the target category of the $\eta_{_{G_{_{\mathbb{C}}}}}$ functor is a hyperk$\ddot{\text{a}}$hler quotient. We conclude outlining the possible extension of this treatment towards a mathematical formulation of magnetic quivers within the context of Coulomb branches of 3D ${\cal N}=4$ quiver gauge theories which will be addressed in an upcoming work by the same author, \cite{VP}. The concluding part of this section recapitulates the essential tools required for performing GIT quotient constructions, highlighting its importance for realising 3D mirror symmetry. In particular, we focus on the calculation of the Hilbert series as specific examples of algebraic varieties of crucial interest in the study of categorical dualities for supersymmetric quiver gauge theories. We conclude showing how Coulomb and Higgs branches from class ${\cal S}$ Hitchin systems carry the same information as the gaugeable algebras leading to SymTFT constructions, discussed in Part III. We conclude outlining interesting open questions and future directions of investigation by the author.

\end{enumerate}

\section{Categorical Dualisability and full extendibility}   \label{sec:cd}
The present section is structured as follows: 

\begin{enumerate}

\item Section \ref{sec:DCFMQS} is devoted to explaining the role played by factorisation homology in describing the invariants of a given TFT. In particular, this prepares the stage for the following sections, where we will be dealing with the applicability of this formalism to the specific case of class ${\cal S}$ theories\footnote{Note the importance of having introduced the notion of contractible cells in section \ref{sec:rqft} in relation to abelianisation and equivariant cohomology.}.

\item In section \ref{sec:334}, we then turn to explaining the relation between categorical dualisability and full-extendibility. In particular, this nicely relates to the notion of Drinfeld centers and gauging categorical structures.

\end{enumerate}

\subsection{Factorisation Homology in context}   \label{sec:DCFMQS}  

The present section is mostly revision of well-known literature\footnote{Specifically, we suggest the interested reader to consult the following reference \cite{Moore:2006dw}.}. It provides a brief mathematical digression, needed for introducing factorisation homology, and its utility for evaluating the index of a 2D TQFT. In doing so, we explain the notions of Hochschild cohomology, stratified spaces, and disk algebras. In the following sections we will see that, combining this formalism to algebro-geometric techniques, leads to interesting new findings.


\subsection*{Cohomology and Homotopy Equivalence}  

In homology theory and algebraic topology, cohomology is a general term for a sequence of abelian groups, usually one associated with a topological space, often defined from a cochain\footnote{Cochains are functions on the group of chains in the homology theory.} complex. Cohomology can be viewed as a way of assigning richer algebraic invariants to a space with respect to homology. 

Some versions of cohomology arise by dualising the homological construction. Some of their properties are formulated in a similar fashion with respect to their homological counterpart. However, cohomology enjoys a richer structure, which is lacking in homology, which we now turn to explain. 

For a given topological space $X$, and a commutative ring, $R$, the cup product is a bilinear map 

\begin{equation}  
H^{^i}(X,R)\ \times \ H^{^j}(X,R)\ \rightarrow\ H^{^{i+j}}(X,R).      
\end{equation}  

This product defines the cohomology ring of $X$ as follows

\begin{equation}  
H^{^{\bullet}}(X,R)\ \equiv\ \bigoplus_{_i} \ H^{^j}(X,R).       
\end{equation}  

It is graded commutative in the sense that, given $u\ \in\ H^{^i}(X,R)$, $v\ \in\ H^{^j}(X,R)$, 

\begin{equation} 
uv\ =\ (-1)^{^{ij}}\ vu.  
\end{equation}

It follows that, if two spaces are homotopy equivalent, their cohomology rings are isomorphic.   

Homotopy equivalence can be expressed in terms of looping by representation spheres. In particular, a homotopy equivalence between topological spaces $X$ and $Y$ is a morphism $f:X\rightarrow Y$ which has a homotopy inverse, hence such that there exists a morphism $g: Y\ \rightarrow X$ and homotopies 

\begin{equation}  
g\ \circ\ f\ \sim\ \mathbf{1}_{_X}\ \ \ \text{and} \ \ \ f\ \circ\ g\ \sim\ \mathbf{1}_{_Y},   
\end{equation}
thereby making homotopy equivalence very similar to the notion of categorical equivalence.

For a given topological space, $X$, elements of $H^{^i}(X)$ can be thought of as represented by codimension-$i$ subspaces of $X$ that can move freely on $X$. For the case at hand, an element of $H^{^i}(X)$ is defined by a continuous map 

\begin{equation}  
f:\ X\ \rightarrow\ M,         
\end{equation}
with pullback 
\begin{equation}  
f^*:\ H^{{\bullet}}(M,R)\ \rightarrow\ H^{\bullet}(X,R),         
\end{equation}
and a closed codimension-$i$ submanifold $N$ of $M$ with an orientation in the normal bundle. As such, the class $f^*([N])\ \in H^{^i}(X)$ can be thought of to lie in the subspace $f^{^{-1}}(N)$ of $X$, since $f^*([N])$ restricts to $0$ in the cohomology of the open subset $X-f^{^{-1}}(N)$. The cohomology class $f^*([N])$ is free to move in $X$ in the sense that $N$ could be replaced by any continuous deformation of $N$ inside $M$.

\subsubsection{Hochschild cohomology and cochain complexes}   

Given a finite group action on a suitably enhanced triangulated category linear over a field, \cite{Perry}, Hochschild cohomology and group actions establish a formula for the Hochschild cohomology of the category of invariants, assuming the order of the group is coprime to the characteristic of the base field.

For a $k$-linear category with $G$-action,

\begin{equation} 
\underset{g\in G}{\bigoplus}\ \phi_{_g}:\ {\cal C}\ \rightarrow\ {\cal C} 
\end{equation}
induces the norm functor 

\begin{equation} 
\begin{aligned}
&N_{_m}: \ {\cal C}_{_G}\rightarrow{\cal C}^{G} \\
&\\
&\ \ \ \ \ \ \ \ x\ \mapsto\ \underset{g\in G}{\sum}\ g(x)       
\end{aligned}
\end{equation}
characterised by the existence of a factorisation property

\begin{equation}  
N_{_m}\circ q: \ {\cal C}\rightarrow {\cal C}^{^G}.  
\end{equation}    

Importantly, $N_{_m}$ is an equivalence.  

If ${\cal C, D}$ are $k$-linear categories, then the $k$-linear exact functors from ${\cal C}$ to ${\cal D}$ form the objects of the $k$-linear category Fun$_{_k}({\cal C},{\cal D})$. 

Let ${\cal C}$ be a $k$-linear category, and let $\phi:\ {\cal C}\rightarrow{\cal C}$ be an endofunctor. The Hochschild cochain complex of ${\cal C}$ with coefficients in $\phi$ is defined as 

\begin{equation}  
HC^{^{\bullet}}({\cal C},\phi)\ \overset{def.}{=}\ \text{Map}_{_{\text{Fun}_{_k}({\cal C,C})}}(\mathbf{1}_{_{\cal C}},\phi)\ \in\ \text{Vect}_{_k}.  
\end{equation}  

The Hochschild cohomology $HH^{^{\bullet}}({\cal C},\phi)$ of ${\cal C}$ with coefficients in $\phi$ is the cohomology of this complex. For the case in which $\phi\equiv\mathbf{1}_{_{\cal C}}$, 

\begin{equation}   
HC^{^{\bullet}}({\cal C})\ \overset{def.}{=}\ HC^{^{\bullet}}({\cal C},\mathbf{1}_{_{\cal C}}) 
\ \ \ ,\ \ \ 
HH^{^{\bullet}}({\cal C})\ \overset{def.}{=}\ HH^{^{\bullet}}({\cal C},\mathbf{1}_{_{\cal C}}) 
\end{equation}   
define the Hochschild cochain complex and the Hochschild cohomology of ${\cal C}$, respectively. The latter comes equipped with a natural algebra structure, induced by composition in $\text{Map}_{_{\text{Fun}_{_k}({\cal C,C})}}(\mathbf{1}_{_{\cal C}},\mathbf{1}_{_{\cal C}})$.

Note that, given a $k$-linear category $\cal C$ with an action by a finite group $G$, there is an induced action of $G\times G$ on $\text{Fun}_{_k}({\cal C,C})$ consisting in the following. An element $(g_{_1},g_{_2})\ \in\ G\times G$ acts on $\text{Fun}_{_k}({\cal C,C})$ by sending $F:\ {\cal C}\rightarrow {\cal C}\ \in\ \text{Fun}_{_k}({\cal C,C})$ to $\phi_{_{g_{_2}}}\circ F\circ \phi^{^{-1}}_{_{g_{_1}}}$. Via the diagonal embedding $G\ \subset\ G\times G$, this restricts to the conjugation action of $G$ on $\text{Fun}_{_k}({\cal C,C})$.   

A main theorem in \cite{Perry} states that, given $\cal C$ a $k$-linear category with an action by a finite group $G$, and assuming the order of $G$ is coprime to the characteristic of $K$, then there is an isomorphism 

\begin{equation}   
HH^{^{\bullet}}({\cal C}^{^G})\ \simeq\ \left(\underset{g\in G}{\bigoplus}\ HH^{^{\bullet}}({\cal C},\phi_{_g})\right)^{^G}, 
\end{equation}
with $\phi_{_g}: {\cal C}\rightarrow {\cal C}$ an autoequivalence corresponding to $g\in G$, and the $G$-action on the right side induced by the conjugation action of $G$ on $\text{Fun}_{_k}({\cal C,C})$. In particular, 

\begin{equation}   
HH^{^{\bullet}}({\cal C})^{^G}\ \hookrightarrow\ HH^{^{\bullet}}({\cal C}^{^G}).
\end{equation}

We will come back to this in section \ref{sec:last} when discussing the GIT quotient and the Kirwan map.

\subsubsection{Factorisation homology of stratified spaces} \label{sec:fhss}


Factorisation homology satisfies a defining property called $\otimes$-excision, which determines $\int_{_M}{\cal A}$ up to equivalence,

\begin{equation} 
\boxed{\ \ \ \int_{_M}{\cal A}\ \simeq\ (H,u_{_{\Sigma}})  \color{white}\bigg]\  \ },   
\label{eq:facthom1111}    
\end{equation}
with such property being a special case of the pushforward property. The RHS of \eqref{eq:facthom1111} consists of the Hilbert space of the TO, $H$, and the ground state degeneracy, $u_{_{\Sigma}}$. \eqref{eq:facthom1111} allows to reduce the computation of the factorisation homology of a surface to that of a lower-dimensional manifold, and, eventually, to that of a 0-dimensional manifold. 

The only known global observable on $\Sigma$ for an anomaly-free TO is the ground state degeneracy. The theory of factorisation homology gives a powerful tool for calculating it. When a closed stratified surface is decorated by anomaly-free topological defects of codimension 0,1,2, the theory of factorisation homology gives us a powerful tool for calculating the degeneracy of the ground state. As an important bi-product, FH provides a powerful tool for calculating global observables on surfaces for anomalous TOs. Indeed, relation between FH and TOs enables to make direct contact with the theoretical physics setup of interest to us.

There is a well-defined fully faithful functor 

\begin{equation}  
\mathfrak{Z}:\ \text{UMTC}^{^{\text{ind}}}\ \rightarrow\ \text{UMTC}  
\label{eq:drinf}
\end{equation} 
acting on objects and morphisms as follows

\begin{equation} 
{\cal C}\ \rightarrow\ \mathfrak{Z}({\cal C})\ \ \ ,\ \ \ _{_{{\cal C}}}{\cal M}_{_{{\cal D}}}\ \mapsto\ \mathfrak{Z}({\cal M})\ \overset{def.}{=}\ \text{Fun}_{_{{\cal C}|{\cal D}}}({\cal M},{\cal M}).     
\end{equation}  

For ${\cal C}$ a UMTC, \eqref{eq:drinf} defines the Drinfeld center of ${\cal C}$, $\mathfrak{Z}({\cal C})$, a canonical braided monoidal category such that 

\begin{equation}   
\mathfrak{Z}({\cal C})\ \simeq\ {\cal C}\ \boxtimes\ \overline{\cal C}.    
\end{equation}

More explicitly, Th.2.5 in \cite{Kong:2013aya} implies that, given ${\cal C}, {\cal D}, {\cal E}$ UMFCs, ${\cal M}$ a multi-fusion ${\cal C}-{\cal D}$-bimodule, and ${\cal N}$ a multi-fusion ${\cal D}-{\cal E}$-bimodule, the assignment 

\begin{equation}  
f\boxtimes_{_{{\mathfrak{Z}({\cal D})}}}g\ \mapsto\ f\boxtimes_{_{\cal D}}g    
\end{equation}   
defines an equivalence between two multi-fusion $\mathfrak{Z}({\cal C})-\mathfrak{Z}({\cal E})$-bimodules:  

\begin{equation}  
\text{Fun}_{_{{\cal C|D}}}({\cal M},{\cal M})\boxtimes_{_{{\mathfrak{Z}({\cal D})}}}\text{Fun}_{_{{\cal D|E}}}({\cal N},{\cal N})\ \simeq\ \text{Fun}_{_{{\cal C|E}}}({\cal M}\boxtimes_{_{{\cal D}}}{\cal N},{\cal M}\boxtimes_{_{{\cal D}}}{\cal N})  
\end{equation}
from which the following two corollaries follow:

\begin{equation}  
{\cal C}\boxtimes_{_{{\mathfrak{Z}({\cal C})}}}{\cal D}^{^{rev}}\ \simeq\ \text{Fun}_{_{H}}({\cal M},{\cal M}) 
\end{equation}  

\begin{equation}  
\boxtimes_{_{({\cal C}_{_{0}},...,{\cal C}_{_{n-1}})}}^{^{\circlearrowright}}({\cal M}_{_{1}},...,{\cal M}_{_{n}})\ \simeq\ \text{Fun}_{_{H}}(\cal P,P), 
\end{equation}  
with ${\cal P}$ a unique unitary category such that this correspondence holds.

Next, consider Mfld$_{_{n}}^{^{or}}$ to be the topological category whose objects are oriented n-manifolds whose objects are oriented n-manifolds without boundary. For any two oriented manifolds $M,N$, the morphism Hom$_{_{\text{Mfld}_{_{n}}^{^{or}}}}(M,N)$ is the space of all orientation-preserving embeddings $e: M\rightarrow N$ endowed with the compact-open topology. 

Denoting with ${\cal M}\text{fld}_{_{n}}^{^{or}}$ the symmetric monoidal $\infty$-category associated to the topological category Mfld$_{_{n}}^{^{or}}$, the symmetric monoidal $\infty$-category Disk$_{_{n}}^{^{or}}$, is the full subcategory of ${\cal M}\text{fld}_{_{n}}^{^{or}}$ whose objects are disjoint union of finitely-many n-dimensional Euclidean spaces equipped with standard orientation. 

If ${\cal V}$ is a symmetric monoidal $\infty$-category, $M$ an oriented manifold, and 

\begin{equation}  
\boxed{\ \ \ {\cal A}:\ \text{Disk}_{_{n}}^{^{or}}\ \rightarrow\ {\cal V} \color{white}\bigg]\ \ }   
\end{equation}  
an n-disk algebra in ${\cal V}$, the factorisation homology of $M$ with coefficients in ${\cal A}$ is an object of ${\cal V}$ given by  

\begin{equation}  
\int_{_{M}}\ {\cal A}\ \overset{def.}{=}\ \text{Colim.}\left(\left(\text{Disk}_{_{n}}^{^{or}}\right)_{_{/M}}\ \rightarrow\ \text{Disk}_{_{n}}^{^{or}}\ \xrightarrow{\cal A}\ {\cal A}\right),  
\end{equation}   
with $\left(\text{Disk}_{_{n}}^{^{or}}\right)_{_{/M}}$ denoting the category of $n$-disks embedded in $M$. If $\mathbf{1}_{_{n}}$ is the trivial $n$-disk albegra assigning to each $\mathbf{R}^{^n}$ $\mathbf{1}_{_{{\cal V}}}\ \in\ {\cal V}$, it follows that 

\begin{equation}  
\int_{_{M}}\mathbf{1}_{_{n}}\ \simeq\ \mathbf{1}_{_{{\cal V}}}.
\end{equation}

The canonical morphism $\mathbf{1}_{_{n}}\rightarrow{\cal A}$ then induces a morphism $\mathbf{1}_{_{{\cal V}}}\rightarrow\int_{_{M}}{\cal A}$, implying the factorisation homology $\int_{_{M}}{\cal A}$ is not just an object of ${\cal V}$, but also comes equipped with a 0-disk algebra structure, the latter playing a crucial role for ground state degeneracy calculations. 

Explicit calculation of factorisation homology as a 0-disk algebra is very useful. Indeed, factorisation homology on a closed stratified surface $\Sigma$ with anomaly-free coefficient ${\cal A}$, $(H,u_{_{\Sigma}})$, exactly gives the ground state degeneracy. of the same surface decorated by anomaly-free topological defects of codimension 0, 1, and 2 that are associated to ${\cal A}$, namely $u_{_{\Sigma}}$.  

\subsection*{Examples of stratified factorisation homology}

Given a stratified 2-disk, $M_{_1}$, with ${\cal A}$ a 2-disk algebra assigned to 2-cells, 1-disk algebras associated to 1-cells, and ${\cal P}$ as the unique 0-algebra assigned to the unique 0-cell, examples of its factorisation homology read as follows

\begin{equation}   
\int_{_{M_{_1}}}{\cal A}_{_1}\ \simeq\ {\cal A}\ \ \ ,\ \ \ \int_{_{M_{_1}\diagdown \{0\}}}{\cal A}_{_1}\ \simeq\ HH_{_{\bullet}}({\cal A})\ \overset{def.}{\equiv}\ {\cal A}\otimes_{_{{\cal A}\otimes{\cal A}^{^{op}}}}{\cal A},
\end{equation}
with $HH_{_{\bullet}}({\cal A})$ denoting the Hochschild cohomology.   

For anomaly-free coefficient systems,

\begin{figure}[ht!]  
\begin{center}    
\includegraphics[scale=1]{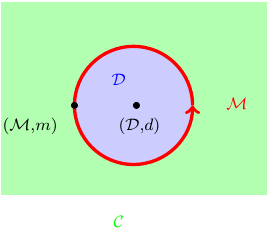} 
\caption{\small A stratified 2-disk with 2-cells labelled by UMTCs ${\cal C}$ and ${\cal D}$. The unique 1-cell is labelled by a closed multi-fusion ${\cal C}-{\cal D}$-bimodule, ${\cal M}$. The two 2-cells are labelled by $({\cal D},d), ({\cal M},m)$, for $d\in{\cal D}$ and $m\in{\cal M}$, respectively.}    
\label{fig:FH}  
\end{center}  
\end{figure}

\begin{equation}   
\boxtimes_{_{({\cal C}_{_{0}},...,{\cal C}_{_{n-1}})}}^{^{\circlearrowright}}({\cal M}_{_{1}},...,{\cal M}_{_{n}})\ \simeq\ \text{Fun}_{_{H}}(\cal P,P).
\end{equation}

The unique 1-cell labelled by a closed multi-fusion ${\cal D}-{\cal C}$-bimodule ${\cal M}$. The two 0-cells are labelled by $({\cal D}, d)$ and $({\cal M},m)$, for $d\in{\cal D}$, and $m\in{\cal M}$, respectively. The closed multi-fusion ${\cal C-D}$-bimodule structure on ${\cal M}$ induces two monoidal functors 
\begin{equation}  
{\cal C}\ \xrightarrow{L}\ {\cal M}\ \xleftarrow{R}\ {\cal D}.  
\end{equation}

Given $(\mathbb{R}^{^2};\Gamma_{_0})$ a stratified 2-disk, the FH reads as follows 

\begin{equation}  
\int_{_{(\mathbb{R}^{^2};\Gamma_{_0})}}({\cal C},{\cal D};{\cal M};({\cal D},d),({\cal M},m))\ \simeq\ ({\cal C},L^{^{\text{V}}}(m\otimes R(d))),  
\label{eq:FH1}  
\end{equation}  
with $L$ the right-adjoint functor of $L$. If ${\cal D}\equiv{\cal C}^{^0}_{_{{\cal A}}}, {\cal M}\equiv{\cal C}_{_{{\cal A}}}$ for a connected commutative separable algebra ${\cal A}$ in ${\cal C}$, $d\equiv\mathbf{1}_{_{{\cal D}}}$, and $m\equiv\mathbf{1}_{_{\cal M}}$, then \eqref{eq:FH1} reduces to   

\begin{equation}  
\int_{_{(\mathbb{R}^{^2};\Gamma_{_0})}}\left({\cal C},{\cal C}^{^0}_{_{{\cal A}}};{\cal C}_{_{{\cal A}}}^{^0};\left({\cal C}_{_{{\cal A}}}^{^0},d\right),\left({\cal C}_{_{{\cal A}}},\mathbf{1}_{_{{\cal C}^{^0}_{_{{\cal A}}}}}\right)\right) \ \simeq\ ({\cal C},{\cal A}).
\label{eq:FH}  
\end{equation}


\subsubsection{Non-contractible loops} \label{sec:2.3}

In presence of non-contractible loops, 

\begin{equation} 
\int_{_{(S^{^2};S^{^1})}}({\cal C},{\cal D};{\cal K};\emptyset)\ \simeq\ (H, \text{Hom}_{_{{\cal C}\boxtimes{\cal D}}}(\mathbf{1}_{_{{\cal C}}\boxtimes_{_{\cal D}}} {\cal A})  
\end{equation}
with

\begin{equation}   
{\cal A}\ \overset{def.}{=}\ (L\ \boxtimes\ R)^{^V} (\mathbf{1}_{_{\cal K}}) 
\end{equation}
a commutative separable algebra in $\bar {\cal C}\ \boxtimes\ {\cal D}$. If ${\cal K}$ is not fusion, then 

\begin{equation} 
\text{dim Hom}_{_{{\cal C}\boxtimes{\cal D}}}(\mathbf{1}_{_{{\cal C}}}\boxtimes \mathbf{1}_{_{{\cal D}}}, {\cal A})\ =\ \text{dim Hom}_{_{k}}(\mathbf{1}_{_{{\cal K}}},  \mathbf{1}_{_{\cal K}})\ >\ 1. 
\end{equation}

When this happens, the corresponding topological order is not stable, and triggers an RG-flow.  

Via factorisation homology, one defines the $W$-matrix, $W^{^{\cal M}}$, for ${\cal M}$ as follows  

\begin{equation}  
(L_{_{\cal M}}\ \boxtimes\ R_{_{\cal M}})^{^V}(\mathbf{1}_{_{{\cal M}}})\ \simeq\ \underset{i\ \in\ O({\cal C}), \ j\ \in\ O({\cal D})}{\bigoplus}\ W^{^{\cal M}}_{_{ij}}\ \ i\ \boxtimes\ j^{^*}  
\end{equation}

\begin{equation}  
(L_{_{\cal M}}\ \boxtimes\ R_{_{\cal M}})^{^V}(\mathbf{1}_{_{{\cal M}\boxtimes_{_{{\cal D}}}{\cal N}}})\ \simeq\ \underset{i\ \in\ O({\cal C}), \ j\ \in\ O({\cal E})}{\bigoplus}\ (W^{^{\cal M}}W^{^{\cal N}})_{_{ij}}\ \ i\ \boxtimes\ j^{^*}, 
\end{equation}  
given $(\mathbf{R}\times S^{^1}; S^{^1}\ \cup\ ...\ \cup\ S^{^1})$ a stratified open cylinder with 2-cells labelled by UMTCs ${\cal C}_{_{0}}, ..., {\cal C}_{_{n}}$, and 1-cells (namely non-contractible loops) labelled by closed multi-fusion ${\cal C}_{_{i-1}}-{\cal C}_{_i}$-bimodules ${\cal M}_{_i}$, $i=1,...,n$, and no 0-cells.

Crucially, the above leads the definition of the ground state degeneracy, namely the index we are looking for

\begin{equation}  
\boxed{\ \ \ \text{dim}\  u \ \overset{def.}{=}\ \text{Tr}\ \left(W^{^{\cal M}}_{_1}W^{^{\cal M}}_{_2}...W^{^{\cal M}}_{_n}\right) \color{white}\bigg] \ }.    
\end{equation}   

From the AGT correspondence, we know that this is equal to the partition function of a class ${\cal S}$ theory. 

As we will see in the upcoming sections, we can map the issue of gluing different class ${\cal S}$ theories separated by a non-invertible defect, \cite{Bashmakov:2022uek,Pasquarella:2023deo}, to that of defining the ground state degeneracy of a 2D TO in presence of non-contractible loops.

\subsection{Dualisability and full-extendibility}   \label{sec:334}  

We now turn to overviewing the relation between categorical dualisability and full-extendibility of a 2D TQFT. As we shall see, when applying this to  a specific algebraic variety, namely the Hilbert series, homological mirror symmetry, in its categorical formulation, results as a biproduct of full-extendibility. 

\subsubsection{Extending TQFTs}

Lurie proved that a TQFT is fully determined by what it assigns to a point. Furthermore, fully extended TQFTs with values in a symmetric monoidal $(\infty, n)$-category, ${\cal C}$, are equivalent to fully dualisable objects in ${\cal C}$. 

For a 2D TQFT, there are two main cases, \cite{lgmtap}:

\begin{enumerate}

\item   The TQFT is fully-extended, in which case the 2-functor reads as follows:

\begin{equation}  
{\cal Z}:\ \text{Bord}_{2,1,0}^{^{\sigma}}\ \rightarrow\ {\cal B},  
\end{equation}
where $\text{Bord}_{2,1,0}^{^{\sigma}}$ denotes the bicategories of points, 1-manifolds with boundaries, and 2-manifolds with corners. The target, ${\cal B}$, is most commonly associated with the bicategory of finite-dimensional $\textbf{k}$-algebras, bimodules, and bimodule maps, Alg$_{_{\textbf{k}}}$. From the cobordism hypothesis, it follows that the extended framed TQFTs with values in Alg$_{_{\textbf{k}}}$are classified by finite-dimensional separable $\textbf{k}$-algebras.

\item  The TQFT is not fully-extended, instead, if

\begin{equation}  
{\cal Z}:\ \text{Bord}_{2,1}^{^{\text{or}}}\ \rightarrow\ {\cal V},  
\end{equation}
with ${\cal V}$ a symmetric monoidal 1-category.

\end{enumerate}

A non-extended 2D TQFT can be extended to the point if there is a symmetric monoidal bicategory, $\cal B$, and an extended TQFT such that

\begin{equation}  
{\cal V}\ \simeq\ \text{End}_{_{\cal B}}\left(\mathbf{1}_{_{\cal B}}\right)\ \ \ \ \text{and} \ \ \ \ {\cal Z}_{_{ne}}\ \simeq\ {\cal Z}\bigg|_{{\text{End}_{_{\text{Bord}_{_{2,1,0}}^{^{\sigma}}(\emptyset)}}}},   
\label{eq:prescr}
\end{equation}
where $\mathbf{1}_{_{\cal B}}\ \in\ {\cal B}$ and $\emptyset\ \equiv\ \mathbf{1}_{_{\text{Bord}_{_{2,1,0}}^{^{\sigma}}(\emptyset)}}$.  

The prescription \eqref{eq:prescr} is clearly not unique. In particular, it depends on the target ${\cal B}$. 

If a non-extended 2D TQFT is a restriction of an appropriate defect TQFT ${\cal Z}_{_{ne}}^{^{\text{def.}}}$, then the non-extended TQFT can be extended to the point as long as we associate it with the bicategory ${\cal B}_{_{{\cal Z}_{_{ne}}^{^{\text{def.}}}}}$.    

\subsection{Comparison with our main example}

We now turn to explaining why this closely resembles the gauging prescription of SymTFTs, \cite{Freed:2022qnc}. In order to do so, we build connection with our treatment and main exaple outlined in Part I and Part II of this work, namely algebraic varieties and ring homologies.


In building the understanding for how the setup of subsection \ref{sec:2.3} is of use for relating 3D mirror symmetry and homological mirror symmetry, we first need a brief mathematical digression, presented in the remainder of this section. Combining these tools with those of section \ref{sec:clt}, will be the main focus of sections \ref{sec:333} and \ref{sec:last}.

\subsubsection{Dualisability in categorical structures}  

A dualisable object in a symmetric monoidal $(\infty,n)$-category, ${\cal C}$, is fully dualisable if the structure maps of the duality unit and counit each admit adjoints, as well as higher-morphisms between them, up to level $(n-1)$. 

By the Cobordism Hypothesis Theorem, Lurie proved that symmetric monoidal $(\infty, n)$-functors out of the $(\infty, n)$-category of cobordisms are characterised by their value at the point, which is a fully-dualisable object.

The notion of dualisability and full dualisability can be extended to the category of chain complexes, where an object is said to be dualisable if and only if it is a bounded chain complex of dualisable modules.

\subsubsection{The case of the Hilbert series} 

We now turn to the higher-categorical structure most relevant for our treatment, whose dualisability is the core topic of the present and following sections.

In \cite{Perry}, the author explores many applications of representation theory of categories to sequences of groups arising in topology, algebra, and combinatorics. 

Given $(V_{_n})_{_{n\in\mathbb{N}}}$ a sequence of representations of groups $G_{_n}$ over a field $k$, it can be lifted to a category ${\cal C}$ whose objects (or isomorphism classes) are indexed by natural numbers, such that the automorphism group of the $n^{th}$ object of ${\cal C}$ is $G_{_n}$. A ${\cal C}$-representation, or ${\cal C}$-module, is a functor ${\cal C}\rightarrow$ Mod$k$. One attempts to use the representation theory of ${\cal C}$ to show that the whole sequence of representations is determined by a finite amount of data, and to discover universal patterns which the sequence must satisfy.

These patterns are often expressed in terms of generating functions. For example, one invariant of a sequence of representations is its sequence of dimensions, (dim $V_{_n})_{_n\in\ \mathbb{N}}$. Such data can be recorded as a generating function, called the Hilbert series of $(V_{_n})_{_{n\in\mathbb{N}}}$

\begin{equation}   
\text{HS}_{_V}(t)\ \overset{def.}{=}\ \underset{n}{\sum}\ \text{dim}_{_k}V_{_n}t^{^n}\ \in\ \mathbb{Z}[[t]].  
\end{equation}

In most cases, theorems from the representation theory of ${\cal C}$ imply that HS$_{_V}(t)$ is rational, with denominator of a certain type. For example, considering the category of finite dimensional vector spaces over $\mathbb{F}_{_q}$ and linear injections between them, if $V$ is a finitely generated $VI_{_q}$ module, and char $k \neq$ char $\mathbb{F}_{_q}$, then HS$(t)$ is rational, with denominator 

\begin{equation} 
w_{_d}(t)\ \overset{def.}{=}\ \overset{d-1}{\underset{j=0}{\prod}}\ \left(1-q^{^j}t\right).  
\label{eq:wp}   
\end{equation}  

Each factor in \eqref{eq:wp} is a Whitney polynomial of the poset of subsets of a $d$-element set.  

Building connection with categorical structures, we the notion of \emph{posets}. In order theory, a poset is a partially-ordered set, namely an arrangement such that, for certain pairs of elements, one precedes the other. The adjective partial denotes the fact that not every pair of elements needs to be comparable; indeed, there may be pairs for which neither element precedes the other. Partial order, thus generalises total orders, in which ever pair is comparable. Given this, one can further define the \emph{order complex} associated to a poset $(S,\le)$ which has the set $S$ as vertices, and the finite chains of $(S,\le)$ as faces. To a poset, one can associate a poset topology, namely a topological structure on the poset of finite chains of $(S,\le)$, ordered by inclusion. Using methods from poset topology, it is possible to construct a chain complex of ${\cal C}$-modules $k_{_d}(V), \forall\ d\in\mathbb{N}$, such that exactness of $k_{_d}(V)$ categorifies the equation 

\begin{figure}[ht!]  
\begin{center} 
\includegraphics[scale=1]{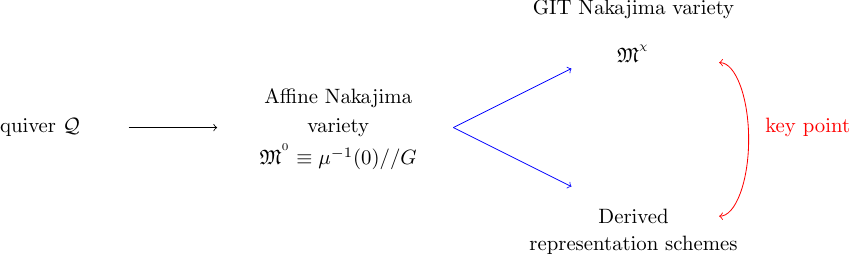} 
\caption{\small The upper and lower blue arrows correspond to geometric and algebraic resolutions of ${\mathfrak{M}^{0}}$, respectively. }  
\label{fig:GIT}  
\end{center}    
\end{figure}

\begin{equation} 
w_{_d}(t)\ \text{HS}_{_V}(t)\ \equiv\ p(t),    
\end{equation} 
for $p(t)$ a polynomial. If $V$ is finitely generated, either $k_{_d}$ or a power of it applied to $V$ is exact. In particular, categorification explains why the denominator \eqref{eq:wp} appears. This goes back to the construction outlined in our main example when integrating over the Kirwan map, \eqref{eq:intkm}.

\subsection*{Poset homology} 

Poset homology are chain complexes classifying rationality.
Extended literature on the subject explains there is a method for assigning a chain complex to a module over a combinatorial category, such that, if the complex is exact, the module has a rational Hilbert series. Furthermore, these complexes are known to satisfy Homology-vanishing theorems. Hasse diagrams, discussed in \cite{Pasquarella:2023exd}, are an example of chain complexes associated to poset representations. Homological mirror symmetry as an equivalence in between Hilber series on two categories, and rationality, are a direct biproduct of the defining features of poset homology, and constitutes a particular example of dualisable categorical structures, \cite{Kontsevich:1994dn}.
   
There is an equivalence of categories between the category of $\mathbb{C}$-linear representations of a quiver ${\cal Q}$ and the category of left $\mathbb{C}{\cal Q}$-modules.

  \subsection*{Moment map and higher homologies}

In general, homologies of derived representation schemes can be highly nontrivial. However, in this particular case, one can identify a necessary and sufficient condition for the vanishing of the higher homologies based on the flatness of $\mu$, \cite{DAlesio:2021hlp}. In particular, in such reference, it was shown that the derived representation scheme DRep$_{_{v,w}}({\cal A})$ has vanishing higher homologies if and only if $\mu^{^{-1}}(0)\ \subset\ M({\cal Q},v,w)$ is a complete intersection, which happens only if the moment map is flat, \cite{mommap}. As explained in \cite{Pasquarella:2023exd,Pasquarella:2023ntw}, the requirement of the algebraic variety to be a complete intersection is crucial for the purpose of our treatment. In particular, it ensures the emergence of a 2-categorical structure, whose importance will be the core topic of sections \ref{sec:333} and \ref{sec:last}.

\section{Drinfeld Centers from Magnetic Quivers}  \label{sec:333}

We now turn to address a key issue highlighted in section \ref{sec:FMT}, namely how to define a fiber functor associated to multiple gaugings not isomorphic to each other. For completeness, prior to explaining how this can be solved combining magnetic quivers and Drinfeld centers\footnote{The definition of the latter will be provided in due course.}, we will first review the features of the setup outlined in section \ref{sec:symtft} needed in the present treatment. This section is structured into three parts:   

\begin{itemize} 

\item Relying upon the description of the HB and CBs as algebraic varieties (cf. section \ref{sec:2.2}), we explain how gauging can be related to the poset ordering leading to the construction of Hasse diagrams, thanks to the unifying role of the moment map.

\item  We then turn to explaining how the identification of such moment map ensures the quiver gauge theory enjoys a generalisation of homological mirror symmetry, with the latter corresponding to the presence of a Drinfeld center and associated fiber functor for a given 2-categorical structure, related to Rozansky-Witten theory, \cite{Rozansky:1996bq}.

\item We conclude opening a connection between the topics outlined in the present work and those of \cite{Teleman:2014jaa, CT}.

\end{itemize}  

The relation between posets, complex cohomology, and Drinfeld centers, will be explored in greater mathematical detail in section \ref{sec:last}. This relation is intrinsic in this section, thanks to the connections outlined in section \ref{sec:mq}; however, we are confident that a more mathematical treatment could also lead to further interesting insights that are worth exploring.

\begin{figure}[ht!]   
\begin{center}
\includegraphics[scale=0.6]{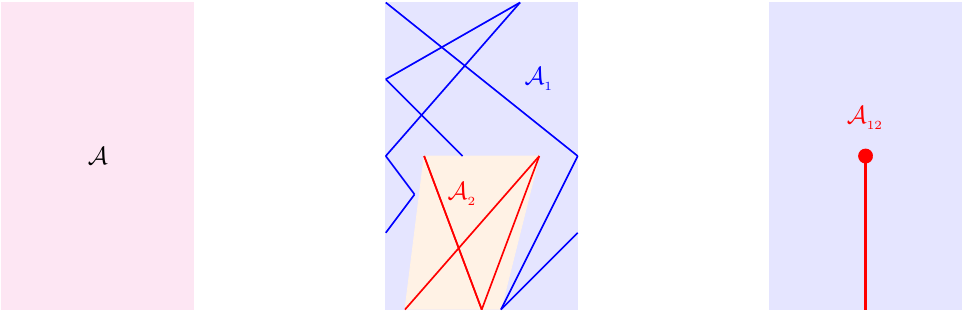}     
\caption{\small Condensing two different subalgebras, ${\cal A}_{_1}, {\cal A}_{_2}\ \subset\ {\cal A}$, the resulting theory corresponds to one with a changed phase with a gauged defect resulting from a relative gaugeable algebra, ${\cal A}_{_{12}}$ ending in the bulk. The defect at the endpoint is nontrivial, and can therefore be thought of as a Hom$(\mathbb{1}_{_{{\cal C}}},{\cal A}_{_{12}})$.}  
\label{fig:2codensationsTJF1}  
\end{center} 
\end{figure} 

\subsection{Gauging and Hasse diagrams}  

We first start by recalling the configuration of interest outlined in section \ref{sec:symtft}, motivating the importance of the Drinfeld center in geometric representation theory, \cite{CT}.

\subsection*{Fiber functors}

    \begin{figure}[ht!]   
\begin{center}
\includegraphics[scale=0.7]{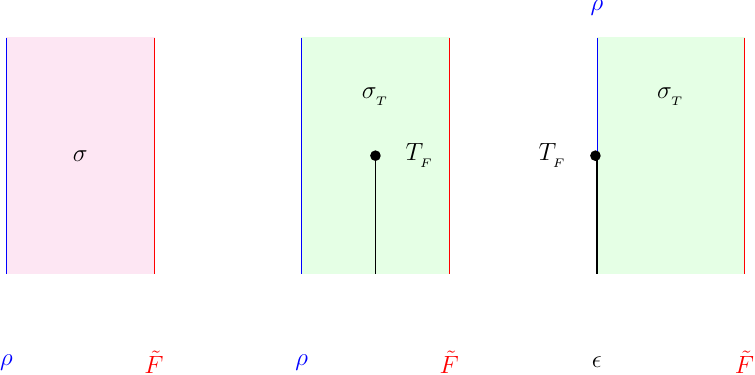}  \ \ \ \ \ \ \ \ \ \ \ \ \ \ \ \ \ \ 
\includegraphics[scale=0.8]{fnf.pdf} 
\caption{\small Adaptation of Freed-Moore-Teleman to the case involving twisted gauged defects. The figure on the far right corresponds to the case of interest to us, namely a configuration involving two different absolute 4D gauge theories separated by a defect. As we shall see, such defect is intrinsically non-invertible, corresponding to the presence of a relative uncondensed subalgebra, dressing ${\cal N}_{_3}$ in a nontrivial way. In higher-categorical terms, it corresponds to a fusion tensor category implementing the morphisms between the operators charged under the gauged symmetry.}
\label{fig:FrredMooreTeleman extended}  
\end{center} 
\end{figure}

\begin{figure}[ht!]     
\begin{center}
\includegraphics[scale=1]{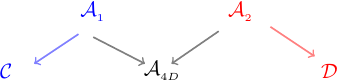}    
\caption{\small This picture reproduces an example of a correspondence, explained in section \ref{sec:2}.}  
\end{center} 
\end{figure}

In our previous work, \cite{Pasquarella:2023deo}, we already argued that, unlike figure \ref{fig:codensationTJF}, the third picture from the left in figure \ref{fig:FrredMooreTeleman extended} does not admit a straightforward expression for the fiber functor as \eqref{eq:functor}. Nevertheless, we claimed that a composite fiber functor could still be assigned thanks to the identification of the underlying algebraic structure of the composite theories

\begin{equation}
\boxed{ \color{white}{bla} \color{black}\large _{_{{\cal A}_{_1}}}{\cal A}_{_{{\cal A}_{_2}}}\ \overset{def.}{\equiv}\ {\cal A}_{_1}\ \otimes_{_{\cal A}}\ {\cal A}_{_2}\ \ \color{white}\bigg]\  }
\end{equation}

Our proposal in \cite{Pasquarella:2023deo} was that the corresponding fiber functor should really be defining the partition function of a 3D theory, whose identification we anticipated being the core topic of the present work. The remainder of this section is devoted to prove that the resulting 3D theory is a 3D ${\cal N}=4$ quiver gauge theory whose HB, when viewed as a complete intersection algebraic variety, enables to define the Drinfeld center of its underlying 2-categorical structure. In doing so, we need to completely specify its 1- and 2-morphisms, namely the generators and the relations in between them, both needed in the evaluation of the HB Hilbert series. As explained in section \ref{sec:DCFMQS}, the HS is nothing but the sum of gauge-invariant algebraic operators living in the chiral ring of the theory.

The Drinfeld center is defined as the 0$^{th}$-Hochschild cohomology, namely the endomorphisms of the identity element of a given category, ${\cal C}$:

\begin{equation}   
\boxed{\color{white}{blan} \color{black}\mathfrak{Z}({\cal C})\ \color{black}\overset{def.}{=}\ \text{End}(\mathbf{1}_{_{{\cal C}}})\ \equiv\ \text{Hom}(\mathbf{1}_{_{{\cal C}}},\mathbf{1}_{_{{\cal C}}})\ \equiv\ HH^{^{{\bullet}\  0}}({\cal C})\ \ \large\color{white}\bigg]\  }
\end{equation}

The existence of the Drinfeld center and that of the fiber functor are mutually related by definition, since 

\begin{equation}   
\boxed{\color{white}{blan} \color{black}{\cal F}:\ \mathfrak{Z}({\cal C})\ \longrightarrow\ {\cal C}\ \ \large\color{white}\bigg]\  }. 
\end{equation}

From what we have just outlined, it clearly follows that, the issue of identifying the Drinfeld center is equivalent to that of assigning a fiber functor projecting to the category ${\cal C}$. For the purpose of our treatment, namely building connection with geometric representation theory, ${\cal C}$ is really meant to be the category of representations, with the latter being described in terms of 2D TFTs, namely Lagrangian submanifolds living in a symplectic manifold $X$, playing the role of boundary conditions for 3D Rozansky-Witten theory, $RW_{_X}$. the meaning of this will be explained in section \ref{sec:homms}.

\subsection*{Relation to Hasse diagrams} 

The relation in between gauging and the construction of the Hasse diagram basically follows from the underlying role played by the moment map. In the former, it comes hand-in-hand with the definition of the fiber functor, meaning a partition function can be assigned to the resulting absolute theory, whereas, in the context of quiver gauge theories, the flatness of the moment map ensures a Hilbert series can be given, thereby fully specifying the generators and relations leading to the Hasse diagram construction itself. Importantly, the poset structure of the Hasse diagram admits a complex cochain description, thereby opening the possibility of studying this from a more mathematically rigorous point of view.

\subsection{Drinfeld centers and mirror symmetry}  \label{sec:homms}

As a key result of our work, we therefore wish to highlight the following statement as being applicable to the theories specified in the previous section \ref{sec:mq}, namely quiver gauge theories with 8 supercharges, 
   \medskip    
   \medskip
\color{blue}

\noindent\fbox{%
    \parbox{\textwidth}{%
   \medskip    
   \medskip
   \begin{minipage}{20pt}
        \ \ \ \ 
        \end{minipage}
        \begin{minipage}{380pt}
      \color{black}  \underline{Key point:} identifying the Drinfeld center is equivalent to assigning a Hilbert series with irreducible representations, i.e. specifying the cones in the Hasse diagram, and their intersection.
        \end{minipage}   
         \medskip    
   \medskip
        \\
    }%
}
 \medskip    
   \medskip     \color{black}

   The remainder of the present section is devoted to proving such assertion. In doing so, we mostly refer to \cite{Teleman:2014jaa, CT}, which was one of the first motivations of this work. A more detailed mathematical description of the correspondence with such references, as well as with \cite{Gonzalez:2023jur}, is the core topic of the remaining sections. For the moment, we wish to highlight that such a correspondence exists, and can be understood from a mathematical physicist's perspective. As a bi-product, we show how magnetic quivers can be combined with higher categories to extend the setup of  \cite{Teleman:2014jaa, CT} for describing higher dimensional QFTs admitting a quiver gauge theory description.

Prior to doing so, we will first briefly recapitulate the notion of homological mirror symmetry, introduced in section \ref{sec:2.1}

   \subsection*{Homological mirror symmetry}

For our purposes, we are mostly interested in a geometrical representation theory realisation of homological mirror symmetry, for the same motivations outlined in \cite{Teleman:2014jaa}. As previously explained in \cite{CT1}, 3D mirror symmetry can be rephrased in terms of topological representation theory thanks to the Drinfeld center, \cite{CT}. We now briefly overview such argument, in order to make contact with the setting we are dealing with.
For the case in which the two Lagrangians are related by mirror symmetry, such as is the case between Neumann and Dirichlet boundary conditions, the LHS of figure \ref{fig:Drinfeld} is such that the intersection is unable to account for the Drinfeld in a straightforward manner, signalling the presence of a non-invertible defect separating the resulting absolute theories. 

\begin{equation} 
\boxed{ \color{white}{bla} \color{black} \text{Vect}<G>\ \otimes_{_{\mathfrak{Z}}}\ \text{Rep}(G)\ \equiv\ \text{Vect}\ \ \color{white}\bigg]\  }
\end{equation}

    \medskip    
   \medskip

\begin{equation}  
\begin{aligned}
&\vcenter{\hbox{\includegraphics[scale=0.6]{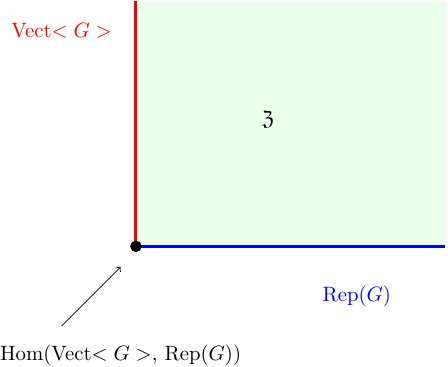} }}\\ 
\end{aligned}
\qquad\qquad\qquad
\begin{aligned}
&\vcenter{\hbox{
\includegraphics[scale=0.8]{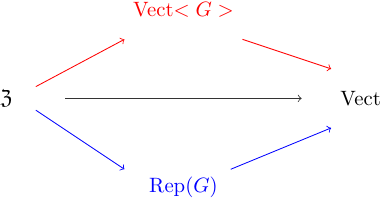}  }}\\
\end{aligned}  
\nonumber
\end{equation}

\begin{figure}[ht!]    
\begin{minipage}[c]{1\textwidth}
\caption{\small The black arrow in the diagram on the RHS denotes the fiber functor mapping from the Drinfeld center to the category of representation. Correspondingly, on the LHS, the Drinfeld center is the green shaded region whose boundaries are the regular and trivial representations, denoted in blue and red, respectively. For the case in which the two are related by ordinary mirror symmetry, their intersection contains the same information as $\mathfrak Z$, and therefore a fiber functor which is fully forgetful can be assigned.}
\label{fig:Drinfeld}
\end{minipage} 
\end{figure}

\subsection{Drinfeld centers from magnetic quivers of 3D ${\cal N}=4$ gauge theories}

In this concluding section, we combine the tools encountered throughout our treatment, drawing additional major conclusions, as well as setting the stage for the core topic of the next section, first presented by the author in \cite{Pasquarella:2023exd}.  

\subsection*{2-fiber products} 

Some preliminary mathematical tools. If $\cal C$ is a 2-category of categories, then there is a notion of a 2-fiber product, $\cal C$$_{_1}\ \times_{_{\cal D}}\ \cal C$$_{_2}$, also denoted by a diagram 

\begin{equation}  
{\cal C}_{_{1}}\ \xrightarrow{f}\ {\cal D}\ \xleftarrow{g}\ {\cal C}_{_2}  
\end{equation}   
is the category of triples $(c_{_1}, c_{_2},\phi)$, where $c_{_1}\in{\cal C}_{_1}$, $c_{_2}\in{\cal C}_{_2}$, and $\phi:f(c_{_1})\ \xrightarrow{\sim}\ g(c_{_2})$ is an isomorphism of their images $\cal D$. For an object $P\in\cal C$, a 2-fiber product $X$$\ \times_{_S}\ $$Y$ associated to the diagram 

\begin{equation}  
X\ \xrightarrow{f}\ S\ \xleftarrow{g}\ Y  
\end{equation} 
is a quadruple $(P,p,q,\phi)$, with 1-morphisms $p:P\rightarrow X, q:P\rightarrow Y$, and a 2-isomorphism

\begin{equation}  
\phi: f\ \circ\ p\ \simeq\ g\ \circ\ q  
\end{equation}  
such that $\forall Z\in\cal C$, the natural functor     

\begin{equation} 
\text{Hom}_{_{\cal C}}\left(Z,P\right)\ \rightarrow\ \text{Hom}_{_{\cal C}}\left(Z,X\right)\times_{_{\text{Hom}_{_{\cal C}}(Z,S)}}\ \text{Hom}_{_{\cal C}}(Z,Y)  
\end{equation}    
is an equivalence of categories.

\subsection*{Rozansky-Witten Theory and the Drinfeld center}   

One of the examples encountered in the previous section, namely $SU(2)$ with $N_{_f}=2$, is the simplest case involving two cones with nontrivial intersection in its Hasse diagram. As such, it is an example where ordinary mirror symmetry in the sense of \eqref{eq:perfectmirror} does not apply for the 3D ${\cal N}=4$ theory arising under dimensional reduction. Nevertheless, a suitable mirror dual can still be assigned once the generators of the two cones featuring in the Hasse diagram, together with their intersection, are correctly accounted for in the calculation of the Hilbert series (HS), i.e. the partition function counting the gauge-invariant operators living in the chiral ring.  
In the previous section we explained how the magnetic quiver prescription enables to identify the HS correctly for cases where ordinary mirror symmetry does not apply. 

To make contact with the calculation of Drinfeld centers, let us remind ourselves of the case of 4D $SU(2)$ ${\cal N}=2$, first studied by Seiberg and Witten within the context of its 3D reduction. Its low-energy limit is described as a sigma-model in the space of vacua, corresponding to the Atiyah-Hitchin moduli space of charge 2 monopoles. The resulting 3D $SU(2)$ theory was then identified with the Rozansky-Witten theory (RW) of the hyper-kah$\ddot{\text{a}}$ler Atiyah-Hitchin manifold. Later developments of this by Kapustin, Rozansky, \cite{Kapustin:2009uw}, and later with Saulina, \cite{Kapustin:2008sc}, addressed their 2-category of branes, with the latter containing smooth holomorphic Lagrangians, $L$.

\begin{figure}[ht!]     
\begin{center}
\includegraphics[scale=0.75]{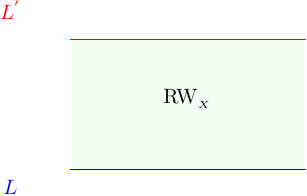}  
\ \ \ \ \ \ \ \    \ \ \ \ \ \ \ \ \ 
\includegraphics[scale=0.75]{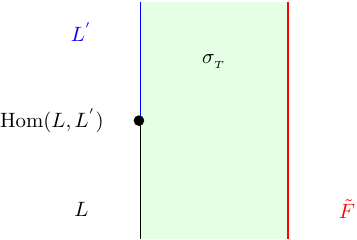}  
\caption{\small Adaptation of the RW$_{_X}$ theory (on the LHS) to the quiche of a relative field theory (RHS).}  
\label{fig:RW}  
\end{center} 
\end{figure}

In the RW model, one must geometrically describe two functors from the 2-category of linear categories with $G$-action to linear categories: 

\begin{enumerate}  

\item the forgetful functor, keeping track of the underlying category (describing the pre-gauged TQFT), represented by the regular representation of $G$,

\item the invariant category, generating the gauged TQFT, associated to the trivial representation of $G$. 

\end{enumerate}   

This already encodes the idea of mirror symmetry, whose generalisation is realised thanks to the composite fiber functor described in \cite{Teleman:2014jaa}, and overviewed in the concluding part of this section in relation to our analysis.

The objects in the 2-category of KRS are branes in the BFM space whose boundaries are Lagrangians $L,L^{^{\prime}}\ \subset\ X$, $X$ a symplectic manifold, specifying the RW$_{_X}$ according to the choice of $X$. $L,L^{^{\prime}}$ are 1-morphisms generating the theory upon acting on the trivial theory   

\begin{equation}  
\boxed{\ \ \  L,L^{^\prime}:\ \text{Id}\ \longrightarrow\ RW_{_X}\color{white}\bigg]\ \ }.    
\end{equation}   

\begin{figure}[ht!]
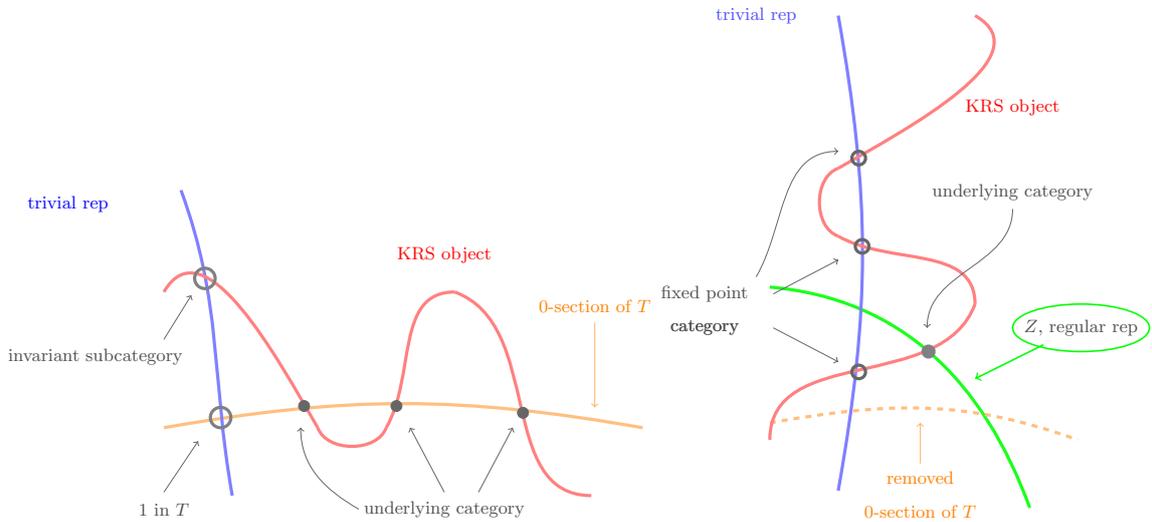
     
\begin{center}
\includegraphics[scale=0.75]{ct1.pdf} 
\includegraphics[scale=0.75]{ct2.pdf} 
\caption{\small The BFM space for an abelian (LHS) and non-abelian (RHS) gauge theory, \cite{Teleman:2014jaa}. The KRS objects living in them are the objects in the 2-category referred to in the text and the trivial and regular representations are the Lagrangian submanifolds $L,L^{^{\prime}}$. Their intersection are Hom-spaces. }  
\label{fig:ctct}  
\end{center} 
\end{figure} 

The 2-morphisms are denoted by Hom$(L,L^{^{\prime}})$. For the case in which $L, L^{^{\prime}}$ are related by ordinary mirror symmetry, the 2-morphisms encode all the information, and the 2-category can be completely reconstructed from them. For the case of non-ablelian gauge groups, a generalisation of the notion of mirror symmetry is needed. Practically, this means we need to specify, both, the 1-morphisms and the 2-morphisms to reconstruct the bulk of the BFM space, and completely describe the KRS branes living in them.

In connection with our previous work, \cite{Pasquarella:2023deo}, it is instructive to visualise what we have just said as shown in figure \ref{fig:RW}. Fully specifying the topological boundary conditions leads to an absolute theory. In case Hom$(L,L^{^\prime})$ leads to a non-invertible defect, the fiber functor, hence the Drinfeld center in figure \ref{fig:Drinfeld} needs a 2-fiber product structure to be fully specified, as also explained in \cite{Pasquarella:2023deo}. In particular, this highlights the importance of specifying the entire underlying 2-categorical structure, namely the Lagrangian submanisfolds as well as the homomorphisms between them. This ensures the objects in the 2-category of KRS, can be fully specified. From the arguments outlined in the preliminary part of this work, we are therefore led to conclude that the configuration on the RHS of figure \ref{fig:RW} is dual to another theory with the same Drinfeld center and only one Lagrangian submanifold on the topological boundary condition, thereby providing a sample realisation of generalised homological mirror symmetry.


\subsection*{Symplectically induced representations}

In conclusion, we outline more explicitly the connection with the work of \cite{Teleman:2014jaa}, setting the stage for a more detailed mathematical treatment, which is the main focus of \cite{Pasquarella:2023exd}.

\begin{figure}[ht!]      
\begin{center}
\includegraphics[scale=0.85]{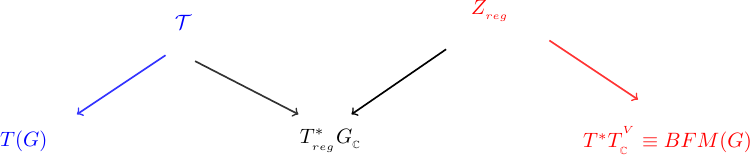}
\caption{\small The composition of fiber functors associated to double symplectic inductions leading to the regular and trivial representations.}
\label{fig:diagram}  
\end{center}
\end{figure}

One of the main aims of \cite{Teleman:2014jaa} is that of identifying a preferred family of simple representations, foliating the BFM space, ultimately describing the space of possible representations. The prescription outlined in such reference is motivated by the realisation of homological mirror symmetry. The construction of the preferred family of foliations also requires making use of a 2-fiber product, defined as follows

\begin{equation}
\boxed{ \large\ \ \ {\cal T}\ \times_{_{T^{^*}_{_{reg}}G_{_{\mathbb{C}}}}}\ Z_{_{reg}}\ \ \color{white}\bigg] } 
\label{eq:composite}
\end{equation}
with ${\cal T}$ and $Z_{_{reg}}$ denoting the trivial and regular fibers, respectively. \eqref{eq:composite} results from the composition outlined in figure \ref{fig:diagram}, where

\begin{equation}     
BFM(G)\ \overset{def.}{=}\ Z_{_{reg}}/G_{_{\mathbb{C}}}\ \ \ \ \ \ ,\ \ \ \ \ T(G)\ \overset{def.}{=}\ (N, \chi) \backslash\backslash T^*G_{_{\mathbb{C}}} // (N, \chi)\ \equiv\ N\ \backslash\ {\cal T}\ /N,
\end{equation}    
with

\begin{equation} 
\chi:\ n\ \longrightarrow\ \mathbb{C}^{^{\times}}
\end{equation}  
the regular character. Every such representation is symplectically induced from a 1D representation of a certain Levi subgroup $G$. As such, the preferred family of foliations resulting from \eqref{eq:composite} is the Fukaya category of a flag variety of $G$.

Given the considerations made when describing figure \ref{fig:RW} in the context of RW-theory, we therefore notice that the aim of describing objects in the bulk of figure \ref{fig:ctct} is equivalent to that of defining the Drinfeld center for two absolute theories separated by a non-invertible defect, as encountered in \cite{Pasquarella:2023deo}. Our main assertion is that, thanks to the prescription outlined in section \ref{sec:2.2}, in presence of multiple symplectically-induced representations, the resulting composite fiber functor defines the Drinfeld center of a mirror 3D ${\cal N}=4$ gauge theory, resulting from the union of multiple Coulomb branches of magnetic quivers associated to 3D ${\cal N}=4$ theories, thereby constituting a generalised realisation of homological mirror symmetry.

A more mathematical reformulation of this is the core topic of upcoming sections.

\section*{Key points} 

We can summarise the main results of sections \ref{sec:cd} and \ref{sec:333} as follows: 

\begin{itemize} 

\item The 2-categorical structure description of the HB Hasse diagram of quiver gauge theories with 8 supercharges is equivalent to that of KRW. 
Importantly, this follows from the dimensional reduction of 4D ${\cal N}=2$ $SU(2)$ gauge theory with $N_{_f}=2$, as the most basic case with HB Hasse diagram consisting of two intersecting cones. 

\item Importantly, the flatness of the moment map, ensuring the vanishing of higher homology, \cite{DAlesio:2021hlp}, corresponds to the possibility of fully specifying the HS associated to the HB of a given theory. We emphasise that this is equivalent to stating the existence of a fiber functor in a 2-category, ultimately leading to the definition of a partition function of an absolute theory, according to the prescription of \cite{Freed:2022qnc}. 

\item The identification of the MQs whose CBs equal the HB of the original theory, corresponds to the identification of the Drinfeld center of a 3D theory, with the latter being a RW theory, $RW_{_X}$, whose BCs are Lagrangian submanifolds within a certain symplectic manifold, $X$. The complete reconstruction of the BFM space, where the objects of the 2-category live, corresponds to defining the Drinfeld center w.r.t. the given Lagrangian submanifolds and their intersection, i.e. the 1- and 2-morphisms characterising the 2-category in question. 

\item The fiber functor defining the partition function of interest is really obtained as a 2-fiber product, and is therefore compatible with the result of \cite{Teleman:2014jaa}, motivated by generalising homological mirror symmetry.

\end{itemize}

\section{Moore-Tachikawa varieties: beyond duality}     \label{sec:last}

We now have all the tools needed for addressing an open question raised in section \ref{sec:clt}, namely extending Moore-Tachikawa varieties beyond categorical duality. In doing so, we will be exploiting tools from algebraic geometry and higher-category theory.    

In particular, we wish to stress the importance of dealing with the hyperk$\ddot{\text{a}}$hler structure of the target 2-category of 2D TFTs describing Moore-Tachikawa varieties, and how this is related to the emergence of new higher-categorical structures requiring generalisation of the ordinary homological mirror symmetry/3D mirror symmetry correspondence. 

The starting point is the following relation in between class ${\cal S}$ theories and the gluing of their AGT dual 2D TFTs, \cite{Moore:2011ee},

\begin{equation}  
\boxed{\ \ \ S_{_L}\ \times_{_{G,q}}\ S_{_R}\ =\ S\left(C_{_L}\ \times \ C_{_R}, D_{_{LR}}\right)\ \color{white}\bigg] },     
\label{eq:cclS}    
\end{equation}  
where $(C_{_L}, D_{_L}), (C_{_R}, D_{_R})$ are decorated 2-surfaces. The AGT correspondence states the equivalence holds in general. However, Moore and Tachikawa pointed out some interesting shortcomings for the case in which the 2-surface $C$ is a sphere with two punctures. In such case, they claim that one does not achieve a genuine 4D class ${\cal S}$ theory. The main aim of the present section is that of proving that this issue could be overcome by making use of a generalised notion of homological mirror symmetry/3D mirror symmetry correspondence, which, from the perspective of Moore-Tachikawa varieties, amounts to accounting for the hyperk$\ddot{\text{a}}$hler structure of the target 2-category of the 2D TFT AGT-dual to the composite class ${\cal S}$ theory on the LHS of \eqref{eq:cclS}.

Prior to delving to delving into explaining a possible way of constructing a 2D TFT whose target category is a hyperk$\ddot{\text{a}}$hler quotient, we briefly overview the work of \cite{Moore:2011ee} who first pointed out the hyperk$\ddot{\text{a}}$hler structure needed generalisation of the target category of the 2D TFT $\eta_{_{G_{_{\mathbb{C}}}}}$.  We briefly overview some specific examples they outlined, explaining the relation with our treatment in Part I when dealing with star-shaped quivers. 

Upon removing the identity from the source category, a morphism is now defined as a pair $(C,a)$, when $C$ is an orientable 2-manifold with boundaries, and $a$ is a positive real number which can be thought of as the area of $C$.   

The composition of the two morphisms adds the area. For example, we see that   

\begin{equation}           
\underset{a\rightarrow 0}{\lim}\ (W,a)\ =\ \mathbb{C}^{^2}\ \otimes\ \mathbb{C}^{^2}\ \otimes\ \mathbb{C}^{^2},       
\end{equation}   
which is equivalent to the Coulomb branch encountered in Part I. Similarly, for $\mathfrak{g}=A_{_2}$, $\eta_{_{A_{_2}}}(W,a)$ is equivalent to the minimal nilpotent orbit of E$_{_6}(\mathbb{C})$ as the hyperk$\ddot{\text{a}}$hler manifold in the $\underset{a\rightarrow 0}{\lim}$. At finite $a$, the hyperk$\ddot{\text{a}}$hler metric is only invariant under $SU(3)^{^3}$ and not the whole $E_{_6}$.

\subsection{Categorical structure without duality}   \label{sec:newcat}

As explained in section \ref{sec:clt}, removing the identity element in Bo$_{_2}$ requires having to introduce at least two different functors $\eta_{_{G_{_{\mathbb{C}}}^{^{\prime}}}}, \eta_{_{G_{_{\mathbb{C}}}^{^{\prime\prime}}}}$, whose action on the circle and its bordism reads as follows 

\begin{equation}   
\eta_{_{G_{_{\mathbb{C}}}^{^{\prime}}}}\left(\ S^{^1}\ \right)\ \equiv\ G_{_{\mathbb{C}}}^{^{\prime}}\ \ \ , \ \ \ \eta_{_{G_{_{\mathbb{C}}}^{^{\prime\prime}}}}\left(\ S^{^1}\ \right)\ \equiv\ G_{_{\mathbb{C}}}^{^{\prime\prime}}   
\label{eq:morph0}  
\end{equation} 

\begin{equation}   
\eta_{_{G_{_{\mathbb{C}}}^{^{\prime}}}}\left(\ U\ \right)\ \equiv\ U_{_{G_{_{\mathbb{C}}}^{^{\prime}}}} \equiv \  G_{_{\mathbb{C}}}^{^{\prime}}\times S_{_n}\ \subset \ G_{_{\mathbb{C}}}^{^{\prime}}\times\mathfrak{g}_{_{\mathbf{C}}}^{^{\prime}}\ \simeq \ T^{^*}G_{_{\mathbf{C}}}^{^{\prime}},       
\label{eq:morph1}  
\end{equation}

\begin{equation}   
\eta_{_{G_{_{\mathbb{C}}}^{^{\prime\prime}}}}\left(\ U\ \right)\ \equiv\ U_{_{G_{_{\mathbb{C}}}^{^{\prime\prime}}}} \equiv \  G_{_{\mathbb{C}}}^{^{\prime\prime}}\times S_{_n}\ \subset \ G_{_{\mathbb{C}}}^{^{\prime\prime}}\times\mathfrak{g}_{_{\mathbf{C}}}^{^{\prime\prime}}\ \simeq \ T^{^*}G_{_{\mathbf{C}}}^{^{\prime\prime}},  
\label{eq:morph2}   
\end{equation}

\begin{equation}   
\eta_{_{G_{_{\mathbb{C}}}^{^{\prime}}}}, \left(\ V\ \right)\ \equiv\ V_{_{G_{_{\mathbb{C}}}^{^{\prime}}}} \overset{def.}{\equiv} \  T^{^*}G_{_{\mathbb{C}}}^{^{\prime}},      
\label{eq:id1}  
\end{equation}

\begin{equation}   
\eta_{_{G_{_{\mathbb{C}}}^{^{\prime\prime}}}}\left(\ V\ \right)\ \equiv\ V_{_{G_{_{\mathbb{C}}}^{^{\prime\prime}}}} \overset{def.}{\equiv} \  T^{^*}G_{_{\mathbb{C}}}^{^{\prime\prime}},  
\label{eq:id2}   
\end{equation} 
where we are assuming

\begin{equation}   
\mathfrak{g}_{_{\mathbf{C}}}^{^{\prime}}\ \cap\ \mathfrak{g}_{_{\mathbf{C}}}^{^{\prime\prime}}\ \neq\ \{ \emptyset \}\   \ , \ \ \text{and}\ \ \   \mathfrak{g}_{_{\mathbf{C}}}^{^{\prime}}\ \cup\ \mathfrak{g}_{_{\mathbf{C}}}^{^{\prime\prime}}\ \equiv\ \mathfrak{g}_{_{\mathbf{C}}}, 
\label{eq:fa}
\end{equation}

\begin{equation}   
\mathfrak{g}_{_{\mathbf{C}}}^{^{\prime}}\ \not\subset\ \mathfrak{g}_{_{\mathbf{C}}}^{^{\prime\prime}}\    \ , \ \ \text{and}\  \  \ \ \mathfrak{g}_{_{\mathbf{C}}}^{^{\prime\prime}}\ \not\subset\ \mathfrak{g}_{_{\mathbf{C}}}^{^{\prime}}.    
\label{eq:sa}
\end{equation} 

\eqref{eq:fa} and \eqref{eq:sa} imply that the two subalgebras involved, $\mathfrak{g}_{_{\mathbf{C}}}^{^{\prime}}, \mathfrak{g}_{_{\mathbf{C}}}^{^{\prime\prime}}$ are associated to different subgroups, $G_{_{\mathbb{C}}}^{^{\prime}}, G_{_{\mathbb{C}}}^{^{\prime\prime}} $, and that the identity elements differ, $T^{^*}G_{_{\mathbf{C}}}^{^{\prime}}\ \neq\ T^{^*}G_{_{\mathbf{C}}}^{^{\prime\prime}}$, even under reparametrisation of the Reimann surface. For each one of the 2D TFTs, $\eta_{_{G_{_{\mathbb{C}}}^{^{\prime}}}}, \eta_{_{G_{_{\mathbb{C}}}^{^{\prime\prime}}}}$ one could use the formalism of \cite{Moore:2006dw,Moore:2011ee}, describing two different class ${\cal S}$ theories, both descending from 6D ${\cal N}=(2,0)$ by dimensionally reducing on a Riemann surface without reparemetrisation invariance. However, given the assumption that $(G_{_{\mathbb{C}}}^{^{\prime}}, \mathfrak{g}_{_{\mathbf{C}}}^{^{\prime}}), (G_{_{\mathbb{C}}}^{^{\prime\prime}}, \mathfrak{g}_{_{\mathbf{C}}}^{^{\prime\prime}}) $ can be embedded in a unique $(G_{_{\mathbb{C}}}, \mathfrak{g}_{_{\mathbb{C}}})$, it is natural to ask what should the triple on the RHS of \eqref{eq:HB1} be for the resulting theory to be absolute?  

We know that the fiber functor and Drinfeld centers for a given absolute theory are defined as follows  

\begin{equation}{\cal F}\ :\ \mathfrak{Z}\left(\text{Bo}_{_2}\right)\ \rightarrow\ \text{Bo}_{_2}, 
\label{eq:F}
\end{equation} 
         
      \begin{equation}  
\mathfrak{Z}\left(\text{Bo}_{_2}\right)\equiv\text{End}_{_{\text{Bo}_{_2}}}\left(T^{^*}G_{_{\mathbf{C}}}\right)\equiv\text{Hom}_{_{\text{Bo}_{_2}}}\left(T^{^*}G_{_{\mathbf{C}}},T^{^*}G_{_{\mathbf{C}}}\right),    
\label{eq:Z}
\end{equation}  
both of which crucially rely upon the presence of an identity in the source and target of $\eta_{_{G_{_{\mathbb{C}}}}}$. Apparently, we run into a contradiction, since the identity element $T^{^*}G_{_{\mathbf{C}}}$ has been removed by assumption, thereby implying $\mathfrak{Z}\left(\text{Bo}_{_2}\right)$ cannot be defined in the ordinary way.   On the other hand, one could define the new identity as being a composite object defined in the following way

      \begin{equation}  
\boxed{\ \ \ T^{^*}\tilde G_{_{\mathbf{C}}}\ \overset{def.}{=}\ T^{^*}G_{_{\mathbf{C}}}^{^{\prime}}\ \otimes_{_{T^{^*}G_{_{\mathbb{C}}}}}\ T^{^*}G_{_{\mathbf{C}}}^{^{\prime\prime}}\color{white}\bigg]\ \ },    
\label{eq:newid}
\end{equation} 
and therefore our proposal for \eqref{eq:F} and \eqref{eq:Z} reads as follows

\begin{equation}
\boxed{\ \ \ {\cal F}\ :\ \mathfrak{Z}\left(\tilde{\text{Bo}_{_2}}\right)\ \rightarrow\ \tilde{\text{Bo}_{_2}}\color{white}\bigg]\ \ }, 
\label{eq:Fnew}
\end{equation} 
         
      \begin{equation}  
\boxed{\ \ \ \mathfrak{Z}\left(\tilde{\text{Bo}_{_2}}\right)\ \equiv\ \text{End}_{_{\tilde{\text{Bo}_{_2}}}}\left(T^{^*}G_{_{\mathbf{C}}}^{^{\prime}}\ \otimes_{_{T^{^*}G_{_{\mathbb{C}}}}}\ T^{^*}G_{_{\mathbf{C}}}^{^{\prime\prime}}\right)\color{white}\bigg]\ \ },    
\label{eq:Znew}
\end{equation}
with 

\begin{equation}  
\tilde{\text{Bo}_{_2}}\ \overset{def.}{=}\ \text{Bo}_{_2}\ /\ V.
\end{equation}  

In the following subsection, we will explain why \eqref{eq:Fnew} and \eqref{eq:Znew} are reasonable proposals for defining an absolute theory in absence of a categorical duality structure.

\subsection{Interesting shortcomings}

As explained in \cite{Pasquarella:2023deo}, the loss of reparametrisation invariance of the Riemann surface signals the presence of an intrinsically-non-invertible defect between different class ${\cal S}$ theories. We will now show that the setup described in section \ref{sec:newcat} is equivalent to that of \cite{Pasquarella:2023deo}.

The starting point in our argument is the conjecture \eqref{eq:HBB} and \eqref{eq:HBtrue}. In absence of categorical duality of, both, source and target, the 2D TFT associated to the maximal dimensional Higgs branch of \cite{Moore:2011ee} is no longer associated to a moment map that is equivalent for all the constituent $S^{^1}$s, thereby violating the conjecture made by \cite{Moore:2011ee}. Explicitly,

\begin{equation}   
\boxed{\ \ \ \ W_{_{G_{_{\mathbb{C}}}}}\ \neq\  \mu^{^-1} /G^{^3}\color{white}\bigg]\ \ } . 
\label{eq:HB}  
\end{equation}  

This is because the algebraic variety associated to the Higgs branch $W_{_{G_{_{\mathbb{C}}}}}$ is not a hyperk$\ddot{\text{a}}$ler quotient. In particular, it is associated to a non-primitive ideal. 

Expanding further on this topic, the crucial point is that, when giving up the duality propriety, there is no longer the identity element in the source category, $T^{^*}G_{_{\mathbf{C}}}$, but, rather, there is one identity for each underlying constituent, \eqref{eq:morph1} and \eqref{eq:morph2}.  

In order to determine how the resulting composite identity element should look like, we need to briefly recall what was outlined in section \ref{sec:od}. Given two different morphisms, and taking their composition

\begin{equation}  
T^{^*}G_{_{\mathbf{C}}}^{^{a^{\prime}}}\ \circ\ T^{^*}G_{_{\mathbf{C}}}^{^a}\equiv\left(T^{^*}G_{_{\mathbf{C}}}^{^a}\ \times\ T^{^*}G_{_{\mathbf{C}}}^{^{a^{\prime}}} \right)\dslash G_{_{\mathbf{C}}},  
\label{eq:compths}   
\end{equation}   
we know that, under the duality assumption, the axiom \eqref{eq:composition} 
comes with two moment maps, \eqref{eq:etaGC1}, each one describing the embedding of the individual morphisms within the algebra of the mother theory. As explained in section \ref{sec:od}, one actually uses the mutual relation in between such moment maps to prove that $T^{^*}G_{_{\mathbf{C}}}$ behaves as the identity. As also claimed in the previos section, \eqref{eq:compths} expresses the need to define a Drinfeld center for the composite system made up of two class $\cal S$ theories. In presence of reparametrisation invariance, such theories can be thought of as being separated by an invertible defect, from which \eqref{eq:compths} can be reduced to a statement of S-duality. 

On the other hand, in absence of reparametrisation-invariance, the resulting class ${\cal S}$ theories would, by definition, be separated by an intrinsically non-invertible defect, with the latter being responsible for the lack of reparametrisation invariance of the Riemann surface \cite{Moore:2006dw}.

Our main question is to find the Drinfeld center for a given Bo$_{_2}$       

\begin{equation}  
{\cal F}:\ \mathfrak{Z}(\text{Bo}_{_2})\ \rightarrow\ \text{Bo}_{_2} 
,
\label{eq:fiber}  
\end{equation}  
and we know that, to a given fiber functor, ${\cal F}$, there is an associated moment map 

\begin{equation}  
\mu:\ \mathfrak{G}\ \rightarrow\ {\cal A}_{_c}.     
\end{equation}  
with $\mathfrak{G}$ being Bo$_{_2}$ in this case, and ${\cal A}$ the algebra of invertible topological defects that projects to the identity $T^{^*}G_{_{\mathbb{C}}}$ under complete gauging of the theory. This is equivalent to stating that $T^{^*}G_{_{\mathbb{C}}}$ is the identity element once having projected over ${\cal A}_{_c}$

\begin{equation}  
{\cal A}\ \dslash_{\mu}\ \mathfrak{G}\ \ \ \text{with}\ \ \ \mu:\ \mathfrak{G}\ \rightarrow\ {\cal A}_{_c}  
\end{equation}   
choosing the definition of the identity in the following way 

\begin{equation}  
T^{^*}G_{_{\mathbb{C}}}\ \simeq\ \mathbf{1}_{_{G_{_{\mathbb{C}}}}}\ \equiv\ {\cal A}\ \dslash_{\mu}\ \mathfrak{G}.    
\end{equation} 

For the purpose of our work, $\mathfrak{G}\ \overset{def.}{=}\ \text{Bo}_{_2}$, therefore    

\begin{equation}  
T^{^*}G_{_{\mathbb{C}}}\ \simeq\ \mathbf{1}_{_{G_{_{\mathbb{C}}}}}\ \equiv\ {\cal A}\ \dslash_{\mu}\ \text{Bo}_{_2}
\end{equation}  
However, in the case of section \ref{sec:newcat}, there are two different moment maps involved, one for each choice of conformal structure on the Riemann surface, that are not mutually related by the moment map constraint following from the axiom \eqref{eq:composition}. We therefore need to find the moment map (and corresponding gauge group $\tilde G_{_{\mathbb{C}}}$)

\begin{equation}  
\tilde\mu:\ \tilde{\mathfrak{G}}\ \rightarrow\ \tilde{\cal A}_{_c},        
\end{equation}
whose identity element

\begin{equation} 
\tilde{\cal A}_{_c}\ \overset{def.}{=}\ {\cal A}_{_1}\ \otimes_{_{T^{^*}G_{_{\mathbb{C}}}}}{\cal A}_{_{2}}    
\end{equation}  
constitutes the identity of the composite theory. If $T^{^*}G_{_{\mathbb{C}}}$ is the identity that has been removed from a particular source category, it still exists, but is no longer the identity present in $\tilde{\text{Bo}}$$_{_2}$ of a given $\tilde\eta_{_{G_{_{\mathbb{C}}}}}$. From the RHS of \eqref{eq:compths}, the identity of $\tilde{\text{Bo}}_{_2}$ therefore reads

      \begin{equation}  
      T^{^*}\tilde G_{_{\mathbf{C}}}\ \overset{def.}{=}\ 
T^{^*}G_{_{\mathbf{C}}}^{^{\prime}}\ \otimes_{_{T^{^*}G_{_{\mathbb{C}}}}}\ T^{^*}G_{_{\mathbf{C}}}^{^{\prime\prime}},    
\label{eq:newidnow}
\end{equation} 
such that its Drinfeld center can be determined.

Defining $\tilde\mu:\ \tilde{\text{Bo}}_{_2}\rightarrow\ \tilde{\cal A}$ as the moment map associated to the composite theory, the corresponding fiber functor can be explicitly rewritten as follows  

\begin{equation}  
\boxed{\ \ \ {\cal F}:\ \mathfrak{Z}\left(\tilde\eta_{_{\tilde G_{_{\mathbb{C}}}}}^{^{-1}}\left(T^{^*}\tilde G_{_{\mathbb{C}}}\right)\right)\ \rightarrow\ \tilde\eta_{_{\tilde G_{_{\mathbb{C}}}}}^{^{-1}}\left(T^{^*}\tilde G_{_{\mathbb{C}}}\right)\color{white}\bigg]\ \ }.   
\label{eq:fiber}  
\end{equation}  
ultimately enabling us to reformulate the problem of finding the Drinfeld center to that of identifying ${\cal F}$ for a given $\mu$.

\section*{Key points}    

The main points are the following:  

\begin{itemize}  

\item Lack of reparametrisation-invariance of bordism operators signals the presence of intrinsic non-invertible defects separating different class ${\cal S}$ theories.

\item Defining the Drinfeld center for a system of composite class ${\cal S}$ theories separated by non-invertible defects constitutes a nontrivial generalisation of an S-duality statement.

\end{itemize}

\subsection{Back to the identity from moduli spaces}  \label{sec:3ch}

In this concluding section, we gather the key tools outlined in this work, highlighting some main features leading towards the relation between homological mirror symmetry and 3D mirror symmetry. 

3D mirror symmetry is defined as a duality exchanging Coulomb and Higgs branches, \cite{Intriligator:1996ex}. Making use of their algebro-geometric definition, briefly overviewed in section \ref{sec:mq}, together with our previous work, \cite{Pasquarella:2023ntw}, we will be arguing that \emph{ordinary} 3D mirror symmetry, formulated as a 1-to-1 correspondence between a Coulomb and a Higgs branch, can be generalised to cases involving more theories (namely more Coulomb branches for a given Higgs branch). Specifically, we will be referring to the case of supersymmetric quiver gauge theories admitting a magnetic quiver description, and whose Higgs branch splits into two cones. This will ultimately provide further evidence to the claims made in \cite{Pasquarella:2023exd}.

As we shall see, the quantity that will enable us to probe such duality structure is the categorical interpretation of Hochschild cohomologies of block algebras, \ref{sec:fhss}.  

The present section is structured into three parts:

\begin{enumerate}

\item At first, we briefly overview the definition of block algebras, their associated Hoschschild cohomology, \cite{Ba}, and the relation with the Drinfeld center, the latter following from our preliminary discussion in section \ref{sec:334}.

\item  We then turn to Coulomb and Higgs branches arising from Hitchin system for class ${\cal S}$ theories, \cite{Neitzke:2014cja}. In particular, we highlight the fact that determining the Hitchin base, namely the Coulomb branch of the moduli space of vacua of class ${\cal S}$ theories, from the Higgs branch, is equivalent to specifying the identity in the embedding category, in complete analogy with the criterion outline din section \ref{sec:DCFMQS} and \ref{sec:334}.

\item Ultimately, we combine the tools outlined in the remainder of the present treatment, together with our previous work, \cite{Pasquarella:2023ntw}, explaining how in presence of algebraic varieties related to Moore-Tachikawa varieties without categorical duality, admit a generalised notion of 3D mirror symmetry thanks to the analogy with algebraic gauging in SymTFT constructions leading to intrinsically-non-invertible defects.

\end{enumerate}

\subsubsection{Hilbert series of Hochschild cohomology of block algebras}

In this first subsection, we will briefly overview a key result of \cite{Ba}, combining it with material outlined in section \ref{sec:DCFMQS}.

Let $p$ be a prime, $k$ an algebraically closed field of characteristic $p$, $G$ a finite group, and $B$ a block algebra of $kG$. A defect group of $B$ is a minimal subgroup $P$ of $G$ such that $B$ is isomorphic to a direct summand of $B\otimes_{_{kP}}B$ as a $B-B$-bimodule. The defect groups of $B$ form a $G$-conjugacy class of $p$-subgroups of $G$, and the defect of $B$ is the integer $d(B)$ such that $p^{^{d(B)}}$ is the order of the defect group of $B$.  

Donovan's conjecture predicts that there should only be finitely-many Morita equivalence classes of block algebras of Hochschild cohomology algebras of block algebras with a fixed defect $d$. For any integer $n\ \in\ \mathbb{N}$, dim $HH^{^n}(B)$ is bounded in terms of the defect groups of the block algebra $B$. 

The first result in \cite{Ba} adds to this the fact that there is a bound on the degrees of the generators and relations of the Hochschild cohomology $HH^{\bullet}(B)$, which only depends on the defect of $B$. In \cite{Ba}, this comes in three theorems, ultimately claiming that, the Hilbert series of the Hochschild cohomology of a block and its defect determine each other up to finitely many possibilities. In particular, 

\begin{equation} 
HH^{^{{\bullet}\ 0}}(B)\ \equiv\ \mathfrak{Z}(B),    
\end{equation}
where $\mathfrak{Z}(B)$ is the Drinfeld center.

\subsubsection{Hitchin systems and mirror quivers} 

We now turn to the more specific setup of interest to us, namely that of class ${\cal S}$ theories. In particular, we are interested in their moduli space of vacua, as well as their description in terms of 2D TQFTs, thanks to the AGT correspondence.

\subsection*{Hitchin systems}  

A Hitchin systems is an integrable system depending on the choice of a complex reductive group and a compact Riemann surface. It lies at the interface of algebraic geometry, Lie algebra theory, and integrable systems. It also plays a crucial role in the geometric Langlands program.  

The Hitchin equations consist in a system of PDEs on the Riemann surface, $C$, concerning a triple $(E, D, \varphi)$, \cite{Neitzke:2014cja}, where: 

\begin{enumerate} 

\item $E$ is a $G$-bundle on $C$.  

\item $D$ is a $G$-connection in $E$.  

\item $\varphi\ \in\ \Omega^{^1}\ (\text{End}\ E)$.  

\end{enumerate}

\subsection*{Spectral curves and Hitchin fibrations}  

The Higgs branch moduli space, ${\cal M}_{_H}$, is an example of one such complex integrable system. In particular, it is a fibration over a complex base space, ${\cal B}$, where the generic fiber is a compact complex torus. 

For gauge groups $G=SU(N)$ or $G=PSU(N)$, given a Higgs bundle $(E_{_h}, \phi)$, then the eigenvalues of $\phi$ in the standard representation of $G$ give an $N$-sheeted branch cover of $C$, \cite{Neitzke:2014cja},

\begin{equation}  
\Sigma\ \overset{def.}{=}\ \left\{\left(z\in C,\lambda\in T^{^*}_{_z}C\right):\text{det}(\phi(z)-\lambda)=0\right\}\ \subset\ T^{^*}C,      
\end{equation}    
which is the spectral curve corresponding to the Higgs bundle, $\left(E_{_h}, \phi\right)$.

The moduli space of harmonic bundles carries a natural family of complex structures $J_{_{\zeta}}$, parameterised by $\zeta\ \in\ \mathbf{CP}^1$. This follows form the fact that ${\cal M}$ carries a natural hyperk$\ddot{\text{a}}$hler metric, \cite{Neitzke:2014cja}. To see this, fix a $G$-bundle $E$. Then let ${\cal C}$ denote the space of pairs $(D,\varphi)$ without imposing the Hitchin equations. By definition, ${\cal C}$ is an infinite-dimensional affine space, with a natural hyperk$\ddot{\text{a}}$hler structure. Moreover, ${\cal C}$ is naturally acted on by the gauge group ${\cal G}$. This action preserves the hyperk$\ddot{\text{a}}$hler structure and has a moment map $\Vec{\mu}$. The Hitchin equations imply that all 3 components of $\Vec{\mu}$ must vanish. Denoting by ${\cal G}$ the smooth sections of Aut $E$, we therefore have

\begin{equation}  
{\cal M}\ \equiv\ \mu^{^{-1}}(0)/{\cal G},     
\end{equation}   
namely the hyperk$\ddot{\text{a}}$hler quotient 

\begin{equation}  
{\cal M}\ \equiv\ {\cal C}\dslash {\cal G},   
\label{eq:hkq}   
\end{equation}  
implying ${\cal M}$ is in turn hyperk$\ddot{\text{a}}$hler. Every hyperk$\ddot{\text{a}}$hler manifold carries a canonical family of complex structures parametrised by $\zeta\  \in\ \mathbf{CP}^1$, in this case denoted by $J_{_{\zeta}}$.  

Let ${\cal B}$ denote the space of all $N$-sheeted branched covers $\Sigma\ \subset T^* C$ of $C$. In this case, ${\cal B}$ is a finite-dimensional complex vector space. Passing from the Higgs bundle to $\Sigma$ gives a projection known as the Hitchin fibration 

\begin{equation}  
{\cal M}_{_H}\ \rightarrow\ {\cal B}, 
\end{equation}
where ${\cal B}$ is the Hitchin base. 

As thoroughly explained in \cite{Neitzke:2014cja}, ${\cal B}$ is the Coulomb branch of the moduli space of vacua of the class ${\cal S}$ theory in question, conventionally denoted $X_{\mathfrak{g}}(C)$. At any nonsingular point in the Coulomb branch, there is an electromagnetic charge lattice 

\begin{equation} 
\Gamma\ \rightarrow\ {\cal B}_{_{reg}}\ =\ {\cal B}/{\cal B}_{sing}.  
\end{equation} 

At every point of base, $u\ \in\ {\cal B}$, there is a spectral curve, $\Sigma_{_u}$, thereby determining the local lattice system as the first cohomology, $H_{_1}(\Sigma_{_u},\mathbf{Z})$.

\subsubsection{Looking for the identity}  

In this concluding part of the section, we are now ready to gather all the ingredients outlined in the previous part of this work, to show how it relates to the formalism outlined in \cite{Pasquarella:2023ntw}. 

In doing so, we mostly need to recall the notion of the identity in Moore-Tachikawa varieties, and their generalisation for the case in which the target 2-category is a hyperk${\dddot{\text{a}}}$hler quotient

      \begin{equation}  
      T^{^*}\tilde G_{_{\mathbf{C}}}\ \overset{def.}{=}\ 
T^{^*}G_{_{\mathbf{C}}}^{^{\prime}}\ \otimes_{_{T^{^*}G_{_{\mathbb{C}}}}}\ T^{^*}G_{_{\mathbf{C}}}^{^{\prime\prime}}.
\label{eq:newidnow}
\end{equation} 

As explained in \cite{Pasquarella:2023ntw}, in such case, the identity is naturally removed by construction, and this clearly fits with taking the categorical quotient according to the GIT prescription, \eqref{eq:hkq}. From the work of Moore and Tachikawa, we also know that such categories are non-dualisable, and, consequently, not associated to a fully-extended TQFT.  

As explained in section \ref{sec:334}, full-extendibility can be recovered only upon specifying the identity element in the fully extended mother category from which the GIT quotient descends, \eqref{eq:prescr}. In particular, as suggested from \eqref{eq:newidnow}, the identity itself might be a composite object, as the one arising when performing double gauging of a SymTFT leading to non-invertible defects, \cite{Pasquarella:2023deo}.  

This partial gauging and gluing of different theories is precisely of the type occurring in the context of supersymmetric quiver gauge theories whose Higgs branch admits a magnetic quiver description, \cite{Cabrera:2019izd}, as well as 3D mirrors of Sicilian theories, \cite{Benini:2010uu}. We plan to report a more detailed treatment for these specific examples in due course, \cite{VP}.

At first, we briefly overviewed cochain level theories as the most important generalisation of the open and closed TFT construction, emphasising its relation to the SymTFT construction leading to absolute theories. Mostly relying upon \cite{Moore:2006dw}, we highlighted the importance of the construction of cobordism operators, emphasising their dependence on the conformal structure of the Riemann surface. In section \ref{sec:BMTCs} we then turned to the discussion of a particular 2D TFT valued in a symmetric monoidal category, namely the maximal dimension Higgs branch of class ${\cal S}$ theories. After briefly reviewing the properties outlined in \cite{Moore:2011ee}, we propose their generalisation for the case in which the target category of the $\eta_{_{G_{_{\mathbb{C}}}}}$ functor is a hyperk$\ddot{\text{a}}$hler quotient. We concluded outlining the possible extension of this treatment towards a mathematical formulation of magnetic quivers within the context of Coulomb branches of 3D ${\cal N}=4$ quiver gauge theories which will be addressed in an upcoming work by the same author.

\subsection{Where to next?}   \label{sec:6}

With the discussion outlined in Part II, III and V, we have shown the importance of abelianisation in geometric representation theory. Specifically, we highlighted the common ground between \cite{Freed:2022qnc,Dimofte:2018abu,Teleman:2014jaa}. 

As explained at the very beginning of our work, one of our key motivations is the need to deepen the mathematical toolbox adopted by Phenomenologists in answering open questions in Particle Physics. Recent developments in the Pheno community have started relying upon the calculation of the Hilbert series, most importantly \cite{Graf:2022rco,Bento:2023owf,Anisha:2019nzx}, successfully encoding key fundamental processes, such as the Higgs mechanism. The crucial point of our analysis, though, is that, albeit Hilbert series calculations are indeed very useful for keeping track of the degrees of freedom and symmetries of a given theory, in principle they are not enough to determine the exact operator content of the theory in question. 

On the other hand, ring homologies, on which Hilbert series are calculated, encode crucial properties of the QFT at hand, enabling a much deeper understanding of its underlying categorical structure. The key take-home message of this work is the crucial role played by abelianisation for identifying the underlying mathematical structure of a given QFT. This concept will be investigated further for the specific case of the SM and its extensions in future work, \cite{VP}. 

Additional complementary directions addressed by the same author will be developing ring homologies for Moore-Tachikawa varieties beyond categorical duality\footnote{Thereby following up from the analysis carried out in \cite{Pasquarella:2023ntw}.}, and understanding their implications for Koszul duality. We plan to report of any advancements in this regard in the near future.

\subsection{Conclusions and Outlook} 

With this work, the aim of the author was that of continuing from past developments, bridging the gap in between the Pure Mathematics and the Theoretical Particle Physics. The purpose of this thesis was mainly:  

\begin{itemize}  

\item   To illustrate (some) key fundamental problems in Theoretical Particle Physics necessitating further mathematical understanding.   

\item   Opening the scene to well-established techniques in representation theory and categorical algebraic geometry to help in this regard.    

\item   For those who are not experts in the field, serve as an oversimplified explanation of the powerful formalism put forward by BFN, and connecting it with Moore-Tachikawa varieties.   

\item Motivating Phenomenologists to engage more with Pure Mathematicians.

\end{itemize}

The main objective of this Part was that of explaining the higher categorical structures that are needed for describing the invariants associated to specific supersymmetric quiver gauge theories, with a particular focus on dualities and their mutual relations in terms of higher-categories. We plan of reporting on any future developments in this regard in the near future, \cite{VP}.

\section{Conclusions and Outlook of Part V}

Part V of this work is structured as follows: 

\begin{enumerate}

\item After a preliminary list of definitions, section \ref{sec:cd} introduces factorisation homology and its applicability within the context of class ${\cal S}$ theories by means of the AGT correspndence. The section ends recapitulating the interplay between dualisability and full-extendibility of TQFTs, with a particular focus of algebraic varieties. This follows the treatment of section \ref{sec:clt}, which briefly overviewed cochain level theories as the most important generalisation of the open and closed TFT construction. Mostly relying upon \cite{Moore:2006dw}, we highlight the importance of the construction of the cobordism operator, highlighting its dependence on the conformal structure of the Riemann surface. We then turn to the discussion of a particular 2D TFT valued in a symmetric monoidal category, namely the maximal dimension Higgs branch of class ${\cal S}$ theories.

 \item Section \ref{sec:333} explains the relation in between the Drinfeld center and magnetic quivers, combining sections \ref{sec:clt} and \ref{sec:symtft}.

\item In section \ref{sec:last} we propose the generalisation of Moore-Tachikawa varieties for the case in which the target category of the $\eta_{_{G_{_{\mathbb{C}}}}}$ functor is a hyperk$\ddot{\text{a}}$hler quotient. We conclude outlining the possible extension of this treatment towards a mathematical formulation of magnetic quivers within the context of Coulomb branches of 3D ${\cal N}=4$ quiver gauge theories which will be addressed in an upcoming work by the same author, \cite{VP}. The concluding part of this section recapitulates the essential tools required for performing GIT quotient constructions, highlighting its importance for realising 3D mirror symmetry. In particular, we focus on the calculation of the Hilbert series as specific examples of algebraic varieties of crucial interest in the study of categorical dualities for supersymmetric quiver gauge theories. We conclude showing how Coulomb and Higgs branches from class ${\cal S}$ Hitchin systems carry the same information as the gaugeable algebras leading to SymTFT constructions, discussed in Part III. We conclude outlining interesting open questions and future directions of investigation by the author.

\end{enumerate}

\part{Conclusions and Outlook}

\section{Conclusions and Outlook}   

String theory is a promising candidate unifying theory of fundamental interactions. In its fermionic formulation, it involves ten dimensions, of which nine are spatial and one is temporal. 
Crucially, such formulation strongly relies upon Supersymmetry (SUSY), which is a correspondence in between bosonic and fermionic particles. 

According to present-day evidence, SUSY is certainly broken in our world, and plenty of effort has been made in identifying traces of such symmetry breaking from the observational side, eventually leading to the extension of the Standard Model (SM) by encoding additional particles and interactions.  
Understanding how the Standard Model can be embedded in String Theory is certainly a step forward towards identifying patterns of SUSY breaking in our world. 
Albeit being broken, SUSY certainly plays a crucial role in understanding which String Theory model is most appropriate for the SM to arise as an effective field theory (EFT), precisely by studying its breaking.

String theory vacua, where the SM would eventually be embedded, are parametrised by the so-called moduli space, and the latter is known to admit a categorical algebro-geometric description. Crucially, great advancements in the latter took place simulatneously to the birth and development of SUSY and String Theory. In recent years, joint collaboration in between pure mathematicians and mathematical physicists, primarily Freed, Hopkins, Lurie, Moore, Segal, and Teleman, shed light to new interesting perspectives which happen to be particularly relevant for the objectives outlined above. The main feature of their approach is that of applying the powerful tool of higher-categorical structures within the context of gauge theories and their supersymmetric counterparts, with the ultimate aim being that of explaining what is the underlying mathematical structure of a quantum field theory (QFT). Their work is based upon identifying underlying topological structures from which partition functions, and correlation functions can be derived. Furthermore, their formalism happens to be particularly efficient for keeping track of the spectrum (i.e. the particle-field content) of a given QFT, and how the latter changes under gauging of the global symmetries of the theory one started from.

Up to now, most of the mathematical control in describing higher-dimensional field theories has been for descendants of maximally-supersymmetric 6D ${\cal N}=(2,0)$ superconformal field theories (SCFTs), from which a rich web of dualities arises upon dimensional reduction to, either, 4 or 3 spacetime dimensions. Such descendants feature fewer supersymmetries with respect to their non-Lagrangian parent theory. In particular, the most studied are 4D ${\cal N}=2$ (also known as class ${\cal S}$ theories), and 3D ${\cal N}=4$ SCFTs, exhibiting exactly half of the amount of supersymmetry of their parent 6D theory.  

From the work of Alday, Gaiotto and Tachikawa (AGT), it is known that class ${\cal S}$ theories can be equally described in terms of a 2D topological field theory (TFT) living on the Riemann surface on which dimensional reduction has been performed. Such a 2D TFT is defined to be a functor in between 2-categories. Such 2-categories were first studied by Moore and Tachikawa, which, in turn, relied upon the pioneering work of Moore and Segal. The main point of their work is that, for a certain class ${\cal S}$ theory to be uniquely defined (i.e. for its spectrum of operators to be uniquely determined), one needs to fully specify the source and target 2-categories of the 2D TFT corresponding to the given 4D ${\cal N}=2$ SCFT in question. In technical terms, the objects of the source 2-category are circles, and the 1-morphisms between them are bordisms. The target 2-category is obtained by acting with the functor on these objects and 1-morphisms, resulting in the gauge group of the 4D dual theory, and the bordism operators (also known as the Moore-Segal bordism operators), with the latter being related to algebraic varieties defining moduli spaces in algebraic geometry, i.e. the Higgs and Coulomb branches.

Higgs and Coulomb branches of supersymmetric gauge theories are fascinating objects to explore, since they shed light on dualities characterising the theory they are associated with. From a pure mathematical point of view, they can be realised in terms of the so-called geometric invariant theory (GIT) construction, and quotienting. For the case of 3D ${\cal N}=4$ SCFTs, exchange in between Higgs and Coulomb branches corresponds to a statement of 3D mirror symmetry between two theories (first studied by Intriligator and Seiberg from a pure theoretical-physics point of view). This 1-to-1 correspondence extends to the calculation of invariants, most importantly the Hilbert Series, accounting for all the gauge-invariant operators of the theory.

However, as also pointed out by Moore, Segal and Tachikawa, their analysis lacked a rigorous mathematical treatment, most importantly for the case in which the target category of the functor is a hyperk$\ddot{\text{a}}$hler quotient, which is indeed the case when describing algebraic varieties such as the Higgs branch of a given theory. The work we have conducted so far provides a step forward towards achieving a rigorous mathematical formulation of such 2-categorical structures for the reason that we will now outline.

As previously explained, TFTs play a crucial role in bridging the gap between QFT and Mathematics. Because of this, the formalism outlined above suggests to always recast a given QFT to a TFT description. For example, 3D ${\cal N}=4$ SCFTs can be turned into 3D TFTs by performing a topological twist. Unlike ordinary 3D TFTs, which are usually trivial in nature in the sense that there are no local degrees of freedom, topologically-twisted TFTs come with nontrivial boundaries, called Lagrangian submanifolds. Importantly, the latter geometrise the boundary conditions of the field content of the theory in question, and, upon studying the intersection (i.e. the 1-morphisms) in between these Lagrangian submanifolds, one can reformulate the statement of 3D mirror symmetry of the parent 3D ${\cal N}=4$ SCFT as that of homological mirror symmetry, which is essentially a duality in between 2-categories. The steps described so far clearly show the advantage of bringing the description down to that the 2-category of Lagrangian submanifolds, which is exactly what we need to keep track of the field content of the theory we started with.  

The 3D mirror symmetry/homological mirror symmetry correspondence has been extensively studied for the specific case in which the correspondence in between the Coulomb and Higgs branch in 3D is 1-to-1.  However, recent developments within the context of supersymmetric quiver gauge theories, mostly led by Hanany and collaborators, have shown that there are cases in which Higgs branches of fermionic theories can be broken down into several constituents, each one associated with the Coulomb branch of a certain 3D ${\cal N}=4$ SCFT (a.k.a. the cone structure), with the latter being described by the so-called magnetic quivers. Such correspondence was primarily identified by studying the Hilbert series as a geometric invariant. In one of our recent works \cite{Pasquarella:2023exd}, we explained how the emergence of this multi-cone structure in the Higgs branch is equivalent to a generalised statement of 3D mirror symmetry, in turn descending form relaxing constraining assumptions on which homological mirror symmetry is based. As a follow up, in \cite{Pasquarella:2023ntw} we proposed a generalisation of the Moore-Tachikawa varieties for the case in which the target category of the 2D TFT is a hyperk$\ddot{\text{a}}$hler quotient. As proved in the former of these two papers, achieving this requires generalising the bordism operators of Moore and Segal to the case involving lack of reparametrisation-invariance on the Riemann surface, ultimately enabling to relate this to the issue of defining a Drinfeld center for composite class ${\cal S}$ theories. 

So far, our main focus has been on class ${\cal S}$ theories, and 3D ${\cal N}=4$ SCFTs, both featuring 8 supercharges. 
To what extent the generalised 3D mirror symmetry/homological mirror symmetry correspondence can be extended to algebraic varieties associated with theories featuring different amounts of supersymmetries is part of my currently ongoing research. The ultimate aim would be to extend this formalism to quiver gauge theories describing the SM and its possible extensions, eventually shedding light on the mathematical description of SUSY breaking mechanisms leading to observable EFTs.

This work explored the importance of 2D structures towards describing higher-dimensional quantum field theories (QFTs). In doing so, we relied upon the joint role played by dualities and categories.

The present work was structured in six parts:

\begin{enumerate} 

\item At first, we introduced the notion of relative QFTs, with detailed explanation for the case of sigma models, gauge theories, and 6D SCFTs. We then turned to the main case of interest to us, namely Coulomb branches of star-shaped quiver gauge theories, introducing the Kirwan map, and the notion of abelianisation, naturally paving the way to the introduction of category theory, which played a crucial role in the remainder of our work.

\item Part II introduced the main ingredients of our treatment, namely categories and dualities. After having introduced higher-categorical structures, we provided two specific examples that were recursively encountered throughout our treatment, specifying how they are realised in the main example outlined in Part II thanks to the work of Moore and Tachikawa, \cite{Moore:2011ee}. Category theory is particularly useful for describing dualities, which is another crucial ingredient characterising this thesis. The dualities we encountered are: 

\begin{itemize}

\item Homological mirror symmetry.

\item 3D mirror symmetry. 

\item The Alday-Gaiotto-Tachikawa (AGT) correspondence.

\item Holographic duality. 

\end{itemize}   

Apart from revising well-known results in the literature, we highlight the importance of going beyond full categorical dualisability in the context of Moore-Tachikawa varieties, to which Part V was devoted. In doing so, we related their formalism to that of SymTFT constructions of Freed, Moore and Teleman, \cite{Freed:2022qnc}, briefly overviewed in Part III.

\item  Part III explained how QFTs arise from knot invariants. Specifically, we explained the Symmetry Topological Field Theory (SymTFT) construction of Freed, Moore, and Teleman. In particular, we focussed on the dimensional reduction of non-Lagrangian theories, specifically 6D ${\cal N}=(2,0)$ SCFTs, \cite{Witten:1995zh,Witten:2007ct}, and the categorical structure of the resulting class ${\cal S}$ theories, \cite{Pasquarella:2023deo}. Making use of the AGT correspondence, \cite{Alday:2009aq}, we turned to a description in terms of 2D topological theories defined as functors between 2-categories. 

This Part mostly centered around: 

\begin{itemize} 

\item The algebro-geometric formulation of moduli spaces, and their construction by means of the Geometric Invariant Theory (GIT) construction. Importantly, we relate this construction to the abelianisation procedure outlined in our main example in section \ref{sec:main}, paving the way to further analysis in Part V.

\item Turning from QM to Symplectic Geometry by means of the WKB approximation. 

\end{itemize} 

Both topics have thoroughly been addressed in the literature, and we therefore devoted some sections in the appendix to explaining them in greater depth.

\item Part IV built from open questions pointed out in the first paper by the author and her supervisor, \cite{DeAlwis:2019rxg}. In addressing this topic, we restrict to lower-dimensional setups in order to deal with open issues usually arising in higher-dimensional settings, extensively addressed in the literature, \cite{Coleman:1980aw,Brown:1988kg,Fischler:1990pk}. In doing so, we built connection with setups involving the emergence of islands, as well as recent developments in the context of von Neumann algebras for calculating generalised entropies in gravitating systems and QFT, \cite{JM2, SWH, Banks:1983by, Giddings:1995gd, EW2, EW3, Witten:2021jzq}. The interesting connection with the remainder of our work is twofold: firstly, the need to resort to a certain dual picture in order to interpret the analytic results of the calculations performed; secondly, the richer structure resulting from the lower-dimensional analysis, emphasises the role played by internal structures in higher-dimensional theories, therefore suggesting alternative perspectives might be needed in order to address other open issues in QFT.

\item  Part V results from the merge of three single-authored papers, \cite{Pasquarella:2023exd,Pasquarella:2023ntw, Pasquarella:2023vks}. The key findings in this Part are:

\begin{itemize}

\item Moore-Tachikawa varieties without dualities are naturally realised in supersymmetric quiver gauge theories, and we therefore put forward a correspondence in between such theories, magnetic quivers, and Drinfeld centers, \cite{Pasquarella:2023exd}. As a result of the formalism adopted (namely the definition of Moore-Segal bordism operators, \cite{Pasquarella:2023ntw,Moore:2006dw}) applied to Moore-Tachikawa varieties without categorical duality, \cite{Moore:2011ee}, we provide further evidence of our conjectured relation in between homological mirror symmetry and 3D mirror symmetry, first presented in \cite{Pasquarella:2023exd}.

\item  The last paper written by the author prior to submitting this Thesis, \cite{Pasquarella:2023vks}, nicely ties together the topics covered in \cite{Pasquarella:2023exd,Pasquarella:2023ntw}, while also preparing the stage for the future research the author will be conducting as a postdoc\footnote{Partly outlined in \cite{Pasquarella:2024mlr}.}. In particular, this part takes a more algebro-geometric approach in explaining the mathematical structure underlying supersymmetric quiver gauge theories, with a particular focus on dualities and their mutual relations in terms of higher-categories. As will be explained in depth, the crucial role is played by factorisation homology, \cite{FHP}. We will not attempt to provide an exhaustive definition of it in this introduction, and, instead, refer the reader to the related section in the core of this work. What we can anticipate, though, is that factorisation homology is related with the counting of ground state degeneracy, and, when applied to the AGT correspondence, provides a tool for evaluating the 4-sphere partition function of class ${\cal S}$ theories. Specifically, as we shall see, factorisation homology allows us to relate categorical dualisability and full-extendibility\footnote{And is therefore in line with the connection between dualities and categorical structures.}. Within the context of non-Lagrangian theories, such as 6D ${\cal N}=(2,0)$ SCFTs this is particularly relevant, since it introduces the notion of ring objects, of which Hilbert series are one of the examples of most interest to us. As an algebraic variety, we focus on the calculation of the Hilbert series on Coulomb and Higgs branches, with the latter being related by the geometric invariant theory (GIT) quotient construction, \cite{Deligne-Mumford}. As such, the GIT quotient provides a recipe for realising 3D mirror symmetry; we therefore show how it can be suitably extended for cases where mirror duals are not necessarily in a 1-to-1 correspondence. In relation to \cite{Cabrera:2019izd}, this nicely fits with the magnetic quiver prescription, for describing the Higgs branch of certain supersymmetric quiver gauge theories with non-primitive ideals\footnote{Namely the 2-cone structure, \cite{Ferlito:2016grh}.}. We then show how the 2D description of disk algebras can be suitably adopted for describing Hitchin systems of class ${\cal S}$ theories.  We concluded outlining future directions that are currently under investigation by the same author.  

\end{itemize}

\item Part V closes outlining the main achievements of the present work, with an outlook on some key questions the author wishes to address in her future research.

\end{enumerate}

We believe our treatment opens the way to further investigating the mutual relation in between homological mirror symmetry and 3D mirror symmetry with the ultimate aim of aiding at understanding the underlying mathematical structure of QFT. These are currently under investigation by the same author, and we plan to report about any advancement in the field in the near future.



\section*{Acknowledgements}    

These acknowledgements are way far from reflecting my sincere gratitude towards the many people who trusted my passion, instinct, and dedication towards research. The names listed below are only part of the ones I wish to thank, and to whom I will be forever grateful. 

First and foremost, my supervisor, Professor Fernando Quevedo. To summarise my working experience with him, I would say our collaboration has been a unique blend of creativity and perseverance, from my side, and patience and wisdom, from his. Thank you for following my progress towards becoming an independent researcher: it has been an honour to be working with you.

I also wish to thank Ahmed Almheiri, Shanta de Alwis, Tudor Dimofte, I$\tilde{\text{n}}$aki Garc\'ia Etxebarria, Victor Gorbenko, David Jordan, Yolanda Lozano, Juan Maldacena, Gregory Moore, Hiraku Nakajima, Carlos Nu$\tilde{\text{n}}$ez, Fernando Quevedo, Jorge Santos, Nathan Seiberg, Marcus Sperling, Constantin Teleman, Mark Van Raamsdonk, and Edward Witten for insightful discussions and questions raised at different stages of this work.

A special acknowledgement goes to, both, the Hodge Institute in Edinburgh, and the Yau Center for Mathematical Sciences, Tsinghua University in Beijing, for their hospitality, where parts of this work were completed and first presented. In particular, many thanks to my respective hosts, Tudor Dimofte, David Jordan, Nicolai Reshetikhin and Shing-Tung Yau.  

I also wish to thank the organisers of all the events that I attended throughout my PhD, allowing this work to be presented at different stages, and the partial support by the STFC Consolidated HEP theory grant \\
ST/T000694/1 through DAMTP towards covering tuition fees. 

Being part of the DAMTP HEP Group made me feel at home while abroad, and I will be forever grateful for their warm welcome, which has certainly stimulated my research advancements in many ways. Many thanks to all other members of the HEP-Gang.
\medskip

Last but not least, I will be forever grateful to the place that raised me during the first six years at university, the beautiful Trieste, and to the sailor whose dream made mine come true.

\appendix

\section{Loop groups}   \label{sec:loop}

A \emph{loop group} is a group of loops in a topological group $G$ with multiplication defined pairwise. It is a group of continuous mappings from a manifold $M$ to a topological group $G$. Assuming $M=S^1$, the loop group is therefore formally defined as follows  

\begin{equation}  
LG\ \overset{def.}{=}\ \{\gamma:S^1\rightarrow G\ |\ \gamma\in\ C(S^1,G)\}.   
\end{equation}

Parametrising $S^1$ with $\theta$, we get that 

\begin{equation} 
\gamma:\ \theta\ \in\ S^1\ \mapsto\ \gamma(\theta)\ \in \ G , 
\end{equation}     
from which multiplication in $LG$ is defined as follows  

\begin{equation} 
\left(\gamma_{_1}\gamma_{_2}\right)(\theta)\ =\ \gamma_{_1}(\theta)\gamma_{_2}(\gamma). 
\end{equation}

Associativity follows from the associativity property in $G$. The inverse os given by 

\begin{equation}  
\gamma^{-1}:\ \gamma^{-1}(\theta)\ \equiv\ \gamma(\theta)^{-1},   
\end{equation}   

and the identity is defined as follows   

\begin{equation}    
\begin{aligned}
e:\ \theta\ \mapsto\ e\ \in\ G.
\end{aligned} 
\end{equation}

An important example of loop groups is the group of based loops on $G$, $\Omega G$. It is the kernel of the evaluation map

\begin{equation}    
\begin{aligned}
e_{_1}:\ &LG\ \rightarrow\ G  \\
&\ \gamma\ \mapsto\ \gamma(1),  
\end{aligned} 
\end{equation}
and as such it is a normal subgroup of $LG$. $G$ can also be embedded in $LG$, as the subspace of constant loops, leading to the split-exact sequence   

\begin{equation}   
1\ \rightarrow\ \Omega G\ \rightarrow\ LG\ \rightarrow\ G\ \rightarrow \ 1.
\end{equation}  

The space $LG$, therefore naturally splits as a semi-direct product   

\begin{equation} 
LG=\Omega G\rtimes G.    
\end{equation}

\section{Supersymmetric Field Theories}

\subsection{Physical Lagrangians}

A theory admitting a Lagrangian formulation, requires the Lagrangian to be physical, namely it must be:   

\begin{enumerate}

\item real

\item local  

\item Poincare'-invariant.  

\end{enumerate}  

The Poincare' group, $P^{^n}$ is a subgroup of the global symmetry group of the theory, and, according to the Coleman-Mandula Theorem, the whole symmetry group of the theory takes the following form  

\begin{equation}   
P^{^n}\ \times\ G,   
\label{eq:timesproduct}
\end{equation}  
with $G$ denoting a compact Lie group. By definition, symmetries give rise to conserved quantities. For the case at hand, \eqref{eq:timesproduct} leads to conservation of energy and momentum, for the Poincare' group, and electric charges and quantum numbers for the compact Lie group.  

The constraints of the Coleman-Mandula Theorem pertinent to bosonic theories can be relaxed in presence of fermions by extending the Poincare' group to the super-Poincare' group, $P^{^{n|s}}$, which is a super Lie group\footnote{Here, $P^{^{n|s}}$ denotes a supersymmetric extensions of the Poincare' algebra $P^{^n}\equiv\text{Lie}(P^{^n})$, whose odd part has dimension $s$.}. The supersymmetric extension of the Coleman-Mandula Theorem, results in stating that the only allowed super Lie groups are of the kind

\begin{equation}   
P^{^{n|s}}\ \times\ G.
\label{eq:timesproduct}
\end{equation}  

For the sake of completeness, we will briefly outline the main definitions of the terminology adopted in this section, \cite{Freed:1999mn}. 

\begin{figure}[ht!]  
\begin{center}     
\includegraphics[scale=1]{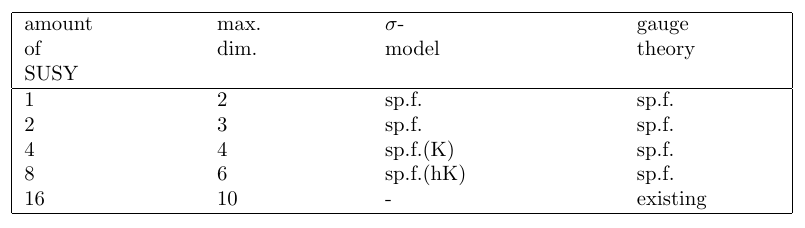}    
\label{fig:susytable}
\caption{\small}  
\end{center}    
\end{figure}  

A \emph{super Lie group} is a group object in the category of supermanifolds. A \emph{super Lie algebra} is a $\mathbb{Z}/2\mathbb{Z}$-graded vector space

\begin{equation}  
\mathfrak{g}=\mathfrak{g}^{^0}\oplus\mathfrak{g}^{^1},     
\end{equation}
equipped with a bracket operation  

\begin{equation}   
[.,.]\ :\ \mathfrak{g}\otimes\mathfrak{g}\ \rightarrow\ \mathfrak{g},     
\end{equation}
which is skew-symmetric and satisfies the Jacobi identity, with

\begin{equation}  
[\mathfrak{g}^{^1},\mathfrak{g}^{^1}]\ \in\ \mathfrak{g}^{^0}.  
\end{equation}   

The role of the super Poincare' group is that of restricting the possible vector spaces (i.e. particle content) and masses that can occur in a given theory. Mathematically speaking, an irreducible representation of $P^{^{n|s}}$ may be realised as the space of sections of a homogeneous $\mathbb{Z}/2\mathbb{Z}$-graded vector bundle over a mass shell. The extension to the case of interacting theories follows suite by noticing that one may use the free-field approximation at any vacuum, giving rise to a certain $P^{^{n|s}}$ representation, which, in turn, follows from the free theory constraints.

\subsection{UV/IR flows and their parametrisation} 

\medskip  

\medskip

\medskip  

\medskip

\subsection{The Witten index} \label{sec:wittenindex}    

\medskip  

\medskip

Supersymmetric theories are characterised by the presence of supersymmetric charge operators, ${\cal Q}$, such that bosons, $|B>$, and fermions, $|F>$, of energy $E$, are transformed into each other 

\begin{equation}  
{\cal Q}|B>=\sqrt{E\ }|F>  
\ \ \ 
,    
\ \ \ 
{\cal Q}|F>=\sqrt{E\ }|B>. 
\label{eq:fb}   
\end{equation} 

\eqref{eq:fb} is equivalent to the statement that non-zero energy states come in Bose-Fermi pairs. This simple statement suggests how to establish whether SUSY is broken or not, namely by counting the zero-energy states. The key element for doing so was first proposed by Witten in \cite{Witten:1982df}, and reads

\begin{equation}  
{\cal I}\overset{def.}{=}\text{Tr}(-1)^{F}\ e^{-\beta H}  
\label{eq:WI}   
\end{equation}
with $F, \beta$ and $H$ denoting the fermion number, the inverse temperature and the energy of the sate, respectively. Due to \eqref{eq:fb}, \eqref{eq:WI} is topological in nature and, by definition, only depends on the number of ground states. In particular, if ${\cal I}\neq 0$, the theory is certainly supersymmetric. Furthermore, it is independent w.r.t. $\beta$ or any other parameter of the theory. Interestingly, though, there is a way in which ${\cal I}$ can be effectively influenced by the features of a specific theory, namely if the number of ground states changes. As we shall see, one way of achieving this is via \emph{spectral flow}, \cite{Schwimmer:1986mf}. The latter will be the main focus of the next sections. However, for consistency, we already provide some qualitative arguments for its importance already at this stage.  

Essentially, the spectral flow is implemented by an element of the superconformal algebra, enabling to interpolate between states defined as path-integrals with antiperiodic (fermionic) and periodic (bosonic) boundary conditions, i.e. between the NS- and R-sectors of a certain SUSY theory. Because of this, the action of the spectral flow is equivalent to changing the ground state degeneracy of the original SUSY theory.

\medskip  

\medskip

\subsubsection{R-symmetry, supergroups, superalgebras and twisting} \label{sec:sca}   

\medskip  

\medskip

Throughout this work, we will often be referring to many types of dualities. Even though they might seem every different in nature, one of the aims of our analysis will be furthering the mutual relation in between them, highlighting how and what one can tell about the others.  

Among the first examples of dualities relevant for our purposes are holography, \cite{Bigatti:1999dp}, and modular transformations. The former was originally formulated as a correspondence in between type-IIB string theory on an AdS$_5\times S^5$, dual to 4D ${\cal N}=4$ SYM. The holographic dictionary comprises the equivalence of the partition function on the field theory and gravity sides as well as the correspondence in between their symmetries, given by the superalgebra $PSU(2,2|4)$. The bosonic subgroup is given by 

\begin{equation}    
SO(4,2)\times SO(6)\ \subset\ SU(2,2)\times SU(4)\ \subset\ PSU(2,2|4)    
\end{equation}
consistently with the isometries of AdS$_5$ and $S^5$, as well as the conformal symmetry and the R-symmetry of the CFT$_4$. By definition, an $R-symmetry$ is one that does not commute with SUSY. As we shall see, its role will be essential throughout our treatment.

Modular transformations, instead, come in 2 kinds, denoted $S$ and $T$. In string theory setups, they amount to changing the Yang-Mills coupling, $\tau_{_{YM}}$, as follows   

\begin{equation}  
\tau_{_{YM}}\ \overset{S}{\longrightarrow}\ -\frac{1}{\tau_{_{YM}}} 
\ \ \ \ \ \ \ \ \ \ \ 
\tau_{_{YM}}\ \overset{T}{\longrightarrow}\ \tau_{_{YM}}+1    
\end{equation}    

At the level of the objects living in a certain string theory setup, modular transformations enable to map one theory into the other by acting on the objects with $S, T$. In particular, $S$ relates strong and weak coupling, and therefore leads to a similar setup as the spectral flow mentioned earlier on \ref{sec:wittenindex}.   

One way of testing these dualities, is by looking for quantities that are protected by supersymmetry. As also mentioned in \ref{sec:wittenindex}, one example is the matching of the spectrum of states characterising the Dp-branes. For example, the extended (i.e. ${\cal N}>1$) superalgebra of 4D massive particles at rest reads

\begin{equation} 
\{{\cal Q}_{\alpha}^{I},{\cal Q}_{\beta}^{\dag J}\}=2M\delta^{IJ}\delta_{\alpha\beta}+2i Z\Gamma_{\alpha,\beta}^0   
\label{eq:superconfalgebra}
\end{equation}  
with $I, J=1,..., {\cal N}$, labelling the number of supersymmetries, $\alpha,\beta=1,2,3,4$, the components of the Majorana spinor supercharge, and $Z^{IJ}=-Z^{JI}$ denoting the \emph{central charge} matrix. The components of the latter constitute a set of conserved quantities that commute with the other generators of the algebra. They re electric and magnetic charges coupling to the gauge fields belonging to the supergravity multiplet. Under the action of a unitary operator, $U$, it can be brought down to a block-diagonal form. The structure of the superconformal algebra therefore implies that the $2{\cal N}\times2{\cal N}$ matrix on the RHS of \ref{eq:superconfalgebra}    

\begin{equation}  
\left(\ \begin{aligned}
&M\ \ \ Z\\
&Z^{\dag}\ \ \ M\\   
\end{aligned}\ \right)
\end{equation} 
should be positive-definite with nonnegative eigenvalues $M\pm|Z_i|$, whre $i=1,...$, and $|Z_1|\ge|Z_2|\ge...$. This results in the so-called $BPS-bound$ 

\begin{equation}  
\boxed{\ \ 
|Z_1|\ \le M \  \color{white}{\bigg ]}}    
\label{eq:BPS}    
\end{equation} 
Saturation of the BPS bound leads to a supermultiplet shortening, due to the emergence of zeros in the superalgebra when $M\equiv |Z_1|$. The bigger the number of equal central charges, the smaller the amount of SUSY left.
Interestingly, BPS states, i.e. those with $M\equiv |Z_1|$, are stable and independent from the parameters of the theory as long as supersymmetry is preserved, and therefore shows exactly the same property as the Witten index, \eqref{eq:WI}. As already argued in \ref{sec:wittenindex}, this quantity, namely the BPS states, can be used for bridging the study of the theory from weak to strong coupling. 

However, the independence of \eqref{eq:BPS} from the specifics of the theory is lost once another representation becomes degenerate with the BPS multiplet, such as occuring due to the Higgs mechanism.  

For the purpose of our analysis, it is convenient to rewrite the superconformal algebra \eqref{eq:superconfalgebra} in terms of the $R$-charges   

\begin{equation} 
\{{\cal Q}_{\alpha}^{I},{\cal Q}_{\beta}^{\dag J}\}=2M_{\alpha\beta}\delta^{IJ}+2i \omega_{\alpha\beta}\epsilon^{IJ}R
\label{eq:superconfalgebra}
\end{equation}
with $\omega_{\alpha\beta}$ defining the symplectic form and $\epsilon^{IJ}$ the Levi-Civita symbol. For ${\cal N}=2$, $M_{\alpha\beta}\in USp(2|2)$ and $R$ being the $U(1)$ superconformal $R$-symmetry. Given that $R$ defines the eentral extension of the superalgebra, it follows that  

\begin{equation}   
[\ R,{\cal Q}_{\alpha}^{I}\ ]
=   
\epsilon^{IJ}{\cal Q}_{\alpha}^{J}  
\label{eq:noncomm}   
\end{equation}   

The lack of commutability in between the 2 sets of operators belonging to the extended superalgebra plays an essential role in the following analysis. In particular, it ensures that, as long as a relic R-symmetry is preserved, so too is SUSY (either totally or partially). These arguments therefore highlight the role played by the $R$-charges for protecting the index. As we shall see, \eqref{eq:noncomm} is at the heart of the noncommutativity in between 2 operations: the spectral flow and twisting. 

\medskip  

\medskip

\subsubsection{The importance of flavour to extremise \texorpdfstring{${\cal Z}$}{}}    

\medskip  

\medskip

In \cite{Jafferis:2010un, Intriligator:2003jj}, for 2 and 4D respectively, it has been shown that the partition function of a CFT can be determined by means of localisation techniques starting from the knowledge of a UV-Lagrangian of the theory as long as the coupling to the background curvature of the sphere on which localisation is performed preserves the superconformal algebra. Equivalently, this statement amounts to ensuring the superconformal algebra contains the exact $R$-charges of the IR-theory, thereby ensuring that, along the spectral flow, the index accounts for the correct number of ground states of the theory.

The present work aims towards analysing the effect of \emph{partial gauging} on, both, modular and holographic dualities in the attempt to provide a unique framework towards understanding the information loss paradox from a higher-dimensional realisation encompassing string theory and defect field theories. In pursuing such task, we will recast the issue of evaporating black hole microstate counting in the language of fusion of non-invertible defects.

\subsection{SUSY enhancement from branes at orbifolds}    

\medskip  

\medskip

Starting from a minimal model exhibiting the same SUSY enhancement w.r.t. the ABJM-theory, namely ${\cal N}=2$ Chern-SImons theory with gauge group $U(N)\times U(N)$, and $SU(2)\times U(1)$ gloal symmmetry. The field conetent consists of:  

\begin{itemize}   
\item 2 gauge fields: $A_{\mu}, \hat A_{\mu}$  

\item bi-fundamental bosons and fermions, $Z^A, \psi^A$, with $A=1,2$, and their hermitian conjugates.  

\end{itemize}  

Upper and lower indices denote the \textbf{2} and $\bar{\mathbf{2}}$-representations of the global $SU(2)$, respectively. Furthermore, the gauge and matter fields are equipped with gauge group indices, $a,b, \hat a, \hat b$, for $U(N)\times U(N)$. Explicitly, they would read $A^{a}_{\ \ \mu b}\ ,\  A^{\hat a}_{\ \ \mu \hat b}\ , \ Z^{A,a}_{\ \ \ \ \hat b}\ , \ \psi^{A,a}_{\ \ \ \ \hat b}$, etc.

For the given field content, the corresponding ${\cal N}=2$ action reads   

\begin{equation}   
S=\int d^3 x \left({\cal L}_o+{\cal L}_{CS}-V_{ferm}-V_{bos}\right)  
\label{eq:SN2}   
\end{equation}   

The kinetic terms are

\begin{equation}   
{\cal L}_o=\text{Tr}\left(-D_{\mu}Z^{\dag}_{A} D^{\mu}Z^A+i\psi^{\dag A}\gamma^{\mu}D_{\mu}\psi_A\right)  
\end{equation}  

with covariant derivative   

\begin{equation}   
D_{\mu}=\partial_{\mu}+i\left[A_{\mu},\ \right]   
\end{equation}  

whereas the Chern-Simons terms read

\begin{equation}   
{\cal L}_{CS}=\epsilon^{\mu\nu\rho}\text{Tr}\left(A_{\mu}\partial_{\nu}A_{\rho}+\frac{2i}{3}A_{\mu}A_{\nu}A_{\rho}-\left(A_{\mu}\longleftrightarrow\hat A_{\mu}\right)\right)  
\end{equation}   

The total action \eqref{eq:SN2} is invariant under SUSY transformations. In particular, we have that

\begin{equation}   
\delta Z^A=i\epsilon^{\dag}\epsilon^{AB}\psi_B  
\end{equation}   

meaning that supersymmetry interpolates between different representations of the fields. Furthermore, the action has an additional ${\cal N}=2$ SUSY dependent on the gauge group. In order to see this explicitly, we need to introduce a new local operator which enables for the representation to remain the same upon turning bosons to fermions and viceversa. As already shown in the literature, such local operator, $T_{ab}$ plays the role of a monopole operator ensuring gauge invariance of the fundamental fields of the theory. Correspondingly, the SUSY transformation rules turn into 

\begin{equation}   
\delta Z^{A a}_{\ \ b}=iT_{ac}\psi^{\dag A c}_{\ \ b}\epsilon     
\end{equation} 
Having introduced a new operator in the spectrum, gauge invariance is restored ensuring the supersymmetric transformations preserve the representation of the fields.

\medskip  

\medskip

\subsubsection{ABJM and mirror symmetry}    

\medskip  

\medskip 

ABJM is a specific example of generalised Gaiotto-Witten (GW) setups. Indeed, it can be thought of as arising from the RG-flow of a 4D ${\cal N}=4$ SYM on an interval with S-dual BCs. 

However, its 1$^{st}$ formulation, was mainly motivated by the search for an AdS$_4$/CFT$_3$ correspondence, that could potentially be applied to string theory compactification setups. It was therefore As such, it constitutes a concrete setup where non-invertibles can be put to test.

\medskip  

\medskip

\section{Higher-categorical symmetries from topological defects} \label{sec:1.1}   

\medskip  

\medskip

To each $D$-dimensional QFT, ${\cal T}$, it is possible to assign a $(D-1)$-dimensional category, ${\cal C}_{_{{\mathfrak T}}}$, defined as \emph{higher}-category if $D>2$. ${\cal C}_{_{{\mathfrak T}}}$ features $D$-levels: 

\begin{enumerate}  

\item 1$^{st}$ level: 0-morphisms, namely the objects living in ${\cal C}_{_{{\mathfrak T}}}$

\item 2$^{nd}$ level: 1-morphisms, defining the relations between the objects  

\item 3$^{rd}$ level: 2-morphisms between the 1-morphisms.  

\item iteratively up to $D-1$.   

\end{enumerate}

\begin{figure}[ht!]   
\begin{center}
\includegraphics[scale=0.7]{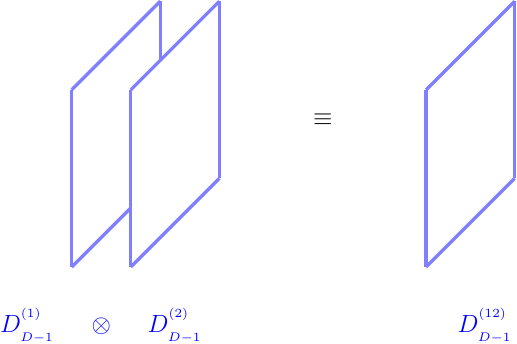}  
\ \ \ \ \ \ \ \ \ \ \ \ \ \ \ \ \ \ \ \ \ 
\includegraphics[scale=0.7]{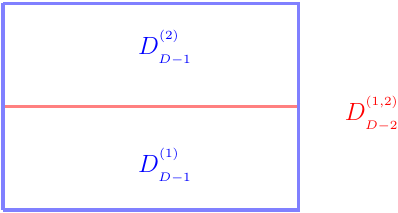}  
\caption{\small Fusion of objects in ${\cal C}_{_{\mathfrak T}}$. }
\label{fig:bfdef}  
\end{center} 
\end{figure}

\begin{figure}[ht!]   
\begin{center}
\includegraphics[scale=0.7]{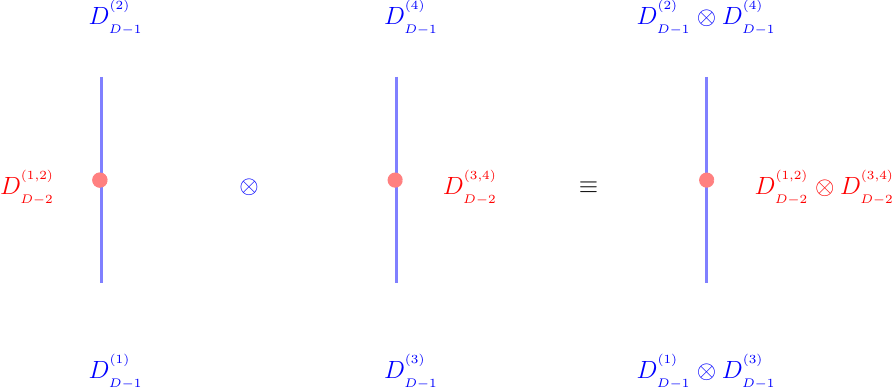}     
\caption{\small Generalised fusion between 1-morphisms.}
\label{fig:newfus}  
\end{center} 
\end{figure}

The objects are codimension-1 topological defects

\begin{equation}  
D_{_{D-1}}\ \overset{def.}{=}\ \bigoplus_{i}\ n_i\ D_{_{D-1}}^{^{(i)}}\ \ \ ,\ \ \ n_i\ge0    
\end{equation}  
where $D_{_{D-1}}^{^{(i)}}$ are codimension-1 topological defects, and $2\le i\ \le D$ denotes the $(i-1)$-morphisms between $(i-2)$-morphisms. $\sum_i n_i$ defining the total number of vacua such that in $n_i$ of them, $D_{_{D-1}}$ behaves as $D_{_{D-1}}^{^{(i)}}$. For the case of a simple object, there is a unique vacuum, and the defect carries a single topological local operator on its worldvolume. 

Under fusion, the objects give rise to a \emph{monoidal} structure   

\begin{equation}  
D_{_{D-1}}^{^{(1)}}\otimes D_{_{D-1}}^{^{(2)}}\ \equiv \ D_{_{D-1}}^{^{(12)}}  
\label{eq:fstr}   
\end{equation} 
as depicted on the LHS of figure \ref{fig:bfdef}.

\begin{figure}[ht!]   
\begin{center}
\includegraphics[scale=0.7]{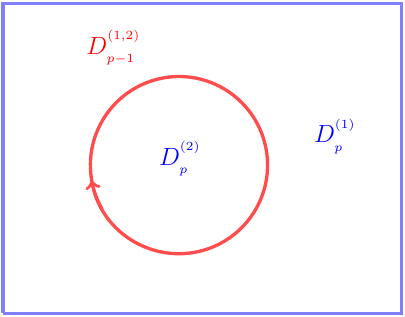}   
\ \ \ \ \ \ \ \ \ \ \ \ \ \ \ \ \ \ 
\includegraphics[scale=0.7]{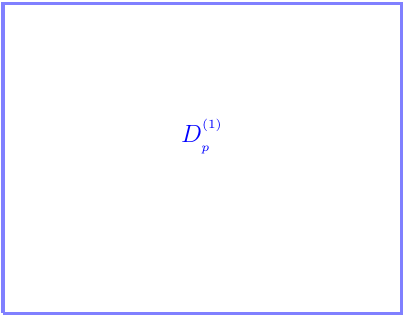}  
\caption{\small If $D{_{p}}^{^{(1)}}$ and $D{_{p}}^{^{(2)}}$ are related by gauging, wrapping the $D_{_{p-1}}^{^{(1,2)}}$ defect around an $S^{^{p-1}}$, bridges the LHS and the RHS of this figure. The algebra can therefore be placed on the defect for defining interfaces.}
\label{fig:contrloop}  
\end{center} 
\end{figure}

The fusion structure \eqref{eq:fstr} descends to a fusion structure of objects of a higher-category of symmetries localised along $D_{_{D-1}}$  

\begin{equation}  
D_{_{D-2}}^{^{(1,2)}}\otimes_{_{D_{_{D-1}}^{^{(2)}}}} D_{_{D-2}}^{^{(2,3)}}\ \equiv \ D_{_{D-2}}^{^{(2,3)}}\circ   D_{_{D-2}}^{^{(1,2)}}
\label{eq:fstr1}   
\end{equation}

For the fusion \eqref{eq:fstr1} to define a 1-morphism, ${\cal C}_{_{\mathfrak T}}$ is required to admit the existence of 2-morphisms, namely the theory has to be, at least, 3D. However, this does not prevent the possibility of defining other types of fusion rules, independently of the existence of higher-morphisms, such a as the one depicted in figure \ref{fig:newfus}.

\subsection{Generalised gauging and localised symmetries}    

\medskip  

\medskip

Let us consider the defects $D_{_{p}}^{^{(1)}}, D_{_{p}}^{^{(2)}}, D_{_{p-1}}^{^{(1,2)}},$ with the latter being defined such that, when wrapping an $S^{^{D-1}}$, $D_{_{p}}^{^{(2)}}$ vanishes. This corresponds to saying that $D_{_{p}}^{^{(1)}}$ and $D_{_{p}}^{^{(2)}}$ are related by gauging. Practically,  this relation implies that, for any given representative of $D_{_{p}}^{^{(1)}}$, we obtain a representative of $D_{_{p}}^{^{(2)}}$ upon gauging the worldvolume theory of $D_{_{p}}^{^{(1)}}$. Furthermore, all topological subdefects of $D_{_{p}}^{^{(2)}}$ can be obtained from those of $D_{_{p}}^{^{(1)}}$.      

This observation enables one to define a generalised gauging construction for any subcategory ${\cal C}_{_{{\mathfrak T}, D_{p}}}\ \subset\ {\cal C}_{_{\mathfrak T}}$. W.r.t. the original categorical theory ${\cal C}_{_{\mathfrak T}}$, the $D_{_{p}}$-defect is a $(D-p-1)$-morphism. The objects of the subcategory ${\cal C}_{_{{\mathfrak T}, D_{p}}}$ are $(D-p)$-morphisms of ${\cal C}_{_{\mathfrak T}}$ from $D_{_{p}}$ to itself. Its 1-morphisms are $(D-p+1)$-morphisms of ${\cal C}_{_{\mathfrak T}}$.   The fusion structure $\otimes$ of ${\cal C}_{_{{\mathfrak T}, D_{p}}}$ descends from $\otimes_{_{D_{p}}}$ on ${\cal C}_{_{\mathfrak T}}$, and, consequently, its fusion rules as well.

For $p=2$, the gauging is performed by acting with an algebra, $A^{^{(1,2)}}$, living in the 1-category ${\cal C}_{_{{\mathfrak T}, D_{2}^{(1)}}}$. The algebra itself consists of the following elements

\begin{equation}  
A^{^{(1,2)}}  
\ 
\overset{def.}{=}\ \left\{A_{_{1}}^{^{(1,2)}}, A_{_{0}}^{^{i}}\right\}    
\ \ \ 
,    
\ \ \     
i\equiv \alpha,\beta,\gamma,\delta  
\label{eq:defectalgebra}   
\end{equation}   
namely an object\footnote{Constituting a gaugeable 1-form symmetry. This is because the algebra described here is an example of an algebra that can be projected to the identity to achieve a gauged theory, namely quotienting the higher-category of topological defects in the SymTFT of the QFT in question, \cite{TJF}.} 

\begin{equation}   
A_{_{1}}^{^{(1,2)}}\ \overset{def}{=}\ D_{_{1}}^{^{(1,2)}}\otimes D_{_{1}}^{^{(2,1)}}    
\end{equation}
and canonical morphisms  

\begin{equation}  
\begin{aligned}
&A_{_{0}}^{^{\alpha}}\ :\ A_{_{1}}^{^{(1,2)}}\otimes A_{_{1}}^{^{(1,2)}}\ \rightarrow\ A_{_{1}}^{^{(1,2)}}\ \ \ \ \ \ , \ \ \ \ \ \ A_{_{0}}^{^{\beta}}\ :\ A_{_{1}}^{^{(1,2)}}\ \rightarrow\ A_{_{1}}^{^{(1,2)}}\otimes A^{^{(1,2)}}\ \\   
\\
&A_{_{0}}^{^{\gamma}}\ :\ A_{_{1}}^{^{(1,2)}}\ \rightarrow\ 1_{D_{2}^{^{(1)}}}\ \ \ \ \ \ \ \ \ \  \ \ \ \ \ \ \ , \ \ \ \ \ \  A_{_{0}}^{^{\delta}}\ :\ 1_{D_{2}^{^{(1)}}}\ \rightarrow\ A_{_{1}}^{^{(1,2)}}.   
\\
\\
\end{aligned}
\end{equation}

\medskip  

\medskip

\subsection*{Category of lines after gauging}    

\medskip  

\medskip  

After gauging, the resulting defect can be expressed in terms of \eqref{eq:defectalgebra} 

\begin{equation}  
D_{2}^{^{(2)}}\ \equiv\ \frac{D_{2}^{^{(1)}}}{A^{^{(1,2)}}}.     
\end{equation}

The gauging of ${\cal C}_{_{{\mathfrak T}, D_{2}^{(1)}}}$ is performed upon inserting a mixture of topological defects, constructed from $A^{^{(1,2)}}$, along the entire domain of the line defect $D_{_{1}}^{^{(1,2)}}$.   

For the case of arbitrary $p$, the gauging is described by a $(p-1)$-algebra in the $(p-1)$-category ${\cal C}_{_{{\mathfrak T}, D_{p}^{(1)}}}$   

\begin{equation}  
A^{^{(1,2)}}     
\ 
\overset{def.}{=}\ \left\{A_{_{p-1}}^{^{(1,2)}}, A_{_{p-2}}^{^{i}}\right\}    
\ \ \ 
,    
\ \ \     
1\le\ i\le\ p-1
\label{eq:defectalgebrap}   
\end{equation}   
where the morphisms (denoted by $i$) describe all possible creation and annihilation channels. $D_{p}^{^{(2)}}$ is obtained by placing a mixture of elements of $A_{_{p}}^{^{(1,2)}}$ along $D_{p}^{^{(1)}}$.

\medskip  

\medskip

\subsection{Fusion rules}    

\medskip  

\medskip  

The importance of having introduced the algebra to implement the gauging, manifests when looking at the fusion rules between the defects characterising a given theory. For ease of notation, we will now argue on the basis of a category ${\cal C}_{_{\text{id}, {\mathfrak T}}}$, and the fusion of 2D objects

\begin{equation}
D_{2}^{^{(O)}}\otimes D_{2}^{^{(O^{\prime})}}  \ = \ \bigoplus_{i} D_{2}^{^{(i)}},
\label{eq:fd}   
\end{equation}   
with $O, O^{\prime}$ denoting different orbits. The line operators describing the 1-morphisms

\begin{equation}
D_{2}^{^{(O)}}\otimes D_{2}^{^{(O^{\prime})}}  \ \rightarrow \  D_{2}^{^{(i)}}
\end{equation}   
arrange into orbits, $O^{\prime\prime}$, under the action of the gauge group $G$, such that the fusion can be re-expressed in terms of them 

\begin{equation}
D_{2}^{^{(O)}}\otimes D_{2}^{^{(O^{\prime})}}  \ = \ \bigoplus_{O^{\prime\prime}} D_{2}^{^{(O^{\prime\prime})}}
\end{equation} 
where the RHS is a representative of the equivalence class of the fusion on the LHS in the quotient theory ${\mathfrak T}/G$. The generalised gauging on top of each $D_{2}^{^{(O^{\prime\prime})}}$ term is encoded by an algebra $A^{^{(O^{\prime\prime}),(OO^{\prime})}}$ in the 1-category ${\cal C}_{_{\text{id}, {\mathfrak T}/G,D_{2}^{^{(O^{\prime\prime})}}}}$, with object   

\begin{equation}
A_{_{1}}^{^{(O^{\prime\prime}),(OO^{\prime})}}  \ \equiv \  \bigoplus_{i} n_{i} D_{1}^{^{(i)}}  
\end{equation}      
where $n_{i}$ denotes the dimension of the vector space constituted by local operators living at the end of $D_{1}^{^{(i)}}$ along $D_{1}^{^{(OO^{\prime}),(O^{\prime\prime})}}$. Consequently, after gauging, the fusion of surface defects, \eqref{eq:fd}, turns into   

\begin{equation}
D_{2}^{^{(O)}}\otimes D_{2}^{^{(O^{\prime})}}  \ = \ \bigoplus_{O^{\prime\prime}} \frac{D_{2}^{^{(O^{\prime\prime})}}}{A_{_{1}}^{^{(O^{\prime\prime}),(OO^{\prime})}}}
\end{equation} 
with the dependence on the algebra describing the gauging on top of each elementary surface term. This is referred to as the \emph{dressing} of the topological operators. Similar considerations can be made for the case of line defects.

For the purpose of this work, we will be mostly focussing on duality, triality and $N$-ality defects. By definition, the first 2 are associated to theories with $\mathbb{Z_{_{N}}^{^(1)}}$ 1-form symmetries, whereas the latter emerge in presence of a $\mathbb{Z}_{_{N}}^{^{(1)}}\times\mathbb{Z}_{_{N}}^{^{(1)}}$ symmetry. It is worth mentioning already at this stage that their fusion rules are basically providing a generalisation of \eqref{eq:defectalgebra}. Specifying to the case of (3+1) TQFTs, the fusion of 3D duality/triality defects, ${\cal N}({\cal M}_{3})\overset{def.}{=}D_3({\cal M}_{3})A_3$ reads  

\begin{equation}  
{\cal N}({\cal M}_{3})\times\bar{\cal N}({\cal M}_{3})\ \equiv\ {\cal C}({\cal M}_{3})  
\label{eq:squareof}   
\end{equation}   
with ${\cal C}({\cal M}_{3})$ denoting the \emph{gauging defect} defined as 

\begin{equation}   
{\cal C}({\cal M}_{3})\ \overset{def.}{=}\ \frac{1}{{|H^0({\cal M}_{3}, G)|}}\ \sum_{\Sigma_2\ \in\ H_2 ({\cal M}_{3}, G)}  L(\Sigma_2)
\end{equation}  

\begin{figure}[ht!]   
\begin{center}
\includegraphics[scale=0.9]{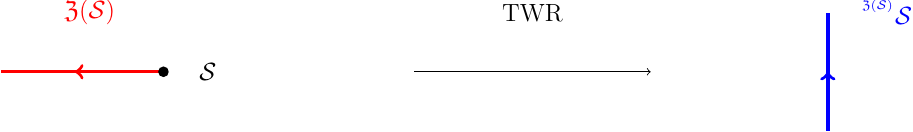}     
\caption{\small TWR as an example of holographic duality.}
\label{fig:mtc}  
\end{center} 
\end{figure} 

From \eqref{eq:squareof}, ${\cal N}({\cal M}_{3})$ is often referred to as the square root of the gauging. As a further major consideration worth mentioning is that the quantum dimensions of the duality and triality defects on $S^3$ coincide.

\section{Topological Wick rotation and holographic dualities}

The AdS/CFT correspondence is an example of holographic duality. Two main features characterising such correspondence are that: 

\begin{enumerate} 

\item Gauge symmetries in AdS $\leftrightarrow$ global symmetries of the CFT. 

\item Correlation functions are only defined when boundary conditions of the AdS theory are specified.

\end{enumerate}

These same properties are shared by other holographic dualities. A new type of holographic duality is the one given by topological Wick rotation.

The holographic dictionary encoded in figure \ref{fig:mtc} reads as follows: 

\begin{enumerate} 

\item an $N+1$ dimensional topological order with a gapped boundary is dual to an $N$ dimensional liquid with an internal symmetry of finite type. ${\cal S}$ is the category of topological defects in the bulk. On the RHS it is the superselection sector of states. $\mathfrak{Z}({\cal S})$ is the category of topological defects in the bulk and naturally acts on ${\cal S}$. $\mathfrak{Z}({\cal S})$ is the category of topological operators (non-local operators invariant under LOA).

\item $^{^{\mathfrak{Z}({\cal S})}}{\cal S}$ is an \emph{enriched category}.

\end{enumerate}     

Consider gapped/gapless edges of a 2+1 dimensional topological order $({\cal C},c)$, with $a,b,c\in{\cal S}$ labelling the topological line defects. The macroscopic observables of the 1+1 dimensional edge form an enriched fusion category $^{^{{\cal B}}}{\cal S}$. The spaces of fields living on the 0 dimensional defect junction $\{M_{_{a,b}}\}$ lead to the OPE

\begin{equation}  
M_{_{b,c}}\ \otimes_{_{\mathbb{C}}}\ M_{_{a,b}}\ \xrightarrow{\ \ \circ\ \ }\ M_{_{a,c}} 
\end{equation}

$\{M_{_{a,b}}\}_{_{a,b\in{\cal S}}}$ is similar to a category $\{\text{hom}(a,b)\}_{_{a,b\in{\cal C}}}$. $M_{_{a,b}}\ \circ\ \in\ {\cal B}$ defines an enriched fusion category.nIntroducing an abstract MTC ${\cal B}$ and a braided equivalence $\phi:\ \text{Mod}_{_{V}}\ \xrightarrow{\ \ \simeq\ \ }\ {\cal B}$, $(\phi,\ ^{^{{\cal B}}}{\cal S})$ can encode different enrichments $^{^{\text{Mod}_{_{V}}}}_{_{\phi}}{\cal S}$. Correlation functions and OPEs among defect fields can be recovered from the triple $(V,\phi, ^{^{\cal B}}{\cal S})$. The gapped/gapless boundaries of a 2+1 dimensional topological order $({\cal C}, c)$ can be completely characterised or classified by the triple $(V,\phi, ^{^{\cal B}}{\cal S})$, where:   

\begin{enumerate}

\item $V$ is the chiral/non-chiral symmetry which  

\begin{itemize}  

\item for a chiral gapless boundary, corresponds to a rational VOA of central charge $c$  

\item for a non-chiral gapless boundary, it is a rational full field algebra  

\item when $V\equiv \mathbb{C}$, it is described by a gapped boundary. 

\end{itemize}

\item $\phi:\ \text{Mod}_{_{V}}\ \xrightarrow{\ \ \simeq\ \ }\ {\cal B}$ is a braided equivalence. 

\item ${\cal S}$ is a fusion category equipped with a braided equivalence ${\cal C}\boxtimes\bar {\cal B}\simeq\mathfrak{Z}({\cal S})$, which determines $^{^{\cal B}}{\cal S}$ via the so-called \emph{canonical construction}.

\end{enumerate}

All gapped/gapless edges of 2+1 dimensional topological orders can be determined by performing a TWR and adding the chiral/non-chiral symmetry data by hand $(V,\text{Mod}_{_{V}}\ \xrightarrow{\ \ \phi\ \ }\ {\cal B}, ^{^{\cal B}}{\cal S})$. $^{^{\cal B}}{\cal S}$ is the \emph{topological skeleton} (encoding the topological or categorical data), whereas $(V,\phi)$ constitutes the local quantum symmetry of the edge (e.g. the VOA or the full field algebra or the conformal net of local operator algebras (LOA)). $^{^{\cal B}}{\cal S}$ is an \emph{abstract enriched FC}, acquiring precise meaning only once ${\cal B}$ is identified with $\text{Mod}_{_{V}}$ via the braided equivalence $\phi:\ \text{Mod}_{_{V}}\ \xrightarrow{\ \ \simeq\ \ }\ {\cal B}$, suggesting that a gapped/gapless quantum liquid $\chi$ can be described by a 2-step data $\chi\equiv(\chi_{_{lqs}}, \chi_{_{sk}})$, since the same $^{^{\cal B}}{\cal S}$ can be realised by different gapped phases. Examples of these constructions are 2D WZW CFTs from 3D Chern-Simons theories. Constructing correlators in 2D CFTS from 3D TFTs. There is a correspondence between Lagrangian algebras in 3D anyon gauging and modular-invariant partition functions in 2D CFTs. Dualities are explicitly constructed from classical statistical models. Correlators mapping 3D TQFTs to 2D RCFTs. String-net construction of 2D correlators.  

\begin{figure} [ht!]  
\begin{center}
\includegraphics[scale=0.75]{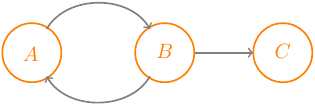}  
\caption{\small }  
\label{fig:FNQV}  
\end{center}  
\end{figure}

\underline{\ Theorem\ } (Kong, Huang)

There is a holographic duality between $(\mathfrak{Z}({\cal S}), {\cal S})$ and $^{^{\cal B}}{\cal S}$. An N+1 dimensional topological order with higher-dimensional symmetry ${\cal R}$ can be completely characterised:   

\begin{itemize}  

\item a fusion n-category ${\cal A}$ equipped with a braided embedding ${\cal R}\ \hookrightarrow\ \mathfrak{Z}_{_{1}}({\cal A})$   

\item a braided equivalence $\phi:\ \mathfrak{Z}_{_{1}}({\cal R})\ \xrightarrow{\ \ \simeq\ \ }\ \mathfrak{Z}_{_{1}}({\cal A})$.  

\end{itemize} 

Generalising ${\cal R}$ to any fusion $(N-1)$-category, one obtaine=s a characterisation of $N$ dimensional gapped quantum liquids by an $N+1$ dimensional topological order with gapped boundary. TWR suggests that this holographic duality is actually functorial. TWR gives an equivalence between the category of $N+1$ dimensional non-chiral topological orders with gapped boundary and the category of $N$ dimensional topological liquids

\begin{equation}   
(\mathfrak{Z}_{_{1}}({\cal A}), {\cal A})  \ \ \longrightarrow\ ^{^{\mathfrak{Z}_{_{1}}({\cal A})}}{\cal A}
\end{equation}   

This holographic dictionary automatically includes a dictionary map of dualities because many dualities can be realised by invertible DWs, e.g. e-m, and K-W duality.  

One of the main goals behind all these studies is the mathematical theory of higher-dimensional CFTs.

\section{Quiver gauge theories  } \label{sec:gth}

\subsection{Graphs}   

A graph, $\Gamma$, gives a category, ${\cal C}_{_{\Gamma}}$, with 3 objects: $A, B$, and $C$. ${\cal C}_{_{\Gamma}}(A,B)$ is just the single arrow from $A$ to $B$.

\subsection{Dynkin diagrams}

\begin{figure} [ht!]  
\begin{center}
\includegraphics[scale=0.6]{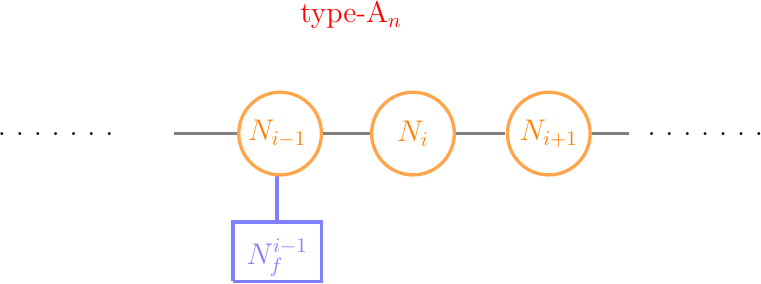}  
\ \ \ \ \ \ \ 
\includegraphics[scale=0.6]{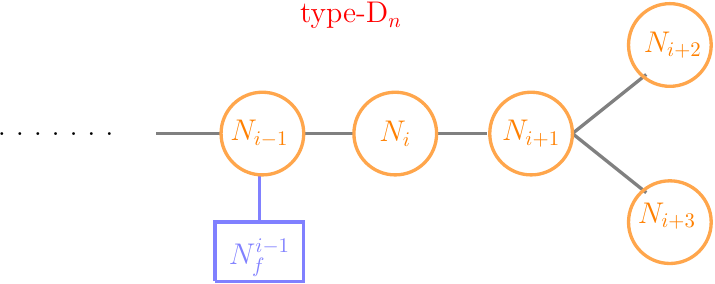}
\caption{\small Examples of type-A and type-D quivers.}  
\label{fig:FNQV}  
\end{center}  
\end{figure}

\begin{figure} [ht!]  
\begin{center}
\includegraphics[scale=0.6]{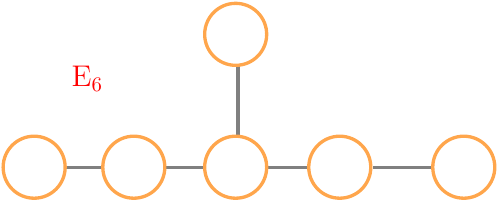}  
\ \ \ \ \ \ \ 
\includegraphics[scale=0.6]{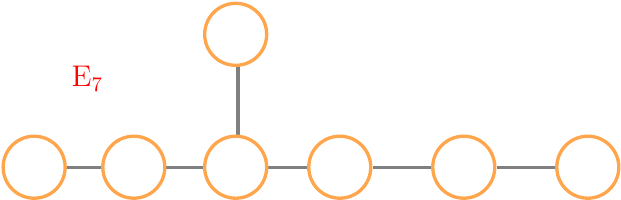}
\caption{\small }  
\label{fig:whatever}
\end{center}  
\end{figure}

\begin{figure} [ht!]  
\begin{center}
\includegraphics[scale=0.6]{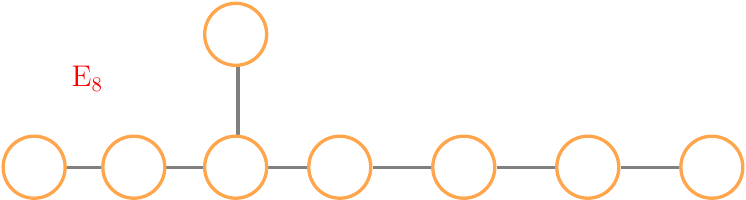}
\caption{\small Examples of type-E quivers.}  
\label{fig:FNQV}  
\end{center}  
\end{figure}

The main focus of our analysis will be the understanding of the behaviour of 4D ${\cal N}=2$ theories. In doing so, we wish to bring the attention or the reader towards the construction of the 4D theory itself, which already accounts for very useful insight towards understanding major issues we wish to address.   

The 4D ${\cal N}=2$ theories that are usually being dealt with, are assumed to possess $S$-duality as inherited from their mother theory, namely 6D ${\cal N}=(2,0)$ SCFT theories, upon dimensional reduction and topological twist on a Riemann surface $\Sigma_2$. Albeit not being this the unique way of deriving such 4D theories, yet they provide the most generic way, which, in turn, allows for a plethora of different descending theories. These 6D SCFTs are very particular UV theories. They were discovered as the decoupling limit of type-IIB on an orbifold singularity. The resolution of the singularity leads to a 6D SCFT theory with characterised according to the ADE classification, with A, D, E denoting the corresponding types of Dynking diagrams.

\subsection{Framed Nakajima quiver varieties}  \label{sec:2.1}

In this first subsection, we briefly overview the features of framed Nakajima quiver varieties, \cite{DAlesio:2021hlp}, and their geometric and algebraic resolutions. 

\begin{figure} [ht!]  
\begin{center}
\includegraphics[scale=1]{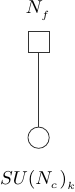}  
\caption{\small An example of a framed Nakajima quiver variety. This will be a prototypical example of the theories we will be addressing in the present work, namely for quiver gauge theories with 8 supersymmetries. The unframed quiver would be the gauge node without matter. Here we have set $G_{_v}\equiv SU(N)$. However, in principle, other choices could have been taken.}  
\label{fig:FNQV}  
\end{center}  
\end{figure}

\subsection*{Geometric (Nakajima) resolution}

Quiver varieties are varieties of quiver representations of a quiver: one fixes a vector space at each vertex, then considers the linear space of representations associated to each arrow of the quiver linear map. A framed version of this was initially introduced by Kronheimer and Nakajima, amounting to doubling the set of vertices, and drawing a new arrow from each new vertex to its corresponding old one. An example of the resulting quiver (which is of the prototypical type of interest for us) is depicted\footnote{The nomenclature featuring in the framed quiver will be explained in section \ref{sec:2.2}, but we emphasise that this is standard notation in the quiver literature.} in figure \ref{fig:FNQV}.

Framed representations appear naturally in ADHM quiver construction of self-dual or anti-self-dual YM on $S^{^4}$. This is also interesting from the point of view of representation theory of Lie algebras because dimension vectors of the framed vertices appear as highest weights of the representations. There are different notions of framing. Using Nakajima's version for quiver varieties, the framed quiver is doubled, meaning each arrow gets doubled by an arrow that goes in the opposite direction. The linear space of representations becomes a linear cotangent bundle 

\begin{equation}
    M({\cal Q},v,w)\ \overset{def.}{=}\ T^{^*} L({\cal Q}^{^{\text{fr}}},v,w),   
\end{equation}
where ${\cal Q}, {\cal Q}^{^{fr}},v,w$ respectively denote the original quiver, the framed quiver, the number of original vertices, and the number of framed vertices.
The gauge group is a general linear group on the original vertices  $G\equiv G_{_v}$, and there is a moment map 

\begin{equation}   
\boxed{\ \ \ \ \ \mu:\ M({\cal Q},v,w)\ \rightarrow\ \mathfrak{g}^{^*} \color{white}\bigg] \ \ }.   
\end{equation}  

Nakajima quiver varieties are Hamiltonian reductions of the following action 

\begin{equation} 
G\  \circlearrowright\ M({\cal Q},v,w),   
\end{equation}
and come into two types:  

\begin{enumerate} 

\item the \emph{affine} Nakajima variety, defined as the partial character variety

\begin{equation}   
\mathfrak{M}^{^0}({\cal Q},v,w)\ \equiv\ \mu^{^{-1}}(0)//G,  
\label{eq:first}
\end{equation} 

\item \emph{quasi-projective} 

\begin{equation}   
\mathfrak{M}^{^{\chi}}({\cal Q},v,w)\ \equiv\ \mu^{^{-1}}(0)//_{_{\chi}}G.   
\label{eq:second}
\end{equation} 

\end{enumerate}

The gauge group by which we take the quotient is   

\begin{equation}  
G\ \equiv\ G_{_v}\ \overset{def.}{=}\ \prod_{a}\ GL_{_{v_{a}}} (\mathbb{C})\ \subset G_{_{v}}\ \times G_{_{w}}   
\end{equation}

For any choice of the nontrivial character    

\begin{equation}  
\chi:\ G\ \rightarrow\ \mathbb{C}^{^{\times}} 
\end{equation}   
there is a proper Poisson morphism 

\begin{equation} 
p: \mathfrak{M}^{^{\chi}}({\cal Q},v,w)\ \rightarrow\ \mathfrak{M}^{^0}({\cal Q},v,w),  
\end{equation}  
which is as a symplectic resolution of the singularities of $\mathfrak{M}^{^0}$.

In both cases, \eqref{eq:first} and \eqref{eq:second}, $\mu^{^{-1}}(0)$ denotes the fiber of zero through the moment map. The latter can be used as a representation scheme for the path algebra, ${\cal A}$, modulo the ideal, ${\cal I}_{_{\mu}}$,

\begin{equation}   
{\cal A}\ \overset{def.}{=}\  \mathbb{C}\overline{{\cal Q}^{^{fr}}}/{\cal I}_{_{\mu}},   
\end{equation}
of the framed doubled quiver 

\begin{equation}   
\mu^{^{-1}}(0)\ \equiv\ \text{Rep}_{_{\mathbb{C}^{^v}\otimes\mathbb{C}^{^w}}}({\cal A})\ \overset{def.}{=}\  \text{Rep}_{_{v,w}}({\cal A}),   
\end{equation} 
where we made use of the following shorthand notation

\begin{equation}   
\mathbb{C}^{^v}\ \overset{def.}{=}\ \bigoplus_{a}\ \mathbb{C}^{^{v_{_a}}}\ \ \ , \ \ \ \mathbb{C}^{^w}\ \overset{def.}{=}\ \bigoplus_{a}\ \mathbb{C}^{^{w_{_a}}},   
\end{equation}

For the purpose of our work, the main result of \cite{DAlesio:2021hlp} is being able to relate such varieties to derived representation schemes, with the latter being obtained by means of an alternative resolution of the affine Nakajima quiver variety. The underlying reason for the importance of this resides with it being a sample realisation of homological mirror symmetry\footnote{We will be explaining this in due course.}. 

Prior to explaining the alternative resolution leading to derived representation schemes, we wish to emphasise that the correspondence that we are looking for, summarised in figure \ref{fig:GIT}, \cite{DAlesio:2021hlp}, emerges from comparing invariants, and the required conditions for them to match at the two endpoints of the red arrow. From the symplectic resolution point of view, this requires flatness of the moment map, as explained in \cite{DAlesio:2021hlp}, following \cite{mommap}. We refer the interested reader to such references for more detailed explanation regarding the definition and properties of flat moment maps. For our purposes, the crucial point is that when the invariants in the symplectic and derived scheme description match, we are dealing with a complete intersection\footnote{The meaning of the latter will be explained in section \ref{sec:2.2}.}.

\subsection*{Algebraic resolution}   

\begin{figure}[ht!]
\begin{center}
\includegraphics[scale=1.1]{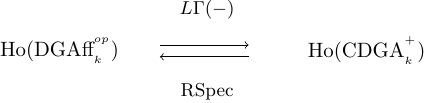} 
\caption{}
\label{}
\end{center}
\end{figure}

The technique of algebraic resolution was first put forward by \cite{Berest}, and consists in resolving the singularities of the representation schemes by introducing homological algebra. 

Prior to explaining derived representation schemes, let us first briefly overview the notion of a differential graded (dg)-schemes. A dg-scheme is a pair $X\overset{def.}{=}\left(X_{_o},{\cal O}_{_{X,\bullet}}\right)$, with $X_{_o}\equiv M({\cal Q},v,w)$ denoting the vector space of linear representations, and ${\cal O}_{_{X,\bullet}}$ a sheaf of dg-algebras such that their zeroth homology reads as follows 

\begin{equation}   
\pi_{_0}(X)\ \overset{def.}{\equiv}\ \text{Spec}\left(H_{_0}\left({\cal O}_{_{X,\bullet}}\right)\right)\ \equiv\ \mu^{^{-1}}(0). 
\end{equation}

The derived representation scheme of the relative algebra\footnote{Associated to the choice of a path in the framed quiver.}, ${\cal A}$, in a vector space $V$, is the object, \cite{DAlesio:2021hlp},  

\begin{equation} 
\text{DRep}_{_V}({\cal A})\ \in\ \text{Ho}\left(\text{DGAff}_{_k}\right), 
\end{equation}
obtained applying the following composition of functors  

\begin{equation} 
\text{DRep}_{_V}(-):\ \text{Alg}_{_S}\ \longrightarrow\ \text{Ho}(\text{DGA}_{_S}^{^+})\ \xrightarrow{L(-)_{_V}}\ \text{Ho}(\text{CDA}_{_k}^{^+})\ \xrightarrow{\text{RSpec}}\ \text{Ho}(\text{DGAff}_{_k}),
\label{eq:totalmap}    
\end{equation}
where   

\begin{equation}   
L(-):\ \text{Ho}\left(\text{DGA}_{_S}\right)\ \rightarrow\ \text{Ho}\left(\text{CDGA}_{_S}\right)  
\end{equation}
is the derived representation functor acting on the homotopy category, Ho, whose homology is the representation homology of a differential graded algebra, ${\cal A}$, 

\begin{equation}    
H_{_{\bullet}}({\cal A},V)\ \overset{def.}{=}\ H_{_{\bullet}}\left(L({\cal A})_{_V}\right).  
\end{equation}

On the other hand, RSpec in \eqref{eq:totalmap} denotes the derived spectrum, defining an equivalence on the homotopy categories

Altogether, \eqref{eq:totalmap} leads to 
\begin{equation} 
\text{DRep}_{_V}({\cal A})\ =\ \text{RSpec}\left(L({\cal A}_{_V})\right) =\ \text{Rep}_{_V}\left({\cal A}_{_{cof}}\right)
\label{eq:DRS}   
\end{equation}
with 

\begin{equation} 
{\cal A}_{_{cof}}\ \overset{\sim}{\twoheadrightarrow}\ {\cal A}
\end{equation}
a cofibrant replacement. Different choices of cofibrant replacements lead to different models of \eqref{eq:DRS}. For the specific case of framed Nakajima quiver varieties, the derived representation scheme reads, \cite{DAlesio:2021hlp},  

\begin{equation} 
\text{DRep}_{_{v,w}}({\cal A})\ =\ \text{Spec}\left(L({\cal A}_{_{v,w}})\right) \ \in\ H_{_0}\left(\text{DGAff}_{_{\mathbb{C}}}\right), 
\label{eq:DRSDA}   
\end{equation}
with representation homology 

\begin{equation} 
H_{_{\bullet}}({\cal A},v,w)\ \equiv\ H_{_{\bullet}}\left(L({\cal A})_{_{v,w}}\right)\ \in\ \text{CDGA}_{_{\mathbb{C}}}^{^+}, 
\label{eq:DRSDA1}   
\end{equation}
a graded commutative algebra. Denoting by $\{U_{_{\lambda}}\}$ the irreducible representations of $G$, \eqref{eq:DRSDA1} explicitly reads as follows 

\begin{equation} 
H_{_{\bullet}}({\cal A},v,w)\ \overset{def.}{=}\ \underset{\lambda}{\bigoplus}\ \text{Hom}_{_G}\left(U_{_{\lambda}}, H_{_{\bullet}}({\cal A},v,w)\right)\otimes U_{_{\lambda}}, 
\label{eq:DRSDA2}   
\end{equation}  
with $\text{Hom}_{_G}\left(U_{_{\lambda}}, H_{_{\bullet}}({\cal A},v,w)\right)$ denoting its isotypical components, namely modules over

\begin{equation} 
H_{_0}({\cal A},v,w)^{^G}\ \equiv\ {\cal O}\left(\mu^{^{-1}}(0)\right)^{^G}.
\label{eq:DRSDA3}   
\end{equation}

For each irreducible representation of $G_{_v}$, the Euler character is defined as follows  

\begin{equation}  
\begin{aligned}
\chi_{_T}^{^{\lambda}}({\cal A},v,w)\ &\overset{def.}{=}\ \overset{\infty}{\underset{i=0}{\sum}}\ (-1)^{^i}\left[\text{Hom}_{_G}\left(U_{_{\lambda}}, H_{_i}({\cal A},v,w)\right)\right]\\
&=\overset{\infty}{\underset{i=0}{\sum}}\ (-1)^{^i}\left[H_{_i}(L({\cal A}))_{_{\lambda,v,w}}^{^{G_{_v}}}\right]\ \ \ \in\ \ K_{_T}\left(\mathfrak{M}^{^0}\right),  
\end{aligned}
\end{equation}
where $T\overset{def.}{=}T_{_w}\times T_{_l}$ accounts for the framing, and 

\begin{equation}  
H_{_i}(\pi):\ H_{_i}\left({\cal A}_{_{cof}}\right)\ \rightarrow\ H_{_i}({\cal A})  
\end{equation}  
is an isomorphism such that ${\cal A}_{_{cof}}$ has no higher homologies, namely, 

\begin{equation}  
\begin{cases} 
&H_{_0}(\pi):\ H_{_i}\left({\cal A}_{_{cof}}\right)\ \xrightarrow{\sim}\ {\cal A}\\    
&\\    
&H_{_i}\left({\cal A}_{_{cof}}\right)\ \equiv\ 0, \ \forall i\ge1.\\
\end{cases}
\end{equation} 
and 

\begin{equation} 
\pi:\ {\cal A}_{_{cof}}\ \longrightarrow\ {\cal A}   
\end{equation} 
defines an acyclic fibration in DGA$_{_{\mathbb{C}}}^{^+}$.

\subsection*{Hilbert series}

One of the key results of \cite{DAlesio:2021hlp} is that agreement in between invariants calculated from the Nakajima and derived representation schemes requires flatness of the moment map on the symplectic side, and vanishing higher homologies in the derived case. Importantly for us, such agreement involves integrated characters, also known as the Hilbert series (HS). The latter is a map which reads

\begin{equation} 
\text{HS}_{_T}:
\begin{cases} 
&K_{_T}\left(\mathfrak{M}^{^0}\right)\ \longrightarrow\ K_{_T}(\text{pt}),\  \text{if $\mathfrak{M}^{^0}$ is compact}\\ 
&\\ 
&K_{_T}\left(\mathfrak{M}^{^0}\right)\ \longrightarrow\ \text{Frac}(T),\  \text{otherwise}.\\ 
\end{cases}
\end{equation}   

From this, the resulting Weyl integral formula reads 

\begin{equation} 
\text{HS}_{_T}\left(\chi_{_T}^{^G}({\cal A},v,w)\right)\ \equiv\ \frac{1}{|G|}\ \int_{_G}dg\ \text{HS}_{_{G\times T}}\left(\chi_{_T}^{^G}({\cal A},v,w)\right).
\label{eq:hs}  
\end{equation}

Upon choosing the variables in the maximal torus of the gauge group, $x\ \in\ T_{_v}\ \subset\ G$, and equivariant variables $t\equiv\ (a,l)\ \in\ T\ \equiv\ T_{_w}\times T_{_l}$, \eqref{eq:hs} can be re-expressed as follows

\begin{equation} 
\text{HS}_{_T}\left(\chi_{_T}^{^G}({\cal A},v,w)\right)\ \equiv\ \frac{1}{|W|}\ \int_{_{T_{_v}}}dx\ WF(x)\ \frac{\ \underset{i}{\prod}\ \left(1-l_{_1}l_{_2}b_{_i}\right)\ }{\underset{j}{\prod}\ \left(1-a_{_j}\right)}, 
\label{eq:hs1}  
\end{equation}   
with $b_{_i}, a_{_i}$ respectively denoting the weights of $M({\cal Q},v,w)^{^*}$ and $\mathfrak g$.

\section{Derivation of generalised black hole (BH) entropy}

\subsection{"Replica trick" derivation}  

\begin{figure}[h!] 
  \begin{center} 
  \includegraphics[scale=0.7]{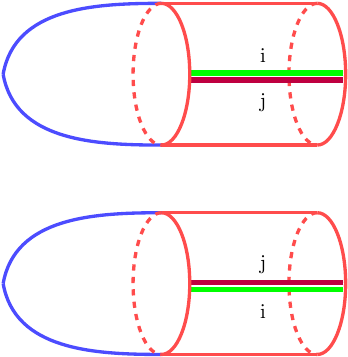}
\ \ \ \ \ \ \ \ \ \ \ \ \ \ \ \ \ 
\includegraphics[scale=0.7]{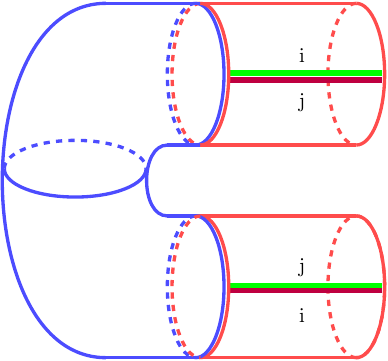}
\caption{} 
\end{center} 
\end{figure}

 \begin{figure}[h!] 
  \begin{center} 
  \includegraphics[scale=0.7]{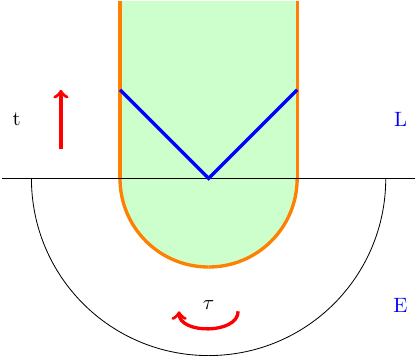}
\ \ \ \ \ \ \ \ \ \ \ \ \ \ \ \ \ 
\includegraphics[scale=0.7]{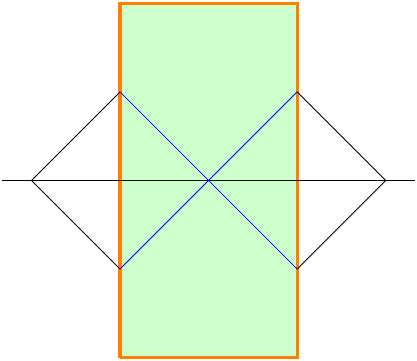}
\caption{} 
\end{center} 
\end{figure}

\subsection{Rènyi and von Neumann entropies} 

Given 2 systems A and B both equipped with a field theory which, for the latter it is assumed to be, in turn, coupled to JT-gravity,

\begin{equation} 
S_{B} 
\ 
= 
\ 
\ln Z_{B} 
\ 
= 
\ 
\frac{\phi_{o}}{4\pi}\left[\int_{B}R+2\int_{\partial B}K\right]+\frac{\Phi}{4\pi}\int_{B}(R-\Lambda)+\frac{\Phi_{b}}{2\pi}\int_{\partial B}K+\ln Z_{CFT}[g] 
\end{equation}
the Rènyi entropies read 

\begin{equation} 
S 
\ 
= 
\ 
\text{tr}\rho_{A}^{n} 
\ 
= 
\ 
\sum_{i_{1}...i_{n}}p_{i_{1}}...p_{i_{n}}<\psi_{i_{1}}|\psi_{i_{2}}><\psi_{i_{2}}|\psi_{i_{3}}>...<\psi_{i_{n}}|\psi_{i_{1}}> 
\ \ \ 
. 
\end{equation} 

\begin{figure}[h!] 
\begin{center} 
\includegraphics[scale=0.8]{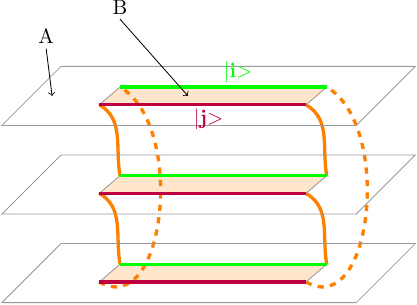}  
\ \ \ 
\includegraphics[scale=0.5]{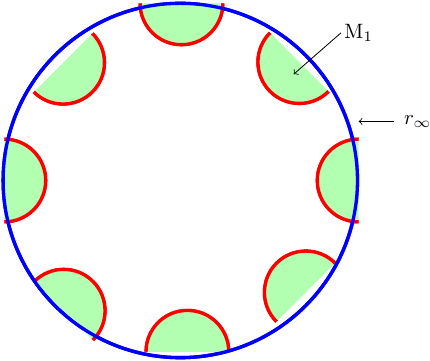}    
\ \ \ 
\includegraphics[scale=0.5]{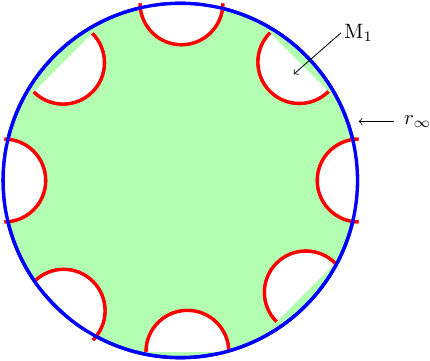}    
\caption{Taking n-copies of the gravitating universe over which the path integral is performed. The pictures at the centre and on the RHS, instead, show the completely disconnected and completely connected saddles, the latter being associated to Euclidean wormholes.} 
\end{center} 
\end{figure} 

The path integral is assumed to be performed over all the possible topologies consistent with the same boundary conditions. 

\begin{equation} 
\text{tr}\rho_{A}^{n} 
\ 
= 
\ 
\frac{1}{Z_{1}^{n}}\left[e^{-I_{grav}[M_{disc}]}Z_{CFT}[M_{disc}]+e^{-I_{grav}[M_{conn}]}Z_{CFT}[M_{conn}]\right] 
\ \ \ 
, 
\end{equation} 

with 

\begin{equation} 
M_{1} 
\ 
= 
\ 
e^{-I_{grav}[M_{conn}]} Z(\beta)  
\end{equation} 
being the saddle point geometry for the gravitating system $B$. 

$$  
\Rightarrow 
\ \ \ 
\text{tr}\rho_{A}^{n} 
\ 
= 
\ 
\frac{1}{Z_{1}^{n}}\left[e^{-nI_{grav}[M_{1}]}\sum_{i}e^{-n\beta E_{i}}+e^{-I_{grav}[M_{conn}]}Z_{CFT}[M_{conn}]\right] 
\ 
= 
$$ 

$$ 
= 
\ 
\frac{1}{Z_{1}^{n}}\left[Z(n\beta)+e^{-I_{grav}[M_{conn}]}Z_{CFT}[M_{conn}]\right] 
\ \ \ 
. 
$$ 

\begin{figure}[h!] 
\begin{center}
\includegraphics[scale=1.2]{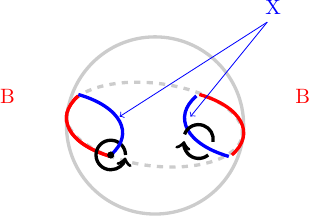} 
\ \ \ \ \ \ \ \ \ \ \ 
\includegraphics[scale=1.2]{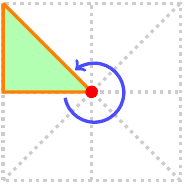} 
\caption{These pictures shows the emergence of fixed points under $Z_{n}$-symmetry of the boundary. On the right, the fixed point is depicted in red. Under the action of the generators of $Z_{n}$-symmetry, the green subregion is mapped to equivalent subregions. The nontrivial fixed point indicates the emergence of a nontrivial boundary for the CFT, which in turn carries d.o.f. associated to the degeneracy of the ground state, as already outlined in previous subsection. }
\end{center} 
\end{figure}

To build $M_{conn}$, we define $C_{x}$ as a potion of a Cauchy slice of $B$ with linear size 2$\pi$ x. Then, by gluing cyclically n-disconnected components of $M_{disc}$ along $C_{x}$ forming a ${\textbf{Z}}_{n}$-symmetric Euclidean wormhole topology obeying the gravitational e.o.m., the Rènyi entropy can be rewritten as follows  

\begin{equation} 
\text{tr}\rho_{A}^{n} 
\ 
= 
\ 
\frac{Z(n\beta)}{Z_{1}^{n}(\beta)}\left[Z(n\beta)+e^{-I_{grav}[M_{conn}]}Z_{CFT}[M_{conn}]\right] 
\end{equation}

In this subsection, we apply the machinery outlined in the previous pages for deriving the explicit expression of the generalised entanglement entropy, showing the emergence of the contribution of the quantum extremal surface resulting from the quotiented boundary geometry. 

As already mentioned, AdS/CFT and, more generally, the Holographic Principle, claims that 

\begin{equation} 
Z_{bdry}(M,J)
\ 
= 
\ Z_{bulk}(M,J) 
\ \ \ 
, 
\end{equation} 
i.e., the bulk string partition function is equivalent to the generating functional of its boundary CFT dual. 

Expecting a UV completion of the bulk theory, i.e. not pure gravity. However, when $G_{N}$ is small, it is possible to approximate it as 

\begin{equation} 
Z_{bulk} 
\ 
= 
\ 
\sum_{g} e^{-I(g)}\ Z_{matter}(g) 
\label{eq:PF}    
\end{equation} 
where $g$ is a classical background with geometry $g$ and 

\begin{equation} 
\frac{\delta}{\delta g}\left[ e^{-I(g)}\ Z_{matter}(g)  \right] 
\ 
= 
\ 
0 
\ \ \ 
\Longleftrightarrow 
\ \ \ 
R_{\mu\nu}\ -\ \frac{1}{2}g_{\mu\nu}R 
\ 
= 
\ 
8\pi G<T_{\mu\nu}> 
\ \ \ 
. 
\label{eq:extr}
\end{equation} 

The partition function is determined by the saddle point extremising (\ref{eq:extr}). In general, these solutions are hard to find even for AdS$_{3}$, and even more for higher-dimensions. Finding $g$ for arbitrary $n$ turns out to be possible through the aid of the replica trick, which in turn relies upon the assumption that the boundary has a $Z_{n}$-symmetry, which in turn corresponds to a cyclic permutation of the $\hat\rho$ featuring in the boundary path-integral. 

Assumptions: 

\begin{enumerate} 
\item{in the dominant saddle, this $Z_{n}$-symmetry is also a symmetry of the bulk, i.e. not spontaneously-broken} 
\item{the dominant saddles are all we need, i.e. it provides a good analytic continuation, but there might be jumps in $n$ leading to discontinuities. } 
\end{enumerate}

As a matter of fact, we can therefore restrict our treatment to \emph{quotiented geometries}. In addition, we need to take into account the \emph{fixed points}\footnote{uniquely identified} of the quotieneted geometry (i.e. the branch cuts). In the bulk a surface with a similar behaviour must also be defined, such that the bulk geometry is smooth everywhere except at the fixed points. 

Quotienting by $Z_{n}$, we end up with a conical singularity. Writing up the semi-classical action in terms of the data of the quotiented geometry, equation (\ref{eq:PF}) becomes 

\begin{equation} 
Z_{n} 
\ 
= 
\ 
e^{-I_{n}(g^{(n)},X^{(n)})}\ Z_{matter,n}(g^{(n)},X^{(n)}) 
\ \ \ 
,     
\ \ \ 
Z_{matter,n} 
\ 
= 
\ 
\text{Tr}_{_{\bar b}}(\rho_{b}^{n}) 
\ \ \ 
. 
\end{equation} 

The full unquotiented geometry is a saddle-point of the gravitational action, therefore satisfies the extremisation 

\begin{equation} 
\frac{\delta}{\delta g}\left[ e^{-I(g)}\ Z_{matter}(g)  \right] 
\ 
= 
\ 
0 
\label{eq:c} 
\end{equation} 

\begin{equation} 
\Longleftrightarrow 
\ \ \ 
\frac{\delta}{\delta g^{(n)}}\left[ e^{-I(g^{(n)})}\ Z_{matter,n}(g^{(n)})  \right] 
\ 
= 
\ 
0 
\ \ \ 
,
\ \ \ 
\frac{\delta}{\delta X^{(n)}}\left[ e^{-I_{n}(g^{(n)})}\ Z_{matter,n}(g^{(n)})  \right] 
\ 
= 
\ 
0 
\ \ \ 
. 
\label{eq:constr}
\end{equation} 

A possible solution to (\ref{eq:constr}) is of the form 

\begin{equation} 
Z_{n} 
\ 
= 
\ 
\exp\left[-nI_{1}(g^{(n)})-(n-1)\frac{A(X^{(n)})}{4G_{N}}\right]\ \text{Tr} (\rho_{b}^{n}) \ \ \ 
. 
\end{equation} 

At this point, we can ease the assumption that $n$ is an integer, in particular $n=1+\epsilon$, with $\epsilon<<1$. Working to leading-order in $\epsilon$, the partition funcion becomes 

$$
Z_{1+\epsilon} 
\ 
= 
\ 
e^{-I(g^{(1)})}\ Z_{matter}\ [1-\epsilon I_{1}(g^{(1)})-\epsilon\frac{\partial g^{(n)}}{\partial N}\frac{\delta I_{1}(g^{(1)}}{\delta g^{(n)}}-\epsilon\frac{A(X^{(1)})}{4G_{N}}\ +\ 
$$

\begin{equation} 
+ 
\ 
\frac{\epsilon}{Z_{matter}}\frac{\partial g^{(n)}}{\partial N}\frac{\delta Z_{matter}(g^{(1)}}{\delta g^{(n)}}\ +\ \frac{\epsilon}{Z_{matter}}\text{Tr}(\rho_{b}\text{ln}\rho_{b})] 
\ \ \ 
. 
\label{eq:1e}
\end{equation} 

At $n=1$, i.e. with one copy of the geometry, the action does not depend on the fixed point and $\text{Tr} \rho_{b}$ doesn't depend on $b$. However, the location of $X^{(1)}$ is not determined at $n=1$, but at leading-order-$\epsilon$ corrections. (\ref{eq:1e}) can be further simplified by applying the semi-classical Einstein equations that read 

\begin{equation} 
\frac{\delta I^{(1)}}{\delta g} \ +\ \frac{1}{Z_{matter}}\frac{\delta Z_{matter}}{\delta g} \ 
= 
\ 
0  
\ \ \ 
, 
\end{equation} 

therefore, we are left with 

\begin{equation} 
Z_{1}^{1+\epsilon}
\ 
= 
\ 
e^{-I_{1}(g^{(1)})}\ Z_{matter}\left[1-\epsilon\ I_{1}(g^{(1)})\ +\ \epsilon\ \text{ln}\ \left(Z_{matter}\right) \right] 
\ \ \ 
. 
\end{equation} 

For the von Neumann entropy 

\begin{equation}
\frac{Z_{1+\epsilon}}{Z_{1}^{1+\epsilon}} 
\ 
= 
\ 
1\ -\ \epsilon\frac{A(X)}{4G_{N}}\ +\ \epsilon\ \text{Tr}\ \left(\hat\rho_{b}\ \text{ln}\ \rho_{b}\right)-\epsilon\ \text{Tr}\left(\hat\rho_{b}\ \text{ln}\ Z_{matter}\right)
\end{equation}

\begin{equation} 
= 
\ 
1\ -\ \epsilon\frac{A(X)}{4G_{N}}\ +\ \epsilon\ \text{Tr}\ \left(\hat\rho_{b}\ \text{ln}\ \hat\rho_{b}\right)
\end{equation} 

\begin{equation}
\Rightarrow 
\ \ \ 
S 
\ 
= 
\ 
\lim_{\epsilon\rightarrow\ 0}\ \left[ -\ \frac{1}{\epsilon}\ \ln\ \frac{Z_{1+\epsilon}}{Z_{1}^{1+\epsilon}}\right] 
\ 
= 
\ 
\frac{A(X)}{4G_{N}}\ +\ S_{bulk}(X) 
\ \ \ 
. 
\label{eq:1l} 
\end{equation} 

For $n=1$, $X$ is nothing but the quantum extremal surface satisfying (\ref{eq:c}). Furthermore, among all possible $X$s choose the smallest. This two-step (extr-min) procedure is what defines the quantum extremal surface prescription. Notice that the deep point of the replica trick consists in treating $n$ as an arbitrary parameter and expanding it around $n=1$, such that $g^{(1)}$ parametrises the background metric without backreactions dominates the path-integral. On the other hand, $Z_{matter}$ across the whole derivation, is really meant to be the full path-integral for the matter content. It is therefore takes into account of all fluctuations in the matter sector, which is what enables to achieve the 1-loop correction to the pure gravity result in (\ref{eq:1l}).

\section{Topological Quantum Field Theories (TQFTs)}  \label{sec:TQFTs}

The present section introduces topological field theories and their String Theory counterparts, namely Topological String Theory. The reason for introducing this is multifold. As already mentioned in the introduction, they constitute an essential tool for probing the mathematical structure of QFTs. The most important advancements in the field took place in 2 and 3-dimensions. Most importantly, the pioneering works of Atiyah, Schwarz, Segal and Witten\footnote{The pioneering works on the subject are the following \cite{Atiyah:2021hsc,Atiyah:2018ijp,Atiyah:2016vlv,Atiyah:2010qs,Atiyah:2007cka,Atiyah:2004jv,Segal:2021mpr,Segal:2010hbt}.}. Sticking to such lower-dimensional setups might seem reductive, especially when having in mind higher-dimensional field theories. However, the great advantage of the 2 and 3D settings have proved to give rise to an extremely rich variety of studies, whose findings can be interestingly applied to higher dimensions as well. As mentioned in the Introduction, the AGT correspondence is a clear example of the utility of 2D TFTs for describing 4D ${\cal N}=2$ SCFTs\footnote{We will come back to this in greater detail in Part \ref{sec:V}.}.

\subsection{Attempts and advancements}

Topological field theories (TFTs) are a bridge between QFT and maths. In gauge theory and mathematical physics, a TQFT is a QFT calculating topological invariants. They are related to knot theory, the theory of 4-manifolds, and the theory of moduli spaces in algebraic geometry \cite{Donaldson,Kontsevich:1994dn,Atiyah:2018ijp,N6}. 

\begin{figure}[h!]    
\begin{center}    
\includegraphics[scale=1]{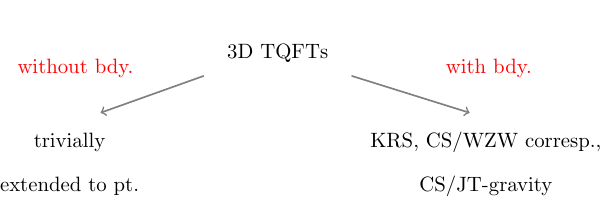}    
\caption{\small }    
\label{fig:plot2f1} 
\end{center}      
\end{figure}

The adjective topological refers to the fact that correlation functions are independent from the spacetime metric. On Minkowski spacetime, topological theories are trivial when contracted to a point. Because of this, they need to be placed on a Riemann surface instead. They come in two main types:   

\begin{enumerate}

\item Schwarz-type TQFTs, e.g. the BF-model

\begin{equation}  
S=\int_MBF,  
\end{equation}
where $B$ is an auxiliary field, whereas $F$ is the derivative of a connection, $A$. This action is explicitly topological, given that the metric does not feature at all. The alternative formulation by Schwarz reads as follows    

\begin{equation}  
S=\int_MA\ \wedge\ dA. 
\end{equation}
An example of this is Chern-Simons theory, applied to knot invariants, to which we shall turn in the following section.

\item   Witten-type TQFTs [topological 4D Yang Mills theory in 4D]. For a given 2-form, $B$, the corresponding action reads as follows

\begin{equation}  
S=\int_MB\ \wedge\ \delta B, 
\end{equation}

\begin{itemize}

\item  $S$ has a symmetry, i.e. $\delta S=0$ under it. 

\item   $\delta^2=0$, i.e. the symmetry is exact. 

\item  There are observables, $\{{\cal O}_{_i}\}$, such that $\delta {\cal O}_{_i}=0$.  

\item   $T^{\alpha\beta}=\delta G^{\alpha\beta}$ for any arbitrary tensor $G^{\alpha\beta}$.

\end{itemize}

\end{enumerate}

2D TFTs are mostly related to mirror symmetry. 
On the other hand, 3D TFTs are much less understood, apart from Chern-Simons theory, which has no nontrivial local observables given that there are no boundary conditions. The only topological observables are Wilson loops (localised on 1D closed submanifolds). Kapustin, Rozansky and Saulina (KRS) were the first to have introduced nontrivial boundary conditions for a 3D TFT. This theory as a 3D sigma-model, first introduced by Rosanzky and Witten (a.k.a. the RW-model). The target of the TFT is required to be hyperk$\ddot{\text{a}}$hler, $X$, or, more generically, a holomorphic symplectic manifold. Upon dimensional reduction on an $S^1$, it yields the B-model.

A physical motivation for studying the RW-model is its relation with 3D mirror symmetry. In particular, understanding the RW-model, is a step forward towards understanding 3D mirror symmetry.

As we will see in greater detail, the boundaries of KRS are Langrangian submanifolds of $X$, with the latter building relation with derived algebraic geometry.

\subsection{The Atiyah-Segal axioms} 

The Atiyah-Segal axioms are particularly useful for understanding the mathematical structure of Schwarz-type QFTs. The key point of such axioms is that a TQFT is a functor from a bordism category to the category of vector spaces.

\subsubsection{Segal's Conformal Field Theory (CFT) axioms}

The object of Segal's work, is that of presenting a mathematical formulation of CFT$_{2}$s, and describing basic examples already known from previous literature. Furthermore, he also presents five examples of areas of pure mathematics where CFTs play a crucial role:

\begin{enumerate}  

\item The monster group. 

\item The representation theory of loop groups, and the group Diff$(S^1)$ of diffeomorphisms on the circle.

\item How representations of Diff$(S^1)$ are related to the geometry of moduli spaces of the Riemann surfaces. (Mumford's classification of holomorphic line bundles descends from this.)   

\item Some of V. Jones' new representations of braid groups, and his classification of subfactors in von Neumann algebras.    

\item Elliptic cohomoogy theory of Lendweber-Stong, \cite{DonZagier}.

\end{enumerate}

\subsubsection{The Atiyah-Segal axioms}   

The key point of such axioms is that a TQFT is a functor from a bordism category to the category of vector spaces.   

Take a commutative ring, $\Lambda$, with the identity, $\mathbf{1}$. Then, a $D$-dimensional TQFT\footnote{Where $D=d+1$, $d$ denoting the number of spatial dimensions.} is defined from $\Lambda$ according to the following axioms:  

\begin{itemize}   

\item  \underline{The Homotopy axiom} A finitely-generated $\Lambda$-module, ${\cal Z}(\Sigma)$ associated to each oriented closed smooth $d$-dimensional manifold, $\Sigma$. 

\item \underline{The Additive axiom} An element ${\cal Z}(M)\in{\cal Z}(\partial M)$ associated to each oriented smooth $(d+1)$-dimensional manifold $M$. 

\item ${\cal Z}$ is functorial with respect to orientation-preserving diffeomorphisms of $\Sigma$ and $M$. 

\item ${\cal Z}$ is involutory, i.e. ${\cal Z}(\Sigma^*)={\cal Z}(\Sigma)^*$.  

\item ${\cal Z}$ is multiplicative. 

\item ${\cal Z}(\emptyset)=\Lambda$ for the $d$-dimensional manifold, and ${\cal Z}(\emptyset)=\mathbf{1}$ for the $(d+1)$-dimensional empty manifold.  

\item \underline{The Hermitian axiom} ${\cal Z}(M^*)=\overline{{\cal Z}(M)}$.   

\end{itemize}

In practice, $\Sigma$ and ${\cal Z}(\Sigma)$ respectively correspond to the physical space and the Hilbert space of the QFT, with associated vanishing Hamiltonian, $H=0$. The latter property follows from the topological nature of the theory in question, implying there is no real dynamics in the $\Sigma\ \times \ I$-cylinder. However, there can still be nontrivial propagation from $\Sigma_{_o}$ to $\Sigma_{_I}$ through an intermediate, topologically nontrivial, manifold $M$ such that   

\begin{equation}  
\partial M\ =\ \Sigma_{_o}^{*}\ \cup\ \Sigma_{_I}.   
\end{equation}

If $\partial M\equiv\Sigma$, then the distinguished vector in the Hilbert space ${\cal Z}(M)$ is the vacuum state defined by $M$. For a closed manifold $M$, ${\cal Z}(M)$ is nothing but the partition function, namely the expectation value of the identity operator.

The additive axiom can be understood by taking into account the fact that finite dimensional vector spaces for a compact closed category. For genuine TFTs, all spaces of quantum states are finite-dimensional, hence one could equivalently consider the linear dual spaces, implying  the propagator (or correlator) map

\begin{equation}   
Z(M):\ Z\left(\partial_{_{in}}M\right)\ \rightarrow\ Z\left(\partial_{_{out}}M\right)  
\end{equation}  
becomes a linear map  

\begin{equation}   
\mathbb{C}\ \rightarrow\ Z\left(\partial_{_{out}}M\right) \ \otimes\ Z\left(\partial_{_{in}}M\right)^{^*}\ =\ Z(\partial M).  
\end{equation}

\subsection{Schwarz-type TQFTs}

\subsubsection{Chern-Simons (CS) Theory} 

A 3D TQFT of Schwarz-type, developed by Edward Witten, and named after Chern and Simons who introduced the Chern-Simons 3-form. The latter is defined as follows: given a manifold and a Lie algebra valued 1-form, $A$, over it, one can define a family of $p$-forms as

\begin{equation}  
d\omega_{_{2k-1}}\ =\ \text{Tr}\left(F^{^k}\right),     
\label{eq:CSF}
\end{equation}
where the right-hand-side is proportional to the $k^{th}$ Chern-character of the connection $A$. For the specific case where $p=3$, \eqref{eq:CSF} reads  

\begin{equation}  
\text{Tr} \left[F\wedge A-\frac{1}{3}A\wedge A\wedge A\right]\ =\ \text{Tr} \left[dA\wedge A+\frac{2}{3}A\wedge A\wedge A\right].  
\end{equation}

In gauge theory, the integral of the Chern-Simons form is a global geometric invariant, modulo addition of an integer.

\subsection{Topological String Theory}

Topological String Theory (TST) is a version of String Theory that follows the idea of TQFTs. There are 2 main versions of it, conventionally referred to as the topological A- and B-model.  

Calculations in TST encode all holomorphic quantities that are protected by spacetime SUSY. Various such calculations are related to CS-theory, Gromov-Witten (GW)-invariants, mirror symmetry, and the Geometric Langlands Program, just to name a few. 

The operators in a TST represent the algebra of operators in the full string theory preserving a certain amount of SUSY.

Abelian gauge theories usually appear in String and M-theory, where abelian means that, in these theories, the space of gauge inequivalent fields forms an abelian group.  

A central extension of a group $G$ by an abelian group $A$ is a group $\tilde G$ such that

\begin{equation}
0\ \rightarrow\ A\ \rightarrow\ \tilde G\ \rightarrow\ G\ \rightarrow\ 0,     
\end{equation}
with $A$ being in the center of $\tilde G$. $\tilde G$ can be thought of as a set of pairs $(g,a)$ with twisted multiplication

\begin{equation}
\left(g_{_1}, a_{_1}\right) \left(g_{_2}, a_{_2}\right)    = \left(g_{_1}g_{_2}, c\left(g_{_1},g_{_1}\right) a_{_1} a_{_2}\right).
\end{equation}   

From a purely mathematical perspective, it is natural to question about the implications of this for the formulation of topological D-branes in terms of derived categories.

\subsubsection{The A- and B-model}  

\begin{itemize}

\item{\underline{The A-model}} 

The topological A-model comes with a 6-dimensional K$\ddot{\text{a}}$hler target space. Such theory is equipped with fundamental strings wrapping around 2D holomorphic curves, as well as D2-branes wrapping Lagrangian submanifolds of spacetime. The worldvolume theory on a stack of $N$ $D2$-branes is the string field theory of the open strings of the A-model,namely a $U(N)$-CS-theory.   

The fundamental strings may, in turn, end on the D2-branes. Additional (coisotropic) branes in various dimensions might also be added [Kapustin, Orlov].

\item{\underline{The B-model}}  

The B-model also contains fundamental strings. However, scattering amplitudes here only depend on the complex structure, and are completely independent w.r.t. the K$\ddot{\text{a}}$hler structure. Furthermore, the theory admits D(-1)-,D1-,D3-, and D5-branes, wrapped around holomorphic 0, 2, 4, and 6-submanifolds, respectively.

\end{itemize}

Mirror symmetry relates the amplitudes calculated on the two sides.

\section{QFT and the Jones polynomial}

\subsection{QFT and the Jones polynomials}

Among the many useful applications of CS-theory as a 3D TQFT of Schwarz-type, is that it interprets the Jones polynomials (JPs).  

JPs assign a Laurent polynomial in $t^{^{1/2}}$ to any knot, $V(knot)$. Each JP can thus be completely determined by the unknot, $V(unknot)=1$, and the Skein relations

\begin{equation}   
\left(t^{^{1/2}}-t^{^{-1/2}}\right)\ V(L_{_{o}})\ =\ t^{^{-1}}\ V\left(L_{_+}\right)-t\ V\left(L_{_-}\right),
\end{equation}  
where $L_{_{\pm}}, L_{_o}$ denote knots differing in between them only at a point, as shown in figure \ref{fig:Skrel}.

\begin{figure}[ht!]  
\begin{center}    
\includegraphics[scale=1]{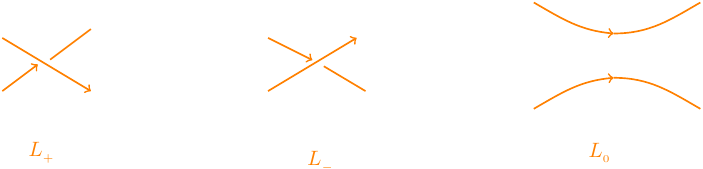}  
\label{fig:Skrel}  
\caption{\small}  
\end{center}  
\end{figure}

\subsection{Schemes or algebraic stacks}  

A scheme is a mathematical structure that enlarges the notion of an algebraic variety, accounting for multiplicities, and allowing varieties over any commutative ring. Scheme theory was originally introduced by A. Grothendieck in 1960 as a key tool for solving open issues in algebraic geometry.  

A scheme is a topological space, as well as a ringed space which is locally a spectrum of a commutative ring.    

A morphisms of schemes can be used for defining algebraic surfaces. In many cases, the family of all varieties of a given type can itself be viewed as a variety of schemes.

\subsection{Morse Theory and Classifying Spaces}

The main idea of Morse Theory is to associate to a Morse functor

\begin{equation}    
f:\ M\rightarrow\mathbf{R}   
\end{equation}
on a closed Riemannian manifold, $M$, a category, ${\cal C}_{_f}$, whose objects are the critical points of $f$, whereas the morphisms, $\overline{\cal M}(a,b)$, between two critical points, $a$ and $b$, are piece-wise flow lines of the gradient flow of $f$ connecting $a$ and $b$. One can recover from the topological category ${\cal C}_{_f}$ the topology of $M$. As such, ${\cal M}(a,b)$ denotes the the moduli space of lines from $a$ to $b$.

From local data, one cannot tell whether a given critical point is required by the topology or if, instead, it is removable. For the system in question to be nontrivial, we need to perform calculations that are sensitive of the presence of more than one critical point on the manifold in question. For this to be the case, we need to ensure that the potential energy in the problem, $V(\phi)$, has more than one minimum, specifically, one for each critical point, allowing for the possibility of tunnelling from one critical point to another. 

Tunnelling can be calculated in the WKB-approximation, namely by means of instantons (using Witten's terminology), \cite{Witten:1982im}. Tunnelling effects often remove spurious degeneracies arising in perturbation theory, as is precisely the case here. From the instanton analysis, we know that the relevant instantons or tunnelling paths are the paths of steepest descent leading from one critical point to another. As such, they are therefore the solutions of the following equation    

\begin{equation}  
\frac{d\phi}{d\lambda} \ =\ \gamma^{^{ij}}\frac{\partial h}{\partial\phi_{_j}}, 
\end{equation}  
where $\lambda$ denotes time in the supersymmetric nonlinear sigma-model, assuming $\phi$ is only a function of time.

The instanton calculation shows that the only relevant solutions of this equation are the ones connecting two critical points whose Morse indices differ by unity. 

Instanton solutions or tunnelling paths in a given theory are extrema of this Lagrangian, written with a Euclidean metric, neglecting the fermionic content. The relevant action would therefore read 

\begin{equation}    
\begin{aligned} 
S&=\frac{1}{2}\ \int d\lambda\ \left(\gamma^{^{ij}}\frac{d\phi^{^i}}{d\lambda}\frac{d\phi^{^j}}{d\lambda}+t^{^2}\gamma^{^{ij}}\frac{\partial h}{\partial\phi^{^i}}\frac{\partial h}{\partial\phi^{^j}}\right)  \\
&=\frac{1}{2}\ \int d\lambda\ \bigg|\frac{d\phi^{^i}}{d\lambda}\pm t \gamma^{^{ij}}\frac{\partial h}{\partial\phi^{^j}}\bigg|^{^2}\ \mp\ t\int d\lambda\frac{dh}{d\lambda}.
\end{aligned}
\end{equation}

The minimum action paths between any two critical points $a$ and $b$ are paths of steepest descent, and the action for such paths reduces to 

\begin{equation}   
S=t\ \bigg|\ h(B)-h(A)\ \bigg|. 
\end{equation}

An example of a classifying space for the $\infty$-cyclic group $G$ is the circle as $X$. Moduli space is essentially a synonym for representing object and classifying space. Usually, the term classifying space is used in topology, whereas moduli space is used in algebraic geometry.

\subsection{Donaldson invariants}

Donaldson theory (DT), \cite{Donaldson}, is the study of the topology of 4-manifolds using moduli spaces of anti-self-dual instantons. Many of the theorems in DT can be proved by a much more convenient formalism, namely that of Seiberg-Witten (SW) Theory. 

Even though this has been amply studied in the literature, there are still some open questions that remain to be addressed, phrased in terms of the following conjectures: 

\begin{enumerate}

\item  The \emph{Witten conjecture}\footnote{The Witten conjecture was later extended by Virasoro.}: in Algebraic Geometry, it is related to the intersection numbers of stable classes on the moduli space of curves. In its formulation, it is a special case of a more general relation between integrable systems of Hamiltonian PDEs, and the geometry of certain 2D TFTs, also known as cohomological field theories\footnote{The latter were first studied by Kontsevich and Manin.}.

\item  The \emph{Atiyah-Floer conjecture}: connecting the instanton Floer homology with the Lagrangian intersection Floer homology\footnote{The latter is, in turn, related to homological mirror symmetry.}.

\end{enumerate}

Witten showed that Donaldson invariants are correlation functions in a supersymmetric gauge theory, and that the infinite-dimensional Morse theory developed by Floer is part of a 2D QFT.   

Mathematical gauge theory was later revolutionised when Seiberg and Witten deduced the long-distance behaviour of the supersymmetric gauge theory calculating the Donaldson invariants. Donaldson invariants are topological in nature, and are therefore scale-independent. Because of this, they found an alternative way of calculating them, this time by means of the so-called \emph{Seiberg-Witten equations}.

The original motivation that led Witten to propose such conjecture was the existence if two different 2D models of quantum gravity that should be exhibiting the same partition function. These are: on the one hand, the intersection of number on the moduli stack of algebraic curves, and, on the other, it is the ln of of the T-function of the KdV-hierarchy, with $T$ a translation operator such that 

\begin{equation}  
T(g)(x)=g(x+1).   
\end{equation}

\subsection{Seiberg-Witten theory}

A Seiberg-Witten theory is an ${\cal N}=2$ supersymmetric gauge theory whose low-energy effective action is exact in the massless limit. Such theory is also referred to as ${\cal N}=2$ SYM theory, whose field content is a single ${\cal N}=2$ vector supermultiplet, in complete analogy with the single vector gauge field (or connection) in ordinary YM theory. Its corresponding Lagrangian density in 2D reads

\begin{equation}  
{\cal L}_{_{SYM}}=\text{Im}\text{Tr}\left(\frac{1}{4\pi}\int d^{^2}\theta d^{^2}\vartheta\ {\cal F}(\Psi)\right),  
\end{equation}
with $\theta,\vartheta$ denoting the coordinates for the spinor directions of superspace, whereas ${\cal F}$ is the prepotential. The minimal theory is defined by

\begin{equation}  
{\cal F}(\Psi)=\frac{1}{2}\tau\Psi^{^2},    
\end{equation}
with $\tau$ the complex coupling constant.

\subsection{From CS to Morse theory}

\subsubsection{Morse Theory}

Morse Theory relates to the description of closed geodesics on smooth Riemannian manifolds. Since the first studies of R. Bott, the theory underwent several reformulations, which we will briefly overview in this section. The crucial feature of Morse Theory is its relation to the so-called critical-point theory.

The main idea is that of associating a Morse function 

\begin{equation} 
f:\ M\ \rightarrow\ \mathbf{R}    
\end{equation}  
on a closed Riemannian manifold, $M$, a category, ${\cal C}_{_f}$, whose objects, $a, b$, are the critical points of $f$, whereas the morphisms between two critical points of $f$ are denoted by $\overline{\cal M}(a,b)$. The latter are piecewise flow lines of the gradient flow of $f$ connecting $a$ and $b$. One can recover the topology of $M$ from the category ${\cal C}_{_f}$. ${\cal M}(a,b)$ is the moduli space of flow-lines from $a$ to $b$. This setup can be rephrased in terms of the space of loops, \ref{sec:loop}.

Morse Theories are classified in terms of: 

\begin{enumerate}

\item Their dimensionality, which can either be finite or infinite.

\item The role of symmetry groups.

\item Their connection to higher or lower-dimensional TQFTs.

\end{enumerate}

Its finite formulation, due to Bott, followed from approximating the description in terms of a finite set of polygons (i.e. piecewise geodesics). Furthermore, he restricted his studies on manifolds with Lie group acting on the manifold while preserving the Morse function, also known as \emph{equivariant} Morse Theory. 

On the other hand, Morse Theory in its infinite-dimensional version has many fascinating applications in lower-dimensional topology, to which we will come back momentarily. 

\subsubsection{Relation to Moduli Spaces of SQFTs}

The most important feature of Morse Theory is its applicability to extended setups going beyond the work of Atiyah and Bott on Riemann surfaces. 

Take a compact Lie group $G$, and $M$ a smooth manifold. A $G$-connection on $M$ gives a notion of parallel transport along paths, which can be considered as an element of $G$ itself. For example, if $M\equiv S^{^2}$, then parallel transport on a half-circle defines a map

\begin{equation}  
S^{^1}\ \longrightarrow \ G,  
\end{equation} 
relating the loop space of $G$ and connections on the Riemann surfaces, the latter being the original setup investigated by Atiyah and Bott, namely the equivariant Morse Theory of the Yang-Mills (YM) functional

\begin{equation} 
YM(A)=\frac{1}{2}\int_{_M}\left|F_{_A}\right|^{^2}\ d^{^D}x.  
\label{eq:YMAB}
\end{equation}  

To determine the critical points of \eqref{eq:YMAB}, one needs to solve the associated Euler-Lagrange equation 

\begin{equation}  
d_{_A}^{^*}F_{_A}=0.   
\end{equation}  

The minima of the energy functional \eqref{eq:YMAB} are flat connections with zero-curvature, which , from the Narasimhan-Seshadri Theorem corresponds to the space of stable bundles. Later developments in 4D by Atiyah, \cite{Atiyah:2021hsc}, Bott, \cite{N1}, and Donaldson, \cite{Donaldson}, eventually lead to the study of instanton equations in 4D, further developed by Atiyah, Hitchin, and Singer, \cite{Hitchin:1999at}.

A great turnaround came when Floer started investigating 3D manifolds, \cite{V2}, in which case the functional to be extremised is CS theory. The critical points of the functional are still flat connections on the given Riemannian manifold, but a crucial difference arises with respect to the 2D and 4D cases: the number of negative eigenvalues of the Hessian at the critical points is infinite-dimensional, implying traditional Morse Theory could not be applied straightforwardly. Importantly, Floer realised that one could define a \emph{relative index} in between two critical points, by making use of a curve interpolating in between them. In such way, one is able to define chain complex, whose homology defines the Floer homology of the 3-manifold in question.  

Furthermore, the latter can be related to the Donaldson invariants of a 4D manifold whose boundary is the 3D manifold whose Floer homology has just been defined. Indeed, Floer homology groups and Donaldson invariants follow some of the defining properties of a TQFT, such as the gluing property.

\begin{figure}[ht!]   
\begin{center}   
\includegraphics[scale=0.7]{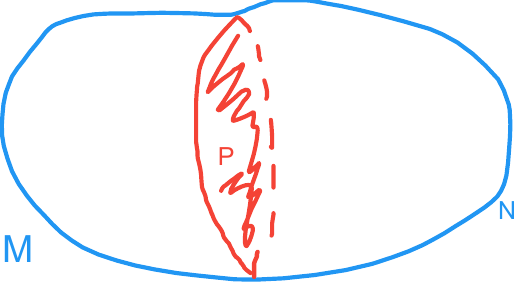}   
\label{fig:gluing}  
\caption{\small} 
\end{center}  
\end{figure}

2 main lines of development in Morse Theory, \cite{Witten:1982im}:

\begin{enumerate}

\item Witten was first motivated by supersymmetry models in lower-dimensional QFT.

\item The second line of development has been, instead, that of understanding that Morse functions decompose smooth manifolds into simple constituents.

\end{enumerate}  

Mathematical constructions of TQFTs usually start with algebraic data which is then assigned to elementary bordisms between manifolds of fixed dimension. Invariants are obtained by decomposing arbitrary bordisms into elementary ones by means of Morse functions. The fact that this defines an invariant follows from comparing the results achieved from different Morse functions.  

For the case of a 2D TFT, such algebraic data is a commutative Frobenius ring, first studied by Moore and Segal.

\subsection*{SQFTs from Morse Theory}

We now briefly outline how Witten's approach shed light on the relation in between supersymmetric QFT (SQFT) and Morse Theory, from a mathematical perspective\footnote{See appendix \ref{sec:gth} for descriptions of gauge theories in mathematical language.}. 

Let us start by taking a smooth compact manifold $M$, and a smooth function  

\begin{equation}
f:M\ \longrightarrow\ \mathbb{R}.   
\end{equation} 

Then, introducing a deformation

\begin{equation}
0\ \longrightarrow\ \Omega^{^1}(X)\ \xrightarrow{d+t\epsilon(df)}\ \Omega^{^2}(X)\ \  \ \xrightarrow{d+t\epsilon(df)}\ \Omega^{^3}(X)\ \xrightarrow{d+t\epsilon(df)}\ ...
\end{equation} 
of the ordinary de Rham complex, if $f$ is a Morse function, and $M$ is endowed with a Riemannian metric, then, in the $\underset{t\rightarrow\infty}{\lim}$, there is a close relation between the spectrum of the associated Laplace operator, $\Delta_{_t}$, and the critical points of $f$. From a physical point of view, these correspond to the different vacuum states of the quantum system. Quantum tunnelling in between them can be encoded as a differential on the $\mathbb{Z}$-graded  abelian group generated by the vacuum states, or the critical points, namely where the grading is given by the Morse index. Smale was the the first to realise this from a pure mathematical perspective, but Witten was the first to start the physical studies behind this, with the relevant system in question being that of supersymmetric quantum mechanics (SQM).  

Once having a consistent understanding of SQM, one could lift the description by, either, increasing the number of spacetime dimensions, and/or the amount of supersymmetries at play, and subsequently studying its impact on the associated Morse Theory.  

For the sake of completeness, we briefly outline a few examples of the statement we have just made.

\begin{enumerate}
    \item Starting from SQM, doubling the amount of supersymmetry while preserving the dimensionality implies removing flows between the critical points of the associated Morse function. Upon raising the theory to 2D QFT (namely Landau-Ginzburg theory), one recovers nontrivial interpolating soliton solutions, with the latter being related to Picard-Lefschetz theory.  

\item For the case of 4D supersymmetric gauge theories, instead, one can study the tunnelling between vacua on a 3D manifold in terms of Floer homology.

\item In higher dimensions, with increasing amount of supersynmetry, namely 5D and 6D SCFTs, Witten proposed a new definition of the Jones polynomial and its categorification in terms of Khovanov homology. In this case, the critical point equations are special cases of the Kapustin-Witten equations introduced in the context of the Geometric Langlands Program (GLP), whereas the flow equations turn out being new equations in a 5D mathematical gauge theory.

\end{enumerate}

\subsubsection{The Cobordism Hypothesis}

More generally to this has been the formulation of the so-called \emph{cobordism hypothesis}. The Cobordism Hypothesis was first formulated by Baez and Dolan, and later proved by Hopkins and Lurie in 2D, and eventually in $D>$2 from Lurie.

The Cobordism Hypothesis states that, given a symmetric monoidal $(\infty,n)$-category, ${\cal C}$, which is fully dualisable and every $k$-morphism, for $1\le k\le n-1$, there is a bi-jection between the ${\cal C}$-valued symmetric monoidal functors of the cobordism category and the objects of ${\cal C}$.  

The motivation for this is a statement we have already introduced in our treatment, namely that symmetric monoidal functors from the cobordism category correspond to TQFTs. In practice, the Cobordism Hypothesis is very similar in nature to the Eilenberg-Steenrod axioms for homology theories, with the latter essentially stating that, a homology theory is uniquely determined by its value from the point. In the cobordism case, this corresponds to saying that the bijection between ${\cal C}$-valued symmetric monoidal functors and the objects of ${\cal C}$ is uniquely determined by its value for the point.

\section{4D descendants of 6D \texorpdfstring{${\cal N}=(2,0)$}{} SCFTs}    \label{sec:6d4d}

\subsection{Class \texorpdfstring{${\cal S}$}{}}   \label{sec:4.1}

First, we briefly overview the properties of class ${\cal S}$ theories, the calculation of the dimension of the Coulomb and Higgs branches, as well as the contribution from punctures to the 4D central charges, and how 3D SCFTs can be accommodated on nontrivial punctures. Then, we recall the basic motivations underlying the 2D/4D correspondence resulting from dimensional reduction of 6D ${\cal N}=$(2,0) SCFTs, needed in core sections of the present work. We end the section by recalling that the 3D theories resulting by twisted compactification on the non-invertible defects of 4D ${\cal N}$=4 SYM are always of the ABJ-type, namely gauge theories with $U(M)\times U(N)$ gauge group, \cite{Kaidi:2022uux}.

\subsection*{\texorpdfstring{${\cal N}$}{}=2 data}    

\medskip  

\medskip 

\begin{enumerate}   
\item The moduli space: Coulomb, Higgs and mixed branches.

\item The charge lattice, $\Gamma$, with 

\begin{equation}   
\text{dim}\ \Gamma\ \equiv\ 2r+f  
\end{equation}  
where $r$ denotes the rank of the gauge group and $f$ the rank of the gauge flavour symmetry. BPS states sre points in $\Gamma$.  

\item An antisymmetric electromagnetic inner product on $\Gamma$  
\begin{equation} 
\circ: \Gamma\times\Gamma\ \longrightarrow\ \mathbb{C} 
\label{eq:emip}   
\end{equation}   
\item A central charge function  

\begin{equation} 
Z_{_{u}}:\Gamma\ \longrightarrow\ \mathbb{C}    
\label{eq:ZU}  
\end{equation}  
where $u$ denotes a generic point in the Coulomb branch.
\end{enumerate} 

Such data is conventionally encoded in a family of Riemann surfaces, $\{\Sigma_{_{u}}\}$, known as \emph{Seiberg-Witten curve}, which varies over the Coulomb branch, and whose homology lattice is identified with $\Gamma$. By definition, \eqref{eq:ZU} is linear in the charges, and can therefore be expressed as a 1-form integral of the \emph{Seiberg-Witten differential}, $\lambda_{_{u}}$   

\begin{equation}   
Z_{_{u}}(\gamma)\ \overset{def.}{=}\ \int_{_{\gamma}}\lambda_{_{u}}  
\ \ \ \ 
,    
\ \ \ \ 
\forall\ \gamma\ \in\ \Gamma   
\end{equation}  

Consequently, the electro-magnetic inner product \eqref{eq:emip} is naturally identified with the intersection of 1-cycles. Jointly, $(\Sigma_{_{u}}, \lambda_{_{u}})$ define the complete solution to the low-energy physics. On the Coulomb branch, this description is provided by a special K\"ahler geometry.

\medskip  

\medskip

\subsubsection*{BPS quivers}    

\medskip  

\medskip 

The purpose of the present work work requires restricting to \emph{complete theories}. These include:   
\begin{itemize}  

\item  Theories associated to triangulation surfaces. 

\item  The quivers $E_{_{n}}, \hat E_{_{n}}, \hat\hat E_{_{n}}$. 

\item The Derksen-Owen quivers, $X_{_{6}}, X_{_{7}}$.

\end{itemize}  

The natural relation between quivers and triangulations is manifest in terms of BPS state counting. 

Assuming the theory of interest admits a set of $2r+f$ hypermultiplett states, $\{\gamma_{_{i}}\}$, forms a positive integral basis for all particles, such as is the case for class ${\cal S}$ theories, a BPS quiver is constructed by drawing a node $\forall \gamma_{_{i}}$, and an arrow in between them, $\gamma_{_{i}}\rightarrow\gamma_{_{j}}$ if $\gamma_{_{i}}\circ\gamma_{_{j}}>0$. If a site of the charge lattice 

\begin{equation}  
\gamma\ \equiv\ \sum_{i}n_{_{i}}\gamma_{_{i}}  \ \in\ \Gamma  
\end{equation}   
is occupied by a BPS state, it is possible to determine its degeneracy and spin. The intarections of the BPS elementary particles are encoded in the 4-supercharge quiver QM of the given configuration, i.e. a 1D supersymmetric gauge theory with gauge and matter groups  

\begin{equation}  
G\overset{def.}{=}\prod_{\text{nodes}}U(n_{_{i}})\ \ \ , \ \ \ \text{Matter}\overset{def.}{=}\bigoplus_{\text{arrows}} B_{_{ij}}^{^a} 
\end{equation}
where $B_{_{ij}}^{^a}$ denotes the bifundamental field charged under $G$. To find the BPS states, we need to study the moduli space of supersymmetric ground states, ${\cal M}_{_{\gamma}}$,

\begin{equation}   
{\cal M}_{_{\gamma}}\ \overset{def.}{=}\ \left\{R\equiv\left\{B_{_{ij}}^{^{a}}:\ \mathbb{C}^{^{n_i}}\rightarrow\mathbb{C}^{^{n_i}}\right\}\ \bigg|\ \frac{\partial {\cal W}}{\ \partial B_{_{ij}}^{^{a}}\ }\ \equiv\ 0, R\  \Pi-\text{stable}\ \right\}\bigg/\ \prod_{i}\ Gl\left(n_{_{i}}, \mathbb{C}\right)  
\label{eq:modsp}   
\end{equation}

Hence, ${\cal M}_{_{\gamma}}$ is the space of solutions to the $F$-term equations 

\begin{equation}  
\frac{\partial {\cal W}}{\ \partial B_{_{ij}}^{^{a}}\ }\ \equiv\ 0  
\end{equation}  
subject to an additional stability constraint, also known as \emph{$\Pi$-stability}, modulo the action of the complexified gauge group $Gl\left(n_{_{i}}, \mathbb{C}\right)$. $\Pi$-stability is the relevant condition for describing moduli space. In quiver representation theory, it means that, a representation $R$ of a quiver $Q$ is a complex vector space $\mathbb{C}^{^{n_{i}}}$ $\forall$ node $i$ and a linera map $B_{_{ij}}^{^{a}}: \mathbb{C}^{^{n_{i}}}\rightarrow \mathbb{C}^{^{n_{j}}}$ $\forall$ arrow $a$ from $i$ to $j$, corresponding to a choice of expectation values of matter fields in the corresponding quiver QM. 

A subrepresentation $S$ of $R$ is a choice of vector subspaces $\mathbb{C}^{^{m_{i}}}\ \subset\ \mathbb{C}^{^{n_{i}}}$ for each node, and maps $b_{_{ij}}^{^{a}}: \mathbb{C}^{^{m_{i}}}\rightarrow \mathbb{C}^{^{m_{j}}}$ for each arrow, such that all diagrams look like the following 

\begin{figure}[ht!]   
\begin{center}
\includegraphics[scale=1]{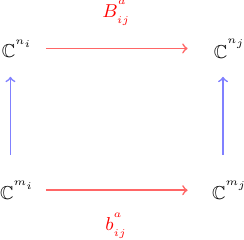}   
\caption{\small}
\label{eq:mew}   
\end{center}
\end{figure}

For a given charge associated to a representation $R$ 

\begin{equation}   
\gamma_{_{R}}\ \equiv\ \sum_{j}n_{_{j}}\gamma_{_{j}}  
\label{eq:pi}  
\end{equation}  
the stabilisation condition corresponds to

\begin{equation}   
\text{arg} Z_{_{u}}(S)\ <\ \text{arg}Z_{_{u}}(R)  
\end{equation}
for $S$ a subrepresentation of $R$. Upon identifying $S$ with a parton state, $\Pi$-stability corresponds to forbidding decay channels of the kind $\gamma_{_{R}}\rightarrow\gamma_{_{S}}$, from which the definition \eqref{eq:modsp} results. 

The nodes of a BPS quiver are  always $\Pi$-stable representations, due to the absence of destabilising subrepresentations. In terms of \eqref{eq:pi}, this corresponds to setting $n_{_{i}}\equiv\ \delta_{_{ij}}$, meaning each note contributes with a multiplicity 1 hypermultiplet BPS state. States of higher spin or multiplicity do not turn into nodes in the quiver.

If ${\cal M}_{_{\gamma}}$ is nonempty, there is a BPS particle in the spectrum with charge $\gamma$. Since ${\cal M}_{_{\gamma}}$ has 4 supercharges, it is a K\"ahler manifold, and its cohomology forms representations of Lefschetz $SU(2)$. If ${\cal M}_{_{\gamma}}$ is a point, it defines a hypermultiplet. If, instead, ${\cal M}_{_{\gamma}}\simeq\mathbb{P}^{^{1}}$, it defines a vector multiplet.  

BPS-wall crossing arises from the $\Pi$-stability condition if at some point $u$ in ${\cal M}_{_{\gamma}}$, $Z_{_{u}}(S)\equiv Z_{_{u}}(R)$, with $S$ a subrepresentation of $R$.

\medskip  

\medskip

\subsection{Vacuum structure: Coulomb, Higgs and mixed branches}

\medskip  

\medskip    

The full moduli space is unknown in most cases. For this reason, in the present work we will only be focussing on class ${\cal S}$-theories obtained by twisted compactification on a Riemann surface of 6D ${\cal N}$=(2,0) on $A_{_{N-1}}$, with generic moduli space

\begin{equation}   
{\cal M}\ \overset{def.}{=}\ \bigcup_{\alpha}\ C_{_{\alpha}}\ \times\ H_{_{\alpha}}  
\end{equation}   
with $C_{_{\alpha}}, H_{_{\alpha}}$ denoting the \emph{Coulomb} and \emph{Higgs branches} of the theory. 

The former is defined by the Seiberg-Witten curve

\begin{equation}   
x^{^{N}}+\sum_{i=2}^{N}\phi_{_{i}}(z)x^{^{N-i}} = 0  
\end{equation}  
where $z$ is the coordinate of the base Riemann surface, and $x$ is the one on the fiber of the cotangent bundle $K\equiv T^*C$. $\phi_{_{i}}(z)$ is a section of the line bundle, $K^{^{\otimes i}}$, namely it is the $i^{^{th}}$ differential on the Riemann surface $C$. In absence of punctures, it s contribution to the dimension of the Coulomb branch reads  

\begin{equation}  
\text{dim}(\phi_{_{i}})\ \equiv\ \text{dim} H^{^{0}}\left(C, K^{^{\otimes i}}\right)\ =\ (2i-1)(g-1)  
\end{equation}  

The general expression in presence of regular punctures reads

\begin{equation}  
\text{dim}_{_{{\mathbb{C}}}} C_{_{\alpha}}\ \equiv\ (k-1)g + \sum_{s=1}^{k}(n_{_{s}}^2-1)(g-1)+\sum_{p}\text{dim}_{_{\mathbb{C}}} C_{_{\alpha}}(p)    
\end{equation}     
where $k$ denotes the number of elements featuring in the partition $X\equiv\left[n_{_{1}},...,n_{_{k}}\right]$, and

\begin{equation}  
\text{dim}_{_{{\mathbb{C}}}} C_{_{\alpha}}(p)\ \equiv\ \sum_{s=1}^{k} d(Y_{_{s}}^{^{D}})\ =\    \sum_{s=1}^{k}\sum_{i=2}p_{_{i}}\equiv \frac{1}{2}\sum_{s=1}^{k} \left(N^2-\sum_{a=1}^{\ell}m_{_{a}}^2\right)    
\end{equation}   
where $Y^{^{D}}$ denotes the local contribution from a regular puncture. 

The total dimension along the Higgs branch, instead, reads   

\begin{equation}  
\text{dim}_{_{{\mathbb{H}}}} H_{_{\alpha}}\ \equiv\ k-1 + \sum_{p}\text{dim}_{_{\mathbb{H}}} H_{_{\alpha}}(p)   
\end{equation}   
with

\begin{equation}   
\text{dim}_{_{\mathbb{H}}} H_{_{\alpha}}(p)
\equiv d(Y^{^{\prime}})-d(Y)  
\end{equation}

\begin{figure}[ht!]   
\begin{center}  
\includegraphics[scale=0.7]{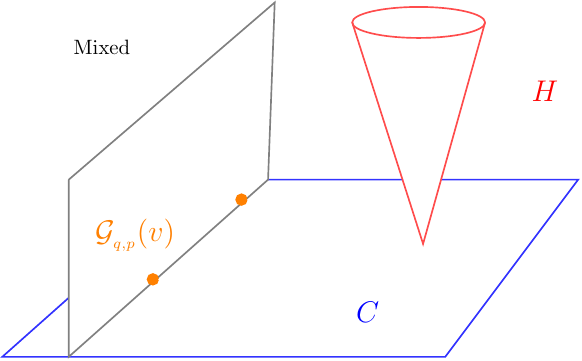}  
\caption{\small The vacuum structure of ${\cal N}=2$ class ${\cal S}$ theories, comprising a Coulomb ($C$), Higgs ($H$), and mixed branch.}    
\label{fig:mbs}  
\end{center}    
\end{figure}

Upon choosing a partition of $N$: $X\equiv\left[n_{_{1}},...,n_{_{k}}\right]$, the maximally factorised SW curve can be expressed as follows

\begin{equation}   
\prod_{s=1}^k\left(x^{^{n_{_{s}}}}+\sum_{i=2}^{n_{_{s}}}\phi_{_{s,i}}(z)x^{^{n_{_{s}}-i}}\right) = 0  
\end{equation}

\subsection{Metrics}\label{sec:metrics}

We now briefly overview the definition of the metric on the Riemann surface, following the work by Gaiotto, Moore, and Nietzke. 

Take ${\cal B}$ to be the Coulomb branch of a certain 4D ${\cal N}=2$ supersymmetric gauge theory. At any point $u\in{\cal B}$, the gauge group is broken to a maximal torus $U(1)^{^r}$. Denoting by $\hat\Gamma_{_u}$ the charge lattice equipped with a pairing $<,>$, with radical being the flavour charge lattice, $\left(\Gamma_{_{flav}}\right)_{_u}$, the quotient charge sublattice of gauge charges can be defined as follows 

\begin{equation}  
\Gamma_{_u}\ \overset{def.}{=}\ \hat\Gamma_{_u}\bigg/\left(\Gamma_{_{flav}}\right)_{_u},     
\end{equation}  
with rank$\Gamma_{_u}=2r$.

The central charge homomorphism is 

\begin{equation}  
Z:\ \hat\Gamma_{_u}\ \longrightarrow\ \mathbb{C}.  
\end{equation}

Then, for every particle, one can define $Z_{_{\gamma}}(u)$, with $\gamma$ a local section of $\hat\Gamma$, as denoting the central charge of a particle of charge $\gamma$.   

For a theory placed on $\mathbb{R}^{^3}\times S^{^1}$, with $R$ denoting the radius of the $S^{^1}$, for $R<<1$, the 4D theory is effectively 3D, and its moduli space is locally a product of hyperk$\ddot{\text{a}}$hler manifolds

\begin{equation} 
{\cal M}_{_{tot}}\ =\ \text{HB}_{_{4D}}\times \text{CB}_{_{3D}}  
\end{equation}

At generic nonsingular points of the 3D CB, the HB is absent, so one can just restrict to the 3D CB 

\begin{equation}  
{\cal M}\ \overset{def.}{=}\ \text{CB}_{_{3D}},  
\end{equation}
such that one can define the map  

\begin{equation}   
\pi:\ {\cal M}\ \longrightarrow\ {\cal B}.    
\end{equation}

That the metric on ${\cal M}$ is hyperk$\ddot{\text{a}}$hler follows from the theory being supersymmetric.  

The main idea proposed by \cite{Gaiotto:2011tf,Gaiotto:2010okc,Gaiotto:2010okc} is that of describing the metric by introducing holomorphic Darboux coordinates for ${\cal M}$ (when viewing the latter as a holomorphic symplectic manifold).

Fixing $U\subset{\cal B}$ over which $\hat\Gamma$ can be trivialised, the holomorphic Darboux coordiantes, $\chi_{_{\gamma}}^{^{\vartheta,u_{_o}}}$, with $u_{_o}\in U\subset {\cal B}$, are labelled by sections $\gamma$ of $\hat\Gamma$ over $U$. These are functions on $\pi^{^{-1}}(U)\times\mathbb{H}_{_{\vartheta}}$, where

\begin{equation}  
\mathbb{H}_{_{\vartheta}}\overset{def.}{=}\ \bigg\{\zeta:\vartheta-\frac{\pi}{2}<\text{arg}\ \zeta<\vartheta+\frac{\pi}{2}\bigg\}. 
\end{equation}

By construction, these coordinates, $\chi_{_{\gamma}}^{^{\vartheta,u_{_o}}}$ experience a jump at specific values of $\vartheta$, and the latter are the phases of the central charges of the BPS states in the vacuum labelled by $u_{_o}$. Furthermore, these jumps are determined by the gauge charges of the BPS states.

such that the following properties are satisfied:  

\begin{enumerate}  

\item $
\chi_{_{\gamma+\gamma^{^{\prime}}}}=\chi_{_{\gamma}}\chi_{_{\gamma^{^{\prime}}}}. $

\item $\forall\ \xi\in\mathbb{C}^{^{\times}}$, $\chi_{_{\gamma}}(.;\xi)\in\mathbb{C}^{^{\times}}$, and is holomorphic.

\item $\{\chi_{_{\gamma}},\chi_{_{\gamma^{^{\prime}}}}\}\ =\ <\gamma,\gamma^{^{\prime}}>\chi_{_{\gamma}}\chi_{_{\gamma^{^{\prime}}}}.  
$

\item $\forall (u,\theta)\in{\cal M}$, $\chi_{_{\gamma}}(u,\theta;\zeta)$ is holomorphic in $\zeta$.

\item $\chi_{_{\gamma}}(.;\zeta)=\overline{\chi_{_{-\gamma}}\left(.;-\frac{1}{\zeta}\right)}$.

\item $\underset{\zeta\rightarrow 0}{\lim}\ \chi_{_{\gamma}}(u,\theta;\zeta)\exp\left[-\frac{\pi RZ_{_{\gamma}}(u)}{\zeta}\right]$ exists.

\item $\chi_{_{\gamma}}^{^{\vartheta,u_{_o}}}$ features discontinuities at pairs $(\vartheta, u_{_o})$ for which there is some $\gamma_{_{BPS}}$ with 

\begin{equation}   
\text{arg}(-Z_{_{BPS}}(u_{_o}))=\vartheta.  
\end{equation}

\item Fixing $\vartheta_{_o}\in\mathbb{R}/2\pi\mathbb{Z}$, $u_{o}\in{\cal B}$, one can define 

\begin{equation}  
\text{S}_{_{\vartheta, u_{_o}}}\overset{def.}{=}\prod_{_{\gamma_{_{BPS}}:\text{arg}(-Z_{_{\gamma_{_{BPS}}}}(u_{_o})=\vartheta_{_o}}}\text{K}_{_{\vartheta, u_{_o}}}^{^{\Omega(\gamma_{_{BPS}};u_{_o})}},   
\end{equation}
with    

\begin{equation}  
\text{K}_{_{\vartheta, u_{_o}}}:\ \chi_{_{\gamma}}\ \mapsto\ \chi_{_{\gamma}}\left(1-\sigma(\gamma_{_{BPS}})\chi_{_{\gamma_{_{BPS}}}}\right)^{^{<\gamma,\gamma_{_{BPS}}>}},   
\end{equation}

\begin{equation}  
\Rightarrow \ \left(\underset{\vartheta\rightarrow\vartheta_{_o}^{^+}}{\lim}\chi_{_{\gamma}}^{^{\vartheta,u_{_o}}}\right)\ =\ \text{S}_{_{\vartheta_{_o}, u_{_o}}}\left(\underset{\vartheta\rightarrow\vartheta_{_o}^{^-}}{\lim}\chi_{_{\gamma}}^{^{\vartheta,u_{_o}}}\right).
\end{equation}

\item When $\zeta\in\mathbb{H}_{_{\vartheta}}$, then  

\begin{equation}  
\chi_{_{\gamma}}\ \sim\ \chi_{_{\gamma}}^{^{approx.}}.    
\end{equation}

\end{enumerate}  

If we now consider a point $u\in{\cal B}$ and two different phases $\vartheta_{_{\pm}}$, such that $\vartheta_{_+}-\vartheta_{_-}<\pi$, then, the two coordinate systems are related as follows

\begin{equation}  
\chi_{_{\gamma}}^{^{\vartheta_{_+},u}}=\text{S}(\vartheta_{_-},\vartheta_{_+};u)\ \chi_{_{\gamma}}^{^{\vartheta,u}},     
\end{equation}
with 

\begin{equation}  
\text{S}(\vartheta_{_-},\vartheta_{_+};u)\ \overset{def.}{=}\ \prod_{_{\gamma_{_{BPS}}:\vartheta_{_-}<\text{arg}(-Z_{_{\gamma_{_{BPS}}}}(u)<\vartheta_{_+}}}\text{K}_{_{\vartheta, u_{_o}}}^{^{\Omega(\gamma_{_{BPS}};u)}},   
\end{equation}  
where the product decomposition is uniquely defined.

\subsection{The Alday-Gaiotto-Tachikawa (AGT) correspondence: a brief overview}    \label{sec:4.22}

\medskip  

\medskip

This can be generalised to arbitrary class ${\cal S}$-theories of $\mathfrak{su}$(2)-type: the 4D ${\cal Z}_{_{S^{^1}\times S^{^{3}}}}$ associated to a Riemann surface of genus $g$ and $n$-punctures, where we place the holonomy $\left(a_{_{i}},\frac{1}{a_{_{i}}}\right)\ \in\ SU(2)_{_{i}}$ for the flavour symmetry for the $i^{^{th}}$ puncture, corresponds to the 2D $q$-deformed T=Yang-Mills theory on the same $\Sigma_{_{g}}$, where the punctures are described as holes with holonomy $\left(a_{_{i}},\frac{1}{a_{_{i}}}\right)\ \in\ SU(2)$, thereby resulting in the following

\begin{equation}  
{\cal Z}_{_{4D, g,n}}(a_{_{i}})\ \equiv\ \frac{\prod_{i}K(a_{_{i}})}{K_{_{o}}^{^{2g-2+n}}}\ {\cal Z}_{_{2D,g,n,A\equiv0}}(a_{_{i}})  
\end{equation}   
where the prefactor can be reabsorbed under suitable redefinition of the handling operator. This correspondence was first found in \cite{Alday:2009aq}.

\medskip  

\medskip 

\underline{\ Undeformed 2D YM\ }    

\medskip  

\medskip

For a gauge theory on a cylinder, with holonomy 

\begin{equation}  
U\overset{def.}{=}\ {\cal P}\ \text{exp}\int_{_{0}}^{^{L}}dx^{^{1}}A_{_{1}} 
\ \ \ 
, \ \ \ 
U\ \xrightarrow{\ \color{white}{}\color{black} \ G\ \color{white}{}\color{black}\ }\ g U g^{^{-1}}\ \ \ , \ \ \ g\in G    
\end{equation}   

\begin{figure}[ht!]  
\begin{center}
\includegraphics[scale=0.8]{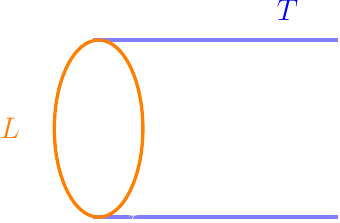}  
\caption{\small}  
\label{eq:cylinder}  
\end{center}   
\end{figure}
the partition function behaves as

\begin{equation}  
{\cal Z}_{_{A+A^{\prime}}}(U)\ \equiv\ e^{^{-Ac_{_{2}}}}\ {\cal Z}_{_{A^{\prime}}}(U)  
\end{equation}   
with $c_{_{2}}$ the operator acting on $\chi_{_{R}}(U)$ by $c_{_{2}}(R)$. From this follows that ${\cal Z}_{_{A\equiv o}}(U)$ is enough to determine ${\cal Z}$ for arbitrary $A$.

\begin{equation}  
{\cal Z}_{_{A\equiv o}}(U)\ \equiv\ \alpha\ \delta(U) 
\end{equation} 
with $\delta(U)$ denoting the $\delta$-function at the ideal of $G$ and  

\begin{equation}
\delta(U)\equiv\sum_{R}\ d_{_{R}}\chi_{_{R}}(U)  \ \ \ ,\ \ \ d_{_{R}}\equiv \int_{_{G}}dU\delta(U)\chi_{_{R}}(U)\equiv\text{Tr}\mathbb{1}\equiv \text{dim} R 
 \end{equation}

 \begin{equation}    
Z_{_{A}}(U)\equiv\alpha\sum_{R}e^{^{-Ac_{_{2}}(R)}}\ \text{dim} R\ \chi_{_{R}}(U)
\end{equation} 

For a 2D surface with 2 boundary holonomies, $U,V$, instead, we have

 \begin{equation}    
Z_{_{A}}(U,V)\equiv\sum_{R}e^{^{-Ac_{_{2}}(R)}}\  \chi_{_{R}}(U) \chi_{_{R}}(V)   
\end{equation} 

Any 2D surface can be cut into pieces, each one being a sphere with 3 holes. This way, we can recover the cylinder by gluing a disk to a hole of a 3-holes sphere. The partition function of the latter reads  

 \begin{equation}    
Z_{_{A}}(U,V,W)\equiv\frac{1}{\alpha}\sum_{R}\frac{e^{^{-Ac_{_{2}}(R)}}}{\text{dim} R}\  \chi_{_{R}}(U) \chi_{_{R}}(V)  \chi_{_{R}}(W)    
\end{equation} 

Proceeding iteratively, the final result for an arbitrary number of holonomies becomes

 \begin{equation}    
Z_{_{A}}(U_{_{i}})\equiv\alpha^{^{2-2g-n}}\sum_{R}e^{^{-Ac_{_{2}}(R)}}\  \frac{\ \prod_{i}\chi_{_{R}}(U_{_{i}})\  }{\ \ ( \text{dim} R)^{^{2g-2+n}}\ \ }   
\label{eq:totZ}   
\end{equation}

\medskip  

\medskip 

\underline{\ $q$-deformed 2D YM\ }    

\medskip  

\medskip 

The arguments outlined in the previous page are proper to the undeformed 2D YM theory.
In order to build the correspondence with the 4D theory, as outlined in \cite{Alday:2009aq}, the 2D theory needs to be deformed. This takes place as follows: quantising $A$ into faces, $f$, to each edge $e$ one assigns a dynamical variable $U_{_{e}}\in G$. The total partition function of the system can therefore be expressed as follows

\begin{equation}  
{\cal Z}_{_{tot}}\ \equiv\ \int\prod_{e}\ dU_{_{e}}\ \prod_{f}\ {\cal Z}_{_{f}} 
\end{equation}  
with 

\begin{equation}    
{\cal Z}_{_{f}}\ \equiv\ \sum_{R}e^{^{-A_{_{f}}c_{_{2}}(R)}}\text{Tr}_{_{R}}\prod_{e} U_{_{e}}    
\end{equation}     

The deformation occurs upon requiring $U_{_{e}}\ \in\ G_{_{q}}$, with $G_{_{q}}$ being the \emph{quantum group} obtained from the ordinary gauge group $G$ upon making its matrix entries noncommutative. For the case of $SU(2)_{_{q}}$

\begin{equation}   
U_{_{i}}^{^{j}}\ \overset{def.}{=}\ \left(\begin{matrix} \alpha\ \ \ \ \ \ \ \beta\\
\\
\gamma\ \ \ \ \ \ \delta\\ 
\end{matrix} \right)\ \ \ \ \ \ \ \ \ \ , \ \ \ \ \ \ \ \ \ \bar U_{_{_i}}^{^{j}}\ \overset{def.}{=}\ \left(\begin{matrix} \delta\ \ \ \ \ \  -q^{1/2}\gamma\\  
\\ 
-q^{^{1/2}}\beta\ \ \ \ \ \ \alpha\\ 
\end{matrix} \right)
\end{equation} 
\medskip 

\begin{equation}  
\Rightarrow \ \ \ \alpha\beta\ \equiv\ q^{^{1/2}}\beta\alpha\ ,\ \alpha\delta-q^{^{1/2}}\gamma\beta \equiv \delta\alpha-q^{^{-1/2}}\gamma\beta \equiv 1  
\end{equation}

\begin{equation}  
\Rightarrow \ \ \ {\cal D}^{^{i\bar i}}U_{_{i}}^{^{j}}\bar U_{_{\bar i}}^{^{\bar j}} \ \equiv\ D^{^{j\bar j}}\ \neq\ \delta^{^{j\bar j}}    
\end{equation}   

\begin{equation}  
{\cal D}^{^{i\bar i}}\ \overset{def.}{=}\ \left(\begin{matrix} q^{^{-1/2}}\ \ \ \ \ 0\\
\\    
0\ \ \ \ \ \ \ q^{^{1/2}}\\ 
\end{matrix} \right)\ \ \ \ \ \ \ \ \ \ \Rightarrow\ \ \ \ \ \ \ \ \ \delta^{^{i\bar i}}\delta_{_{i\bar i}}\ \equiv\ 2\ \ \ \ \ \ \Rightarrow\ \ \ \ \boxed{\ \ \ D^{^{i\bar i}}\delta_{_{i\bar i}} \equiv q^{^{1/2}}+q^{^{-1/2}}  \color{white}\Bigg]\color{black}\ \ }
\label{eq:qd}   
\end{equation} 

The boxed expression in \eqref{eq:qd} defines the \emph{quantum dimension} of the 2D representation of $SU_{_{q}}(2)$ and is basically the character, $\chi_{_{SU(2)}}[q]$. For the $SU(N)$ case,   \eqref{eq:qd} generalises as follows  

\begin{equation}  
\text{dim}_{_{q}}R\ \overset{def.}{=}\ \text{Tr}_{_{q}}\text{diag}\ \left(q^{^{(N-1)/2}}, q^{^{(N-3)/2}},...,q^{^{(1-N)/2}}\right)   
\end{equation} 

Upon substituting this to the dimension of the representation in \eqref{eq:totZ},

 \begin{equation}    
\boxed{\ \ \ \ Z_{_{A,g}}(U_{_{i}})\equiv\alpha^{^{2-2g-n}}\sum_{R}e^{^{-Ac_{_{2}}(R)}}\  \frac{\ \prod_{i}\chi_{_{R}}(U_{_{i}})\  }{\ \ ( \text{dim}_{_{q}} R)^{^{2g-2+n}}\ \ }  \color{white}\Bigg]\color{black}\ \ }
\label{eq:totZ1}   
\end{equation}

The key point to notice is the fact that \eqref{eq:totZ1} can be calculated for a surface with arbitrary number of holes just by knowing ${\cal Z}_{_{cyl}}, {\cal Z}_{_{3-holes}}$ and suitable gluing. This statement can be readily understood by saying that correlation functions of a given theory can be determined just by knowing 2- and 3-point functions.  

For example, a 4-punctured sphere is obtained by connecting 2 3-punctured spheres. 

\begin{figure}[ht!]  
\begin{center}   
\includegraphics[scale=0.6]{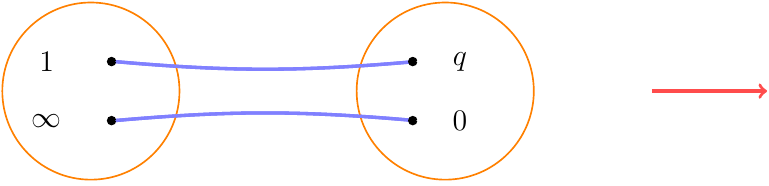}   \ \ \ \ \ \ 
\includegraphics[scale=0.6]{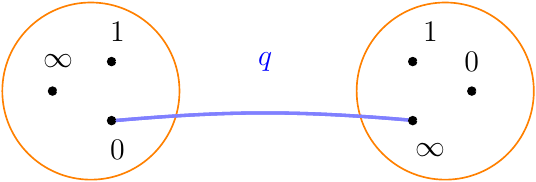}      
\caption{\small}  
\label{eq:cylinder1}  
\end{center}   
\end{figure}

\medskip  

\medskip 

\underline{\ 4D class ${\cal S}$ theories of $SU(2)$-type\ }    

\medskip  

\medskip 

In a similar fashion, they can also be constructed from 3-punctured spheres and cylinders. In the 2D case, we saw how complex numbers are assigned to 2D surfaces. Once more, the main building blocks will be cylinders and punctured surface, with an $SU(2)$ flavour symmetry assigned to each puncture.

\begin{figure}[ht!]  
\begin{center}   
\includegraphics[scale=0.6]{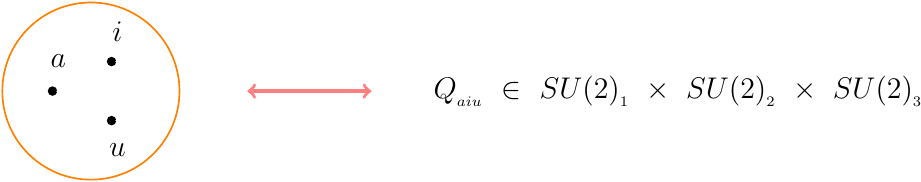}   
\caption{\small The 3-punctured sphere is equivalent to the ${\cal N}=1$ chiral multiplet in the $(2,2,2)$ representation of $SU(2)_{_{1}}\times SU(2)_{_{2}}\times SU(2)_{_{3}}$, constituting a half hypermultiplet.}  
\label{eq:cylinder1}  
\end{center}   
\end{figure}    
The cylinder is associated to an ${\cal N}$=2 vector multiplet with gauge group $SU(2)$ whose complexified coupling is   

\begin{equation}  
\tau\ \overset{def.}{=}\ \frac{4\pi i}{g^{^2}}\ +\ \frac{\theta}{2\pi}  \ \ \ , \ \ \ g\ \in\ \mathbb{R}\ \ ,\ \ \theta\sim\theta+2\pi
\end{equation}   

Given a collection of 3-punctured spheres, joining 2 of them with a cylinder, leads to a punctured torus. Parametrically, this corresponds to taking a cylinder with parameter $\tau$ and making the following identification 

\begin{equation}   
zw\ \equiv\ e^{^{2\pi i\tau}}\ \equiv\ q  
\end{equation}  
with $z, w$ indicating the positions of the 2 joined punctures. From the 4D g auge theory perspective, this corresponds to coupling 2 $SU(2)$ flavour symmetries to a dynamical $SU(2)$ gauge field with coupling constant $\tau$. Once having joined the 2 punctures, their flavour symmetry groups are identified, therefore the hypermultiplet $Q_{_{aiu}}$ can be relabelled into triplets and singlets, $\{A_{_{Iu}}\}, \{H_{_{u}}\}$, defining an ${\cal N}$=2 SU(2) gauge theory with enhanced ${\cal N}$=4 SUSY, corresponding to ${\cal N}$=4 SYM, namely a theory exhibiting S-duality. 

S-duality states that, a theory with gauge group $G$ and coupling constant $\tau$ is equivalent to the theory with dual gauge group $G^{^{\text{V}}}$ and coupling constant $-\frac{1}{\tau}$. For the case of $SU(2)$, $G\equiv G^{^{\text{V}}}$.    

For ${\cal N}$=2 Lagrangian theories, the partition function can be calculated performing a path integral on $S^{^{1}}\times S^{^{3}}$. In addition to the gauge holonomies parameterising the BPS configurations on which integration is performed, there might also be additional flavour holonomies. As previously mentioned, the holonomy $g\ \in\ G$ around $S^{^{1}}$ can be expressed as 

\begin{equation}   
g\ \equiv\ (z_{_{1}}, ... z_{_{r}})\ \in\ U(1)^{^{r}}\ \in G  
\end{equation}  

For an ${\cal N}=2$ hypermultiplet consisting of ${\cal N}=1$ chiral multiplets in a representation $R$ of a symmetry $G$, the partition function is

\begin{equation}   
{\cal Z}_{_{R}}(q,g)\ \equiv\ \prod_{n\ge0}\prod_{w}\ \frac{1}{1 1-q^{^{n+1/2}}z^{^{w}}\ }  
\end{equation}
with $w\equiv (w_{_{1}},...w-_{{r}})$ the weights of $R$ and 

\begin{equation}   
z^{^{w}}\ \equiv\ \prod_{i}\ z_{_{i}}^{^{w_{_{i}}}}    
\end{equation}

A 3-punctured sphere, the partition function on $S^{^{1}}\times S^{^{3}}$ with holonomies of $SU(2)_{_{i}}$ $\left(a_{_{i}},\frac{1}{a_{_{i}}}\right) \in\ SU(2)_{_{i}}$

\begin{equation}   
{\cal Z}(a_{_{1}},a_{_{2}},a_{_{3}})\ \equiv\ \frac{\ K(a_{_{1}})\ K(a_{_{2}})\ K(a_{_{3}})\ }{K_{_{o}}}\ \sum_{n\ge0}\ \frac{\chi_{_{n}}(a_{_{1}})\ \chi_{_{n}}(a_{_{2}})\ \chi_{_{n}}(a_{_{3}})\ }{\chi\left(q^{1/2}\right)}
\end{equation}  
where the RHS is equivalent to the amplitude of a $q$-deformed $SU(2)$ YM theory on a 3-punctured sphere.

\medskip 

\medskip

\subsection{3D SCFTs on punctures }   \label{sec:4.3}   

\medskip  

\medskip

\underline{\ Higgs branch of $T^{^{\rho}}[\mathfrak{g}]$\ }

\medskip  

\medskip 
Given a 4D ${\cal N}$=4 SYM with gauge group $G$ on $\frac{1}{2}$-space $\mathbb{R}^{^{2,1}}\times\mathbb{R}^{^{\ge0}}$, obtained by dimensional reduction of a 6D ${\cal N}$=(2,0) SCFT, admits $\frac{1}{2}$-BPS boundary conditions for the adjoint fields specified by a homomorphism  

\begin{equation}  
\rho:\mathfrak{su}(2)\rightarrow \mathfrak{g}.  
\end{equation}

The adjoint orbit for each element $e\ \in\mathfrak{g}$, $O_{_{e}}^{^{\mathfrak{g}}}$, is such that

\begin{equation}  
O_{_{e}}^{^{\mathfrak{g}}}\overset{def.}{=} G_{_{\mathbb{C}}}\cdot\ e.  
\end{equation} 

The classification of such $\rho$s, up to conjugation, is equivalent to that of nilpotent elements in $\mathfrak{g}$. When $\mathfrak{g}\equiv \mathfrak{su}(N)$, a nilpotent orbit is specified by the size of the Jordan blocks

\begin{equation}  
n\equiv n_{_{1}}+...+n_{_{k}}  
\end{equation} 
or, equivalently, by the partition, $[n_{_{i}}]$ of $n$. Similar considerations hold for $\mathfrak{g}\equiv \mathfrak{so}(n), \mathfrak{sp}(n)$. 

The boundary conditions for ${\cal N}$+4 SYM are given by Neumann boundary conditions for $\Phi_{_{1,2,3}}$ and Dirichlet BCs for $\phi_{_{4,5,6,}}$  

\begin{equation}  
\Phi_{_{1,2,3}}(s)\ \sim\ \frac{\rho(\tau_{_{1,2,3}})}{s}\ \ \ \ ,\ \ \ \ \Phi_{_{4,5,6}}(s)\bigg|_{_{s\equiv 0}}\ \equiv 0  
\end{equation}

Under S-duality of the 4D bulk, the gauge group $G\rightarrow G^{^{\text{V}}}$, and at $s\equiv0$ lives a 3D SCFT, $T^{^{\rho}}[\mathfrak{g}]$, with $ G^{^{\text{V}}}$-flavour symmetry on the Coulomb branch, coupled to the restriction of the bulk gauge fields to the boundary.

For $\rho\equiv0$, the Coulomb and Higgs branches of $T[\mathfrak{g}]$ are ${\cal N}_{_{\mathfrak{g}^{^{\text{V}}}}},{\cal N}_{_{\mathfrak{g}}}$,  with the latter defining the nilpotent cone of $\mathfrak{g}$, namely the subset of $\mathfrak{g}$ consisting of its nilpotent elements. Its dimension reads   

\begin{equation}  
\text{dim}_{_{\mathbb{C}}}\ {\cal N}_{_{\mathfrak{g}}}\ \equiv\ \text{dim}\  G-\text{rank}\  G  
\end{equation}  

Note that this is always an even number due to the fact that ${\cal N}_{_{\mathfrak{g}}}$ is a hyperh\"ahler cone.

More generally, turning on some Higgs vevs $e$, the low-energy limit of the theory at this point would feature singular moduli along the direction orthogonal to $O_{_{e}}$, whereas those along the nilpotent orbit will be smooth. Hence, their contribution as free hypermultiplets to the total Higgs branch quaternionic dimension of $T^{^{\rho}}[\mathfrak{g}]$ reads   

\begin{equation}
\text{dim}_{_{\mathbb{H}}}\ H\left(T^{^{\rho}}[\mathfrak{g}]\right)\ \equiv\ \frac{1}{2}\left(\text{dim}\  G-\text{rank}\  G -\text{dim}_{_{\mathbb{C}}}O_{_{e}}\right) 
\end{equation}

For classical Lie algebras, $\mathfrak{g}=\mathfrak{su}(N), \mathfrak{so}(2N), \mathfrak{sp}(N), \mathfrak{so(2N+1)}$, $O_{_{e}}$ is labelled by $[n_{_i}]$ and its dimensions are known.

\medskip  

\medskip

\underline{\ The Coulomb branch and the Spaltenstein map\ }

\medskip  

\medskip

\begin{equation}  
\text{CB}\left(T^{^{\rho}}[\mathfrak{g}]\right)\ \subset\ \text{CB}\left(T[\mathfrak{g}]\right)\  
\end{equation}   
implying the LHS is the union of closures of nilpotent orbits in $\mathfrak{g}^{^{\text{V}}}$. For $\mathfrak{g}$ a classical Lie algebra, $T^{^{\rho}}[\mathfrak{g}]$ can be constructed as a braneweb.  

The Spaltstein map leads to 

\begin{equation}  
d:\ {\cal N}_{_{\mathfrak{g}}}/G\ \longrightarrow {\cal N}_{_{\mathfrak{g}^{^{\text{V}}}}}/G
\end{equation}   
hence acting on the orbits as follows 

\begin{equation}
d(O_{_{p}})\ \equiv\ O_{_{q}}  \ \ \ ,\ \ \ q\equiv p^{^{\dag}}\ \ \text{for}\ \ \mathfrak{su}(N)
\end{equation}

Importantly, 

\begin{equation}   
\text{if}\ \ \ O\ge O^{^{\prime}}\ \ \ \Rightarrow\ \ \ d(O)\le d(O^{^{\prime}})  \ \ \ \text{and}\ \ \ \text{CB}\left(T^{^{\rho}}[\mathfrak{g}]\right)\ \subset\ \text{CB}\left(T^{^{\rho^{\prime}}}[\mathfrak{g}]\right)\
\end{equation}
if $T^{^{\rho^{\prime}}}[\mathfrak{g}]$ can be Higgsed to $T^{^{\rho}}[\mathfrak{g}]$. Rephrasing this in terms of the mixed branches of a given $T[\mathfrak{g}]$: given the maximum Coulomb branch ${\cal N}_{_{\mathfrak{g}^{^{\text{V}}}}}$ and a maximum Higgs branch ${\cal N}_{_{\mathfrak{g}}}$, the mixed branches defining the total vacuum moduli space of $T[\mathfrak{g}]$ is 

\begin{equation}  
C\times H\ \equiv\ \left\{(x,y)\ \in\ {\cal N}_{_{\mathfrak{g}^{^{\text{V}}}}}\times {\cal N}_{_{\mathfrak{g}}}\ \bigg|\ d(O_{_{x}})\ge\ O_{_{y}}\ \ \text{and}\ \ d(O_{_{y}})\ge O_{_{x}}\ \right\}  
\end{equation}
such that, upon assigning a vev $e\ \in\ {\cal N}_{_{\mathfrak{g}}}$ to the Higgs branch, the Coulomb branch is restricted to the closure of $d(O_{_{e}})$. 

\medskip  

\medskip

\subsection{The Geometric Langlands Program (GLP)}

The category of modules of a vertex operator algebra (VOA) is semisimple if and only if the VOA is rational. Any QFT has a trivial line operator. In a topological twist of a supersymmetric theory, it corresponds to an object $\mathbb{1}$ in the differential graded (dg) category. A dg-category is an additive category whose morphism spaces are dg vector spaces. Each space Hom$(L_{_{i}},L_{_{j}})$ has a cohomological $\mathbb{Z}$-grading and a nilpotent differential of degree 1, ${\cal Q}$, behaving as a derivation on a composition of morphisms. Equivalence relations are imposed by taking the ${\cal Q}$-cohomology.

A dg category, ${\cal D}$, can often be represented as a dg enhancement of the bounded derived category $D^{^{b}}({\cal C})$ of an abelian category ${\cal C}$

\begin{equation} 
{\cal D}\simeq D^{^{b}}({\cal C})    
\end{equation}   

If ${\cal D}\equiv D^{^{b}}({\cal C})$ is represented as a derived category, then the complex describing local operators at a junction of objects in it are

\begin{equation} 
\text{Hom}_{_{{\cal D}}}(L, L^{\prime})=\text{Hom}_{_{{\cal C}}}(L, L^{\prime})  
\end{equation}  

The degree of the complex corresponds to U(1) R-charge of the local operator. $R$-charges manifesting in terms of higher-extension groups follows from having chosen to represent the intrinsic category of line operators ${\cal D}$ as an enhancement of $D^{^{b}}({\cal C})$. 

The category of lines in 3D TQFTs is expected to be a dg braided tensor category equipped with 

\begin{enumerate}

\item fusion rules 

\begin{figure}[ht!]   
\begin{center}
\includegraphics[scale=0.8]{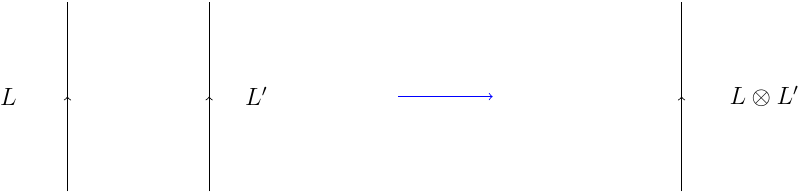}  
\caption{\small  Fusion in ${\cal C}_{_{line}}$.}
\label{fig:fusion1}  
\end{center} 
\end{figure}

\item and braiding 

 \begin{figure}[ht!]   
\begin{center}
\includegraphics[scale=0.8]{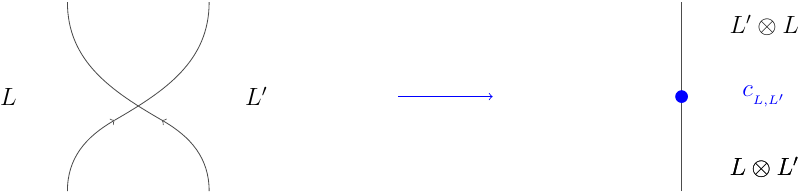}  
\caption{\small  Braiding in ${\cal C}_{_{line}}$.}
\label{fig:br}  
\end{center} 
\end{figure} 

\end{enumerate}  

Non-semisimplicity affects the structure of state spaces on surfaces $\Sigma_{_{g}}$, which are closely related to the category of line operators. In a TQFT with semisimple dg category of line operators, ${\cal D}$, the torus Hilbert space si given by the Grothendieck group $K_{_{o}}({\cal D})$, with a basis labelled by simple objects. For higher genus, the Hilbert spaces are given by the fusion algebra of simple objects, with quantum dimensions provided by the Verlinde formula. When ${\cal D}$ is not semisimple, state spaces are more complicated, in particular, for the torus, it is given by the Hoschild homology of ${\cal D}$, $HH_{_{\circ}}({\cal D})$.  

Admitting a semisimple category of line operators is a strong constrain to impose on a 3D TQFT; one such example being Chern-Simons theory with compact gauge group £G£ and level $k$. Indeed, CS is a TQFT of cohomological type, whose category of line operators ${\cal D}$ is finite semisimple, with trivial dg structure. The simple objects defining it are Wilson lines labelled by irreps of $G$, $\{S_{_{i}}\}$. There are only finitely many of them , due to an equivalence imposed by large gauge transformations. Furthermore, there are no local operators to define gauge-invariant junctions among different Wilson lines, namely, there are no local bulk operators apart from the identity, $\mathbb{1}$. On the other hand, topological twists of supersymmetric theories typically admit non-finite and non-semisimple categories of line operators. 

3D ${\cal N}=$4 gauge theories admit 2 different topological twists:

\begin{enumerate}

\item a reduction of Witten's 4D ${\cal N}=$2 Donaldson twist (i.e. the $A$-twist)  

\item an intrinsically 3D twist (the $B$-twist).

\end{enumerate}   

The two are related to the 2D $A/B$-twists under dimensional reduction on an $S^{^1}$. They also arise in 4D ${\cal N}=4$ SYM from the Kapustin-Witten $A/B$-twists. In such setups, semisimplicity is rare. Looking for a 3D TQFT that matches the structure of the CGP TQFT described above based on the non-semisimple category $U_{_{q}}(\mathfrak{g})$-mod at an even root of unity $q\equiv e^{^{i\pi/k}}$. The theory is CS-like, containing a subset of line operators labelled by the same irreps of $G$ at level $k$, matching the modules of $U_{_{q}}(\mathfrak{g})$ surviving semisimplicity. 

The starting point ia the Gaiotto-WItten 3D ${\cal N}=$ 4 SCFTs, $T[G]$, defined as the S-duality interface of 4D ${\cal N}=4$ SYM. $G$ is a compact simple Lie group depending on the complexified Lie algebra, $\mathfrak{g}$. The theory has $\tilde G\times \tilde G^{^{\text{V}}}$ flavour symmetry, with $\tilde G, \tilde G^{^{\text{V}}}$ denoting the simply-connected forms of $G$ and its Langlands dual, respectively. The two factors of the flavour group act on the Coulomb and Higgs branches of the moduli space of vacua of $T[G]$, which are Langlands dual nilpotent cones  

\begin{equation}   
{\cal M}_{_{Coul}}[T[G]]\simeq {\cal N}^{^{\text{V}}}\in\ \mathfrak{g}^{^{\text{V}}} 
\ \ \ ,\ \ \ 
{\cal M}_{_{Higgs}}[T[G]]\simeq {\cal N}\in\ \mathfrak{g}    
\end{equation}

Requiring that $k\in\mathbb{Z}\simeq H^{^{4}}(B\tilde G)$, $k\ge h$, and $G\equiv\tilde G$ is simply connected, gauging the simply-connected $\tilde G$ symmetry of $T[G]$ by introducing a 3D ${\cal N}=2$ gauge multiplet and a SUSY CS term at level $K$, the resulting theory 

\begin{equation} 
{\cal T}_{_{\tilde G, k}}\ \overset{def.}{=}\ T[G]/G_{_{k}}  
\end{equation}   
which is still a 3D ${\cal N}=$ 4 SUSY theory, with WIlson line operators in the $G$-representation and $G^{^{\text{V}}}$ global symmetry. However, it is not topological due to the presence of superconformal matter in $T[G]$. To solve this issue, a topological twist is needed

\begin{equation}   
{\cal T}_{_{\tilde G, k}}^{^{A}}\ \sim\ \text{CS}[G_{_{k-h}}]\ \otimes \ T[G]^{^{A}}    
\end{equation}   
with resulting Hilbert space

\begin{equation}   
{\cal H}_{_{{\cal T}_{_{\tilde G, k}}^{^{A}}}}(\Sigma)\ \sim\ {\cal H}_{_{G_{_{k-h}}}}(\Sigma)\ \otimes \ {\cal H}_{_{{\cal N}^{^{\text{V}}}}}(\Sigma)   
\end{equation}

  \subsection*{4D construction and 6D relations}

The Gaiotto-Witten theories originate from 6D ${\cal N}=$(2,0) SCFTs of ADE-type, $\mathfrak{g}$, on a product of a 3-manifold and a twisted cigar, $M\times D^{^2}\times_{_{q}}S^{^{1}}$, where $q\equiv e^{^{2\pi\Psi^{^{\text{V}}}}}$. The theory is topologically twisted along $M$ and holomorphically topologically-twisted on $D^{^2}\times_{_{q}}S^{^{1}}$. At $\partial D^{^{2}}$, one places BCs labelled by a completely flat connection, ${\cal A}$, on $M$. Compactifying on the cigar circle and the noncontractible $S^{^{1}}$ in various orders, the resulting theory is 4D YM on $M\times \mathbb{R}_{_{+}}$ with different BCs. 

As analysed by Kapustin and Witten within the context of the Geometric Langlands program, there is a prescription describing how to embed calculations in CS theory into GL-twisted 4D ${\cal N}=$4 SYM. A CS calculation on a 3-manifold ${\cal M}_{_3}$ maps to a 4D gauge theory calculation ${\cal M}_{_3}\times\mathbb{R}^{^+}$ with a specific BC deforming the standard Neumann BCs. The conventional notation for such generalised BCs is $B_{_{(p,q)}}$. The analytically-continued CS level $k$ is related to the coupling $\Psi$ of the GL-twisted 4D theory as 

\begin{equation}  
k+h\equiv\Psi  \ \overset{def.}{=}\ \frac{\theta}{2\pi}+\frac{4\pi i}{g_{_{YM}}^{^2}}\ \frac{t-t^{^-1}}{t+t^{^-1}}
\end{equation}
where $t\equiv\pm i$ corresponds to $\Psi\equiv\tau$.

\begin{figure}[ht!]   
\begin{center}
\includegraphics[scale=1]{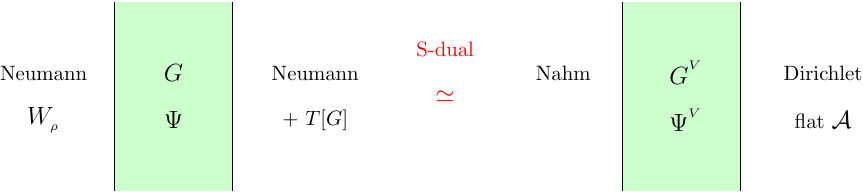}  
\caption{\small  }
\label{fig:tg4D}  
\end{center} 
\end{figure}

\begin{figure}[ht!]   
\begin{center}
\includegraphics[scale=1]{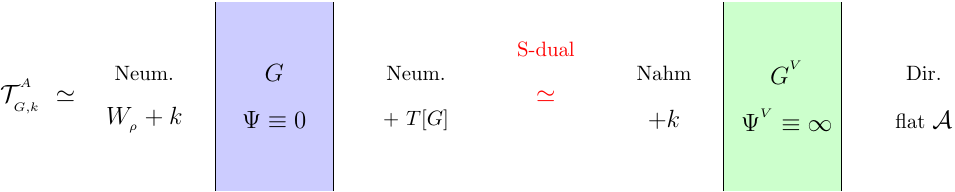}  
\caption{\small  }
\label{fig:tg4D1}  
\end{center} 
\end{figure}

\begin{figure}[ht!]   
\begin{center}
\includegraphics[scale=0.8]{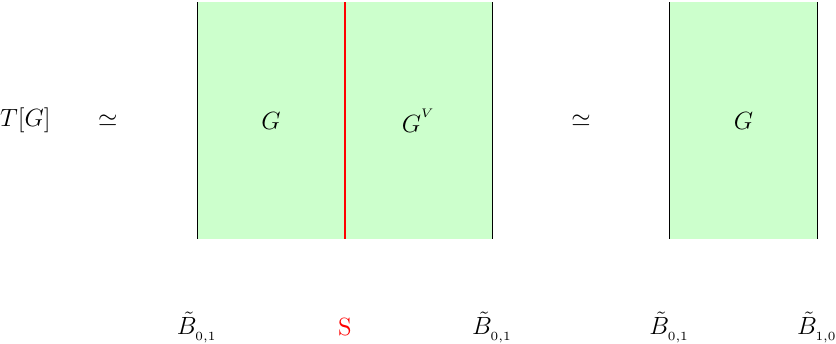}  
\caption{\small  $\tilde B_{_{1,0}}$ and $\tilde B_{_{0,1}}$ denote Neumann and Dirichlet boundary conditions, respectively. They are related by the action of the S-duality wall.}
\label{fig:tg1}  
\end{center} 
\end{figure}

\begin{figure}[ht!]   
\begin{center}
\includegraphics[scale=0.8]{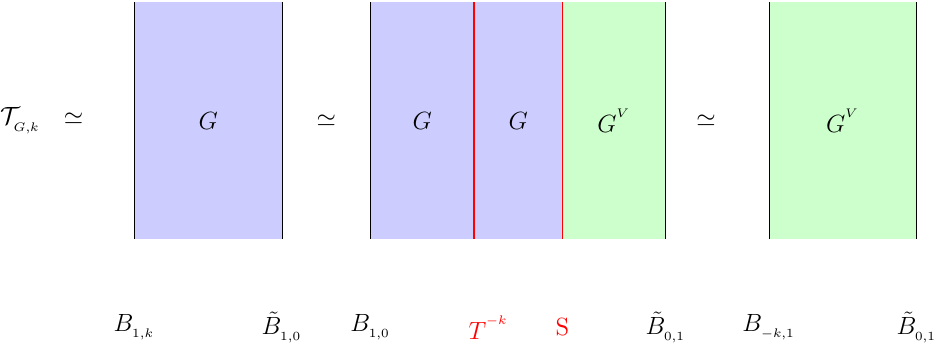}  
\caption{\small  Gauging the $G$-flavour symmetry of $T[G]$ at CS level $k$, corresponds to replacing one of the Dirichlet boundary conditions with a deformed Neumann BC $B_{_{1,k}}$, obtained by placing an $T^{^{-k}}S$ interface between ordinary Neumann and Dirichlet BCs.}
\label{fig:tg2}  
\end{center} 
\end{figure}

When $G\equiv (P)SU(N)$, the configuration of figure \ref{fig:tg2} can be further lifted to branewebs in type-IIB string theory. 

\begin{figure}[ht!]   
\begin{center}
\includegraphics[scale=0.8]{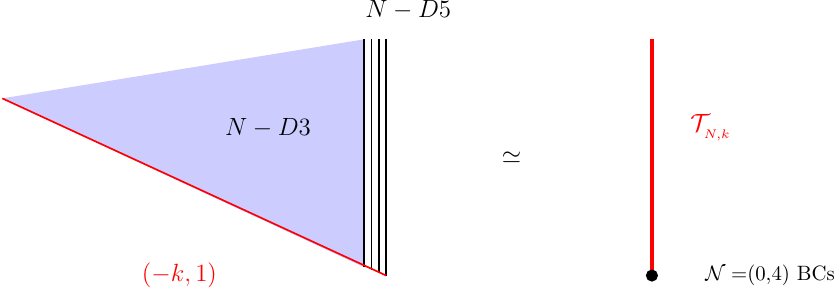}  
\caption{\small  }
\label{fig:tg3}  
\end{center} 
\end{figure}

  Physically, given an object $L$ in ${\cal C}_{_{ab}}$, one may perform a path-integral on a solid torus $D^{^2}\times S^{^1}$ with line operator $L$ inserted at its core to produce a state in the torus state space. This defines a map 

\begin{equation}  
\text{Ob}({\cal C}_{_{ab}})\ \xrightarrow{\ \kappa\ }\ {\cal H}(T^{^2})   
\end{equation} 
with the following properties: 

\begin{enumerate}

\item  it is linear 

\begin{equation} 
\kappa(A\oplus B)\equiv \kappa(A)\oplus \kappa(B)  
\end{equation}  

\item  \begin{equation} 
\kappa(A\otimes B)\equiv \kappa(A)\cdot \kappa(B)  
\end{equation}  

\item  

it factors through the Grothendieck ring $K_{_{o}}({\cal C}_{_{ab}})$ because the states on the RHS don't see the difference between direct sums of linear operators $A\oplus B$ and nontrivial extensions (i.e. bound states) 

\item images of $\kappa$ belong to the cohomological degree-0 subspace of the state space ${\cal H}(T^{^2})$. 

\end{enumerate}   

All this, together with the isomorphism ${\cal H}(T^{^2})\simeq HH_{_{o}}({\cal C})$, implies that $\kappa$ induces the map 

\begin{equation}  
K_{_{o}}({\cal C})\ \rightarrow\  HH_{_{o}}({\cal C})\ \subset\ {\cal H}(T^{^2})  
\end{equation}

  \subsection*{Identity stalk and the flag manifold}

  \begin{equation}  
  D^{^b}({\cal C}_{_{1}}^{^{ab}})\equiv{\cal C}_{_{1}}\ \simeq\ \text{Vect}^{^{\oplus 2}}\ \oplus\ [\text{Coh}(T^*[2]\mathbb{P}^{^1})]^{^\oplus k-1}  
  \end{equation}
where each of the two $\text{Vect}$s is an abelian semisimple category with a single simple object, whereas, $[\text{Coh}(T^*[2]\mathbb{P}^{^1})]^{^\oplus k-1}$, is a dg enrichment of the derived category. The direct sum  signals that there are no morphisms between objects belonging to the two summands. The abelian and derived categories decompose as a direct sum of $K+1$ blocks 

  \begin{equation}  
{\cal C}_{_{1}}^{^{ab}}\ \simeq\ B_{_{k}}^{^{ab+}}\oplus B_{_{k}}^{^{ab-}}\oplus \ \bigoplus_{j=1}^{k+1} \ B_{_{j}}^{^{ab}}  
  \end{equation}
  where $B_{_{k}}^{^{ab\pm}}$ denote the semisimple subcategories generated by the simple projectives $S_{_{k}}^{^{\pm}}$, wheras each $B_{_{j}}^{^{ab}}$ consists of a subcategory generated by the pair $(S_{_{j}}^{^+}, S_{_{k-j}}^{^-})$.

\section{Categories, knots, and entanglement}\label{sec:5}

\subsection{Higher-categorical structure of (linked minimal models) knots}    \label{sec:5.1}

This appendix contains additional background material underlying section \ref{sec:1.1}. In particular, the main focus is that of identifying fusion spaces associated with knots, \cite{Huang:2021gsc}. A concrete calculation of knot multiplicity shows that the knot complement of a trefoil knot can store quantum information. Also spiral maps enabling to understand consistency relations for torus knots and spiral fusions of fluxes.  

\emph{Immersion} of $A$ in $M$ is defined by triples $(A_+, A, \mathfrak{p}_{_{A}})$. The projection $\mathfrak{p}_{_{A}}:A\rightarrow M$ may or may not be an embedding. In particular, the Hilbert space induced by $M$ on $A$ might be smaller w.r.t. the one with which $A$ is already equipped. Pictorially, this would appear as an overlap of $A$ on $M$. At the level of the Hilbert space, the overlap corresponds to a double copy in the tensor product structure. The correct definition can be achieved by suitably lifting the refined state on local disks contained in $\tilde\Gamma(A_+)$.

The information convex set, $\Sigma(\Omega)$, of a 2-hole disk can be completely characterised. Any element belonging to it can be expressed as   

\begin{equation}  
\rho\ =\ \bigoplus_{a,b,c\ \in\ {\cal C}}\ p_{_{ab}}^{^{c}}\ \rho^{^{abc}}  
\end{equation}   
with ${\cal C}$ denoting the set of topological charges, $\{p_{_{ab}}^{^{c}}\}$ a probability distribution, and $\{\rho^{^{abc}}\}$ a set of mutually orthogonal quantum states. For each triple $(a,b,c)$, $\{\rho^{^{abc}}\}$ defines the state space of a finite-dimensional Hilbert space, whose dimension is given by the fusion multiplicities of the underlying anyon theory. $(a,b)$ label the charges associated with the holes, and $c$ the total charge of the disk.

The \emph{information convex set} for an immersed region is defined as $\Sigma(\Omega)=\{\rho_{_{\Omega}}\}$ such that 

\begin{enumerate}  
\item   $\rho_{_{\Omega}}\equiv\text{Tr}_{_{\Omega_+\diagdown \Omega}}\ \rho_{_{\Omega_+}}$.
\item $\rho_{_{\Omega_+}}$ is consistent with $\sigma_{_{\omega}}^{^{\mathfrak{p}}}$\  $\forall\ \omega\ \in\ \tilde\Gamma(\Omega_+)$.    
\end{enumerate}   

Elements of the information convex sets are closed under the merging operation as long as a certain mild condition is satisfied. The \emph{merging theorem} states that immersed spaces, $T_1, T_2$, with isomorphic ICSs, share the same number of elements. From this follows that, under the composition of density matrices belonging to each one of them, $\rho_{_{ABC}}\ \in\ \Sigma(ABC),\ \lambda_{_{BCD}}\ \in\ \Sigma(BCD)$, the resulting merged state reads 

\begin{equation}   
\tau\ \ \overset{def.}{=}\ \ \rho \bowtie \lambda \ \ \ 
, \ \ \ 
\tau\ \in\ \Sigma(ABCD)  
\end{equation}

For the case in which $\Omega$ is \emph{sectorisable}, it ICS can be expressed in the following way  

\begin{equation}  
\Sigma(\Omega)\equiv\ \bigg\{\ \sum_{i}\ p_{_{I}}\rho^{^{i}}_{_{\Omega}}\ \bigg|\ \sum_{_{i}}p_{_{i}}=1, p_{_{i}}\ge0\bigg\}    
\end{equation}  
with $\{\rho^{^{i}}_{_{\Omega}}\}$ denoting the set of mutually orthogonal density matrices. The set of labels $i$ forms the \emph{set of superselection sectors}, ${\cal C}_{_{\Omega}}$, with $i=1$ defining the \emph{vacuum} sector. For any given sectorisable region $\Omega$ embedded in $S^3$ or $B^3$, $\rho_{_{\Omega}}^{^{1}}\equiv\sigma_{_{\Omega}}$ is unique.   

The \emph{quantum dimension} can therefore be defined as follows  

\begin{equation}  
d_{_{i}}\ \overset{def.}{=}\ \exp\ \left(\frac{S\left(\rho_{_{\Omega}}^{^{i}}\right)-S\left(\sigma_{_{\Omega}}^{^{*}}\right)}{2}\right)   
\end{equation}    
where $\sigma_{_{\Omega}}^{^{*}}$ denotes the vacuum state.

A set of \emph{superselection sectors} can be assigned to any sectorisable region. Conventionally, we will be denoting them as $\{{\cal C}_{_{I}}\}$, with the subscript $I$ specifying the sector. Among the simplest choices one could make are: sphere shells, ${\cal C}_{_{point}}$, solid tori, ${\cal C}_{_{flux}}$, and torus shells, ${\cal C}_{_{Hopf}}$. For the purpose of this work, the latter happens to be of particular importance. However, as we shall be arguing throughout, all 3 can be thought of as subcategories defining an $n$-category associated to an $(n+1)$-dimensional theory.

\begin{figure}[ht!]   
\begin{center}  
\includegraphics[scale=0.5]{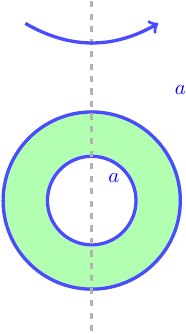}  
\ \ \ \ \ \ \ \ \ \ \ \ \ \ \ \ \ \ \ \ \ \ \ \ \  \ \ \ \ \ \ 
\includegraphics[scale=0.9]{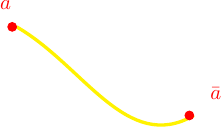}  
\caption{\small A sphere shell, (on the LHS). On the RHS, the topological charges of the 2 connected components are mapped to a particle/antiparticle pair joined by a string.}  
\label{fig:sectreg1}  
\end{center}  
\end{figure} 

\begin{figure}[ht!]   
\begin{center}  
\includegraphics[scale=0.5]{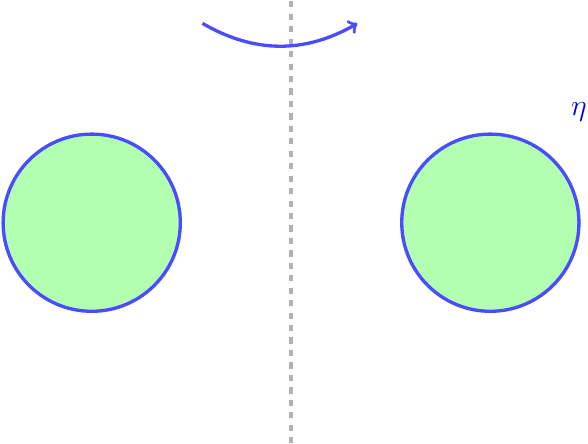}  
\ \ \ \ \ \ \ \ \ \ \ \ \ \ \ \ \ \ \ \ \ \ \ \ \  \ \ \ \ \ \ 
\includegraphics[scale=0.9]{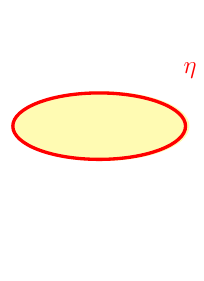}  
\caption{\small A solid torus, $\mathbb{T}$, (on the LHS). On the RHS, a visualisation of a flux operator operator, $\eta$.}  
\label{fig:sectreg1}  
\end{center}  
\end{figure} 

\begin{figure}[ht!]   
\begin{center}  
\includegraphics[scale=0.5]{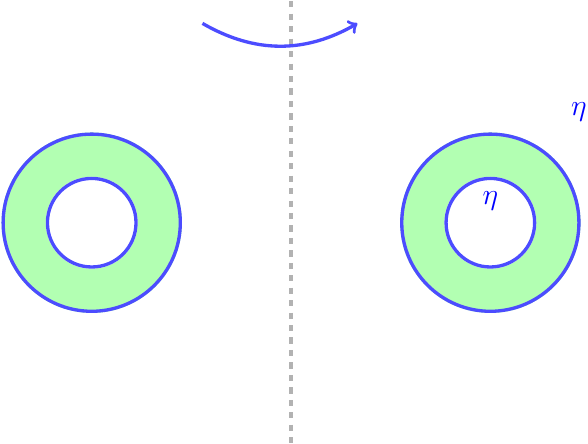}  
\ \ \ \ \ \ \ \ \ \ \ \ \ \ \ \ \ \  
\includegraphics[scale=0.8]{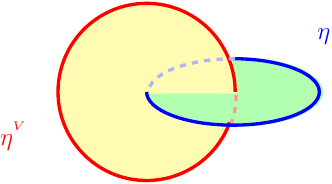}  
\caption{\small A torus shell, $\mathbb{T}$, (on the LHS). On the RHS, a visualisation of the Hopf link, resulting from the mutual intersection of $\eta$ and $\eta^{^{\text{V}}}$.}  
\label{fig:sectreg1}  
\end{center}  
\end{figure}

Its corresponding superselection sector type is the Hopf fibration, ${\cal C}_{_{Hopf} } \equiv\{1, \eta, \zeta, ...\}$. Superselection sectors are associated with different classes of loop excitations. Each one of them can be created by applying a suitable type of membrane operator. The description in terms of the information convex set is somehow dual to that of the excitations, since it focuses on their complement when embedded in an $S^3$. 

For a torus shell, $\Sigma(\mathbb{T})$ is a simplex with extreme points $\{\rho_{_{\mathfrak{T}}}^{^{\eta}}\}_{\eta\ \in\ {\cal C}_{_{Hopf}}}$. The set of Hopf excitations admit a natural decomposition 

\begin{equation}  
{\cal C}_{_{Hopf}}\ \equiv\ \bigcup_{\mu\ \in\ {\cal C}_{_{flux}}}\ {\cal C}_{_{Hopf}}^{^{[\mu]} }  
\label{eq:emb}   
\end{equation}

$\Sigma(\mathbb{T})$ comes equipped with nontrivial automorphisms, such as the one responsible for permuting the consituting loops in the Hopf link of its complement

\begin{equation} 
v\ :\ {\cal C}_{_{Hopf}}\ \xrightarrow{\color{white}aaaa\color{black}}\ {\cal C}_{_{Hopf}}   \ \ \ , \ \ \ v(\eta)\ \equiv \eta^{^{\text{V}}}   
\end{equation}  

The set of shrinkable loops are not linked to any other by definition, and therefore provide a natural embedding for the fluxes

\begin{equation}   
{\cal C}_{_{flux}}\ \overset{\varphi}{\hookrightarrow}\ {\cal C}_{_{loop}}\ \equiv\ {\cal C}_{_{Hpof}}^{^{[1}}\ \subset\ {\cal C}_{_{Hopf}} 
\end{equation}   

The embedding \eqref{eq:emb} implies the matching of the quantum dimensions

\begin{equation}  
\sum_{a\ \in\ {\cal C}_{_{loop}}}\ d_{_{a}}^2\ \equiv\ \sum_{\mu\ \in\ {\cal C}_{_{flux}}}\ d_{_{\mu}}^2   
\end{equation}    
as well as the presence of a well defined map ${\cal C}_{_{loop}}\rightarrow{\cal C}_{_{flux}}$. Indeed, given a torus shell $\mathbb{T}$, and $T\subset\mathbb{T}$ a solid torus, an extreme point $\rho_{_{\mathbb{T}}}^{^{\eta}}\ \in\ \Sigma(\mathbb{T})$, 

\begin{equation}   
\text{Tr}_{_{\mathbb{T}/T}}\rho_{_{\mathbb{T}}}^{^{\eta}}   \equiv \rho_{_{T}}^{^{\mu}} \ \in\ \Sigma(T)    
\end{equation}

  \subsection*{Fusion rules and knot complements}

  \medskip 

  \medskip 

\begin{figure}[ht!]   
\begin{center}  
\includegraphics[scale=0.5]{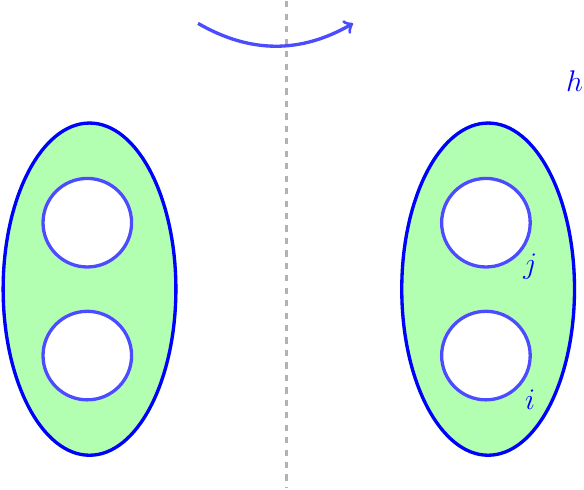}  
\ \ \ \ \ \ \ \ \ \ \ \ \ \ \ \ \ 
\includegraphics[scale=0.7]{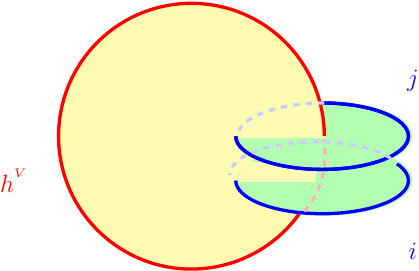}  
\caption{\small On the LHS, a solid torus with 2 tori removed. On the RHS, the fusion of 2 loops in ${\cal C}_{_{Hopf}}^{^{[mu]}}$, from which the multiplicities $\{N_{_{ij}}^{^{h}}\}$ can be extracted.}    
\label{fig:fusion}  
\end{center}  
\end{figure}

The Hilbert space theorem states that, 

\begin{equation}   
\Sigma_{_{I}}(\Omega)\ \simeq\ S\left(\mathbb{V}_{_{I}}\right)   
\label{eq:HST}   
\end{equation}
where $I$ labels the superselection sector and $S\left(\mathbb{V}_{_{I}}\right)$ defines the state space of a finite-dimensional Hilbert space $\mathbb{V}_{_{I}}$. From \eqref{eq:HST}, it follows that the information convex set is completely characterised by the \emph{fusion multiplicity}

\begin{equation}   
N_{_{I}}\ \equiv\ \text{dim}\ \mathbb{V}_{_{I}}   
\end{equation}   
For $N_{_{I}}=0$, $\Sigma_{_{I}}(\Omega)$ is empty, whereas for $N_{_{I}}=1$ it contains a single element, which is an isolated fixed point. For $N_{_{I}}=2$ it is isomorphic to  the Bloch ball and contains an infinite number of continuously parameterised extreme points. When $N_{_{I}}>1$, $\Omega$ can store quantum information.

the information convex sets of 3D regions with boundaries can be associated with a set of fusion spaces. The fusion of 2 loops, depicted on the RHS of figure \ref{fig:fusion}, enables to derive the \emph{fusion multiplicities}, $\{N_{_{ij}}^{^{h}}\}$,

\begin{equation}  
h\times i\equiv\ \sum_j\ N_{_{ij}}^{^{h}} j    
\ \ \ 
,   
\ \ \ 
h,i,j \ \in\ {\cal C}_{_{Hopf}}^{^{[\mu]}}   
\end{equation}

The \emph{complement} of an unknot in $B^3$ is a shrinkable loop at a poin, with \emph{shrinking rule} 

\begin{equation}  
\ell\ \equiv\ \sum_a N_{_{\ell}}^{^{a}}\ a  
\ \ \ 
,   
\ \ \ 
\ell \ \in\ {\cal C}_{_{loop}}\ \ \ 
,   
\ \ \ 
a \ \in\ {\cal C}_{_{point}}   
\end{equation}

 \begin{figure}[ht!]   
\begin{center}  
\includegraphics[scale=0.7]{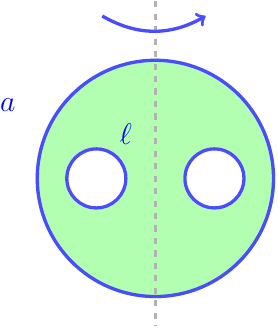}  
\ \ \ \ \ \ \ \ \ \ \ \ \ \ \ \ \ 
\includegraphics[scale=0.7]{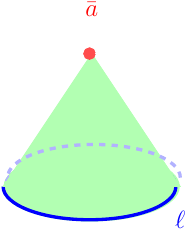}  
\caption{\small $B^3\diagdown T$ on the LHS, with its corresponding shrinking rule on the right.}    
\label{fig:fusion1}  
\end{center}  
\end{figure}

The next essential ingredient that needs to be introduced is the \emph{knot complement}, which plays an important role in their classification. For the purpose of our treatment, their key role resides in the fact that their information convex sets are characterised by nontrivial fusion data. For a given knot, $K$, we will denote its complement on the 3-sphere as 

\begin{equation}  
\Omega_{_{K}}\ \overset{def.}{=}\ S^3/K  
\end{equation}   
with a thickened boundary $\partial\Omega_{_{K}}$, i.e. a knotted embedded torus shell. It therefore follows that 

\begin{equation}  
{\cal C}_{_{\partial\Omega_{_{K}}}}\ \simeq\ {\cal C}_{_{Hopf}}  
\end{equation}   
namely the 2 sets are identical up to permutations. The information convex set of the knot complement, $\Sigma(\Omega_{_{K}})$, is a convex hull of the subset $\Sigma_{_{\zeta}}(\Omega_{_{K}})$, with $\zeta\ \in\ {\cal C}_{_{Hopf}}$. In general, the set of superselection sectors that can exist on a knot $K$ of an $S^3$ alone is a subset of ${\cal C}_{_{Hopf}}$, implying that the knot excitations are a small number w.r.t. the Hopf link. However, they are more coherent, and therefore are able to store quantum information. The proof of this last statement was carried out in \cite{Huang:2021gsc} and requires some additional elements we will now present.

\medskip 

\medskip

  \subsection{Spiral maps, fusion multiplicities and holomorphic blocks}\label{sec:5.2}

\medskip 

\medskip     

The \emph{knot multiplicity}, $N_{_{\zeta}}(K)$ is the dimension of the Hilbert space of $\Sigma_{_{\zeta}}\left(\Omega_{_{K}}\right)$   

\begin{equation}  
\boxed{\ \ \ N_{_{\zeta}}(K)\ \overset{def.}{=}\ \text{dim}\ \mathbb{V}_{_{\zeta}}\left(\Omega_{_{K}}\right) \color{white}\bigg]\color{black} \ \ }    
\label{eq:km}   
\end{equation}   

For the unknot, \eqref{eq:km} reduces to 

\begin{equation}  
N_{_{\eta}}(\text{unknot})\ \overset{def.}{=}\ \sum_{\mu\ \in\ {\cal C}_{_{flux}}}\ \delta_{_{\eta, \varphi(\mu)}}   \ \ \ , \ \ \ \forall \eta\ \in\ {\cal C}_{_{Hopf}}    
\label{eq:km1}   
\end{equation} 
whereas for a general $(p,q)$-knot it reads as follows    
\begin{equation}  
\boxed{\ \ \ N_{_{\zeta}}^{^{\eta}}(p,q)\ \overset{def.}{=}\ \text{dim}\ \mathbb{V}_{_{\zeta}}^{^{\eta}}\left(T\diagdown K_{_{p,q}}\right) \color{white}\bigg]\color{black} \ \ \ , \ \ \ \eta, \zeta\ \in\ {\cal C}_{_{Hopf}}  \ \ } 
\label{eq:km2}   
\end{equation} 
with $\eta, \zeta$ labelling the unknotted (outer) and knot boundaries, respectively.

  \begin{figure}[ht!]   
\begin{center}  
\includegraphics[scale=0.5]{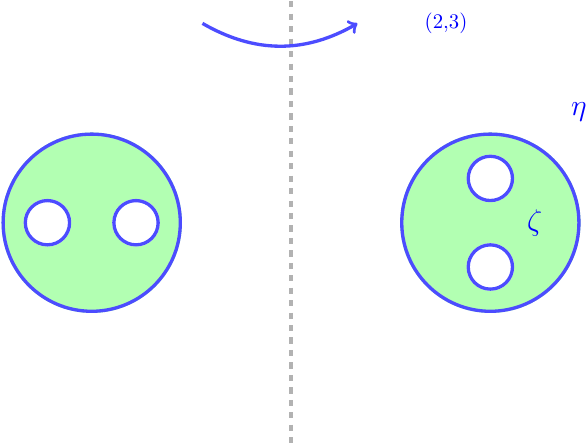}  
\caption{\small Example of $S^3\diagdown K(2,3)$. With $\eta, \zeta$ denoting the unknotted and knotted boundaries, respectively.}    
\label{fig:tmt}  
\end{center}  
\end{figure}  

$\Sigma(\Omega)$ is completely determined by the \emph{set of extreme points}, $\text{ext}(\Sigma(\Omega))$. In order to define the latter, some preliminary definitions are of order. 

\begin{figure}[ht!]   
\begin{center}  
\includegraphics[scale=1]{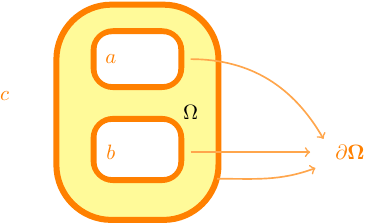}  
\caption{\small $\Omega$ is not sectorisable, but $\partial\Omega$ is, since it has 3 connected components, labelled $i=a,b,c$. Each one of them corresponds to a different superselection sector, collectively denoted ${\cal C}_{_{\partial\Omega}}$.}  
\label{fig:Omega}  
\end{center}  
\end{figure} 

For regions that are not sectorisable, such as the one depicted in figure \ref{fig:Omega}, we can distinguish between $\Omega$ and its boundary defined as follows 

\begin{equation} 
\boxed{\ \ \ \partial\Omega\ \overset{def.}{=}\ {\cal M}_{_{D-1}}\ \times I  \color{white}\bigg]\color{black}\ \ }  
\end{equation}   
where ${\cal M}_{_{D-1}}$ is a $(D-1)$-dimensional closed manifold, and $I$ an interval. $\Sigma(\partial\Omega)$ is a \emph{simplex} since $\partial\Omega$ is sectorisable, namely it is possible to assign to it a finite set of superselection sectors labelled by ${\cal C} _{_{\partial\Omega}}$. If $\partial\Omega$ contains multiple connected components, each component is sectorisable.  

Under the action of partial tracing, every extreme point of $\Sigma(\Omega)$ reduces to one of $\Sigma(\partial\Omega)$.

Equipped with this machinery, it is now possible to define the \emph{extreme point criterion}, stating that, for a given immersed $\Omega$, $\rho_{_{\Omega}}\ \in\ \Sigma(\Omega)$ is an extreme point if and only if

\begin{equation}    
\left(S_{_{\Omega}}+S_{_{\Omega\diagdown\partial\Omega}}-S_{_{\partial\Omega}}\right)_{_\rho}\ \equiv\ 0  
\end{equation}    

Furthermore, $\forall i\ \in\ {\cal C}_{_{\partial\Omega}}$, corresponding to a nonempty $\Sigma_{_{i}}(\Omega)$, the \emph{quantum dimension} is defined as    

\begin{equation}   
\ln\ d_{_{i}}\ \overset{def.}{=}\ S\left(\rho_{_{\Omega}}^{^{i,e}}\right)-S\left(\rho_{_{\Omega}}^{^{1,e}}\right)   
\end{equation}    
with $\rho_{_{\Omega}}^{^{i,e}}$ an extreme point of $\Sigma_i(\Omega)$, and $\rho_{_{\Omega}}^{^{1,e}}$ the vacuum reference state. 

Once having identified the quantum dimensions associated to the different superselection sectors, we can calculate their \emph{fusion multiplicities}, $N_{ab}^{c}$,

\begin{equation}  
d_{_{a}}\ d_{_{b}}\ =\ \sum_{c}\ N_{ab}^{c}\ d_{_{c}}  
\end{equation}
with $(a,b,c)$ denoting the different topological charges involved in the fusion process.

\subsection{The total quantum dimension}  

\medskip 

\medskip

The \emph{total} quantum dimensions can therefore be extracted as follows:

\begin{enumerate}   

\item for points and fluxes

\begin{equation}  
{\cal D}\ \overset{def}{=}\sqrt{\sum_{_{a\ \in\ {\cal C}_{_{point}}}}\ d_{_{a}}^2\ \  }\ \equiv\ \sqrt{\sum_{_{\mu\ \in\ {\cal C}_{_{flux}}}}\ d_{_{\mu}}^2\ \  }
\end{equation}    

\item for a knotted torus, instead, the definition can be relaxed, turning the equality into an upper bound on the topological entanglement entropy   

\begin{equation}  
\sum_{_{\zeta\ \in\ {\cal C}_{_{Hopf}}}}\ N_{\zeta}(K)\ d_{_{\zeta}}\ \equiv\ \sum_{_{\overset{\mu,\nu\ \in\ {\cal C}_{_{flux}}}{t_p(\mu)\equiv t_q(\nu)}}}\ \frac{d_{_{\mu}}^2 d_{_{\nu}}^2 }{d^2_{_{t_p(\mu)}}} \ \le\ {\cal D}^4
\end{equation}    

For the case of $K$ with arbitrary $(p,q)$

\begin{equation}  
\sum_{_{\zeta\ \in\ {\cal C}_{_{Hopf}}}}\ N_{\zeta}^{\eta}(p,q)\ d_{_{\zeta}}\ \equiv\ \sum_{_{\overset{\mu,\nu\ \in\ {\cal C}_{_{flux}}}{t_p(\mu)\equiv t_{(p,q)}(\nu)}}}\ \frac{d_{_{\varphi^{\text{V}}}} d_{_{\eta}} }{d^2_{_{t_p(\mu)}}} \ \le\ {\cal D}^4  
\label{eq:fm}   
\end{equation}
where 

\begin{equation}  
N_{\zeta}^{^{\eta}}(p,q)\ \equiv\ \sum_{_{\mu  \in\ {\cal C}_{_{flux}}}} N^{^{\eta}}_{\zeta\varphi^{^{\text{V}}}(\mu)}      
\label{eq:SSN}   
\end{equation}
\end{enumerate}

Importantly, the embedding operation     

\begin{equation}   
\varphi^{^{\text{V}}}\ :\ {\cal C}_{_{flux}}\ \hookrightarrow\ {\cal C}_{_{Hopf}}
\end{equation}
really consists of 2 consecutive parts, namely 

\begin{equation}  
\varphi\ :\ {\cal C}_{_{flux}}\ \hookrightarrow\ {\cal C}_{_{loop}}\ \subset\ {\cal C}_{_{Hopf}}
\end{equation}   
and 

\begin{equation} 
\eta\ \rightarrow\ \eta^{^{\text{V}}}   
\end{equation}
from which follows that \eqref{eq:SSN} is equivalent to a gauging procedure.

In a 4D TQFT, the entanglement entropy is given by the area-law plus a topological subleading correction  

\begin{equation}  
S_{_{EE}}\ =\ \alpha\ \ell\ +\ \ln{\cal D}  
\end{equation}  
where $\alpha$ is a constant, $\ell$ is the area of the surface on which $S_{_{EE}}$ is being calculated, and ${\cal D}$ is the quantum dimension. Following the Casini-Huerta-Myers prescription, the latter corresponds to the so-called boundary free energy of the 3D boundary to the 4D TQFT. It is a universal quantity, and its importance is crucial for the study of holographic embeddings of such theories.

For $\Sigma(\Omega_{_{K}})$, with corresponding maximum entropy state $\rho_{_{\Omega_{_{K}}}}^{^{*}}$, the entropy w.r.t. the reference state can be extracted from the fusion multiplicities \eqref{eq:fm} and therefore reads

\begin{equation} 
S\left(\rho_{_{\Omega_{_{K}}}}^{^{*}}\right)-S\left(\sigma_{_{\Omega_{_{K}}}}^{^{}}\right)\ =\ 
\ln\left(\sum_{_{\zeta\ \in\ {\cal C}_{_{Hopf}}}}\ N_{\zeta}^{\eta}(p,q)\ d_{_{\zeta}}\right)   
\label{eq;fm1}   
\end{equation}

[analogy with Janus configurations]

\section{Sigma-models}  \label{sec:sigmamodels}

\subsection{The sigma-model}  

By definition, \cite{Freed:1999mn}, a sigma-model is a  field theory describing the field as a point particle confined to move on a fixed manifold. The latter could either be a Riemannian manifold, a Lie group, or a symmetric space. 

When coupling it to a gauge field, we get the Landau-Ginzburg (LG)-model, \cite{LandauGinzburg}. 

The Lagrangian density for the sigma-model can be expressed in many different ways. One of the simplest expressions is the following   

\begin{equation}   
{\cal L}=\frac{1}{2}\sum_{i=1}^{n}\sum_{_j=1}^{n}\ g_{_{ij}}(\phi)\partial^{^{\mu}}\phi_{_i}\partial_{_{\mu}}\phi_{_j}, 
\end{equation}
where $g_{_{ij}}(\phi)$ denotes the metric tensor on the field space $\phi\ \in\ \Phi$, where $\Phi$ can be any Riemannian manifold.

In a more fully-geometric notation, it can be written as a fiber bundle with fibers $\Phi$ over a differentiable manifold $M$. Given a section 

\begin{equation}  
\phi:\ M\ \rightarrow\ \Phi,  
\end{equation}
fix a point $x\ \in\ M$. Then, the pushforward at $x$ is a map of tangent bundles 

\begin{equation}  
d_{_{x}}\phi:\ T_{_{x}}M\ \rightarrow\ T_{_{\phi(x)}}\Phi,  
\end{equation}  
taking 

\begin{equation}  
\partial_{_{\mu}}\ \mapsto\ \frac{\partial \phi^{^i}}{\partial x^{^{\mu}}}\partial _{_i}.  
\end{equation}  

The sigma model action then reduces to   

\begin{equation}  
S=\frac{1}{2}\ \int_{_M}d\phi\ \wedge\ \star\ d\phi. 
\label{eq:clsm}
\end{equation}

Among the most important interpretation of the classical sigma-model is that of non-interacting QM. Taking $\Phi\equiv \mathbf{C}$, \eqref{eq:clsm} becomes 
\begin{equation}  
S=\frac{1}{2}\ <<\phi\ ,\ \Delta\ \phi>>, 
\label{eq:clsm1}
\end{equation}
with

\begin{equation}
   \phi:\ M\rightarrow\mathbb{C} 
\end{equation}
interpreted as a wavefunction, whose Laplacian corresponds to its associated kinetic energy. Hence, the classical sigma-model on $\mathbf{C}$ can be interpreted as the QM of a free particle.

\subsection{Riemann surfaces}

The geodesic structure of a Riemannian manifold is described by the Hamilton-Jacobi equations. As previously mentioned, the $M$ and $\Phi$ defining the sigma-model can both be taken to be Riemannian. 

The cotangent bundle $T^*\Phi$ can always be locally trivialised, namely  
\begin{equation}
   T^*\Phi\bigg|_{_U}\ \simeq\ U\times\ \mathbf{R}^{^n}.
\end{equation}   

Given $g_{_{ij}}$ on $\Phi$, the Hamiltonian function is defined as follows 

\begin{equation}
H(q,p)\ \overset{def.}{=}\ \frac{1}{2}\ g^{^{ij}}(q)\ p_{_i}p_{_j},
\end{equation} 
from which the Hamilton-Jacobi equations define the geodesic (or Hamiltonian) flow on $\Phi$ 

\begin{equation}   
\dot q^{^i}=\frac{\partial H}{\partial p_{_i}}\ \ \ ,\ \ \ \dot p_{_i}=-\frac{\partial H}{\partial q^{^i}}.  
\end{equation} 

The Hamiltonian description therefore enables us to interpret the sigma-model as the gluing of two energy functionals, namely the momenta in $T^*\Phi$ and $T^*M$.

\subsection{Moduli spaces}

In algebraic geometry, a moduli space is a geometric space (a scheme or an algebraic stack) whose points represent algebro-geometric objects of some kind, or their isomorphic classes. The main motivation for introducing moduli spaces is that of finding solutions of geometric problems.

A scheme is a mathematical structure that enlarges the notion of an algebraic variety, accounting for multiplicities, and allowing varieties defined on any commutative ring. Scheme theory was introduced by A. Grothendieck in 1960 for addressing deep problems in algebraic geometry. 

A scheme is a topological space together with commutative rings for all of its open sets, which arises from gluing together spectra of commutative rings along their open subsets, i.e. ringed space which is locally a spectrum of a commutative ring.

Algebraic surfaces can be studied making use of morphisms of schemes. In many cases, the family of all varieties of a given type can itself be viewed as a variety of schemes, known as moduli space.

\subsection{The nonlinear sigma-model} 

Considering Minkowski spacetime, and denoting with ${\cal F}$ the space of fields, $L$ the Lagrangian, and $\gamma$ the variational 1-fomr, ${\cal M}$ the space of classical solutions, and $\omega$ the local symplectic form. In its classical formulation, the Lagrangian encodes a classical Hamiltonian system once a specified time direction has been selected, thereby breaking Poincare' invariance. 

Upon taking a real-valued scalar field 

\begin{equation}   
\phi: M^{^n}\ \rightarrow\ \mathbb{R}.     
\end{equation}  

The kinetic term in the Lagrangian would therefore read  

\begin{equation} 
\begin{aligned}
L_{_{kin}} &=\frac{1}{2}|d\phi|^{^2}\left|d^{^n}x\right| \\    
&=\frac{1}{2}g^{^{\mu\nu}}\partial_{_{\mu}}\phi\partial_{_{\nu}}\phi \left|d^{^n}x\right| \\  
&=\frac{1}{2}\left[(\partial_{_0}\phi)^{^2}-\sum_{_{i=1}}^{^{n-1}}(\partial_{_i}\phi)^{^2}\right]\left|d^{^n}x\right|.  
\label{eq:lkin}
\end{aligned}   
\end{equation}    

Adding a potential term of degree $\le 2$ of the kind

\begin{equation}  
V: \mathbb{R}\ \longrightarrow\ \mathbb{R}   
\end{equation}  
to \eqref{eq:lkin}, and assuming $V$ is bounded below by choosing suitable coefficients of th e polynomial, then one can eliminate the linear and constant terms by assuming the minimum of $V$ is at the origin with $V=0$ there. The total Lagrangian therefore becomes  

\begin{equation}  
L=\left(\frac{1}{2}|d\phi|^{^2}-\frac{m^{^2}}{2}\phi^{^2}\right)\left|d^{^n}x\right|.  
\end{equation}

In non-free theories, the potential is no longer quadratic. The simplest example of an interacting theory is the case of the quartic potential. In either case, in a pure mathematical formulation, the non-linear $\sigma$-model is characterised by the following data:  

\begin{enumerate}

\item A Riemannian manifold, $X$. 

\item   A potential energy function  

\begin{equation}    
V:\ X\ \longrightarrow\ \mathbb{R}.   
\end{equation}   

\end{enumerate}  

In this formulation, the space of fields is th emapping space    

\begin{equation}  
{\cal F}=\left(\text{Map}(M^{^n}, X\right).  
\end{equation}

The total Lagrangian therefore reads

\begin{equation}  
L=\left[\frac{1}{2}|d\phi|^{^2}-\phi^{^*}V\right]\left|d^{^n}x\right|,      
\end{equation}   
from which the energy density reads

\begin{equation}  
{\cal E}(\phi)=\left[\frac{1}{2}|\left|\partial_{_0}\phi\right|^{^2}+\sum_{_{1=1}}^{^{n-1}}\frac{1}{2}\left|\partial_{_i}\phi\right|^{^2}+V\right]\left|d^{^{n-1}}x\right|,      
\end{equation}

The space of solutions, namely the moduli space of vacua, ${\cal M}_{_vac}$, is 

\begin{equation}   
{\cal M}_{_vac}\ \subset\ {\cal F}{\cal E}_{_N},  
\end{equation}
with ${\cal F}{\cal E}_{_N}$ denoting the space of static fields of finite energy. 

For the case of a real scalar field with potential $V$, one usually assumes that the minimum value is at $V=0$, and therefore that the solutions lie at

\begin{equation}  
{\cal M}\ \overset{def.}{=}\ V^{^{-1}}(0).    
\end{equation}  

For example, if $X\equiv\mathbb{R}$, and $V=0$, the resulting theory is that of a massless real scalar field. For this case, one finds that ${\cal M}_{_{vac}}\simeq\mathbb{R}$. On the other hand, for the case of a scalar field with potential 

\begin{equation}   
V(\phi)=\frac{m^{^2}}{2}\phi^{^2}   
\end{equation}  
${\cal M}_{_{vac}}$ is a single point $\phi=0$. For a quartic potential, instead, ${\cal M}_{_{vac}}$ is constituted by two points, i.e. the values of $\phi$ where $V$ is minimised.

\subsection{Principle bundles and connections} 

Formulating gauge theories in Lagrangian formulation that is suitable for a QFT, some preliminary differential geometry notatioons are needed.  

The initial ingredients we start from are a manifold, $M$, and a Lie group, $G$. Then, a \emph{principal G-bundle}, $P\rightarrow M$ is defined as a manifold $P$ on which $G$ acts freely on the right with quotient 

\begin{equation} 
P/G\simeq M,  
\end{equation}  
such that there exist local sections. If $P^{^{\prime}}, P$ are principal $G$-bundles over $M$, then an isomorphism  

\begin{equation}  
\varphi^{^{\prime}}:P^{^{\prime}}\ \longrightarrow\ P,     
\end{equation}   
is a smooth diffeomorphism which commutes with $G$, and induces the identity map on $M$. If $P\equiv P^{^{\prime}}$, such automorphism defines a gauge transformation of $P$.

For every $M$, there is a category of principal $G$-bundles and isomorphisms. 

A \emph{connection} on a principal $G$-bundle $\pi: P\rightarrow M$ is a $G$-invariant distribution in $TP$ which is transverse to the vertical distribution kerd$\pi$. A connection can therefore be expressed as a 1-form $A\in\Omega^{^1}(P;\mathfrak{g})$ whose value at $p$ is the projection

\begin{equation}    
T_{_p}P\ \longrightarrow\ V_{_p}\simeq\mathfrak{g},      
\end{equation}   
with kernel $H_{_p}$.

Connections form a category, arranging in a set of equivalence classes under isomorphisms.

\subsection{Mathematical Gauge theory}  

The prototypical example of a gauge theory is Maxwell's theory. In its quantised formulation, one needs to introduce Dirac's charge quantisation, replacing the group $\mathbb{R}$ by the compact group $\mathbb{Z}/2\mathbb{Z}$. This statement can be generalised by replacing $\mathbb{Z}/2\mathbb{Z}$ with any compact Lie group $G$. 

Hence, we are drawn to the conclusion that the Lagrangian formulation of the field theory in question is simply based on the following data:   

\begin{enumerate}  

\item A compact Lie group with Lie algebra, $\mathfrak{g}$.

\item  A bi-invariant inner product on $\mathfrak{g}$, $<.,.>$.

\end{enumerate}

From this, the total Lagrangian for the gauge field reads

\begin{equation}  
L=-\frac{1}{2}\left<F_{_A}\wedge\star F_{_A}\right>  
\end{equation}
from which the energy density reads   

\begin{equation}  
{\cal E}(A)=\left[\ \sum_{_{\mu<\nu}}]\frac{1}{2}\left|F_{_{\mu\nu}}\right|^{^2}\right]\ \left|d^{^{n-1}}x\right|,      
\end{equation}  
whose moduli space is simply 

\begin{equation}  
{\cal M}_{_{vac}}=\text{pt},       
\end{equation} 
where the calculation is understood to be performed on the space of equivalence classes of connections.

\subsection{The gauged sigma-model}  

This is a very general bosonic field theory, combining pure gauge theory with the $\sigma$-model, both of which are greatly used in QFT. The data specifying the model in question are: 

\begin{enumerate}

\item A Lie group, $G$, with Lie algebra, $\mathfrak{g}$. 

\item A bi-invariant scalar product, $<,>$, on $\mathfrak{g}$.    

\item A Riemannian manifold, $X$, on which $G$ acts by isometries.

\item A potential function   

\begin{equation}    
V:\ X\ \longrightarrow\ \mathbb{R},     
\end{equation}
invariant under $G$.  

\end{enumerate}  

The space of fields, ${\cal F}$, consists of pairs 

\begin{equation} 
{\cal F}\ \overset{def.}{=}\{(A,\phi)\},    
\end{equation}  
with $A$ denoting a collection of principal $G$-bundles $P\rightarrow M$, and $\phi$ a section of the associated bundle over $P\times_{_G}X\rightarrow M$. In most cases, it is convenient to view $\phi$ as an equivariant map 

\begin{equation}    
\phi:\ P\ \longrightarrow\ X.   
\end{equation}

The space of fields can be thought of as a category, with an isomorphism 

\begin{equation}    
\varphi:\ P^{^{\prime}}\ \longrightarrow\ P     
\end{equation}
of principal bundles, inducing an isomorphism of fields  

\begin{equation}     
\left(A^{^{\prime}},\phi^{^{\prime}}\right)\ \longrightarrow\ (A,\phi).  
\end{equation}    

The global symmetry group of the theory is the subgroup of isometries of $X$ commuting with the $G$-action, and preserving the potential, $V$. 

The Lagrangian density, therefore reads  

\begin{equation}    
{\cal L}=-\frac{1}{2}|F_{_A}|^{^2}+\frac{1}{2}|{\cal D}_{_A}\phi|^{^2}-\phi^{^*}V,  
\label{eq:Lagrangiandensity}    
\end{equation}  
where ${\cal D}_{_A}$ denotes the covariant derivative, coupling $\phi$ with $A$. The energy density of the pair $(A,\phi)$, can readily be extracted from \eqref{eq:Lagrangiandensity}, and reads  

\begin{equation}   
{\cal E}(A,\phi)=\left[\ \sum_{_\mu<\nu}\frac{1}{2}|F_{_{\mu\nu}}|^{^2}+\sum_{_{\mu}}\frac{1}{2}\left|\left(\partial_{_A}\right)_{_{\mu}}\phi\right|^{^2}+\phi^{^*}V\ \right]\ \left|d^{^{D-1}}x\right|,     
\label{eq:energyd}   
\end{equation} 
where $D$ denotes the total number of spacetime dimensions. 

Assuming $V$ has a minimum at 0, one can look for soulitions with 0-energy. By direct inspection of \eqref{eq:energyd}, it is possible to deduce that:   

\begin{enumerate}  

\item  The first term implies that $A$ is flat, hence, up to equivalence, it is the trivial connection with zero-curvature.

\item   The second term tells us that $\phi$ has to be a constant for it to be a zero-energy solution.

\item   The third term constraints the constant value of $\phi$ to lie in the $V^{^{-1}}(0)$ set.

\end{enumerate}   

A trivial connection $A$ has a group of automorphisms that are isomorphic to $G$, i.e. the group of global gauge transformations. For a vacuum solution, $\phi$ must be a constant function with values in $V^{^{-1}}(0)$, and, therefore, $V^{^{-1}}(0)$ inherits the same $G$-action from $\phi$. Hence, the moduli space of vacua is a subquotient space of $X$  

\begin{equation}   
\boxed{\ \ \ {\cal M}_{_{vac}}\ \overset{def.}{=}\ V^{^{-1}}(0)\bigg/G\color{white}\bigg]\ \ }.  
\end{equation}

For the case of certain supersymmetric field theories, $X$ is K$\ddot{\text{a}}$hler of hyperk$\ddot{\text{a}}$hler, and $V^{^{-1}}(0)$ is the norm square of an appropriate moment map, in which case ${\cal M}_{_{vac}}$ reduces to a K$\ddot{\text{a}}$hler of hyperk$\ddot{\text{a}}$hler quotient (with the latter actually being the setting where this had first been studied).

\section{(A)dS\texorpdfstring{$_3/T\bar T$}{}-deformed CFTs }\label{sec:TT}

As previously argued, the entropy of pure AdS is divergent, compatibly with the fact that bulk reconstruction enables to probe the IR bulk at arbitrary depth by evaluating $S_{_{EE}}$ on ever-increasing subregions of the conformal boundary.

Within the AdS/CFT correspondence, finiteness of bulk-reconstruction can be achieved by shifting the conformal boundary at a finite cutoff by deforming the CFT Lagrangian density by means of an irrelevant operator. In 2D, this is known as the $T\bar T$-deformation, initially introduced by Zamolodchikov, \cite{BB901}. The RG-flow equations driven by this operator interpolate between the UV and the IR. 

For the 2D dual to an (A)dS$_3$, \cite{BB70, BB48}, the action associated to the deformed theory reads

\begin{equation}     
S_{T\bar T}^{^{AdS}}
\ 
= 
\ 
S_{CFT}+\lambda\int d^{2} x\ \sqrt{-\gamma} \ T\bar T     
\label{eq:TTADS}    
\end{equation}

\begin{equation}    
S_{T\bar T}^{^{dS}}
\ 
= 
\ 
S_{CFT}+\lambda\int d^{2} x\ \sqrt{-\gamma} \  \left[\ T\bar T +\frac{1}{2\lambda^2}\  \right]
\label{eq:TTDS}   
\end{equation}  
respectively, with

\begin{equation}  
T\bar T\ \overset{\text{def.}}{=}\ \frac{1}{8}(T^{ab}T_{ab}-(T^{a}_{a})^{2}) 
\end{equation}

The deformation of the CFT$_2$ Lagrangian induces a change in the partition functions,  which, for AdS$_3$ reads 

\begin{equation} 
\delta \ln {\cal Z} 
\ 
= 
\ 
\ln {\cal Z}_{T\bar T}-\ln {\cal Z}_{CFT} 
\ 
= 
\ 
-\frac{1}{2}\ \int d^{2}x\ \sqrt{\gamma}\ <T^{ab}>\ \delta \gamma_{ab} 
\label{eq:zeta}     
\end{equation} 
and \emph{trace flow equations}, \cite{BB901}, 

\begin{equation} 
<T^{\ a}_{\ a}> 
\ 
= 
\ 
-\frac{c}{24}{\cal R}-\frac{\lambda}{4}\left(\ <T^{ab}><T_{ab}>-<T^{a}_{a}>^{2}\ \right)      
\end{equation} 
where the 1$^{st}$ term defines the trace anomaly associated with a CFT$_{2}$.

The first application of these deformations for describing the gluing of bulk spacetimes from a holographic perspective, was carried out by H. Verlinde et al. in \cite{BB70} for AdS$_{3}$ spacetimes, and further developed by Silverstein et al. \cite{BB48} for local realisations of dS$_3$s in an asymptotically AdS$_3$ background, with the latter relying upon the dS$_{_{D+1}}$/dS$_{_{D}}$-correspondence (dS/dS for brevity). The fact that the latter defines a duality in between 2 dS theories follows from the fact that the deformed Lagrangian driving the RG-flow contains an additional term w.r.t. the AdS$_3$/CFT$_2$ case which is playing the role of a positive 2D cosmological constant. We will come back to this in greater detail when deriving the IR $S_{_{EE}}$s in both setups. Before doing so, we will briefly outline the key features characterising the dS/dS-correspondence, motivating that the gluing of different spacetimes in the IR can be naturally accounted for in both correspondences alike.

According to the dS/dS correspondence, \cite{BB22}, a D+1-dimensional dS spacetime with curvature radius $1/H_{D+1}$ is equivalent to 2 interacting CFTs of D-dimemsions with a cutoff $H_{D+1}$ and coupled to gravity (cf. figure \ref{fig:dSdS}).

\begin{figure}[ht!]        
\begin{center}    
\includegraphics[scale=1.5]{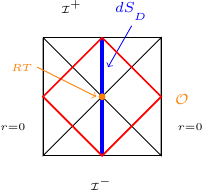} 
\ \ \ \ \ \ 
\includegraphics[scale=1.5]{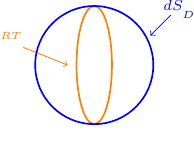}        
\caption{\small{The conformal boundary is depicted in blue. The RT-surface is still a codimension-2 surface, in complete analogy with the AdS/CFT setup. For $D=2$, it reduces to 2 antipodal points on an $S^{1}$.  } }   
\label{fig:dSdS} 
\end{center}    
\end{figure} 

The analogy in between the 2 AdS/CFT and the dS/dS correspondences already shows at the level of the metric, which read\footnote{$\sin$ is for dS and $\sinh$ for AdS.}

\begin{equation} 
  ds^{2}_{_{D+1} }
  \ 
  = 
  \ 
 dw^{2}+\sin(\text{h})^{2}\left(w\ H\right)\left[-d\tau^{2}+\frac{1}{H^{2}}\cosh^{2}\left(\tau\ H\right)\ d\Omega_{_{D-1}}^{2}\right] 
\end{equation}    
with the key difference being the change in the scale factor $\sin\rightarrow\sinh$, implying that the 2 match in the IR-region of the bulk, i.e. for small values of the arguments of the trigonometric/hyperbolic functions. From the CFT point of view this corresponds to the fact that the CFT on the dS/dS side has to be deformed by an irrelevant operator.

\medskip 

\medskip

\subsection{\texorpdfstring{${\cal Z}_{S^{2}}^{^{\ \ T\bar T}}$}{ZSTbT} and  \texorpdfstring{${S}_{_{T\bar T}}$ }{2}  for AdS }  

\medskip 

\medskip

For the case of a CFT$_{2}$, the $S^{2}$-partition function can be determined upon integrating (\ref{eq:zeta}) over the metric defined by the line element 

\begin{equation} 
ds^{2} 
\ 
= 
\
 r^{2}(d\theta^{2}+\sin^{2}\theta\ d\phi^{2}) 
\end{equation}   
 
The partition function changes under a change in $r$, \cite{BB70, BB48}, hence, from (\ref{eq:zeta})

\begin{equation}    
\frac{\partial\ \ln {\cal Z}_{S^{2}}^{^{\ \ AdS,  T\bar T}} }{\partial r} 
\ 
= 
\ 
-\frac{1}{r}\ \int d^{2}x\ \sqrt{\gamma}\ T^{a}_{a} 
\ 
= 
\ 
-\frac{1}{r}\int\ d\theta\ d\phi\ \sin\theta\ r^{2}\ \frac{2}{\lambda}\left(1-\sqrt{1+\frac{c\lambda}{12 r^{2}}}\right)     
\nonumber    
\end{equation}   

\begin{equation}   
\Rightarrow 
\ \ \ 
\frac{\partial\ \ln {\cal Z}_{S^{2}}^{^{\ \ AdS,  T\bar T}} }{\partial r} 
  \ 
= 
\ 
-\frac{8\pi}{\lambda}\ \left(r-\sqrt{r^{2}+\frac{c\lambda}{12 }}\right)    
\label{eq:t2}   
\end{equation}

\begin{equation}    
\Rightarrow 
\ \ \ 
\ln {\cal Z}_{S^{2}}^{^{\ \ AdS,  T\bar T}} 
  \ 
= 
\ 
-\frac{8\pi}{\lambda}\ \int \ dr\ \left(r-\sqrt{r^{2}+\frac{c\lambda}{12 }}\right)     
\ 
= 
\ 
\frac{8\pi}{\lambda}\ \left[\ -\frac{r^{2}}{2}+\frac{c\ \lambda}{12}\int\ dx\ \ \cosh^{2}x\ \right] 
\nonumber 
\end{equation}
where $r(x)\overset{\text{def.}}{=}\sqrt{\frac{c\lambda}{12}}\ \sinh x$; the partition function becomes

\begin{equation} 
\ln {\cal Z}_{S^{2}}^{^{\ \ AdS, T\bar T}} 
  \ 
= 
\ 
\pi \frac{c}{3} \sinh^{-1}\ \left(\sqrt{\frac{12}{c\lambda}}\ r\ \right)-\frac{4\pi\  r^{2}}{ \lambda}\left[\ 1-\ \sqrt{\frac{c\lambda}{12\  r^{2}}+1}\ \right]    
\label{eq:AdSPF2}    
\end{equation} 
up to an overall integration constant. The entanglement entropy can be determined from (\ref{eq:t2}) and (\ref{eq:AdSPF2}), \cite{BB70},

\begin{equation} 
\boxed{\ \ S_{_{T\bar T}} 
\ 
= 
\ 
\left(1-\frac{r}{2}\frac{\partial}{\partial r}\right) \ln {\cal Z}_{S^{2}}^{^{\ \ AdS, T\bar T}} 
\ 
= 
\ 
\pi \frac{c}{3}\  \sinh^{-1}\ \left(\sqrt{\frac{12}{c\lambda}}\ r\ \right)    \ \ } 
\label{eq:finalresult1} 
\end{equation}

For later purposes, it is important to highlight that, \eqref{eq:AdSPF2} and \eqref{eq:finalresult1} differ by means of the following term 

\begin{equation}    
-\frac{r}{2}\frac{\partial}{\partial r}  \ln {\cal Z}_{S^{2}}^{^{\ \ AdS, T\bar T}} 
=    
\frac{4\pi\  r^{2}}{ \lambda}\left[\ 1-\ \sqrt{\frac{c\lambda}{12\  r^{2}}+1}\ \right]    
\label{eq:diffttb} 
\end{equation}    
and therefore, one could argue that, under a suitable choice of $\lambda$, \eqref{eq:diffttb} can be set to vanish. In such case, $S_{_{T\bar T}}$ would be uniquely determined by $\ln{\cal Z}_{S^{2}}^{^{\ \ AdS, T\bar T}}$. We will be coming back to the importance of this observation later on in section \ref{sec:5.4}.

For $r>>\sqrt{\lambda c}$,

\begin{equation} 
\begin{aligned}
\sinh^{-1} x 
& 
=  
\ln\ \left|\ x\ +\ \sqrt{1+x^{2}}\ \right | 
\ 
= 
\
\ln\ \left|\ x\ +\ x\ \sqrt{1+\frac{1}{x^{2}}}\ \right | 
\ 
\sim \nonumber\\ 
&    
\sim 
\ln\ \left|\ x\ \right| +\left(1+\frac{1}{2 x^{2}}\right) 
\ 
= 
\ 
\ln\ \left|\ x\ \right| +\frac{2x^{2}+1}{2 x^{2}} 
\end{aligned}  
\end{equation}
and (\ref{eq:finalresult1}) therefore turns into

\begin{equation}    
S_{_{T\bar T}} 
\ 
= 
\ 
\frac{c}{3}\ \ln\ \left|\ \sqrt{\frac{12}{c\lambda}}\ r\ \right|\ \ +  \frac{c}{3}\ \frac{2\left(\frac{12}{c\lambda}\ r^{2}+1\right)}{2 \frac{12}{c\lambda}\ r^{2}} \ - \ \frac{c}{3}\ \frac{1}{\frac{24}{c\lambda}\ r^{2}}    
\label{eq:lee1} 
\end{equation}   

By further substituting $r(x)\overset{\text{def.}}{=}\sqrt{\frac{c\lambda}{12}}\ \sinh x$, (\ref{eq:lee1}) can be brought down to the more familiar expression

\begin{equation} 
S_{_{T\bar T}} 
\ 
= 
\ 
\frac{c}{3}\ \ln\ \left|\ \sinh x\right|\ \ +   \frac{c^{2}\lambda}{36}\ \coth^{2}x \ - \ \frac{c}{6}\ \frac{1}{ \sinh^{2}x }   
\label{eq:lee2} 
\end{equation}    
where the 1$^{st}$ term is the entanglement entropy for an interval of length $r$ at finite temperature $\beta\sim\sqrt{12/c\lambda}$. Equation (\ref{eq:lee2}), to leading order in the coupling $\lambda$, therefore reads

\begin{equation} 
S_{_{T\bar T}} 
\ 
= 
\ 
\frac{c}{3}\ \ln\ \left|\ \sqrt{\frac{12}{c\lambda}}\ r\ \right|\ \ +  \frac{c^{2}\ \lambda}{72}\ \frac{1}{ r^{2}} \ + \ \frac{c}{3}  
\label{eq:lee3} 
\end{equation}

From (\ref{eq:lee1}) we can now see that in the UV-limit, i.e. as we take $\underset{r\rightarrow 0}{\lim} S_{_{T\bar T}}$, the first two terms cancel each othert out, thereby preventing the entanglement entropy form diverging. This can be checked explicitly through de l'Hopitàl's rule for the $\infty-\infty$ case.    
\medskip 

\medskip


\subsection{ \texorpdfstring{${\cal Z}_{S^{2}}^{^{\ \ T\bar T}}$ }{2} and  \texorpdfstring{${S}_{_{T\bar T}}$ }{2}    for dS }  

\medskip 

\medskip   

For dS$_{3}$/dS$_{2}$, the corresponding geodesic in the space of 2D theories is given by, \cite{BB47}, 

\begin{equation} 
\boxed{\ \ \ \frac{\delta{\cal L}}{\delta \lambda} 
\ 
= 
\ 
2\  T\bar T-\frac{1}{ \lambda^{2}}  \textcolor{white}{\Biggl[} \color{black}\ \ } 
\end{equation} 
which differs from the AdS case by means of the 2$^{nd}$ term, taking into account that the CFT$_{2}$ is now coupled to gravity. The corresponding change to the sphere partition function now reads

\begin{equation} 
\frac{\partial}{\partial \lambda}\ \ln {\cal Z}_{S^{2}}^{^{\ \ dS, T\bar T}} 
\ 
= 
\ 
-2\int\ d^{2}x\ \sqrt{\gamma}\ <T\bar T> +\ \frac{1}{ \lambda^{2}}\ \int\ d^{2}x\ \sqrt{\gamma} 
\ 
= 
\ 
-8\pi r^{2}\ <T\bar T> +\frac{4\pi\ r^{2}}{ \lambda^{2}} 
\end{equation} 

From the factorisation of the $T\bar T$ expectation value, 

\begin{equation} 
<T\bar T> 
\ 
= 
\ 
-\frac{1}{4}<T^{\theta}_{\theta}>^{2} 
\ 
= 
 \ 
 -\frac{1}{4\ \lambda^{2}}\ \left(\frac{\lambda c}{12r^{2}}-2\sqrt{\frac{\lambda c}{12r^{2}}-1}\right) 
 \end{equation}

 \begin{equation} 
 \Rightarrow 
 \ \ \ 
\frac{\partial}{\partial \lambda}\ \ln {\cal Z}_{S^{2}}^{^{\ \ dS,  T\bar T}} 
\    
= 
\    
\frac{2\pi\ r^{2}}{\lambda^{2}}\ \left(\frac{\lambda c}{12r^{2}}-2\sqrt{\frac{\lambda c}{12r^{2}}-1}\right) +\frac{4\pi\ r^{2}}{\lambda^{2}} 
\end{equation} 
which can be integrated w.r.t. $\lambda$ resulting in 

\begin{equation}      
 \ln {\cal Z}_{S^{2}}^{^{\ \ dS,  T\bar T}} 
\ 
= 
\ 
\pi\frac{c}{6}\ \ln\left|\ \frac{\lambda c}{12 r^{2}}\ \right| -\frac{4\pi\ r^{2}}{\lambda}- 4\pi\ r^{2}\ \int d\lambda\ \frac{1}{\lambda^{2}} \sqrt{\frac{\lambda c}{12r^{2}}-1}     
\end{equation}        

The last term can be integrated by parts 

\begin{equation}   
\begin{aligned}  
- 4\pi\ r^{2}\ \int d\lambda\ \frac{1}{\lambda^{2}} \sqrt{\frac{\lambda c}{12r^{2}}-1}    
& 
= 
- 4r^{2} \left(\ \-\frac{1}{\lambda}\  \sqrt{\frac{\lambda c}{12r^{2}}-1}  +\int\ d\lambda\ \frac{1}{2\lambda}\ \frac{1}{ \sqrt{\frac{\lambda c}{12r^{2}}-1}  }\ \frac{c}{12 r^{2}} \right)    
= \nonumber\\ 
&    
= 
- 4\pi\ r^{2} \left(\ \-\frac{1}{\lambda}\  \sqrt{\frac{\lambda c}{12r^{2}}-1}  +\frac{c}{12 r^{2}}\ \tan^{-1}\ \left(\frac{1}{\sqrt{\frac{\lambda c}{12r^{2}}-1}  }\right) \right)    
\end{aligned}  
\end{equation}

Gathering all the terms, the partition function thus reads 

\begin{equation} 
\ln {\cal Z}_{S^{2}}^{^{\ \ dS, T\bar T}} 
\ 
= 
\ 
\pi \left[\frac{c}{3}\ \ln\left|\ \sqrt{\frac{\lambda c}{12 }}\frac{1}{r}\ \right| -\frac{4\ r^{2}}{\lambda}\left(1-  \sqrt{\frac{\lambda c}{12r^{2}}-1} \right) +\frac{c}{3}\ \tan^{-1}\ \left(\frac{1}{\sqrt{\frac{\lambda c}{12r^{2}}-1}  }\right)\right] 
\label{eq:dSPF1}    
\end{equation} 
from which the entanglement entropy thus follows, \cite{BB47}, 

\begin{equation}    
 \boxed{\ \ S_{_{T\bar T}} 
 \ 
 = 
 \ 
\left(1-\frac{r}{2}\frac{\partial}{\partial r}\right) \ln {\cal Z}_{S^{2}}^{^{\ \ dS,  T\bar T}} 
 \ 
 = 
 \ 
 \pi\left[\frac{c}{6}+\frac{c}{3}\ \ln\left|\ \sqrt{\frac{\lambda c}{12 }}\frac{1}{r}\ \right|+\frac{c}{3}\ \tan^{-1}\ \left(\frac{1}{\sqrt{\frac{\lambda c}{12r^{2}}-1}  }\right)   \right] \ \ } 
\label{eq:dSEE11}    
\end{equation}

Notice that \eqref{eq:dSEE11} and \eqref{eq:dSPF1} differ by means of the following terms

\begin{equation}  
-\frac{r}{2}\frac{\partial}{\partial r}  \ln {\cal Z}_{S^{2}}^{^{\ \ dS,  T\bar T}} 
=   
-\frac{c}{6}+\frac{4\ r^{2}}{\lambda}\left(1-  \sqrt{\frac{\lambda c}{12r^{2}}-1} \right) 
\end{equation}
which, as also argued for the AdS case, \eqref{eq:diffttb}, can be set to vanish under suitable choice of $\lambda$.

\medskip 

\medskip


 \subsection{Comparing \texorpdfstring{$S_{_{T\bar T}}$}{} for AdS and dS} 
 
 
 \medskip 

\medskip   

Comparing the expressions for the partition functions on the dS and AdS side, i.e. (\ref{eq:dSPF1}) and (\ref{eq:AdSPF2}) respectively, we can rewrite the $\sinh^{-1}$ term in the latter as a $\ln$,

\begin{equation}  
\sinh^{-1} \left(\ \sqrt{\frac{12}{c\lambda}}\ r\ \right) 
\ 
= 
\ 
\ln\ \left|\sqrt{\frac{12}{c\lambda}}\ r+\sqrt{\frac{12 r^{2}}{c\lambda}+1}\ \right| 
\ 
= 
\ 
\ln\ \left|\sqrt{\frac{12}{c\lambda}}\ r\ \left(1+\sqrt{\frac{c\lambda}{12 r^{2}}+1}\ \right) \right| 
\nonumber     
\end{equation}      
which is reminiscent of the usual structure of $S_{_{EE}}$ for a finite-T CFT. The AdS partition function therefore becomes

\begin{equation}
\ln {\cal Z}_{S^{2}}^{^{\ \ AdS, T\bar T}} 
  \ 
= 
\ 
\pi \frac{c}{3} \ln\ \left|\sqrt{\frac{12}{c\lambda}}\ r\ \left(1+\sqrt{\frac{c\lambda}{12 r^{2}}+1}\ \right) \right| +\frac{4\pi}{ \lambda}\left[\ r\ \sqrt{r^{2}+\frac{c\lambda}{12}}\  -r^{2}\ \right] 
\label{eq:AdSPF3}    
\end{equation} 
suggesting the factor of $\sqrt{\frac{12}{c\lambda}}=\frac{1}{\epsilon}$ can be interpreted as the cutoff at which the dS can be glued in the IR of the original AdS. By means of such regularisation, (\ref{eq:dSPF1}) can be rewritten in terms of $\epsilon$ as follows

\begin{equation} 
\ln {\cal Z}_{S^{2}}^{^{\ \ dS,  T\bar T}} 
\ 
= 
\ 
\pi\left[\frac{c}{3}\ \ln\left|\ \sqrt{\frac{\lambda c}{12 }}\frac{1}{\epsilon}\ \right| -\frac{4r^{2}}{\lambda}\left(\ 1- \frac{1}{\pi}\ \sqrt{\frac{\lambda c}{12r^{2}}-1}\ \right) +\frac{c}{3}\ \tan^{-1}\ \left(\frac{1}{\sqrt{\frac{\lambda c}{12r^{2}}-1}  }\right)   \right]  
\label{eq:dSPF2}    
\end{equation}

Notice that only the $\epsilon$-dependent term diverges one upon taking the UV-limit with $\underset{r\rightarrow 0}{\lim}$, i.e. when probing short-distance scales. From the redefinition (\ref{eq:dSPF2}),  the entanglement entropy for the $T\bar T$ deformed dS-theory now reads, \cite{BB47}, 
 
 \begin{equation} 
 \boxed{\ \ S_{_{T\bar T}} 
 \ 
 = 
 \ 
\left(1-\frac{r}{2}\frac{\partial}{\partial r}\right) \ln {\cal Z}_{S^{2}}^{^{\ \ dS,  T\bar T}} 
 \ 
 = 
 \ 
\pi\left[\frac{c}{6}+ \frac{c}{3}\ \ln\left|\ \sqrt{\frac{\lambda c}{12 }}\frac{1}{\epsilon}\ \right|+\frac{c}{3}\ \tan^{-1}\ \left(\frac{1}{\sqrt{\frac{\lambda c}{12r^{2}}-1}  }\right)   \right] \ \ } 
\label{eq:dSEE1}    
\end{equation} 

In the work of Silverstein et al., \cite{BB47}, the 2$^{nd}$ term in (\ref{eq:dSEE1}) is set to vanish by appropriately tuning the value of the CFT parameters and consequently of their corresponding bulk duals.
For our purposes, though, this cutoff dependence plays a key role, and will therefore be kept explicit. As a further remark, notice that, for the case in which the dS well approximates the original AdS, the universal parts of the entanglement entropies on the AdS and dS side agree, since the  $\tan^{-1}$ term in (\ref{eq:dSEE1}) can be rewritten as $\sinh^{-1}$ by making the argument imaginary

\begin{equation}  
 \tan^{-1}\ \left(\frac{1}{\sqrt{\frac{\lambda c}{12r^{2}}-1}  }\right) 
 = 
 \ 
\sinh^{-1}\ \left(\sqrt{\frac{12}{c\lambda}}\ r\ \right)   
\end{equation}      
thereby proving that, under a suitable definition of $\epsilon$, dS can arise from the RG-flow of a bulk AdS. We will be coming back to this when comparing localisation calculation with the actions for vacuum transitions. 
  
\medskip 

\medskip   


 \subsection{Cutoff dependence of \texorpdfstring{$S_{_{EE}}$}{}}

\medskip 

\medskip   

The divergence of $S_{_{EE}}$ for pure AdS when evaluated at $z=0$ follows from removing the UV-cutoff $\epsilon$ acting as a regulator, \cite{CC}. When deforming the CFT with an irrelevant operator, $S_{_{EE}}$ will still be dependent on the cutoff. However, its explicit dependence can be removed by appropriately tuning the parameters in the UV and IR of the theory, thereby leading to finite $S_{_{EE}}$s.

\medskip 

\medskip

\subsection{Key points of \texorpdfstring{$T\bar T$}{}-deformations}

\medskip 

\medskip   

\begin{itemize}

\item  They constitute an example of UV/IR-mixing, and therefore provide an interesting setup where to address unitarity issues similar to the ones arising within the context of the information loss paradox.

\item dS can arise as the IR-limit of an RG-flow driven by an irrelevant operator starting from a UV-theory dual to AdS.

\item Localisation techniques play a key role in evaluating the partition function for a given CFT. Given its key role in determining $S_{_{EE}}$, the extremisation procedure rooted in the formalism is compatible with the method defining $S_{_{TOT}}$ by means of the FMP method. It is therefore reasonable to argue, as we shall see in section \ref{sec:5}, that these quantities must be proportional to each other. 

\item   The cutoff radius ensures the finiteness of $S_{_{T\bar T}}$ on both sides, explicitly featuring only universal terms.  

\item   According to the dimensionality of the CFTs being involved, the behaviour of the universal terms is predicted by the $g,c$ and $a-theorems$.   

\item   The main advantage of having introduced $\lambda\neq 0$ is that, in the limit $r/\sqrt{c\lambda }\rightarrow 0$, $S_{EE}\rightarrow 0$ and does not diverge upon taking the UV-limit, unlike the usual CFT result. This proves that, removing the regulator, unitarity is still preserved. As we will see, removing the CFT-cutoff is equivalent to taking the limit of the vanishing cosmological constant in the dS$\rightarrow$ dS transition in absence of a black hole.

\end{itemize}

\section{Von Neumann algebras (VNAs)}

\subsection{Types of VNAs}

A von Neumann algebra (VNA) is a $*$-subalgebra $M$ of ${\cal B}({\cal H})$ that coincides with its bicommutant  

\begin{equation}  
M=M^{^{\prime\prime}}.   
\end{equation}  

A VNA $M\subset {\cal B}({\cal H})$ is called a factor if $M\cap M^{^{\prime}}=\mathbb{C}\mathbf{1}_{_{{\cal B}({\cal H})}}$.  

VN factors can be of different types:  

\begin{enumerate}  
\item I$_{_n},$ or I$_{_{\infty}}$, classified by their dimension.  

\item II$_{_1}$ or II$_{_{\infty}}$, with 

\begin{equation}  
II_{_{\infty}}\simeq II_{_1}\otimes {\cal B}({\cal H}).  
\end{equation}   

\item III$_{_{\lambda}}$ with $\lambda\in[0,1]$.
\end{enumerate} 

For the case of type-II$_{_1}$, a \emph{projection} in a VNA $M$ is an element $p\in M$ such that

\begin{equation}   
p^{^*}=p=p^{^2}.   
\end{equation}

The type of factor depends on the existence of small projections. Specifically, a linear functional $\tau$ on a VNA $M$ is  

\begin{itemize}  
\item Positive, if $\tau(x^{^*}x)\ge 0 \forall x\in M$.  

\item Faithful, if $\tau(x^{^*}x)=0\Rightarrow x=0$.  

\item Stable, if $\tau (1)=1$.  

\item Tracial, if $\tau(xy)=\tau(yx)\  \forall x,y\in M$. A tracial state is simply referred to as trace.  

\end{itemize}

\subsection{Standard form of a II\texorpdfstring{$_{_1}$}{} factor}   

Let $M$ be a factor with continuous and faithful trace $\tau$. The pairing

\begin{equation}  
<x,y>=\tau(y^{^*}x)\ \in\ \mathbb{C}  
\end{equation}   
defines an inner product on $M$. Let $L^{^2}(M,\tau)$ be the associated completion of $M$. Then there is an embedding 

\begin{equation}  
\begin{aligned}
&M\ \longrightarrow\ L^{^2}(M)  \\
&x\ \mapsto\ \hat x\\   
&1\ \mapsto\ \Omega.
\end{aligned}
\end{equation}  

$M$ is represented on a Hilbert space $L^{^2}(M)$ by considering   

\begin{equation}  
\pi_{_\tau}(x)\hat y=\hat{xy}\equiv x\hat y  
\end{equation}  
and extending it to a morphism    

\begin{equation}    
\pi_{_{\tau}}:\ M\ \longrightarrow\ {\cal B}\left(L^{^2}(M)\right).  
\end{equation}

This representation is called the standard form of $M$, implying that   

\begin{equation}    
\hat x=\hat{x1}=x\hat 1=x\Omega,   
\end{equation}    
such that

\begin{equation}    
\tau(x)=<x\Omega,\Omega. 
\end{equation}

\subsection{Representations and von Neumann dimensions}   

A representation of a factor $M$ is a Hilbert space ${\cal H}$ with a structure of an $M$-module. 

There is a Theorem, stating that any representation ${\cal H}$of a II$_{_1}$ factor $M$ is equivalent to 

\begin{equation}    
p\left(L^{^2}(M)\otimes\ell^{^2}(\mathbb{N})\right),
\end{equation}  
where $p$ is a projection 

\begin{equation}    
p=vv^*    \ \in\ (M\times 1)^{\prime}
\end{equation} 
whose trace is the size of the representation ${\cal H}$ 

\begin{equation}    
\text{dim}_{_M}{\cal H}\ \equiv\ \text{Tr}\left(vv^*\right)\ \in\ [0,\infty].
\end{equation} 

\subsection*{Properties of the VN dimension}  

If $M$ is a type-II$_{_1}$ factor, then 

\begin{equation}    
\text{dim}_{_M}\bigoplus_{_{j\in J}}{\cal H}_{_j}\ \equiv\ \sum_{_{j\in J}}\text{dim}_{_M}{\cal H}_{_j}.
\end{equation}   

If $p\in M^{\prime}$ is a projection, then   

\begin{equation}    
\text{dim}_{_M}L^{^2}(M)_{_p}\ =\ \tau(p).
\end{equation} 

Representations are classified by their $M$-dimension, and it therefore follows that   

\begin{equation}    
\text{dim}_{_M}{\cal H}\ =\ \text{dim}_{_M}{\cal K}\ \ \ \iff\ \ \ {\cal H}\ \simeq\ {\cal K}.
\end{equation} 

\subsection{The Jones index} 

Let $N\subset M$ be type II$_{_1}$-factors. Then $N$ is a representation on $L^{^2}(M)$. 

The \emph{Jones index} of the subfactor $N\subset M$ is defined as follows  

\begin{equation}  
[M:N]\ \overset{def.}{=}\ \text{dim}_{_N}L^{^2}(M).   
\end{equation} 

If $N$ is represented on ${\cal H}$ with dim$_{_N}{\cal H}<\infty$, then  

\begin{equation}  
[M:N]\ =\ \frac{{dim}_{_N}{\cal H}}{\text{dim}_{_M}{\cal H}}.   
\end{equation}

\section{Other appendices to Part IV}


\subsection{Twist operators and Rényi entropies }  \label{subsec:3.2}   


The definition of entanglement entropy involves a path integral over a Riemann surface, where the boundary conditions dictated by the twist field operators\footnote{Twist field operators are located at the endpoints of the intervals w.r.t. which $S_{_{EE}}$ is being evaluated.}, ${\cal T}$, encode the nontrivial topological features of the theory, \cite{CC}. This is of particular interest whenever dealing with theories with no 1:1 correspondence between the solutions to the e.o.m. and the original Lagrangian density. 

A similar issue arises when performing coordinate transformations in 2D gravity theories, such as maximal extensions, ultimately leading to a rich web of dualities. The bulk-boundary correspondece of AdS/CFT, as well as the factorisation problem, provide further examples. Before delving into a more detailed analysis of concrete examples, we wish to highlight first the importance of one major quantity relating all these topics together, which in turn prompted the question of understanding its relation with non-invertible symmetries, namely the R\'enyi entropies.

For later convenience, we will briefly overview the main features of the arguments outlined by Cardy and Calabrese for proposing this quantity as the suitable candidate to study CFT$_2$s in all their richness.  In doing so, we will also recapitulate how the emergence of the island becomes manifest from a pure $CFT/BCFT$ calculation as the change in dominance of connected/disconnected channels within correlation functions of twist operators, stressing the importance of the fact that $S_{_{EE}}$ for a single interval in a BCFT shares the same behaviour as that of 2 intervals on a CFT.  
\medskip    

\underline{  4-point functions in a $CFT$\ \ } 

\medskip

In this digression, we briefly outline the derivation of $S_{_{EE}}$, following the techniques of \cite{CC}.
In presence of conformal symmetry breaking, $<T^{\mu}_{\mu}>\neq 0$, and the boundary-to-boundary propagator picks up a correction of the kind $<{\cal T}{\cal T} T_{\mu}^{\mu}>$. The map from the Riemann surface\footnote{... where the $n$-point functions of the twist operators are evaluated.} to the complex plane is defined as, \cite{CC},

\begin{equation}
  z 
  \ 
  = 
  \ 
  \xi^{1/n} 
\ 
= 
  \ 
  \left(\frac{w-u}{w-v}\right)^{1/n} 
  \ \ \ 
  \Rightarrow 
  \ \ \ 
  {\cal R}_{n,1}\rightarrow\mathbb{C} 
\end{equation} 
and the trace of the energy-momentum tensor transforms accordingly as follows    
  
\begin{equation}
  T(w) 
  \ 
  = 
  \ 
  \left(\frac{dz}{dw}\right)^{2}T(z)+\frac{c}{12}\{z,w\} 
  \label{eq:Scw} 
\end{equation} 
with the 2$^{nd}$ term in \ref{eq:Scw} denoting the \emph{Schwarzian derivative},  from which its expectation value becomes

\begin{equation}
  <T(w)\underset{{\ \cal R}_{n,1} }{>\ \ \ }  
  \ 
  = 
  \ 
    <T(z)\underset{\mathbb{C} }{>\ \ \ }\ +\ \frac{c(1-n^{-2})}{24}\frac{(v-u)^{2}}{(w-u)^{2}(w-v)^{2}} 
   \frac{ <{\cal T}_{n}(u,0)\tilde{\cal T}_{n}(v,0)T(w)\underset{{\cal L}^{(n)}, \mathbb{C} }{>\ \ \ } }{<{\cal T}_{n}(u,0)\tilde{\cal T}_{n}(v,0)>}
\end{equation}
with twist operators ${\cal T}_{n}, \tilde{\cal T}_{n}$ being primary fields of conformal dimension $\Delta_{n},\bar \Delta_{n}$ determined from the 2-point function 
    
    \begin{equation} 
   <{\cal T}_{n}(u,0)\tilde{\cal T}_{n}(v,0)>_{{\cal L}^{(n)},\mathbb{C}}
    \ 
    =    
    \ 
     |u-v|^{-2\Delta_{n}} 
     \ \ \ 
     \Rightarrow 
     \ \ \ 
     \Delta_{n} 
     \ 
     = 
     \ 
     \frac{c}{12}\left(n-\frac{1}{n}\right) 
     \ \ \ 
     . 
     \end{equation}

Following the work of Cardy and Calabrese, \cite{CC}, the 4-point function of the twist operators characterising the Riemann surface on which a field theory lives, defines the partition function for 2 disjoined intervals on the FT side. The $n^{th}$-\emph{Rènyi entropy} is defined as

\begin{equation}      
S^{(n)}_{_{EE}} 
\ 
\overset{\text{def.}}{=}    
\ 
\text{Tr}(\rho_{A})^{n}   
\ 
= 
\ 
\frac{{\cal Z}_{n}  }{{\cal Z}^{n}}    
\ \ \ 
, 
\end{equation}    
with 

\begin{equation}    
\begin{aligned}
 {\cal Z}_{n}    \ 
 &    \overset{\text{def.}}{=}      \      
<{\cal T}(z_{1})\bar{\cal T}(z_{2})\bar{\cal T}(z_{3}){\cal T}(z_{4})> _{cyl} 
\ 
= \nonumber\\    
&    
=    
\left(\frac{2\pi}{\beta}\right)^{\frac{c}{3}\left(n-\frac{1}{n}\right)} |w_{1}w_{2}w_{3}w_{4}|^{\frac{c}{12}\left(n-\frac{1}{n}\right)}<{\cal T}(w_{1})\bar{\cal T}(w_{2})\bar{\cal T}(w_{3}){\cal T}(w_{4})> _{\mathbb{C}} 
\label{eq:2ch} 
\end{aligned}
\end{equation}    
with the complex coordinates $w_{i}$ being related to the spacetime coordinates $(t,r)$ via the following change of variables   

\begin{equation}
\begin{aligned}
&w_{1} 
\ 
= 
\ 
e^{-2\pi(r_{I}+t_{I})/\beta}e^{i\pi/2 } 
\ \ \ 
, 
\ \ \ 
w_{2} 
\ 
= 
\ 
e^{2\pi(r_{I}-t_{I})/\beta}e^{i\pi/2 } 
\nonumber\\ 
&     
w_{3} 
\ 
= 
\ 
e^{-2\pi(r_{R}+t_{R})/\beta}e^{-i\pi/2 } 
\ \ \ 
, 
\ \ \ 
w_{4} 
\ 
= 
\ 
e^{2\pi(r_{R}-t_{R})/\beta}e^{-i\pi/2 } 
\end{aligned}  
\end{equation}

Given a subregion $A$ obtained by partial tracing over the density matrix of the total system, the Rényi entropies $S^{(n)}$, the entanglement entropy $S_{_{EE}}$, the reduced denisty matrix $\rho_{_{A}}$ and the partition function ${\cal Z}$ are related according to the following 

\begin{equation} 
\boxed{\ \ S_{A}^{^{EE}}  
\ 
\overset{\text{def.}}{=}     
\ 
\underset{n\rightarrow1}{\lim}\  S_{A}^{(n)} 
\ 
= 
\ 
-\underset{n\rightarrow1}{\lim}\ \left[\ \frac{1}{1-n}\ln\text{Tr}\rho_{A}^{n}\ \right]     
\ 
= 
\ 
-\underset{n\rightarrow1}{\lim}\  \frac{\partial}{\partial n}\ \text{Tr}\rho_{A}^{n} 
\ 
= 
\ 
-\underset{n\rightarrow1}{\lim}\ \frac{\partial}{\partial n}\ \frac{{\cal Z}_{n}(A)}{{\cal Z}^{n}} \ \ }        
\label{eq:ententr} 
\end{equation} 

\medskip  

\medskip

\medskip  

\medskip   
\newpage    

\subsection*{Key points:}

\begin{itemize}  

\item $n$ is assumed to be analytically continued in order for the derivative to be defined in \eqref{eq:ententr}.   

\item The need to consider 1 interval rather than 2 for the case of a BCFT, suggests one should be looking for a more generic approach towards describing the behaviour of the system for $n\ \in \mathrm{R}$.  

\item The twist operators are playing the same role as the non-invertible defects of a QFT featuring global internal symmetries.

\end{itemize}

\subsection{\texorpdfstring{$S_{tot}^{^{\ \text{AdS$\rightarrow$AdSBH}}}$}{}}\label{sec:B}      


In this appendix we will briefly argue how to deal with transitions involving a black hole only on one side of the brane. At the level of the Lagrangian density, this amounts to turning on the mass deformation on, either, the interior or the exterior vacua. The procedure to obtain the total action turns out to be a combination of the cases outlined in sections \ref{sec:2} and \ref{sec:3}, and we therefore quote the key steps, referring the reader to the aforementioned parts of this work for further details and intermediate steps. 

The extrinsic curvatures on the 2 sides of the brane would now be

\begin{equation} 
\frac{\phi^{\prime}}{L}\bigg|_{\pm} 
\ 
= 
\ 
\frac{\Lambda_{+}-\Lambda_{-}\mp\frac{\kappa^{2}}{4}}{\kappa}\ \phi_{b}+\frac{{\cal B}}{\kappa} 
\nonumber 
\end{equation}

Assuming the black hole lies in the inner spacetime\footnote{Labelled by the ``-''  sign.}, the turning point can be determined from the potential

\begin{equation} 
\phi_{1,2} 
\ 
= 
\ 
-\frac{\left(\Lambda_{+}-\Lambda_{-}\ -\frac{\kappa^{2}}{4}\right)\ {\cal B}}{\kappa}\ \phi_{o}^{2}\ \left[1\pm\sqrt{1-\frac{4\kappa^{2}\left(\frac{{\cal B}^{2}}{\kappa^{2}}+{\cal C}\right)}{\left(\Lambda_{+}-\Lambda_{-}-\frac{\kappa^{2}}{4}\right)^{2}\ {\cal B}^{2}\ \phi_{o}^{2}}}\right]   
\nonumber 
\end{equation}

The bulk horizon contribution will only come from the AdSBH, 

\begin{equation} 
S_{hor} 
\ 
= 
\ 
\frac{\eta}{G}\frac{{\cal B}}{\Lambda_{-}}\left[\ 1-\sqrt{1-\frac{4{\cal C}\Lambda_{-}}{{\cal B}^{2}}}\ \right]
\nonumber    
\end{equation} 
whereas the brane action would now read 

\begin{equation} 
\begin{aligned}
S_{brane} 
&
= 
&
\frac{2\eta}{G}\int_{\phi_{1}}^{\phi_{2}}d\phi_{b} \left[\cosh^{-1}\left[\frac{\left(\Lambda_{+}-\Lambda_{-}+\frac{\kappa^{2}}{4}\right)\phi_{b}+{\cal B}}{\kappa\sqrt{-{\cal C}-{\cal B}\phi_{b}-\Lambda_{-}\phi_{b}^{2}}}\right]-\cosh^{-1}\left[\frac{\left(\Lambda_{+}-\Lambda_{-}-\frac{\kappa^{2}}{4}\right)\phi_{b}+{\cal B}}{\kappa\sqrt{-{\cal C}-\Lambda_{+}\phi_{b}^{2}}}\right]\right]
\nonumber \\
&
=
&         
\frac{2\eta}{G}\ \left[\   
-\frac{{\cal B}}{2\Lambda_{-}}\  \ln\left|\frac{y_{2}}{y_{1}}\sqrt{\frac{1-y_{1}^{2}}{1-y_{2}^{2}}}\right| -\phi_{2}\ln\left|\frac{A_{2}\ y_{2}\sqrt{1-y_{4}^{2}}}{A_{4}\ y_{4}\sqrt{1-y_{2}^{2}}}\right|+\phi_{1}\ln\left|\frac{A_{2}\ y_{1}\sqrt{1-y_{3}^{2}}}{A_{4}\ y_{3}\sqrt{1-y_{1}^{2}}}\right| +\  \right.    
\nonumber\\ 
&& 
- 
\left.\frac{1}{2}\sqrt{\frac{\frac{{\cal B}^{2}}{4\Lambda_{-}}-{\cal C}}{\Lambda_{-}}}\ \ln\ \left|\frac{1-y_{2}}{1+y_{2}}\frac{1+y_{1}}{1-y_{1}}\right| +\mu_{+}\ \ln\ \left|\frac{1-y_{4}}{1+y_{4}}\frac{1+y_{3}}{1-y_{3}}\right| \  \right]   
.
\label{eq:2VR}  
\end{aligned}
\end{equation}

\subsection{No obstructions from the Generalised Second Law (GSL)}

 Wormhole traversability has been amply studied in the last 2 decades thanks to the development of the AdS/CFT correspondence, \cite{JM}. The birth of an inflating universe from an asymptotically flat background has been argued to violate wormhole traversability as a consequence of the  \emph{generalised 2$^{nd}$ law} (GSL), \cite{BBAW}. However, we would like to highlight that the results obtained in the present paper by means of the Hamiltonian formalism show that this does not hold in our case for several reasons that we will now outline. In light of the closing remarks in section \ref{sec:4}, the total action for the transition can be recast in the form of the difference of 2 $S_{_{EE}}^{T\bar T}$. Relying upon the holographic understanding of such kinds of theories, we can deduce that the transition is only involving a subregion of the entire spacetime, and is, therefore, compatible with the traversable wormhole picture first outlined by Gao, Jafferis and Wall (GAO) by means of double-trace deformations, \cite{BBAW2}.  In summary, in our setup, there is no obstruction to up-tunnelling due to the following features: 

\begin{enumerate} 

\item Transitions are local

\item The instanton is singular and the metric is degenerate

\item The transition amplitude is related to the Entanglement entropy, $S_{_{EE}}$, rather than the Von Neumann entropy. This justifies the necessity for topology change to take place whenever describing vacuum transitions. Notice that, in the Hamiltonian formalism, transitions are only described on spatial slices and not on the entire manifold. For wormhole saddles of the kind expected to emerge in the path integral of quantum gravity, the entire spacetime should really be understood of as being multiply-connected. 

\end{enumerate} 

As long as the manifold is simply connected, the action is related to a \emph{monotonic} $S_{_{EE}}$, and therefore behaves as the ordinary von Neumann entropy of a bipartite system. In any other case, though, such as for when describing up-tunnelling, the von Neumann entropy needs to be upgraded to $S_{_{EE}}$, meaning spacetime is multiply-connected. The singular metric involved in up-tunnelling is unsuitable for applying the GSL, given that the latter takes into account the causal dependence in between different spacetimes at ${\cal I}^{\pm}$. But this is different from the local transitions we are dealing with. The Hamiltonian constraint defines the energy of the system, i.e. not of the entire spacetime, but only that of the local subregion involved in the transition. 

\subsection*{Singular instantons}\label{sec:4.2}        
The present appendix is meant to be a short digression providing further insight into the singular instanton solution obtained by FGG. As also argued by the authors in \cite{Farhi:1989yr}, there are 2 possible ways in which the singularity might arise: 
\begin{enumerate}

\item Either in the past of the spacetime, or

\item In the wormhole configuration mediating the tunnelling process (double analytic continuation). 

\end{enumerate} 

The former can be avoided via the quantum tunnelling interpolating between 2 classical solutions, defined as \emph{type-a} and \emph{type-b} trajectories. The latter is unavoidable given the change in sign of the lapse function in the forbidden region, as already noticed in \cite{Fischler:1990pk}.

\begin{equation}    
\begin{aligned}    
&\text{type-a} 
\\ 
\\ 
\\ 
\\ 
\\ 
\\ 
\\ 
\\ 
&\text{type-b} 
\\ 
\\ 
\\ 
\\ 
\end{aligned}    
\ \   
\begin{aligned} 
\text{dS}   
\ \ \ \ \ \ \ \ \ \ \ \  \ \ \ \ \ \ \ \ \ \ \ \ \ \ \           \text{Schw    }    \ \ \ \ \ \ \ \ \    
\\
\includegraphics[scale=0.5]{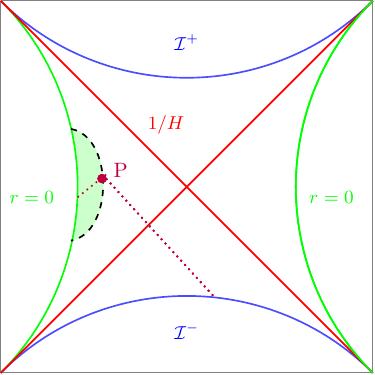}   
\ \ \ \ 
\includegraphics[scale=0.5]{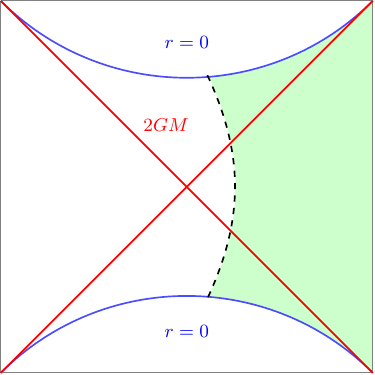}   
\\    
\\     
\includegraphics[scale=0.5]{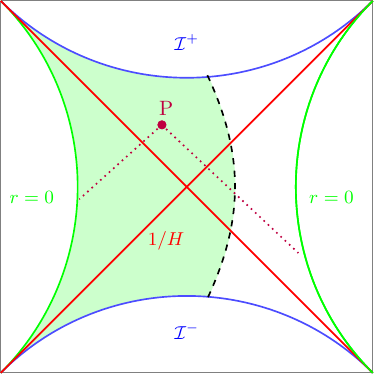}   
\ \ \ \ 
\includegraphics[scale=0.5]{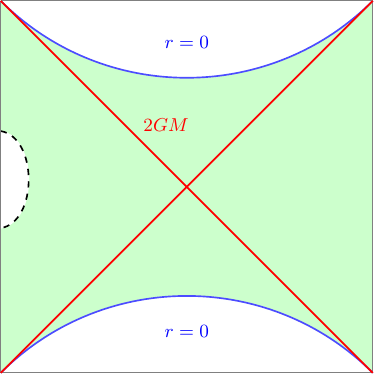}   
\end{aligned}       
\ \ 
\vcenter{\hbox{\includegraphics[scale=0.45]{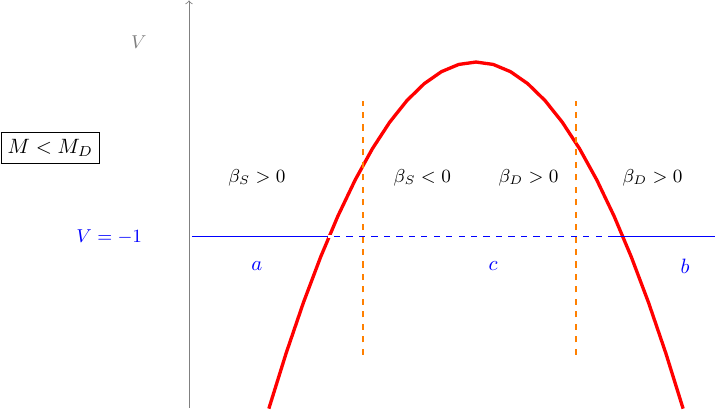} } }      
\nonumber    
\end{equation}

For the quantum process to take place, and effectively interpolate in between the 2 classical trajectories depicted in the figure above, the brane's mass, $M$, must be at least smaller w.r.t, $M_{S}$. i.e. the Schwarzschild mass. This ensures that the wall's trajectory intersects the horizon and the extrinsic curvature changes sign. In particular, \cite{Farhi:1989yr} choose $M<M_{D}$. The reason for this is that, being $M$ smaller than any other scale of the theory, it can be tuned to be arbitrarily small, ultimately enabling to circumvent the emergence from the initial singularity\footnote{\cite{Farhi:1989yr} also argue in favour of this by means of the so-called trapped and anti-trapped surfaces, with the latter being those to which the Penrose Theorem applies.} as we will now briefly outline. Given a 2-surface, $P$, its past light-cone will inevitably intersect the singularity with at least one light ray. However, in the weak gravity limit, it is possible to create a bubble of suitable radius to be at the 1$^{st}$ turning point of the potential barrier, treating the nucleation process as a pure quantum fluctuation of a flat background.
  
\subsection{Roots of a quartic}\label{sec:D}   

In this appendix we outline the procedure for determining the extremising value of $\bar\rho$ for an (A)dS$_2\rightarrow$(A)dS$_2$ in the thin-wall approximation, \eqref{eq:dstodstw}, outlined in section \ref{sec:thinwall}. Upon extremising $B_{_{TOT}}$ w.r.t. $\bar\rho$, we end up with a polynomial with all even powers of $\bar\rho$. In particular, redefining $x\overset{def.}{=}\bar\rho^2$, it can be brought down to the following form

\begin{equation} 
x^4+a_3x^3+a_2x^2+a_1x+a_o =0 
\label{eq:quart1}    
\end{equation}    

By further redefining $y\overset{def.}{=}x+\frac{a_3}{4}$, \eqref{eq:quart1} becomes   

\begin{equation} 
y^4+py^2+qy+r =0 
\label{eq:quart2}  
\end{equation} 

Adding and subtracting $y^2u+\frac{u^2}{4}$, \eqref{eq:quart2} becomes   

\begin{equation} 
\left(y^2+\frac{1}{2}u\right)^2-\left[(u-p)x^2-qx+\left(\frac{u^2}{4}-r\right)\right]  =0  
\end{equation}

After a suitable manipulation, it is possible to factorise it into the product of 2 quadratics subject to satisfying a cubic constraint equation, namely

\begin{equation} 
q^2=4(u-p)\left(\frac{u^2}{4}-r\right) 
\ \ \ \text{with}\ \ \ 
\begin{cases} 
p\overset{def.}{=}a_2-\frac{3}{8}a_3^2\\  
q\overset{def.}{=}a_1-\frac{1}{2}a_2a_3+\frac{1}{8}a_3^3  \\  
r\overset{def.}{=}a_o-\frac{1}{4}a_1a_3+\frac{1}{16}a_2a_3^2-\frac{3}{256}a_3^4    
\end{cases}
\end{equation}
The corresponding solution $u_o=t_o-\frac{p}{3}$, with 

\begin{equation} 
t_o\overset{def.}{=}\frac{2}{3}\sqrt{p^2+12r\ }\ \cos\left[\frac{1}{3}\cos^{-1}\left(\frac{2p^3+9(3q^2-8rp)}{18}\left(\frac{3}{p^2+12r}\right)^{3/2}\right) \right]
\end{equation}   

Once having obtained such factorisation, the corresponding value of the turning point can be extracted. In the 4D case, extremisation leads to a relation which is equivalent to the junction conditions in BT, and given that the turning point can be analytically computed in such case, it also corresponds to the value of the bubble radius extremising $B_{tot}$. In this case, though, it is not so straightforward, therefore we need to calculate it explicitly from the extremising procedure. However, we anticipate that \eqref{eq:dstodstw} prior to extremisation already shares the same structure as the total bounce for type-1 instantons as found by BT in 2D. The turning point reads

\begin{equation}   
\bar\rho^2 
=
-\frac{2A_3+A_5}{2}\left[1\pm\sqrt{1-\frac{4\left(A_3^2+A_4+A_5A_6\right)}{(2A_3+A_5)^2}\ }\right]    
\end{equation} 

with

\begin{equation}   
A_3\overset{def.}{=}\frac{a_3}{4}
\ \ , \ \  
A_4\overset{def.}{=}\frac{1}{2}\left(t_o-\frac{p}{3}\ \right)   
\ \ , \ \  
A_5\overset{def.}{=}\sqrt{t_o-\frac{p}{3}\ }    
\ \ , \ \  
A_6\overset{def.}{=}-\frac{1}{2}\sqrt{\frac{\left(t_o-\frac{p}{3}\right)^2-4r}{t_o-\frac{4p}{3}}\ }       
\end{equation}


\section{Geometric Invariant Theory (GIT)}

Delving in the explicit description of the duality structure arising from class ${\cal S}$ theories, to which Part \ref{sec:V} is devoted, requires some key preliminary algebro-geometric tools, specifically the definition of Coulomb and Higgs branches from geometric invariant theory, \cite{Nakajima:2022sbi,Gonzalez:2023jur,Braverman:2016pwk,Braverman:2016wma,Teleman:2022oiu,DP2021,Berest,Teleman:2018wac,GIT,Braverman:2017ofm}.  

This appendix recapitulates the essential tools required for performing GIT quotient constructions, highlighting its importance for realising 3D mirror symmetry. In particular, we focus on the calculation of the Hilbert series as specific examples of algebraic varieties of crucial interest in the study of categorical dualities for supersymmetric quiver gauge theories.

\medskip 

\medskip 

\begin{figure}[ht!]  
\begin{center}   
\includegraphics[scale=1]{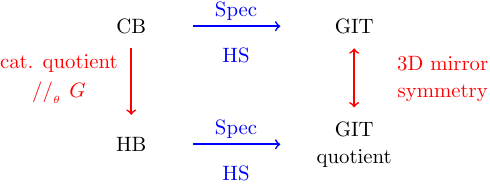}    
\caption{\small Key algebraic varieties and dualities between invariants.}   
\label{fig:SMSM1}  
\end{center}  
\end{figure}

\subsection{Equivariant cohomology}

Equivariant cohomology, \cite{BFM}, also known as Borel cohomology, is a cohomology theory from algebraic topology which applies to topological spaces with a group action. It can be viewed as a generalisation of group cohomology. More explicitly, the equivariant cohomology ring of a space $X$ with action of a topological group $G$ is defined as the ordinary cohomology ring with coefficient ring $\Lambda$ of the homotopy quotient EG$\times_{_G}$X  

\begin{equation}  
H^{^{\bullet}}_{_G}(X;\Lambda)\ =\ H^{^{\bullet}}\left(EG\times_{_G} X;\Lambda\right).    
\end{equation}  

\subsection*{The Kirwan map}  

The Kirwan map, first introduced in \cite{FK}, states 

\begin{equation}  
H^{^{\bullet}}_{_G}(M)\ \rightarrow\ H^{^{\bullet}}\left(M//_{_p} G\right), 
\end{equation}  
where $M$ is a Hamiltonian $G$-space, i.e. a symplectic manifold acted on by a Lie group $G$ with a moment map 

\begin{equation}     
\mu:\ M\ \rightarrow\ \mathfrak{g}^{*}.  
\end{equation}   

$H^{^{\bullet}}(M)$ is an equivariant cohomology ring of $M$, i.e. the cohomology ring of the homotopy quotient $EG\times_{_G}M$ of $M$ by $G$.

\begin{equation}  
M//_{_p}G\ =\ \mu^{^{-1}}(p)/G  
\end{equation}   
is the symplectic quotient of $M$ by $G$ at a regular central value $p\in Z(\mathfrak{g}^{*})$ of $\mu$. It is defined as the map of equivariant cohomology induced by the inclusion 

\begin{equation}    
\mu^{^{-1}}(p)\ \hookrightarrow\ M   
\end{equation}
followed by the canonical isomorphism  

\begin{equation}  
H^{^{\bullet}}_{_G}\left(\mu^{^{-1}}(p)\right)\ =\ H^{^{\bullet}}\left(M//_{_p} G\right).
\end{equation} 

\subsection{The GIT construction}    

Geometric invariant theory (GIT), \cite{Deligne-Mumford} is a recipe for constructing quotients in algebraic geometry. An affine GIT quotient, or affine geometric invariant theory quotient of an affine scheme $X=\text{Spec} {\cal A}$ with an action by a group scheme $G$, is the affine scheme Spec$({\cal A}^{^G})$, namely the prime spectrum of the ring of invariants of ${\cal A}$, and is denoted by $X//G$.

The input of the GIT construction consists of:

\begin{enumerate} 

\item A complex reductive group, $G$.

\item A representation of $G$  

\begin{equation} 
X \ =\ \mathbf{A}^{^n}.  
\end{equation}

\item A character

\begin{equation}  
\theta:\ G\ \rightarrow\ G.   
\end{equation}

\end{enumerate}  

GIT takes this data and gives an open subset of $X$, namely the semistable locus  

\begin{equation}    
X_{_{\theta}}^{^{ss}}(G)\ \overset{\text{open}}{\subset}\ X, 
\end{equation}
from which one can define the stack quotient as follows 

\begin{equation}    
X//_{_{\theta}}(G)\ \overset{def.}{\equiv}\ \left[X_{_{\theta}}^{^{ss}}(G)\ /\ G\right]. 
\label{eq:ssq}
\end{equation}  

As \eqref{eq:ssq} suggests, a GIT quotient is therefore a categorical quotient of the locus of semistable points, i.e. the quotient of the semistable locus.

Examples of GIT are Grassmannians, whose definition goes as follows. Take $G$ a complex (connected) reductive group. Let ${\cal O}$ denote the formal power series ring $\mathbf{C}[[z]]$ and ${\cal K}$ its fraction field $\mathbf{C}((z))$. The Grassmannian is defined by the group quotient 

\begin{equation}   
\text{Gr}_{_G}\ \overset{def.}{=}\ G_{_{\cal K}}/G_{_{\cal O}},   
\end{equation}
where $G_{_{\cal K}}$ and $G_{_{\cal O}}$ denote the groups of ${\cal K}$ and ${\cal O}$-valued points of $G$, respectively. More formally, Gr$_{_G}$ is defined as being the moduli space of pairs $({\cal P}, \varphi)$ of a $G$-bundle, ${\cal P}$, on the formal disk 

\begin{equation} 
D\ \overset{def.}{=}\ \text{Spec}({\cal O}), 
\end{equation}
and its trivialisation $\varphi$ over the punctured disk   

\begin{equation} 
D^{^*}\ \overset{def.}{=}\ \text{Spec}({\cal K}). 
\end{equation}    

Similarly, for the case of a complex representation, \textbf{N}, the moduli space is now parameterised by a triple

\begin{equation} 
{\cal R}\ \overset{def.}{=}\ ({\cal P}, \varphi, s),   
\end{equation}  
where $({\cal P}, \varphi)$ is in Gr$_{_G}$, and $s$ is a section of an associated vector bundle ${\cal P}_{_N}={\cal P}\times_{_G}\textbf{N}$, such that it is sent to a regular section of the trivial bundle under $\varphi$.

For a given triple ${\cal R}$, there is a corresponding projection

\begin{equation} 
\pi:\ {\cal R}\ \rightarrow \text{Gr}_{_G},     
\end{equation} 
where Gr$_{_G}$ is the affine Grassmannian, and $G$ is a complex reductive group. There is a natural commutative ring object, ${\cal A}$, in the derived $G_{_{\cal O}}$-equivariant constructible category on Gr$_{_G}$, $D_{_G}\left(\text{Gr}_{_G}\right)$, such that $H_{_{\bullet}}^{^{G_{_{\cal O}}}}\left({\cal R}\right)$ can be constructed from $H_{_{\bullet}}^{^{G_{_{\cal O}}}}\left(\text{Gr}_{_G},{\cal A}\right)$ equipped with a commutative ring structure.

In presence of a collection of such commutative ring objects, $\{{\cal A}_{_i}\}$, in $D_{_G}\left(\text{Gr}_{_G}\right)$, one can perform a \emph{gluing construction} by defining a new commutative ring object as 

\begin{equation}   
\iota _{_{\Delta}}\left(\ \boxtimes\ {\cal A}_{_i}\right),  
\end{equation}
where 

\begin{equation}  
\iota _{_{\Delta}}:\ \text{Gr}_{_G}\ \rightarrow\ \prod_{_i} \text{Gr}_{_G}  
\end{equation}
is the diagonal embedding. One such example are star-shaped quivers, \cite{Benini:2010uu}.

Its equivariant Borel-Moore homology, \cite{BWB}, can be defined by the $G_{_{\cal O}}$-equivariant Borel-Moore homology of   

\begin{equation}  
{\cal R}_{_{\le\lambda}}\ \equiv\ {\cal R}\ \cap\ \pi^{^{-1}}\left(\bar{\text{Gr}}_{_G}^{^{\lambda}}\right)     
\end{equation}    
in such a way that an embedding

\begin{equation}  
{\cal R}_{_{\mu}}\ \hookrightarrow\ {\cal R}_{_{\lambda}}   
\end{equation}  
will induce a map   

\begin{equation}  
H_{_{\bullet}}^{^{G_{_{\cal O}}}}\left({\cal R}_{_{\le\mu}}\right) \ \rightarrow\ H_{_{\bullet}}^{^{G_{_{\cal O}}}}\left({\cal R}_{_{\le\lambda}}\right) 
\end{equation}
for $\mu\le\lambda$.    

More generally, the euivariant Borel-Moore homology can be expressed as a graded sum 

\begin{equation}  
H_{_{\bullet}}^{^{G_{_{\cal O}}}}\left({\cal R}\right) \ \simeq\ \underset{\gamma}{\bigoplus}\  H_{_{\bullet}}^{^{G_{_{\cal O}}}}\left({\cal R}^{^{\gamma}}\right), 
\end{equation}
with ${\cal R}^{^{\gamma}}$ the connected component corresponding to $\gamma\ \in\ \pi_{_1}(G)$.

\subsection{Moduli spaces: Coulomb and Higgs branches }  
The Coulomb branch is an affine algebraic variety whose coordinate ring is the equivariant Borel-Moore homology group, $H_{_{\bullet}}^{^{G_{_o}}}(\cal R)$ of a certain space ${\cal R}$, defining the variety of triples.

The Coulomb branch is defined as, \cite{Braverman:2016wma}, 

\begin{equation}    
{\cal M}_{_C}\ \overset{def.}{=}\ \text{Spec}\ {\cal A} (G, \mathbf{N})\ \equiv\ \text{Spec}\left(H_{_*}^{^{G_{_o}}}({\cal R}), *\right).     
\end{equation} 

Similarly, one can define the quantised Coulomb branch as follows  

\begin{equation}    
{\cal M}_{_C}\ \overset{def.}{=}\ \text{Spec}\ {\cal A}_{_{\hbar}} (G, \mathbf{N})\ \equiv\ \text{Spec}\left(H_{_*}^{^{G_{_o}\ \rtimes\mathbf{C}^{^{\times}}}}({\cal R}), *\right),     
\end{equation} 
which, in terms of quantum homology, can also be re-expressed as follows,

\begin{equation} 
{\cal M}_{_C}\ \overset{def.}{=}\ \text{Spec}\left(QH_{_G}^{^{\bullet}}\left(G;HF^{^{\bullet}}(X)\right)\right).     
\end{equation}

On the other hand, the Higgs branch is defined as  the spectrum of the categorical quotient, namely, \cite{Braverman:2016wma},

\begin{equation} 
{\cal M}_{_H}\ \overset{def.}{=}\ \text{Spec}\left(QH^{^\bullet}\left(X//G\right)\right)\ \simeq\ QH^{^{\bullet}}_{_{LG}}(X).  
\end{equation}

\subsection{The monopole formula}

The monopole formula contains all the gauge-invariant chiral operators that acquire a non-zero expectation value\footnote{Their specific VEV depends on their respective dimensions and quantum numbers.} along the CB. It is particularly useful for investigating expected properties of Coulomb branches, and reads as follows   

\begin{equation}  
P_{_t}^{^{G_{_{\cal O}}}}({\cal R})\ =\ \sum_{_k}\ t^{^{-k}}\ \text{dim} \ H_{_k}^{^{G_{_{\cal O}}}}({\cal R}).    
\label{eq:MF}   
\end{equation}

As already remarked in \cite{Braverman:2016wma}, \eqref{eq:MF} is the same as the Hilbert series on the Coulomb branch of a 3D ${\cal N}=4$ supersymmetric quiver gauge theory associated with the pair $(G_{_{\mathbf{C}}}, \mathbf{N}\ \oplus\ \mathbf{N}^{^*})$.

\section{Magnetic quivers} \label{sec:magneticquivers}

\subsection{Magnetic Quivers: a unifying framework}   \label{sec:mq}  

This section outlines some further preliminary background tools needed in Part \ref{sec:V} for addressing the formulation of mirror symmetry from the perspective of geometric representation theory, \cite{Pasquarella:2023exd}. Mostly inspired by \cite{Teleman:2014jaa} and further upcoming work by Teleman, \cite{CT}, we argue how constructions as the ones addressed in \cite{Pasquarella:2023deo}, involving multiple gauging, can still be assigned a fiber functor, as in the Freed-Moore-Teleman symmetry TFT (SymTFT) setup, \cite{Freed:2022qnc}; however, such fiber functor should not be associated to either of the absolute theories separated by the non-invertible defect, but, rather to a 3D theory whose partition function correctly accounts for the symplectic projection leading to either of the two absolute theories.

Our approach consists in a combination of the following: 

\begin{enumerate} 

\item Defining a fiber functor for different absolute theories connected by non-invertible defects, \cite{Pasquarella:2023deo}. 

\item Relating mirror symmetry with the identification of a Drinfeld center, \cite{CT}. 

\item Recent advancements in the understanding of the Higgs branch (HB) structure of higher-dimensional quiver gauge theories with 8 supercharges by means of Coulomb branches (CBs) of magnetic quivers (MQs) associated to 3D ${\cal N}=4$ gauge theories, \cite{Cabrera:2019izd,Bourget:2019aer,Bourget:2021siw,Bourget:2023cgs,Bourget:2019rtl,Ferlito:2016grh}. 

\end{enumerate}

In the current and next sections we mostly outline how the topics listed above combine together, and in the following sections we will provide a more detailed mathematical correspondence with the work of \cite{Teleman:2014jaa}, highlighting how his constructions can effectively be generalised to higher dimensions and be naturally embedded in the setup of \cite{Freed:2022qnc} precisely thanks to the understanding in terms of magnetic quivers of 3D ${\cal N}=4$ theories addressed here.  

One could see our proposal as further supporting the idea that higher-categorical symmetries probe representation theory structures.

In the present section, we review the correspondence between geometric and algebraic resolutions of framed Nakajima quiver varieties, \cite{DAlesio:2021hlp,Braverman:2016pwk,Braverman:2016wma}, highlighting it as an interesting example of homological mirror symmetry. In particular, we emphasise the property the moment map and higher homologies need to satisfy to ensure agreement in between the calculation of the two Hilbert series. We conclude the section with a brief overview of Hasse diagram constructions via magnetic quivers for quiver gauge theories with 8 supercharges, highlighting an interesting 2-categorical structure arising when dealing with complete intersections.

Section \ref{sec:mq} is meant to provide a brief overview of some key elements we will be using throughout our treatment, emphasising the ones that are needed for building the connection with the work of \cite{Teleman:2014jaa, CT}\footnote{Throughout the entire treatment, we will be assuming the basic knowledge of higher categories and ADE quivers. We refer the reader to the extended literature on both topics for detailed definitions and examples. Specific additional tools will be explained in due course when needed.}, which will be the core focus of section \ref{sec:333}. The first part of the present article is structured as follows: 

\begin{itemize}  

\item  At first we review the correspondence between geometric and algebraic resolutions of framed Nakajima quiver varieties, \cite{DAlesio:2021hlp,Braverman:2016pwk,Braverman:2016wma}, highlighting it as an interesting example of homological mirror symmetry\footnote{The main explanation for this statement will become manifest in Part III of our treatment.}. In particular, we emphasise the property the moment map and higher homologies need to satisfy to ensure agreement in between the two resolutions. Consequently, this is mapped to an equivalence in between the Hilbert series resulting from summing over characters.

\item We then turn to a brief overview of Higgs and Coulomb branches\footnote{Simply denoted by HB and CB for convenience.} of quiver gauge theories as algebraic varieties, pointing out the role of magnetic quivers (MQs) for describing the HB of a quiver gauge theory with 8 supercharges and arbitrary flavour, \cite{Cabrera:2019izd,Bourget:2019aer,Bourget:2021siw,Bourget:2023cgs,Bourget:2019rtl,Ferlito:2016grh}. In doing so, we highlight the importance of the role played by, both, the Hilbert series (HS) and highest weight generating function (HWG) in identifying different intersecting cones in the HB Hasse diagram, \cite{Cabrera:2019izd,Bourget:2019aer,Bourget:2021siw,Bourget:2023cgs,Bourget:2019rtl,Ferlito:2016grh}, suggesting a 2-categorical structure.  

\end{itemize}

In section \ref{sec:333} we explain how gauging can be related to the poset ordering leading to the construction of Hasse diagrams, thanks to the unifying role of the moment map. We then explain how the identification of such moment map ensures the quiver gauge theory enjoys a generalised notion of homological mirror symmetry, with the latter corresponding to the presence of a Drinfeld center and a corresponding fiber functor for a 2-categorical structure, related to Rozansky-Witten theory, \cite{Rozansky:1996bq}. We conclude the section highlighting connections between the topics outlined in the present work and those of \cite{Teleman:2014jaa, CT}, thereby opening the scene to the more mathematical treatment to which Part III is devoted.

As we shall see, the underlying motivation for building the connections explained in section \ref{sec:333} is the importance of identifying a unifying framework for realising mirror symmetry and its generalisations.

\subsection{Intersecting cones} \label{sec:2.2}  

From the more mathematically-inclined overview of section \ref{sec:2.1}, we know how to obtain a configuration admitting homological mirror symmetry, and the conditions that must be satisfied for the Hilbert series to match on the GIT and derived representation scheme side. For the purpose of what we will be addressing in section \ref{sec:3}, we will now overview a setup where homological mirror symmetry can be described in an interestingly generalised and controlled manner, suggesting this as a probe for investigating its underlying mathematical structure. 

\begin{figure}[ht!]     
\begin{center}
\includegraphics[scale=1]{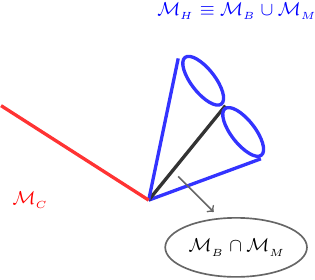} 
\caption{\small Moduli space with Coulomb branch (CB), ${\cal M}_{_C}$, Higgs branch (HB), ${\cal M}_{_H}\equiv{\cal M}_{_B}\cup{\cal M}_{_M}$, with its nontrivially intersecting mesonic and baryonic branch, denoted by ${\cal M}_{_B}$ and ${\cal M}_{_M}$, respectively.}  
\label{}  
\end{center} 
\end{figure} 


\subsection*{Higgs and Coulomb branches as algebraic varieties}

The main focus of this section are quiver gauge theories with 8 supercharges, classical flavour and gauge groups, whose vacuum moduli space is related to the nilpotent orbits, ${\cal O}_{_{\rho}}$, of classical Lie algebras, $\mathfrak{g}$. For a given gauge group $G$, the nilpotent orbits of an algebra $\mathfrak{g}$ are defined by equivalence classes of nilpotence conditions on representation matrices, \cite{Hanany:2019tji}. From the theoretical physics point of view, the nilpotent conditions describe the way in which the scalar fields in the $F$-term equations vanish at the  SUSY vacuum, and can be specified by a quiver gauge theory. 

According to the Jacobson Morozov Theorem, these nilpotent orbits are 
in 1 to 1 correspondence with equivalence classes of embeddings of $\mathfrak{su}(2)$ into $\mathfrak{g}$. Each such embedding 

\begin{equation}   
\rho:\mathfrak{su}(2)\ \rightarrow\ \mathfrak{g}  
\end{equation}   
is a homomorphism, labelled, by, either, the partition of the representation of $G$, or by a characteristic, using Dynking labels to specify the mapping of the roots and weights of $\mathfrak{g}$ onto $\mathfrak{su}(2)$.

A \emph{Slodowy slice}, $S_{_{\rho}}\overset{def.}{\equiv}{\cal O}_{_{\rho}}^{^{\perp}}$, is a space transverse to a nilpotent orbit, and therefore commuting with it, while living within the adjoint orbit of the ambient group $G$. These transverse spaces can be further restricted to their intersections with the closure of any enclosing nilpotent orbit $\overline{\cal O}_{_{\sigma}}$, thereby leading to spaces labelled by pairs of nilpotent orbits, whose elements are \emph{Slodowy interesections} 

\begin{equation}    
S_{_{{\sigma,\rho}}}\ \overset{def.}{\equiv}\ \overline{\cal O}_{_{\sigma}}\ \cap\ S_{_{\rho}}  
\label{eq:SlInt}    
\end{equation}    
encoding, as particular examples, nilpotent orbits, Slodowy slices and Kraft-Procesi transitions\footnote{The connection between the 3D boundary conditions on type-II brane systems in 4D ${\cal N}=4$  CFTs and Slodowy intersections was highlighted in \cite{Gaiotto:2008ak}.}. 

\eqref{eq:SlInt} are the algebraic varieties of which a Hilbert series can be calculated\footnote{Hilbert series for Slodowy intersections can also be constructed by means of purely group theoretic methods, making use of localisation formulae related to the Hall Littlewood polynomials.}, as we will explain later on. Prior to turning to that we outline some preliminary definitions that will be recursively used throughout the remainder of this work.

The \emph{nilcone} is defined as

\begin{equation}    
\overline{\cal O}_{_{max}}\ \overset{def.}{\equiv}\  {\cal N}.    
\end{equation} 
with dimension

\begin{equation}    
\left|{\cal N}\right|\ \equiv\ \left|\mathfrak{g}\right|-\text{rank}\left[\mathfrak{g}\right].       
\end{equation}

Nilpotent orbits can be arranged as a \emph{Hasse diagram} according to the inclusion relations of their closures:  

\begin{equation}    
{\cal N}\ \equiv\ \overline{\cal O}_{_{max}}\ \supset\  \overline{\cal O}_{_{sub-reg}}\ \supset\ ...\ \overline{\cal O}_{_{min}}\ \supset\ \overline{\cal O}_{_{trivial}}\ \equiv\ \{0\}.    
\end{equation}  

From the definition of the Slodowy intersection, it follows that

\begin{equation}    
S_{_{{{\cal N},trivial}}}\ \equiv\ \overline{\cal O}_{_{max}}   \equiv\ {\cal N}
\end{equation}

By means of such terminology, the definition of the \emph{Higgs} and \emph{Coulomb branches} can be expressed as follows
\begin{equation}    
\overline{\cal O}_{_{max}}\   \overset{def}{\equiv}\ \text{HB}\left[{\cal M}_{_A}(\rho,0)\right] 
\ \ \ , \ \ \      
S_{_{{{\cal N},\rho^{^T}}}}\   \overset{def}{\equiv}\ \text{CB}\left[{\cal M}_{_A}(\rho,0)\right] 
\ \ \ , \ \ \     
{\cal M}_{_A}(\rho,0)\   \overset{def}{\equiv}\ {\cal L}_{_{A}}\left(\rho^{^T}\right)
\end{equation}  
with ${\cal M}_{_A}(\rho,0)$ denoting a single-flavoured linear quiver.

\subsubsection{Magnetic quivers and Hasse diagrams}  

Having said this, we now briefly overview the prescription for constructing magnetic quivers and Hasse diagrams from quiver subtraction, \cite{Cabrera:2018ann}\footnote{Such terminology is explained in due course. We refer to the extensive literature, especially the one in the references, for more detailed explanations as well as several examples explicitly analysed.}. The procedure can be summarised by the following steps:  

\begin{enumerate}  

\item Start from a certain quiver gauge theory with 8 supercharges. Focussing on SQCD theories, namely a 3-parameter family of theories labelled by $(N_{_c}, N_{_f}, k)$, denoting number of colours, flavours, and Chern-Simons level, respectively. As a framed Nakajima quiver variety, we assign to it an electric quiver of the kind \ref{fig:FNQV}. Its HB is an algebraic variety associated to a 5-brane web configuration. 

\item Identify the number of maximal decompositions of the 5-brane web. 

\item To each maximal decomposition associate the magnetic quiver (MQ) of a 3D ${\cal N}=4$ gauge theory.  

\item The HB of the SQCD theory is preserved under dimensional reduction, and is equivalent to the union of the CBs of the MQs identified in step 3.    

\item Construct the Hasse diagram associated to the HB by implementing quiver subtraction\footnote{We are assuming familiarity with the notion of ADE quivers. We refer to the extended literature on the topic if needed.} on the MQs.  

\end{enumerate}   

Most importantly, this procedure generalises the following relation:  

\begin{equation}  
\boxed{\ \ \ \ \text{HB}\left(T\right)\ \equiv\ \text{CB}\left(T^{^{\text{V}}}\right)\color{white}\bigg]\ \ }, 
\label{eq:mirrordual}  
\end{equation} 
with $T$ a 3D ${\cal N}=4$ quiver gauge theory, and $T^{^{\text{V}}}$ its mirror dual. \eqref{eq:mirrordual} can also be re-expressed in terms of the electric and magnetic quivers 

\begin{equation}  
\boxed{\ \ \ \ \text{HB}\left(\text{EQ}\right)\ \equiv\ \text{CB}\left(\text{MQ}\right)\color{white}\bigg]\ \ }. 
\label{eq:mirrordualquivers}  
\end{equation} 

The advantage of the procedure put forward by \cite{Cabrera:2019izd,Bourget:2019aer,Bourget:2021siw,Bourget:2023cgs,Bourget:2019rtl,Ferlito:2016grh} is that, when $T$ does not admit a unique $T^{^{\text{V}}}$, its \emph{generalised} mirror symmetric theory can still be recovered by specifying the MQs, or, equivalently, the generators of the cones featuring in the Hasse diagram, together with their intersections. For concreteness, we briefly overview some key examples of the procedure summarised above, encompassing the main features we will be needing to consider in section \ref{sec:3}.

\subsubsection{Rank-1 and rank-2 examples}

The examples reproduced in this subsection are for 5D ${\cal N}=1$ SQCD, with varying parameters $(N_{_c},N_{_f},k)$. Notice that, for $SU(2)$, $k=0$, reason why it is conventionally omitted in the quiver representation. However, for $N_{_c}>2$ it might be $k\neq0$, reason why in such case we specify this in the corresponding quiver.

For each case we draw brane intersections (NS5s-branes and (p,q)-branes are always present). When adding flavour, D5-branes also contribute to the intersection. On the right of each brane intersection, we show the Hasse diagram constructed by implementing quiver subtraction on the magnetic quiver. Conventionally, the node at the bottom of the Hasse diagram denotes the original magnetic quiver, prior to implementing any quiver subtraction. At each step of the iteration, we draw a vertical line, ending at another node, which therefore denotes a reduced magnetic quiver. When a bifurcation takes place, it signals that there are multiple possible quiver subtractions that could be performed. This corresponds to the emergence of multiple cones in the Hasse diagram. In case no bifurcations are present, it means the cone is unique.

\subsection*{5D\ \ SU(N\texorpdfstring{$_{_c})$, $N_{_f}=0$}{}}

\medskip 

\medskip

\begin{equation} 
\begin{aligned}
&\vcenter{\hbox{\includegraphics[scale=0.8]{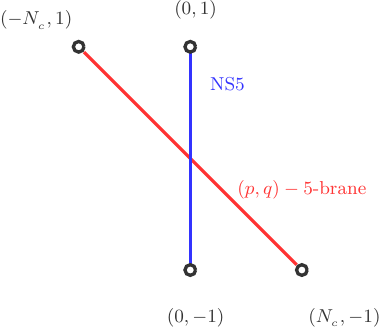} }}\\ 
\end{aligned}
\qquad\qquad\qquad\qquad\qquad  
    \begin{aligned}
        &\vcenter{\hbox{\includegraphics[scale=1]{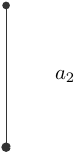}}}\\
    \end{aligned}  
\label{}  
\nonumber
\end{equation}

\begin{figure}[ht!]    
\begin{minipage}[c]{1\textwidth}
\caption{\footnotesize In absence of flavour degrees of freedom, the 5-braneweb decomposition is unique (LHS) leading to a HB Hasse diagram characterised by a unique cone (RHS).}     \label{fig:2p}   
\end{minipage}    
\end{figure}

\medskip 

\begin{equation}  
\text{HB}_{_{\infty}}^ {\ \text{5D}}\ \left(\ \raisebox{-30pt}{\includegraphics[ scale=0.75]{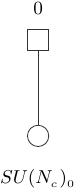}}\ \right) \ \equiv\ \mathbb{C^{^3}}/\mathbb{Z}_{_{N_{_c}}}\ \equiv \ \text{CB}^{\ \text{3D}}\ \left(\ \raisebox{-15pt}{\includegraphics[scale=0.75]{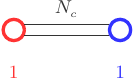}}\ \right) 
\end{equation}

\medskip 

\medskip 

\subsection*{5D\ \  SU(2),\  N\texorpdfstring{$_{_f}=2$}{}}

\medskip 

\medskip

\begin{figure}[ht!]    
\begin{minipage}[c]{1\textwidth}
\caption{\small 5-braneweb with added flavour (LHS) and HB Hasse diagram with two cones.}  
\label{fig:22}  
\end{minipage}   
\end{figure}

\begin{figure}[ht!]     
\begin{center}
\includegraphics[scale=0.6]{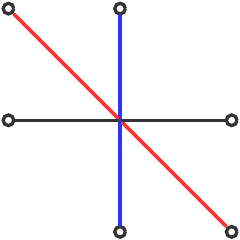}  
\ \ \ \ \ \ \ \ \ \ \ \ \ \ \ \ \ \ \ \ 
\includegraphics[scale=0.6]{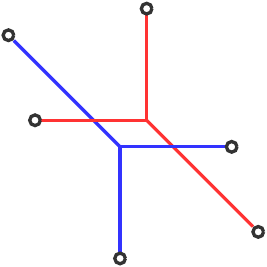}  
\caption{\small Adding flavour, there are two possible decompositions of the braneweb. Each decomposition is mapped to a different magnetic quiver. The union of their CBs is equivalent to the HB of the original 5D quiver gauge theory we started from. The intersection of the two cones is given by a single symplectic leaf.}  
\label{fig:2cones}  
\end{center} 
\end{figure}   

\begin{equation}  
\text{HB}_{_{\infty}}^ {\ \text{5D}}\ \left(\ \raisebox{-30pt}{\includegraphics[scale=0.75]{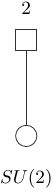}}\ \right) \ \equiv \ \text{CB}^{\ \text{3D}}\ \left(\ \raisebox{-23pt}{\includegraphics[scale=0.6]{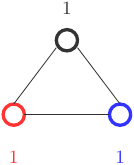}}\ \right) \ \bigcup\ \text{CB}^{\ \text{3D}}\ \left(\ \raisebox{-10pt}{\includegraphics[scale=0.6]{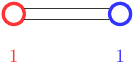}}\ \right) 
\end{equation}

\medskip 

\medskip

\subsection*{5D \ \ SU(2), \ N\texorpdfstring{$_{_f}=4$}{}}

\medskip 

\medskip 

\begin{equation}  
\begin{aligned}
&\vcenter{\hbox{\includegraphics[scale=0.5]{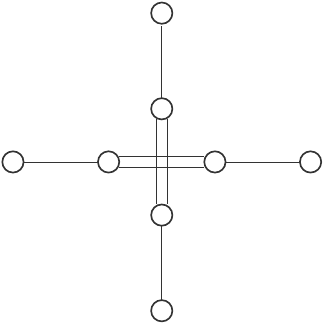}}}
\end{aligned}
\qquad\qquad\qquad\qquad   
\begin{aligned}
&\vcenter{\hbox{
\includegraphics[scale=1]{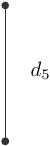}  }} 
\end{aligned}
\nonumber    
\end{equation}

\begin{figure}[ht!]
\begin{center} 
\caption{\small 5-braneweb and corresponding Hasse diagram for 5D $SU(2)$, $N_{_f}=4$.}  
\label{fig:11}  
\end{center} 
\end{figure}

\begin{equation}  
\text{HB}_{_{\infty}}^ {\ \text{5D}}\ \left(\ \raisebox{-30pt}{\includegraphics[scale=0.75]{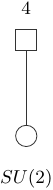}}\ \right) \ \equiv\ \text{CB}^{\ \text{3D}}\ \left(\ \raisebox{-25pt}{\includegraphics[scale=0.6]{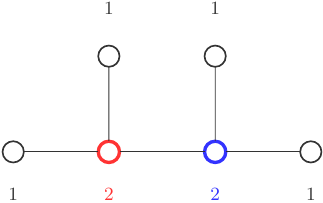}}\ \right) 
\end{equation}

\medskip 

\medskip 

\subsection*{5D \ \ SU(3)\texorpdfstring{$_0$, \ N$_{_f}=6$}{}}

\begin{equation}   
\begin{aligned}
&\vcenter{\hbox{\includegraphics[scale=0.6]{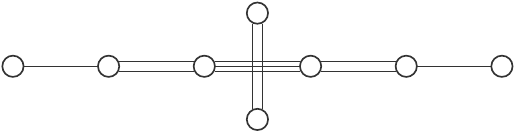} }}\\  
\end{aligned}
\qquad\qquad    
\begin{aligned}
&\vcenter{\hbox{\includegraphics[scale=0.8]{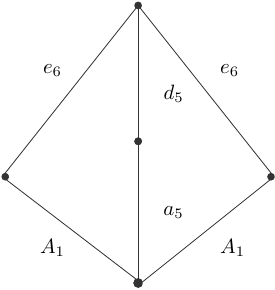}  }}\\  
\end{aligned}
\nonumber
\end{equation}

\begin{figure}[ht!]    
\begin{minipage}[c]{1\textwidth}
\caption{\small 5-braneweb for 5D $SU(3)_0$, $N_{_f}=6$. The cone is unique since $k=0$.}  
\label{fig:44}  
\end{minipage} 
\end{figure}

\begin{equation}  
\text{HB}_{_{\infty}}^ {\ \text{5D}}\ \left(\ \raisebox{-30pt}{\includegraphics[scale=0.75]{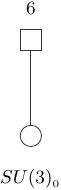}}\ \right) \ \equiv\ \text{CB}^{\ \text{3D}}\ \left(\ \raisebox{-25pt}{\includegraphics[scale=0.55]{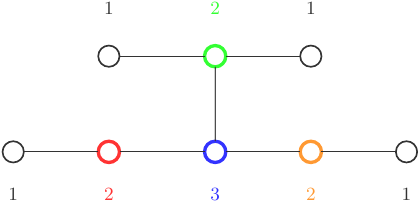}}\ \right) 
\end{equation}

\medskip 

\medskip 
\subsection*{5D \ \ SU(6)\texorpdfstring{$_2$}, \ N\texorpdfstring{$_{_f}=8$}{}}  
\medskip 

\medskip

\begin{figure}[ht!]     
\begin{center}
\includegraphics[scale=0.6]{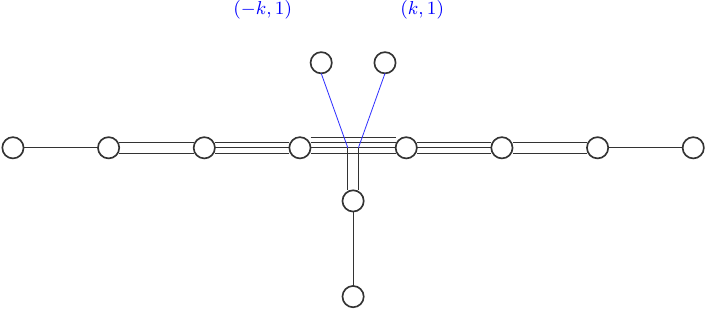}  \ \ \ \ \ \ \ \ \ \ \ \ 
\includegraphics[scale=0.6]{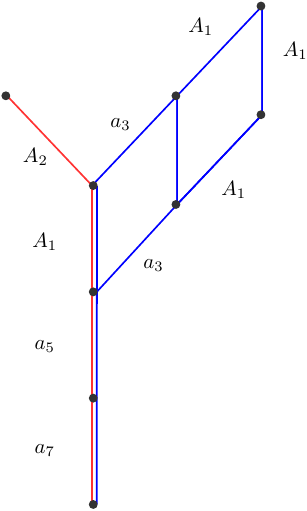}  
\caption{\small Adding a CS level $k\neq0$ leads the coexistence of two maximal decompositions of the 5-braneweb. The corresponding Hasse diagram is therefore constituted by 2 intersecting cones. The intersection in this case goes along the vertical direction where the blue and red lines are parallel to each other.}  
\label{fig:33}  
\end{center} 
\end{figure}

\begin{equation}  
\begin{aligned}
\text{HB}_{_{\infty}}^ {\ \text{5D}}\ \left(\ \raisebox{-30pt}{\includegraphics[scale=0.75]{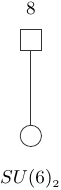}}\ \right) \ &\equiv \ \text{CB}^{\ \text{3D}}\ \left(\ \raisebox{-25pt}{\includegraphics[scale=0.35]{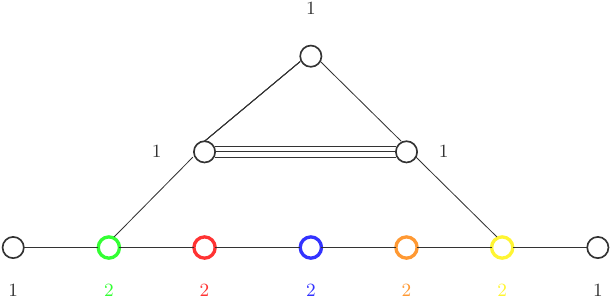}}\ \right) \ \bigcup\ 
\text{CB}^{\ \text{3D}}\ \left(\ \raisebox{-20pt}{\includegraphics[scale=0.35]{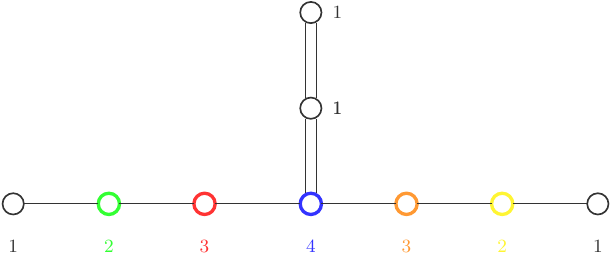}}\ \right) 
\end{aligned}
\end{equation}

From the point of view of the HB of quiver gauge theories with 8 supercharges in 5, 4, and 3 dimensions share a similar structure. In particular, the ramification of the HB due to the presence of multiple cones is preserved when considering the derived category of BPS objects in the lower-dimensional theory.

\subsection*{The 2-categorical structure of complete intersections}  
As previously mentioned, there are quiver gauge theories with 8 supercharges that do not admit a mirror symmetric 3D ${\cal N}=4$ theory description. The reason for this can be recast to the fact that its HB is not a single kyperk$\ddot{\text{a}}$ler cone, but, rather, the union of two, with nontrivial intersection, such as the cases shown in figure \ref{fig:22} and \ref{fig:33}. 

For classical gauge groups $G\equiv U(N), Sp(N)$ or $USp(2N), SO(2N)$, the moduli space is a complete intersection, meaning it is an algebraic variety whose dimension, $d$, is given by the following expression

\begin{equation}  
d\ \overset{def.}{=}\ g-r,     
\end{equation}  
where $g, r$ respectively denote the generators and relations between them. 

For the case in which a 3D ${\cal N}=4$ gauge theory, $T$, with given gauge group and suitable choice of matter content, admits a mirror dual theory, $T^{^{\text{V}}}$, the Hilbert series evaluated on the Coulomb branch (CB) of the former equals that of the Higgs branch (HB) of the latter,   

\begin{equation} 
\text{HS}\left[\text{CB}\  \left(T\right)\right]\ \equiv\ \text{HS}\left[\text{HB} \ \left(T^{^{\text{V}}}\right)\right],   
\label{eq:perfectmirror}   
\end{equation}     
hence   

\begin{equation}  
\text{HB}\left(\text{EQ}\right)\ \equiv\ \text{CB}\left(\text{MQ}\right). 
\label{eq:pm1}
\end{equation} 

Perfect realisation of ordinary mirror symmetry in \eqref{eq:perfectmirror} follows from the HB and CB of the dual theory both being hyperk$\ddot{\text{a}}$hler quotients. Recent developments in terms of magnetic quivers and Hasse diagrams highlighted the fact that setups where \eqref{eq:perfectmirror} effectively takes place involve Hasse diagrams featuring a single cone, such as in figure \ref{fig:11} and \ref{fig:44}. This feature naturally emerges from the HS calculation in the sense that \eqref{eq:HS1} would only account for the contribution from the generators of the unique cone involved in the construction of the diagram. On the other hand, for theories whose Hasse diagram is the union of two intersecting cones, \eqref{eq:HS} splits into terms associated to the generators of each individual cone, together with additional subtractive terms accounting for the nontrivial intersection in between them. 

Extended calculation of HS for arbitrary gauge theories with 8 supercharges has been the main focus of several recent works, mostly \cite{Bourget:2023cgs,Bourget:2019rtl,Ferlito:2016grh}. As shown in \cite{Bourget:2019rtl}, there are some cases in which one could trade a cone for an intersection, thereby indicating that the generators and the relations in between them encode the same information. This turns out to be of particular importance for our purposes, in particular in building the correspondence\footnote{A more detailed explanation of this will be provided in the next section, when comparing with \cite{Teleman:2014jaa}.} with the Drinfeld center\footnote{As stated in the introduction, the focus of \cite{Pasquarella:2023exd} was outlining the main proposal of connecting Drinfeld centers and magnetic quivers from a theoretical point of view, while reserving a more formal mathematical treatment of the same correspondence to followup work by the same author, \cite{Pasquarella:2023ntw, Pasquarella:2023vks}.}.   

Within the context of quiver gauge theories with 8 supercharges, the Hilbert series (HS) is a partition function counting chiral gauge-invariant operators, encoding the variety of vacua generated by such operators. The highest weight generating function (HWG), encodes the same information in a more succinct way, that is more useful to be dealing with for cases involving higher rank, \cite{Ferlito:2016grh}.   

The general expression reads

\begin{equation}  
\begin{aligned}
\text{HS}_{_{N_{_{f}}}}(t;x_{_1},,...x_{_k})\ &\equiv\ \underset{j}{\sum}\ f_{_j}(x_{_1},,...x_{_k})\ t^{^j},
\end{aligned}   
\label{eq:HS1}   
\end{equation}
with $t$ denoting a fugacity for the highest weight of the $SU(2)_{_R}$ $R$-symmetry group providing the grading for the ring of functions, while $f_{_i}(x_{_1},,...x_{_k})$ are sums of characters for irreducible representations of the global symmetry group. From this follows the equivalence with \eqref{eq:hs} and \eqref{eq:hs1}. In the notation of \cite{Ferlito:2016grh}, this reads

\begin{equation}  
\begin{aligned}
\text{HS}_{_{N_{_{f}}}}(t;x_{_1},,...x_{_k})\ &\equiv\ \int d\mu_{_{G}}\ \text{HS}\left(\frac{\mathbb{C}[{\cal Q}, \tilde{\cal Q}]}{\sqrt{\text{$F$-terms}\ }}\right) \ \equiv \text{HS}\left(\frac{\mathbb{C}[M, {\cal Q}, \tilde{\cal Q}]}{\sqrt{\text{Rel.s}\ }}\right)
\end{aligned}   
\label{eq:HS}   
\end{equation} 
with $M, {\cal Q}$ denoting mesons and baryons, respectively.

For the case of 4D ${\cal N}=2$ SQCD with gauge group $G=SU(N_{_c})$ and $N_{_f}$ hypermultiplets in the fundamental representation of $SU(N_{_c})$, the calculation of \ref{eq:HS} requires: 

\begin{enumerate} 

\item solving the $F$-term equations, setting the derivatives of the superpotential to zero 

\item identifying the gauge-invariant operators. 

\end{enumerate}

For ${\cal N}=2$ SU(2) $N_{_f}=2$, the HS is a sum of rational functions, \cite{Ferlito:2016grh},

\begin{equation}  
\begin{aligned}
\text{HS}(t;x_{_1},x_{_2})\ &\equiv\ \int d\mu_{_{SU(2)}}\ F(t,z,x_{_1},x_{_2})\\  
\ &\equiv\ \text{HS}(\mathbb{C}^{^2}/\mathbb{Z}_{_2};t,x_{_1})+\text{HS}(\mathbb{C}^{^2}/\mathbb{Z}_{_2};t,x_{_2})-1 
\label{eq:hs3}
\end{aligned}
\end{equation}
with the first two terms corresponding to the two cones and the last term denoting the intersection in between them at the origin, as shown in figure \ref{fig:22}. When expanding it in terms of powers of $t$, we get the Plethystic logarithm (PL),

\begin{equation}  
\begin{aligned}
\text{PL}(t;x_{_1},x_{_2})\ &\equiv\ ([2;0]+[0;2])t^{^2}-([2;2]+2[0;0])t^{^4}+...
\label{eq:hs4}
\end{aligned}
\end{equation}
with $[;]$ denoting the characters of the corresponding representations of $SO(4)$.
The character corresponding to a certain representation can be encoded in the corresponding Dynkin label. We can therefore choose a set of fugacities to keep track of them, enabling us to rewrite \eqref{eq:hs3} and \eqref{eq:hs4} as a highest weight generating function 

\begin{equation} 
\text{HWG}\left(t;\mu_{_1},\mu_{_2}\right)\ \equiv\ \text{PE}\left[\left(\mu_{_1}^{^2}+\mu_{_2}^{^2}\right)t^{^2}-\mu_{_1}^{^2}\mu_{_2}^{^2}t^{^4}\right],
\end{equation}
with the first two terms corresponding to the generators of the two cones, and the last (subtractive) term to their intersection in the Hasse diagram.

\section*{Key points} 

The main message of this section can be summarised as follows: 

\begin{itemize} 

\item Quiver gauge theories with 8 supercharges are characterised by a structure fully specified by knowing the generators and intersections of the cones in their HB Hasse diagram. The latter is obtained by implementing quiver subtraction on MQs associated to maximal decompositions of the 5-brane webs associated to the electric quiver. As such, it admits a 2-categorical structure description.

\item The moment map plays a key role in ensuring the identification of the underlying MQs. The flatness condition it is required to satisfy defines the HS, hence the HWG function, \cite{DAlesio:2021hlp}. Combining the works of \cite{Cabrera:2019izd,Bourget:2019aer,Bourget:2021siw,Bourget:2023cgs,Bourget:2019rtl,Ferlito:2016grh}, this constitutes a sample realisation of homological mirror symmetry. 

\item In presence of more than one cone in the Hasse diagram, the identification of the MQs whose CBs' union equals the HB of the original theory, is a generalised statement of homological mirror symmetry.

\end{itemize}  

We wish to emphasise that, while most of the above statements have been individually addressed in the literature, to the best of our knowledge, their combined application towards identifying the Drinfeld center in connection to CBs constitutes an original approach, to which Part \ref{sec:V} is devoted.  

\cleardoublepage
\phantomsection


\end{document}